%% file: english.tex
\documentclass[a4paper,12pt]{report}
\newcommand{\printVersion}{} % comment it to have boxes around hyperlinks
\usepackage[utf8]{inputenc}

\usepackage{amsmath,amssymb,amsthm,
stmaryrd,bbm,color,
stmaryrd,
xargs,upgreek,
subfigure,figs
}
\usepackage[Q= yes,pverb-linebreakchar={~}]{examplep} %for verbatim
\usepackage{graphicx}

\usepackage{tikz}
\usetikzlibrary{decorations} 
\usetikzlibrary{decorations.pathmorphing} 
\usetikzlibrary{calc}

\usepackage{makeidx}
\makeindex

\include{MyUtf8}

\numberwithin{equation}{chapter}
\renewcommand{\thechapter}{\Roman{chapter}}
\renewcommand{\theequation}{\thechapter.\arabic{equation}}

\addtocounter{secnumdepth}{1}

\newcounter{appendices}

\PassOptionsToPackage{hyphens}{url}
\usepackage[breaklinks]{hyperref}

\usepackage{mysplit}

\begin{category}{%
\addcontentsline{toc}{chapter}{%
References%
}%
\label{sec:biblio}
\addcontentsline{lot}{chapter}{References}%
Articles written during This PhD}
  \begin{category}{Not yet published in a peer-reviewed journal}
    \SBentries{2010arXiv1007.1770K,Alexandrov:2011aa}
  \end{category}
  \begin{category}{Accepted for publication}
    \SBentries{Gromov:2011cx}
  \end{category}
  \begin{category}{Published}
    \SBentries{Gromov:2010km,Kazakov:2010iu}    
  \end{category}
\end{category}

\makeatletter
\newif\if@display

\everydisplay{\@displaytrue}
\everymath{\@displayfalse}

\@displayfalse

\makeatother

\newcommand{\fat}{\Pv{fat}}
\newcommand{\hook}{\Pv{hook}}
\newcommand{\nested}{\Pv{nested}}
\newcommand{\nesting}{\Pv{nesting}}
\newcommand{\jacobi}{\Pv{Jacobi}}
\newcommand{\ppath}{\Pv{path}}
\newcommand{\level}{\Pv{level}}
\newcommand{\KPhi}{\Pv{KP hierarchy}}
\newcommand{\vdm}{\Pv{Vandermonde}}
\newcommand{\her}{\Pv{hermitian}}
\newcommand{\pau}{\Pv{Pauli}}
\newcommand{\anz}{\Pv{ansatz}}
\newcommand{\cannnot}{\Pv{cannot}}
\newcommand{\NuMr}{\Pv{numerical}}
\newcommand{\Wronskian}{\Pv{Wronskian}}
\newcommand{\zerm}{\Pv{zero-mode}}
\newcommand{\lbd}{\Pv{labeled}}
\newcommand{\lbg}{\Pv{labeling}}
\newcommand{\faml}{\Pv{family}}
\newcommand{\below}{\Pv{below}}
\newcommand{\sugr}{\Pv{super-group}}
\newcommand{\sugrs}{\Pv{super-groups}}
\newcommand{\Sugrs}{\Pv{Super-groups}}
\newcommand{\cd}{\Pv{co-derivative}}
\newcommand{\cdrs}{\Pv{co-derivatives}}
\newcommand{\yn}{\Pv{Young}}
\newcommand{\rp}{\Pv{representation}}
\newcommand{\rhs}{\Pv{right-hand-side}}
\newcommand{\lhs}{\Pv{left-hand-side}}
\newcommand{\another}{\Pv{another}}
\newcommand{\Ysys}{\Pv{Y-system}}
\newcommand{\YsE}{\Pv{Y-system equation}}
\newcommand{\Yfs}{\Pv{Y-functions}}
\newcommand{\Yf}{\Pv{Y-function}}
\newcommand{\TBAE}{\Pv{TBA-equations}}
\newcommand{\op}{\Pv{operator}}

\newcommand{\ops}{\Pv{operators}}

\newcommand{\ing}{\Pv{integrable}}
\newcommand{\cds}{\Pv{spin chain}}

\newcommand{\csds}{\Pv{spin chains}}
\newcommand{\Csds}{\Pv{Spin chains}}
\newcommand{\Ham}{\Pv{Hamiltonian}}
\newcommand{\Matm}{\Pv{Mathematica}}
\newcommand{\PCM}{\Pv{principal chiral model}}
\newcommand{\SYM}{\Pv{super Yang-Mills}}
\newcommand{\idest}{\Pv{i.e.}}
\newcommand{\PCF}{\Pv{principal chiral model}}
\newcommand{\Ydag}{\Pv{{\yn} diagram}}
\newcommand{\Topr}{\Pv{{\Top}-{\op}}}
\newcommand{\Toprs}{\Pv{{\Top}-{\op}s}}
\newcommand{\Tf}{\Pv{{\Tfu}-function}}
\newcommand{\Tfs}{\Pv{{\Tfu}-functions}}
\newcommand{\bTfs}{\Pv{{\bTft}-functions}}
\newcommand{\upTfs}{\Pv{{\upTft}-functions}}
\newcommand{\riTfs}{\Pv{{\riTft}-functions}}
\newcommand{\wTfs}{\Pv{{\wTft}-functions}}

\newcommand{\tauf}{\Pv{\ensuremath{\tau}-function}}
\newcommand{\taufs}{\Pv{\ensuremath{\tau}-functions}}
\newcommand{\qfs}{\Pv{{\qfu}-functions}}
\newcommand{\qf}{\Pv{{\qfu}-function}}
\newcommand{\hQfs}{\Pv{{\hfu}-functions}}
\newcommand{\uqfs}{\Pv{{\upqfqq}-functions}}
\newcommand{\upfs}{\Pv{{\uppfqq}-functions}}
\newcommand{\pfs}{\Pv{{\pfu}-functions}}
\newcommand{\Qfs}{\Pv{{\Qfu}-functions}}
\newcommand{\cQfs}{\Pv{{\cqfu}-functions}}
\newcommand{\Qf}{\Pv{{\Qfu}-function}}
\newcommand{\Qopr}{\Pv{{\Qop}-{\op}}}
\newcommand{\Qoprs}{\Pv{{\Qop}-{\op}s}}

\newcommand{\Roprs}{\Pv{{\Rop}-{\op}s}}
\newcommand{\Sma}{\Pv{{\Smat}-matrix}}
\newcommand{\MID}{\Pv{main identity on {\cd}s}}
\newcommand{\MMID}{\Pv{Main identity on {\cd}s}}
\newcommand{\TBA}{\Pv{thermodynamic Bethe \anz}}
\newcommand{\Bafra}{\Pv{the linear system (\ref{eq:BT1},~\ref{eq:BT2})
  which defines the Bäcklund transform}}

\newcommand{\ADF}{\Pv{AdS/CFT}}
\newcommand{\thesis}{\Pv{thesis}}
\newcommand{\PhD}{\Pv{PhD}}

\usepackage{empheq}

\include{MyNotations}

\newtheorem{statmt}{Statement}

\title{{% 
    Modèles intégrables et dualité AdS/CFT}}
\author{% 
S. Leurent}
\begin{document}

%\maketitle
\begin{titlepage}
  \includegraphics[width=.3\textwidth]{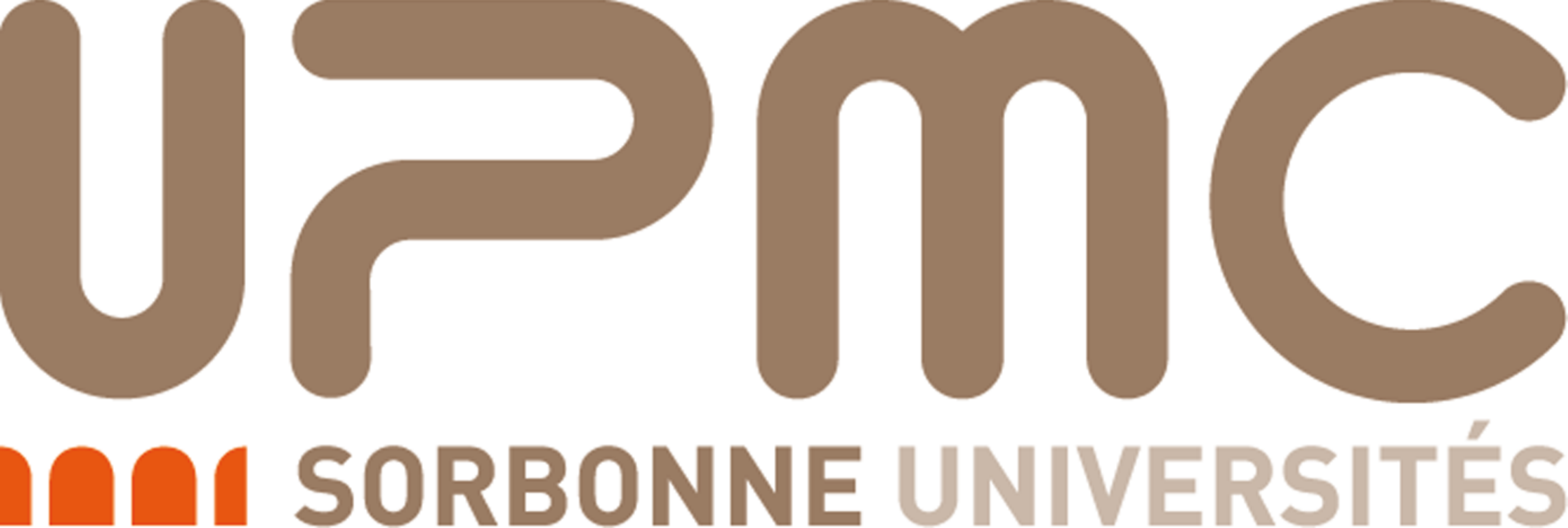} \hspace{\stretch{1}}
  \includegraphics[width=.2\textwidth]{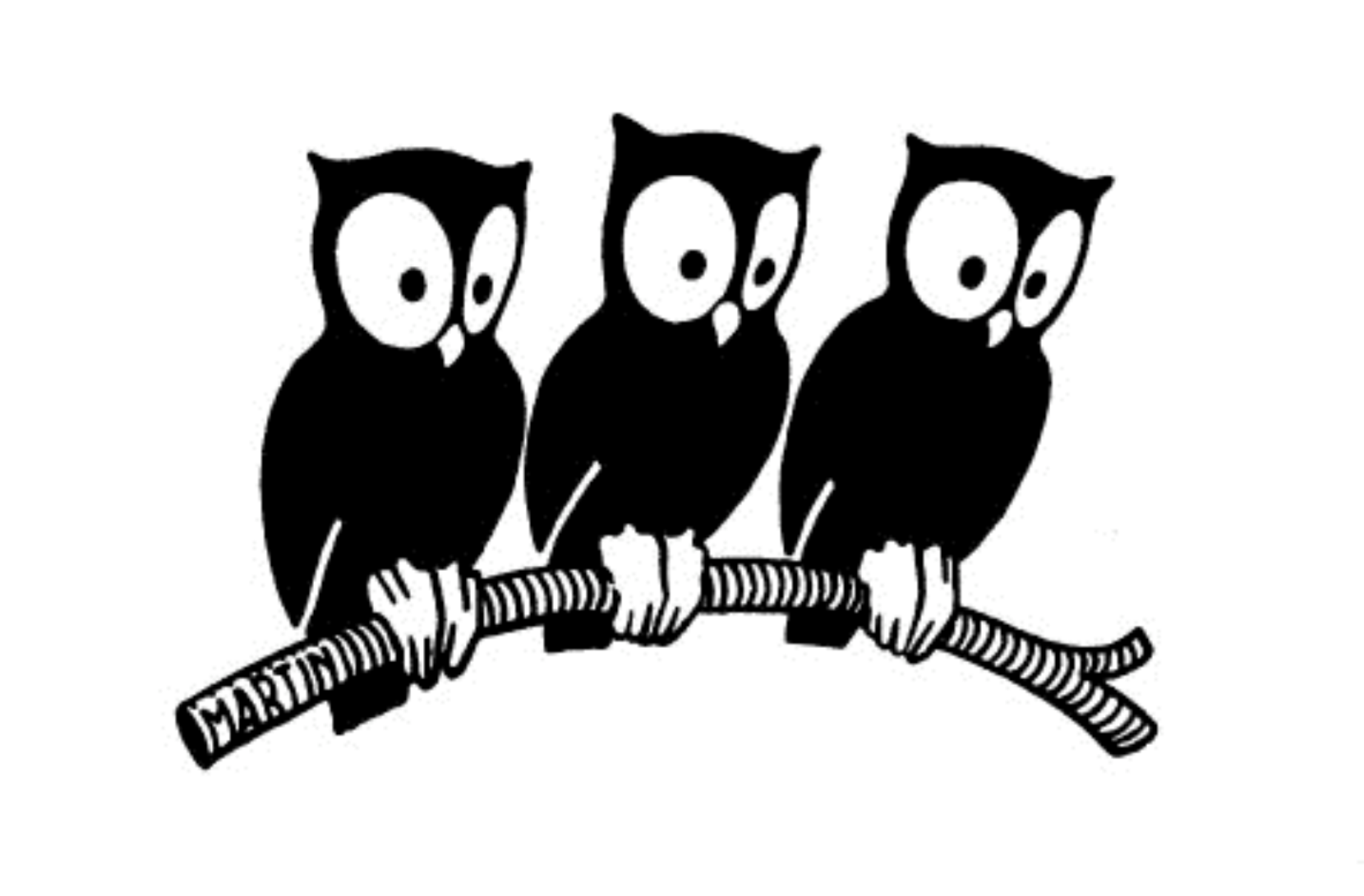} \hspace{\stretch{1}}
  \includegraphics[width=.2\textwidth]{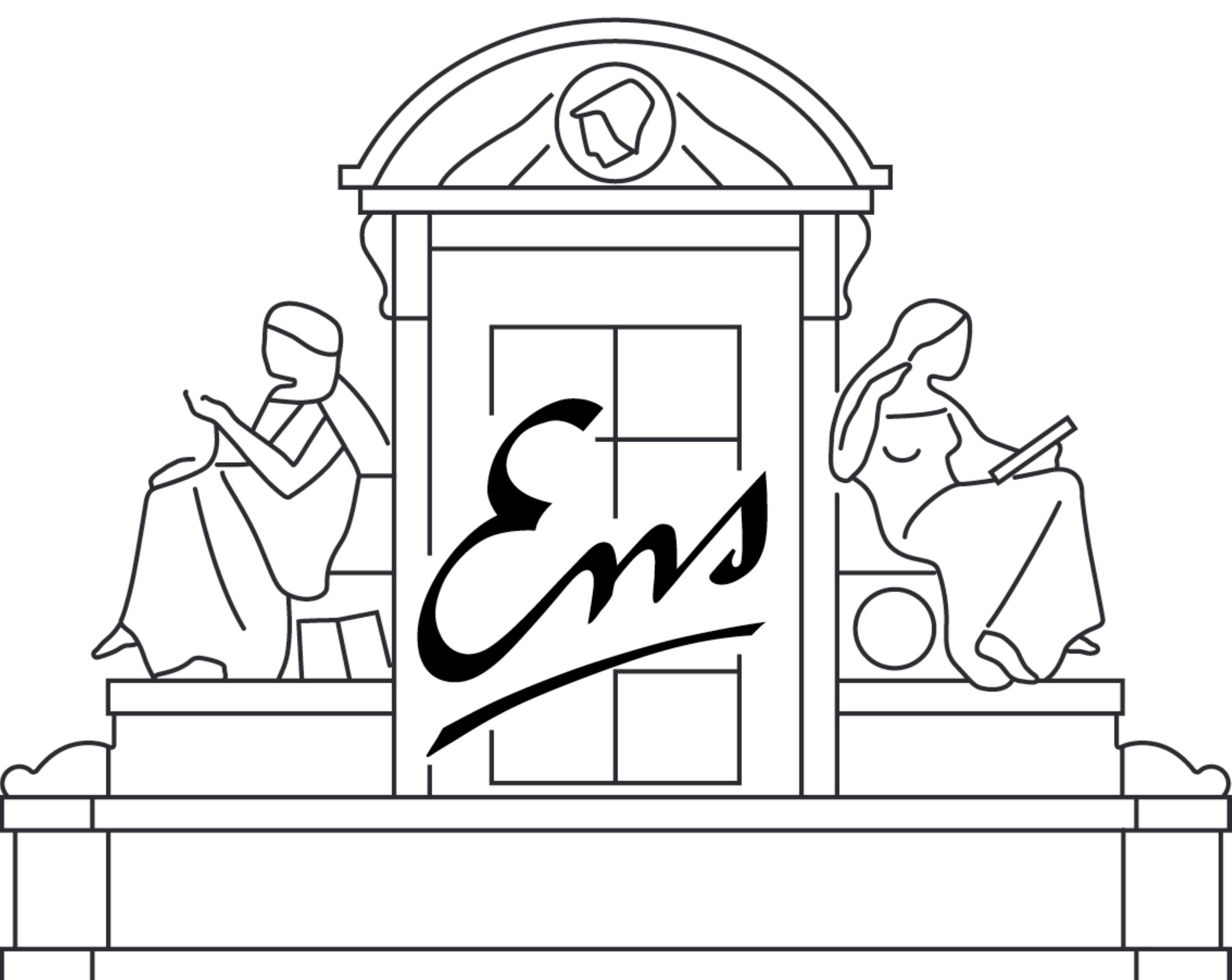}
  \begin{center}
    \vspace{1cm}
    {\Large \textsc{Thèse de doctorat de \\ l'université Pierre et
        Marie Curie}}\\
\vspace{1cm}
    {\large Spécialité : Physique théorique}\\
\vspace{2cm}
    {\large Présentée par}\\
{\Large Sébastien LEURENT}\\
\vspace{2cm}
    {\large pour obtenir le grade de}\\
    {\Large \textsc{Docteur de l'université Pierre et
        Marie Curie}}\\
  \end{center}
\vspace{2cm}
{\large Sujet :} {\Large \textbf{Systèmes intégrables et dualité
    AdS/CFT}}
\vspace{1cm}\\
{\large Soutenance le 20 juin 2012, devant le jury composé de}\\
\begin{center}
  \begin{tabular}{ll}
    Romuald \textsc{Janik}&Rapporteur\\
    Vladimir \textsc{Kazakov}&Directeur de thèse\\
    Gregory \textsc{Korchemsky}&Rapporteur\\
    Arkady \textsc{Tseytlin}&Examinateur\\
    Konstantin \textsc{Zarembo}&Examinateur\\
    Jean-Bernard \textsc{Zuber}&Examinateur\\
  \end{tabular}
\end{center}
 ~\\
Thèse délivrée par l'\textit{UPMC}, Place Jussieu, 75005 Paris \textit{FRANCE},\\
suite à des recherches effectuées au \textit{LPT-ENS}, 24 rue Lhomond, 75005 Paris \textit{FRANCE}.
\end{titlepage}
 \addtocounter{page}{1}

\include{francais}

\include{prelim}

\chapter%
{Integrability and Bethe ansatz}
\addcontentsline{lot}{chapter}{I. Integrability and Bethe ansatz}
\label{part:introduction}

\include{intro}

\chapter{{\Qop} {\ops} for {\csds}}
\addcontentsline{lot}{chapter}{II. {\Qop} {\ops} for {\csds}}
\label{part:qoperatorsspin}

\include{spinchains}

\chapter{Thermodynamic Bethe Ansätze and Y-systems}
\label{cha:ansatzs-de-bethe}
\addcontentsline{lot}{chapter}{III. Thermodynamic Bethe Ansätze and Y-systems}

\include{TBA}

\chapter{FiNLIE for the AdS/CFT Duality}
\addcontentsline{lot}{chapter}{IV. AdS/CFT Duality}
\label{cha:dualite-adscft}

\include{AdsCFT}

\include{concl}

 \setcounter{chapter}{\value{appendices}}
 \renewcommand{\thechapter}{\Alph{chapter}}
 \renewcommand\theHchapter{\Alph{chapter}}
\rstchpapp

\chapter{%
Introduction to representations of matrix groups
}
\addcontentsline{lot}{chapter}{A. Introduction to representations of matrix groups}
\label{sec:elements-de-theorie}

\include{groups}

\chapter{Properties of {\cdrs}}
\addcontentsline{lot}{chapter}{B. Properties of {\cdrs}}
\label{app:cod}

\include{diagrammatics}

\pagebreak 
\printindex

\pagebreak 
\bibliographystyle{MyAlpha}
\bibliography{Biblio}

\end{document}

%% file: MyUtf8.tex
\usepackage[utf8]{inputenc}
\usepackage[T1]{fontenc}
\usepackage{amssymb}
\usepackage{amsmath}

\DeclareMathOperator{\guilleright}{\mbox{\guillemotright}}
\DeclareMathOperator{\guilleleft}{\mbox{\guillemotleft}}

\DeclareUnicodeCharacter{00A1}{\mbox{\textrm{\textexclamdown}}}
\DeclareUnicodeCharacter{00B7}{\ensuremath{\cdot}}
\DeclareUnicodeCharacter{00BA}{\ensuremath{^o}}
\DeclareUnicodeCharacter{00D7}{\ensuremath{\times}}
\DeclareUnicodeCharacter{00F6}{\"o}

\DeclareUnicodeCharacter{00C5}{\c{S}R}
\DeclareUnicodeCharacter{0552}{\c{S}R}
\DeclareUnicodeCharacter{015E}{\c{S}}

\DeclareUnicodeCharacter{0278}{\ensuremath{\phi}}

\DeclareUnicodeCharacter{02B0}{\ensuremath{^h}}
\DeclareUnicodeCharacter{02B2}{\ensuremath{^j}}
\DeclareUnicodeCharacter{02B3}{\ensuremath{^r}}
\DeclareUnicodeCharacter{02B7}{\ensuremath{^w}}
\DeclareUnicodeCharacter{02B8}{\ensuremath{^y}}
\DeclareUnicodeCharacter{02E1}{\ensuremath{^l}}
\DeclareUnicodeCharacter{02E2}{\ensuremath{^s}}
\DeclareUnicodeCharacter{02E3}{\ensuremath{^x}}

\DeclareUnicodeCharacter{0393}{\ensuremath{\Gamma}}
\DeclareUnicodeCharacter{0394}{\ensuremath{\Delta}}
\DeclareUnicodeCharacter{0398}{\ensuremath{\Theta}}
\DeclareUnicodeCharacter{039B}{\ensuremath{\Lambda}}
\DeclareUnicodeCharacter{03A0}{\ensuremath{\Pi}}
\DeclareUnicodeCharacter{03A3}{\ensuremath{\Sigma}}
\DeclareUnicodeCharacter{03A8}{\ensuremath{\Psi}}
\DeclareUnicodeCharacter{03A9}{\ensuremath{\Omega}}

\DeclareUnicodeCharacter{03B1}{\ensuremath{\alpha}}
\DeclareUnicodeCharacter{03B2}{\ensuremath{\beta}}
\DeclareUnicodeCharacter{03B3}{\ensuremath{\gamma}}
\DeclareUnicodeCharacter{03B4}{\ensuremath{\delta}}
\DeclareUnicodeCharacter{03B5}{\ensuremath{\epsilon}}
\DeclareUnicodeCharacter{03B6}{\ensuremath{\zeta}}
\DeclareUnicodeCharacter{03B7}{\ensuremath{\eta}}
\DeclareUnicodeCharacter{03B8}{\ensuremath{\theta}}
\DeclareUnicodeCharacter{03BA}{\ensuremath{\kappa}}
\DeclareUnicodeCharacter{03BB}{\ensuremath{\lambda}}
\DeclareUnicodeCharacter{03BC}{\ensuremath{\mu}}
\DeclareUnicodeCharacter{03BD}{\ensuremath{\nu}}
\DeclareUnicodeCharacter{03C0}{\ensuremath{\pi}}
\DeclareUnicodeCharacter{03C3}{\ensuremath{\sigma}}
\DeclareUnicodeCharacter{03C4}{\ensuremath{\tau}}
\DeclareUnicodeCharacter{03C6}{\ensuremath{\varphi}}
\DeclareUnicodeCharacter{03C7}{\ensuremath{\chi}}
\DeclareUnicodeCharacter{03C9}{\ensuremath{\omega}}

\DeclareUnicodeCharacter{05D0}{\ensuremath{\aleph}}

\DeclareUnicodeCharacter{2022}{\ensuremath{\bullet}}
\DeclareUnicodeCharacter{202F}{\,}
\DeclareUnicodeCharacter{2070}{\ensuremath{^0}}
\DeclareUnicodeCharacter{00B9}{\ensuremath{^1}}
\DeclareUnicodeCharacter{00B2}{\ensuremath{^2}}
\DeclareUnicodeCharacter{00B3}{\ensuremath{^3}}
\DeclareUnicodeCharacter{2074}{\ensuremath{^4}}
\DeclareUnicodeCharacter{2075}{\ensuremath{^5}}
\DeclareUnicodeCharacter{2076}{\ensuremath{^6}}
\DeclareUnicodeCharacter{2077}{\ensuremath{^7}}
\DeclareUnicodeCharacter{2078}{\ensuremath{^8}}
\DeclareUnicodeCharacter{2079}{\ensuremath{^9}}
\DeclareUnicodeCharacter{207A}{\ensuremath{^+}}
\DeclareUnicodeCharacter{207B}{\ensuremath{^-}}
\DeclareUnicodeCharacter{207C}{\ensuremath{^=}}
\DeclareUnicodeCharacter{207F}{\ensuremath{^n}}
\DeclareUnicodeCharacter{2080}{\ensuremath{_0}}
\DeclareUnicodeCharacter{2081}{\ensuremath{_1}}
\DeclareUnicodeCharacter{2082}{\ensuremath{_2}}
\DeclareUnicodeCharacter{2083}{\ensuremath{_3}}
\DeclareUnicodeCharacter{2084}{\ensuremath{_4}}
\DeclareUnicodeCharacter{2085}{\ensuremath{_5}}
\DeclareUnicodeCharacter{2086}{\ensuremath{_6}}
\DeclareUnicodeCharacter{2087}{\ensuremath{_7}}
\DeclareUnicodeCharacter{2088}{\ensuremath{_8}}
\DeclareUnicodeCharacter{2089}{\ensuremath{_9}}

\DeclareUnicodeCharacter{2102}{\ensuremath{{\mathbb{C}}}}
\DeclareUnicodeCharacter{2115}{\ensuremath{{\mathbb{N}}}}
\DeclareUnicodeCharacter{211D}{\ensuremath{{\mathbb{R}}}}
\DeclareUnicodeCharacter{2124}{\ensuremath{{\mathbb{Z}}}}
\DeclareUnicodeCharacter{2148}{\ensuremath{{\mathbbm{i}}}}
\DeclareUnicodeCharacter{1D540}{\ensuremath{{\mathbb{I}}}}
\DeclareUnicodeCharacter{1D7D7}{\ensuremath{{\mathbbm{1}}}}

\DeclareUnicodeCharacter{2190}{\ensuremath{\leftarrow}}
\DeclareUnicodeCharacter{2191}{\ensuremath{\uparrow}}
\DeclareUnicodeCharacter{2192}{\ensuremath{\rightarrow}}
\DeclareUnicodeCharacter{2193}{\ensuremath{\downarrow}}
\DeclareUnicodeCharacter{219D}{\ensuremath{\leadsto}}
\DeclareUnicodeCharacter{21A0}{\ensuremath{\twoheadrightarrow}}
\DeclareUnicodeCharacter{21A3}{\ensuremath{\rightarrowtail}}
\DeclareUnicodeCharacter{21A6}{\ensuremath{\mapsto}}
\DeclareUnicodeCharacter{21D0}{\ensuremath{\Leftarrow}}
\DeclareUnicodeCharacter{21D1}{\ensuremath{\Uparrow}}
\DeclareUnicodeCharacter{21D2}{\ensuremath{\Rightarrow}}
\DeclareUnicodeCharacter{21D3}{\ensuremath{\Downarrow}}
\DeclareUnicodeCharacter{2200}{\ensuremath{\forall}}
\DeclareUnicodeCharacter{2203}{\ensuremath{\exists}}
\DeclareUnicodeCharacter{2205}{\ensuremath{\emptyset}}
\DeclareUnicodeCharacter{2207}{\ensuremath{\nabla}}
\DeclareUnicodeCharacter{2208}{\ensuremath{\in}}
\DeclareUnicodeCharacter{2209}{\ensuremath{\not\in}}
\DeclareUnicodeCharacter{2217}{\ensuremath{\ast}}
\DeclareUnicodeCharacter{2223}{\ensuremath{\mid}}
\DeclareUnicodeCharacter{2225}{\ensuremath{\parallel}}
\DeclareUnicodeCharacter{2227}{\ensuremath{\wedge}}
\DeclareUnicodeCharacter{2228}{\ensuremath{\vee}}
\DeclareUnicodeCharacter{223C}{\ensuremath{\sim}}
\DeclareUnicodeCharacter{2243}{\ensuremath{\simeq}}
\DeclareUnicodeCharacter{2245}{\ensuremath{\cong}}
\DeclareUnicodeCharacter{2248}{\ensuremath{\approx}}
\DeclareUnicodeCharacter{224A}{\ensuremath{\approxeq}}
\DeclareUnicodeCharacter{2250}{\ensuremath{\doteq}}
\DeclareUnicodeCharacter{225C}{\ensuremath{\triangleq}}
\DeclareUnicodeCharacter{2260}{\ensuremath{\neq}}
\DeclareUnicodeCharacter{2261}{\ensuremath{\equiv}}
\DeclareUnicodeCharacter{2264}{\ensuremath{\leq}}
\DeclareUnicodeCharacter{226B}{\ensuremath{\gg}}
\DeclareUnicodeCharacter{227A}{\ensuremath{\prec}}
\DeclareUnicodeCharacter{227B}{\ensuremath{\succ}}
\DeclareUnicodeCharacter{2270}{\ensuremath{\nprec}}
\DeclareUnicodeCharacter{2281}{\ensuremath{\nsucc}}
\DeclareUnicodeCharacter{2283}{\ensuremath{\supset}}
\DeclareUnicodeCharacter{2286}{\ensuremath{\subseteq}}
\DeclareUnicodeCharacter{2291}{\ensuremath{\sqsubseteq}}
\DeclareUnicodeCharacter{2295}{\ensuremath{\oplus}}%⊕
\DeclareUnicodeCharacter{2296}{\ensuremath{\ominus}}
\DeclareUnicodeCharacter{2297}{\ensuremath{\otimes}}%⊗
\DeclareUnicodeCharacter{22A2}{\ensuremath{\vdash}}
\DeclareUnicodeCharacter{22A4}{\ensuremath{\top}}
\DeclareUnicodeCharacter{22A5}{\ensuremath{\bot}}
\DeclareUnicodeCharacter{22A8}{\ensuremath{\models}}
\DeclareUnicodeCharacter{22A9}{\ensuremath{\Vdash}}
\DeclareUnicodeCharacter{22B8}{\ensuremath{\multimap}}
\DeclareUnicodeCharacter{22C6}{\ensuremath{\star}}
\DeclareUnicodeCharacter{22C8}{\ensuremath{\bowtie}}
\DeclareUnicodeCharacter{22EE}{\ensuremath{\vdots}}

\DeclareUnicodeCharacter{2423}{\,}
\DeclareUnicodeCharacter{2588}{\ensuremath{\blacksquare}}
\DeclareUnicodeCharacter{25A1}{\ensuremath{\Box}}
\DeclareUnicodeCharacter{25B7}{\ensuremath{\rhd}}
\DeclareUnicodeCharacter{25C1}{\ensuremath{\lhd}}
\DeclareUnicodeCharacter{25C7}{\ensuremath{\diamond}}
\DeclareUnicodeCharacter{25CB}{\ensuremath{\circ}}
\DeclareUnicodeCharacter{25CF}{\ensuremath{\bullet}}
\DeclareUnicodeCharacter{25EF}{\ensuremath{\bigcirc}}

\DeclareUnicodeCharacter{27E8}{\ensuremath{\langle}}
\DeclareUnicodeCharacter{27E9}{\ensuremath{\rangle}}
\DeclareUnicodeCharacter{27EA}{\ensuremath{\guilleleft}}
\DeclareUnicodeCharacter{27EB}{\ensuremath{\guilleright}}

\DeclareUnicodeCharacter{1D44E}{\textsl{a}}
\DeclareUnicodeCharacter{1D44F}{\textsl{b}}
\DeclareUnicodeCharacter{1D452}{\textsl{e}}
\DeclareUnicodeCharacter{1D454}{\textsl{g}}
\DeclareUnicodeCharacter{1D456}{\textsl{i}}
\DeclareUnicodeCharacter{1D459}{\textsl{l}}
\DeclareUnicodeCharacter{1D45B}{\textsl{n}}
\DeclareUnicodeCharacter{1D45F}{\textsl{r}}
\DeclareUnicodeCharacter{1D461}{\textsl{t}}
\DeclareUnicodeCharacter{1D466}{\textsl{y}}

%% file: MyNotations.tex
\newcommand{\Pv}[1]{#1}

\definecolor{darkgreen}{rgb}{0,0.5,0}

\ifdefined\printcolor
\newcommand{\colorprint}[2]{#1}
\else
\newcommand{\colorprint}[2]{#2}
\fi

\ifdefined\printVersion
\hypersetup{pdfborder= 0 0 0}
\fi

\renewcommand{\leq}{\leqslant}
\renewcommand{\geq}{\geqslant}

\newcommand{\forget}[1]{}

\newcommand{\Dnothing}{D}
\newcommand{\Lnothing}{L}
\newcommand{\Mnothing}{M}
\newcommand{\Pnothing}{P}
\newcommand{\Tnothing}{T}
\newcommand{\gnothing}{g}
\newcommand{\jnothing}{j}
\newcommand{\nnothing}{n}
\newcommand{\mnothing}{m}

\def\clap#1{\hbox to 0pt{\hss#1\hss}}
\def\mathllap{\mathpalette\mathllapinternal}
\def\mathrlap{\mathpalette\mathrlapinternal}
\def\mathclap{\mathpalette\mathclapinternal}
\def\mathllapinternal#1#2{%
           \llap{\(\mathsurround=0pt#1{#2}\)}}
\def\mathrlapinternal#1#2{%
           \rlap{\(\mathsurround=0pt#1{#2}\)}}
\def\mathclapinternal#1#2{%
           \clap{\(\mathsurround=0pt#1{#2}\)}}

\newcommand{\indown}{\Pv{
    \begin{tikzpicture}
      \node[rotate=-90] at (0,0) {\(\in\)};
    \end{tikzpicture}
}}
\newcommand{\inleft}{\Pv{\ensuremath{%\not
    \begin{tikzpicture}
      \node[rotate=180] at (0,0) {\ensuremath{\in}};
    \end{tikzpicture}}
}}

\newcommand{\cb}[1]{\Pv{\ensuremath{\left[ #1\right]}}}
\newcommandx*{\scomm}[5][1=0.33cm,2=0pt,usedefault]{\Pv{\ensuremath{{
\raisebox{#2}{\resizebox{0.28cm}{#1}{
\ensuremath{\llparenthesis}}}\: #4\:,\:#5 {{
\raisebox{#2}{\resizebox{0.28cm}{#1}{
\ensuremath{\rrparenthesis}}}_{#3}}}}}}}
\newcommandx*{\comm}[2][1=0.33cm,2=0pt,usedefault]{
\scomm [#1] [#2] - }
\newcommandx*{\acomm}[2][1=0.33cm,2=0pt,usedefault]{
\scomm [#1] [#2] + }

\newcommand{\ibrapket}[2]{\Pv{\ensuremath{\left\langle\{#1\}\middle|#2\right\rangle}}}
\newcommand{\nndisp}[1]{{\color{gray}(\color{black}}#1{\color{gray})\color{black}}}
\newcommand{\nbranket}[2]{\braket{\nndisp{#1}}{\nndisp{#2}}}
\newcommand{\nbraket}[2]{\braket{\nndisp{#1}}{{#2}}}
\newcommand{\vac}{\Pv{\ensuremath{|\{\quad\!\}\rangle}}}
\newcommand{\pket}[1]{\Pv{\ensuremath{\left|#1\right \rangle}}}%impulsion
\newcommand{\iket}[1]{\Pv{\ensuremath{\left|\{#1\}\right \rangle}}}%position
\newcommand{\nket}[1]{\ket{\nndisp{#1}}}
\newcommand{\nbra}[1]{\bra{\nndisp{#1}}}
\newcommand{\ket}[1]{\Pv{\ensuremath{\left|#1\right \rangle}}}
\newcommand{\bra}[1]{\Pv{\ensuremath{\left\langle#1\right|}}}
\newcommand{\braket}[2]{\Pv{\ensuremath{\left\langle#1\middle|#2\right\rangle}}}
\newcommand{\sket}[1]{\Pv{\ensuremath{|#1\rangle}}}

\newcommand{\phie}{\Pv{\ensuremath{\left<\phi,  e\right>}}}
\newcommand{\ninter}[2]{\Pv{\ensuremath{\llbracket #1,#2\rrbracket}}}
\newcommand{\Hilb}{\Pv{\ensuremath{\mathcal{H}}}}
\newcommand{\Hilbl}{\Pv{\ensuremath{{\scalebox{0.7}{\ensuremath{\mathcal{H}}}}}}}

\newcommand{\Hami}{\Pv{\ensuremath{\mathrm{H}}}}
\newcommand{\HK}{\Pv{\ensuremath{\mathbb{L}}}}%fat hook
\newcommand{\Tk}{\Pv{\ensuremath{\mathbb{T}}}}%T hook
\newcommand{\St}{\Pv{\ensuremath{\mathbb{S}}}}%strip
\newcommand{\Sh}{\Pv{\ensuremath{\mathbbm{w}}}}%half strip
\newcommand{\Yl}[1]{\Pv{\ensuremath{\left[#1\right]_\Yfu}}}% to
\newcommand{\Tl}[1]{\Pv{\ensuremath{\left[#1\right]_\Tft}}}% to
\newcommand{\lcds}{\Pv{\ensuremath{\mathrm{L}}}}
\newcommand{\LF}{\Pv{\textrm{L}}}%space's length for field theories
\newcommand{\RF}{\Pv{\textrm{R}}}% length of the other dimension of the torus
\newcommand{\spi}{\Pv{\ensuremath{\mathrm{i}}}}
\newcommand{\spj}{\Pv{\ensuremath{\mathrm{j}}}}
\newcommand{\spk}{\Pv{\ensuremath{\mathrm{k}}}}
\newcommand{\spl}{\Pv{\ensuremath{\mathrm{l}}}}
\newcommand{\spm}{\Pv{\ensuremath{\mathrm{m}}}}
\newcommand{\spn}{\Pv{\ensuremath{\mathrm{n}}}}
\newcommand{\su}{\Pv{\ensuremath{\mathtt{u}}}}
\newcommand{\sv}{\Pv{\ensuremath{\mathtt{v}}}}
\newcommand{\us}{\Pv{\ensuremath{\mathsf{u}}}}
\newcommand{\vs}{\Pv{\ensuremath{\mathsf{v}}}}
\newcommand{\xs}{\Pv{\ensuremath{\mathsf{x}}}}
\newcommand{\ys}{\Pv{\ensuremath{\mathsf{y}}}}
\newcommand{\zs}{\Pv{\ensuremath{\mathsf{z}}}}
\newcommand{\ii}{\Pv{\ensuremath{i}}}
\newcommand{\jj}{\Pv{\ensuremath{j}}}
\newcommand{\kk}{\Pv{\ensuremath{k}}}
\newcommand{\lL}{\Pv{\ensuremath{l}}}
\newcommand{\ivp}{\Pv{\ensuremath{\mybold i}}}
\newcommand{\jvp}{\Pv{\ensuremath{\mybold j}}}
\newcommand{\kvp}{\Pv{\ensuremath{\mybold k}}}

\newcommand{\irt}{\Pv{\ensuremath{\mathsf i}}}
\newcommand{\jrt}{\Pv{\ensuremath{\mathsf j}}}
\newcommand{\krt}{\Pv{\ensuremath{\mathsf k}}}
\newcommand{\lrt}{\Pv{\ensuremath{\mathsf l}}}
\newcommand{\mrt}{\Pv{\ensuremath{\mathsf m}}}
\newcommand{\nrt}{\Pv{\ensuremath{\mathsf n}}}
\newcommand{\jlvl}{\Pv{\ensuremath{\mathbf j}}}
\newcommand{\klvl}{\Pv{\ensuremath{\mathbf k}}}
\newcommand{\mlvl}{\Pv{\ensuremath{\mathbf m}}}
\newcommand{\nlvl}{\Pv{\ensuremath{\mathbf n}}}
\newcommand{\iq}{\Pv{\ensuremath{i}}}
\newcommand{\jq}{\Pv{\ensuremath{j}}}

\newcommand{\bjq}{\Pv{\ensuremath{\mybold j}}}
\newcommand{\bkq}{\Pv{\ensuremath{\mybold k}}}
\newcommand{\dkp}{\Pv{\ensuremath{2 ~% 
\kappa
      ~ \pi}}}
\newcommand{\tk}{\Pv{\ensuremath{\mathbf t}}}
\newcommand{\mmass}{m}
\newcommand{\jmathvp}{\Pv{\ensuremath{\mybold \jmath}}}
\newcommand{\mm}{\Pv{\ensuremath{m}}}
\newcommand{\nn}{\Pv{\ensuremath{n}}}
\newcommand{\uu}{\Pv{\ensuremath{u}}}
\newcommand{\MM}{\Pv{\ensuremath{\mathrm{M}}}}

\newcommand{\NN}{N}
\newcommand{\HH}{H}
\newcommand{\RR}{R}
\newcommand{\vv}{v}
\newcommand{\Mp}{\Pv{\ensuremath{M}}} % number of particules in Bethe
\newcommand{\Mpo}{\Pv{\ensuremath{\mathbf{M}}}} % operatorial number of particules 
\newcommand{\DD}{\Pv{\ensuremath{D}}}
\newcommand{\oD}{D}
\newcommand{\coordi}{\Pv{\ensuremath{i}}}
\newcommand{\coordj}{\Pv{\ensuremath{j}}}
\newcommand{\coordk}{\Pv{\ensuremath{k}}}
\newcommand{\coordl}{\Pv{\ensuremath{l}}}
\newcommand{\coordm}{\Pv{\ensuremath{m}}}
\newcommand{\coordn}{\Pv{\ensuremath{n}}}
\newcommand{\Lop}{\Pv{\ensuremath{L}}}
\newcommand{\Ltw}{\Pv{\ensuremath{\mathcal{L}_g}}}
\newcommand{\Top}{\Pv{\textrm{T}}}
\newcommand{\Tfu}{\Pv{\textrm{\it{T}}}}
\newcommand{\Qfu}{\Pv{\textrm{\it{Q}}}}
\newcommandx*{\rT}[4][1=\su,2=a,3=s,4={},usedefault]{\gT[#1][#2,#3][#4]}
\newcommandx*{\lT}[3][1=\su,2=\lambda,3={},usedefault]{\gT[#1][(#2)][#3]}
\newcommandx*{\tT}[3][1=\su,2=\tk,3={},usedefault]{\Top_{#3}(#1,#2)}
\newcommandx*{\gT}[3][1=\su,2=(\lambda),3={},usedefault]{\Pv{\ensuremath{\Top^{#2}_{#3}(#1)}}}
\newcommandx*{\gQ}[2][1=\su,2={},usedefault]{\gQn[(#1)][#2]}%Q
\newcommandx*{\gQn}[2][1={},2={},usedefault]{\Pv{\ensuremath{\Qop_{{#2}}{#1}}}}%Q operator
\newcommandx*{\gQf}[2][1=\su,2={},usedefault]{\Pv{\ensuremath{\Qfu_{{#2}}(#1)}}}%Q
\newcommandx*{\fQ}[3][1=\us,2={\mlvl},3={},usedefault]{\Pv{\ensuremath{\fQf_{[{#2}]}^{#3}(#1)}}}
\newcommandx*{\vf}[1][1={},usedefault]{\Pv{\ensuremath{\varphi^{#1}}}}
\newcommandx*{\ffQ}[3][1=\us,2={\mlvl},3={(\Rg)},usedefault]{\Pv{\ensuremath{\fQf_{{#2}}^{#3}\left(#1\right)}}}%
\newcommand{\fQf}{Q}%Q-polynomials for a field thory

\newcommandx*{\gpf}[2][1={\{\iq\}},usedefault]{\Pv{\ensuremath{\pfu_{{#1}}^{#2}}}}%q
\newcommandx*{\gpfs}[2][1={\iq},2={},usedefault]{\gpf[#1]{[#2]}}
\newcommandx*{\gps}[2][1={\iq},2={},usedefault]{\gpf[\{#1\}]{#2}}
\newcommandx*{\gpe}[1][1={},usedefault]{\gpf[\emptyset]{#1}}
\newcommandx*{\gpss}[2][1={\iq},2={},usedefault]{\gpf[\{#1\}]{[#2]}}
\newcommandx*{\gpes}[1][1={},usedefault]{\gpf[\emptyset]{[#1]}}
\newcommand{\pfu}{\Pv{\textrm{\sl p}}}%Q-fun for a field thory
\newcommand{\pbfu}{{\tilde {\pfu}}}%Q-fun for a field thory
\newcommandx*{\gpbf}[2][1={\{\iq\}},usedefault]{\Pv{\ensuremath{\pbfu_{{#1}}^{#2}}}}%q
\newcommandx*{\gpbfs}[2][1={\iq},2={},usedefault]{\gpbf[#1]{[#2]}}
\newcommandx*{\gpbs}[2][1={\iq},2={},usedefault]{\gpbf[\{#1\}]{#2}}
\newcommandx*{\gpbe}[1][1={},usedefault]{\gpbf[\emptyset]{#1}}
\newcommandx*{\gpbss}[2][1={\iq},2={},usedefault]{\gpbf[\{#1\}]{[#2]}}
\newcommandx*{\gpbes}[1][1={},usedefault]{\gpbf[\emptyset]{[#1]}}

\newcommandx*{\cqf}[2][1={\{\iq\}},usedefault]{\Pv{\ensuremath{\cqfu_{{#1}}^{#2}}}}%q
\newcommandx*{\cqfs}[2][1={\iq},2={},usedefault]{\cqf[#1]{[#2]}}
\newcommandx*{\cqs}[2][1={\iq},2={},usedefault]{\cqf[\{#1\}]{#2}}
\newcommandx*{\cqe}[1][1={},usedefault]{\cqf[\emptyset]{#1}}
\newcommandx*{\cqss}[2][1={\iq},2={},usedefault]{\cqf[\{#1\}]{[#2]}}
\newcommandx*{\cqes}[1][1={},usedefault]{\cqf[\emptyset]{[#1]}}
\newcommand{\cqfu}{\Pv{\ensuremath{\mathcal{Q}}}}%Q-fun for a field
\newcommand{\cqbfu}{\Pv{\ensuremath{{{\tilde {\cqfu}}}}}}%same tilded
\newcommandx*{\cqbf}[2][1={\{\iq\}},usedefault]{\Pv{\ensuremath{\cqbfu_{{#1}}^{#2}}}}%q
\newcommandx*{\cqbfs}[2][1={\iq},2={},usedefault]{\cqbf[#1]{[#2]}}
\newcommandx*{\cqbs}[2][1={\iq},2={},usedefault]{\cqbf[\{#1\}]{#2}}
\newcommandx*{\cqbe}[1][1={},usedefault]{\cqbf[\emptyset]{#1}}
\newcommandx*{\cqbss}[2][1={\iq},2={},usedefault]{\cqbf[\{#1\}]{[#2]}}
\newcommandx*{\cqbes}[1][1={},usedefault]{\cqbf[\emptyset]{[#1]}}

\newcommandx*{\hqf}[2][1={\{\iq\}},usedefault]{\Pv{\ensuremath{\hfu_{{#1}}^{#2}}}}%q
\newcommandx*{\hqfs}[2][1={\iq},2={},usedefault]{\hqf[#1]{[#2]}}
\newcommandx*{\hqs}[2][1={\iq},2={},usedefault]{\hqf[\{#1\}]{#2}}
\newcommandx*{\hqe}[1][1={},usedefault]{\hqf[\emptyset]{#1}}
\newcommandx*{\hqss}[2][1={\iq},2={},usedefault]{\hqf[\{#1\}]{[#2]}}
\newcommandx*{\hqes}[1][1={},usedefault]{\hqf[\emptyset]{[#1]}}
\newcommand{\hfu}{\ensuremath{\hat {\qfu}}}%Q-fun for a field thory

\newcommandx*{\bgqf}[2][1={\{\iq\}},usedefault]{\Pv{\ensuremath{\bqfu_{{#1}}^{#2}}}}%q
\newcommandx*{\bgqfs}[2][1={\iq},2={},usedefault]{\bgqf[#1]{[#2]}}
\newcommandx*{\bgqs}[2][1={\iq},2={},usedefault]{\bgqf[\{#1\}]{#2}}
\newcommandx*{\bgqe}[1][1={},usedefault]{\bgqf[\emptyset]{#1}}
\newcommandx*{\bgqss}[2][1={\iq},2={},usedefault]{\bgqf[\{#1\}]{[#2]}}
\newcommandx*{\bgqes}[1][1={},usedefault]{\bgqf[\emptyset]{[#1]}}
\newcommand{\bqfu}{\Pv{\ensuremath{\overline{\qfu}}}}%Q-fun for a field thory

\newcommandx*{\gqf}[2][1={\{\iq\}},usedefault]{\Pv{\ensuremath{\qfu_{{#1}}^{#2}}}}%q
\newcommandx*{\gqfs}[2][1={\iq},2={},usedefault]{\gqf[#1]{[#2]}}
\newcommandx*{\gqs}[2][1={\iq},2={},usedefault]{\gqf[\{#1\}]{#2}}
\newcommandx*{\gqe}[1][1={},usedefault]{\gqf[\emptyset]{#1}}
\newcommandx*{\gqss}[2][1={\iq},2={},usedefault]{\gqf[\{#1\}]{[#2]}}
\newcommandx*{\gqes}[1][1={},usedefault]{\gqf[\emptyset]{[#1]}}
\newcommand{\qfu}{\Pv{\textrm{\sl q}}}%Q-fun for a field thory
\newcommand{\qbfu}{{\tilde {\qfu}}}%Q-fun for a field thory
\newcommandx*{\gqbf}[2][1={\{\iq\}},usedefault]{\Pv{\ensuremath{\qbfu_{{#1}}^{#2}}}}%q
\newcommandx*{\gqbfs}[2][1={\iq},2={},usedefault]{\gqbf[#1]{[#2]}}
\newcommandx*{\gqbs}[2][1={\iq},2={},usedefault]{\gqbf[\{#1\}]{#2}}
\newcommandx*{\gqbe}[1][1={},usedefault]{\gqbf[\emptyset]{#1}}
\newcommandx*{\gqbss}[2][1={\iq},2={},usedefault]{\gqbf[\{#1\}]{[#2]}}
\newcommandx*{\gqbes}[1][1={},usedefault]{\gqbf[\emptyset]{[#1]}}

\newcommandx*{\hbqfs}[2][1={\iq},2={},usedefault]{\hbqf[#1]{[#2]}}
\newcommandx*{\hbqf}[2][1={\{\iq\}},usedefault]{\Pv{\ensuremath{\hbqfw_{{#1}}^{#2}}}}%white
\newcommandx*{\hbqs}[2][1={\iq},2={},usedefault]{\hbqf[\{#1\}]{#2}}
\newcommandx*{\hbqss}[2][1={\iq},2={},usedefault]{\hbqf[\{#1\}]{[#2]}}
\newcommand{\hbqfw}{\hat{\mathbf q}}%Q-fun for a field thory

\newcommandx*{\hbqbfs}[2][1={\iq},2={},usedefault]{\hbqbf[#1]{[#2]}}
\newcommandx*{\hbqbf}[2][1={\{\iq\}},usedefault]{\Pv{\ensuremath{\hbqbfw_{{#1}}^{#2}}}}%white
\newcommandx*{\hbqbs}[2][1={\iq},2={},usedefault]{\hbqbf[\{#1\}]{#2}}
\newcommandx*{\hbqbss}[2][1={\iq},2={},usedefault]{\hbqbf[\{#1\}]{[#2]}}
\newcommand{\hbqbfw}{\hat{\overline{\mathbf q}}}%Q-fun for a field thory

\newcommandx*{\bqfs}[2][1={\iq},2={},usedefault]{\bqf[#1]{[#2]}}
\newcommandx*{\bqf}[2][1={\{\iq\}},usedefault]{\Pv{\ensuremath{\bqfw_{{#1}}^{#2}}}}%white
\newcommandx*{\bqs}[2][1={\iq},2={},usedefault]{\bqf[\{#1\}]{#2}}
\newcommandx*{\bqss}[2][1={\iq},2={},usedefault]{\bqf[\{#1\}]{[#2]}}
\newcommand{\bqfw}{{\mathbf q}}%Q-fun for a field thory

\newcommandx*{\bqbfs}[2][1={\iq},2={},usedefault]{\bqbf[#1]{[#2]}}
\newcommandx*{\bqbf}[2][1={\{\iq\}},usedefault]{\Pv{\ensuremath{\bqbfw_{{#1}}^{#2}}}}%white
\newcommandx*{\bqbs}[2][1={\iq},2={},usedefault]{\bqbf[\{#1\}]{#2}}
\newcommandx*{\bqbss}[2][1={\iq},2={},usedefault]{\bqbf[\{#1\}]{[#2]}}
\newcommand{\bqbfw}{\overline{\mathbf q}}%Q-fun for a field thory

\newcommandx*{\wqfs}[2][1={\iq},2={},usedefault]{\wqf[#1]{[#2]}}
\newcommandx*{\wqf}[2][1={\{\iq\}},usedefault]{\Pv{\ensuremath{\qfw_{{#1}}^{#2}}}}%white
\newcommandx*{\wqs}[2][1={\iq},2={},usedefault]{\wqf[\{#1\}]{#2}}
\newcommandx*{\wqss}[2][1={\iq},2={},usedefault]{\wqf[\{#1\}]{[#2]}}
\newcommand{\qfw}{\mathbbm{q}}%Q-fun for a field thory
\newcommand{\qbfw}{{\tilde {\qfw}}}%Q-fun for a field thory
\newcommandx*{\wqbf}[2][1={\{\iq\}},usedefault]{\Pv{\ensuremath{\qbfw_{{#1}}^{#2}}}}%q
\newcommandx*{\wqbs}[2][1={\iq},2={},usedefault]{\wqbf[\{#1\}]{#2}}
\newcommandx*{\wqbfs}[2][1={\iq},2={},usedefault]{\wqbf[#1]{[#2]}}

\newcommandx*{\wpfs}[2][1={\iq},2={},usedefault]{\wpf[#1]{[#2]}}
\newcommandx*{\wpf}[2][1={\{\iq\}},usedefault]{\Pv{\ensuremath{\pfw_{{#1}}^{#2}}}}%white
\newcommandx*{\wps}[2][1={\iq},2={},usedefault]{\wpf[\{#1\}]{#2}}
\newcommandx*{\wpss}[2][1={\iq},2={},usedefault]{\wpf[\{#1\}]{[#2]}}
\newcommand{\pfw}{\mathbbm{p}}%Q-fun for a field thory
\newcommand{\pbfw}{{\tilde {\pfw}}}%Q-fun for a field thory
\newcommandx*{\wpbf}[2][1={\{\iq\}},usedefault]{\Pv{\ensuremath{\pbfw_{{#1}}^{#2}}}}%q
\newcommandx*{\wpbs}[2][1={\iq},2={},usedefault]{\wpbf[\{#1\}]{#2}}
\newcommandx*{\wpbfs}[2][1={\iq},2={},usedefault]{\wpbf[#1]{[#2]}}

\newcommandx*{\rTf}[4][1=\su,2=a,3=s,4={},usedefault]{\gTf[#1][#2,#3][#4]}
\newcommandx*{\gTf}[3][1=\su,2=(\lambda),3={},usedefault]{\Pv{\ensuremath{\Tfu^{\,#2}_{#3}\left(#1\right)}}}
\newcommandx*{\las}[2][1=a,2=s,usedefault]{\Pv{\ensuremath{\la_{[#1,#2]}}}}
\newcommandx*{\pg}[2][1=g,2=\lambda,usedefault]{\Pv{\ensuremath{\pge_{#2}(#1)}}}
\newcommandx*{\pe}[2][1=e_{\alpha,\beta},2=\lambda,usedefault]{\Pv{\ensuremath{\pee_{#2}(#1)}}}
\newcommandx*{\Y}[3][1=a,2=s,3={},usedefault]{\Pv{\ensuremath{\Yfu_{#1,#2}^{#3}}}}
\newcommandx*{\Yb}[3][1=a,2=s,3={},usedefault]{\Pv{\ensuremath{\Ybfu_{#1,#2}^{#3}}}}
\newcommandx*{\subT}[3][1=a,2=s,usedefault]{\Pv{\ensuremath{\Tft_{#1,#2}^{\,#3}}}}
\newcommandx*{\T}[3][1=a,2=s,3={},usedefault]{\subT[#1][#2]{#3}}
\newcommandx*{\Ts}[3][1=a,2=s,3={},usedefault]{\subT[#1][#2]{[#3]}}
\newcommandx*{\tilT}[3][1=a,2=s,3={},usedefault]{\Pv{\ensuremath{\tilTft_{#1,#2}^{\,#3}}}}
\newcommandx*{\breT}[3][1=a,2=s,3={},usedefault]{\Pv{\ensuremath{\breTft_{#1,#2}^{\,#3}}}}
\newcommandx*{\sga}[2][1={},2={},usedefault]{\Pv{\ensuremath{{\sqrt{g_{#2}}}^{[#1]}}}}
\newcommandx*{\ga}[3][1={},2={},3={},usedefault]{\Pv{\gan[{[#1]}][#2][#3]
  }}
\newcommandx*{\gan}[3][1={},2={},3={},usedefault]{\Pv{\ensuremath{{#3{g}}^{#1}_{#2}}}}
\newcommandx*{\hT}[3][1=a,2=s,3={},usedefault]{\Pv{\ensuremath{\hTft_{#1,#2}^{\,#3}}}}
\newcommand{\Yfu}{\Pv{\ensuremath{Y}}} % Y-function
\newcommand{\Ybfu}{\Pv{\ensuremath{\overline{Y}}}} % Y-function
\newcommand{\Tft}{\Tfu} % T-function in the TBA section
\newcommand{\hTft}{\Pv{\ensuremath{\mathrlap{\hat{\phantom{~\Tfu}}}\Tfu}}}
\newcommand{\tilTft}{\mathrlap{\tilde{\phantom{~\Tfu}}}\Tfu}
\newcommand{\breTft}{\mathrlap{\breve{\phantom{~\Tfu}}}\Tfu}
\newcommand{\bTft}{\Pv{\textrm{\bf{T}}}}
\newcommandx*{\bT}[3][1=a,2=s,3={},usedefault]{\Pv{\ensuremath{\bTft_{#1,#2}^{#3}}}}

\newcommand{\hbTft}{\Pv{\hat{\textrm{\bf{T}}}}}
\newcommandx*{\hbT}[3][1=a,2=s,3={},usedefault]{\Pv{\ensuremath{\hbTft_{#1,#2}^{#3}}}}

\newcommand{\wTft}{\Pv{\ensuremath{\mathbb{T}}}}
\newcommandx*{\wT}[3][1=a,2=s,3={},usedefault]{\Pv{\ensuremath{\wTft_{#1,#2}^{#3}}}}

\newcommand{\hwTft}{\Pv{\ensuremath{\hat{\mathbb{T}}}}}
\newcommandx*{\hwT}[3][1=a,2=s,3={},usedefault]{\Pv{\ensuremath{\hwTft_{#1,#2}^{#3}}}}

\newcommand{\upTft}{\Pv{\ensuremath{\left.\underbracket{\mathrm{T}}\right.}}}
\newcommandx*{\upT}[3][1=a,2=s,3={},usedefault]{\Pv{\ensuremath{\upTft_{\smash{\ensuremath{\raisebox{.2cm}{\ensuremath{_{#1,#2}}}}}}^{\,#3}}}}

\newcommand{\huTft}{\Pv{\ensuremath{\left.\underbracket{\smash{\mathrlap{{\hat{\mathrm{T}}}}}\phantom{T}}\right.}}}
\newcommandx*{\huT}[3][1=a,2=s,3={},usedefault]{\Pv{\ensuremath{\huTft_{\smash{\ensuremath{\raisebox{.2cm}{\ensuremath{_{#1,#2}}}}}}^{\,#3}}}}

\newcommand{\riTft}{\Pv{\ensuremath{\underrightarrow{\mathrm{T}}}}}
\newcommandx*{\riT}[3][1=a,2=s,3={},usedefault]{\Pv{\ensuremath{\riTft_{#1,#2}^{#3}}}}

\newcommand{\hrTft}{\Pv{\ensuremath{\underrightarrow{\hat{\mathrm{T}}}}}}
\newcommandx*{\hrT}[3][1=a,2=s,3={},usedefault]{\Pv{\ensuremath{\hrTft_{#1,#2}^{#3}}}}

\newcommandx*{\hrqfs}[2][1={\iq},2={},usedefault]{\hrqf[#1]{[#2]}}
\newcommandx*{\hrqf}[2][1={\{\iq\}},usedefault]{\Pv{\ensuremath{\hrqfqq_{{#1}}^{#2}}}}%white
\newcommandx*{\hrqs}[2][1={\iq},2={},usedefault]{\hrqf[\{#1\}]{#2}}
\newcommandx*{\hrqss}[2][1={\iq},2={},usedefault]{\hrqf[\{#1\}]{[#2]}}
\newcommandx*{\hrqe}[1][1={},usedefault]{\hrqf[\emptyset]{#1}}
\newcommandx*{\hrqes}[1][1={},usedefault]{\hrqf[\emptyset]{[#1]}}
\newcommand{\hrqfqq}{\Pv{\ensuremath{\underrightarrow{\hat{q}}\!}}}

\newcommandx*{\rbqfs}[2][1={\iq},2={},usedefault]{\rbqf[#1]{[#2]}}
\newcommandx*{\rbqf}[2][1={\{\iq\}},usedefault]{\Pv{\ensuremath{\rbqfqq_{{#1}}^{#2}}}}%white
\newcommandx*{\rbqs}[2][1={\iq},2={},usedefault]{\rbqf[\{#1\}]{#2}}
\newcommandx*{\rbqss}[2][1={\iq},2={},usedefault]{\rbqf[\{#1\}]{[#2]}}
\newcommandx*{\rbqe}[1][1={},usedefault]{\rbqf[\emptyset]{#1}}
\newcommandx*{\rbqes}[1][1={},usedefault]{\rbqf[\emptyset]{[#1]}}
\newcommand{\rbqfqq}{\Pv{\ensuremath{\underrightarrow{\overline{q}}\!}}}

\newcommandx*{\hrbqfs}[2][1={\iq},2={},usedefault]{\hrbqf[#1]{[#2]}}
\newcommandx*{\hrbqf}[2][1={\{\iq\}},usedefault]{\Pv{\ensuremath{\hrbqfqq_{{#1}}^{#2}}}}%white
\newcommandx*{\hrbqs}[2][1={\iq},2={},usedefault]{\hrbqf[\{#1\}]{#2}}
\newcommandx*{\hrbqss}[2][1={\iq},2={},usedefault]{\hrbqf[\{#1\}]{[#2]}}
\newcommandx*{\hrbqe}[1][1={},usedefault]{\hrbqf[\emptyset]{#1}}
\newcommandx*{\hrbqes}[1][1={},usedefault]{\hrbqf[\emptyset]{[#1]}}
\newcommand{\hrbqfqq}{\Pv{\ensuremath{\underrightarrow{\hat{\overline{q}}}\!}}}

\newcommandx*{\rqfs}[2][1={\iq},2={},usedefault]{\rqf[#1]{[#2]}}
\newcommandx*{\rqf}[2][1={\{\iq\}},usedefault]{\Pv{\ensuremath{\rqfqq_{{#1}}^{#2}}}}%white
\newcommandx*{\rqs}[2][1={\iq},2={},usedefault]{\rqf[\{#1\}]{#2}}
\newcommandx*{\rqss}[2][1={\iq},2={},usedefault]{\rqf[\{#1\}]{[#2]}}
\newcommandx*{\rqe}[1][1={},usedefault]{\rqf[\emptyset]{#1}}
\newcommandx*{\rqes}[1][1={},usedefault]{\rqf[\emptyset]{[#1]}}
\newcommand{\rqfqq}{\Pv{\ensuremath{\underrightarrow{q}\!}}}

\newcommandx*{\rpfs}[2][1={\iq},2={},usedefault]{\rpf[#1]{[#2]}}
\newcommandx*{\rpf}[2][1={\{\iq\}},usedefault]{\Pv{\ensuremath{\rpfqq_{{#1}}^{#2}}}}%white
\newcommandx*{\rps}[2][1={\iq},2={},usedefault]{\rpf[\{#1\}]{#2}}
\newcommandx*{\rpss}[2][1={\iq},2={},usedefault]{\rpf[\{#1\}]{[#2]}}
\newcommandx*{\rpe}[1][1={},usedefault]{\rpf[\emptyset]{#1}}
\newcommandx*{\rpes}[1][1={},usedefault]{\rpf[\emptyset]{[#1]}}
\newcommand{\rpfqq}{\Pv{\ensuremath{\underrightarrow{p}\!}}}

\newcommandx*{\hrpfs}[2][1={\iq},2={},usedefault]{\hrpf[#1]{[#2]}}
\newcommandx*{\hrpf}[2][1={\{\iq\}},usedefault]{\Pv{\ensuremath{\hrpfqq_{{#1}}^{#2}}}}%white
\newcommandx*{\hrps}[2][1={\iq},2={},usedefault]{\hrpf[\{#1\}]{#2}}
\newcommandx*{\hrpss}[2][1={\iq},2={},usedefault]{\hrpf[\{#1\}]{[#2]}}
\newcommandx*{\hrpe}[1][1={},usedefault]{\hrpf[\emptyset]{#1}}
\newcommandx*{\hrpes}[1][1={},usedefault]{\hrpf[\emptyset]{[#1]}}
\newcommand{\hrpfqq}{\Pv{\ensuremath{\underrightarrow{\hat{p}}\!}}}

\newcommandx*{\upbqfs}[2][1={\iq},2={},usedefault]{\upbqf[#1]{[#2]}}
\newcommandx*{\upbqf}[2][1={\{\iq\}},usedefault]{\Pv{\ensuremath{\upbqfqq_{{#1}}^{#2}}}}%white
\newcommandx*{\upbqs}[2][1={\iq},2={},usedefault]{\upbqf[\{#1\}]{#2}}
\newcommandx*{\upbqss}[2][1={\iq},2={},usedefault]{\upbqf[\{#1\}]{[#2]}}
\newcommandx*{\upbqe}[1][1={},usedefault]{\upbqf[\emptyset]{#1}}
\newcommandx*{\upbqes}[1][1={},usedefault]{\upbqf[\emptyset]{[#1]}}
\newcommand{\upbqfqq}{\Pv{\ensuremath{\left.\underbracket{\overline{\mathrm{q}}}\right.\!}}}

\newcommandx*{\huqfs}[2][1={\iq},2={},usedefault]{\huqf[#1]{[#2]}}
\newcommandx*{\huqf}[2][1={\{\iq\}},usedefault]{\Pv{\ensuremath{\huqfqq_{{#1}}^{#2}}}}%white
\newcommandx*{\huqs}[2][1={\iq},2={},usedefault]{\huqf[\{#1\}]{#2}}
\newcommandx*{\huqss}[2][1={\iq},2={},usedefault]{\huqf[\{#1\}]{[#2]}}
\newcommandx*{\huqe}[1][1={},usedefault]{\huqf[\emptyset]{#1}}
\newcommandx*{\huqes}[1][1={},usedefault]{\huqf[\emptyset]{[#1]}}
\newcommand{\huqfqq}{\Pv{\ensuremath{\left.\underbracket{{\hat{\mathrm{q}}}}\right.\!}}}

\newcommandx*{\hubqfs}[2][1={\iq},2={},usedefault]{\hubqf[#1]{[#2]}}
\newcommandx*{\hubqf}[2][1={\{\iq\}},usedefault]{\Pv{\ensuremath{\hubqfqq_{{#1}}^{#2}}}}%white
\newcommandx*{\hubqs}[2][1={\iq},2={},usedefault]{\hubqf[\{#1\}]{#2}}
\newcommandx*{\hubqss}[2][1={\iq},2={},usedefault]{\hubqf[\{#1\}]{[#2]}}
\newcommandx*{\hubqe}[1][1={},usedefault]{\hubqf[\emptyset]{#1}}
\newcommandx*{\hubqes}[1][1={},usedefault]{\hubqf[\emptyset]{[#1]}}
\newcommand{\hubqfqq}{\Pv{\ensuremath{\left.\underbracket{{\hat{\overline{\mathrm{q}}}}}\right.\!}}}

\newcommandx*{\upqfs}[2][1={\iq},2={},usedefault]{\upqf[#1]{[#2]}}
\newcommandx*{\upqf}[2][1={\{\iq\}},usedefault]{\Pv{\ensuremath{\upqfqq_{{#1}}^{#2}}}}%white
\newcommandx*{\upqs}[2][1={\iq},2={},usedefault]{\upqf[\{#1\}]{#2}}
\newcommandx*{\upqss}[2][1={\iq},2={},usedefault]{\upqf[\{#1\}]{[#2]}}
\newcommandx*{\upqe}[1][1={},usedefault]{\upqf[\emptyset]{#1}}
\newcommandx*{\upqes}[1][1={},usedefault]{\upqf[\emptyset]{[#1]}}
\newcommand{\upqfqq}{\Pv{\ensuremath{\left.\underbracket{\mathrm{q}}\right.\!}}}

\newcommandx*{\uppfs}[2][1={\iq},2={},usedefault]{\uppf[#1]{[#2]}}
\newcommandx*{\uppf}[2][1={\{\iq\}},usedefault]{\Pv{\ensuremath{\uppfqq_{{#1}}^{#2}}}}%white
\newcommandx*{\upps}[2][1={\iq},2={},usedefault]{\uppf[\{#1\}]{#2}}
\newcommandx*{\uppss}[2][1={\iq},2={},usedefault]{\uppf[\{#1\}]{[#2]}}
\newcommandx*{\uppe}[1][1={},usedefault]{\uppf[\emptyset]{#1}}
\newcommandx*{\uppes}[1][1={},usedefault]{\uppf[\emptyset]{[#1]}}
\newcommand{\uppfqq}{\Pv{\ensuremath{\left.\underbracket{\mathrm{p}}\right.\!}}}

\newcommand{\leTft}{\Pv{\ensuremath{\underleftarrow{\mathrm{T}}}}}
\newcommandx*{\leT}[3][1=a,2=s,3={},usedefault]{\Pv{\ensuremath{\leTft_{#1,#2}^{#3}}}}

\newcommand{\pge}{\uppi} 
\newcommand{\pee}{\textrm{\boldmath \ensuremath{\pge}}} 
\newcommand{\mybold}[1]{\textrm{\boldmath\ensuremath{#1}}}
 
\newcommand{\PI}{\textrm{\boldmath \ensuremath{\Pi}}}
\newcommand{\BI}{\textrm{B}}
\newcommand{\la}{\Pv{\ensuremath{{\lambda}}}}
\newcommand{\Qop}{\Pv{\textrm{Q}}}
\newcommand{\Rop}{\Pv{\ensuremath{R}}}
\newcommand{\Rg}{\Pv{\ensuremath{R}}}
\newcommand{\RL}{\Pv{\ensuremath{R/L}}}
\newcommand{\Sact}{\Pv{\ensuremath{\mathcal{S}}}} % the S matrix 
\newcommand{\Smat}{\Pv{\ensuremath{\hat S}}} % the action 
\newcommand{\Sscal}{\Pv{\ensuremath{S}}}% the dressing phase
\newcommand{\bSact}{\Pv{\ensuremath{\mybold{\Sact}}}} % the S matrix 
\newcommand{\Sb}{\textrm{\boldmath \ensuremath{\Sscal}}}% the product
\newcommand{\CDD}{\Pv{\ensuremath{\chi_{_{\mathrm{CDD}}}}}}% the CDD
\newcommand{\CDb}{\textrm{\boldmath \ensuremath{\CDD}}}
\newcommand{\Sgrp}[1]{\Pv{\ensuremath{\mathcal{S}^{#1}}}}%the
\newcommand{\Ssph}{S}%the sphere
\newcommand{\Lf}{\Pv{\ensuremath{L}}}
\newcommand{\hD}{\Pv{\ensuremath{\hat{\mathrm{D}}}}}
\newcommand{\hDt}{\hD ~~}
\newcommandx*{\DL}[3][1=\su_{\spi}+\hD,2=1,3=\lcds,usedefault]{\bigotimes_{\spi=#2}^{#3}\left(#1\right)}
\newcommandx*{\DLt}[3][1=\su_{\spi}+\hD,2=1,3=\lcds,usedefault]{\DL[#1][#2][#3] ~~}
\newcommand{\De}{\Pv{\ensuremath{\mathcal{D}}}}
\newcommand{\Dm}{\Pv{\ensuremath{\mathrm{D}}}}
\newcommand{\Kr}{\Pv{\textrm{K}}}%like in GL(K)
\newcommand{\Mr}{\Pv{\textrm{M}}}%like in GL(K|M)
\newcommand{\Np}{\Pv{\textrm{N}}}% like in SU(N) PCF
\newcommand{\Ker}{\Pv{\ensuremath{{K}}}}% a Kernel
\newcommand{\CK}{\Pv{\ensuremath{\mathcal{K}}}}% the Cauchy Kernel
\def\sK{\not{\!\!\;\!\CK}}
\def\pint{-\hskip-0.43cm \int}
\newcommand{\perm}{\Pv{\ensuremath{\mathcal{P}}}}
\newcommand{\proj}{\Pv{\ensuremath{\mathrm{P}}}}
\newcommand{\PP}{P}
\newcommand{\gr}[1]{\Pv{\ensuremath{\mathrm{p}_{#1}}}}
\newcommand{\sg}[1]{\Pv{\ensuremath{(-1)^{\gr {#1}}}}}
\newcommand{\C}{\textrm{C}}
\newcommand{\G}{\Pv{\ensuremath{\mathrm{g}}}}%AdS/CFT coupling

\newcommandx*{\rlb}[3][1={{{,}}},2={},3={},usedefault]{\Pv{\ensuremath{{{(#2
        #1 #3 )}}}}}%label for roots
\newcommandx*{\strb}[2][1=\mlvl,2=\kk,usedefault]{\Pv{\ensuremath{% (#1)_{#2}
[#1,#2]
    }}}
\newcommandx*{\dg}[1][1=\mlvl,usedefault]{\Pv{\ensuremath{d^{\rlb[{#1}]}}}}
\newcommand{\cha}[1]{\chi_{_{#1}}}
\newcommand{\chs}[1]{\chi^{(#1)}}
\newcommandx*{\chas}[2][1=a,2=s,usedefault]{
  \chi^{(#1,#2)}}

\newcommand{\g}[1][{}]{\Pv{\ensuremath{g_{_{#1}}}}}
\newcommand{\gp}[1]{\Pv{\ensuremath{g_{% _{
          #1% }
      }}}}%operator g at a given position

\newcommandx{\Il}[1]{\Pv{\ensuremath{I_{{#1}}}}}

\newcommand{\js}{\Pv{\ensuremath{\mathtt{j}}}}
\newcommand{\sutr}{\Pv{\ensuremath{\mathrm{Str}}}}%supertrace
\newcommand{\sudet}{\Pv{\ensuremath{\mathrm{Sdet}}}}%supertrace
\newcommand{\Max}{\Pv{\ensuremath{\mathrm{max}}}}
\newcommand{\Min}{\Pv{\ensuremath{\mathrm{min}}}}
\newcommandx*{\sht}[2][1={\theta_\nrt},2=N,usedefault]{\Pv{\ensuremath{\mathrm{sinh}\left(\frac{2\pi}{#2} #1\right)}}}
\newcommandx*{\cht}[2][1={\theta_\nrt},2=N,usedefault]{\Pv{\ensuremath{\mathrm{cosh}\left(\frac{2\pi}{#2} #1\right)}}}

\newcommandx*{\VdM}[3][1=\nn,2=1,3=z,usedefault]{\Pv{\ensuremath{\dVdM({#3}_{#2},\cdots,{#3}_{#1})}}}
\newcommand{\dVdM}{\Pv{\ensuremath{\Updelta}}}

\newcommand{\Det}[2]{\Pv{\ensuremath{
\left|\left(#1\right)_{#2}\right|}}}
\newcommand{\BlockDet}[4]{\Pv{\ensuremath{
\left|
  \begin{array}{c}
\left(#1\right)_{#2}\\
\left(#3\right)_{#4}
\end{array}
\right|}}}

\newcommand{\where}{\textrm{where }}
\newcommand{\when}{\textrm{when }}

\newcommand{\ie}{\textrm{i.e.}}
\newcommand{\If}{\textrm{ if }}
\newcommand{\IF}{\textrm{ If }}
\newcommand{\oth}{\textrm{ otherwise }}
\newcommand{\Then}{\textrm{ then }}
\newcommand{\et}{\textrm{and }}
\renewcommand{\And}{\textrm{ and }}
\newcommand{\hence}{\textrm{hence }}
\newcommand{\Or}{\textrm{ or }}
\newcommand{\whereas}{\textrm{whereas }}

\definecolor{lightgray}{rgb}{.85,0.85,0.85}

\newcommand{\framedline}[2]{\lefteqn{\fbox{\ensuremath{\displaystyle 
#1 #2}}}\phantom{\rule{\fboxsep}{0pt}\rule{\fboxrule}{0pt}#1}&\phantom{#2\rule{\fboxsep}{0pt}\rule{\fboxrule}{0pt}}}

 \newcommand{\standardline}[2]{#1 & #2}
\ifdefined\PreviewMacro
\renewcommand{\framedline}[2]{#1 & #2}
\fi

\newcommand{\parDer}[1]{\Pv{\ensuremath{\frac{\partial}{\partial
#1}}}}
\newcommand{\Vect}[1]{\Pv{\ensuremath{\mathrm{Vect}\left\{#1\right\}}}}

\renewcommand{\Im}{\Pv{\ensuremath{\mathrm{Im}}}}
\renewcommand{\Re}{\Pv{\ensuremath{\mathrm{Re}}}}

\newcommand{\PST}{\Pv{\textrm{PSU(2,2\ensuremath{|}4)}}}
\newcommand{\SU}[1]{\Pv{\textrm{SU(#1)}}}
\newcommand{\U}[1]{\Pv{\textrm{U(#1)}}}
\newcommand{\GL}[1]{\Pv{\textrm{GL(#1)}}}
\newcommand{\SL}[1]{\Pv{\textrm{SL(#1)}}}
\newcommand{\GLKM}{\GL{\Kr\ensuremath{|}{\Mr}}}
\newcommand{\Mat}[1]{\Pv{\textrm{M(#1)}}}
\newcommand{\Orth}[1]{\Pv{\textrm{O(#1)}}}

\newcommand{\bR}{ℝ}%\mathbb{R}
\newcommand{\bN}{ℕ}%\mathbb{N}
\newcommand{\bC}{ℂ}%\mathbb{C}
\newcommand{\bZ}{ℤ}%\mathbb{Z}
\newcommand{\bI}{𝕀}%\mathbb{I}
\newcommand{\bi}{ⅈ}%\mathbbm{i}
\newcommand{\bo}{{\ensuremath{{\mathbbm{1}}}}}%\mathbbm{1}

\newcommand{\Zf}{\Pv{\ensuremath{ℤ_4}}}%\mathbb{Z}

\newcommand{\eu}{e^{\partial_\su}}
\newcommand{\ove}{{\overline{\emptyset}}}
\newcommand{\oj}{{\overline{\jmathvp}}}
\newcommand{\Ib}{{{\Pv{\ensuremath{\overline{I}}}}}}
\newcommand{\Ijb}{{{\Pv{\ensuremath{\overline{I\Delta\jvp}}}}}}

\newcommand{\Normed}{\Pv{\ensuremath{\mathcal{N}}}}

\newcommandx*{\Wt}[3][1=z,2=\su,3={},usedefault]{\Pv{\ensuremath{\mathcal{W}_{#3}(#2;#1)}}}

\newcommandx*{\OOp}[2][1=\jvp,2=I,usedefault]{\Pv{\ensuremath{\mathcal{O}_{#2}(#1)}}}

\newcommand{\xx}{\Pv{\ensuremath{x}}}
\newcommand{\hx}{\Pv{\ensuremath{\hat x}}}%physical sheet of "x"
\newcommandx*{\MyLine}[5][1={jj1/ii1},%standard
2={}%dashed
,3=\forget%label
,4={}% Identity
,5={}%labels for lines
]{\raisebox{-5pt}{  \begin{tikzpicture}[scale=.5]
\tikzstyle{bboule} = [circle,scale=0.2,ball color=black]
\node[bboule,below] (jj1) at (0,-.5) {};
\node[bboule,above] (ii1) at (0,.5) {};
#3{\node[above] at (ii1.north) {\(_{\coordi}\)};\node[below] at (jj1.south) {\(_{\coordj}\)};}
\foreach \from/\to in {#1}
     \draw [line width=.25mm%,-% >
     ] (\from) -- (\to);
\foreach \from/\to in {#2}
     \draw [style=densely dotted,line width=.25mm%,-% >
      ]
     (\from) -- (\to);
\foreach \from/\to in {#4}
     \draw [double, line width=.2mm%,-% >
     ] (\from) -- (\to);
\foreach \bla/\alb in {#5}
     \node[below] at (\bla) {\footnotesize \alb};
  \end{tikzpicture}
}}
\newcommandx*{\MyTwoNodes}[5][1={jj1/ii1,jj2/ii2},%standard
2={}%dashed
,3=\forget%label
,4={}% Identity
,5={}%labels for lines
]{\raisebox{-5pt}{  \begin{tikzpicture}[scale=.5]
\tikzstyle{bboule} = [circle,scale=0.2,ball color=black]
\node[bboule] (jj1) at (0,-.5) {};
\node[bboule] (jj2) at (0.5,-.5) {};
\node[bboule] (ii1) at (0,.5) {};
\node[bboule] (ii2) at (0.5,.5) {};
#3{\node[above] at (ii1.north) {\(_{\coordi_1}\)}; \node[above] at (ii2.north)
  {\(_{\coordi_2}\)}; \node[below] at (jj1.south) {\(_{\coordj_1}\)}; \node[below]
  at (jj2.south) {\(_{\coordj_2}\)}; }
\foreach \from/\to in {#1}
     \draw [line width=.25mm%,-% >
     ] (\from) -- (\to);
\foreach \from/\to in {#2}
     \draw [style=densely dotted,line width=.25mm%,-% >
      ]
     (\from) -- (\to);
\foreach \from/\to in {#4}
     \draw [double, line width=.2mm%,-% >
     ] (\from) -- (\to);
\foreach \bla/\alb in {#5}
     \node[below] at (\bla) {\footnotesize \alb};
  \end{tikzpicture}
}}
\newcommandx*{\MyThreeNodes}[4][1={jj1/ii1, jj2/ii2, jj3/ii3},2={},3=\forget,4={}]{\raisebox{-5pt}{  \begin{tikzpicture}[scale=.5]
\tikzstyle{bboule} = [circle,scale=0.2,ball color=black]
\node[bboule] (jj1) at (0,-.5) {};
\node[bboule] (jj2) at (0.5,-.5) {};
\node[bboule] (jj3) at (1,-.5) {};
\node[bboule] (ii1) at (0,.5) {};
\node[bboule] (ii2) at (0.5,.5) {};
\node[bboule] (ii3) at (1,.5) {};
#3{\node[above] at (ii1.north) {\(_{\coordi_1}\)}; \node[above] at (ii2.north)
  {\(_{\coordi_2}\)}; \node[above] at (ii3.north)
  {\(_{\coordi_3}\)}; \node[below] at (jj1.south) {\(_{\coordj_1}\)}; \node[below]
  at (jj2.south) {\(_{\coordj_2}\)};\node[below]
  at (jj3.south) {\(_{\coordj_3}\)}; }
\foreach \from/\to in {#1}
     \draw [line width=.25mm%,-% >
     ] (\from) -- (\to);
\foreach \from/\to in {#2}
     \draw [style=densely dotted,line width=.25mm%,-% >
      ]
     (\from) -- (\to);
\foreach \from/\to in {#4}
     \draw [style=densely dotted,line width=.5mm,color=red
      ]
     (\from) -- (\to);
  \end{tikzpicture}
}}

\newcommand{\fdisp}[1]{\fbox{\ensuremath{\displaystyle{#1}}}}

\newcommand{\st}{\ast}

\newcommand{\staref}[1]{\Pv{Statement \ref{#1}}}
\newcommand{\stapref}[1]{\Pv{Statement \ref{#1} (page \pageref{#1})}}
\newcommand{\figref}[1]{\Pv{figure \ref{#1}}}
\newcommand{\figpref}[1]{\Pv{figure \ref{#1} (page \pageref{#1})}}

\newcommand{\FS}[1][\Np]{{\Pv{\ensuremath{{\lbrack #1 \rbrack}_\DD}}}}
\newcommand{\Ast}[2][]{{\Pv{\ensuremath{\mathtt{A}^{#1}_{#2}}}}}
\newcommand{\Af}[2][]{{\Pv{\ensuremath{\mathcal{A}^{#1}_{#2}}}}}

\newcommandx*{\rf}[2][1={\iq},2={},usedefault]{\Pv{\ensuremath{f_{#1}^{#2}}}}
\newcommandx*{\Pf}[2][1={\iq},2={},usedefault]{\Pv{\ensuremath{P_{#1}^{#2}}}}

\newcommand{\SnS}{S} %nothing
\newcommand{\MnM}{M} %nothing

\def\tth{{\tilde{\theta}}}
\def\bth{{\mybold{\theta}}}

\def\bZc{\Pv{\ensuremath{
{{\resizebox{0.28cm}{0.33cm}{\ensuremath{\hspace{0.03cm}\check {\hspace{-0.03cm}\resizebox{0.14cm}{0.18cm}{\ensuremath{Z}}}}}}}}}}
\def\hbZc{\Pv{\ensuremath{\widehat{Z}}}}

\newcommand{\CF}{\Pv{\ensuremath{\mathcal{F}}}}

\def\rhou{{\rho_{_U}}}
\def\rhob{{\rho_{_b}}}
\def\rhotB{{\tilde\eta_{_b}}}
\def\rhoc{{\eta_{_c}}}
\def\rhocd{{\rho_{_c}}}

\newcommand{\bB}{\Pv{\ensuremath{\mathbf{B}}}}
\newcommand{\tbB}{\Pv{\ensuremath{\tilde{\mathbf{B}}}}}
\newcommand{\tbBb}{\Pv{\overline{\tbB}}}
\newcommand{\bfC}{\Pv{\ensuremath{\mathbf{C}}}}
\newcommand{\bfCb}{\Pv{\ensuremath{\overline{\bfC}}}}

\newcommand{\baQ}{\Pv{\ensuremath{Q}}} % Polynomial whose roots are
\newcommand{\tbaQ}{\Pv{\ensuremath{\tilde Q}}} % Polynomial whose roots are
\DeclareMathOperator{\disc}{disc}
\DeclareMathOperator{\sign}{sign}

\newcommandx*{\GmC}[1][1=\us]{\Pv{\ensuremath{\left[#1\right]_\gamma}}}

\newcommand{\smr}{\Pv{\ensuremath{\mathtt{s}}}}%magic ratio for the
\newcommand{\rmr}{\Pv{\ensuremath{\mathtt{r}}}}%magic ratio for the
\newcommandx*{\args}[2][1=+\frac \bi
2,2={},usedefault]{\Pv{\ensuremath{\left(\us #1 #2 \right)}}}

\newcommand{\JacTruFig}{\colorprint{\begin{figure}[p]}{\begin{figure}} %
  \centering %
  \includegraphics{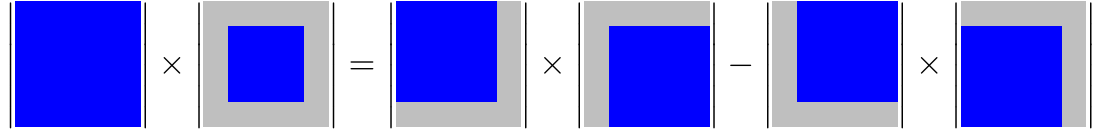} %
  \caption{The {\jacobi} identity on determinants. } %
  \label{fig:{\jacobi}} %
\end{figure} %
}

\newcommand{\FigContour}{\begin{figure} %
\fbox{\begin{minipage}{.95\textwidth} %
  \begin{center} %
   \includegraphics{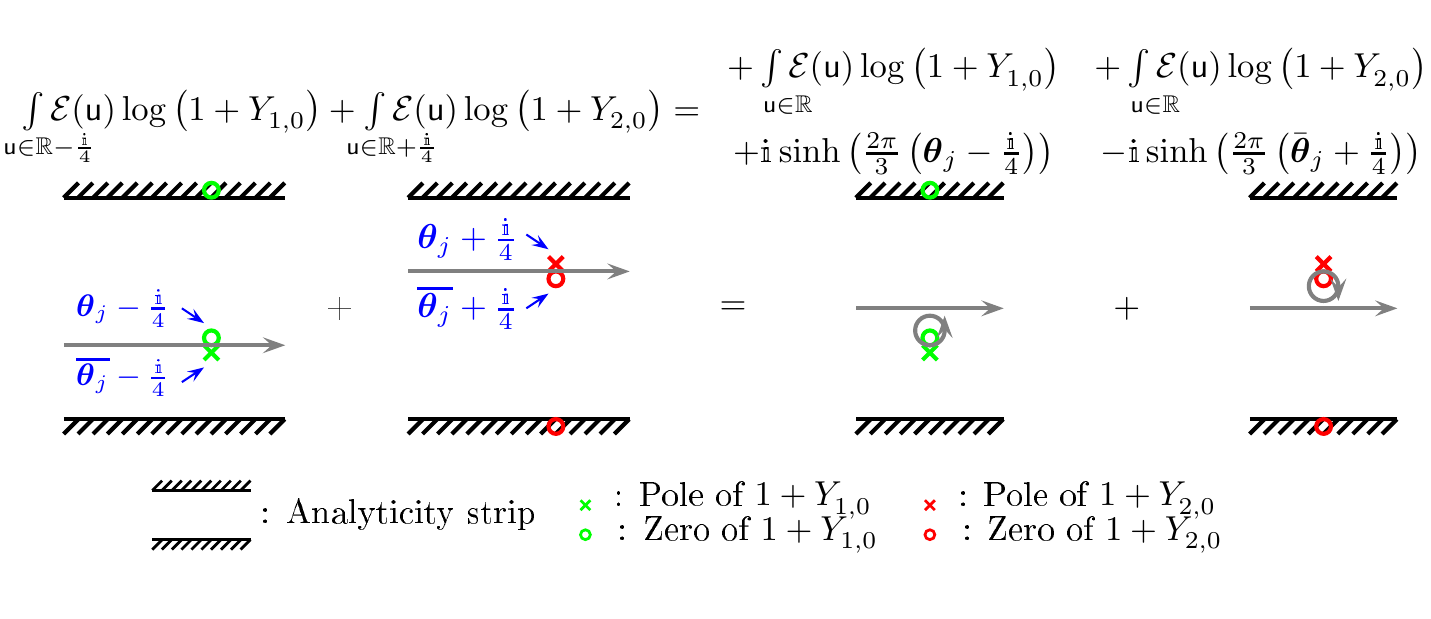} %
   \caption{Choice of contour for the energy of excited states when \(\Np=3\)} %
   \label{fig:contour3} %
  \end{center} %
Illustration of the analyticity of the integrand and of the choice of %
integration contour, in the expression of %
the energy \({E(\LF)=\sum_{a=1}^{\Np-1}\int \mathcal{E}(\us)\log \left( 1 + \Y[][0](\us)         %
\right) \mathrm{d}\us}\) (where \({\mathcal{E}(\us)=-\frac 1 3 %
{\cosh\left(\frac{2 \pi}{3} \us \right)}}\)). This illustration %
corresponds to the case when \(\Im(\bth_\jrt)>0\), and it shows that %
the expressions \eqref{eq:EnergyOdd} and \eqref{eq:oddNU1energy} coincide. %
\end{minipage}}\end{figure} %
} %

\newcommand{\FigEnergiesPCMO}{ %
\colorprint{\begin{figure}[p]}{\begin{figure}} %
\fbox{\begin{minipage}{.95\textwidth} %
  \begin{center} %
   \includegraphics{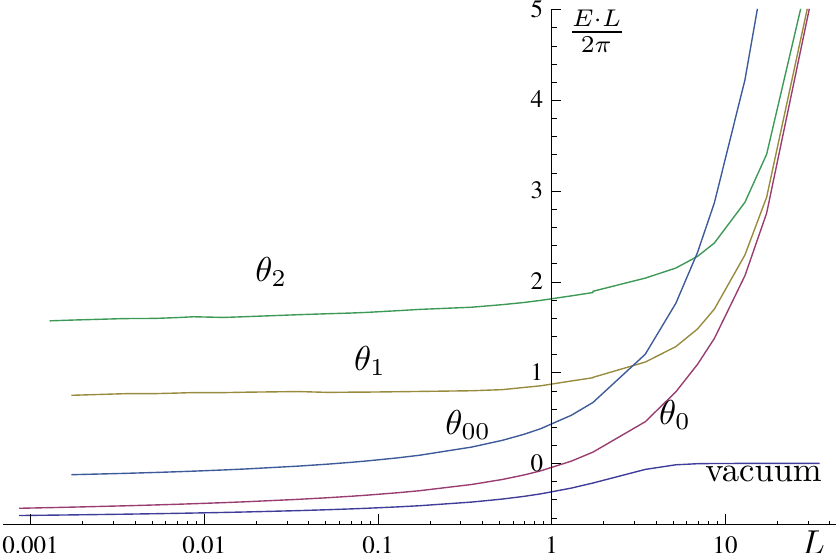} %
  \caption{ Energies of the vacuum and of a few low-energy excited states, as functions of \(\LF\), at \(\Np=3\).} %
  \label{fig:PCMen} %
\end{center} %
This graph shows the energies of a few low-lying states in the \(\SU %
3\times \SU3\) {\PCM}. The state {\lbd} \(\theta_0\) (resp \(\theta_1\) %
resp \(\theta_2\)) is the state with one particle, which has momentum number \(0\) %
(resp \(1\), resp \(2\)). %
When the size \(\LF\) is large, this ``momentum number'' is the integer \(\kappa\) such that %
\index{momentum number} %
in the Bethe equation \eqref{eq:BethePCM2}, we have  \(\dkp=\bi~ \lcds~ %
\sht + \mathrm{log}\left( \Sscal(\theta_\nrt) \right)\). On the other %
hand, the state {\lbd} \(\theta_{0,0}\) is the state with two %
particles, having momentum number \(0\). %

  We see that in the asymptotic limit (\(\LF\to\infty\)), the energy is %
equal to the number of particles, hence \(\frac{E\LF}{2\pi}\) is linear %
in \(\LF\) and looks exponential in the logarithmic scale of %
this figure. By contrast, in the %
\(\LF\to 0\) limit, \(\frac{E\LF}{2\pi}\) goes to a constant, %
equal to \(- \frac 2 3\) plus the sum of the momentum numbers. %}%
\end{minipage} %
} %
\end{figure} %
}

\newcommand{\FigEnergiesPCMT}{ \begin{figure} %
\fbox{\begin{minipage}{.95\textwidth} %
  \begin{center} %
   \includegraphics{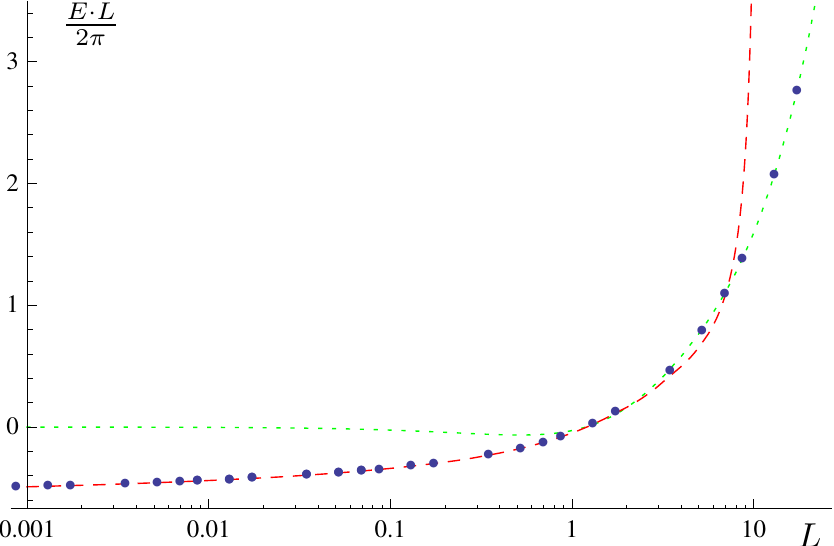}  %
   \caption{Energy of the first excited state.}  %
   \label{fig:MGraph3} %
  \end{center} %
The {\NuMr} energy of the first excited state \(\theta_0\) (blue dots)
is compared to the  %
    analytic expression \eqref{eq:MGLuscher} of  %
    \(E^{mass gap}_{\LF\to\infty}\) (green dotted line and to  %
 the conformal limit \(E_{\theta_0}=E_{vac}+\frac 8 9 \frac {2 \pi}\LF  %
 \frac 1{\log(c/\LF)+\frac 1 2 \log(\log(c/\LF))}\) (see  %
 \cite{2010arXiv1007.1770K}) (red dashed line), where  %
 \(c=12.3\) is   %
chosen to fit the data. %
\end{minipage}}\end{figure} %
}

\newcommand{\XFig}{
\colorprint{ \renewcommand\floatpagefraction{.95}
 \begin{figure}[p]}{ \begin{figure}[h]}
\fbox {\begin{minipage}{.95\textwidth}
  \begin{center}
    \subfigure[The functions \(x({\us})\) and \(1/x({\us})\).]
    {\label{fig:Xriemannl} \includegraphics{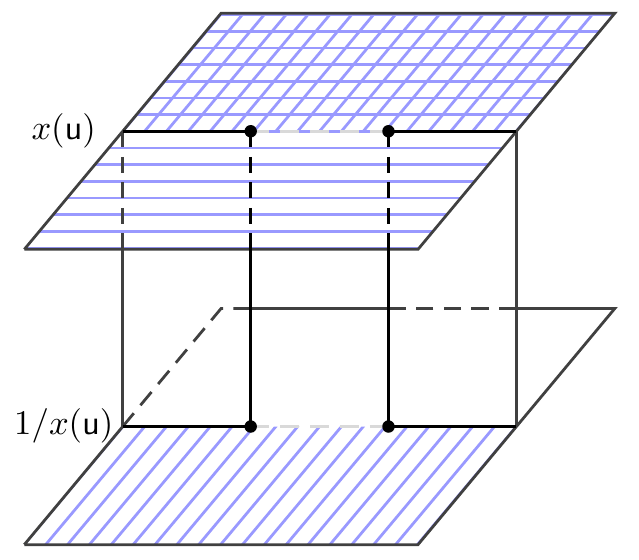}}
    \quad
    \subfigure[The functions \(\hat x({\us})\) and \(1/\hat x({\us})\).]
    {\label{fig:Xriemanns} \includegraphics{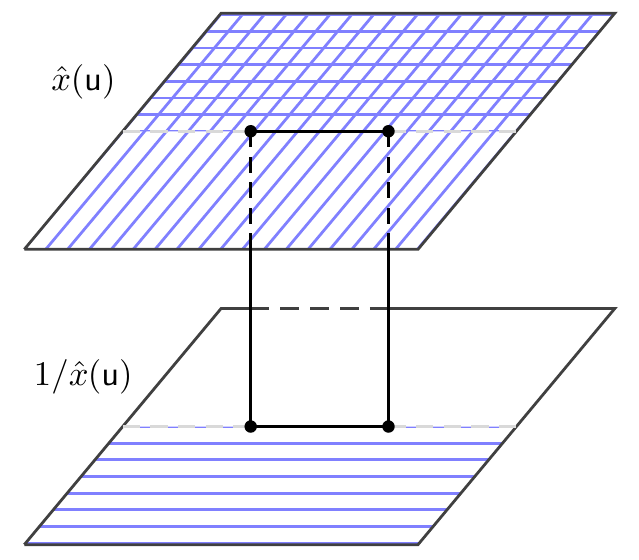}}
    \caption{Different branches of the multivalued function \(x({\us})\).}
    \label{fig:Xriemann}
  \end{center}

  The top sheet on  subfigure \ref{fig:Xriemannl} illustrates the
function \(x(\us)\) % 
(mirror branch) as
defined in \eqref{eq:DefXMir}. It has a cut on 
\({\bZc_0\equiv]-\infty,-2\,\G]\cup [2\,\G,\infty[}\), which connects
it to the function \(1/x(\us)\) (in the sense that %
\(1/x(\us)\) %
is
obtained by analytic continuation of \(x(\us)\) across this cut). 
The function \(\hx(\us)\) (``magic branch'',  top sheet of subfigure
\ref{fig:Xriemanns}) is obtained by
``gluing together'' the upper half plane of \(x(\us)\) with the lower
half plane \(1/x(\us)\) (see  \eqref{eq:DefXPhys}) and it has only a
``short'' 
Zhukovsky cut on \(
{\hbZc}_0\equiv [-2\,\G,2\,\G]\).
\end{minipage}}
\end{figure}
}

\newcommand{\FigNumAdsCFT}{
\begin{figure}
 \centering
\fbox{\begin{minipage}{.95\textwidth}
   \begin{center}
   \subfigure[Middle nodes {\Yfs} (\ensuremath{\Y[a][0]})]
 {\label{fig:Ymid}\includegraphics[width=0.45 \textwidth]{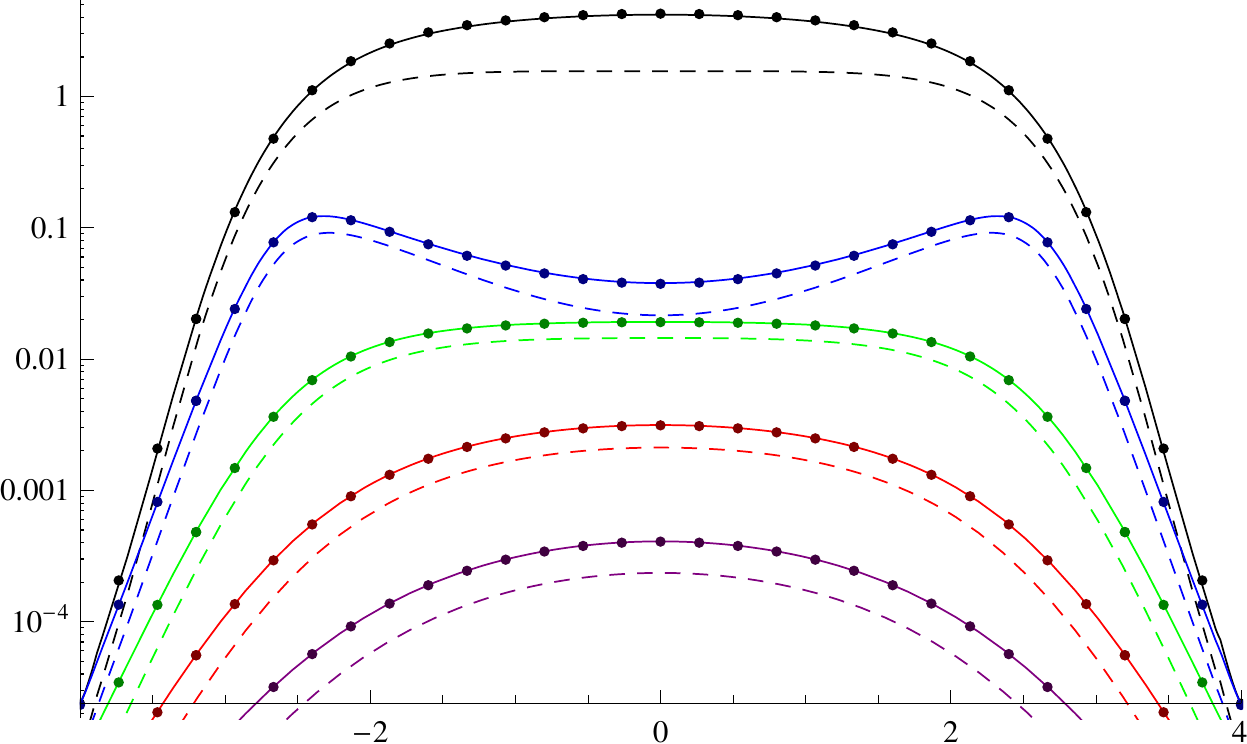}} \qquad
   \subfigure[The {\Yfs} \ensuremath{\Y[1][1]} and \ensuremath{1/\Y[2][2]}]
 {\label{fig:Yf}\includegraphics[width=0.45 \textwidth]{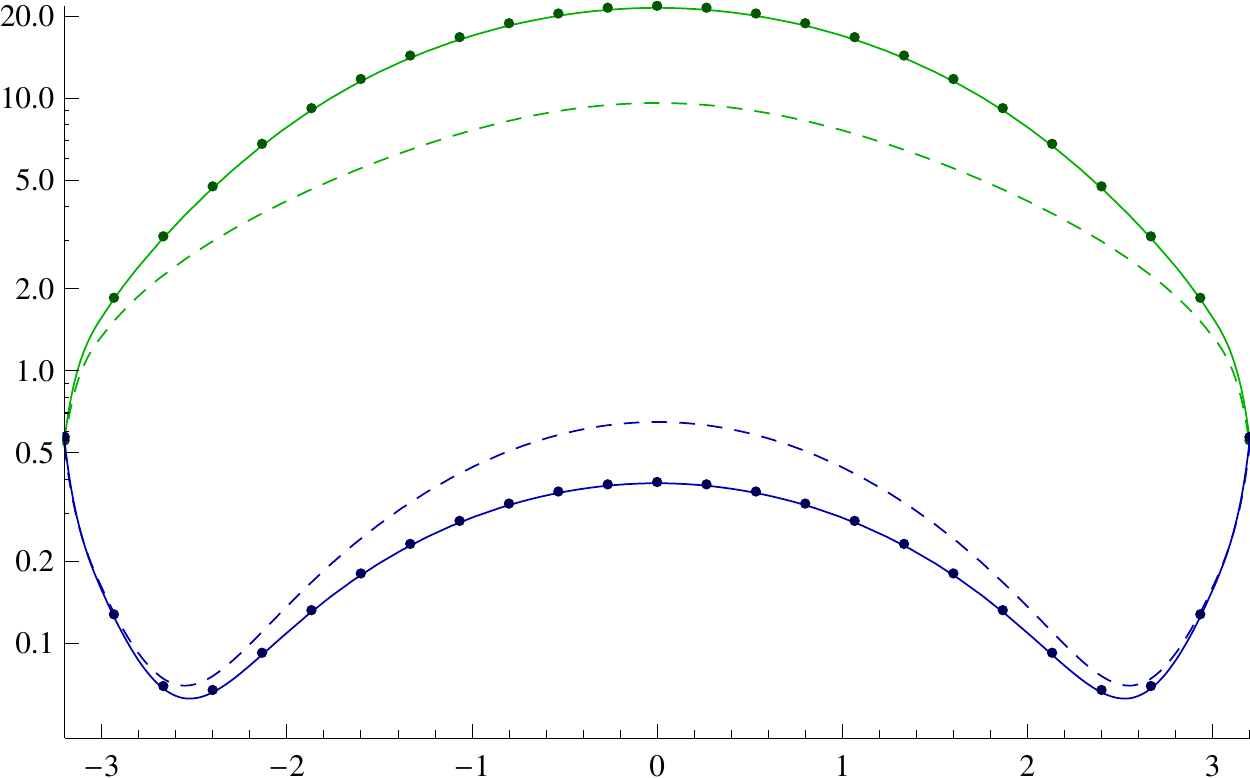}}
 \caption{Numerical {\Yfs} for Konishi state at \ensuremath{\G=1.6}}
  \label{fig:Yfuns}
   \end{center}

 The {\Yfs} obtained from FiNLIE  iterations (dots) are compared
 to the outcome of TBA-iterations  \cite{Gromov:2009zb}
 (solid lines) and to the asymptotic expression (dashed lines).
 The figure~\ref{fig:Ymid} shows \(\Y[1][0]\) (black), \(\Y[2][0]\) (blue), \(\Y[3][0]\)
 (green), \(\Y[4][0]\) (red) and \(\Y[5][0]\) (violet), while the
 figure~\ref{fig:Yf} shows \(\Y[1][1]\) (green) and its continuation
 \(1/\Y[2][2]\) (blue). %}
\end{minipage}}
\end{figure}
}

%% file: francais.tex
\begin{center}
  \textbf{Résumé}
\end{center}

     Cette thèse est consacrée à l'étude de systèmes quantiques
     intégrables tels des chaînes de spins, des théories de champs à
     1+1 dimensions, et la dualité AdS/CFT. Cette dualité AdS/CFT est
     une conjecture, émise à la fin du siècle dernier, qui relie
     notamment le régime non-perturbatif d'une théorie de jauge
     superconforme (nommée \(\mathcal{{\NN}}\)=4 {\SYM}) au régime perturbatif d'une
     théorie de cordes dans un espace à 10 dimensions (de géométrie
     AdS₅\(\times\){\Ssph}⁵).

  Ce manuscrit explore les similarités entre des chaînes de spins
  intégrables et des théories de champs intégrables, tels Super Yang
  Mills. Il commence par une étude approfondie des chaînes de spins
  intégrables pour y construire explicitement un ``flot de Bäcklund'' et
  des ``opérateurs {\Qop}'' polynômiaux, qui permettent de diagonaliser le
  Hamiltonien. Des théories de champs intégrables sont ensuite
  étudiées et des ``fonctions {\Qfu}'' sont obtenues, qui sont l'analogue des
  opérateurs {\Qop} construits pour les chaînes de spins. Il apparaît que
  de nombreuses informations sont contenue dans les propriétés
  analytiques des fonctions {\Qfu}. Cela permet d'aboutir, dans le cadre de
  l'{\anz} de Bethe thermodynamique, à un nombre fini d'équations
  non-linéaires intégrales qui encode le spectre des niveaux d'énergie
  de la théorie considérée (en taille finie). Ce système d'équations
  est équivalent au système infini d'équations, connu sous le nom de
  système Y, qui dans le cas de la dualité AdS/CFT avait été
  conjecturé assez récemment. \\ ~ \\ ~ \\
\textbf{Mots-clé :} {Ans}{atz} de Bethe, Chaînes de spins, Dualité
AdS/CFT, Systèmes intégrables, Théories de champs conformes, Théories
de jauge supersymmétriques

\pagebreak

\renewcommand\listtablename{
  Table des matières}
\listoftables{}
\addcontentsline{lot}{section}{Table des matières}

\chapter*{Remerciements}
\label{cha:remerciements}
\addcontentsline{lot}{section}{Remerciements}

Cette thèse s'est déroulée au LPT de l'ÉNS, sous la direction de
Vladimir Kazakov, qui a su, par
de nombreuses discussions, me communiquer son goût pour la recherche et 
me présenter des problèmes extrêmement intéressants. Que ce soit pour
les collaborations qu'il a initiées, pour les explications qu'il
{\mnothing}'a 
fournies, pour la liberté qu'il {\mnothing}'a toujours donnée, pour les nombreux
voyages au cours de cette thèse, je lui suis extrêmement
reconnaissant pour ce qu'il {\mnothing}'a apporté dans mon travail de thèse,
mais aussi pour l'amitié que nous avons nouée.

Je tiens aussi à remercier Romuald Janik et Gregory Korchemsky qui ont
accepté la lourde tâche d'être rapporteurs, ainsi qu'Arkady Tseytlin,
Konstantin Zarembo et Jean-Bernard Zuber, qui ont bien voulu faire
partie du jury pour cette soutenance.

~

Je tiens aussi à remercier les collaborateurs avec lesquels {\jnothing}'ai eu la
chance de travailler de manière fructueuse sur des sujets variés :
Volodya, tout d'abord, sans lequel je {\nnothing}'aurais probablement jamais
étudié ces sujets, Dima qui a toujours si bien su partager ses
trouvailles comme ses questionnements, mais aussi mes autres
coauteurs Sacha Alexandrov, Nikolay Gromov, Zengo Tsuboi et Anton
Zabrodin, avec lesquels {\jnothing}'ai eu le plaisir de travailler et de me
rendre compte que ma propension naturelle à chercher seul les
réponses à mes questions devait impérativement être complétée par
les discussions fréquentes qui permettent de donner la place
à plusieurs points de vue sur un même problème, d'avoir 
du recul sur
son propre travail et 
de s'enrichir des idées d'autres
chercheurs.

~

J'ai eu la chance de mener à bien cette thèse au sein du Laboratoire
de Physique Théorique de l'ÉNS, laboratoire dont tous les membres ont
été plus accueillants et stimulants les uns que les autres. J'ai eu
la chance de passer presque sept ans au sein de l'ÉNS (comme élève
puis comme thésard et caïman), et ces années
{\mnothing}'apportèrent beaucoup, aussi bien professionnellement
qu'humainement. J'ai tout particulièrement apprécié la vie associative
dense de 
cette école (et de ses alentours), ainsi que la très grande diversité
de parcours des étudiants et la diversité des formations que l'on peut
y suivre. 

Je tiens d'abord à remercier tous ceux qui (tels les
secrétaires et informaticiens par exemple) simplifient tellement la
vie en prenant sur eux de nombreuses préoccupations concrètes. Je
tiens aussi à remercier tous les enseignants, qui {\mnothing}'ont permis de
suivre à l'ÉNS une scolarité alliant une très grande qualité avec une
variété 
extraordinaire de cours. 
Je souhaite aussi remercier les 
enseignants hors du commun que {\jnothing}'ai eu au long de ma scolarité, 
de l'école primaire à la prépa et à l'université. 

Les meilleurs enseignants que {\jnothing}'ai eus étaient bien sûr beaucoup plus que
des formateurs, puisqu'ils {\mnothing}'ont aidé à grandir et {\mnothing}'ont (sup)porté
depuis ma plus tendre enfance, jusqu'à me laisser finalement voler de
mes propres ailes : il s'agit de mes parents qui méritent tout
particulièrement d'être remerciés ici.

~

 Je tiens aussi à remercier les élèves que {\jnothing}'ai côtoyés au long de ma
 scolarité, car la 
réussite scolaire n'est pas purement individuelle, mais résulte de
l'émulation, du travail commun, des discussions avec d'autres
élèves, et d'une atmosphère collective qui permette à la fois le
travail et l'épanouissement personnel.

 Je souhaite aussi remercier les élèves que {\jnothing}'ai eus en faisant mes
 premiers pas dans l'ensei\-gnement, en prépa comme à l'Université puis à
 l'ÉNS. Avoir des élèves attentifs, réactifs et assidus a été un tel
 plaisir, que je fus très heureux d'avoir (je l'espère) contribué à
 leur apprentissage.

~

 Lors de ces années de thèse, {\jnothing}'ai particulièrement apprécié
 l'atmosphère du LPT, ses chercheurs accueillants, ses secrétaires
 efficaces et ses informaticiens (parfois bénévoles) tellement
 dévoués. Ce fut un plaisir de travailler dans le bureau qui fut le
 mien, et je remercie les thésards qui ont eu à me supporter et avec
 lesquels nous avons notamment partagé un engouement commun pour les
 biscuits au beurre, et diverses infusions.

 De nombreux autres amis, connus sur les bancs de l'école ou dans des
 associations ou partis politiques  {\mnothing}'ont aussi permis de traverser
 cette période, et méritent d'être remerciés ici pour leur joie de
 vivre communicative.

~

 Mais bien plus encore, je remercie Céline, qui a été à mes côtés
 quotidiennement, {\mnothing}'a accompagné au cours de ces années et a toujours
 su me signifier son amour par ses nombreuses attentions et tout ce
 qu'elle fit pour moi, se chargeant même de relire cette thèse. Je
 tiens donc à lui témoigner ici mon amour et ma gratitude.

\chapter*{Introduction}
\label{cha:resume}
\addcontentsline{lot}{section}{Introduction}

La physique théorique actuelle est notamment confrontée à deux
défis de taille : les propriétés non-perturbatives de certaines
théories de jauge d'une part, et la description quantique de la
gravité d'autre part. 

Les théories de jauge sont une description quantique de particules en
interaction. Dans ce formalisme, des champs
(comme le champ électromagnétique) sont ``quantifiés'', et leur
quantification fait naturellement apparaître des particules (tels les
photons), comme médiateurs des interactions. Dans ce formalisme quantique,
on ne peut pas considérer qu'une particule ait une trajectoire bien
définie c'est-à-dire qu'on ne peut pas lui associer une position à
chaque instant. Au contraire, pour chaque mesure que l'on effectue, de
nombreuses trajectoires possibles contribuent et leur contributions
s'ajoutent. Parmi les différentes ``trajectoires'' (ou plutôt
devrait-on dire les différentes histoires) qui contribuent, certaines
contiennent des désintégrations de particules, des créations de
nouvelles particules, ou des interactions entre particules. 

Une difficulté de taille de ce formalisme est que chaque processus
physique est décrit par la somme des contributions d'une infinité
d'``histoires'' différentes. La question naturelle qui se pose est de
savoir si cette somme est bien finie. Pour certaines théories, il
existe au moins certains régimes (qualifiés de perturbatifs)  où plus une
histoire fait intervenir de créations ou d'anihilations de particules,
plus sa contribution est faible. Dans ce cas, on peut calculer de
nombreuses propriétés physiques en tronquant la somme pour ne garder
qu'un nombre fini d'événements, faisant intervenir un nombre limité
de créations ou anihilations de particules.  Plus la précision
souhaitée est élevée, plus il faudra considérer de termes.

Pour certains modèles de ce type, par exemple les interactions entre
les quarks (les plus petits constituants connus des noyaux atomiques),
cette approche perturbative permet uniquement de décrire des processus
se déroulant à une énergie suffisante. Il existe des propriétés de ces
théories qui ne sont pas expliquées par cette approche perturbative,
notamment le confinement des quarks à l'intérieur de hadrons, comme
les neutrons et les protons (phénomène qui explique que l'on ne puisse
pas observer un quark ``seul'' mais uniquement des particules
constituées de plusieurs quarks).

~

Un autre défi de taille pour la physique théorique est la description
quantique de la gravité. Le formalisme des théories de jauge
a abouti à d'importants succès, au premier rang desquels se trouve le
modèle standard. Celui-ci décrit toutes les particules observées à ce
jour, et explique trois des quatre interactions fondamentales
(connues) de la nature : l'interaction électromagnétique,
l'interaction ``faible'' (responsable de certaines réactions
nucléaires) et l'interaction ``forte'' (décrivant les interactions
entre les quarks). Seule la gravité n'est pas décrite par ce modèle,
et à ce jour, les théories des cordes sont la façon la plus aboutie de la
décrire de façon quantique. Une différence notoire avec les théories
de jauge mentionnées ci-dessus est que les particules ne sont pas
considérées comme ponctuelles mais 
comme unidimensionnelles (que
l'on peut imaginer comme de petites cordes en mouvement).

~

À la fin du siècle dernier, une dualité a été conjecturée, qui relie
ces deux défis majeurs. Cette dualité fait intervenir d'une part la théorie
de jauge nommée {\SYM}, et caractérisée par de nombreuses
symétries (elle est invariante sous l'effet des transformations
conformes et sous quatre transformations de super-symétrie), 
et d'autre part une théorie des cordes dans
l'espace dix-dimensionel \(AdS_5\times {\Ssph}^5\), produit d'une sphère
par un espace Anti de Sitter. 
Cette dualité
conjecture par exemple que certaines quantités, difficiles à calculer
en
théorie de jauge dans 
le
régime de couplage fort (régime
hautement non-perturbatif),  peuvent être obtenues par un calcul en
théorie de cordes à couplage faible (dans un régime perturbatif de la
théorie des cordes). Inversement, certaines quantités difficiles à
calculer dans le régime de couplage fort de la théorie des cordes
peuvent s'obtenir à partir de calculs perturbatif dans {\SYM}.

Cette dualité, nommée dualité AdS\(_5\)/CFT\(_4\) (ou plus simplement
AdS/CFT) est très puissante car elle relie des calculs perturbatifs
et non-perturbatifs. Elle est cependant difficile à vérifier
concrètement, car il faut explicitement mener à bien un calcul
perturbatif et un calcul non-perturbatif pour vérifier si le résultat
coïncide. Dans cette thèse, nous verrons néanmoins comment calculer
certaines quantités de manière exacte (et non pas perturbative) dans
le cadre de cette dualité et plus généralement dans le cadre de
modèles intégrables. 

~

Comme nous le verrons, l'intégrabilité est un outil puissant permettant
des calculs exacts dans de nombreux modèles jouissant des bonnes
propriétés. Ces modèles, appelés ``modèles intégrables'', sont
notamment caractérisés par un nombre important de charges
conservées. En particulier, nous verrons que des opérateurs nommés
``matrices de transfert'', (ou simplement ``opérateurs {\Top}''),
peuvent être construits pour des chaînes de spins intégrables. Nous
verrons d'ailleurs aussi que des
opérateurs ayant les mêmes propriétés (ils obéissent notamment à
l'équation de 
Hirota) existent pour des théories des champs intégrables, et en
particulier 
pour AdS/CFT. Un des résultats 
de cette thèse est de montrer
que ces opérateurs {\Top} peuvent s'exprimer en termes d'opérateurs
{\Qop} de Baxter, qui sont construits explicitement dans le cas des
chaînes de spins intégrables. Ce résultat s'appuie notamment sur des
arguments combinatoires élémentaires qui seront introduits et motivés.

~

Nous verrons ensuite que de nombreux autres modèles que ces chaînes de
spins sont intégrables, en particulier des théories de champs
bidimensionnelles. Plus précisément, nous verrons que si l'espace est
périodique, mais suffisamment grand, ces modèles peuvent être
exactement résolus grâce à l'{\anz} de Bethe. Une question
intéressante est dès lors de savoir ce qu'il advient pour un espace
plus petit. Cette interrogation sur les corrections de taille finie
trouve notamment une réponse grâce à la méthode d'{\anz} de Bethe
thermodynamique, très utilisée dans cette thèse. Cette méthode est
d'autant plus intéressante qu'elle s'applique notamment à la dualité
AdS/CFT, dont elle permet de calculer exactement le spectre
(c'est-à-dire les niveaux d'énergie de la théorie des cordes, ou les
dimensions des opérateurs de {\SYM}). Nous verrons que
cette méthode d'{\anz} de Bethe thermodynamique donne lieu à des
fonctions {\Top} qui généralisent les opérateurs
des chaînes de spins, et satisfont
l'équation de Hirota.

~

Nous verrons ensuite comment résoudre de manière plus générale
l'équation de Hirota pour certains groupes de (super)-symétrie.  
Cette solution, qui constitue en partie un résultat original de cette
thèse, 
fait intervenir des fonctions {\Qfu} qui
généralisent les opérateurs {\Qop} construits pour les
chaînes de spins. 

{\SnS}'appuyant sur cette solution, un résultat très
important de cette thèse est l'écriture, pour plusieurs modèles
intégrables, d'un système fini d'équations intégrales non linéaires qui
décrit les corrections de tailles finies. Ce résultat 
est
obtenu en trouvant les propriétés analytiques de 
ces fonctions 
  {\Qfu}. Ces propriétés peuvent être obtenues à
partir des équations issues de l'{\anz} de Bethe thermodynamique, ou
elles peuvent être postulées à partir des symétries du modèle et de
considérations physiques. 
Nous verrons comment cette nouvelle méthode peut être appliquée aussi
bien dans le cas du champ chiral principal que pour la dualité
AdS/CFT.
Dans le cas de la dualité AdS/CFT, 
les équations issues de l'{\anz} de Bethe thermodynamique
sont
bien connues et ont été largement étudiées. Nous montrerons qu'une forme
de ces équations permet de prouver les propriétés postulées pour les
fonctions {\Tfu} et {\Qfu}. 
  Réciproquement, ces 
propriétés analytiques
des 
fonctions {\Tfu} et {\Qfu} permettent de démontrer les équations traditionnellement
obtenues à partir de l'{\anz} de Bethe thermodynamique. Cela 
  signifie 
que les conditions d'analyticité sur lesquelles s'appuie notre système
fini d'équations intégrales constitue une nouvelle formulation des
équations d'{\anz} de Bethe thermodynamique, mais qui se ramène
désormais à un nombre fini d'équations.

\chapter*{Résumé détaillé et plan de ce manuscrit}
\label{cha:resume-detaille}
\addcontentsline{lot}{section}{Résumé détaillé}

Ce manuscrit, rédigé essentiellement en anglais, est principalement
divisé en chapitres, mais contient aussi quelques annexes, dont la
lecture n'est pas forcément nécessaire mais donne au lecteur les
éléments pour comprendre les outils utilisés. On pourra noter
l'existence d'un index par mots clés (page \pageref{sec:index}), où
sont aussi indiquées de nombreuses notations utilisées dans ce texte, avec un
renvoi à la page correspondante. Par ailleurs la bibliographie (page
\pageref{sec:biblio}) regroupe les différents articles cités dans ce
manuscrit, par ordre alphabétique. Seuls les articles 
\cite{2010arXiv1007.1770K}, \cite{Gromov:2010km},
\cite{Kazakov:2010iu}, \cite{Gromov:2011cx} et
\cite{Alexandrov:2011aa} sont regroupés a part. Il s'agit des
articles écrits pendant ma thèse, et qui présentent les résultats
exposés dans ce manuscrit.

Le présent manuscrit ne se donne pas uniquement pour objectif de
reproduire le contenu de ces articles, mais aussi
d'introduire les idées ayant mené à ces articles,  
et de justifier de manière détaillée les constructions utilisées. Ce souhait d'écrire une présentation pédagogique des
résultats a parfois abouti à présenter des arguments différents de
ceux donnés dans ces articles, et tous les résultats de ces articles
{\nnothing}'ont pas forcément été reproduits dans ce manuscrit de
façon exhaustive.
Un lecteur souhaitant aller un peu plus loin est donc invité à lire aussi ces
articles en complément du présent manuscrit. 

La structure de ce manuscrit et la suivante :
\paragraph{Chapitre introductif}
\label{sec:chapitre-introductif}

Le premier chapitre de ce manuscrit introduit la notion
\linebreak d'intégrabilité et l'{\anz} de Bethe à partir de l'exemple simple de la chaîne de
spins de Heisenberg, avec conditions de bord périodiques. Nous verrons
dans ce chapitre comment il est 
possible, pour cette chaîne de spins, de déterminer les
états propres du Hamiltonien. Nous verrons que si l'on cherche des
états propres sous la forme de  combinaisons linéaires d'ondes de
spins, alors des équations appelées ``équations de Bethe'' apparaissent
naturellement, et décrivent à quelle condition une combinaison
linéaire d'ondes planes est un état propre du Hamiltonien.

Cet exemple de chaîne de spins,
étudié dans la section \ref{sec:integr-et-},
 permettra 
de montrer ce que l'on entend dans cette thèse par système
intégrable, à savoir un système où les fonctions d'onde des états
propres sont des superpositions d'ondes planes obéissant à des équations
de Bethe. Ces modèles sont exactement solubles, non pas au sens où
l'on connaît une expression complètement explicite des vecteurs
propres et de leurs énergies, mais où l'on sait les construire de
manière exacte si l'on résout une équation (l'équation de Bethe), dont
la forme est la même pour tous les modèles intégrables, mais qui est
en général difficile à résoudre analytiquement. 
Cette équation fait apparaître une matrice {\Smat}, qui dépend du
modèle et caractérise l'interaction de deux particules. Nous verrons
que dans ces modèles intégrables les interactions entre un nombre arbitraire de
particules s'expriment  en termes d'interactions entre deux particules.

Comme indiqué en section \ref{sec:gener-other-integr}, il existe des
théories de champs quantiques,
généralement bidimensionnelles (avec
une dimension spatiale et une dimension temporelle),
 qui sont intégrables au sens
ci-dessus. En revanche, nous
verrons aussi que ces théories ne sont intégrables que si la dimension
spatiale est de grande taille (et périodique). 
Nous nous intéresserons ici aux  corrections de taille finie, pour lesquelles
les niveaux d'énergie peuvent aussi être calculés de manière exacte,
grâce à la méthode dite d'{\anz} de Bethe thermodynamique (présentée dans
le chapitre \ref{cha:ansatzs-de-bethe}).

Enfin la section \ref{sec:dualite-adscft} introduira brièvement la
dualité AdS/CFT, et indiquera pourquoi elle constitue un modèle
intégrable. L'étude des corrections de taille finie de ce modèle fera
l'objet du chapitre \ref{cha:dualite-adscft}.

\paragraph{Chapitre \ref{part:qoperatorsspin}}
  \label{sec:chapter-ref-}

Le chapitre \ref{part:qoperatorsspin} propose une analyse beaucoup plus détaillée d'une
première classe de modèles intégrables, à savoir certaines chaînes de 
spins généralisant la chaîne de spins de Heisenberg étudiée dans le
chapitre introductif \ref{sec:integr-et-}. Ce chapitre \ref{part:qoperatorsspin} montrera
comment obtenir les équations de Bethe et 
le spectre du Hamiltonien pour ces chaînes de spins intégrables, en
introduisant des charges conservées (les opérateurs {\Top} et les
opérateurs {\Qop}) et en définissant un ``flot de Bäcklund''. 

Ces chaînes de spins seront introduites dans la section
\ref{sec:chaines-de-spin}.
Nous y construirons de nombreuses charges
conservées appelées ``opérateurs {\Top}'',
en faisant appel à
des notions de
théorie des groupes et des représentations introduites dans l'annexe
\ref{sec:elements-de-theorie}.  Ces opérateurs {\Top} sont
construits en section \ref{sec:operateurs-t}, puis sont exprimés en
termes d'un opérateur différentiel \(\hD\), dont de nombreuses
propriétés sont données dans l'annexe \ref{app:cod}. Cela permet de
montrer en section \ref{sec:hirota-equat-cher} que ces opérateurs
{\Top} satisfont une équation bilinéaire appelée équation de
Hirota. La preuve de cette équation est donnée en suivant
\cite{2008JHEP...10..050KV}, et repose sur les propriétés
combinatoires de l'opérateur \(\hD\).
 Nous montrerons ensuite comment cette 
équation peut se réécrire comme une identité combinatoire,
 obtenue
au cours de cette thèse, que nous
avons appelée ``Master identity'' dans l'article
\cite{Kazakov:2010iu}, et que nous appellerons ``{\MID}'' dans ce
manuscrit. 

La section \ref{sec:transf-de-backl} 
présentera 
des résultats connus sur les transformées de Bäcklund. Cette section montrera comment exprimer une
solution assez générale de l'équation de Hirota en termes de
``fonctions {\Qfu}''. 
Nous y supposerons
l'existence d'un flot de
Bäcklund polynômial et montrerons que sous cette hypothèse, les
équations de Bethe découlent de conditions d'analyticité. 

La section \ref{sec:expr-diff-Qop} présente des résultats obtenus au
cours de
cette thèse \cite{Kazakov:2010iu}. Il y est démontré explicitement,
pour toutes les chaînes de spins définies en section
\ref{sec:chaines-de-spin} qu'un flot de Bäcklund polynômial
existe. Cette section permet aussi d'écrire explicitement les
opérateurs {\Top} en termes des opérateurs {\Qop}, qui sont eux-mêmes
définis à partir de l'opérateur différentiel \(\hD\). Ils peuvent aussi
être vus comme une certaine limite 
des opérateurs {\Top}.

Enfin la section \ref{sec:quantum-classical} présente un autre
résultat obtenu pendant cette thèse \cite{Alexandrov:2011aa}, à savoir
le lien entre cette construction du flot de Bäcklund et les
constructions de ``fonctions \(\tau\)'' qui décrivent l'intégrabilité
classique.

\paragraph{Chapitre \ref{cha:ansatzs-de-bethe}}
\label{sec:chapitre}

Le chapitre \ref{cha:ansatzs-de-bethe} se concentre sur les
corrections de taille finie : il décrit la résolution de 
 théories
de champs intégrables avec une
dimension spatiale 
de taille  finie et des conditions au bord périodiques. La méthode de résolution, présentée en section
\ref{sec:example-princ-chir} dans le cas du modèle chiral
principal, est l'{\anz} de Bethe thermodynamique, décrit dans la
littérature pour de nombreux modèles. Elle aboutit en général à un
système Y, qui donne lieu à la même équation de Hirota que dans le
chapitre précédent. C'est pourquoi l'on a besoin de décrire de la
façon la plus générale possible les solutions de l'équation de Hirota.

La section \ref{sec:solution-generale-de} décrit donc des solutions
assez générales de l'équation de Hirota, qui prennent exactement la même forme
que les solutions trouvées au chapitre \ref{part:qoperatorsspin}. Une
condition est identifiée (la condition de typicalité) sous laquelle
on peut écrire les fonctions {\Tfu} à partir d'un nombre fini de
fonctions {\Qfu}. Cette expression en termes de fonctions {\Qfu} était
déjà connue dans certains cas
\cite{springerlink:10.1007/s002200050165}, mais constitue pour partie
(pour les ``\(\Tk\)-{\hook}s'') un résultat obtenu dans cette thèse
\cite{Gromov:2010km}. 
  
Enfin la section \ref{sec:application-au-champ} introduit un important
résultat \cite{2010arXiv1007.1770K} 
de cette thèse, à savoir la
possibilité d'écrire le système Y sous la forme d'un nombre fini
d'équations portant sur un nombre fini de densités. De plus ces
équations expriment simplement des conditions d'analyticité assez
naturelles. 

\paragraph{Chapitre \ref{cha:dualite-adscft}}
\label{sec:chapitre-}
Enfin le chapitre \ref{cha:dualite-adscft} conclut ce manuscrit en
s'attaquant à la dualité AdS/CFT. Cette dualité suscite
un fort intérêt scientifique 
du fait des espoirs qu'elle engendre notamment pour
comprendre des théories de champs à un niveau non-perturbatif. 

La méthode d'{\anz} de Bethe thermodynamique a déjà permis  d'obtenir
un système Y pour les niveaux d'énergie de cette dualité. Dans ce chapitre nous montrerons
comment écrire des conditions d'analyticité naturelles sur les
fonctions {\Tfu}, qui sont équivalentes au système Y déjà connu. Ces
résultats \cite{Gromov:2011cx} éclairent singulièrement la nature de ce système Y, en
trouvant de nouvelles symétries qu'il satisfait et qui s'interprètent
physiquement. 

Suivant la même méthode que dans le cas du champs chiral principal,
nous commençons par exprimer
en section \ref{sec:parameterization-qfs-3}
 la solution générale de l'équation de
Hirota en termes de trois fonctions réelles. Les principales équations
et les nouvelles symétries qui caractérisent ces fonctions {\Tfu}
sont ensuite présentées en section \ref{sec:set-equations-1}, avant de
présenter nos résultats numériques.

Enfin une conclusion page \pageref{app:ccl} discute les apports de cette approche et les
questions soulevées par cette thèse.

%%% Local Variables: 
%%% mode: latex
%%% eval: (ispell-change-dictionary "francais") ***
%%% eval: (find-file "english.tex") ***
%%% TeX-master: "english.tex"
%%% End: 

%% file: prelim.tex
\addcontentsline{lot}{part}{Manuscrit (en anglais)}

\chapter*{Abstract}
     This {\thesis} is devoted to the study of {\ing} quantum systems
     such as {\csds}, two-dimensional field theories and the AdS/CFT
     duality.  This AdS/CFT duality is a conjecture, stated in the end
     of the last century, which relates (for instance) the
     non-perturbative regime of a superconformal gauge theory (called
     \(\mathcal{{\NN}}\)=4 {\SYM}) and the perturbative regime of a string
     theory on a 10-dimensional space with the geometry AdS₅\(\times\){\Ssph}⁵. 

  This {\thesis} explores the similarities between {\ing} {\csds}
  and quantum field theories, such as Super Yang Mills. We first study
  {\ing} {\csds} and build explicitly a polynomial ``Bäcklund
  flow'' and polynomial ``{\Qoprs}'', which allow to diagonalize the
  {\Ham}. We then study {\ing} field theories et show how to
  obtain ``{\Qfs}'', analogous to the {\Qoprs} built for spin
  chains. It turns out that several important informations are
  contained in the analytic properties of these {\Qfs}. That
  allows to obtain, in the framework of the thermodynamic Bethe
  {\anz}, a finite number of non-linear integral equations encoding
  the spectrum of the theory which we study. This system of equations
  is equivalent to an infinite system of equations, known as
  ``{\Ysys}'', which had been quite recently conjectured in the case of the
  AdS/CFT duality.

\chapter*{Acknowledgements}
\addcontentsline{toc}{section}{Acknowledgements}

I want to thank very deeply my {\PhD} advisor V. Kazakov for the
way he managed to introduce me to some very interesting problems and
the way he guided me through this couple of years of research. 

I also thank Romuald Janik and Gregory Korchemsky 
who accepted to be the \emph{rapporteurs} for this {\thesis}, and 
Arkady Tseytlin, Konstantin Zarembo and Jean-Bernard Zuber, 
for taking part in the jury. 

I also want to thank my coauthors A. Alexandrov, {\NN}. Gromov,V. Kazakov, 
 Z. Tsuboi, {\Dnothing}. Volin and A. Zabrodin, for the work we did together
during this {\PhD}. I also wish to thank all the AdS/CFT community for
many discussions  (for instance with I. Kostov, {\Tnothing}. {\L}ukowski, {\Dnothing}. Serban,
{\Mnothing}. Staudacher, {\Pnothing}. Vieira and many others).
I also want to thank A. Bobenko,
{\Lnothing}. Fevrier, and W. Krauth for fruitful discussions.

 \pagenumbering{arabic}
\renewcommand\contentsname{Contents}
\tableofcontents
 \addtocounter{page}{-1}
\addcontentsline{toc}{section}{Contents}%
\addcontentsline{lot}{section}{Contents}%
 \addtocounter{page}{1}

\chapter*{Introduction}
\addcontentsline{toc}{section}{Introduction}

Theoretical quantum physics currently faces two very important
challenges: the non perturbative understanding of gauge theories on
the one hand, and the quantum description of gravity on the other hand.

Gauge theories are a quantum description of interacting particles. In
this formalism, some fields (such as the electromagnetic field) are
``quantized'', and their quantization gives rise to particles (such
as the photons), which transmit interactions. In this quantum
formalism, one {\cannnot} say that a particle has a well-defined
trajectory, in the sense that it does not a have a well-defined
position at every time. Instead, in every process which we can
measure, many possible trajectories do contribute, and one should sum
their contributions. In fact, one should even sum over different
histories (which generalize the idea of ``trajectories''), including
some histories which involve creations or annihilations of particles.

We see that this formalism describes the interaction of quantum particles by
summing infinitely many different ``histories''. A natural question
is then whether this sum is finite or not. For some of these theories
(called asymptotically free), there exist at least some regimes
(called perturbative regimes), where an history contributes less and
less as it contains more and more creations or annihilations of
particles. In this regime, it is possible to compute some physical
quantities by truncating the sum, keeping only a finite number of
terms, which involve only a limited number of creations or
annihilations of particles. The better accuracy we wish to obtain, the
more terms we have to keep in the sum.

For some of these models, such as the interactions between quarks (the
smallest known components of atomic nuclei), this perturbative
approach only allows to describes processes which take place at an
high enough energy. There are some properties of these theories which
are not explained by this perturbative approach, such as the
confinement of quarks inside hadrons (such as neutrons and
protons). This confinement is a property of quarks which means that it
is not possible to observe one single quark, and which only allows to
observe particles made of several quarks.

~

Another key challenge of theoretical physics is the quantum
description of gravity. The formalism of gauge theories has already led
to important successes, such as the construction of the ``standard
model''. This model describes all particles that we have observed so
far, and it explains three out of the four (known) fundamental
interactions of 
nature: the electromagnetic interaction, the ``weak'' interaction
(involved in some nuclear reactions) and the ``strong'' interaction
(which describes the interactions between quarks). But gravity does
not fit this picture, and up to now our most successful description of
gravity at the quantum {\level} is given by ``string theories''. A key
difference between gauge theories and string theories is that the
particles are not viewed as point-like, but as one-dimensional
extended objects (which can be viewed as small strings).

~

In the end of the last century, a duality was conjectured, which
connects these two key challenges. On the one hand, this duality
involves the gauge theory called {\SYM}, which exhibits many
symmetries (it is invariant under conformal transformations, and under
four super-symmetric transformations). On the other hand, it involves
a string theory on the ten-dimensional spacetime \(AdS_5\times
{\Ssph}^5\), 
made out of a sphere and an ``Anti de Sitter'' space.
This duality conjectures that for instance, some quantities which are
hard to compute in the strong coupling regime of the gauge theory (in
the non-perturbative regime), can be obtained by a perturbative
computation in the string theory. Conversely, properties of the (non-perturbative)
strong coupling  regime of the string theory can be obtained from the
perturbative regime of the {\SYM} gauge theory.

This duality, called  AdS\(_5\)/CFT\(_4\) (or simply AdS/CFT) is very
powerful because it relates perturbative and non-perturbative
regimes. But this property also makes it very hard to test this
duality explicitly, because in order to check it, we should perform
independently a perturbative and a non-perturbative computation to
compare them. In this {\thesis}, we will see that some quantities can be
computed exactly (in the sense that the computation will not involve
the truncation of a sum, and that it will not be restricted to a
perturbative regime), in the framework of this duality and, more
generally, in the framework of {\ing} systems.

~

As we will see, integrability is a powerful tool which allows to
perform exact computations in several models exhibiting quite specific
features. These models, called ``{\ing} models'', have for
instance an important number of conserved charges. In particular we
will see how to construct, for {\ing} {\csds}, some {\ops}
called ``transfer matrices'' (or simply ``{\Toprs}''). We will also
see that some {\ops} having the same properties (they obey the same
Hirota equation) exist for {\ing} field theories, and in
particular in the case of AdS/CFT.
One of the results of this {\thesis} is that these {\Toprs} can be written
through ``{\Qoprs}'', which we construct explicitly in the case of
{\ing} {\csds}. This result involves elementary combinatorial arguments
which will be introduced and motivated.

~

Next, we will see that several other models than {\csds} (in
particular some two-dimensional field theories) are
{\ing}. More
precisely, we will see that if the space is periodic, but large
enough, these models can be solved exactly by means of the ``Bethe
{\anz}''. Hence we will come to the question of finite size effects,
{\idest} the question whether anything can still be obtained when the space
is smaller. In this {\thesis}, we will answer this question by means of
the {\TBA}, a method which is all the more interesting as it applies
for instance to the AdS/CFT duality. In the framework of this duality,
this method allows to compute exactly ({\idest} non-perturbatively) the
spectrum ({\idest} the energy levels of the string theory, or the scaling
dimensions of the {\ops} of {\SYM}). We will see that this {\TBA}
gives rise to ``{\Tfs}'' which generalize the {\Toprs} of {\csds},
and obey the same Hirota equation.

~

Then, we will construct the general\footnote{More exactly, the
  solution which we will construct is the ``typical solution'' of
  Hirota equation, and we will define what typicality means.} solution
of Hirota equation for several symmetry groups. This result, part of
which is a new result of this {\PhD}, involves ``{\Qfs}'' which
generalize the ``{\Qoprs}'' constructed explicitly for {\csds}.

Using this solution, a very important result of this {\thesis} is the
possibility to write, for several {\ing} field theories, a finite
set of non-linear integral equations (FiNLIE), encoding the finite
size corrections. This FiNLIE is obtained by finding the analytical
properties of these {\Qfs}, and we will see that these analytical
properties can either be conjectured from the symmetries of the model
and from physical considerations, or they can be derived from the {\TBA}.
We will see how this new approach can be applied in the case of the
{\PCM} as well as in the case of the AdS/CFT duality.

In the case of the AdS/CFT duality, the equations arising from the
{\TBA} are well known and have already been quite extensively studied
in the literature. We will see that one form of these equations allows
to prove the analytical properties conjectured for the {\Tfu}- and
{\Qfs}. Conversely these analytical properties of the {\Tfu}- and
{\Qfs} imply the equations which are usually obtained from the {\TBA}.
This means that the analytical conditions giving rise to our FiNLIE
are a new formulation of the {\TBA}'s equations, reduced
to a finite set of equations.

\chapter*{Detailed summary}
\addcontentsline{toc}{section}{Detailed summary}

This manuscript is mainly divided into chapters, but it also contains
two appendices. It is not necessary to read through these appendices,
but it should give some basic tools to the reader, which will allow to
understand the arguments of the main text. One should note the
existence of an index (page \pageref{sec:index}), which lists several
notations and keywords used in the text and which refers to the
corresponding pages. There is also a bibliography, (page
\pageref{sec:biblio}), which lists, in alphabetic order, the articles
cited in this text. Only the articles 
\cite{2010arXiv1007.1770K}, \cite{Gromov:2010km},
\cite{Kazakov:2010iu}, \cite{Gromov:2011cx} and
\cite{Alexandrov:2011aa} are listed separately. They are the articles
written during my {\thesis}, and they give the results written in this
{\thesis}.

The present {\thesis} aims not only at repeating the content of these
articles, but it also aims at motivating the constructions we used,
with a quite high {\level} of details. This aim to write a pedagogical
presentation of these articles sometimes led to showing arguments which
differ from the one exposed in these articles, and all the results of
these articles were not necessarily repeated in this
{\thesis}. Therefore, a reader who wishes to go further than the present
manuscript is invited to read these articles in addition to this {\thesis}.

The structure of this {\thesis} is as follows :
\paragraph{Introductory chapter}

The first chapter of this {\thesis} introduces the notion of
integrability and of Bethe {\anz} from the simple example of the
Heisenberg {\cds} with periodic boundary conditions. In this
chapter we will see how it is possible to find the eigenstates of the
{\Ham} of this {\cds}. We will see that these eigenstates can
be found in the form of linear combinations of planar waves obeying
some equation called the ``Bethe equations'' (these equations are the
conditions under which a combination of planar waves is an eigenstate
of the {\Ham}).

This {\cds} example, studied in section \ref{sec:integr-et-}, will
be used to introduce what we will mean  by an {\ing} system, {\idest} a
system where the wave functions of the eigenstates are given by
superpositions of planar waves obeying some Bethe equations. These
models are exactly solvable, not in the sense that we know a
completely explicit expression of the eigenstates and of their energy,
but rather in the sense that we know how to construct them if we solve
an equation (the Bethe equation), which takes the same form for all
{\ing} models but is in general quite hard to solve analytically.
This equation involves an {\Sma}, which depends on the model and
characterizes the interaction between two particles. We will see that
for these {\ing} models, the interactions between an arbitrary
number of particles can be expressed in terms of successive
interactions between two particles.

As we will see in section \ref{sec:gener-other-integr}, there exist
quantum field theories which are {\ing} in the above sense. These
models are usually two-dimensional (with one
space dimension and one time dimension), and we will see that they are
{\ing} only if the space dimension has a very large size (and is
periodic).
In this {\thesis}, we will be interested in the finite size corrections,
which means the exact ({\idest} non-perturbative) computation of the energy
levels when the size of the space dimension is finite, and we will
study them by means of a method called the {\TBA} (explained in the
chapter \ref{cha:ansatzs-de-bethe}).

Then the section \ref{sec:dualite-adscft} will briefly introduce the
AdS/CFT duality, and explain why it is an {\ing} model. In the
framework of this duality, the finite size corrections will be studied
in the chapter \ref{cha:dualite-adscft}.

\paragraph{Chapter \ref{part:qoperatorsspin}}

The chapter \ref{part:qoperatorsspin} gives a more detailed analysis
of a first {\ing} model, namely a {\cds}, generalizing the
analysis of the Heisenberg {\cds} presented in the introductory
chapter \ref{sec:integr-et-}. This chapter \ref{part:qoperatorsspin}
will show one derivation of the Bethe equations for these {\ing}
{\csds}, obtained by introducing some conserved charges (the
``{\Toprs}'' and the ``{\Qoprs}'') and by defining a ``Bäcklund
flow''.

These {\csds} will be introduced in section
\ref{sec:chaines-de-spin}. In this section, we will construct many
conserved charges called ``{\Toprs}'', using some notions of group
theory and representations, introduced in the appendix
\ref{sec:elements-de-theorie}. These {\Toprs} are constructed in
section \ref{sec:operateurs-t}, and are then expressed through a
differential {\op} \(\hD\). Many important properties of this
{\op} are given in the appendix \ref{app:cod}. This construction
allows to show in section \ref{sec:hirota-equat-cher} that these
{\Toprs} obey a bilinear equation called the Hirota equation. The
proof of this equation is given by following
\cite{2008JHEP...10..050KV}, and it relies on the combinatorial
properties of the {\op} \(\hD\). Next we will see how this equation
can be rewritten as a combinatorial identity obtained in this {\PhD},
which we called ``Master identity'' in the article
\cite{Kazakov:2010iu}, and which we will call ``{\MID}'' in the
present {\thesis}.

The section \ref{sec:transf-de-backl} will then give known results on
the Bäcklund transforms.  This section will show how to express a
quite general solution of the Hirota equation in terms of a finite number
of ``{\Qfs}''. In this section, we will assume that a polynomial
Bäcklund flow exists, and we will show that then, the
Bethe equations follow from the analyticity constraints ({\idest} from the polynomiality).

Next, the section \ref{sec:expr-diff-Qop} gives results obtained in
this {\PhD}. It is explicitly derived, for all the {\csds}
introduced in section  \ref{sec:chaines-de-spin}, that a polynomial
Bäcklund flow exists. This section also allows to explicitly write the
{\Toprs} in terms of some {\Qoprs}, which are themselves defined
through the differential {\op} \(\hD\). These {\Qoprs} can also be
viewed as a quite singular limit of the {\Toprs}.

Finally the section \ref{sec:quantum-classical} sketches {\another} result
of this {\PhD} \cite{Alexandrov:2011aa}, namely the relation between
this construction of the Bäcklund flow and the construction of the
``{\taufs}'' which describe classical integrability.

\paragraph{Chapter \ref{cha:ansatzs-de-bethe}}

The chapter \ref{cha:ansatzs-de-bethe} is focussed on the finite size
corrections: it explains the solution of {\ing} field theories
with a space dimension of size \(\LF<\infty\) and with periodic boundary
conditions. This solution is based on the method called {\TBA},
presented in section \ref{sec:example-princ-chir} for the principal
chiral model. This method is described in the literature for several
{\ing} models, and it usually gives rise to a {\Ysys}, which
implies the same Hirota equation as in the previous chapter. Therefore
we become interested in writing the most general possible solution of
this Hirota equation.

The section \ref{sec:solution-generale-de} describes quite general
solutions of the Hirota equation, and these solutions take exactly the
same form as the solutions found in chapter
\ref{part:qoperatorsspin}. A condition is identified (the typicality
condition) under which the {\Tfs} are expressed in terms of a finite
number of {\Qfs}. This expression was already known in some cases
\cite{springerlink:10.1007/s002200050165}, but part of it (for the
``\(\Tk\)-{\hook}s'') is a new result of this {\PhD}.

Finally, the section \ref{sec:application-au-champ} presents a very
important result of this thesis \cite{2010arXiv1007.1770K}: it is
shown that the {\Ysys} can be recast into a finite set of equations on
a finite set of ``densities''. Moreover these equations simply express
some quite natural analyticity conditions.

\paragraph{Chapter \ref{cha:dualite-adscft}}

 To finish with, the chapter \ref{cha:dualite-adscft} concludes this
 manuscript with the case of the AdS/CFT duality. This duality is 
 a very active field of research because  it gives (for
 instance) very interesting hopes for a non-perturbative
 understanding of field theories.

The  {\TBA} already allowed to obtain a {\Ysys} (described in the
literature \cite{2009PhRvL.103m1601G}) which gives the energy levels
in this AdS/CFT duality. In this chapter, we will show how to write
some
natural analyticity conditions on the {\Tfs}, which are equivalent to
the previously-known {\Ysys}. This result sheds light on the nature of
the {\Ysys}, by finding new symmetries which it obeys, and which have
a very physical interpretation.

Following the same method as in the case of the {\PCM}, we start by
parameterizing in section \ref{sec:parameterization-qfs-3} a general
solution of Hirota equation in terms of three real functions. The main
equations and symmetries which constrain these {\Tfs} are then given
in section \ref{sec:set-equations-1}, where the FiNLIE is derived,
before we show our {\NuMr} results.

Finally, a conclusion (page \pageref{app:ccl}) discusses how our
approach solves the initial problem, and what new questions arise from this.

%%% Local Variables: 
%%% mode: latex
%%% TeX-master: "english.tex"
%%% End: 

%% file: intro.tex
\index{Bethe ansatz|(}

\section{Coordinate Bethe {\anz} for the Heisenberg {\cds}}
\label{sec:integr-et-}

As a first introduction to this manuscript, let us briefly recall the
solution to the so-called Heisenberg ``XXX\(_{1/2}\)'' {\cds}, which
corresponds to a
quantum version of the Ising model.

This chain consists of \(\lcds\) spins, {\lbd} by \(\spi\in
\ninter 1 \lcds\), where we use the notation \(\ninter {{\nn}_1}
{{\nn}_2}\) to denote 
the set \([{\nn}_1,{\nn}_2]\cap ℤ\).
\index{00@\(\ninter {{\nn}_1} {{\nn}_2}\equiv [{\nn}_1,{\nn}_2]\cap ℤ~~\)}
Each spin is in a superposition 
\(\ket{\psi}\in \Hilbl_\spi=ℂ^2\)
of the states \(\ket{↑}\) and \(\ket{↓}\).
The Hilbert space is therefore \(\Hilb= 
{\bigotimes_{\spi=1}^\lcds} \Hilbl_\spi=\left(ℂ^2\right)^{\otimes
  \lcds}\), while 
the {\Ham} is 
\begin{align}
\label{eq:hamilHeisenbergintro}
 \framedline{ \Hami=}{\lcds - 2
\sum_{\spi}
  \perm_{\spi,\spi+1}}\,\,.
\end{align}
It is expressed in terms of a permutation {\op}  \(\perm_{\spi,\spj}\) defined by
\index{Pa (permutation {\op})@\perm (permutation {\op})}
\begin{gather}
\label{eq:DefPerm}  \perm_{\spi,\spj} 
  \left(\ket{\phi_1}\otimes\ket{\phi_2}\otimes\cdots\otimes\ket{\phi_\lcds}\right)=\ket{\phi_{\smash{\tau_{[\spi,\spj]}(1)}}}\otimes\ket{\phi_{\smash{\tau_{[\spi,\spj]}(2)}}}\otimes\cdots\otimes\ket{\phi_{\smash{\tau_{[\spi,\spj]}(\lcds)}}}\\
\where \tau_{[\spi,\spj]}(\spk)=\left\{
  \begin{array}{ll}
\spj&\If \spk=\spi\\
\spi&\If \spk=\spj\\
\spk& \mathrm{otherwise}
\end{array}\right.\,.
\label{eq:DefTAUij0}
\end{gather}
\index{tauij (transposition)@\ensuremath{\tau_{[\spi,\spj]}} (transposition)}
This {\op} exchanges the values of two spins, giving for instance
\(\perm_{1,3}\ket{↓↓↑}=\ket{↑↓↓}\).

More precisely, we will study a {\cds} with periodic boundary
conditions, which means that the {\Ham}
\eqref{eq:hamilHeisenbergintro} is defined as 
\begin{gather}
   \Hami=\lcds - 2
\sum_{\spi=1}^{\lcds-1}
  \perm_{\spi,\spi+1}
-2 \,\perm_{\lcds,1}
\,\,.
\end{gather}

The expression \eqref{eq:hamilHeisenbergintro} may seem 
unusual,
but we will
actually show in section \ref{sec:chaine-de-spin} that it coincides
with the usual ferromagnetic {\Ham} \(\Hami= -\sum_{\spi}
     \vec{\sigma_\spi}\cdot\vec{\sigma_{\spi+1}}\).

The simplest eigenstate of the {\Ham} \eqref{eq:hamilHeisenbergintro} is
the following state, which we will call the vacuum:
\begin{align}
\label{eq:DefVac}
  \vac ~ \equiv&\sket{\underbrace{↓↓↓\cdots↓}_\lcds}\,.
\end{align}
It is clearly an eigenstate, and its energy is \(E_0\equiv-\lcds\) ({\ie} 
it satisfies \({\Hami \vac = - \lcds \vac}\)).
It is  an arbitrary convention to choose a state with all spins down,
while the opposite convention (choosing \(\vac\) with all spins up)
would give the same results.

\paragraph{Single-particle states}
\label{sec:single-part-stat}

The next simplest states one can think of are the states
\begin{align}
\label{eq:1magnon}
  \iket{\spi}\equiv&\sket{\underbrace{↓↓↓\cdots↓}_{\spi-1}↑ \underbrace{↓↓↓\cdots↓}_{\lcds-\spi}}\,.
\end{align}
These states will be viewed as the presence of a ``particle'', called
magnon, \index{Magnon} at site \(\spi\). This magnon physically is 
just
a flip
of one spin 
with respect
to the vacuum \(\vac\). More generally we will
call ``number of particles'' the number of ``spins up'', and this
number turns out to be invariant under \(\Hami\). Therefore
we can look for a basis of eigenstates of \(\Hami\) having
fixed ``number of particles''. The states \eqref{eq:1magnon}
are actually  not eigenstates of \(\Hami\) if
\(\lcds\geq2\) but we will show that some combinations of them are
eigenstates.

To this end, let us recall that the permutation
{\op} \(\perm_{\spi,\spj}\) exchanges the spins at position \(\spi\)
and \(\spj\), hence 
\begin{align}
  \perm_{\spi,\spj}\iket{\spk}=&\iket{\tau_{[\spi,\spj]}(\spk)}\,
\end{align}
where \(\tau_{[\spi,\spj]}\) is defined by \eqref{eq:DefTAUij0}. We can then write
the action of \(\Hami\) on the state
\begin{equation}
  \pket{p}\equiv\sum_{\spj=1}^\lcds e^{\bi p
    \spj}\iket{\spj}\,:
\label{eq:1-part-{\anz}}
\end{equation}
\begin{align}
\Hami \pket{p}=&\lcds \pket{p}-2 \sum_{\spj=1}^{\lcds}\sum_{\spi=1}^{\lcds} 
e^{\bi p  \spj}~ \perm_{\spi,\spi+1} \iket{\spj}
\label{eq:then-hamil-ketp=}
\\=&\lcds \pket{p}-2\left((\lcds-2)\pket{p}+\sum_{\spj=1}^{\lcds} 
e^{\bi p  \spj}\left(\iket{\spj+1}+\iket{\spj-1}\right)\right)
\label{eq:=hui}
\\=&(-\lcds+4-4 \cos(p))\pket{p} - 2\left(e^{\bi \lcds p}-1\right)
\iket{1} 
-2\left(e^{\bi p}-e^{\bi(\lcds+1)
    p}\right)\iket{\lcds}\,,
\label{eq:-left.-}
\end{align}
where ⅈ denotes the imaginary number with imaginary part
equal to one,
\index{ia@\(ⅈ~\equiv~\sqrt{-1}~~\)}
and we identified \(\iket{\lcds+1}=\iket{1}\) in \eqref{eq:=hui}.
From \eqref{eq:-left.-}, we see that \(\pket{p}\) is an (unnormalized)
eigenstate of the {\Ham}
if and only if \(e^{\bi \lcds p}=1\). This condition is simply the
constraint that  
the wave function 
\(\Psi(\spj)=\ibrapket{\spj}{p} = e^{\bi p  \spj}\) has to be periodic with
period \(\lcds\), as imposed by the identification \(\iket{\lcds+1}=\iket{1}\).

This already allowed us to identify  \(\lcds\) ``single-particle''
eigenstates (corresponding to \(p=0,\frac{2 \pi}{\lcds},\frac{4
  \pi}{\lcds}, \cdots,\frac{2(\lcds-1) \pi}{\lcds}\)).
They have energy \((E_0+4-4\cos(p))\), with \(e^{\bi \lcds p}=1\).

\paragraph{Two-particles states}
\label{sec:two-part-stat}

Next, one can consider the following ``two-particles'' states:
\begin{gather}
\label{eq:2magnons}
    \iket{\spj,\spk}\equiv\sket{\underbrace{↓↓↓\cdots↓}_{\spj-1}↑
      \underbrace{↓↓↓\cdots↓}_{\spk-\spj-1}↑\underbrace{↓↓↓\cdots↓}_{\lcds-\spk}}\,,\hspace{3cm}\where
    \spj<\spk\\
\label{eq:2-part-{\anz}}
\pket{p_1,p_2;\Sscal}\equiv\sum_{\spj<\spk} \left(e^{\bi (p_1 \spj + p_2
  \spk)}+\Sscal~ e^{\bi (p_1 \spk + p_2 \spj)}\right) \iket{\spj,\spk}\,.
\end{gather}
The action of \(\Hami\) is a bit harder to compute explicitly than in
(\ref{eq:then-hamil-ketp=}-\ref{eq:-left.-}), but we can 
see
that 
\begin{multline}
  \Hami \iket{\spi,\spj}=\lcds \iket{\spi,\spj} -2\left(
     (\lcds-4)\iket{\spi,\spj}
     +\iket{\spi+1,\spj}+\iket{\spi-1,\spj}
+\iket{\spi,\spj+1}
\rule{0pt}{3.7mm}
  \right.
  \\
  \left.
  \rule{0pt}{3.7mm}
 +\iket{\spi,\spj-1}\right)
 \hspace{2cm}\If 1<\spi<\spj-1<\lcds-1\,.
 \label{eq:-left.r-+ketspi}
\end{multline}
If we do the natural identifications
\(\iket{\spi,\lcds+1}=\iket{1,\spi}\) and
\(\iket{0,\spi}=\iket{\spi,\lcds}\), then 
this equation 
holds
even if \(\spj=\lcds\) or \(\spi=1\). 

From \eqref{eq:-left.r-+ketspi} we can 
see that up to ``boundary''
terms generalizing the last terms of \eqref{eq:-left.-}, 
we
have \(\Hami \pket{p_1,p_2;\Sscal} = (E_0+8-4(\cos(p_1) 
{+}\cos(p_2))) \pket
{p_1,p_2;\Sscal}\), and we therefore expect that by setting these ``boundary''
terms to zero, we will find eigenstates with
energy \(E_{p_1,p_2}\equiv E_0+4-4\cos(p_1)+4-4\cos(p_2)\). 
The boundary terms, which 
are given by
\(\Hami \pket{p_1,p_2;\Sscal} -E_{p_1,p_2} \pket{p_1,p_2;\Sscal}\),
are of two types: 
\begin{itemize}
\item First, the terms {\below} arise from the fact that \eqref{eq:-left.r-+ketspi}
  fails if \(\spj=\spi+1\):
  \begin{align*}
    \left(2(e^{\bi p_1}\Sscal+e^{\bi
        p_2})-(1+\Sscal)\left(1+e^{\bi(p_1+p_2)}\right)\right)\sum_{\spj=1}^\lcds e^{\bi \spj (p_1+p_2)}\iket{\spj,\spj+1}\,.
  \end{align*}
  To have an eigenstate of \(\Hami\), it is necessary that these
  terms vanish, {\ie}
  \begin{align}
\label{eq:Smat}
    \framedline{\Sscal=}{\Sscal(p_2,p_1)\equiv-\frac{1+e^{\bi(p_1+p_2)}-2e^{\bi p_2}}{1+e^{\bi(p_1+p_2)}-2e^{\bi p_1}}}\,\,.
  \end{align}
\item Other terms appear at the boundary of the chain, as in
  \eqref{eq:-left.-}. One can show that they cancel if the proper
  periodicity condition is imposed. This periodicity condition is
  \begin{align}
  \forall 1\leq {\spk}<\lcds,~\Psi(\spk,\lcds)=&\Psi(0,\spk)&\where
  \Psi(\spj,\spk)\equiv& e^{\bi (p_1 \spj + p_2
  \spk)}+\Sscal e^{\bi (p_1 \spk + p_2 \spj)}
\end{align}
and it is solved by 
\begin{align}
\label{eq:QuantCond2}
  \framedline{e^{\bi \lcds p_2}=}{\Sscal}\,\,,& \framedline{e^{\bi \lcds p_1}=}{1/\Sscal}\,\,.
\end{align}
\end{itemize}
One can 
check that the above conditions
(\ref{eq:Smat},~\ref{eq:QuantCond2}) are sufficient conditions, under
which \(\pket{p_1,p_2;\Sscal}\) is an (unnormalized) eigenstate. One can also check (see
for instance \cite{1998cond.mat..9162K}), that
this gives \(\frac{\lcds(\lcds-1)}2\) independent eigenstates of this
form.
This means that all the ``two-particles'' eigenstates of the
{\Ham} are of this form.

\paragraph{{\Mp}-particles states}
\label{sec:mp-particles-states}

More generally, the Bethe {\anz} 
tells that for an
arbitrary number \(\Mp\) of particles, the eigenstates should be looked
for in the form
\begin{gather}
\label{eq:mpart-state-wavefun}
  \pket{p_1,p_2,\cdots,p_\Mp ; \{\mathcal{A}_\sigma\}_{{\sigma\in
      \Sgrp\Mp}}}
\equiv 
\sum_{\qquad\mathclap{1\leq \spj_1<\spj_2<\cdots<\spj_\Mp\leq \lcds}\qquad}
\Psi(\spj_1,\spj_2,\cdots,\spj_\Mp)
\iket{\spj_1,\spj_2,\cdots,\spj_\Mp}\,,
\\
\where 
\Psi(\spj_1,\spj_2,\cdots,\spj_\Mp)
\equiv \sum_{\sigma\in \Sgrp\Mp}
\mathcal{A}_\sigma ~ e^{\bi\sum_\kk p_{\sigma(\kk)} \spj_\kk} \,,
\label{eq:AnsatzNPart}
\end{gather}
where 
\(\Sgrp\Mp\) denotes the
\index{S (symmetric group)@\ensuremath{\mathcal{S}} (symmetric group)}
  set of all permutations of \(\{1,2,\cdots,\Mp\}\).
This 
state
is a linear combination of \(\Mp\) planar waves, and it is parameterized by 
the \(\Mp\) momenta (or impulsions) \(p_\ii\) of these planar waves,
and by the \(\Mp 
!\) coefficients \(\mathcal{A}_{\sigma}\) of the linear combination. 
This definition \eqref{eq:AnsatzNPart} generalizes the special cases 
\eqref{eq:1-part-{\anz}} and \eqref{eq:2-part-{\anz}}
 corresponding to 
 \(\Mp=1\) or \(\Mp=2\), 
written with the normalization choice
 \(\mathcal{A}_{\bo}=1\), where \(\bo\) denotes the identity
permutation.

Like before, we can first see that 
\begin{multline}
  \Hami \iket{\spj_1,\spj_2,\cdots,\spj_\Mp}=
  +\lcds 
\iket{\spj_1,\spj_2,\cdots,\spj_\Mp}
-2\left(\lcds-2\Mp\right)\iket{\spj_1,\spj_2,\cdots,\spj_\Mp}
\\
-2\sum_{\kk=1}^\Mp \left(\rule{0pt}{4.2mm}\iket{\spj_1,\cdots,\spj_{\kk-1},
\spj_\kk+1,\spj_{\kk+1},\cdots,\spj_\Mp} 
\right.
\\
\left.\rule{0pt}{4.2mm}
+
\iket{\spj_1,\cdots,\spj_{\kk-1},
\spj_\kk-1,\spj_{\kk+1},\cdots,\spj_\Mp}\right)
\end{multline}
\begin{equation*}
  \hspace{5cm}\If\spj_1> 1 ~\And~ \forall\kk<\Mp,~ \spj_{\kk+1}>\spj_\kk+1 ~\And~
\spj_\Mp<\lcds\,.
\end{equation*}
This implies that up to ``boundary'' terms, 
\(\Hami \pket{p_1,p_2,\cdots,p_\Mp ; \{\mathcal{A}_\sigma\}}\) is
equal to \linebreak  \(\left(E_0+\sum_{\kk=1}^\Mp4-4\cos(p_\kk)\right)\pket{p_1,p_2,\cdots,p_\Mp
  ; \{\mathcal{A}_\sigma\}}\). 
As a consequence, if this state is an eigenstate, its energy is
\begin{align}
\label{eq:BetheEnergyN}
  E=&E_0+\sum_{\kk=1}^\Mp\left(4-4\cos(p_\kk)\right)\,.
\end{align}
Physically, this expression
means that each particle has an energy \linebreak
\(4-4\cos(p_\kk)=8\sin^2({p_\kk}/2)\). The energy of an ``{\Mp}-particles state''
is simply the vacuum energy \(E_0\), plus the sum of the energies of the
particles.
As in the two-particles case, one can then investigate all the
extra-terms which have to be set to zero in order to obtain an
eigenstate of \(\Hami\):
\begin{itemize}
\item First, some terms arise from the states
  \(\iket{\spj_1,\spj_2,\cdots,\spj_\Mp}\) where there is one \(\kk\) 
  such that \(\spj_{\kk+1} = \spj_\kk+1\). For a given \(\kk\), the
  cancellation of these terms reduces to the constraint
  \begin{align}
\forall &\sigma,&
    \mathcal{A}_{\sigma\circ\tau_{[\kk,\kk+1]}}=&-\frac{1+
      e^{\bi(p_{\sigma(\kk)}+p_{\sigma(\kk+1)})} -2  e^{\bi p_{\sigma(\kk+1)}}}{1+
      e^{\bi(p_{\sigma(\kk)}+p_{\sigma(\kk+1)})} -2  e^{\bi
        p_{\sigma(\kk)}}}  \mathcal{A}_{\sigma} \,.
\label{eq:BetheAmplitudeConstr}
  \end{align}
 These constraints impose that
 \begin{align}
\label{eq:BetheAmplitudeCoefs}
   \mathcal{A}_{\sigma}=&{\NN}~\epsilon(\sigma)\prod_{\substack{\jj<\kk}}
   \left(1+e^{\bi (p_{\sigma(\jj)} + p_{\sigma(\kk)})}-2 e^{\bi p_{\sigma(\kk)}}\right)\,,
 \end{align}
where \({\NN}\) is a \(\sigma\)-independent normalization, and
\(\epsilon(\sigma)\equiv \prod_{\ii<\jj}\frac{\sigma(\ii)-\sigma(\jj)}{\ii-\jj}\) is the signature of the permutation \(\sigma\).
\index{e(s)@\ensuremath{\epsilon(\sigma)} (signature of a permutation)}
\item Then some extra terms arise at the boundaries of the {\cds},
  and their vanishing requires the periodicity of the wave
  function. More explicitly, the requirement
  \(\Psi(\spj_1,\spj_2,\cdots,\spj_{\Mp-1},\lcds)=\Psi(0,\spj_1,\spj_2,\cdots,\spj_{\Mp-1})\)
  is satisfied if the momenta \(p_\jj\) satisfy:
  \begin{align}
\framedline{\forall 1\leq \jj \leq \Mp ,~~ e^{\bi \lcds p_\jj} =}
{\prod_{\kk\neq\jj} \Sscal(p_\jj,p_\kk)}\,\,\,,
\label{eq:QuantBetheN}
  \end{align}
  where \(\Sscal(p_\jj,p_\kk)\) is defined by \eqref{eq:Smat}.
\end{itemize}

One can actually show that finding sets of momenta satisfying the
``Bethe equation'' \eqref{eq:QuantBetheN}
\index{Bethe equations}
 always gives the
wave function of an eigenstate, by plugging these momenta into
\eqref{eq:AnsatzNPart} and \eqref{eq:BetheAmplitudeCoefs}. Moreover
one can show that 
all eigenstates are obtained this way.

The ``Bethe {\anz}'' is the name given to the ``guess'' that
eigenstates should be found in the form \eqref{eq:AnsatzNPart}. Solving
the Bethe equations allows to find exactly the eigenstates, and their energy
\eqref{eq:BetheEnergyN}. This is why we say that the Heisenberg spin
chain is ``{\ing}''. That does not really mean that we know a completely
explicit expression of the 
eigenstates and of their energy, but only that we know
a simple equation \eqref{eq:QuantBetheN} called ``Bethe equation'',
and we know that solving 
this equation solves exactly this model. Solving the Bethe equation
analytically can nevertheless be a difficult task, especially when the
number of particles ({\idest} the number of planar waves, also called magnons)
is large. 

A more detailed account of this Bethe {\anz} can be found in
\cite{springerlink:10.1007/BF01341708,1998cond.mat..9162K}. The method
given above is often called the ``coordinate Bethe {\anz}'', because
it gives an expression for the wave function. There actually
exist other ways to derive
\eqref{eq:BetheEnergyN}, \eqref{eq:QuantBetheN} and
\eqref{eq:Smat}, and one of them, based on the Hirota equation will be
detailed in the chapter \ref{part:qoperatorsspin} of this
manuscript. 
An nice introduction to the Bethe {\anz}, as well as 
several alternative 
methods to obtain the Bethe
equations \eqref{eq:BetheEnergyN}, \eqref{eq:QuantBetheN} and
\eqref{eq:Smat}  are introduced for instance in  
the 
review \cite{Staudacher:2010jz} and the references therein.

The chapter \ref{part:qoperatorsspin} will introduce a 
method
based on the construction of 
a {\faml} of conserved charges called ``{\Toprs}''. We also see how to
construct 
{\ops} called the ``{\Qoprs}'', which
belong to same {\faml} of commuting {\ops}, and
are the building blocks to express {\Toprs}. The chapter
\ref{part:qoperatorsspin} will introduce a few {\csds} which
  generalize the Heisenberg XXX\(_{1/2}\) {\cds} and
for which we can construct these {\ops} explicitly (which is an
original result obtained during this {\PhD}). We will also see how these
{\ops} allow to diagonalize the {\Ham} and recover these Bethe
equations.

\section{Generalization to other {\ing} models}
\label{sec:gener-other-integr}

In the previous section, we saw how the Bethe {\anz} arises for
the Heisenberg {\cds}. Interestingly enough the same procedure can
be used for many other {\ing} models, including {\csds}
with various Hamiltonians, as well as a few two-dimensional quantum
field theories, also called ``\(\sigma\)-models''  (some of them can be
viewed as the limit of spin 
chains when the space is continuous instead of discrete).

As it is argued in \cite{Zamolodchikov1979253}, the wave function of
several field theories can be obtained by a Bethe {\anz} generalizing
the {\anz} of the previous section. The {\anz} is then that the
eigenstates, parameterized by sets of rapidities \(p_\kk\) (or
momenta), have a wave function of the form 
\begin{gather}
\label{eq:FieldBetheAnsatz}
  \Psi(\spj_1,\spj_2,\cdots,\spj_\Mp) \equiv \sum_{\sigma\in
    \Sgrp\Mp} \mathcal{A}_\sigma e^{\bi\sum_\kk p_{\sigma(\kk)}
    \spj_\kk} \,,
\end{gather}
where (by contrast to the previous section) the positions 
\(\spj_1,\spj_2,\cdots,\spj_\Mp\) are not necessarily integers. In
general, the wave function has several components (if the theory
has several different types of particles), and the coefficients 
\(\mathcal{A}_\sigma\) are vectors\footnote{For a field theory with
  \(\Kr\) different types of particles, the analogous of the state
  (\ref{eq:1magnon},\ref{eq:2magnons},\(\ldots\)) 
is the states
\(\iket{\nn_1(\spj_1),\nn_2(\spj_2),\ldots,\nn_\Mp(\spj_\Mp)}\) denoting
the presence of a particle of type \(\nn_1\) at position \(\spj_1\), and 
of a particle of type \(\nn_2\) at position \(\spj_2\), etc.
Then a general state (with \(\Mp\) particles) is written as 
\(\sum \Psi_{\nn_1,\nn_2,\ldots,\nn_\Mp}(\spj_1,\spj_2,\cdots,\spj_\Mp)
\iket{\nn_1(\spj_1),\nn_2(\spj_2),\ldots,\nn_\Mp(\spj_\Mp)}\), where
\(\Psi_{\nn_1,\nn_2,\ldots,\nn_\Mp}(\spj_1,\spj_2,\cdots,\spj_\Mp)\)
denotes the coordinates of \(\Psi(\spj_1,\spj_2,\cdots,\spj_\Mp)\in
\left(\bC^{\Kr}\right)^{\otimes \Mp}\). Hence we see that the
wave function belongs to \(\left(\bC^{\Kr}\right)^{\otimes \Mp}\), and
hence \(\mathcal{A}_\sigma\) also belongs to \(\left(\bC^{\Kr}\right)^{\otimes \Mp}\).
}.

 To write
the {\anz} \eqref{eq:FieldBetheAnsatz}, it is clearly
crucial\footnote{The fact that the space dimension is one-dimensional
  is necessary here to have a sum over permutations. This sum
  corresponds to the fact that if the momenta of the spin waves are numbers
  (and not vectors), then they are necessarily ordered in one out of
  \(\Mp!\) possible ways.
} that the
space is one-dimensional, 
{\idest} that the positions \(\spj_1,\spj_2,\cdots,\spj_\Mp\) are numbers, and
not vectors. It also assumes that the number of particles is
conserved.

In field theories, planar waves describe free particles and the {\anz}
\eqref{eq:FieldBetheAnsatz} only describes a specific domain
in the physical space: the domain where
 the positions
\(\spj_1,\spj_2,\cdots,\spj_\Mp\) are separated by distances large
enough compared to the interaction range.
Describing particles coming close to each other is a more complicated,
but fortunately not necessary task. In fact it is enough to know the
main features of the 
wave function of 
two-particles
states (when the two particles are separated by small distances), and
from  this one can construct the wave function when two particles are
close to each other and all the other ones are far away.
One can then argue that the only properties of two-particles states
which we need to know are encoded into an {\Sma} \(\Smat(p,p')\),
and that the wave function is constrained to obey the equation
\begin{align}
  \forall &\sigma,&
    \mathcal{A}_{\sigma\circ\tau_{[\kk,\kk+1]}}=&-
\Smat(p_{\sigma(\kk+1)},p_{\sigma(\kk)}) \mathcal{A}_{\sigma} \,,
\label{eq:BethField}
\end{align}
which generalizes the equation \eqref{eq:BetheAmplitudeConstr}. One
should note that 
\(\Smat(p,p')\) has to be a matrix 
because
\(\mathcal{A}_{\sigma}\) and
\(\mathcal{A}_{\sigma\circ\tau_{[\kk,\kk+1]}}\) are vectors. 
This
matrix encodes all the information that we need to know about the
behavior of the wave function when two particles are close to each
other (compared to the interaction range).

If this {\anz} holds, then the eigenstates are completely fixed by the
two-particles interactions, encoded into the {\Sma} \(\Smat(p,p')\). 
This {\Sma} {\cannnot} be completely arbitrary, and if \eqref{eq:BethField} has a
solution, then \(\Smat(p,p')\) has to obey the following constraint
\begin{align}
\label{eq:YBfield}
{ A_{\sigma\circ\tau_{[\kk,\kk+1]}\circ\tau_{[\kk-1,\kk]}\circ\tau_{[\kk,\kk+1]}}}
=&
{ A_{\sigma\circ\tau_{[\kk-1,\kk]}\circ\tau_{[\kk,\kk+1]}\circ\tau_{[\kk-1,\kk]}}}
  \\
\hence~~ \Smat(p',p)\cdot \Smat(p'',p)\cdot \Smat(p'',p')=&\Smat(p'',p')\cdot \Smat(p'',p)\cdot \Smat(p',p)
\end{align}
which arises from the fact that \(\tau_{[\kk,\kk+1]} \circ
\tau_{[\kk-1,\kk]} \circ
\tau_{[\kk,\kk+1]} = \tau_{[\kk-1,\kk]} \circ \tau_{[\kk,\kk+1]} \circ
\tau_{[\kk-1,\kk]}\), by denoting \(p=p_{\sigma({\kk}-1)}\),
\(p'=p_{\sigma({\kk})}\) and \(p''=p_{\sigma({\kk}+1)}\). This factorization
formula (illustrated by figure \ref{fig:YBfield})
 actually means that the interaction of three particles is
obtained as a product of two-particles interactions, and  this
product is invariant under a specific reordering.

\begin{figure}
  \centering
\fbox{\begin{minipage}{.95 \textwidth}
    \begin{center}
      \PictYbSma
    \end{center}
  \caption{Illustration of the Yang Baxter factorization formula
    \protect\eqref{eq:YBfield}: the three-points interaction (left)
    can be written in two different ways as a product of two-points
    interactions, and the result has to be the same. The same property
  holds for the interaction of arbitrarily many particles.}
  \label{fig:YBfield}
\end{minipage}}
\end{figure}

We see that this {\anz} puts very strong constraints on the theory,
and it can only work for very specific models. These models should
be two-dimensional (with a one-dimensional space dimension, 
and a time dimension) with a conservation of the
number of particles, and 
a factorization property. It is argued in \cite{Zamolodchikov1979253}
that the {\anz} holds if
the number of conserved charges is infinite.

Investigating the symmetry properties of a model sometimes allows to
put even more constraints and to completely solve it. For instance
for the {\PCM} (which will be introduced in section
\ref{sec:example-princ-chir}),  the integrability can be motivated by
finding an infinite set of conserved charges \cite{Polyakov1977224},
and the symmetries of the model (which is relativistic, has an \(\SU
\Np \times \SU \Np\) symmetry, and obeys ``unitarity'' and ``crossing''
constraints) allow to fix the {\Sma} uniquely
\cite{Zamolodchikov1979253} (it is also discussed in section
\ref{sec:example-princ-chir}).

In the very specific field theories 
(such as the 
 {\PCM})
where it holds, this {\anz}
only describes the wave function when the particles are separated by
large distances (compared to the interaction range). 
It is therefore necessary that the size \(\LF\) of the spatial
dimension is large enough (otherwise the particles {\cannnot} be
separated by long distances).

In this manuscript, we will be interested in the finite size effects
which occur when the size \(\LF\) is not large enough
compared to the interaction range, and the {\anz} above {\cannnot} be
used. In this case we have to use a method called the ``{\TBA}''. This method is explained in the chapter
\ref{cha:ansatzs-de-bethe}, which is more specifically focussed on the
{\PCM}. 

We will see in this section that a trick (sometimes called ``double
Wick rotation'' or ``Matsubara transform'') 
allows to write equations for these finite size effects. For several
models, this trick allows to express the finite size-effects from a
set of non-linear integral equations, which is often infinite, 
and can be reduced\footnote{More precisely, this (usually
    infinite) set of integral equation implies the functional relation
    called {\YsE}. On the other hand, the {\YsE} has to be
    supplemented with analyticity conditions in order to imply the
    original (usually
    infinite) set of integral equation.
} to a functional relation taking the universal
form of a ``{\Ysys}''.
 This system of equations is
tightly related to the Hirota equation \cite{1994IJMPA...9.5215K} found for the
{\csds}.  

An important result of this {\thesis} is that the ``{\Qoprs}'', the
fundamental objects  introduced in section \ref{part:qoperatorsspin}
for {\csds}, have a direct analogue (the
{\qfs}) for {\ing} field theories and 
this
allows to 
solve several Y-systems. These results are presented in chapter
\ref{cha:ansatzs-de-bethe} in the case of the {\PCM},
and the results of \cite{2010arXiv1007.1770K} are presented.

\section{AdS/CFT duality}
\label{sec:dualite-adscft}

The arguments suggested above allow to solve several two-dimensional
gauge theories with one space dimension and one time dimension. 
Many such
models are relativistic and have massive
particles\footnote{The fact that the particles are massive introduces
  a mass scale \(\mmass\) and a length scale \(\frac 1 \mmass\). We have
  seen that this length scale was important because integrability
  comes from the regime where \(\LF\gg \frac 1 \mmass\).}.

On the other hand, the known particle physics (described by the
standard model), is a four-dimensional relativistic gauge theory,
which is asymptotically free. This means that the interactions which
occur above an energy scale are well described by sums of ``Feynman
diagrams'' which correspond to different possible interaction
processes. For instance, the simplest way for two electrons to interact
is by exchanging one photon. But they could as well exchange two
photons, or more. Or an electron could emit a photon which transforms
afterwards into an electron-positron pair which annihilates into photons
that are finally absorbed by the other electron. All the processes which
can happen are described by ``Feynman diagrams'', and 
in general one should sum an infinite series of these processes.
Asymptotic freedom (which occurs for instance in the standard model)
means that above a given energy scale,
the more complicated\footnote{More precisely, the statement
is that the more loops a diagram has, the less it contributes.} a
diagram is, the less it contributes to the sum.
This allows to show that, in order to compute the 
  properties 
of an interaction
to a given accuracy, it is sufficient to keep a finite number of
terms.

We see that in general, for these gauge theories, what we can do is to
write an infinite series (which is an asymptotic expansion), which is
called a perturbative expansion. This captures important physical
properties of the interactions, but it {\cannnot} be used {\below} a given
energy scale. As a consequence, there are some aspects of these gauge
theories that are not captured by this approach.
For instance, one of these non-perturbative aspects is the confinement
of quarks inside hadrons (such as the neutrons and the 
protons), which explains that we {\cannnot} observe an isolated quark, but
only some particles made of multiple quarks.

These questions arise for lots of different gauge theories and we will
see that there exists at least one four-dimensional gauge theory,
the so-called {\SYM} field theory with four supersymmetries
 (\(\mathcal{{\NN}}=4\) SYM) for which some exact computations can be done (as
opposed to the 
perturbative expansion mentioned above). This theory is a conformal
field theory (CFT), which means that it is invariant under several
transformations including dilatations (see {\below}). 

  \paragraph{Conformal invariance and dimension of {\ops}}
  \label{sec:conf-invar-dimens}

Every  field theory describes some fields which are functions of the
positions (these functions are 
{\op}-valued for quantum field theories).
Conformal field theories are invariant under the transformations of
positions which preserves angles ({\idest} these transformations locally
look like compositions of translations, rotations, and dilatations).

When the space has dimension \(\DD>2\), all the conformal
transformations take the form 
\begin{subequations}
\label{eq:xconformaltransfo}
  \begin{align}
    x_\mu&\mapsto x'_\mu=x_\mu +a_\mu+\Omega_{\mu,\nu}x_\nu& \where
    a&\in
    \bR^\DD&\And \omega&\in \Orth \DD\,\\
    x_\mu&\mapsto x'_\mu=\lambda\:x_\mu & \where \lambda&\in \bR\,\\
    \Or x_\mu&\mapsto
    x'_\mu=\frac{\frac{x_\mu}{(x)^2}+\alpha_\mu}{\left(\frac{x_\rho}{(x)^2}+\alpha_\rho\right)^2}
    & \where (x)^2&\equiv x_\nu x_\nu=\|x\|^2&\And \alpha&\in \bR^\DD\,
  \end{align}
\end{subequations}
(or a composition of these three transformations)
where we use Einstein's sum convention in an Euclidean metric (which
means that the repeated indices are summed over, {\idest} that
\(\Omega_{\mu,\nu}x_\nu\) denotes the sum \(\sum_{\nu=1}^\DD\Omega_{\mu,\nu}x_\nu\)).

These transformations locally conserve the ratios of distances ({\idest} the
angles), which means that conformal transformations are the
transformations \(x_\mu \mapsto x'_\mu\) such that there exists a
positive function \(\lambda(x)\) such that\footnote{Here,
  \(\left(\mathrm{d}x'\right)^2\) denotes the bilinear form which is
  formally constructed as \(\mathrm{d}x'_\mu \mathrm{d}x'_\mu = 
\frac{\partial x'_\mu}{\partial x_\nu} \mathrm{d}x_\nu  \frac{\partial x'_\mu}{\partial x_\rho} \mathrm{d}x_\rho\)
} 
\begin{equation}
  \label{eq:defLambaConformal}
  \left(\mathrm{d}x'\right)^2 = \lambda(x)^2 \left(\mathrm{d}x\right)^2\,.
\end{equation}
In conformal fields theories, 
the  coordinates can be transformed as in
\eqref{eq:xconformaltransfo}, and then 
some fields \(\Phi_\ii(x)\) (called primary {\ops}) transform as 
\begin{align}
\label{eq:PhiConformalTransfo}
  \Phi_\ii(x)&\mapsto \Phi_\ii'(x')&\where&\:\Phi_\ii'(x')\equiv
  \lambda(x)^{\Delta_\ii}\:\: \Phi_\ii(x)\,
\end{align}
where \(\Delta_\ii\) is called the conformal dimension of the field
\(\Phi_\ii(x)\), which indicates how \(\Phi_\ii(x)\) is rescaled when the
coordinates are rescaled.

An important information that we want to extract in a field theory
is the correlation functions such as
\(\left\langle \Phi_1(x) \:\Phi_2(y)\right\rangle\), \(\left\langle
  \Phi_1(x) \:\Phi_2(y) \:\Phi_3(z)\right\rangle\), etc. These
correlations functions capture the properties of the quantum
fluctuations, and they are strongly constrained by the above symmetry:
in conformal field theories, they are invariant under the
transformations
(\ref{eq:xconformaltransfo},\ref{eq:PhiConformalTransfo}).

It is then possible to show \cite{Polyakov:1970xd} that  the ``two-points''
and ``three-points'' correlation functions are given by
\begin{gather}
  \label{eq:2points}
  \left\langle \Phi_{\ii}(x)
    \:\Phi_{\jj}(y)\right\rangle=\frac{\delta_{\Delta_{\ii},\Delta_{\jj}}}{\left\|
    x-y\right\|^{2 \Delta_{\ii}}}\,,\\
  \label{eq:3points}
\left\langle
  \Phi_{\ii}(x) \:\Phi_{\jj}(y) \:\Phi_{\kk}(z)\right\rangle=\frac{C_{{\ii},{\jj},{\kk}}}{
\left\| x-y\right\|^{\Delta_{\ii}+\Delta_{\jj}-\Delta_{\kk}} \left\|
  x-z\right\|^{\Delta_{\ii}-\Delta_{\jj}+\Delta_{\kk}} \left\| y-z\right\|^{-\Delta_{\ii}+\Delta_{\jj}+\Delta_{\kk}}
}\,,
\end{gather}
where the expression \eqref{eq:2points} fixes the normalization of the
fields\footnote{If each field \(\Phi_{\kk}\) is multiplied by an arbitrary
  constant, then the relation \eqref{eq:2points} clearly has to be
  multiplied by a constant. The choice to have only
  \(\delta_{\Delta_{\ii},\Delta_{\jj}}\) in the numerator (without any extra
  constant) fixes this degree of freedom in the definition of the
  fields.}. The ``conformal dimensions'' \(\Delta_{\ii}\) and the
``structure constants'' \(C_{{\ii},\jj,\kk}\) are important properties a theory,
as they allow to compute the correlation functions \eqref{eq:2points}
and \eqref{eq:3points}.

In the present manuscript we will specifically focus on the conformal
dimensions of the {\ops} in \(\mathcal{{\NN}}=4\) {\SYM}. We will see
  that they can be obtained using {\ing} properties of this model.

  \paragraph{Integrability in the AdS/CFT duality}
  \label{sec:integr-adscft-dual}

Interestingly enough the {\ing} properties of the {\SYM}
field theory are best understood and tested in the framework of the
``AdS/CFT correspondence''.  This conjectured duality
\cite{Maldacena:1997re,Gubser:1998bc,Witten:1998qj} says that several
quantities (such as correlation functions) that we wish to compute on
one side 
of the duality, for instance in {\SYM}, can be obtained by
computing other quantities in the other side of the duality, for
instance AdS. This ``AdS'' denotes a string theory ({\idest} a quantum
theory of gravity) on a 10-dimensional space-time having the geometry
\(AdS_5\times {\Ssph}^5\), where \(AdS_5\) denotes the 
5-dimensional anti de Sitter space, a curved manifold which is roughly
speaking a multi-dimensional hyperboloid.

Given a quantity that we want to compute (for instance) in
  {\SYM}, it is not easy to understand what computation in the AdS string
  theory is associated to this quantity. Nevertheless,
this 
  duality 
is very interesting because it turns out to
relate the perturbative domain of one model to the non-perturbative
domain of the other model. It means that for instance, classical
string theory is related to deeply non-perturbative gauge theory.

We will not enter deeply into the details of this duality, but we will
simply work with one of the predictions of this duality: the fact that
the energy spectrum of the super-strings is equal to the spectrum of
the conformal dimensions of the {\SYM} {\ops}. 

On the {\SYM} side, integrability was first noticed in
\cite{Lipatov:1993yb}, and in the planar limit (a limit when the rank
of the gauge group is 
large), a mapping was noticed \cite{Faddeev:1994zg,Minahan:2002ve}
between the study 
of the spectrum of the 
conformal dimensions in {\SYM} and an {\SL 2} {\cds}
(see for instance \cite{Minahan:2010js}\footnote{The
  review \cite{Minahan:2010js} is part of the collection
  \cite{Beisert:2010jr} of reviews, which provide an excellent
  introduction to the subject.}).
More precisely, 
it was shown that in this
planar limit 
the only relevant {\ops} are linear combinations of {\ops} of
the form 
\(\mathrm{tr}({\mathcal{O}_{1}},{\mathcal{O}_2}, \cdots, \mathcal{O}_\lcds)\)
where \(\mathcal{O}_1\), \(\mathcal{O}_2\), \(\cdots\), \(\mathcal{O}_\lcds\)
are arbitrary fields.
In general, the {\op} {\(\Phi_{\nn_1,\cdots,\nn_\lcds}\equiv
\mathrm{tr}({\mathcal{O}_{\nn_1}},\cdots,
\mathcal{O}_{\nn_\lcds})\)} 
is not a primary {\op}, which transforms as in
\eqref{eq:PhiConformalTransfo}. Instead it is a linear combination of
primary {\ops}, and the {\ops} of this form transform as 
\(\Phi_A'(x')=
  Z_A^{~B}\: \Phi_B(x)\) where the ``mixing matrix'' \(Z_A^{~B}\) has
  eigenvalues  \(\lambda(x)^{\Delta_\ii}\) where \({\Delta_\ii}\) denotes
  the dimensions of primary {\ops}.
In particular 
 there is a sector called \(\SU 2\)
sector, where only two elementary {\ops} can appear ({\idest} each \(\mathcal{O}_\ii\) is
either equal to \(X\) or \(Y\)). Then the {\op} 
\(\mathrm{tr}(XXYX)\) can be mapped to
the state \(\ket{↑↑↓↑}\) of an \(\SU 2\) {\cds} of size
\(\lcds=4\).
 Then it was shown that for ``long {\ops}'', {\idest} for
composite {\ops} made of the trace 
of the product of many elementary {\ops}, this mapping transforms
  the 
mixing matrix
  into an {\ing} {\cds}
  {\Ham}, and
 some Bethe equations arise 
 that allow to find
 the spectrum of {\SYM}' long {\ops}.

The equations obtained for these ``long {\ops}'' can then be
continued to short {\ops}. This gives a {\Ysys} which was
conjectured in \cite{2009PhRvL.103m1601G} and then understood in terms
of the {\TBA} approach
\cite{2009JPhA...42K5401B,Gromov:2009bc,Arutyunov:2009ur}.
This {\Ysys} was successfully tested in the weak coupling regime, by
comparison with perturbative expansion in {\SYM}
\cite{Janik:2007wt,Heller:2008at,Bajnok:2008bm,Fiamberti:2008sm,Velizhanin:2008pc,Minahan:2009wg,Arutyunov:2010gb,Balog:2010xa},
but also in the strong coupling regime \cite{Gromov:2009tq}.

In this manuscript, we will take these equations as the starting
point for the chapter \ref{cha:dualite-adscft}, and derive a simpler
set of equations \cite{Gromov:2011cx}. This work, performed during this PhD,
is similar in spirit to the analysis of the {\PCM} in the chapter
\ref{cha:ansatzs-de-bethe}, but we will see that the analyticity
conditions are much richer. In particular we will find a new symmetry
of the {\Ysys} (which we call ``quantum-\(\bZ_4\) symmetry'', and
which we interpret from the string theory on \(AdS_5\times {\Ssph}^5\)), and show
how to recast the infinite set of equations arising from the {\TBA}
into a finite set of equations, where the analyticity of several
functions is much better understood than in previous analyses. 

\index{Bethe ansatz|)}

%%% Local Variables: 
%%% mode: latex
%%% TeX-master: "english.tex"
%%% End: 

%% file: spinchains.tex
In this chapter we will prove the integrability of a class of spin
chains which generalize the Heisenberg {\cds} studied in chapter \ref{sec:integr-et-}. 

To do this, we will first construct a {\faml} of commuting {\ops} (and
which also commute with the {\Ham} \(\Hami\)),
called the {\Toprs} \cite{1990JPhA...23.1477B}. We will then show,
following the paper 
\cite{2008JHEP...10..050KV}, that these {\Toprs} obey some fusion
relations governed by the ``Cherednik-Bazhanov-Reshetikhin''
determinant formula \eqref{eq:CBR}
\cite{springerlink:10.1007/BF01077327,1990JPhA...23.1477B}, which can also be recast into
the bilinear form of the Hirota equation \cite{1992PhyA..183..304K,Kuniba:1992qv,springerlink:10.1007/s002200050165,1997JPhA...30.7975T}. The proof of this relies on
combinatorial identities introduced in the appendix
\ref{sec:diagr-expr-co}, and it will conclude the section
\ref{sec:chaines-de-spin}. In the next sections, we will show how
to diagonalize the 
{\Toprs} and the {\Ham}, by writing a ``Bäcklund flow''. 
More precisely we will start by motivating the introduction of the
Bäcklund flow in section \ref{sec:transf-de-backl}, where we will show
(as in \cite{springerlink:10.1007/s002200050165,Zabrodin:1996vm,Kazakov:2007fy,Zabrodin:2007rq})
that if this flow exists and is polynomial,  
then some strong constraints arise that allow to
diagonalize the {\Toprs}.
Finally new results of this {\PhD} will be presented in the
section \ref{sec:expr-diff-Qop},
where this Bäcklund flow is constructed explicitly at the operatorial
{\level},  and an original construction of the so-called
``{\Qoprs}'' is presented for the {\GLKM} {\cds}.

Starting from the introduction of ``{\Qoprs}'' in \cite{Baxter:1972hz}
for the eight-vertex lattice model, some {\Qoprs} have been
constructed for a large variety of {\ing} systems\footnote{For
  instance, constructions of {\Qoprs} for different models are given
  in \cite{Baxter:1972hz,1997CMaPh.190..247B,1999CMaPh.200..297B,2001NuPhB.604..580H,2002NuPhB.622..475B,2002cond.mat..7177F,2003math......6242K,2005NuPhB.709..578K,2005JPhA...38.6641K,2006JPhA...39..S11B,2007CMaPh.272..263B,2007JSMTE..01....5B,2008JPhA...41I5206K,2008NuPhB.805..451B,2009JPhA...42g5204D,2010JPhA...43O5208B,2010JSMTE..11..002B,2011NuPhB.850..148B,Staudacher:2010jz,2011NuPhB.850..175F,2011arXiv1112.3600F,2012arXiv1205.1471T}
.}
The construction given in this {\thesis} for the {\GLKM} {\cds} is
quite different from these 
constructions, 
and allows to define {\Qoprs} directly as {\ops} and to show their
polynomiality.
In particular, this gives {\Wronskian} determinant expressions for the
{\Toprs} in terms of {\Qoprs}. These {\Wronskian} expressions are given in section
\ref{sec:wronsk-expr-qq}, and can also be found in the literature
\cite{1997CMaPh.190..247B,springerlink:10.1007/s002200050165,2008NuPhB.805..451B,Gromov:2010km,2010NuPhB.826..399T}. These
{\Wronskian} expressions are known solutions of the Hirota equation, and
we will prove that 
these expression apply to 
the {\Toprs} of these {\csds}\footnote{One can easily see that
  there also  exists solutions of the Hirota equation which {\cannnot} be
  written as a {\Wronskian} determinant. This will be discussed in
  chapter \ref{cha:ansatzs-de-bethe}, where a sufficient condition
  (called typicality) is given, under which one can write such
  {\Wronskian} determinants.%
}. %

Interestingly, this construction turns out to have very deep
connections \cite{Alexandrov:2011aa} with the ``classical
integrability'', as explained  
in the section \ref{sec:quantum-classical}.%

\section{{\Csds} and {\Toprs}}
\label{sec:chaines-de-spin}

\index{T-operators@{\Toprs}|(}

{\Csds} are particularly simple examples of {\ing} systems.
In this section, we will see how to construct the {\faml} of conserved
charges which will allow to diagonalize the {\Ham}. We will also
see that these charges can be expressed %
  through %
the action of a
``{\cd}'', and we will see that it allows to prove the
``Cherednik-Bazhanov-Reshetikhin'' 
determinant formula \eqref{eq:CBR}.

\subsection{Construction of the {\Toprs}}
\label{sec:operateurs-t}

\subsubsection{Heisenberg {\cds}}
\label{sec:chaine-de-spin}

The ``Heisenberg XXX\(_{1/2}\) {\cds}'' is the simplest {\cds}, and corresponds to a
quantum version of the Ising model. As we already saw in the
introductory section \ref{sec:integr-et-}, its Hilbert space is  \(\Hilb=%
{\bigotimes_{\spi=1}^\lcds} \Hilbl_\spi=\left(ℂ^2\right)^{\otimes
  \lcds}\). In this {\cds}, the interactions are only between nearest
neighbors, and are governed by the {\Ham}
\eqref{eq:hamilHeisenbergintro}, or by the (more physical) expression
{\below} 
(we will show that the two expressions coincide)
\begin{gather}
\label{eq:hamilHeisenberg}
\fdisp
{
 { \Hami=}{%
   {\sum_{\spi}\Hami_{\spi,\spi+1}=-\sum_{\spi}
     \vec{\sigma_\spi}\cdot\vec{\sigma_{\spi+1}}}}
}\\
\where
\vec{\sigma_\spi}\cdot\vec{\sigma_{\spj}}\equiv\sum_{{\coordl}=1}^{3}\sigma_\spi^{({\coordl})}.\sigma_\spj^{({\coordl})}%
\,,%
\qquad\qquad
\sigma_\spi^{({\coordl})}=%
𝕀^{\otimes {\spi}-1}\otimes \sigma^{({\coordl})}\otimes 𝕀^{\otimes {\lcds}-{\spi}}\,.
\end{gather}
In the expression, \(\sigma^{(1)}\), \(\sigma^{(2)}\) and \(\sigma^{(3)}\)
denote {\pau} matrices 
\begin{align}
\label{eq:DefPauli}
  \sigma^{(1)}=&\left(
\begin{array}{cc}
 0 & 1 \\
 1 & 0
\end{array}
\right)\,
&  \sigma^{(2)}=&\left(
\begin{array}{cc}
 0 & -{ⅈ} \\
 {ⅈ} & 0
\end{array}
\right)\,
&  \sigma^{(3)}=&\left(
\begin{array}{cc}
 1 & 0 \\
 0 & -1
\end{array}
\right)
\,,
\end{align}
and \(\bI\) denotes the unity matrix. %
The symbol ⅈ denotes the imaginary number with imaginary part
equal to one.
\index{ia@\(ⅈ~\equiv~\sqrt{-1}~~\)}

The minus sign in front of \(\sum_{\spi}
\vec{\sigma_\spi}\cdot\vec{\sigma_{\spi+1}}\) in the definition of the
{\Ham} means that the {\cds} is ferromagnetic. With this choice,
the state \(\vac\) introduced in \eqref{eq:DefVac} is a state %
of
lowest energy.
To define the {\Ham} \eqref{eq:hamilHeisenberg} completely, a
boundary condition has to be specified. For instance, if the chain is
``open'', the sum \eqref{eq:hamilHeisenberg} runs over \({\spi}\in
\ninter 1 {{\lcds}-1}\). On the contrary, if the chain is periodic,
the sum runs over \({\spi}\in \ninter 1 {\lcds}\), and the
identification
\(\sigma_{{\lcds}+1}^{({\coordl})}=\sigma_{1}^{({\coordl})}\) is used.

As we already saw in the introductory section, 
the proof %
that this {\cds} is
``{\ing}'' is obtained by  %
rewriting %
the {\Ham} 
(and %
\(\vec{\sigma_\spi}.\vec{\sigma_\spj}\)) in terms of the permutation {\op}
\index{Pa (permutation {\op})@\perm (permutation {\op})}
\(\perm_{\spi,\spj}\)
defined by \eqref{eq:DefPerm}. For instance if \(\lcds=2\),
\(\perm_{1,2}\) is the {\op}
defined by
\begin{gather}
  {\perm_{1,2}
  \left(\ket{\phi}\otimes\ket{\psi}\right)=\ket{\psi}\otimes\ket{\phi}},\\
\label{eq:PermFromGen}
\ie~  \perm_{1,2}=\sum_{ \mathclap
    {\alpha
,\beta
\in\{↑,↓\}%
    }}%
e_{\alpha,\beta}\otimes e_{\beta,\alpha} \qquad\qquad \where e_{\alpha,\beta}=\ket\alpha\bra\beta
\end{gather}
One can write for instance
\begin{align}
\label{eq:sigsig1}
  \sigma^{(1)}\otimes \sigma^{(1)} + \sigma^{(2)} \otimes
  \sigma^{(2)}=&
\left(e_{{↓},{↑}} + e_{{↑},{↓}}\right) \otimes \left(e_{{↓},{↑}} +
  e_{{↑},{↓}}\right)\\&~ + \left(ⅈ~e_{{↓},{↑}} -ⅈ~ e_{{↑},{↓}}\right)
\otimes \left(ⅈ~e_{{↓},{↑}} -ⅈ ~e_{{↑},{↓}}\right)\nonumber\\ 
=& 2 \left(\rule{0pt}{3.7mm}e_{{↓},{↑}}\otimes e_{{↑},{↓}}
+ e_{{↑},{↓}} \otimes e_{{↓},{↑}}\right)
\end{align}
\begin{align}
\And \sigma^{(3)}\otimes \sigma^{(3)}=& 2 \left(\rule{0pt}{3.7mm}e_{{↑},{↑}}\otimes
e_{{↑},{↑}} + e_{{↓},{↓}}\otimes e_{{↓},{↓}}\right)
-\bI\,,
\label{eq:sigsigl}
\end{align}
whence we can deduce 
\begin{align}
\label{eq:perm=sigsig01}
  \vec{\sigma}_1.\vec{\sigma}_2=&
2~ \perm_{1,2}-\bI\,.
\end{align}

The same result holds if \(\lcds>2\), and then it reads
\begin{align}
  \perm_{\spi,\spj}=&\frac 1 2
  \left(\vec{\sigma_\spi}.\vec{\sigma_\spj} + \bI\right)\,,
\end{align}
which allows to rewrite the {\Ham} as 
\begin{align}
  \label{eq:HamilHeis}
  {\Hami}=&
-\sum_{\spi}
  \left(2 \perm_{\spi,\spi+1}-𝕀\right)=\lcds - 2 \sum_{\spi}
  \perm_{\spi,\spi+1}\,.
\end{align}
In \eqref{eq:HamilHeis}, \(\lcds\) implicitly denotes the {\op} \(\lcds ~\bI\).

~

We will see in the next %
section
that this property allows to define
a {\faml} of {\ops} commuting with the {\Ham}
\eqref{eq:hamilHeisenberg}. 
Before we construct these {\ops}, let us notice that the Heisenberg
{\cds} can be generalized to spins in a superposition of {\Kr}
states.
In that case the Hilbert space is \(\Hilb=%
{\bigotimes_{\spi=1}^\lcds} \Hilbl_\spi=\left(ℂ^{\Kr}\right)^{\otimes
  \lcds}\), whereas the {\Ham} is
\begin{align}
\label{eq:HaK1}
  \framedline{{\Hami}=}{-\sum_{\spi} \vec{\uplambda_\spi}\cdot\vec{\uplambda_{\spi+1}}
=
-\sum_{\spi}
  \left(2~ \perm_{\spi,\spi+1}-\frac 2 {\Kr} 𝕀\right)}\,\,,
\end{align}
where \(\vec{\uplambda}=
(\uplambda^{(1)},\uplambda^{(2)},\ldots,\uplambda^{({\Kr}^2-1)})\) 
denotes the Gell-Mann matrices, which generalize the {\pau} matrices:
\(\frac{\Kr^2-\Kr}2\) of them are of the form
\(e_{\alpha,\beta}+e_{\beta,\alpha}\) or
\(ⅈ~e_{\alpha,\beta}-ⅈ~e_{\beta,\alpha}\), (where \(e_{\alpha,\beta} =
\ket{\alpha} \bra{\beta}\)), while the other \(\Kr-1\) Gell-Mann matrices
are zero-trace diagonal matrices. The form of these matrices is then
such that the computation (\ref{eq:sigsig1}-\ref{eq:sigsigl})
holds for Gell-Mann matrices as well. That implies
{\begin{align}
\label{eq:perm=lamlam}
  \frac 2 {\Kr} 𝕀 + \sum_{{\coordl}=1}^{{\Kr}^2-1}\frac
  {\uplambda_\spi^{({\coordl})}.\uplambda_\spj^{({\coordl})}}2=& 2
  \perm_{\spi,\spj}\,,%
\end{align}
which generalizes \eqref{eq:perm=sigsig01} to the \({\Kr}\neq 2\) case,
and explains the second equality in \eqref{eq:HaK1}.
}

\subsubsection{Yang-Baxter equation and the construction of conserved charges}
\label{sec:equation-de-yang}

Rewriting the {\Ham} \eqref{eq:hamilHeisenberg} in terms of %
permutation {\op}s will be useful due to the simple algebra
satisfied by these permutation {\op}s: for instance, the relation 
\(\perm_{\spi,\spj}\perm_{\spj,\spk}=\perm_{\spj,\spk}
\perm_{\spi,\spk}\)
allows to check the following equality%
\begin{gather}
\label{eq:RRR=RRR}
\fdisp
{  \Rop_{\spi,\spj}(\su-\sv)\Rop_{\spi,\spk}(\su)\Rop_{\spj,\spk}(\sv)=%
  \Rop_{\spj,\spk}(\sv)\Rop_{\spi,\spk}(\su)\Rop_{\spi,\spj}(\su-\sv)}\\
\where \Rop_{\spm,\spn}(\su)=\su + \perm_{\spm,\spn}\,.
\label{eq:DefR}
\end{gather}
The equality \eqref{eq:RRR=RRR} will be crucial in what follows, and
it is called the ``Yang-Baxter'' identity. More precisely, we will say
that the {\Rop}-matrix defined by \eqref{eq:DefR} 
does satisfy the ``Yang-Baxter'' identity \eqref{eq:RRR=RRR}.

Now, we will see how this identity allows to define a {\faml} of
{\ops} which commute with each other and with the {\Ham}. To
this end, we need to introduce a larger Hilbert space \(\Hilb\otimes
\Hilbl_{a_1}\otimes\Hilbl_{a_2}\otimes\cdots \otimes\Hilbl_{a_\nn}\),
where each
\(\Hilbl_{a_{\spk}}=ℂ^{\Kr}\) is an ``auxiliary'' space.
Here, the symbol \(\Hilb\) denotes the original Hilbert space, \(\Hilb=%
{\bigotimes_{\spi=1}^\lcds} \Hilbl_\spi=\left(ℂ^\Kr\right)^{\otimes
  \lcds}\), whereas the smaller symbol \(\Hilbl\) denotes the smaller spaces
(isomorphic to \(ℂ^{\Kr}\)) which appear in the tensor products. Each
\(\Kr\)-dimensional space \(\Hilbl_{\spi}=ℂ^{\Kr}\) corresponds to one
spin in a superposition of \(\Kr\) states.

We will then show by recurrence that
\begin{gather}
\label{eq:RLL=LLR}
{\Rop_{a_1,a_2}(\su-\sv)
{\Lop}^{(1)}(\su){\Lop}^{(2)}(\sv)=}{{\Lop}^{(2)}(\sv) {\Lop}^{(1)}(\su)
\Rop_{a_1,a_2}(\su-\sv)}
\\
\where
{\Lop}^{(\kk)}(\su)=%
\Rop_{\lcds,a_\kk}(\su)\Rop_{\lcds-1,a_\kk}(\su)\cdots
\Rop_{1,a_\kk}(\su)\,.
\label{eq:Ldef}
\end{gather}
In the definition \eqref{eq:Ldef}, the ``monodromy matrix'' \({\Lop}^{(\kk)}(\su)\)
is an {\op} acting on \(\Hilb\otimes  \Hilbl_{a_\kk}\). By
contrast the {\ops} in \eqref{eq:RLL=LLR} are acting on \(\Hilb\otimes
 \Hilbl_{a_1}\otimes
 \Hilbl_{a_2}\), and in this equation it is implicit that for instance
 \({\Lop}^{(1)}(\su)\) rigorously denotes the  {\op}
 \({\Lop}^{(1)}(\su)\otimes \bI\). 

 \begin{proof}[Proof of the relation \eqref{eq:RLL=LLR}]
The proof relies on the following recurrence relation:
\begin{gather}
\label{eq:RLL=LLRLL}
\Rop_{a_1,a_2}(\su-\sv)
{\Lop}^{(1)}(\su){\Lop}^{(2)}(\sv)=%
{\Lop}_{\lcds,\spi+1}^{(2)}(\sv) {\Lop}_{\lcds,\spi+1}^{(1)}(\su) \Rop_{a_1,a_2}(\su-\sv)
{\Lop}_{\spi,1}^{(1)}(\su){\Lop}_{\spi,1}^{(2)}(\sv)\,,\\
\where {\Lop}_{\spi,\spj}^{(\kk)}(\su)=%
\Rop_{\spi,a_\kk}(\su)\Rop_{\spi-1,a_\kk}(\su)\cdots \Rop_{\spj,a_\kk}(\su)\,,
\end{gather}
where we see that \(  \)\({\Lop}^{(\kk)}(\su)= {\Lop}_{\lcds,1}^{(\kk)}(\su)\).
The initialization of the recurrence  for \(\spi=\lcds\) is trivial,
whereas the case  \(\spi=0\) is the statement \eqref{eq:RLL=LLR} that
we want to  prove. As explained graphically in the figure~\ref{fig:TYB1}, going from
\(\spi\) to \(\spi-1\) in \eqref{eq:RLL=LLRLL} is done by using the Yang-Baxter relation
\eqref{eq:RRR=RRR},
together with the commutation relation 
\begin{gather}
\label{eq:RR=RR}
\comm{\Rop_{\spi,a_1}(\su)}{\Rop_{\spj,a_2}(\sv)}
=%
0%
\hspace{3cm}
\If%
\spi\neq\spj\\
\where \scomm{\pm}{A}{B}\equiv%
A\cdot B ~\pm~ B\cdot A\,.
\end{gather}
\index{01@\ensuremath{\scomm{\pm}{A}{B}\equiv A\cdot B ~\pm~ B\cdot A}}
Indeed, the {\Roprs} have
the form \(\bI\otimes \Rop\otimes \bI\), and the commutation relation
\eqref{eq:RR=RR} is nothing but the statement that
\begin{align}
\label{eq:ComBosTens}
  \comm{A\otimes\bI}{\bI\otimes B}
=&0\,.
\end{align}

\begin{figure}
 \centering
\fbox{\begin{minipage}{.95 \textwidth}
  \begin{center}
    \includegraphics{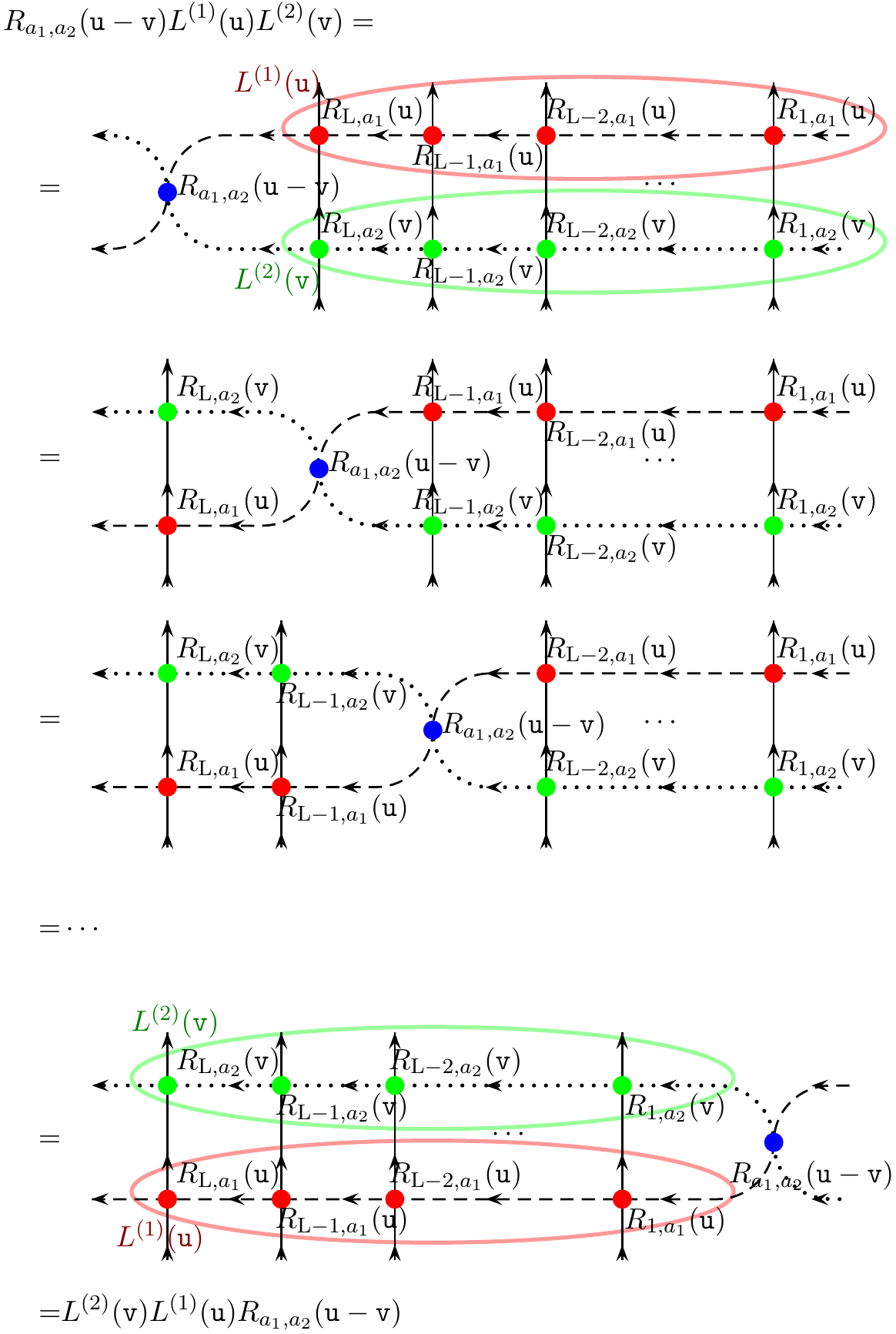}%
    \caption{Illustration of the proof of \protect{~\eqref{eq:RLL=LLR}}}
      \label{fig:TYB1}
  \end{center}
\end{minipage}}
\end{figure}

It can be interesting to write diagrammatically the structure of this
iterative proof, which is done in 
figure~\ref{fig:TYB1}.
In this figure, the vertical lines stand for the Hilbert spaces
\(\Hilbl_1\), \(\Hilbl_2\), \(\ldots\), whereas 
the horizontal lines stand for the auxiliary spaces
 \(\Hilbl_{a_1}\) et \(\Hilbl_{a_2}\). The dots (colored in the online
 version) stand for the {\Roprs} and the order of multiplications is
 given by the arrows. These arrows allow an ambiguity in the order of
 {\ops}, which is fully consistent with \eqref{eq:RR=RR}, and
 crucial for the proof. Graphically, we see that the iterative
 structure of the proof simply consists in moving vertical lines from
 the right side of the blue dot to its left side, and the relation
 \eqref{eq:RRR=RRR} exactly allows to move lines this way across
 each other.
 \end{proof}

~

This equation \eqref{eq:RLL=LLR} turns out to be very important, as it
allows to define a {\faml} of {\ops}, (denoted by \(\gT[][{ }]\)),
which commute with each other:
\begin{align}
\label{eq:commutT}
\framedline{
\comm{ \gT[][{ }]} { \gT[\sv][{ }] }=}
{0} &\where \framedline{\gT[][{ }]\equiv}%
{\mathrm{tr}_{a_\ii}
 {\Lop}^{(\ii)}(\su)}\,\,.
\end{align}
These {\Toprs} (also called ``transfer matrices'') are {\lbd} by
different values of the scalar parameter 
\(\su\), which will be called the ``spectral parameter''.
\index{u(s@\ensuremath{\su} (spectral parameter)}
One should note that in \eqref{eq:commutT}, each \(\gT[][{ }]\) is 
defined by introducing an auxiliary space, and taking a partial trace
defined by \eqref{eq:tracpart} in appendix
\ref{sec:notat-tens-prod}. After this trace, we get an {\op} acting
on the initial Hilbert space \(\Hilb\). The two {\ops} \(\gT[][{ }]\) 
and \(\gT[\sv][{ }]\) are defined by introducing two different auxiliary
spaces (denoted  by \(a_1\) and \(a_2\) in \eqref{eq:RLL=LLR}).

\begin{proof}[Proof of \eqref{eq:commutT}]
  This relation is obtained from \eqref{eq:RLL=LLR}, which implies
  that if
  \(\Rop_{a_1,a_2}(\su-\sv)\) is invertible, {\idest} if
  \(\su-\sv\neq\pm1\), then 
\(
\begin{array}[t]{rl}
  \gT[][{ }]\cdot \gT[\sv][{
  }]=&\mathrm{tr}_{a_1\otimes a_2}
  {\Lop}^{(1)}(\su){\Lop}^{(2)}(\sv)\\
  =&\mathrm{tr}_{a_1\otimes a_2} {\Lop}^{(2)}(\sv){\Lop}^{(1)}(\su)
  =\gT[\sv][{ }]\cdot \gT[][{ }]\,.
\end{array}
\)
    As \(\gT[][{ }]\) is a polynomial in the variable \(\su\), %
    \eqref{eq:commutT} also holds for arbitrary \(\su\) and
    \(\sv\) (by continuation).
\end{proof}

\begin{figure}
 \centering
\fbox{\begin{minipage}{.95\textwidth}
  \begin{center}
    \includegraphics[scale=1]{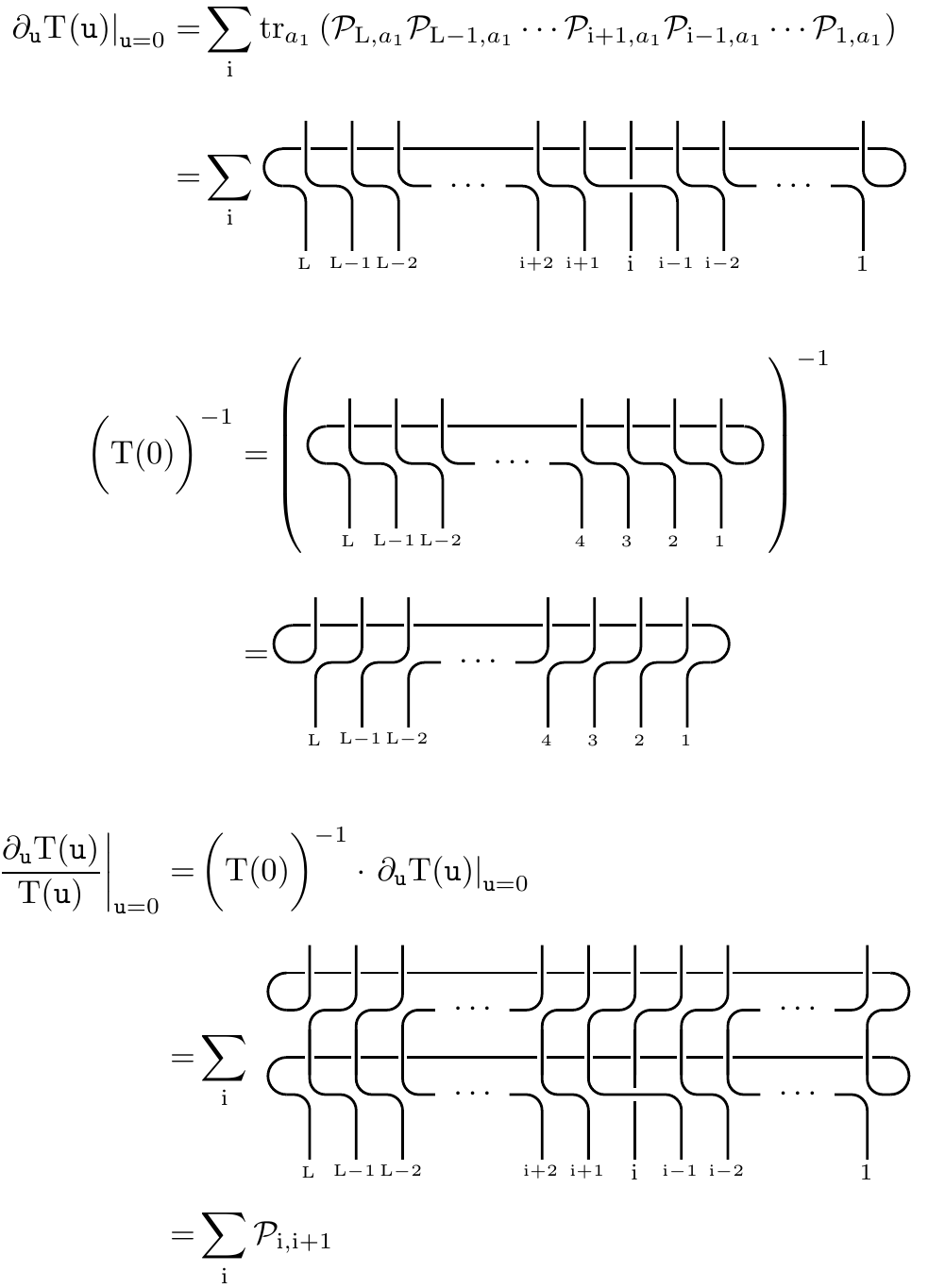}
    \caption{Proof of \protect{~\eqref{eq:Ha=dLogT}}}
      \label{fig:HdT}
  \end{center}
Products of permutations are symbolized by a set of lines. %
If a
line ends at position \(\spk\) in the bottom of a diagram and at position
\(\spl\) in the top of the diagram, then this line stands for a factor
\(\delta_{\coordj_\spk}^{\coordi_\spl}\) in the expression of the
coordinates \(\mathcal{O}_{~{\coordj}_1,{\coordj}_2,\cdots,{\coordj}_\lcds}^{{\coordi}_1,{\coordi}_2,\cdots,{\coordi}_\lcds}\)
  of the {\op} \(\mathcal{O}\) corresponding to this diagram. Vertical lines are associated to the Hilbert
spaces \(\Hilbl_\lcds\), \(\Hilbl_{\lcds-1}\), \(\cdots\),
\(\Hilbl_1\), while the auxiliary space is horizontal. %
With these diagrammatic rules, one immediately notices that for instance
\(\Top(0)^{-1}=\perm_{\sigma_c}\), where \(\sigma_c\) is the cyclic
permutation such that \(\forall \spi>1,~\sigma_c(\spi)=\spi-1\), and
where \(\perm_\sigma\) is defined by \eqref{eq:Pgen} (which
generalizes \eqref{eq:DefPerm} to an arbitrary permutation) in
appendix \ref{sec:notat-tens-prod}.
We finally get \(  \left.\frac{\partial_\su
    \Top(\su)}{\Top(\su)}\right|_{\su=0}=
\sum_{\spi}\perm_{\spi,\spi+1}\), which gives
\eqref{eq:Ha=dLogT}.
\end{minipage}}
\end{figure}

At this point, we have defined a {\faml} of commuting {\ops}
\(\gT[][{ }]\), {\lbd} by an arbitrary \(\su\in ℂ\). 
For periodic {\csds}, we will now see 
that
they are conserved charges, {\ie} that the {\Ham} commutes with
them.
As shown in figure \ref{fig:HdT}, the {\Ham} can be written as
\begin{align}
  \Hami=& - \sum_{\spi}
  \left(2 \perm_{\spi,\spi+1}-\frac 2 {\Kr} 𝕀\right)=\frac 2 {\Kr} \lcds - 2 
\left.\frac{\partial_{\su}
    \gT[][{ }]}{\gT[][{ }]}\right|_{{\su}=0}\\
\framedline{\Hami=}{\frac 2 {\Kr} {\lcds}-2
\left.\partial_{\su} \log \gT[][{ }]\right|_{{\su}=0}}\,\,,
\label{eq:Ha=dLogT}
\end{align}
which immediately implies the commutation
\begin{align}
  \comm{\Hami}{\gT[][{ }]}=&0\,.
\end{align}
Let us note that in \eqref{eq:Ha=dLogT}, the expression
\(\frac{\partial_{\su} \Top({\su})}{\Top({\su})}\) makes sense due to
the commutation relation \eqref{eq:commutT}. Let us also remark that
the identity {\op} \(\bI\) was omitted, like in 
\eqref{eq:HamilHeis}. This omission of implicit identity {\ops}
will be frequent in this manuscript.

We have now shown that for a periodic Heisenberg {\cds},
with {\Ham}  \eqref{eq:hamilHeisenberg}
(or more generally \eqref{eq:HaK1}), there exists a {\faml} of
commuting charges called {\Toprs}. We also showed how the {\Ham}
can be expressed from {\Toprs}. This construction can also be found in  
\cite{Nepomechie:1998jf} or in \cite{Faddeev1996},
where the method of ``algebraic Bethe {\anz}'' is reviewed. This
method allows to derive the Bethe equations \eqref{eq:QuantBetheN} and
to diagonalize the {\Toprs}, to recover the  spectrum
\eqref{eq:BetheEnergyN}. In this manuscript, we will diagonalize
{\Toprs}   through
a different {\ppath}, relying on a ``Bäcklund flow'', introduced
in the next sections.

But in order to proceed with this flow, we actually need %
to define the {\Toprs} in 
a more general context (as in \cite{2008JHEP...10..050KV})
: 
we will deform the periodicity condition (by introducing a twist
\(\g\in GL({\Kr})\)), add  ``inhomogeneities'', and choose more general
auxiliary spaces than \(ℂ^{\Kr}\).

\subsubsection{Inhomogeneous twisted {\cds}}
\label{sec:chain-de-spin}

Let us now define {\Toprs} generalizing the {\op}s we constructed for
the Heisenberg {\cds}.
We will do this by finding more general definitions of {\Roprs} such
that the Yang-Baxter \eqref{eq:RRR=RRR} still holds. This way, all
the results above will immediately apply and define a {\faml} of
commuting {\Toprs}.

\paragraph{Introduction of a twist and of inhomogeneities}
\label{sec:introduction-twist}

The first generalization consists in a modification of the periodicity
condition, and involves a ``twist'' \(\g\in GL({\Kr})\).
In this case, the definition \eqref{eq:Ldef} is replaced with
\begin{align}
\label{eq:DefLtw}
  {\Ltw}^{(\kk)}(\su)=&{\Lop}^{(\kk)}(\su) \cdot \gp {a_\kk}
  =\Rop_{\lcds,a_\kk}(\su)\Rop_{\lcds-1,a_\kk}(\su)\cdots
  \Rop_{1,a_\kk}(\su)\cdot \gp {a_\kk}%
\,,
\end{align}
where \(\gp {a_\kk}\) denotes the {\op} \(𝕀⊗\g\), which
acts on \(\Hilb⊗\Hilbl_{a_{\kk}}\).
Noticing that \(\gp {a_\kk}\) commutes
with \({\Lop}^{(\lL)}(\su)\), as soon as \(\kk\neq\lL\), we get
\begin{align}
  \Rop_{a_1,a_2}(\su-\sv)%
{\Ltw}^{(1)}(\su){\Ltw}^{(2)}(\sv)%
=&\Rop_{a_1,a_2}(\su-\sv)
{\Lop}^{(1)}(\su){\Lop}^{(2)}(\sv)
\cdot \gp {a_1}\cdot \gp {a_2}
\label{eq:RLLt1}
\\
\label{eq:RLLt2}
=&{{\Lop}^{(2)}(\sv) {\Lop}^{(1)}(\su)
\Rop_{a_1,a_2}(\su-\sv)}\cdot \gp {a_1}\cdot \gp {a_2}\\
\label{eq:RLLt3}
=&{{\Lop}^{(2)}(\sv) {\Lop}^{(1)}(\su) \cdot \gp {a_1}\cdot \gp {a_2}\cdot
\Rop_{a_1,a_2}(\su-\sv)}
\\=&{{\Ltw}^{(2)}(\sv) {\Ltw}^{(1)}(\su)
\Rop_{a_1,a_2}(\su-\sv)}\,
\label{eq:RLLt4}
\end{align}
which means that introducing this twist does not break the
 relation \eqref{eq:RLL=LLR}.
The line \eqref{eq:RLLt2} is obtained from \eqref{eq:RLLt1} by simply writing
the equation 
\eqref{eq:RLL=LLR}, whereas \eqref{eq:RLLt3} is obtained by using the
commutation between \(\gp {a_1}\cdot \gp {a_2}\) and
\(\perm_{a_1,a_2}\). 
As before, 
\eqref{eq:RLLt4} implies that the {\Toprs} commute with
each other:
\begin{align}
\label{eq:commutTw}
  \framedline{%
\comm{
      {\Top}(\su)}%
{{\Top}(\sv)} %
=}{0} &\where \framedline{{\Top}(\su)=}%
{\mathrm{tr}_{a_\ii}
 {\Ltw}^{(\ii)}(\sv)}\,\,.
\end{align}
The ``twist'' in the boundary condition gives a different {\Ham},
expressed from
\(\left.\partial_{\su} \log \Top({\su})\right|_{{\su}=0}\). This
{\Ham} is equal to
\begin{align}
\label{eq:Ht=dLogT}
  \Hami=& \frac 2 {\Kr} {\lcds}-2
\left.\partial_{\su} \log \Top({\su})\right|_{{\su}=0}\\
=&\frac 2 {\Kr} {\lcds}-2\left(
\sum_{\spi=1}^{\lcds-1}
  \perm_{\spi,\spi+1}
\right) -2 \perm_{1,{\lcds}}\cdot \gp \lcds^{-1}\cdot \gp 1\,.
\label{eq:Ht=dT2}
\end{align}
The last term, which contains \(\perm_{1,{\lcds}}\), acts only on the
first and the last spins and 
is associated to
the periodicity condition. Only this term is changed with respect to %
\eqref{eq:Ha=dLogT}. 
\begin{proof}[Proof of \eqref{eq:Ht=dT2}]
This relation is checked by exactly the same argument as
\eqref{eq:Ha=dLogT}. Using the notations of figure \ref{fig:HdT}, the
result is obtained by writing
\begin{align}
  \left.\frac{\partial_\su \Top(\su)}{\Top(\su)}\right|_{\su=0}=&
\left(\rule{0pt}{.5cm}\Top(0)\right)^{-1}
\cdot 
  \left.\partial_\su \Top(\su)\right|_{\su=0}
\\=&
\sum_{\spi=1}^{\lcds-1} \raisebox{-1.5cm}{\includegraphics{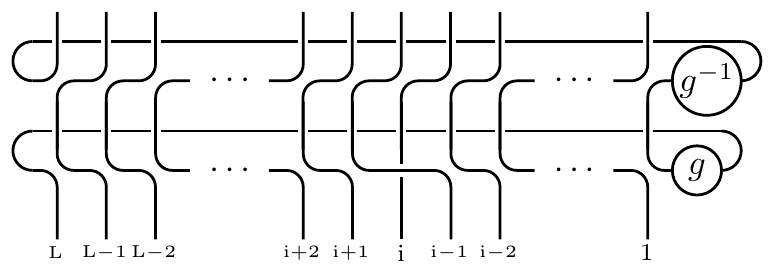}}\nonumber
\\&\qquad\qquad+\raisebox{-1.5cm}{\includegraphics{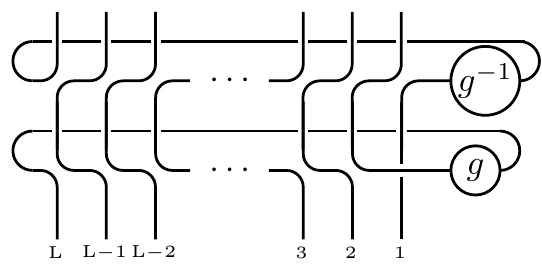}}
\\=&
\sum_{\spi=1}^{\lcds-1}
  \perm_{\spi,\spi+1} + \perm_{1,{\lcds}}\cdot \gp \lcds\cdot \gp 1^{-1}\,.
\end{align}
\end{proof}

A second convenient generalization is obtained by introducing
inhomogeneities \(\theta_\spi\).
\index{ti@\ensuremath{\theta_\spi} (inhomogeneities)}
The definition
\eqref{eq:DefLtw} is then replaced with
\begin{align}
\label{eq:DefLtwIh}
  {\Ltw}^{(\spk)}(\su)=&
  =\Rop_{\lcds,a_\spk}(\su-\theta_\lcds)\Rop_{\lcds-1,a_\spk}(\su-\theta_{\lcds-1})\cdots
  \Rop_{1,a_\spk}(\su-\theta_1)\cdot \gp {a_\spk}%
\,,
\end{align}
and it still gives (by the same arguments as before)
\begin{align}
\label{eq:commutTwIh}
  \framedline{%
\comm
      {{\Top}(\su)} %
{{\Top}(\sv)} %
=}{0} &\where \framedline{{\Top}(\su)=}%
{\mathrm{tr}_{a_\spi}
 {\Ltw}^{(\spi)}(\su)}\,\,.
\end{align}
In this commutation relation \eqref{eq:commutTwIh}, the two {\Toprs}
must be defined with the same torsion {\g}, and the same
inhomogeneities  \(\theta_\spi\), but with two different (but
isomorphic) auxiliary spaces.

\paragraph{General auxiliary space}
\label{sec:gener-auxil-space}

  Finally, one can also define other {\Toprs}
by changing the auxiliary
space. This will give rise to other {\Toprs}, 
which commute with the
{\Toprs} defined in \eqref{eq:commutTwIh}, and are therefore conserved
charges. %
In \eqref{eq:RLL=LLR}, the ``auxiliary'' spaces
\(\Hilbl_{a_{\spk}}=ℂ^{\Kr}\) are isomorphic to the ``physical''
spaces \(\Hilb_{\spi}=ℂ^{\Kr}\) (corresponding to spins in a
superposition of \(\Kr\) different states \(\nket{1}\), \(\nket{2}\),
\(\cdots\), \(\nket{\Kr}\)). By contrast, we will now choose the
auxiliary spaces to be in
a different
 {\rp} of {\GL \Kr}, which means that \(\Hilb_{a_\spk}\) is a
 given vector space%
 , and that there exists a morphism \(\pge: {\GL \Kr} \to
 \GL{\ensuremath{\Hilb_{a_\spk}}}\) such that
\begin{align}
\label{eq:defreprchap2}
  \forall {\g},{\g}'\in \GL \Kr,&~\pg[{\g}\cdot {\g}'][{ }]=\pg[][{ }]\pg[{\g}'][{ }]\,.
\end{align}
We will actually choose a {\rp} characterized by an arbitrary
{\yn} diagram  \(\lambda\)
(see appendix \ref{sec:elements-de-theorie}
for an introduction to representations and {\yn} diagrams), and the
morphism of equation \eqref{eq:defreprchap2} will be denoted
as \(\pg\). %

  Since the ``physical'' spaces \(\Hilbl_\spi\) are in general %
  not isomorphic to the
  ``auxiliary'' ones, the definition \eqref{eq:DefPerm} of the
  permutation {\op}  \(\perm\) does not make sense any more. 
In order to define a new set of {\Toprs} associated to various
representations, let us first define a generalization of the
permutation {\op} and of the {\Rop}-matrix:
\begin{align}
\label{eq:DefRGeneral}
 \Rop_{\spi,\lambda}(\su)=&\su + \perm_{\spi,\lambda}\,,\\
\where \framedline{\perm_{\spi,\lambda}=} {\sum_{1\leq \coordk,\coordl\leq \Kr}
e_{\coordk,\coordl} \otimes
\pe[e_{\coordl,\coordk}]
}\,\,\,.
\label{eq:DefPermGeneral}
\end{align}
In \eqref{eq:DefPermGeneral},
\(e_{\coordk,\coordl}\) is a generator of {\GL{\Kr}} and it acts on
the space \(\Hilbl_\spi\). It can be defined as
\(e_{\coordk,\coordl}=\nket\coordk \nbra\coordl\) (in the basis of
appendix \ref{sec:notat-tens-prod}), or as a matrix with
coefficients \(\left(e_{\coordk,\coordl}\right)_{~\coordj}^{\coordi} =
\delta_{\coordk,\coordi}\delta_{\coordl,\coordj}\) which are all equal
to zero except at position (\(\coordk,\coordl\)). 
By contrast, \(\pe[e_{\coordl,\coordk}]\) denotes the corresponding
generator in the {\rp} \(\lambda\), and it acts on the
auxiliary space. This generator, defined by
\begin{align}
\pge_\lambda\left(  \exp~e_{\coordk,\coordl} \right) =& \exp~\pe
\end{align}
is introduced in more details in the appendix \ref{sec:groupes-de-rang}.

For the %
fundamental\footnote{The fundamental {\rp} is the
  {\rp} given by the vector space \(\Hilb_{a_\spk}=ℂ^{\Kr}\)
and by the morphism \(\pg[][{ }]=\g\).
} {\rp} \(\Hilb_{a_\spk}=ℂ^{\Kr}\) (denoted by 
\(\lambda={\raisebox{-.1cm}{\includegraphics[scale=.3]{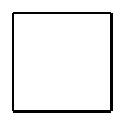}}}\)) 
 we recover the definition
\eqref{eq:DefPerm} of the permutation {\op}.
Indeed, we have 
\begin{gather}
  \sum_{1\leq \coordk,\coordl\leq \Kr}
e_{\coordk,\coordl} \otimes
e_{\coordl,\coordk} \nket{\nn,\mm}=e_{\mm,\nn} \otimes
e_{\nn,\mm} \nket{\nn,\mm}=\nket{\mm,\nn}\,,
\label{eq:PgenFond}\\
\where \nket{\mm,\nn}\equiv \nket{\mm}\otimes\nket{\nn}\,.
\end{gather}

As we will show {\below}, the {\Rop}-matrix defined by
\eqref{eq:DefRGeneral} satisfies 
a Yang-Baxter equation generalizing \eqref{eq:RRR=RRR}:
\begin{align}
\label{eq:YBlambda}
\framedline{  \Rop_{\spi,\lambda}(\su-\sv)\Rop_{\spj,\lambda}(\su)\Rop_{\spi,\spj}(\sv)-}{\Rop_{\spi,\spj}(\sv)\Rop_{\spj,\lambda}(\su)\Rop_{\spi,\lambda}(\su-\sv)=0}\,\,\,.
\end{align}

\begin{proof}
  For ``standard'' permutation {\ops}, we proved the equation
\eqref{eq:RRR=RRR} using the
relation \(\perm_{\spi,\spj}\perm_{\spj,\spk}=\perm_{\spj,\spk}
\perm_{\spi,\spk}\). By contrast, the algebra of generalized
permutations is more subtle: although 
\(\perm_{\spi,\spj}\perm_{\spi,\lambda}\) is equal to
\(\perm_{\spj,\lambda}\perm_{\spi,\spj}\), the product 
\(\perm_{\spi,\lambda}\perm_{\spj,\lambda}\) is in general not equal
to \(\perm_{\spi,\spj}\perm_{\spi,\lambda}\).

Nevertheless, as explained in appendix \ref{sec:elements-de-theorie}
(see \eqref{eq:commgenGLK}, which holds for arbitrary
representations), the commutation relations are the same as for the
fundamental representations. For instance, this implies that
\begin{align}
  \comm{\perm_{\spi,\spj}+\perm_{\spi,\lambda}}{\perm_{\spj,\lambda}}=&0
\label{eq:commp-spj+p-lambd}
\end{align}
holds  even for generalized permutations. 
This allows to prove that the {\lhs} of \eqref{eq:YBlambda},
as a polynomial of \(\su\) and \(\sv\), has all its coefficients equal to
zero. For instance the coefficient of \(\su^0 ~\sv^1\) is zero due to
\eqref{eq:commp-spj+p-lambd}. 
The same argument proves that the coefficient of \(\su^1~\sv^0\) is
zero, whereas the coefficients of \(\su^2~\sv^1\),
\(\su^1~\sv^2\), \(\su^2~\sv^0\), \(\su^0~\sv^2\) and  \(\su^1~\sv^1\)
are trivially zero. Finally, the constant term is
\(\perm_{\spi,\lambda}\perm_{\spj,\lambda} \perm_{\spi,\spj} 
-\perm_{\spi,\spj}\perm_{\spj,\lambda} \perm_{\spi,\lambda}\), and vanishes
due to the relation
\(\perm_{\spi,\spj}\perm_{\spi,\lambda}=\perm_{\spj,\lambda}\perm_{\spi,\spj}\).
\end{proof}

As a consequence of the Yang-Baxter equation \eqref{eq:YBlambda},
we can now define {\Toprs} associated to each {\yn} diagram:
\begin{gather}
\label{eq:DefTtwIh}
\fdisp{  \lT
=%
\mathrm{tr}_{\lambda} \left(
\Rop_{\lcds,\lambda}(\su%
_\lcds)\Rop_{\lcds-1,\lambda}(\su%
_{\lcds-1})\cdots
  \Rop_{1,\lambda}(\su%
  _1)\cdot {\pg}\right)}\,,\\
\where %
\su_\spi\equiv \su-\theta_\spi
\end{gather}
\index{u(s@\ensuremath{\su} (spectral parameter)!ui@\ensuremath{\su_\spi}}
where the partial trace is performed on the auxiliary space, which
corresponds to the {\rp} \(\lambda\).
This generalizes the previous {\Toprs} to the case when
\(\lambda\neq\raisebox{-.1cm}{\includegraphics[scale=.3]{figYdiag_1}}\).

The commutation relation 
\begin{align}
 \framedline{ \forall \su,\sv,\lambda,~~}{\quad   
\comm[0.45cm][-.05cm]{\lT}{\Top(\sv)}
   =0}\,,
\label{eq:TcommLambda}
\end{align}
is then obtained from \eqref{eq:YBlambda}, and it holds for two
{\Toprs} defined with the same twist
\(\g\), and the same inhomogeneities \(\theta_\spi\). They only differ
by the spectral parameter  \(\su\) and the {\rp} \(\lambda\).

The commutation relation \eqref{eq:TcommLambda} ensures that the {\op}s
\(\lT\) are conserved charges, because the {\Ham} is expressed in
terms of the {\op}s  \({\Top}(\su)\).
It is also possible to prove {\another} commutation relation:
\begin{align}
 \standardline{ \forall \su,\sv,%
   s,s'}{\quad %
\comm[0.45cm][-.05cm]{\lT[\su][\overbrace{\raisebox{-.07cm}{\includegraphics[scale=.2]{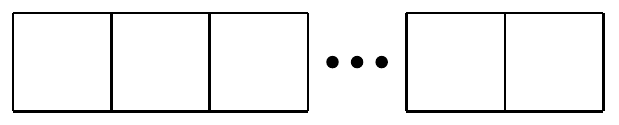}}}^s]}
{\lT[\sv][\overbrace{\raisebox{-.07cm}{\includegraphics[scale=.2]{figYdiag2}}}^{s'}]}
   =0}\,,
\label{eq:TcommSSprime}
\end{align}
where
\(\overbrace{\raisebox{-.07cm}{\includegraphics[scale=.2]{figYdiag2}}}^{s}\)
denotes the {\yn} diagram \(\lambda=(s,0,0,\cdots)\) (see
\eqref{eq:DefYoung}). This diagram corresponds to the {\rp}
obtained by symmetrizing \((ℂ^{\Kr})^{\otimes s}\).

To this end, we should consider the expression \eqref{eq:=GenSym} of the
generators, shown in appendix \ref{sec:elements-de-theorie}. Then, it
is actually possible to rewrite the projector\footnote{As explained in
  the appendix \ref{sec:elements-de-theorie}, 
  this projector
  performs a projection from \((ℂ^{\Kr})^{\otimes s}\) to the
  {\rp}
  \(\underbrace{\raisebox{-.07cm}{\includegraphics[scale=.2]{figYdiag2}}}_{s}\)
  by symmetrizing with respect to all indices.
} \(\proj_{%
  {\includegraphics[scale=.2]{figYdiag2}}%
}\) as a product of fundamental {\Rop}-matrices defined in \eqref{eq:DefR}, as
explained in section 6 of the paper 
\cite{2008JHEP...10..050KV}. Then, the commutation relation
\eqref{eq:TcommSSprime} arises from manipulations of this fundamental
{\Rop}-matrix (see \cite{2008JHEP...10..050KV}). 

Actually we will even show in section \ref{sec:commutation-relation}
that all the {\Toprs} commute with each other:
\begin{align}
 \standardline{ \forall \su,\sv, \lambda, \mu
 }{\quad %
\comm[0.45cm][-.05cm]{\lT[\su][]}
{\lT[\sv][\mu]}
   =0}\,.
\label{eq:TcommLambdaMu}
\end{align}

\index{T-operators@{\Toprs}|)}

\subsection{Differential expression of the {\Toprs}}
\label{sec:expr-diff-des}

In  \cite{2008JHEP...10..050KV}, it was shown that these
{\Toprs} can be expressed in terms of differential {\ops}, by
differentiating with respect to the twist  \(\g\).

In \cite{2008JHEP...10..050KV}, a differential {\op}  {\hD} called
{\cd} is
introduced and defined by
\Pv{\begin{empheq}[box=\fbox]{align}
  \hD \otimes f(\g)&\equiv \parDer {\phi^{^t}} \otimes \left.f\left(e^{\phie%
      } \g\right)\right|_{\phi=0}%
\label{eq:DefD1}\\
&=\sum_{\alpha,\beta}e_{\alpha,\beta}\otimes  \left.\left(
\parDer {\phi_{\beta,\alpha}}f\left(e^{\sum_{\gamma,\delta}
 e_{\gamma,\delta}\phi_{\gamma,\delta}}{\g}\right)
\right)\right|_{\phi\to 0}\,,
\end{empheq}}
\index{M(K)@\ensuremath{\Mat \Kr} : \ensuremath{\Kr\times \Kr} matrices}
where \(\phi\in \Mat \Kr %
\) is a %
\(\Kr\times \Kr\) 
matrix and
{\phie}
denotes \(\sum_{\alpha,\beta} e_{\alpha,\beta}\phi_{\alpha,\beta}\),
where the  \(e_{\alpha,\beta}\) are the generators of \(GL(\Kr)\),
introduced in the appendix \ref{sec:repr-tens-de}
(they are matrices with one single non-zero coefficient at position \(\alpha,\beta\)).
The {\op}  \(\parDer
{\phi^{^t}}\) is a matrix, whose coefficients are differentiations
with respect to the coefficients of the transpose \(\phi^{^t}\) of \(\phi\): 
\begin{align}
\label{eq:DefparDerPhi}
  \parDer {\phi^{^t}}\equiv&\sum_{\alpha,\beta} e_{\alpha,\beta}\parDer {\phi_{\beta,\alpha}}\,.
\end{align}
If  \(f(\g)\) belongs to a space \(E\), we see that 
\(\hD \otimes f(\g)\in \Mat\Kr\otimes E\). In practice \(f(\g)\) will be
a linear {\op} on a Hilbert space \((ℂ^{\Kr})^{\otimes \spi}\):
\begin{align}
  f(\g)~:~&~ (ℂ^{\Kr})^{\otimes \spi}\to (ℂ^{\Kr})^{\otimes \spi}
\\  \hD \otimes f(\g)~:~&~ (ℂ^{\Kr})^{\otimes \spi+1}\to (ℂ^{\Kr})^{\otimes \spi+1}\,,
\end{align}
and we see that in terms of Hilbert space, the {\cd} {\hD}
``adds a spin'' to the {\cds}.
In the particular case when \(f(\g)\) is not an {\op} but a scalar, we
will usually write \(\hDt f(\g)\) instead of \(\hD \otimes f(\g)\).

This definition \eqref{eq:DefD1} is useful to obtain permutation
{\ops}:
\begin{align}
\label{eq:hDrhoG}
  \hD \otimes \pg %
=&\left.\sum_{\alpha,\beta}e_{\alpha,\beta}\otimes \left(
\parDer {\epsilon}\pg[ %
  e^{\epsilon ~
    e_{\beta,\alpha}}%
]\nonumber
\cdot \pg %
\right)\right|_{\epsilon \to 0} \\
=&\left.\sum_{\alpha,\beta}e_{\alpha,\beta}\otimes \left(
\parDer {\epsilon} %
  e^{\epsilon ~
    {\pe[e_{\beta,\alpha}]}}%
\cdot \pg %
\right)\right|_{\epsilon \to 0} \\
=&\left(\sum_{\alpha,\beta}e_{\alpha,\beta}\otimes
    \pe[e_{\beta,\alpha}]\right)
\cdot (\bI\otimes \pg )%
= 
\perm_{\spi,\lambda}\cdot \left(\bI\otimes \pg%
\right)\,.\nonumber
\end{align}

Moreover, 
the usual Leibniz rule (\(({\uu}~{\vv})'={\uu}'~{\vv}+{\uu}~{\vv}'\)) has a generalization for this
{\cd} {\op}: %
\begin{align}
\label{eq:codLeibnitz}  \hD \otimes \left(f_{1}({\g}) \cdot f_{2}({\g})\right)=&\cb{\hD \otimes
  f_{1}(\g)} \cdot \left(\bI \otimes f_{2}(\g)\right) + 
\left(\bI \otimes f_{1}(\g)\right) \cdot \cb{\hD \otimes
  f_{2}(\g) }\,,
\end{align}
Where the ``dot'' symbol (~\(\cdot\)~) denotes the multiplication of
{\ops}, and where \(\hD\) only acts on what is inside the same
brackets \([~~]\).

This implies 
\begin{align}
  \hD\otimes \hD \otimes \pg %
  =& \hD\otimes\left(
    \perm_{2,\lambda}\cdot \left(\bI\otimes \pg %
    \right)
  \right)=\perm_{2,\lambda} 
  \perm_{1,\lambda} \cdot \left(\bI\otimes\bI\otimes \pg %
  \right)\,,
\end{align}
 and more generally
 \begin{align}
\label{eq:LDef=DDD}
\lefteqn{
  (\su_1+\hD)\otimes(\su_2+\hD)\otimes\cdots\otimes(\su_\lcds+\hD)\otimes
\pg %
}\qquad~\nonumber\\
=&\left(\su_\lcds+\perm_{\lcds,\lambda}\right)\cdot
\left(\su_{\lcds-1}+\perm_{\lcds-1,\lambda}\right)\cdots
\left(\su_1+\perm_{1,\lambda}\right) \cdot
\left(\bI^{\otimes\lcds}\otimes \pg %
\right)\\
=&{\Ltw}^{(\lambda)}(\su)\,.
\end{align}

Hence, the {\Toprs}, which are the partial trace of {\Ltw}, can be
written as 
\begin{align}
\label{eq:TfromD}
\framedline{
  \lT %
  =}{\left[\DLt
\cha \lambda(\g)\right]}%
&\where  &
\su_\spi\equiv \su-\theta_\spi\,,
\end{align}
where \(\cha \lambda(\g)\) denotes the character of \(\g\) in the
{\rp} \(\lambda\), {\idest} the trace of \(\pg\) (see appendix
\ref{sec:caracteres}).

An important property that we can notice from this expression
\eqref{eq:TfromD} is that \(\lT\) is polynomial in the variable \(\su\). Let us define
\begin{align}
\label{eq:aDef}
  \framedline{|\lambda|\equiv}{\mathrm{Max}\left\{{\ii}\middle|\lambda_{\ii}>0\right\}}\,\,.
\end{align}
\index{ll@\ensuremath{|\lambda|}}
Then we can see that
for generic \(\g\in\GL \Kr\) (for instance in a vicinity of identity),
this polynomial has degree \(\lcds\) if \(|\lambda|\leq \Kr\), whereas
if \(|\lambda|> \Kr\), the polynomial is identically zero.
Indeed, as
explained in appendix \ref{sec:elements-de-theorie}, \(\cha
\lambda(\g)\) is 
identically zero 
if and only if \(|\lambda|> \Kr\).
This statement %
can be
easily understood as the impossibility to antisymmetrize more than
\(\Kr\) indices if they take values in \(\ninter 1 \Kr\).

The condition \(|\lambda|\leq \Kr\) means that the {\yn} diagram
\(\lambda\) has to lie %
inside
the lattice of \colorprint{\figpref{fig:BosHook}}{\figref{fig:BosHook}},
where the {\yn}
 diagrams are drawn %
as in
 \eqref{eq:DefYoung}, %
 and the constraint \(|\lambda|\leq \Kr\) forces the {\yn}-diagram to
 live inside 
the lattice
 \(\ninter 0 \Kr \times \bN\), where \(\bN\) denotes the
 set \(\{0,1,2,\cdots\}\) of all non-negative integers.
 \index{N@\ensuremath{\bN=\{0,1,2,\cdots\}}}.

\colorprint{\renewcommand\floatpagefraction{.8}
 \begin{figure}[p]}{\begin{figure}}
\colorprint{}{\fbox}{\colorprint{}{\begin{minipage}{.95 \textwidth}}
  \centering
\subfigure[A {\yn} digram %
{\color{white}large phantom }
(here
\(\lambda=(5,3,2,2,0,0,\cdots)\)).
]
{
%
  % \smash
  {
        \begin{tikzpicture}
    \newcommand{\KK}{2.5}
    \newcommand{\UU}{3.2}
    \renewcommand{\RR}{4.2}
    \newcommand{\LL}{-1}
    \newcommand{\rr}{3.9}
    \tikzstyle{axis}=[thick]
    \tikzstyle{diag}=[color=red,very thick]
    \draw[style=axis](\LL,0)--(\RR,0);
    \draw[style=axis](0,0)--(0,\UU);
    \draw[style=axis](0,\KK)--(\rr,\KK);
    \draw[step=.5cm] (0,0) grid (\rr,\KK);
    \node at (-.3,\KK){\(\Kr\)};
    \node at (1.5,-.3){\(\phantom s\)};
    \foreach \n/\l in {{0/2.5},{.5/1.5},{1/1},{1.5/1}}
{
    \fill[black!20] (0,\n) rectangle +(\l,.5);
    \draw[step=.5cm,style=diag] (0,\n) grid +(\l,.5);
}
  \end{tikzpicture}}}\quad \qquad
\subfigure[A rectangular {\yn} diagram %
(here, \(\lambda=(6,6,6,0,0,\cdots)\))]
{
  \begin{tikzpicture}
    \newcommand{\KK}{2.5}
    \newcommand{\UU}{3.2}
    \renewcommand{\RR}{4.2}
    \newcommand{\LL}{-1}
    \newcommand{\rr}{3.9}
    \tikzstyle{axis}=[thick]
    \tikzstyle{diag}=[color=red,very thick]
    \draw[style=axis](\LL,0)--(\RR,0);
    \draw[style=axis](0,0)--(0,\UU);
    \draw[style=axis](0,\KK)--(\rr,\KK);
    \draw[step=.5cm] (0,0) grid (\rr,\KK);
    \fill[black!20] (0,0) rectangle +(3,1.5);
    \draw[step=.5cm,style=diag] (0,0) grid +(3,1.5);
    \node at (-.3,\KK){\(\Kr\)};
    \node at (-.3,1.5){\(a\)};
    \node at (3,-.3){\(s\)};
  \end{tikzpicture}}
  \caption{{\yn} diagrams and \((a,s)\) lattice for {\GL \Kr} (here \(\Kr=5\))}
  \label{fig:BosHook}
\colorprint{}{\end{minipage}}}
\end{figure}

Let us also notice that, in this formalism, the {\Toprs} are obtained
by the action of 
{\cdrs} on characters. 
The expression \eqref{eq:TfromD} of {\Toprs} has the specificity that
is starts from characters (obtained from a trace), and then spins are
``created'' by the action of {\cd}. This is conceptually quite
different from usual definitions like \eqref{eq:DefTtwIh}, where one
should first multiply {\Rop}-matrices corresponding to each spin, and
take the trace afterwards.
As a consequence, several non-trivial properties of the
representations become irrelevant because the only thing we have to know
about representations is their character.

\paragraph{Symmetry group}
\label{sec:symmetry-group}

In this construction, one point which may look surprising is that for
the \(\SU 2\) Heisenberg {\cds}, we introduce a twist \(\g\in\GL 2\)
and various representations of \(\GL 2\). Actually choosing \(\GL \Kr\)
instead of \(\SU \Kr\) simply makes the structure a little more general,
and \(\SU \Kr\) can be obtained by restricting the authorized values of
\(\g\) to \(\SU \Kr\). Then all the representations of \(\GL \Kr\) would be
replaced by representations of \(\SU\Kr\). 

To elaborate a little more, let us notice that using the generators
\(\uplambda^{({\coordl})}\) of \(\SU\Kr\) ({\idest} the Gell-Mann matrices), we
could define an ``{\SU \Kr} {\cd}'' as \linebreak \({\left.\sum_\coordl
\uplambda^{({\coordl})}\otimes \parDer {\phi_\coordl}f\left(e^{-\bi \sum
    \phi_\coordk \uplambda^{({\coordk})}}\right)\right|_{\phi\to 0}}\), where \(\phi\) is
vector with \(\Kr^2-1\) components. Then the formula \eqref{eq:hDrhoG}
would become (for the {\rp} \(\mu\))
\begin{align}
  \hD \otimes \pg[][\mu] 
=&\left.
\sum_\coordl
\uplambda^{({\coordl})}\otimes 
\left(
\parDer {\epsilon}\pg[ 
  e^{-\bi \epsilon ~ \uplambda^{({\coordl})}}%
][\mu]\nonumber
\cdot \pg[][\mu] %
\right)\right|_{\epsilon \to 0} \\
=&-\bi \left(\sum_\coordl \uplambda^{({\coordl})}\otimes
    \pe[\uplambda^{({\coordl})}][\mu]\right)
\cdot (\bI\otimes \pg[][\mu] )
\end{align}

But one can show that \(\left(\sum_\coordl \uplambda^{({\coordl})}\otimes
    \pe[\uplambda^{({\coordl})}][\mu]\right)=4 \perm_{\spi,\mu}- \frac
  4 {\Kr} 𝕀\), which means that this ``{\SU \Kr} {\cd}'' obeys
  exactly the same algebra as the {\GL\Kr} {\cd} up to
  multiplicative factors \(-4 \bi\) and additive terms proportional to
  \(\bI\).

  This means that the analysis we will perform applies as
  efficiently to the group {\SU \Kr} (or even {\SL \Kr}) as to {\GL\Kr}.

\subsection{Generalization to {\sugrs}}
\label{sec:superspins}

{\Sugrs}, such as {\GLKM} introduced in appendix
\ref{sec:gener-glkens}, are groups of ``matrices'' such that the
property 
\eqref{eq:ComBosTens} sometimes holds up to a sign, {\idest} there exist
``anticommuting'' objects such that %
\((A\otimes\bI)\cdot(\bI\otimes B)
= -(\bI\otimes B)\cdot(A\otimes\bI)\). For these groups,
the construction of the {\Rop} matrix in
equation \eqref{eq:DefR} (or more generally \eqref{eq:DefRGeneral})
has to be modified so that the Yang-Baxter equation %
still
holds. %
For these
{\sugrs}, the ``physical'' Hilbert spaces
\(\Hilbl_\spi\) are linear combinations of \(\nket{1},\nket{2},\cdots,
\nket{\Kr+\Mr}\), where \(\nket{1},\nket{2},\cdots,
\nket{\Kr}\) are ``commuting'' and \linebreak \(\nket{\Kr+1},\nket{\Kr+2},\cdots,
\nket{\Kr+\Mr}\) are ``anti-commuting''.
This means that for instance \(\left(\nbra{\ii}\otimes
  \nbra{\jj}\right)\cdot \left(\nket{\kk}\otimes
  \nket{\lL}\right) = \pm \nbranket{\ii}{\kk} \nbranket{\jj}{\lL}\) where
the sign is plus if \(\Min(\jj,\kk)\leq \Kr\) ({\idest} if either \nket{\jj} or
\nket{{\kk}} is ``commuting'') and minus otherwise.
The signs can be summarized by saying that both \(\nket\ii\) and \(\nbra \ii\)
have the grading
\(\gr{\coordi}\in \bZ/2\bZ\) defined by:
\begin{align}
{\sg {\coordi}}=&1&\If&\coordi\in\ninter 1 \Kr\\
{\sg {\coordi}}=&-1&\If&\coordi\in\ninter {\Kr+1} \Mr\,.
\end{align}
For arbitrary objects \(A\) and \(B\) with well-defined gradings 
\(\gr A\) and \(\gr B\), we have
\(\left(A\otimes \bI \right)\cdot\left(\bI \otimes B
\right)=A\otimes B = (-1)^{\gr A \gr B}\left(\bI \otimes B \right)\cdot\left(A\otimes
  \bI \right)\). 
For usual matrix groups such as \GL \Kr, all gradings are zero and
there is %
no
such sign.
The %
simplest way to define a permutation {\op} would be to keep the definition
\eqref{eq:PermFromGen} unchanged. Let us see how it would then act on
\(\ket{\coordi}\otimes\ket{\coordj}\):
\begin{align}
{\left( \sum_{1\leq \coordk,\coordl\leq \Kr+\Mr}
  \ket \coordk \bra \coordl%
\otimes
  \ket \coordl \bra \coordk%
\right)\cdot
\ket{\coordi}\otimes\ket{\coordj}}%
\nonumber %
=&\sum_{1\leq \coordk,\coordl\leq \Kr+\Mr}
(-1)^{\left(\gr{\coordk}+\gr\coordl\right)\gr \coordi} \left(
  \ket \coordk \bra \coordl%
\ket{\coordi}\right)\otimes
\left(
  \ket \coordl \bra \coordk%
\ket{\coordj}\right)\\
=&(-1)^{\left(\gr{\coordj}+\gr\coordi\right)\gr
  \coordi}\ket{\coordj}\otimes \ket{\coordi}\,,
\label{eq:=badsuperperm2}
\end{align}
using
the fact that the grading of \(\ket \coordl \bra \coordk\)
is \(\gr{\coordk}+\gr\coordl\).%

Finally, one can check that if the permutation {\op} was %
defined as \linebreak \(\perm_{1,2}=\sum_{1\leq \coordk,\coordl\leq \Kr+\Mr}
  \ket \coordk \bra \coordl%
\otimes
  \ket \coordl \bra \coordk\), then due to the sign in
  \eqref{eq:=badsuperperm2} the Yang-Baxter equation
  \eqref{eq:RRR=RRR} would fail. 
To ensure that 
the Yang-Baxter equation
still holds,
  it is actually sufficient to change the
  signs in the definition of \perm, which gives the definition
  \begin{align}
\label{eq:superPerm}
    \perm_{1,2}=&
\sum_{1\leq \coordk,\coordl\leq \Kr+\Mr}
{\sg  \coordl}
  \ket \coordk \bra \coordl
\otimes
  \ket \coordl \bra \coordk\,.
  \end{align}
This equation gives very naturally
\begin{align}
  \perm_{1,2}\cdot
\ket{\coordi}\otimes\ket{\coordj} =& (-1)^{\gr \coordi \gr \coordj}
\ket{\coordj}\otimes\ket{\coordi} \,,
\end{align}
which is the natural generalization of \eqref{eq:DefPerm} when some
vectors are anti-commuting.

For more spins, one gets for instance %
\begin{align}
  {\perm_{1,3}\cdot
\ket{\coordi}\otimes\ket{\coordj}\otimes\ket{\coordn}}%
=&\left(\sum_{1\leq \coordk,\coordl\leq \Kr+\Mr}
{\sg  \coordl}
  \ket \coordk \bra \coordl
\otimes\bI\otimes
  \ket \coordl \bra \coordk
\right)
\ket{\coordi}\otimes\ket{\coordj}\otimes\ket{\coordn}
\\=&
\left(\sum_{1\leq \coordk,\coordl\leq \Kr+\Mr}
(-1)^{\gr \coordl+\left(\gr \coordl+\gr \coordk\right)\left(\gr
    \coordi + \gr \coordj \right)}
  (\ket \coordk \braket \coordl \coordi)
\otimes\ket \coordj\otimes
  (\ket \coordl \braket \coordk \coordn)
\right)\\=&
 (-1)^{\gr \coordi+\left(\gr \coordi+\gr \coordn\right)\left(\gr
    \coordi + \gr \coordj \right)} \ket \coordn \otimes \ket \coordj
\otimes \ket \coordi
\label{eq:superm3}
\end{align}

To show that \eqref{eq:RRR=RRR} holds with the definition
\eqref{eq:superPerm}, it is sufficient to prove that
\(\perm_{\spi,\spj}\perm_{\spj,\spk}=\perm_{\spj,\spk}
\perm_{\spi,\spk}\) holds. %
For instance for three spins:
\begin{align}
  \perm_{1,2}\perm_{2,3} \ket{\coordi,\coordj,\coordk}=&
(-1)^{\gr \coordk(\gr
  \coordi+\gr\coordj)}\ket{\coordk,\coordi,\coordj}=
\perm_{2,3}\perm_{1,3} \ket{\coordi,\coordj,\coordk}\,,
\end{align}
where we used \eqref{eq:superm3}. It is not complicated to deduce that
\(\perm_{\spi,\spj}\perm_{\spj,\spk}=\perm_{\spj,\spk}
\perm_{\spi,\spk}\) also holds for more spins.

Finally with this definition of the permutation {\op}, the
construction of section \ref{sec:chaines-de-spin} still gives a
{\faml} of commuting {\Toprs}.

This definition of the permutation {\op}, introduced here to
  reproduce the Yang-Baxter identity \eqref{eq:RRR=RRR}, actually also
  allows to define representations associated to {\Ydag}s (see
  appendix \ref{sec:gener-glkens}), and to define {\Toprs} associated
  to arbitrary {\Ydag}s.%

In order to also write the expression of {\Toprs} in terms of
{\cdrs}, as in section \ref{sec:expr-diff-des}, we then have to
incorporate the same sign into the definition of the {\cd}, by
replacing \eqref{eq:DefparDerPhi} with
\begin{align}
    \parDer {\phi^{^t}}\equiv&\sum_{\alpha,\beta} {\sg  \beta} e_{\alpha,\beta}\parDer {\phi_{\beta,\alpha}}\,.
\end{align}
The introduction of this sign is such that the relation \(  \hD
\otimes \pg %
  =
\perm_{\spi,\lambda}\cdot \left(\bI\otimes \pg %
\right) \)
holds by the same argument as %
in \eqref{eq:hDrhoG}. %

Therefore, the relation \eqref{eq:TfromD} holds for {\sugrs} as
well, and we will now be indistinctly working with either a
{\sugr}, or a more ``standard'' matrix group such as \GL \Kr.

This construction gives a set of commuting {\ops}, which we want to
interpret as conserved quantities of a given model. For this we can
for instance define the {\Ham}
\begin{align}
  \Hami=& \frac 2 {\Kr+\Mr} \lcds - 2 
\left.\partial_{\su} \log 
\gT[][
\raisebox{-.07cm}{\includegraphics[scale=.2]{figYdiag_1}}]
\right|_{{\su}=0}\,,
\label{eq:Ha=dLogTsu}
\end{align}
where
\(\gT[][\raisebox{-.07cm}{\includegraphics[scale=.2]{figYdiag_1}}]\) is
the {\Topr} corresponding to the fundamental {\rp} (this
{\op} was denoted \(\Top(\su)\) in the section \ref{sec:equation-de-yang}).

In the case when all inhomogeneities are set to zero, it can be
rewritten as
\begin{equation}
  \Hami=\frac 2 {\Kr+\Mr} 𝕀- 2\left( \sum_{\spi=1}^{\lcds-1}
\perm_{\spi,\spi+1}\right) -2 \perm_{1,{\lcds}}\cdot \gp \lcds\cdot \gp 1^{-1}\,,
\end{equation}
exactly like in \eqref{eq:Ht=dT2}.
This {\Ham} corresponds to interactions between nearest neighbors,
because it is the sum over \(\spi\) of an {\op} (the generalized
permutation) acting only on the spins \(\spi\) and \(\spi+1\). We will
see that with this {\Ham}, 
the {\cds} is {\ing}, due to the relation \eqref{eq:Ha=dLogTsu}
between \(\Hami\) and \(\gT[][
\raisebox{-.07cm}{\includegraphics[scale=.2]{figYdiag_1}}]\).
For this {\Ham}, the
{\Toprs} are conserved quantities, because they commute with 
\(\gT[][
\raisebox{-.07cm}{\includegraphics[scale=.2]{figYdiag_1}}]\).

An important difference with the bosonic case ({\idest} the {\GL \Kr} spin
  chain of section \ref{sec:operateurs-t}) is that the character
\(\cha \lambda ({\g})\) is nonzero when \(\lambda_{\Kr+1}\leq \Mr\)~\cite{Deguchi:1991aq}. This
condition reduces to the condition \(|\lambda|\leq \Kr\) if \(\Mr=0\), but
if \(\Mr\neq 0\), it gives rise to the lattice of %
\colorprint{\figpref{fig:FatHook}}{\figref{fig:FatHook}}, where we see that the {\yn} diagrams are forced to
lie inside a lattice, having the shape of a {\fat} letter {\Lnothing}. This lattice
will be called a {\fat} {\hook} in this manuscript.

\colorprint{\begin{figure}[p]}{\begin{figure}}
  \centering
\colorprint{}{\fbox}{\colorprint{}{\begin{minipage}{.95 \textwidth}}
    \begin{center}
      \subfigure[A {\yn} digram %
{\color{white}large phantom }
      (here
      \(\lambda=(7,6,4,4,3,1,0,0,\cdots)\)).]  {
        \vspace{-2cm}
        \begin{tikzpicture}
          \newcommand{\KK}{2} \renewcommand{\MM}{1.5}
          \newcommand{\UU}{4.2} \renewcommand{\RR}{4.2}
          \newcommand{\LL}{-1} \newcommand{\rr}{3.9}
          \renewcommand{\mm}{3.9} \tikzstyle{axis}=[thick]
          \tikzstyle{diag}=[color=red,very thick]
          \draw[style=axis](\LL,0)--(\RR,0);
          \draw[style=axis](0,0)--(0,\UU);
          \draw[style=axis](\MM,\KK)--(\rr,\KK);
          \draw[style=axis](\MM,\KK)--(\MM,\mm); \draw[step=.5cm]
          (0,0) grid (\rr,\KK); \draw[step=.5cm] (0,\KK) grid
          (\MM,\mm); \node at (-.3,\KK){\(\Kr\)}; \node at
          (\MM,-.3){\(\Mr\)}; \foreach \n/\l in
          {{0/3.5},{.5/3},{1/2},{1.5/2},{2/1.5},{2.5/.5}} {
            \fill[black!20] (0,\n) rectangle +(\l,.5);
            \draw[step=.5cm,style=diag] (0,\n) grid +(\l,.5); }
        \end{tikzpicture}}\quad\qquad \subfigure[A rectangular {\yn} diagram
      (\(\lambda=(7,7,0,0,0,\cdots)\)).]  {
        \vspace{-2cm}
        \begin{tikzpicture}
          \newcommand{\KK}{2} \renewcommand{\MM}{1.5}
          \newcommand{\UU}{4.2} \renewcommand{\RR}{4.2}
          \newcommand{\LL}{-1} \newcommand{\rr}{3.9}
          \renewcommand{\mm}{3.9} \tikzstyle{axis}=[thick]
          \tikzstyle{diag}=[color=red,very thick]
          \draw[style=axis](\LL,0)--(\RR,0);
          \draw[style=axis](0,0)--(0,\UU);
          \draw[style=axis](\MM,\KK)--(\rr,\KK);
          \draw[style=axis](\MM,\KK)--(\MM,\mm); \draw[step=.5cm]
          (0,0) grid (\rr,\KK); \draw[step=.5cm] (0,\KK) grid
          (\MM,\mm); \node at (-.3,\KK){\(\Kr\)}; \node at
          (\MM,-.3){\(\Mr\)}; \fill[black!20] (0,0) rectangle
          +(3.5,1.); \draw[step=.5cm,style=diag] (0,0) grid +(3.5,1.);
          \node at (-.3,\KK){\(\Kr\)}; \node at (-.3,1.){\(a\)}; \node
          at (3.5,-.3){\(s\)};
        \end{tikzpicture}} \vspace{-.5cm}
    \end{center}
  \caption{{\yn} diagrams and \((a,s)\) lattice for
    {\GL{\Kr\ensuremath{|}\Mr}} (here \(\Kr=4\) and \(\Mr=3\))} 
  \label{fig:FatHook}
\colorprint{}{\end{minipage}}}
\end{figure}

\subsection[Hirota equation and CBR determinant formula]{Hirota equation and Cherednik-Bazhanov-Reshetikhin determinant formula}
\label{sec:hirota-equat-cher}

In this section, we will introduce 
the ``fusion relations'' between the {\Toprs}
corresponding to various representations.
We will introduce the
results obtained in \cite{2008JHEP...10..050KV} (in particular the
derivation of the CBR
determinant formula \eqref{eq:CBR} found in \cite{springerlink:10.1007/BF01077327,1990JPhA...23.1477B}), and briefly introduce the proof
of these results.

First let us introduce a particular {\faml} of {\Toprs}
corresponding to ``rectangular'' {\yn} diagrams: in view of the
parameterization \eqref{eq:DefYoung} let us define %
\begin{align}
\label{eq:DefRecT}
\framedline{\rT\equiv\lT[\su][\las%
]}{\qquad\quad%
  \where
  \las
  \equiv
(\underbrace{s,s,\ldots,s,s}_{\ensuremath{a\textrm{ 
        times}}},0,0,\ldots)}
\end{align}
which corresponds to a rectangular {\yn} diagram of horizontal size \(s\) and
vertical size \(a\). %

As explained in appendix \ref{sec:elements-de-theorie}, the character
of a generic group element \({\g}\in{\GLKM}\) is nonzero if and only if \(\lambda\)
obeys the condition \(\lambda_{\Kr+1}\leq \Mr\). Using
\eqref{eq:TfromD} we deduce that {\gT} is a  nonzero polynomial of the
variable \(\su\) for {\Ydag}s if and only if \(\lambda_{\Kr+1}\leq \Mr\).
For rectangular representations, that means that {\rT} is nonzero if
and only if \linebreak
\((a,s)\in \HK(\Kr,\Mr)\), where the ``{\fat} {\hook}'' \(\HK(\Kr,\Mr)\) is defined by
\begin{align}
\label{eq:DefHK}
  \HK(\Kr,\Mr)\equiv&\left\{(a,s)\in\bN\times\bZ \middle|
  \begin{array}{ll}
    s\geq 0 &\And 0\leq a\leq \Kr\\
    &\Or\\
    a\geq 0 &\And 0\leq s\leq \Mr\\
    &\Or\\
    a=0
  \end{array}
\right\}\,.
\end{align}
\index{L(K,M)@\ensuremath{\HK(\Kr,\Mr)}}
This lattice of authorized values of \((a,s)\) is shown in the %
\figpref{fig:BosHook}
for \(\GL \Kr={\GL{\Kr\ensuremath{|}{0}}}\) and in the
\figpref{fig:BosHook} for \(\GL{\Kr\ensuremath{|}{\Mr}}\).
Moreover, one should note that %
the representations associated to
\((a=0;~ s\in \bZ)\) or to \((s=0;~ a\in \bN)\) are identical (they correspond
to \(\lambda=(0,0,0,\cdots)\)), which means that
\(\rT[][0]=\rT[][][0]=\rT[][0][0]\). In the present case, this
{\rp} is associated to the {\Topr}
\(\rT[][0][0]=\prod_{\spi=1}^\lcds \su_\spi\).

An important relation satisfied by {\Toprs}
is then the
following determinant relation (which is called the
``Cherednik-Bazhanov-Reshetikhin'' formula \cite{springerlink:10.1007/BF01077327,1990JPhA...23.1477B}, and which we will prove below) 
\begin{align}
\label{eq:CBR}
\framedline{  \lT %
  =}{\frac{\Det{\rT[\su+1-\ii][1][\lambda_\jj+\ii-\jj]}{1\leq\ii,\jj\leq
    |\lambda|}}{\prod_{{\coordk}=1}^{|\lambda|-1}
\rT[\su-{\coordk}][0][0]}}\,\,,
\end{align}
where \(|\lambda|\) is defined by \eqref{eq:aDef}.
\index{CBR formula}
\index{Cherednik-Bazhanov-Reshetikhin|see{CBR}}
\index{Bazhanov-Reshetikhin|see{CBR}}

The numerator of the {\rhs} is the determinant of the
\(|\lambda|\times|\lambda|\) matrix whose coefficients are the commuting
{\ops} \(\rT[\su+1-\ii][1][\lambda_\jj+\ii-\jj]\), where \(\rT[][1]\)
corresponds to the symmetric {\rp} \(\las [1]=\underbrace{\includegraphics[scale=.3]{figYdiag2}}_{s}\).

This equation generalizes the relation
\begin{align}
  \label{eq:Weyl02}
  \cha \lambda({\g})=&%
\Det{
\chs {{\lambda_\ii+\jj-\ii}}
({\g})}
   {1\leq \ii,\jj\leq|\lambda|}\,, 
\end{align}
on characters (see appendix
\ref{sec:caracteres}), where \(\chs {s}\) denotes the character
associated to the {\rp} \(\las[1]\). It 
allows to express an arbitrary
{\Topr} in terms of the {\ops} {\rT[\su][1]}. That will allow to
prove the commutation relation \eqref{eq:TcommLambdaMu}, but also to
show the Hirota equation \eqref{eq:Hirota}, found in \cite{1992PhyA..183..304K,Kuniba:1992qv,springerlink:10.1007/s002200050165,1997JPhA...30.7975T}.

Though this determinant relation was proven in \cite{2008JHEP...10..050KV},
it will be helpful to recall this proof, which extensively relies on
the ``{\cd}'' formalism.
The proof is done in two main steps: the first step shows that the
{\rhs} of \eqref{eq:CBR} is a polynomial (by showing the vanishing
of specific minors \eqref{eq:KV4.3} of the determinant), and the second step
checks that it really coincides with {\lT}.

\subsubsection{Proof of the CBR formula : Part one}
\label{sec:proof-cbr-formula}

Let us start by proving that the {\rhs} of \eqref{eq:CBR} is
polynomial:

By definition, all the {\rT[\sv][1]} inside %
the determinant are polynomial functions of the variable
{\sv}. Moreover, the denominator is a polynomial, which can be written
explicitly due to the relation %
\begin{align}
\label{eq:TT00}
  \rT[\su][0][0]\equiv&\prod_{\spi=1}^\lcds \su_\spi=\prod_{\spi=1}^\lcds (\su-\theta_\spi)\,.
\end{align}
\index{ti@\ensuremath{\theta_\spi} (inhomogeneities)}
\index{u(s@\ensuremath{\su} (spectral parameter)!ui@\ensuremath{\su_\spi}}
This relation is nothing but the definition \eqref{eq:TfromD}, where
the {\rp} associated to an empty {\yn} diagram
\(\las[0][0]%
=\las[a][0]%
=\las[0][s]%
\) has the character
\(\cha {{\las[1][0]%
}}(\g)=1\), as it can be seen from the relation
\eqref{eq:SymCharac} in the appendix \ref{sec:repr-tens-de}.

Hence in order to prove that the {\rhs} of \eqref{eq:CBR} is polynomial,
it is sufficient to prove that it has no pole, {\idest} that the determinant
is zero\footnote{Rigorously, %
the argument works
  under the condition that the denominator only has simple
  zeroes {\idest} under the condition that %
  the
  \(\theta_{\spi}-\theta_{\spj}\)'s are not integer. We will explain
  later in the text how to deal with this constraint. %
} when
\(\su=\theta_\spi+{\coordk}\) for arbitrary \(\spi\in\ninter 1 \lcds, {\coordk}\in
\ninter 1 {|\lambda|-1}\).
To do this, we can expand the determinant with respect to the two
successive lines at position \(\coordk\) and \(\coordk+1\). That gives  a
sum of terms
of the form \(\left(\rT[\theta_{\spi}+1][1][s_1]
\rT[\theta_{\spi}][1][s_2]-\rT[\theta_{\spi}+1][1][s_2-1]
\rT[\theta_{\spi}][1][s_1+1]\right)\cdot
\Dm(\coordk,\coordk+1;{\jj}_0,{\jj}_1)\), where \({\Dm}(\coordk,\coordk' ;
\coordl,\coordl')\) denotes the \(\left(|\lambda|-2\right) \times \left(|\lambda|-2\right)\) minor
obtained by removing the lines \(\coordk\) and \(\coordk'\) and the
columns \(\coordl\) and \(\coordl'\) from the determinant, and where we introduced
\(s_1=\lambda_{{\jj}_0}+\coordk-{\jj}_0\)
and \(s_2=\lambda_{{\jj}_1}+\coordk+1-{\jj}_1\).

In order to prove that this numerator is zero when
\(\su=\theta_\spi+{\coordk}\), we will prove 
the relation
\begin{align}
\label{eq:KV4.3}
 \framedline{ \rT[\theta_{\spi}+1][1][s_1]\cdot
\rT[\theta_{\spi}][1][s_2]-\rT[\theta_{\spi}][1][s_1+1]\cdot
\rT[\theta_{\spi}+1][1][s_2-1] =}{0}\,\,\,, 
\end{align}
which %
means that all
the terms of the 
determinant (expanded with respect to the columns \(\coordk\)
and \(\coordk+1\)) vanish. %
That will prove that the {\rhs} of 
\eqref{eq:CBR} has no pole, and is indeed a polynomial function of the variable \(\su\).

In order to prove \eqref{eq:KV4.3}, it is very convenient to introduce
the generating series of \(\rT[\su][1]\) and to rewrite \eqref{eq:KV4.3}
as the equivalent statement
\begin{gather}
\label{eq:WKV4.3}
\Wt[][\theta_{\spi}+1]\cdot \Wt[y][\theta_{\spi}]-%
\frac y z
\Wt[z][\theta_{\spi}]\cdot \Wt[y][\theta_{\spi}+1]=0\\
\where
~~\fdisp{\Wt\equiv\sum_{s=0}^{\infty}\rT[\su][1]z^s}{}=\left[%
\DLt
w(z)\right]\,,
\end{gather}
where \(w(z)\equiv \sum_{s=0}^{\infty} z^s \chs s\) is defined by
\eqref{eq:DefWz} in the appendix 
\ref{sec:YoungTab} introducing the representations associated
to given {\yn} diagrams.

To prove the relation \eqref{eq:WKV4.3} for arbitrary
inhomogeneities \(\theta_\spj\),
we will first prove it in the simplest case
: when all inhomogeneities are equal to zero. In that case, the
relation that we have to prove is simply
\begin{align}
\label{eq:DD.1+DD=1+DD.DD}
 z \left[(1+\hD)^{\otimes \lcds} w(z)\right] \cdot
    \left[\hD^{\otimes \lcds} w(y)\right] =& 
 y  \left[\hD^{\otimes \lcds} w(z)\right] \cdot \left[(1+\hD)^{\otimes \lcds} w(y)\right]
\end{align}

\begin{proof}[Proof of \eqref{eq:DD.1+DD=1+DD.DD}]
  Using the appendix \ref{sec:diagr-expr-co}, %
  we will now see that \eqref{eq:DD.1+DD=1+DD.DD} has a \linebreak remarkably
  simple %
  proof, which relies on a diagrammatic expression of {\ops} like \linebreak
  \({\left[(1+\hD)^{\otimes \lcds} w(z)\right]}\). For instance, one can
  show the relation  \(\hDt w(\xx)=\frac {\g \xx}{1- \g\xx} w(\xx)\), which can
  be graphically represented as
  \begin{align}
    \hDt w(\xx)=& \MyLine
    w(\xx)\,.
  \end{align}
  where the line \(\MyLine\) stands for the {\op} \(\frac {\g \xx}{1-
    \g\xx}\).  Next, one computes \linebreak \(\hD\otimes \hDt w(\xx)=\left( \frac
    {\g \xx}{1- \g \xx} \otimes \frac {\g \xx}{1- \g \xx}
    +\perm_{1,2}(\frac {1}{1- \g \xx} \otimes \frac {\g \xx}{1- \g
      \xx})\right)w(\xx)\), which can be written diagrammatically as
  \begin{align}
    \label{eq:DiagDDw} \hD\otimes \hDt w(\xx)=&\left(\MyTwoNodes+
      \MyTwoNodes[{jj2/ii1}][{jj1/ii2}] \right)w(\xx)\,,
  \end{align}
  where \(\MyLine[{}][{jj1/ii1}]\) denotes the {\op} \(\frac {1}{1- \g
    \xx}\). More details about this diagrammatic can be found in the
  appendix \ref{sec:diagr-expr-co}, including a very simple pattern to
  write \(\hD^{\otimes\lcds} w(\xx)\): one should write a term for
  each permutation, and dash all the lines going up to the right.  For
  instance, this rule gives
  \begin{align}
    \hD^{\otimes 3} w(\xx)=&\left(\MyThreeNodes +
      \MyThreeNodes[{jj1/ii1,jj3/ii2}][{jj2/ii3}]
      +\MyThreeNodes[{jj3/ii3,jj2/ii1}][{jj1/ii2}]
      +\MyThreeNodes[{jj2/ii2,jj3/ii1}][{jj1/ii3}]
      +\MyThreeNodes[{jj3/ii2,jj2/ii1}][{jj1/ii3}]
      +\MyThreeNodes[{jj3/ii1}][{jj1/ii2,jj2/ii3}] \right) w(\xx)\,.
    \label{eq:Diag3Dw}
  \end{align}
In this expression each picture 
is a graphical
{\rp} of a given {\op}: 
for instance \(\MyThreeNodes\) stands for
the {\op} \(\frac {\g \xx}{1- \g\xx}\otimes \frac {\g \xx}{1-
  \g\xx}\otimes \frac {\g \xx}{1- \g\xx}\), whereas
\(\MyThreeNodes[{jj3/ii3,jj2/ii1}][{jj1/ii2}]\) stands for 
the {\op} \(\perm_{1,2}\cdot\left(
\frac {1}{1- \g\xx}\otimes \frac {\g \xx}{1-
  \g\xx}\otimes \frac {\g \xx}{1- \g\xx}
\right)\). These representations of {\ops} will be
called \(\hD\)-diagrams, because they arise from the Leibniz rule when the
effect of successive {\cdrs} is computed.
\index{D-diagrams@\ensuremath{\hD}-diagrams}

  One can also show that a very similar diagrammatic rule can be used
  to compute \(\left(1+\hD\right)^{\otimes 3} w(\xx)\): in this case,
  one should dash all the lines which are either vertical, or going up
  to the right. For instance, one gets
  \begin{align}
    \left(1+\hD\right)^{\otimes 3} w(\xx)=&\left(
      \MyThreeNodes[][{jj1/ii1,jj2/ii2,jj3/ii3}] +
      \MyThreeNodes[{jj3/ii2}][{jj1/ii1,jj2/ii3}]
      +\MyThreeNodes[{jj2/ii1}][{jj3/ii3,jj1/ii2}]
      +\MyThreeNodes[{jj3/ii1}][{jj2/ii2,jj1/ii3}]
      +\MyThreeNodes[{jj3/ii2,jj2/ii1}][{jj1/ii3}]
      +\MyThreeNodes[{jj3/ii1}][{jj1/ii2,jj2/ii3}] \right) w(\xx)\,.
    \label{eq:diad31+DD}
  \end{align}

  Let us then consider the {\op} \(\left[\left(1+\hD\right)^{\otimes
      3} w(\xx)\right] \cdot \perm_{\sigma_c}\) where \(\sigma_c\) is
  the cyclic permutation \(\sigma_c(1)=\lcds\),
  \(\sigma_c(\spi+1)=\spi\).  For an arbitrary {\op} \(\mathcal{O}\),
  the coordinates of the product \(\mathcal{O}\cdot \perm_{\sigma_c}\)
  are easily obtained as
  \begin{align}
    \left(\mathcal{O}\cdot
      \perm_{\sigma_c}\right)_{~{\coordj}_1,{\coordj}_2,\cdots,{\coordj}_\lcds}^{{\coordi}_1,{\coordi}_2,\cdots,{\coordi}_\lcds}=&
    \mathcal{O}_{~{\coordj}_2,{\coordj}_3,\cdots,{\coordj}_\lcds,{\coordj}_1}^{{\coordi}_1,{\coordi}_2,~\,\cdots\,~,{\coordi}_\lcds}\,.
  \end{align}  

As a consequence, if the {\op} \(\mathcal{O}\) corresponds to a given
\(\hD\)-diagram (as introduced above, or with more details in the
appendix \ref{sec:diagr-expr-co}), then the {\op}
\(\mathcal{O}\cdot \perm_{\sigma_c}\) corresponds to {\another}
\(\hD\)-diagram, obtained by 
  (cyclicly) shifting
 to the left the lower
  dots of each \(\hD\)-diagram, 
  to get (for instance for \(\lcds=3\))
  \begin{gather}
    \label{eq:1+DLwP}
      \left[\left(1+\hD\right)^{\otimes 3} w(\xx)\right] \cdot
      \perm_{\sigma_c}%
    =\left(
      \MyThreeNodes[{}][{jj1/ii2,jj2/ii3}][\forget][{jj3/ii1}] +
      \MyThreeNodes[{jj2/ii2}][{jj1/ii3}][\forget][{jj3/ii1}]
      +\MyThreeNodes[{jj1/ii1}][{jj2/ii3}][\forget][{jj3/ii2}]
      +\MyThreeNodes[{jj2/ii1}][{jj1/ii2}][\forget][{jj3/ii3}]
      +\MyThreeNodes[{jj2/ii2,jj1/ii1}][][\forget][{jj3/ii3}]
      +\MyThreeNodes[{jj2/ii1}][{jj1/ii3}][\forget][{jj3/ii2}] \right)
    w(\xx)\,.
  \end{gather}%
\colorprint{\JacTruFig \renewcommand\floatpagefraction{.5}}{}%
  This coincides exactly with \eqref{eq:Diag3Dw} up to the fact that
  one line (emphasized, in red in the online version) is dashed
  instead of solid. Indeed, both \eqref{eq:diad31+DD} and
  \eqref{eq:1+DLwP} are sums over all permutations \(\sigma\in
  \Sgrp \lcds\), with a specific dashing rule: in
  \eqref{eq:diad31+DD}, the solid lines are the lines going up to the
  left. After multiplication by the permutation, these lines go up
  either vertically or to the left, and the fact that they are solid
  consistently reproduces \eqref{eq:Diag3Dw}. The same argument holds
  for dashed lines, except for the line connected to the
  bottom-right-dot in \eqref{eq:diad31+DD}. Thus, for arbitrary
  \(\lcds\geq 1\), \(\left[\left(1+\hD\right)^{\otimes \lcds}
    w(\xx)\right] \cdot \perm_{\sigma_c}\) coincides with
  \(\hD^{\otimes \lcds} w(\xx)\) up to the fact that in every \(\hD\)-diagram,
  the line connected to the bottom-right-dot is dashed instead of
  solid. If we recall that solid (resp dashed) lines stand for \(\frac
  {\g~\xx}{1-\g~\xx}\) (resp \(\frac {1}{1-\g~\xx}\)), one gets the
  statement
  \begin{align}
    \label{eq:leftl-lcds-wxxr}
    \left[\left(1+\hD\right)^{\otimes \lcds} w(\xx)\right] \cdot
    \perm_{\sigma_c} \cdot \left(\bI^{\otimes (\lcds-1)}\otimes \g ~
      \xx\right)=&\hD^{\otimes \lcds} w(\xx)\,.
  \end{align}

  The same arguments allow to prove that
  \begin{align}
    \label{eq:leftb-lcds-1otim}
    \left(\bI^{\otimes (\lcds-1)}\otimes \g ~ y \right)^{-1} \cdot
    \perm_{\sigma_c}^{-1} \cdot \left[ \hD^{\otimes \lcds} w(y)\right]
    =&\left(1+\hD\right)^{\otimes \lcds} w(y)\,.
  \end{align}
  Multiplying \eqref{eq:leftl-lcds-wxxr} by
  \eqref{eq:leftb-lcds-1otim} gives exactly
  \eqref{eq:DD.1+DD=1+DD.DD}.
\end{proof}

\begin{proof}[Proof of \eqref{eq:WKV4.3}]
  As shown in the section
  \ref{sec:bilinear-identities} (in appendix
  \ref{app:cod}), a simple recurrence allows to show that the identity
  \eqref{eq:DD.1+DD=1+DD.DD} implies the more general relation
  \eqref{eq:MINoNest}, which reads
  \begin{gather}
    \label{eq:MID0Nest}
    \Wt[z][\su+1]\cdot \Wt[y][\su]-%
    \frac y z
    \Wt[z][\su]\cdot \Wt[y][\su+1]%
    =(1-\frac y z)\Wt[y,z][\su+1]\cdot \left[\prod_{\spi=1}^\lcds
      \su_\spi\right]\,,
    \\
    \where ~~%
    {\Wt[y,z]\equiv}%
    {\left[%
        \DLt w(y)w(z)\right]}\,.
  \end{gather}
  If we recall that \(\su_\spi\equiv \su-{\theta_\spi}\), this
  relation immediately implies \eqref{eq:WKV4.3}.
\end{proof}

This proves the vanishing
\eqref{eq:KV4.3} of the determinant \eqref{eq:CBR} at the 
zeroes of the denominator, because all terms in its expansion with
respect to the 
lines \(\coordk\) and \(\coordk+1\) vanish.

If the denominator only has simple zeroes, this is enough to show that
the  {\rhs} of \eqref{eq:CBR} is indeed polynomial in the variable \({\su}\).

On the other hand, the denominator has multiple zeroes only if there
exist some inhomogeneities \(\theta_{\spi}\) and \(\theta_{\spj}\) such
that \(\theta_\spi-\theta_\spj\) is an integer.
If this is the case,
the zeroes of the denominator are easily transformed
into simple zeroes
 by adding small perturbations to the
\(\su_\spi\)'s. %
When these perturbations are removed, several
zeroes of the denominator collide (to form a zero of multiplicity
greater than one), while the same number of zeroes collide in the
numerator, giving rise to a zero with (at least) the same
multiplicity. This shows that the {\rhs} of \eqref{eq:CBR} is indeed
polynomial in the variable \({\su}\), even when the denominator has zeroes
with multiplicities.

\subsubsection{Proof of the CBR formula : part two}
\label{sec:proof-cbr-formual}

Having proven that the {\rhs} of 
\eqref{eq:CBR} is a polynomial (as a function of the spectral
parameter \(\su\)), the next step is
now
 to show that it coincides exactly
with the {\lhs}. To this end, %
the %
denominator in \eqref{eq:CBR} can be explicited
from \eqref{eq:TT00} and then incorporated into the determinant:
\begin{align}
{  \frac{\Det{\rT[\su+1-\ii][1][\lambda_\jj+\ii-\jj]}{1\leq\ii,\jj\leq
    a}}{\prod_{{\coordk}=1}^{a-1}
\rT[\su-{\coordk}][0][0]}}%
=&
\Det{
  \DLt[1+\frac 1 {\su_\spi +1-\spi}\hD] \chs {\lambda_\jj+\ii-\jj}(\g)
}{1\leq\ii,\jj\leq
    a}\cdot\prod_{\spi=1}^{\lcds}\su_\spi\,,
  \label{eq:toCBR}
\end{align}
where \(\chs s\) is the character of the symmetric {\rp},
introduced in \eqref{eq:DefWz}
\index{cs@\ensuremath{\chs s \equiv \chas [1]}}.

This expression can be expanded in each variable \(\su_\spi\) around
\(\su_\spi\to\infty\). %
For instance, when 
\(\su_1 \to \infty \), \eqref{eq:toCBR} becomes
\begin{multline}
{  \frac{\Det{\rT[\su+1-\ii][1][\lambda_\jj+\ii-\jj]}{1\leq\ii,\jj\leq
    a}}{\prod_{{\coordk}=1}^{a-1}
\rT[\su-{\coordk}][0][0]}}%
\\=
  (\su_1+\hD)~~\Det{
  \DLt[1+\frac 1 {\su_\spi +1-\spi}\hD][2] \chs {\lambda_\jj+\ii-\jj}(\g)
}{1\leq\ii,\jj\leq
    a}\cdot\prod_{\spi=2}^{\lcds}\su_\spi %
+\mathcal{O}\left(\frac 1 {\su_1}\right)\,.
\label{eq:totoCBR}
\end{multline}
\begin{proof}[Proof of \eqref{eq:totoCBR}]
  In the determinant on the {\rhs} of \eqref{eq:toCBR},
the term of degree \(0\) in \(\su_1\) is
  simply %
  \(\Det{\DLt[1+\frac 1 {\su_\spi +1-\spi}\hD][2] \chs
  {\lambda_\jj+\ii-\jj}(\g) }{1\leq\ii,\jj\leq a}\), 
obtained %
  by neglecting all terms with \(\frac 1 {\su_1}\). %
After
multiplication by \(\prod_{\spi=1}^{\lcds}\su_\spi\), this gives the
term of degree \(1\) in \({\su}_1\). The next term is obtained by recalling
that a determinant is a sum (running over permutations) of products of
coefficients. For each term of this sum, the coefficient of \(\frac 1
{\su_1}\) is obtained by keeping a \(\frac {\hD} {\su_1}\) in one (and
only one) of the factors. %
This
prescription exactly
coincides with the {\cd} of
  \(\frac 1 {\su_1}\Det{\DLt[1+\frac 1 {\su_\spi +1-\spi}\hD][2] \chs
  {\lambda_\jj+\ii-\jj}(\g) }{1\leq\ii,\jj\leq a}\),
expressed through the Leibniz rule.
\end{proof}

Moreover, %
we can note that due to the
polynomiality of the {\lhs}, the term \(\mathcal{O}\left(\frac
  1 {{\su}_1}\right)\) in \eqref{eq:totoCBR} is necessarily equal to
zero. We can then reproduce the 
  argument to expand the result \eqref{eq:totoCBR} around \(\su_2\to \infty\), and
  iterate up to \(\su_\lcds\to\infty\).  After these iterations, we get

\begin{align}
{  \frac{\Det{\rT[\su+1-\ii][1][\lambda_\jj+\ii-\jj]}{1\leq\ii,\jj\leq
    a}}{\prod_{{\coordk}=1}^{a-1}
\rT[\su-{\coordk}][0][0]}}=%
\DLt 
\Det{
\chs {\lambda_\jj+\ii-\jj}(\g)
}{1\leq\ii,\jj\leq
    a}\,.
  \label{eq:CBRr}
\end{align}
Finally, the Weyl formula \eqref{eq:Weyl02} allows to write the
{\rhs} of \eqref{eq:CBRr} as \linebreak \(\left[\DLt
\cha \lambda(\g)\right]\), which is equal to \(\lT\). As a consequence,
\eqref{eq:CBRr} is exactly the CBR formula \eqref{eq:CBR}.

\subsubsection{Fusion rule and commutation relation \texorpdfstring{\protect{\eqref{eq:TcommLambdaMu}}}{}}
\label{sec:commutation-relation}

As explained in the construction of the model, the {\Toprs} obey the
commutation \eqref{eq:TcommSSprime}\footnote{One should remember that
  \(\lT[\su][\overbrace{\raisebox{-.07cm}{\includegraphics[scale=.2]{figYdiag2}}}^s]
  = \rT[][1]\)}. This means that all the {\Toprs}
in the right-hand side of \eqref{eq:CBR} commute with each other, and
therefore, \eqref{eq:CBR} implies the general commutation relation 
\begin{align}
   \forall& \su,\sv, \lambda, \mu,&
\framedline{\comm[0.45cm][-.05cm]{\lT[\su][]}
{\lT[\sv][\mu]}}
   {=0}\,.
\label{eq:TcommLambdaMu2}
\end{align}

But of course, the CBR determinant formula \eqref{eq:CBR} tells much
more than just a commutation relation: it tells how to express the
{\Toprs} for an arbitrary {\rp} in terms of {\Toprs} for simpler
{\rp}s corresponding to {\yn}-diagrams with one single
row. This result is often called a ``fusion rule''.

Moreover, we will now show (in section \ref{sec:cbr-determ-form})
that
when restricted to 
rectangular representations \eqref{eq:DefRecT}, this
CBR determinant formula 
\eqref{eq:CBR} %
is equivalent to
the following bilinear
identity, called the Hirota Identity \cite{1992PhyA..183..304K,Kuniba:1992qv,springerlink:10.1007/s002200050165,1997JPhA...30.7975T}:
\begin{align}
  \label{eq:Hirota}
\rT[\su+1] \cdot \rT = \rT[\su+1][a+1]\cdot\rT[][a-1]+\rT[\su+1][][s-1]\cdot\rT[][][s+1]
\end{align}
\index{Hirota equation}
This identity involves the product of commuting {\ops}, and it
occurs frequently for {\ing} models. What we will show in the next
sections is that this identity allows to diagonalize the {\Toprs} and
to recover the spectrum of the theory.

Moreover, as we have already seen, the {\Toprs} are nonzero only
inside the lattice \(\HK(\Kr,\Mr)\) of \figpref{fig:FatHook}.
In the next section we will
investigate some properties of the solutions of Hirota equation on
this lattice. An explicit proof of these properties will then be given in
section \ref{sec:expr-diff-Qop}, where fundamental quantities of
interests, called {\Qoprs}, will be defined.

\subsection{{\jacobi} identity and bilinear equations}
\label{sec:jacobiident-bilin}

Let us now show in what sense the Hirota equation \eqref{eq:Hirota} is
equivalent to the CBR formula \eqref{eq:CBR}. To do this we will use
an important tool, which is the 
{\jacobi} identity%
. It is a general identity on determinants and allows to prove an
equivalence between bilinear relations and determinant formulae.

We will see that this {\jacobi} identity allows to prove on the one hand that
the CBR formula \eqref{eq:CBR}, once restricted to rectangular
representations, is equivalent to the Hirota equation
\eqref{eq:Hirota}. On the other hand, we will %
see that the CBR formula is also
equivalent to {\another} bilinear relation, which we will call the {\MID}.

This %
will involve the
minors of an
arbitrary determinant. Let us define
\begin{align}
  {\De}(\kk,\lL ;
    \mm,\nn)=\Det{a_{\coordi,\coordj}}{\substack{\kk\leq\coordi\leq\lL\\
\mm\leq\coordj\leq\nn}}\,.
\end{align}
Then the {\jacobi} identity is the very general statement that for any
coefficients \(a_{\coordi,\coordj}\) (which are either numbers, or
{\ops} commuting with each other), 
\begin{multline}
  \label{eq:{\jacobi}}
{\De}(1,\nn;1,\nn){\De}(2,\nn-1;2,\nn-1)=
{\De}(1,\nn-1;1,\nn-1){\De}(2,\nn;2,\nn)\\ -
{\De}(1,\nn-1;2,\nn){\De}(2,\nn;1,\nn-1)\,.
\end{multline}
\colorprint{}{\JacTruFig}%
This identity is represented graphically in
\colorprint{\figpref{fig:{\jacobi}}}{\figref{fig:{\jacobi}}},
where the matrix with coefficients \((a_{{\coordi},{\coordj}})\) is represented by a
blue square. Its minors are denoted by a square where some lines and
columns are grayed-out. If the matrix is of size \(2\times 2\), we can
recognize the usual definition of the determinant.

\subsubsection{CBR determinant formula and Hirota equation}
\label{sec:cbr-determ-form}

Let us start by illustrating this in the case of Hirota equation:
first, we restrict the CBR determinant formula to the rectangular
{\rp} \(\las\) defined in \eqref{eq:DefRecT}:
\begin{gather}
\label{eq:CBRrect}
  \rT 
   =\frac{
 \Det{\rT[\su+1-\ii][1][s+\ii-\jj]}{1\leq\ii,\jj\leq
     a}}{\prod_{{\coordk}=1}^{a-1}
 \rT[\su-{\coordk}][0][0]}\,,
  \\{\ie\quad} \frac{\rT }{\rT [][0][0]}=
  \Det{\frac{\rT[\su+1-\ii][1][s+\ii-\jj]}{\rT[\su+1-\ii][0][0]}}{1\leq\ii,\jj\leq
      a}
\,.
\end{gather}
We then choose the coefficients 
\begin{align}
a_{\coordi,\coordj}=  &
\frac{\rT[\su+1-\coordi][1][s+\coordi-\coordj]}{\rT[\su+1-\coordi][0][0]}
\end{align}
and write the {\jacobi} identity \eqref{eq:{\jacobi}} for \(\nn=a+1\):
\begin{align}
  \frac{\rT[][a+1] }{\rT [][0][0]}   \frac{\rT[\su-1][a-1] }{\rT
    [\su-1][0][0]}
=&\frac{\rT}{\rT [][0][0]} \frac{\rT[\su-1]}{\rT [\su-1][0][0]} 
-  \frac{\rT[][][s-1] }{\rT [][0][0]} \frac{\rT[\su-1][][s+1] }{\rT [\su-1][0][0]}\,.
\end{align}
This way, we see that the Hirota equation \eqref{eq:Hirota} is a direct
consequence of the CBR determinant formula \eqref{eq:CBR} (or actually
its restriction \eqref{eq:CBRrect} to rectangular {\yn} diagrams).

Interestingly enough, we can also go the other way round, and show
that the Hirota equation \eqref{eq:Hirota} implies the rectangular CBR
formula \eqref{eq:CBRrect}, under the condition that \(\rT=0\) if \(a<0\),
that \(\rT[][0]=\rT[][0][0]\) is not identically zero, and that the
solution is ``typical'' in a sense which will be explained {\below}.
 Indeed, if we assume that
\eqref{eq:Hirota} holds, then its restriction to \(a=1\) gives
\begin{align}
\rT[][2] = \frac{
  \begin{vmatrix}
    \rT[\su][1]&\rT[\su][1][s-1]\\\rT[\su-1][1][s+1]&\rT[\su-1][1]
  \end{vmatrix}
}{\rT[\su-1][0][0]}\,,
\label{eq:CBRrect2}
\end{align}
which gives the \((a=2)\) case of the rectangular CBR
formula \eqref{eq:CBRrect}. %

Then one can iteratively express \(\rT\) for increasing values of \(a\).
For instance if we plug the expression \eqref{eq:CBRrect2} of
\(\rT[][2]\) into the Hirota equation, we get
\begin{align}
\label{eq:CBRa3}
\rT[][3] =& \frac{
  \begin{vmatrix}
    \rT[\su][2]&\rT[\su][2][s-1]\\\rT[\su-1][2][s+1]&\rT[\su-1][2]
  \end{vmatrix}
}{\rT[\su-1][1]}
\end{align}
if \(\rT[\su-1][1]\) is non-zero. We can then plug the expression
\eqref{eq:CBRrect2} to get
\begin{align}
\label{eq:CBRa3develop}
\rT[][3]
=&
\frac{ 
\begin{vmatrix}
    \rT[\su][1]&\rT[\su][1][s-1]\\\rT[\su-1][1][s+1]&\rT[\su-1][1]
  \end{vmatrix}
\cdot \begin{vmatrix}
    \rT[\su-1][1]&\rT[\su-1][1][s-1]\\\rT[\su-2][1][s+1]&\rT[\su-2][1]
  \end{vmatrix}
}
{{\rT[\su-1][1]}{\rT[\su-1][0][0]\rT[\su-2][0][0]}}\nonumber\\
&\quad -
\frac{ 
\begin{vmatrix}
    \rT[\su][1][s-1]&\rT[\su][1][s-2]\\\rT[\su-1][1]&\rT[\su-1][1][s-1]
  \end{vmatrix}
\cdot \begin{vmatrix}
    \rT[\su-1][1][s+1]&\rT[\su-1][1]\\\rT[\su-2][1][s+2]&\rT[\su-2][1][s+1]
  \end{vmatrix}
}
{{\rT[\su-1][1]}{\rT[\su-1][0][0]\rT[\su-2][0][0]}}
\\
=&\frac{\rT[\su-2][1][s+1] \left(\rT[][1][s-2] \rT[\su-1][1][s+1]-\rT[\su-1][1][s-1]
   \rT[][1]\right)}{{\rT[\su-1][0][0]\rT[\su-2][0][0]}}\nonumber\\
&~~+\frac{\rT[\su-2][1] \left(\rT[\su-1][1] \rT[][1]-\rT[][1][s-1]
   \rT[\su-1][1][s-1]\right)}{{\rT[\su-1][0][0]\rT[\su-2][0][0]}}\nonumber\\
&~~+\frac{\rT[\su-2][1][s+2]\left(\rT[\su-1][1][s-1] \rT[][1][s-1]-\rT[][1][s-2] \rT[\su-1][1]\right)
   }{{\rT[\su-1][0][0]\rT[\su-2][0][0]}}\\
=&\frac{\left|%
\begin{array}{ccc}
 \rT[\su ][1][s] & \rT[\su ][1][s-1] & \rT[\su ][1][s-2] \\
 \rT[\su -1][1][s+1] & \rT[\su -1][1][s] & \rT[\su -1][1][s-1] \\
 \rT[\su -2][1][s+2] & \rT[\su -2][1][s+1] & \rT[\su -2][1][s]
\end{array}
\right|}{\rT[\su-1][0][0]\rT[\su-2][0][0]}
\,,\label{eq:finallyCBR3}
\end{align}

The equation \eqref{eq:finallyCBR3} obtained this way is  %
exactly the \((a=3)\) case of the rectangular CBR
formula \eqref{eq:CBRrect}. The reason why the result coincides with
\eqref{eq:CBRrect} is simply that \eqref{eq:CBRrect} satisfies the
Hirota equation. Then a simple recurrence shows that if the Hirota
equation \eqref{eq:Hirota} holds, then one gets iteratively the
rectangular CBR formula \eqref{eq:CBRrect} when \(a\geq 2\). The case
\(a=1\) of \eqref{eq:CBRrect} is also
trivially true because it reduces to \(\rT [][1]=\rT [][1]\).

This proof that 
the
bilinear equation \eqref{eq:Hirota} is equivalent to the
determinant expression \eqref{eq:CBRrect}
holds
 provided the recurrence
sketched above never involves a division by a {\Topr} which is
identically zero. 
Let us now %
see how to deal with this constraint: 
as we will see,
a correct statement is that
the %
``typical'' solutions of the Hirota equation \eqref{eq:Hirota} are
given by the determinant expression \eqref{eq:CBRrect}.
In this statement, a ``typical solution'' of the Hirota equation
\index{Typical (solution)}
\eqref{eq:Hirota} is a solution \(\rT[][][][(0)]\) of Hirota equation such that, for every
small perturbation
\(\rT[][1][][(\epsilon)]=\rT[][1][][(0)]+\mathcal{O}(\epsilon)\) of \(\rT[][1][][(0)]\), there
exists a solution \(\rT[][][][(\epsilon)]\) of Hirota equation such that
\(\rT[][][][(0)]=\lim_{\epsilon\to 0} \rT[][][][(\epsilon)]\) and
  such that
  \(\rT[][a][][(\epsilon)]=0\) if \(a<0\). I will
not enter into the details here, but it is easy to show that for a
``typical 
solution'' of the Hirota equation, for every \((a,s,\su)\) one can find a small
perturbation \(\rT[][1][][(\epsilon)]\) such that in the vicinity of
\(\epsilon=0\), \(\rT[\su-1][a-2][]\neq 0\) and the Hirota equation allows to
express \(\rT[][][]\) and to proceed with the recurrence.

~

To conclude this remark about typical solutions, let us give an
example of a non-typical solution of Hirota equation:
\begin{align}
\label{eq:HirNonTyp1}
  \rTf=&1&&\If (a,s)\in\HK(1,0)~\textrm{ or if }~ a=4 \And s\geq 0\,,\\
  \rTf=&0&&\oth\!\!.
\label{eq:HirNonTyp2}
\end{align}
Then one possible choice of perturbation \(\rTf[][1][][(\epsilon)]\) is
given by
\begin{align}
  \rTf[][1][][(\epsilon)]=&\chas[1](\g[(\epsilon)])&\where
  \g[(\epsilon)]\equiv&
  \mathrm{diag}(1,\epsilon,\epsilon,\cdots,\epsilon)\in {\GL 5}\,.
\end{align}
For this choice of perturbation, if a solution \(\rTf[][][][(\epsilon)]\)
exists for all \(a\), then we can show by recurrence (with the
arguments above) that for \(a\leq 5\),
we get
\begin{align}
    \rTf[][][][(\epsilon)]=&\chas[](\g[(\epsilon)])\,,
\end{align}
and (for \(a\leq 5\)), this recurrence never involves a division by
zero. Then we see that for \(s\geq 1\), \(\lim_{\epsilon\to 0}
\rTf[][4][][(\epsilon)] =0 \neq \rTf[][4]\), which proves that the
solution (\ref{eq:HirNonTyp1},\ref{eq:HirNonTyp2}) of Hirota equation
is a non-typical solution, which is why it does not satisfy the CBR
determinant formula \eqref{eq:CBRrect}.

\subsubsection{{\MMID}}
\label{sec:mid}

To construct the {\Qoprs} and the Bäcklund flow, we will need a
combinatorial identity on {\cdrs}, which reads as follows:
\begin{align}
\label{eq:MIDww}
\lefteqn{(z_1-z_\nn)  \Wt[%
z_1,\cdots , z_\nn
  ][\su+1] \cdot
\Wt[%
z_2,\cdots,z_{\nn-1}
]}\nonumber\phantom{\where\quad }%
& \\
&\qquad=~ z_1 \Wt[z_1%
,\cdots,z_{\nn-1}][\su+1] \cdot
\Wt[z_2%
,\cdots,z_\nn]\nonumber\\
&\qquad ~~~-z_\nn \Wt[z_1%
,\cdots,z_{\nn-1}] \cdot
\Wt[z_2%
,\cdots,z_\nn][\su+1]\\
\where\quad &{\Wt[z_\lL,\cdots,z_\mm]\equiv}{\left[%
\DLt
 w(z_\lL)w(z_{\lL+1})w(z_{\lL+2})\cdots w(z_\mm)\right]}\,,
\end{align}
where \(w(z)\equiv \sum_{s=0}^{\infty} z^s \chs s\) is the generating
series of symmetric characters.
This formula holds for arbitrary \(\g\in {\GL{\Kr\ensuremath{|}\Mr}}\),
 \(\lcds\geq 0\), \(\nn\geq 2\), \(\su\in\bC\), \(\{\theta_\spi\}\in\bC^\lcds\) and \(\{z_{\ii}\}\in\bC^\nn\).
It generalizes the identity \eqref{eq:MID0Nest} to the \(\nn\geq 2\)
case.

To prove this identity, let us first use the 
{\jacobi}
identity to prove %
that \eqref{eq:MIDww} is  equivalent to the
determinant expression
\begin{gather}
\label{eq:MIDdet}
{  \Wt[z_1,z_2,\cdots,z_\nn]}
= \frac 1 {\prod_{\coordk=1}^{\nn-1}
    \Wt[\emptyset][\su%
    -\coordk]}
  \frac{\Det{z_\coordj^{1-\coordk}
{\Wt[z_\coordj][\su+1-\coordk]}%
}{1\leq \coordj,\coordk\leq \nn}}{
\VdM
}\, \\
\where \Wt[\emptyset]=\prod_{\spi=1}^{\lcds}\su_\spi\qquad \And \quad
\VdM[b][a]=\Det{z_\coordj^{a-\coordk}
 }{a\leq \coordj,\coordk\leq b}
\,.
\end{gather}
\begin{proof}
  This equivalence between \eqref{eq:MIDdet} and \eqref{eq:MIDww} is
  proven by the same means as the equivalence between
  \eqref{eq:CBRrect} and \eqref{eq:Hirota}, and we will just sketch it
  here.

  First, one proves that the determinant \eqref{eq:MIDdet} satisfies
  the equation \eqref{eq:MIDww}. For this, one writes the {\jacobi}
  identity \eqref{eq:{\jacobi}} for the coefficients
  \(a_{\coordj,\coordk}= z_\coordj^{1-\coordk} \frac
  {\Wt[z_\coordj][\su+1-\coordk]}{\Wt[\emptyset][\su+1-\coordk]}
  \). For instance the minor \({\De}(2,\nn-1;2,\nn-1)\) is %
  equal
  to \( \frac{\Wt[z_2,\cdots,z_{\nn-1}][\su-1]}{\Wt[\emptyset][\su-1]}
  \VdM[\nn-1][2]
  / \prod_{\coordj=2}^{\nn-1} z_\coordj\). By writing carefully all
  terms of the {\jacobi} identity, and using the following property of
  the {\vdm} determinant \(\VdM\):
  \begin{align}
    \VdM[\nn-1][2]\VdM=&\left(\frac 1{z_\nn}-\frac
      1{z_1}\right)\VdM[\nn-1]\VdM[][2] \,,
  \end{align}
  we exactly obtain \eqref{eq:MIDww}.

To finish the proof of the equivalence between \eqref{eq:MIDww}
and \eqref{eq:MIDdet}, one proves that \eqref{eq:MIDww} implies
\eqref{eq:MIDdet} by a recurrence over the
number \(\nn\) of parameters \(z_1,\cdots,z_\nn\). Indeed, \eqref{eq:MIDww}
allows to express \(\Wt[z_1,z_2,\cdots,z_\nn]\) in terms of \(\Wt[J]\), for
different strict subsets \(J\) of \(\{z_1,z_2,\cdots,z_\nn\}\).
Exactly like in (\ref{eq:CBRa3develop}-\ref{eq:finallyCBR3}), one can
compute explicit expressions, but it is not necessary, since the
{\jacobi} identity ensures that the outcome will be exactly
\eqref{eq:MIDdet}.
 This recurrence %
 completes the proof of the equivalence between
\eqref{eq:MIDww} and \eqref{eq:MIDdet}.
\end{proof}

Now, in order to finish with the proof of \eqref{eq:MIDww}, we just
have to prove the determinant expression \eqref{eq:MIDdet}. %
Let us
present two different proofs: first a simple proof which reproduces
the arguments used in section \ref{sec:hirota-equat-cher} to prove
the CBR formula. The second proof will show that the equality
\eqref{eq:MIDdet} itself
is 
actually
equivalent to the CBR formula. 

\begin{proof}[First proof of \eqref{eq:MIDdet}]
A simple way to show this relation
is by first checking that the
{\rhs} is polynomial, and then by expanding it. This proof is
exactly the same as the proof given in section
\ref{sec:hirota-equat-cher} for the CBR formula 
\eqref{eq:CBR}, and it relies on the fact that when \(\su=\theta_\spi+{\coordk}\) (for arbitrary \(\spi\in\ninter 1 \lcds, {\coordk}\in
\ninter 1 {\nn-1}\)), the minors associated to the lines \(\coordk\) and
\(\coordk+1\) vanish due to the identity \eqref{eq:WKV4.3}. Then the
expansion of the determinant around \(\su_\spi=\infty\) is performed
exactly like in \eqref{eq:totoCBR}. It gives
\begin{multline}
  \frac 1 {\prod_{\coordk=1}^{\nn-1}
    \Wt[\emptyset][\su%
    -\coordk]}
  \frac{\Det{z_\coordj^{1-\coordk}
{\Wt[z_\coordj][\su+1-\coordk]}%
}{1\leq \coordj,\coordk\leq \nn}}{
\VdM
}\\=\DLt 
  \frac{\Det{z_\coordj^{1-\coordk}
w(z_\coordj)
}{1\leq \coordj,\coordk\leq \nn}}{
\VdM
}=\Wt[%
z_1,\cdots , z_\nn
  ]\,.\qedhere
\end{multline}
\end{proof}

Another proof of the {\MID}, written in \cite{Kazakov:2010iu}, sheds
more light into the relation between \eqref{eq:MIDdet} and the CBR
formula \eqref{eq:CBR}. Understanding this proof will also be
interesting for further generalizations of this result.

\begin{proof}[Second proof: Equivalence between \eqref{eq:MIDdet} and \eqref{eq:CBR}]
  Let us expand the quantity \linebreak \(\Wt[z_1,z_2,\cdots,z_\nn] \cdot 
\VdM\) in powers of \(z_1,z_2,\cdots,z_\nn\):
\begin{multline}
\label{eq:pMIDzExpL}
  \Wt[z_1,z_2,\cdots,z_\nn] \cdot 
\VdM\\= \DLt \sum_{\sigma \in \Sgrp {\nn}}
\sum_{%
(s_1,s_2,\ldots,s_\nn)\in \bN^\nn
} \epsilon(\sigma)
\prod_{\kk=1}^{\nn} \chs {s_\kk}(\g) z_\kk^{s_\kk+1-\sigma(\kk)}
\end{multline}
where 
\(\epsilon(\sigma)\equiv
\prod_{\ii<\jj}\frac{\sigma(\ii)-\sigma(\jj)}{\ii-\jj}\) is the
signature of the permutation \(\sigma\). In \eqref{eq:pMIDzExpL}, we
can see that the coefficient of \(\prod_{\kk=1}^\nn z_\kk^{\lambda_\kk+1-\kk}\)
is
\begin{multline}
  \DLt \sum_{\sigma \in {\Sgrp \nn}}
\epsilon(\sigma)
\prod_{\kk=1}^{\nn} \chs {
\lambda_\kk+\sigma(\kk)-\kk
}(\g) %
\\=
\DLt
\Det{\chs {
\lambda_\kk+{\jj}-\kk
}(\g)}{1\leq {\jj},\kk\leq |\lambda|}%
= \gT\,.
\end{multline}

 By comparison, we can expand
\(%
{\Det{z_\coordj^{1-\coordk}
{\Wt[z_\coordj][\su+1-\coordk]}%
}{1\leq \coordj,\coordk\leq \nn}
} %
\), which reads
\begin{multline}
{\Det{z_\coordj^{1-\coordk}
{\Wt[z_\coordj][\su+1-\coordk]}
}{1\leq \coordj,\coordk\leq \nn}
}
 \\
=
\sum_{\sigma \in {\Sgrp \nn}}
  \sum_{
(s_1,s_2,\ldots,s_\nn)\in \bN^\nn
} \epsilon(\sigma)
\prod_{\kk=1}^{\nn} \rT[\su+1-\sigma(\kk)][1][s_\kk] z_\kk^{1-\sigma(\kk)+s_\kk}\,,
\end{multline}
where the coefficient of \(\prod_{\kk=1}^\nn z_\kk^{\lambda_\kk+1-\kk}\) is
\begin{gather}
  \sum_{\sigma \in {\Sgrp \nn}}
\epsilon(\sigma)
\prod_{\kk=1}^{\nn} \rT[\su+1-\sigma(\kk)][1][
\lambda_\kk+\sigma(\kk)-\kk
]%
=
\Det{\rT[\su+1-\sigma(\kk)][1][
\lambda_\kk+{\jj}-\kk
]}{1\leq {\jj},\kk\leq |\lambda|}\,.
\end{gather}

Then we immediately see that the equality (of the coefficient of a
given degree in each \(z_\kk\) in) \eqref{eq:MIDdet} is simply
equivalent to the CBR formula \eqref{eq:CBR}, which was already proven
earlier in the text.
\end{proof}

To finish this section, let us note that due to the diagrammatic
expressions given in appendix \ref{app:cod}, the {\MID}
\eqref{eq:MIDww} can also be written in the following, slightly
stronger form:
\begin{empheq}[box=\fbox]{multline}
\label{eq:MIDPi}
\lefteqn{(z-t) 
\left[\DLt[{\su}_\spi+1+\hD] 
  w(z) w(t) \PI \right]\cdot
\left[\DLt \PI \right]}%
\\
=~ z \left[\DLt[{\su}_\spi+1+\hD] 
  w(z) \PI \right]\cdot
\left[\DLt w(t) \PI \right]%
\\
-t 
\left[\DLt
  w(z) \PI \right]\cdot
\left[\DLt[{\su}_\spi+1+\hD]  w(t) \PI \right]\,,
\end{empheq}
\begin{gather}
 \where\PI=\prod_{\kk=1}^\nn \left(w(z_\kk)\right)^{a_\kk}\,, 
\label{eq:MIDPiDef}
\end{gather}
for \(\nn\) arbitrary pairs of numbers \((z_\kk,a_\kk)\). 

\begin{proof}
  First, if all \(a_\kk\) are equal to \(1\), then \eqref{eq:MIDPi} 
  is exactly the identity \eqref{eq:MIDww} written for 
  \(\tilde \nn=\nn+2\), \(\tilde z_1=z\), \(\tilde z_{\tilde\nn}=t\), and
  \(\tilde z_\kk=z_{\kk-1}\) for \(\kk=2,3,\cdots \nn+1\).

  Next if all the powers \(a_\kk\) in \eqref{eq:MIDPiDef} are non-negative
  integers, then   \(\PI\) can be written as
  \begin{align}
    \PI=&\prod_{\kk=1}^{\sum{a_{\coordi}}} w(\tilde z_{\kk})&
    \where \tilde z_{\kk}\equiv& z_{\Max\{{\jj}|\sum_{{\coordi}\leq {\jj}}a_{\coordi}\leq \kk\}}\,.
  \end{align}
  Therefore, the case when all the powers \(a_\kk\) are non-negative
  integer reduces to 
  the case when they are all equal to \(1\), which reduces to
  \eqref{eq:MIDww}.

  Finally, it is easy to see from the diagrammatic introduced in
  appendix \ref{app:cod} that \(\frac 1 {w(z)^a} \left[\DLt f({\g}) \left(w(z)^a\right)\right]\) is a
  polynomial in the variable \(a\). As a consequence, if
  \eqref{eq:MIDPi} holds when the powers \(a_\kk\) are non-negative
  integer, then it holds for arbitrary powers \(a_\kk\).
\end{proof}

\subsection{Conservation of the number of particles}
\label{sec:cons-numb-part}

The CBR formula \eqref{eq:CBR} which we have just shown also allows to
find simple eigenspaces of all {\Toprs}. To this end it is enough to
find some spaces which are stable under all \(\rT[][1]\), and the CBR
formula \eqref{eq:CBR} will then imply that these spaces are stable
under all {\Toprs}.

As we have seen for instance in \eqref{eq:Diag3Dw} (see appendix
\ref{sec:diagr-expr-co} for more details),
\(\Wt\equiv\sum_{s=0}^{\infty}\rT[\su][1]z^s\) can be 
written as a sum of \(\hD\)-diagrams. To each \(\hD\)-diagram is associated an
expression of the form \(\perm_\sigma\cdot
\left(\bigotimes_{\spi=1}^\lcds \mathcal{O}_\spi\right)\), where
\(\sigma \in \Sgrp \lcds\) is a permutation, and
\(\mathcal{O}_\spi\) is an {\op} (equal to either \(\frac 1 {1-{\g}~z}\),
\(\frac {{\g}~z}{1-{\g}~z}\) or \(\su_\spi ~\bI\)), which %
commutes with
\({\g}\). As a consequence,
\begin{align}
\label{eq:commWfg}
  \comm[0.45cm][-.05cm]{(f({\g}))^{\otimes\lcds}}{\Wt}=&0\,,
\end{align}
for any analytic function \(f\).
\begin{proof}
  First, each {\op} \(\mathcal{O}_{\spi}\) is diagonal is the same basis
  as \({\g}\) (and \(f({\g})\)). That is why
  \(\comm[0.65cm][-.15cm]{(f({\g}))^{\otimes\lcds}}{\left(\bigotimes_{\spi=1}^\lcds
      \mathcal{O}_\spi\right)}=0\). Next one sees that
  \((f({\g}))^{\otimes\lcds}\) commutes with any permutation {\op}
  \(\perm_\sigma\), so that finally it commutes with each \(\perm_\sigma\cdot
\left(\bigotimes_{\spi=1}^\lcds \mathcal{O}_\spi\right)\), and with
their sum \(\Wt\).
\end{proof}

If we denote by \(\left({\vv}_\jvp\right)_{1\leq\jvp\leq \Kr+\Mr}\)
the eigenvectors of the twist \({\g}\), 
and by \(\left(x_\jvp\right)_{1\leq\jvp\leq \Kr+\Mr}\) the
corresponding eigenvalues,
then we can notice that
\begin{align}
  (f({\g}))^{\otimes\lcds}\ket{{\vv}_{\jvp_1},{\vv}_{\jvp_2},\cdots,{\vv}_{\jvp_\lcds}}=&
\left(\prod_{\spi=1}^{\lcds}f(x_{\jvp_\spi})\right)\ket{{\vv}_{\jvp_1},{\vv}_{\jvp_2},\cdots,{\vv}_{\jvp_\lcds}}\,.
\end{align}
For fixed \(\Mp_1\), \(\Mp_2\), \(\cdots\), \(\Mp_{\Kr+\Mr}\), we can
introduce the sets
\begin{align}
\label{eq:DefEMMMM}
  E_{\Mp_1,\Mp_2,\cdots,\Mp_{\Kr+\Mr}}\equiv&
  \Vect{\ket{{\vv}_{\jvp_1},{\vv}_{\jvp_2},\cdots,{\vv}_{\jvp_\lcds}}\middle|
\forall %
\kvp
,~ \Mp_\kvp=\sum_{\spi=1}^\lcds \delta_{\jvp_\spi,\kvp}
 }\,.
\end{align}
These
spaces are the sets of states having a fixed number \(\Mp_\kvp\) of spins
pointing in each direction \(\ket {{\vv}_\kvp}\). In the spirit of the
introductory section \ref{sec:integr-et-}, they are the sets of
states having a fixed number \(\Mp_\kvp\) of particles of each type.
One can
see that for any state \(\ket\phi \in   E_{\Mp_1,\Mp_2,\cdots,\Mp_{\Kr+\Mr}} \), 
\begin{align}
  (f({\g}))^{\otimes\lcds}\ket\phi
=&\prod_{\jvp=1}^{\Kr+\Mr} f(x_\jvp)^{\Mp_\jvp}\ket\phi\,.
\end{align}
It is also possible to promote the numbers \(\Mp_\kvp\) into
{\ops} \(\Mpo_\kvp\) having eigenvalue \(\Mp_\kvp\) on the
eigenspace \(E_{\Mp_1,\Mp_2,\cdots,\Mp_{\Kr+\Mr}}\). Then the {\op}
\((f({\g}))^{\otimes\lcds}\) is equal to \(\prod_{\jvp=1}^{\Kr+\Mr} f(x_\jvp)^{\Mpo_\jvp}\).

The relation \eqref{eq:commWfg} shows that all {\Toprs}
commute with \((f({\g}))^{\otimes\lcds}\) (in addition to commuting
with each other), %
and if we assume that the eigenvalues
\(x_\jvp\) are distinct, we can choose some functions \(f_\kvp\) such
that \(f_\kvp(x_\jvp)=1+\delta_{\jvp,\kvp}\), {\idest} such that
\((f_\kvp({\g}))^{\otimes\lcds} = 2^{\Mpo_\kvp}\). In this case we see
that the {\Toprs} commute with \(2^{\Mpo_\kvp}\) and hence they commute
with each \({\Mpo_\kvp}\). In other words, the sets
\(E_{\Mp_1,\Mp_2,\cdots,\Mp_{\Kr+\Mr}}\) 
 are therefore stable under all the {\Toprs}.

If the eigenvalues of \(\g\) are not distinct, one can show that
the spaces \(E_{\Mp_1,\Mp_2,\cdots,\Mp_{\Kr+\Mr}}\) are still stable under
all {\Toprs}, provided \(\g\) is diagonalizable. To show this, we can
for instance show that \(E_{\Mp_1,\Mp_2,\cdots,\Mp_{\Kr+\Mr}}\) is
stable under the {\op} \(\perm_\sigma\cdot
\left(\bigotimes_{\spi=1}^\lcds \mathcal{O}_\spi\right)\) associated
to each \(\hD\)-diagram.

Therefore, we have seen that the spaces
\(E_{\Mp_1,\Mp_2,\cdots,\Mp_{\Kr+\Mr}}\) are stable under 
all {\Toprs}, {\idest} that all the {\Toprs} commute with the {\op}
\(\Mpo_\jvp\). This means that the number of ``particles'' of each type
is invariant under the action of all {\Toprs}, and therefore it is preserved under
the action of the {\Ham} as well.

This remark will be useful in section \ref{sec:expr-diff-Qop}, because
it allows to build some {\ops} which trivially commute with all
{\Toprs} (for instance, this will be the case of the {\op} \(\BI_I\)
which we will define in \eqref{eq:DefPIi}).

\section{Bäcklund transform and Bethe equations}
\label{sec:transf-de-backl}
As we saw in the previous section, the ``2\(^{\textrm{nd}}\)'' Weyl
character formula \eqref{eq:Weyl02}, satisfied by characters (which are
the {\Toprs} of a {\cds} with length \(\lcds=0\)), can be
generalized to \(\su\)-dependent {\Toprs} by introducing the correct
shift. This gives the CBR formula \eqref{eq:CBR}, and we 
also saw that (for rectangular representations \eqref{eq:DefRecT}),
this CBR formula is equivalent to the Hirota equation
\eqref{eq:Hirota}, a bilinear equation on {\Toprs}. 

In this section and the next section, we will show how to perform the
same program for the ``1\(^{\textrm{st}}\) Weyl formula'' {\below} (see
appendix \ref{sec:caracteres})
\begin{align}
\label{eq:Weyl01}
  \cha \lambda({\g})=&\frac{ \Det%
{
      x_\jvp^{\lambda_\ii+\Kr-\ii}
}
    {1\leq \ii,\jvp\leq\Kr} }
  {\Det%
{
      x_\jvp^{\Kr-\ii} 
}%
{1\leq
      \ii,\jvp\leq\Kr}}\,,
\end{align}
where \(x_1\), \(x_2\), \(\cdots\), \(x_\Kr\) denote the eigenvalues of
\(\g\in\GL \Kr\).

First we will see that this formula gives rise to a
bilinear identity on characters of \(\GL \Kr\) as compared to \(\GL
{\Kr-1}\). We will see how this bilinear identity generalizes to {\Toprs}
(or to their eigenvalues) by adding a dependence on the spectral parameter
\(\su\) \cite{1992PhyA..183..304K,Kuniba:1992qv,springerlink:10.1007/s002200050165,1997JPhA...30.7975T}. Then we will see that if this bilinear identity is satisfied,
it %
gives an
expression for {\Toprs}, %
out of which the spectrum of the theory can be obtained.
In particular we will obtain some Bethe equations
\cite{Zabrodin:2007rq,Lai1974,PhysRevB.12.3795,Babelon:1981un,Kulish:1983rd,Kazakov:2007fy} 
 which
  generalize the equations of section \ref{sec:integr-et-}.
 Finally, we
will prove in the next section \ref{sec:expr-diff-Qop} that these
bilinear relations are indeed satisfied by polynomial {\ops}.

In order to do this, we will restrict, from now on, to the case when
\(\g\) is diagonalizable and has distinct non-zero eigenvalues. 
\index{xj@\ensuremath{x_{\jvp}}(eigenvalue of the twist)}
The eigenvalues of
\(\g\in {\GL{\Kr\ensuremath{|}\Mr}}\) will be denoted as
  \((x_1,x_2,\cdots,x_{\Kr+\Mr})\), and they are supposed to be
  distinct and non-zero.
The case when \(\g\) is not diagonalizable or when several eigenvalues
are 
equal %
can in principle be obtained as a limit, but this point
will not be discussed much in the present manuscript.

\index{T-functions@{\Tfs}}
The present section reviews results which were known before the start
of the {\PhD}. In this section, we will not use any explicit
definition
of the {\Toprs} obtained in the previous section, but
instead we will consider some {\Tfs}, which are functions of the
spectral parameter \(\su\) and obey the Hirota equation.  We will assume
that a polynomial\footnote{
We will call ``Bäcklund flow''  
any
solution of
the equations (\ref{eq:BT1},~\ref{eq:BT2}) given in the first
subsection \ref{sec:intr-backl-flow}%
.
} ``Bäcklund flow''  
 exists, which gives rise to
so-called {\Qfs}.

\index{Q-functions@{\Qfs}}
The {\Tfs} and {\Qfs} will be denoted by slant letters 
as opposed to the vertical letters for {\Top}- and {\Qoprs}. What we
will see in the next section \ref{sec:expr-diff-Qop} is that 
some
{\Qoprs} can be explicitly constructed and a polynomial Bäcklund
\index{Q-operators@{\Qoprs}}
flow can be constructed at the {\level} of {\ops}. Then we will be
able to identify the {\Qfs} (and the {\Tfs}) of the present section
with the eigenvalues of the {\Qoprs} (and the {\Toprs}). It will then
be obvious that writing an equation on the {\Toprs} or on the {\Tfs}
is strictly equivalent, because the {\Tfs} are the eigenvalues of the
{\Toprs}, which commute with each other. For instance the equation
\begin{align}
  \label{eq:HirotaTf}
\rTf[\su+1] \rTf = \rTf[\su+1][a+1]\rTf[][a-1]+\rTf[\su+1][][s-1]\rTf[][][s+1]\,,
\end{align}
\index{Hirota equation}
will mean that the eigenvalues of {\Toprs} obey the same Hirota equation as the
{\ops} themselves. %
In \eqref{eq:HirotaTf},
 the symbols like \(\rTf[\su+1]\),  \(\rTf\) etc. denote the eigenvalues
 of the {\op}s \(\rT[\su+1]\),  \(\rT\) etc.,
but these eigenvalues all correspond to the
 same eigenspace (we use the fact that the {\Toprs} commute with each
 other, so that they have common eigenstates).

For the moment, we will introduce a 
``Bäcklund flow'', for some %
{\Tfs}, which %
obey the Hirota equation \eqref{eq:HirotaTf}.
For \(\lcds>0\), we will
assume (but not prove yet) the existence of this polynomial Bäcklund
flow, and we will restrict %
to results which were 
known \cite{springerlink:10.1007/s002200050165,Zabrodin:1996vm,Kazakov:2007fy,Zabrodin:2007rq}
 before the start of this {\PhD}, as opposed
to the results of the sections \ref{sec:expr-diff-Qop} and
\ref{sec:quantum-classical}, which are original results of this
thesis \cite{Kazakov:2010iu,Alexandrov:2011aa}.  

 This Bäcklund flow, %
 can %
for instance
 be found in \cite{springerlink:10.1007/s002200050165} (for usual
 matrix groups), in \cite{Kazakov:2007fy} (for super groups) and in
 \cite{Zabrodin:2007rq} (for twisted super-spin chains), and is the
 starting point for 
original works of this {\PhD}
\cite{Kazakov:2010iu,Alexandrov:2011aa}, including the explicit
operatorial constructions of the {\Top}- and {\Qoprs} given in the
next section \ref{sec:expr-diff-Qop}.

\subsection{Introduction of the Bäcklund flow}
\label{sec:intr-backl-flow}

As written in appendix \ref{sec:caracteres}, the characters of the {\GL
\Kr} group are expressed through the Weyl formula
\eqref{eq:Weyl01}. 
The idea behind the Bäcklund flow will be to write this expression for
the character of
\begin{align}
  \g[I]\equiv&\mathrm{diag}\left((x_{\ivp})_{{\ivp}\in I}\right)\in
  \GL{|I|}\,,&
\where &|I|=\mathrm{Card}\{I\}\,,
\end{align}
for an arbitrary subset \(I\) of \(\ninter 1 {\Kr}\). For such a set we
will denote
\begin{align}
\label{eq:BarDef}
  \overline I\equiv & \ninter 1 \Kr \setminus I\,,&\And&&
  \ove\equiv&\ninter 1 \Kr\,.
\end{align}
\index{Ibar@\ensuremath{\overline I\equiv \ninter 1 {\Kr+\Mr} \setminus I}}
\index{0ove@\ensuremath{\ove \equiv \ninter 1 {\Kr+\Mr}}}

Given an ordering \(({\ivp}_1,{\ivp}_2,\cdots, {\ivp}_\Kr)\) of
\(\{{\ivp}_1,{\ivp}_2,\cdots,{\ivp}_\Kr\}=\ninter 1 \Kr\), we will be interested in an
undressing procedure
\begin{gather}
\label{eq:undressg}
  \begin{array}{ccccccc}
    \g [{I_\Kr}]&
\rightsquigarrow
&\g [{I_{\Kr-1}}]&\rightsquigarrow&\g [{I_{\Kr-2}}]&\cdots&\g [{\emptyset}]\\
     \indown&&\indown&&\indown&&\indown\\
{\GL \Kr}&\supset&{\GL{\Kr-1}}&\supset&{\GL{\Kr-2}}&\cdots&\{\bo\}
  \end{array}\\[.5cm]
  \where I_\nlvl\equiv \{{\ivp}_1,{\ivp}_2,\cdots {\ivp}_\nlvl\}\,.
\end{gather}
This procedure gradually decreases the rank of the group, and it is
dependent on the ordering \(({\ivp}_1,{\ivp}_2,\cdots, {\ivp}_\Kr)\) of the set
\(\ninter 1 \Kr\). This ordering, which governs the undressing
procedure, will be called a ``nesting path''.\index{Nesting ! nesting path} 
For a given set \(I\subset \ninter 1 \Kr\), the number \(\Kr-|I|\) will be
called the ``nesting level''.\index{Nesting ! nesting level}  It is
the number of steps, in the undressing procedure \eqref{eq:undressg},
which are needed to reach \(\g [I]\) by starting from \(\g=\g[\ove]\).

The ``Hasse diagram''\cite{2010NuPhB.826..399T} (see figure
\ref{fig:Hasse}) shows all 
the possible sets \(I\subset \ove\), with lines connecting each set to
its subsets. On this diagram, each {\nesting} {\ppath} is one (out of
\(\Kr !\)) {\ppath} connecting \(\ove=\ninter 1 \Kr\) to \(\emptyset\).

\begin{figure}
\fbox{\begin{minipage}{.95\textwidth}
  \centering
  \begin{tikzpicture}[scale=4]
\tikzstyle{np} = [red,line width=.55mm]
 \node (i) at (0.,-0.) {\(\emptyset\)};
 \node (i1i) at (-0.75,--0.612372) {{\{1\}}};
 \node (i2i) at (-0.25,--0.612372) {{\{2\}}};
 \node (i3i) at (0.25,--0.612372) {{\{3\}}};
 \node (i4i) at (0.75,--0.612372) {{\{4\}}};
 \node (i1i2i) at (-1.02062,--1.22474) {{\{1, 2\}}};
 \node (i1i3i) at (-0.612372,--1.22474) {{\{1, 3\}}};
 \node (i1i4i) at (-0.204124,--1.22474) {{\{1, 4\}}};
 \node (i2i3i) at (0.204124,--1.22474) {{\{2, 3\}}};
 \node (i2i4i) at (0.612372,--1.22474) {{\{2, 4\}}};
 \node (i3i4i) at (1.02062,--1.22474) {{\{3, 4\}}};
 \node (i1i2i3i) at (-0.75,--1.83712) {{\{1, 2, 3\}}};
 \node (i1i2i4i) at (-0.25,--1.83712) {{\{1, 2, 4\}}};
 \node (i1i3i4i) at (0.25,--1.83712) {{\{1, 3, 4\}}};
 \node (i2i3i4i) at (0.75,--1.83712) {{\{2, 3, 4\}}};
 \node (i1i2i3i4i) at (0.,--2.44949) {{\{1, 2, 3, 4\}}};
 \draw (i) -- (i1i);
 \draw[np] (i) -- (i2i);
 \draw (i) -- (i3i);
 \draw (i) -- (i4i);
 \draw (i1i) -- (i1i2i);
 \draw (i1i) -- (i1i3i);
 \draw (i1i) -- (i1i4i);
 \draw[np] (i2i) -- (i1i2i);
 \draw (i2i) -- (i2i3i);
 \draw (i2i) -- (i2i4i);
 \draw (i3i) -- (i1i3i);
 \draw (i3i) -- (i2i3i);
 \draw (i3i) -- (i3i4i);
 \draw (i4i) -- (i1i4i);
 \draw (i4i) -- (i2i4i);
 \draw (i4i) -- (i3i4i);
 \draw (i1i2i) -- (i1i2i3i);
 \draw[np] (i1i2i) -- (i1i2i4i);
 \draw (i1i3i) -- (i1i2i3i);
 \draw (i1i3i) -- (i1i3i4i);
 \draw (i1i4i) -- (i1i2i4i);
 \draw (i1i4i) -- (i1i3i4i);
 \draw (i2i3i) -- (i1i2i3i);
 \draw (i2i3i) -- (i2i3i4i);
 \draw (i2i4i) -- (i1i2i4i);
 \draw (i2i4i) -- (i2i3i4i);
 \draw (i3i4i) -- (i1i3i4i);
 \draw (i3i4i) -- (i2i3i4i);
 \draw (i1i2i3i) -- (i1i2i3i4i);
 \draw[np] (i1i2i4i) -- (i1i2i3i4i);
 \draw (i1i3i4i) -- (i1i2i3i4i);
 \draw (i2i3i4i) -- (i1i2i3i4i);
\end{tikzpicture}
  \caption{Hasse diagram for {\GL 4}. In red (and thicker), the {\nesting} {\ppath}
    \((2,1,4,3)\) is highlighted.%
  }
  \label{fig:Hasse}
\end{minipage}}
\end{figure}
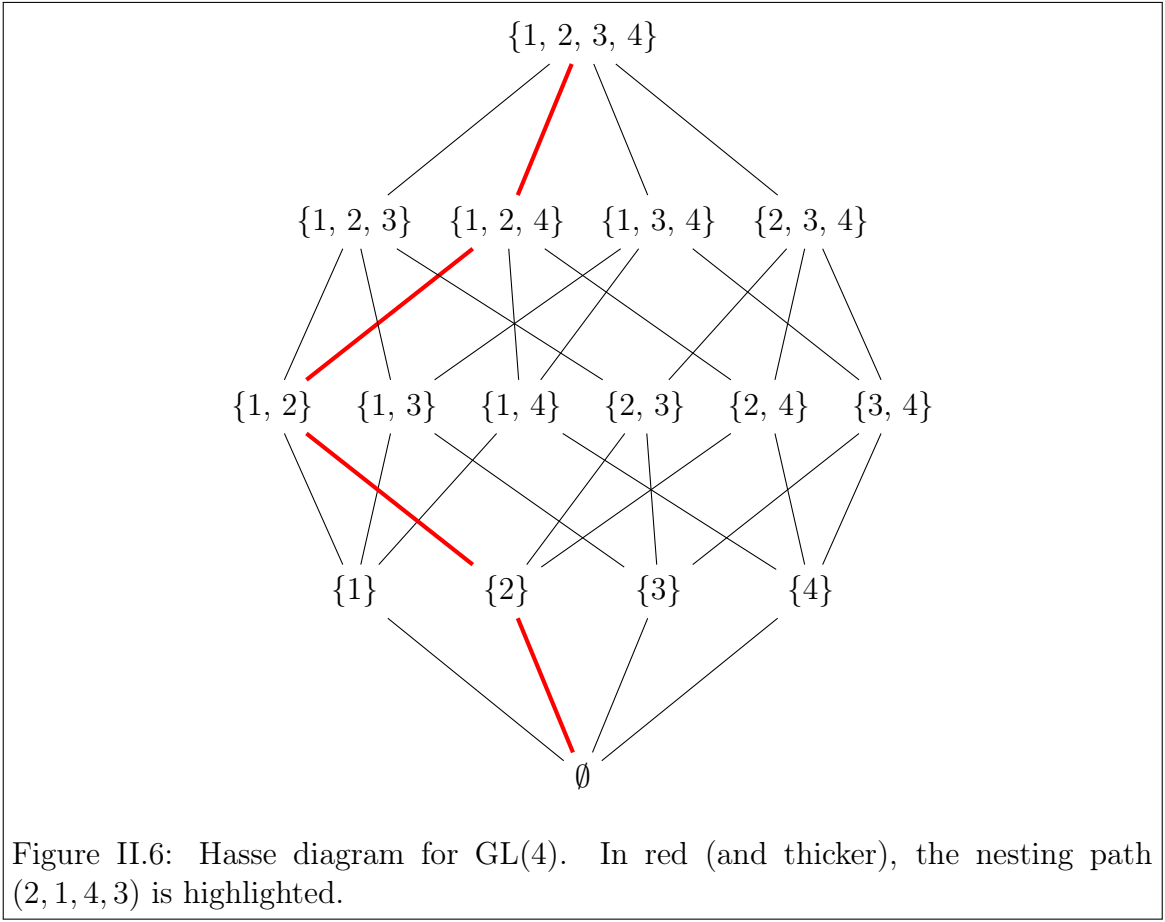

For the character \(\chas \equiv \cha {\las}\)
associated to rectangular {\rp}s, 
the formula \eqref{eq:Weyl01} can be written at an arbitrary {\nesting}
{\level} and it reads
\begin{align}
\label{eq:Weyl1rI}
  \chas %
(\g[I])=&\frac{
\BlockDet{x_\jvp^{s+|I|-\ii}}{\substack{1\leq
    \ii\leq a\\\jvp\in I}}
{x_\jvp^{|I|-\ii}}{\substack{a+1\leq \ii\leq |I|\\\jvp\in I}}
}
  {\Det
{ %
x_\jvp^{|I|-\ii} 
}%
{\substack{1\leq
      \ii\leq|I|\\\jvp\in I}}}\,,
\end{align}
where the numerator is the determinant of a matrix made of two blocks of
respective size \(a\times\Kr\) and \((\Kr-a)\times \Kr\). %
This expression is valid for \(0\leq a\leq |I|\), whereas \( \cha
{\las}(\g[I])=0\) if \(a>|I|\).

Writing Plücker identities (a generalization of the Jacobi identity \cite{springerlink:10.1007/s002200050165}) %
for the determinant
\eqref{eq:Weyl1rI}, one gets
\begin{empheq}[left=\empheqlbrace]{align}
  \label{eq:chibacklund2}
  \chas[a+1] (\g[I,\jvp]) \chas(\g[I])=& \chas(\g[I,\jvp])\chas[a+1]
  (\g[I])
  +x_\jvp \chas[a+1][s-1](\g[I,\jvp]) \chas[][s+1](\g[I]) \,,\\
  \label{eq:chibacklund1}
  \chas[][s+1] (\g[I,\jvp]) \chas(\g[I])=&
  \chas(\g[I,\jvp])\chas[][s+1] (\g[I]) +x_\jvp \chas[a+1](\g[I,\jvp])
  \chas[a-1][s+1](\g[I]) \,,
\end{empheq}
where \(I,\jvp\) denotes the set \(I\cup \{\jvp\}\), for an arbitrary
\(\jvp\in \ninter 1 \Kr\setminus I\).

An interesting particular case of \eqref{eq:chibacklund2}, when \(a=0\)
is the relation
\begin{align}
\label{eq:TQ-char}
  \chs{s} (\g[I])=&
  \chs{s}(\g[I,\jvp]) -x_\jvp \chs{s-1}(\g[I,\jvp]) \,,
\end{align}
were \(\chs{s}\equiv\chas[1]\) denotes the character associated to the
symmetric {\rp} \(\lambda=(s,0,0,\cdots)\).

If we remember the expression of the generating series %
(see appendix \ref{sec:caracteres})
\begin{align}
\label{eq:GenSerI}
  w_{I}(z)\equiv&\sum_{s\geq 0} z^s \chs s (\g [I])=
  \prod_{\jvp\in  I}\frac 1 {1- x_\jvp ~z}\,,
\end{align}
the relation \eqref{eq:TQ-char} actually reduces to the simple
statement that 
\begin{align}
  w_{I}(z)\equiv&\left(1-x_\jvp~z\right) ~ w_{I,\jvp}(z).
\end{align}

The relations (\ref{eq:chibacklund2},~\ref{eq:chibacklund1}) are
written for the characters of {\GL \Kr}. For 
{\GL{\Kr\ensuremath{|}\Mr}}, the equations
(\ref{eq:chibacklund2},~\ref{eq:chibacklund1}) hold provided
\(\sg \jvp =1 %
\). Moreover, the generating series
\eqref{eq:GenSerI} becomes
\begin{align}
\label{eq:GenSerIs}
  w_{I}(z)\equiv&\sum_{s\geq 0} z^s \chs s (\g [I])=
\sudet\left( \frac 1 {1-
        {\g}~ z}\right)
=
\prod_{\substack{\jvp\in  I}}\left(1-x_\jvp~z\right)^{{\sg  \jvp}}\,
\end{align}
where \(\sudet\left( \frac 1 {1-
        {\g}~ z}\right)\) is the super-determinant of 
\(\frac 1 {1-
        {\g}~ z}\) (see appendix \ref{sec:gener-glkens}).

The generalization of \eqref{eq:TQ-char} is then
\begin{gather}
\label{eq:TQ-su-char}
  \chs{s} (\g[I])=
  \chs{s}(\g[I\Delta \jvp]) -x_\jvp \chs{s-1}(\g[I \Delta \jvp]) \,.\\
\where I\subset \ninter 1 {\Kr+\Mr} \And
\left\{
  \begin{array}{ll}
    {\sg  \jvp}=+1,\quad\jvp\not\in I& \And~~  I\Delta \jvp\equiv
    I\cup\{\jvp\}\\
\qquad\textrm{or}\\
    {\sg  \jvp}=-1, \quad\jvp\in I& \And~~  I\Delta \jvp\equiv I\setminus\{\jvp\}
  \end{array}
\right.
\label{eq:DeltaSugr}
\end{gather}

For these {\sugrs}, the undressing procedure (along a given {\nesting}
{\ppath} \linebreak \(({\ivp}_1,{\ivp}_2,\cdots,{\ivp}_{\Kr+\Mr})\)) can become for
instance 
\begin{gather}
\label{eq:undresssug}
  \begin{array}{ccccccc}
    \g [{I_{\Kr+\Mr}}]&
\rightsquigarrow
&\g [{I_{\Kr+\Mr-1}}]&\rightsquigarrow&\g [{I_{\Kr+\Mr-2}}]&\cdots&\g [{\emptyset}]\\
     \indown&&\indown&&\indown&&\indown\\
{\GL {\Kr\ensuremath{|}\Mr}}&\supset&{\GL{\Kr-1\ensuremath{|}\Mr}}&\supset&{\GL{\Kr-1\ensuremath{|}\Mr-1}}&\cdots&\{\bo\}
  \end{array}\\[.5cm]
  \where I_\nlvl\equiv \{{\ivp}_1,{\ivp}_2,\cdots {\ivp}_\nlvl\}\,.
\end{gather}
At every step, either \(\Kr\) or \(\Mr\) is decreased by one, 
and \eqref{eq:undresssug} involves the inclusions
\({\GL
  {\Kr\ensuremath{|}\Mr}}\supset{\GL{\Kr-1\ensuremath{|}\Mr}}\supset{\GL{\Kr-1\ensuremath{|}\Mr-1}}\)
which
correspond to the case \(\sg {{\ivp}_{\Kr+\Mr}}=1\) and \(\sg {{\ivp}_{\Kr+\Mr-1}}=-1\).

We will also call {\nesting} {\level} (associated to a set \(I\subset \ninter
1 {\Kr+\Mr}\)) the integer \linebreak \(\Kr+\Mr-|I|\), which is the number
\index{Nesting ! nesting level} 
of steps necessary to reach \(\g[I]\) from \(\g=\g[\ove]\) following the
procedure \eqref{eq:undresssug}. We also define the complement of a
set (as a generalization of \eqref{eq:BarDef}) by
\begin{align}
\label{eq:SuBarDef}
  \overline I\equiv & \ninter 1 {\Kr+\Mr} \setminus I\,,&\And&&
  \ove\equiv&\ninter 1 {\Kr+\Mr}\,.
\end{align}
\index{Ibar@\ensuremath{\overline I\equiv \ninter 1 {\Kr+\Mr} \setminus I}}
\index{0ove@\ensuremath{\ove \equiv \ninter 1 {\Kr+\Mr}}}

In \eqref{eq:TQ-su-char}, we see that the case \(\sg \jvp=1\)
corresponds to the transformation \(\GL
{\kk\ensuremath{|}\mm}\rightsquigarrow \GL {\kk-1\ensuremath{|}\mm}\),
where \(\GL {\kk-1\ensuremath{|}\mm}\) 
corresponds to \(I\) on the {\lhs} and \(\GL
{\kk\ensuremath{|}\mm}\)
corresponds to \(I\Delta\jvp\equiv I\cup\{\jvp\}\)
 on the {\rhs}. By
contrast, the case \(\sg \jvp=-1\) corresponds to the transformation \(\GL
{\kk\ensuremath{|}\mm}\rightsquigarrow \GL {\kk\ensuremath{|}\mm-1}\),
where \(\GL {\kk\ensuremath{|}\mm}\) 
corresponds to \(I\)
 on the {\lhs} and \(\GL
{\kk\ensuremath{|}\mm-1}\rightsquigarrow\)
corresponds to \(I\Delta\jvp\equiv I\setminus\{\jvp\}\) on the {\rhs}.

With the notations \eqref{eq:DeltaSugr}, the generalization of
(\ref{eq:chibacklund2},~\ref{eq:chibacklund1}) to {\sugrs} is
simply 
\begin{empheq}[left=\empheqlbrace]{align}
\label{eq:chibacklund2s}
  \chas[a+1] (\g[I\Delta\jvp]) \chas(\g[I])=&
  \chas(\g[I\Delta\jvp])\chas[a+1] (\g[I])\nonumber\\&\qquad\qquad
  +x_\jvp \chas[a+1][s-1](\g[I\Delta\jvp]) \chas[][s+1](\g[I]) \,,\\
\label{eq:chibacklund1s}
  \chas[][s+1] (\g[I\Delta\jvp]) \chas(\g[I])=&
  \chas(\g[I\Delta\jvp])\chas[][s+1] (\g[I])\nonumber\\&\qquad\qquad
  +x_\jvp \chas[a+1](\g[I\Delta\jvp]) \chas[a-1][s+1](\g[I]) \,.
\end{empheq}

We have already shown that the {\Tfs}, which depend
non-trivially on the spectral parameter \(\su\), generalize characters
in %
a way which satisfies the Hirota equation 
\eqref{eq:Hirota}. We should therefore generalize
(\ref{eq:chibacklund2s},~\ref{eq:chibacklund1s}) in a way which is
consistent with this Hirota equation.

The correct generalization of
(\ref{eq:chibacklund2s},~\ref{eq:chibacklund1s}) is then \cite{springerlink:10.1007/s002200050165,Zabrodin:1996vm,Kazakov:2007fy,Zabrodin:2007rq}
\begin{empheq}[left=\empheqlbrace]{align}
\label{eq:BT1}
  \rTf[\su][a+1][][I\Delta\jvp]\rTf[][][][I] -
  \rTf[][][][I\Delta\jvp]\rTf[][a+1][][I] =& x_{\jvp} \rTf[\su+1][a+1][s-1][I\Delta\jvp]
  \rTf[\su-1][][s+1][I]\,,  \\
\rTf[][][s+1][I\Delta\jvp]\rTf[][][][I] -
\rTf[][][][I\Delta\jvp]\rTf[][][s+1][I]
=&x_{\jvp} \rTf[\su+1][a+1][][I\Delta\jvp] \rTf[\su-1][a-1][s+1][I]\,,
\label{eq:BT2}
\end{empheq}
where the ``{\nested} {\Tfs}'' \(\rTf[][][][I]\) (defined by this
equation) generalize the characters 
\(\chas(\g[I])\).  

In the case \(\sg \jvp=+1\) (corresponding to a transformation
\(\GL{\kk\ensuremath{|}\mm} \rightsquigarrow
\GL{\kk-1\ensuremath{|}\mm}\)), we will say that if
 we know a function
\(\rTf[][][][I\Delta \jvp] = \rTf[][][][I,\jvp]\) (where
\(I\raisebox{-4pt}{\inleft}\!\!\!\!\!\!\!\!/~ \jvp\) and
\(I,\jvp\equiv I\cup\{\jvp\}\))
  which satisfies the Hirota equation
\eqref{eq:Hirota}, then any solution \(\rTf[][][][I]\) of
(\ref{eq:BT1},~\ref{eq:BT2}) %
is a
``Bäcklund transformed'' 
of \(\rTf[][][][I,\jvp]\).

On the other hand, in the case \(\sg \jvp=-1\) (which corresponds to a
transformation 
\(\GL{\kk\ensuremath{|}\mm} \rightsquigarrow
\GL{\kk\ensuremath{|}\mm-1}\)), we will say that if
we know a function
\(\rTf[][][][J]\equiv\rTf[][][][I,\jvp]\) (
where \(I\raisebox{-4pt}{\inleft}\!\!\!\!\!\!\!\!/~ \jvp\) and 
\(J\equiv I,\jvp\equiv I\cup\{\jvp\}\))
which
satisfies the Hirota equation \eqref{eq:Hirota},
then any solution \(\rTf[][][][J\Delta {\jvp}]=\rTf[][][][I%
]\) of
(\ref{eq:BT1},~\ref{eq:BT2}) %
is
a ``Bäcklund transformed'' 
of \(\rTf[][][][I,\jvp]\).

In \cite{Zabrodin:2007rq}, (where the function \(\rTf[][][][\{1,2,\cdots,\kk,\Kr+1,\Kr+2,\Kr+\mm\}]\) is
denoted as \linebreak \({{\Tnothing}_{{\kk},{\mm}}(a,s,s-a+2\su)       %
}\)), it is shown that if these Bäcklund transforms exist, then
\begin{itemize}%
\item If the function \(\rTf[][][][I,\jvp]\) satisfies the Hirota
  equation, then its Bäcklund transformed also satisfies Hirota
  equation. Symmetrically, if the Bäcklund transformed \(\rTf[][][][I]\)
  of a function \(\rTf[][][][I,\jvp]\) satisfies the Hirota equation, then
  \(\rTf[][][][I,\jvp]\)  also satisfies the Hirota equation.
\item If \(\rTf[][][][I\Delta\jvp]\) is zero outside \(\HK(\kk,\mm)\)
  (defined by \eqref{eq:DefHK}), then %
  one can show that
  \(\rTf[][][][I]\) is zero outside \(\HK(\kk',\mm')\), where %
  \begin{align}
    \left\{
      \begin{array}{c}
        \kk'=\kk-1\\
        \mm'=\mm
      \end{array}
    \right.
&&\textrm{or}&&
    \left\{
      \begin{array}{c}
        \kk'=\kk\\
        \mm'=\mm+1
      \end{array}
    \right.\,\,.
  \end{align}
We will be interested in solutions of (\ref{eq:BT1},~\ref{eq:BT2})
such that \(\kk'=\kk-1\) and \(\mm'=\mm\) if \({\sg  \jvp}=1\), whereas \(\kk'=\kk\)
and \(\mm'=\mm+1\) if \({\sg  \jvp}=-1\). Then the Bäcklund transformed of
a {\Tf} lives in a ``{\fat} {\hook}'' with one less row (if \({\sg 
  \jvp}=1\)) or one less column (if \({\sg 
  \jvp}=-1\)) than the original {\Tf} (see figure \ref{fig:Backlund}).

For the characters this statement about the size of the ``{\fat}
{\hook}'' is trivial, but for the {\Tfs}, it is an important statement
which is not completely obvious because we do not use any explicit expression
of the {\Tfs}, but simply the fact that \(\rT[][][][\ove]\)  obeys the
Hirota equation on the ``{\fat} {\hook}'' \(\HK(\Kr,\Mr)\).

 This statement about the size of the ``{\fat} {\hook}'' justifies
 that, for the {\Tfs}, the Bäcklund flow can be interpreted as the sequence
 of inclusions \eqref{eq:undresssug}.
\end{itemize}

\begin{figure}
  \centering
\raisebox{-1cm}{\begin{tikzpicture}
    \newcommand{\KK}{1}
    \renewcommand{\MM}{.5}
    \newcommand{\UU}{2.7}
    \renewcommand{\RR}{2.2}
    \newcommand{\LL}{-.6}
    \newcommand{\rr}{1.9}
    \renewcommand{\mm}{2.4}
    \tikzstyle{axis}=[thick]
    \tikzstyle{diag}=[color=red,very thick]
    \draw[style=axis](\LL,0)--(\RR,0);
    \draw[style=axis](0,0)--(0,\UU);
    \draw[style=axis](\MM,\KK)--(\rr,\KK);
    \draw[style=axis](\MM,\KK)--(\MM,\mm);
    \draw[step=.5cm] (0,0) grid (\rr,\KK);
    \draw[step=.5cm] (0,\KK) grid (\MM,\mm);
    \node at (-.3,\KK){\(2\)};
    \node at (\MM,-.3){\(1\)};
  \end{tikzpicture}}
\(\rightsquigarrow\)
\raisebox{-1cm}{    \begin{tikzpicture}
    \newcommand{\KK}{.5}
    \renewcommand{\MM}{.5}
    \newcommand{\UU}{2.7}
    \renewcommand{\RR}{2.2}
    \newcommand{\LL}{-.6}
    \newcommand{\rr}{1.9}
    \renewcommand{\mm}{2.4}
    \tikzstyle{axis}=[thick]
    \tikzstyle{diag}=[color=red,very thick]
    \draw[style=axis](\LL,0)--(\RR,0);
    \draw[style=axis](0,0)--(0,\UU);
    \draw[style=axis](\MM,\KK)--(\rr,\KK);
    \draw[style=axis](\MM,\KK)--(\MM,\mm);
    \draw[step=.5cm] (0,0) grid (\rr,\KK);
    \draw[step=.5cm] (0,\KK) grid (\MM,\mm);
    \node at (-.3,\KK){\(1\)};
    \node at (\MM,-.3){\(1\)};
  \end{tikzpicture}}
\(\rightsquigarrow\)
\raisebox{-1cm}{    \begin{tikzpicture}
    \newcommand{\KK}{.5}
    \renewcommand{\MM}{0}
    \newcommand{\UU}{2.7}
    \renewcommand{\RR}{2.2}
    \newcommand{\LL}{-.6}
    \newcommand{\rr}{1.9}
    \renewcommand{\mm}{2.4}
    \tikzstyle{axis}=[thick]
    \tikzstyle{diag}=[color=red,very thick]
    \draw[style=axis](\LL,0)--(\RR,0);
    \draw[style=axis](0,0)--(0,\UU);
    \draw[style=axis](\MM,\KK)--(\rr,\KK);
    \draw[style=axis](\MM,\KK)--(\MM,\mm);
    \draw[step=.5cm] (0,0) grid (\rr,\KK);
    \draw[step=.5cm] (0,\KK) grid (\MM,\mm);
    \node at (-.3,\KK){\(1\)};
    \node at (\MM,-.3){\(\phantom{1}\)};
  \end{tikzpicture}}
\(\rightsquigarrow\)
\raisebox{-1cm}{    \begin{tikzpicture}
    \newcommand{\KK}{0}
    \renewcommand{\MM}{0}
    \newcommand{\UU}{2.7}
    \renewcommand{\RR}{2.2}
    \newcommand{\LL}{-.6}
    \newcommand{\rr}{1.9}
    \renewcommand{\mm}{2.4}
    \tikzstyle{axis}=[thick]
    \tikzstyle{diag}=[color=red,very thick]
    \draw[style=axis](\LL,0)--(\RR,0);
    \draw[style=axis](0,0)--(0,\UU);
    \draw[style=axis](\MM,\KK)--(\rr,\KK);
    \draw[style=axis](\MM,\KK)--(\MM,\mm);
    \draw[step=.5cm] (0,0) grid (\rr,\KK);
    \draw[step=.5cm] (0,\KK) grid (\MM,\mm);
    \node at (\MM,-.3){\(\phantom{1}\)};
  \end{tikzpicture}}

  \caption{Bäcklund flow for a {\nesting} {\ppath} of {\GL
      {2\ensuremath{|}1}}. This %
    shows the successive \((a,s)\)-lattices for a {\nesting} {\ppath}
    corresponding to
    \({\GL
      {2\ensuremath{|}1}} \supset {\GL{1\ensuremath{|}1}} \supset
    {\GL{1%
      }} \supset \{\bo\}\), as in \eqref{eq:undresssug}.}
  \label{fig:Backlund}
\end{figure}

This Bäcklund flow is then often called ``undressing procedure'':
from a
solution of Hirota on a given lattice
\(\HK(\kk,\mm)\), it %
produces a
solution of Hirota on a smaller and smaller lattice.
On the last step,
the lattice \(\HK(0,0)\) is trivial and the {\Tf}
\(\rTf[][][][\emptyset]\) is a constant (independent of \(\su\), \(a\), and \(s\))
inside \(\HK(0,0)\).

In the section \ref{sec:expr-diff-Qop} we will see how to explicitly
construct the \(\rTf[][][][I]\) for all \(I\), and we will check that they are
polynomial and satisfy the Hirota equation. Before this construction is
presented, let us now see how it will be used: the ``dressing
procedure'' will allow to recover the spectrum of {\Toprs} from the
existence of this polynomial Bäcklund flow.

\subsection{Bethe equations and energy spectrum}
\label{sec:bethe-equat-energy}

In this section, we will assume that for all subsets \(I\subset \ninter
1 {\Kr+\Mr}\), the {\Tfs} satisfying (\ref{eq:BT1},~\ref{eq:BT2}) are 
polynomial functions of the spectral parameter \(\su\).
We can then define some {\Qfs}, which are {\Tfs} associated to empty
{\yn} diagrams:
\begin{align}
  \framedline{\gQf[][I]\equiv}{\rTf[][0][0][I]}\,\,.
\end{align}
\index{Q-functions@{\Qfs}}
This defines \(2^{\Kr+\Mr}\) different {\Qfs}, which are all polynomial
in the spectral parameter \(\su\).

We will also assume that the function 
\(\gQf[][\emptyset]\) is independent of \(\su\) (see also section
\ref{sec:solution-generale-de} for an interpretation of this
constraint in terms a physical gauge).

We will show that under these hypotheses, we will be able to recover
the Bethe equations (derived in section \ref{sec:integr-et-} in
the case of the XXX\(_{1/2}\) Heisenberg {\cds}), and to
express
the spectrum of the model.

In the next section \ref{sec:expr-diff-Qop}, we will construct an
explicit realization of this Bäcklund flow, and prove that the
polynomiality condition is satisfied (and that \(\gQ[][\emptyset]\) is independent of \(\su\)).

\subsubsection{``Dressing'' procedure and {\Qfs}}
\label{sec:dressing-procedure}

For characters, we noticed that the relation \eqref{eq:TQ-su-char} was
tightly connected to the expression \eqref{eq:GenSerIs} of the
generating series of symmetric characters. Let us now generalize this
to {\Tfs}: the restriction of \eqref{eq:BT1} to \(a=0\) reads \index{T-functions@{\Tfs}!TQ-relation}
\begin{align}
\label{eq:TQs}
  \rTf[\su][1][][I\Delta\jvp]\gQf[][I] -
  \gQf[][I\Delta\jvp]\rTf[][1][][I] =& x_{\jvp} \rTf[\su+1][1][s-1][I\Delta\jvp]
  \gQf[\su-1][I]\,,
\end{align}
which will be called the ``{\Tfu}{\Qfu}-relation''. It can be rewritten 
in terms of a generating series %
if we
define
\begin{align}
\label{eq:DefWIfun}
 \framedline{\Wt[][][I]\equiv}{\sum_{s=0}^\infty z^s \rTf[][1][][I]}\,\,.
\end{align}
Then, the equation \eqref{eq:TQs} is equivalent to
\begin{align}
\label{eq:TQWs}
\Wt[][][I\Delta\jvp]\gQf[][I] -
  \Wt[][][I]\gQf[][I\Delta\jvp] =& x_{\jvp}~z~ \Wt[][\su+1][I\Delta\jvp]
  \gQf[\su-1][I]\,,
\end{align}
or equivalently
\begin{align}
  \label{eq:TQOp}
\Wt[][][I]=&%
\frac{\gQf[][I]}{\gQf[][I\Delta\jvp]}
\left(
1-x_\jvp~z~\frac{\gQf[\su-1][I]}{\gQf[\su][I]} e^{\partial_\su}
\right)
\Wt[][][I\Delta\jvp]\,,
\end{align}
where %
the {\op} 
\(\eu\) is defined by
\begin{align}
\left[\eu f(\su)\right]=&\sum_{\nn=0}^\infty\left[\frac{\partial_\su^\nn}{{\nn}!}f(\su)\right]=f(\su+1).
\end{align}
We see that 
\begin{align}
  \left[\eu f(\su) {\tilde f}(\su)\right]=&f(\su+1){\tilde f}(\su+1)=f(\su+1)\left[\eu {\tilde f}(\su)\right]\,
\end{align}
so that \(\eu\) actually obeys the rule
\begin{align}
  \eu f(\su)=&f(\su+1)\eu\,.
\end{align}
\index{edu@\ensuremath{e^{\partial_\su}}}

If we recall the definition \eqref{eq:DeltaSugr} of \(I\Delta {\jvp}\),
\eqref{eq:TQWs} can be rewritten as
\begin{align}
  \label{eq:TQsugrOp1}
&\quad\Wt[][][I,\jvp]=\OOp \Wt[][][I]\\[3mm]
  \label{eq:TQsugrOpD}
\where \OOp=&
\left\{
  \begin{array}{ccc}
    \left(
1-x_\jvp~z~\frac{\gQf[\su-1][I]}{\gQf[\su][I]} e^{\partial_\su}
\right)^{-1}
\frac{\gQf[][I,\jvp]}{\gQf[][I]}&\If {\sg \jvp}=1\\[.3cm]
\frac{\gQf[][I,\jvp]}{\gQf[][I]}
\left(
1-x_\jvp~z~\frac{\gQf[\su-1][I,\jvp]}{\gQf[\su][I,\jvp]} e^{\partial_\su}
\right)
&\If {\sg \jvp}=-1
  \end{array}
\right.\,.
\end{align}
In the case \({\sg \jvp}=-1\), this expression is obtained from \eqref{eq:TQWs} by the
substitution \(I\to I',\jvp\) and \(I\Delta \jvp\to I'\).  In \eqref{eq:TQsugrOpD}, the
{\op}  
\(\left(
1-x_\jvp~z~\frac{\gQf[\su-1][I]}{\gQf[\su][I]} e^{\partial_\su}
\right)^{-1}
\frac{\gQf[][I,\jvp]}{\gQf[][I]}\) is defined by
\begin{align}
\left(
1-x_\jvp~z~\frac{\gQf[\su-1][I]}{\gQf[\su][I]} e^{\partial_\su}
\right)^{-1}
&\frac{\gQf[][I,\jvp]}{\gQf[][I]}\equiv\sum_{\nn=0}^{\infty}
\left(x_\jvp~z~\frac{\gQf[\su-1][I]}{\gQf[\su][I]}
  e^{\partial_\su}\right)^\nn \frac{\gQf[][I,\jvp]}{\gQf[][I]}\\
=&\sum_{\nn=0}^{\infty}
\frac{\gQf[\su-1][I]}{\gQf[\su+\nn-1][I]}\frac{\gQf[\su+\nn][I,\jvp]}{\gQf[\su+\nn][I]}
\left(
x_\jvp~z~ e^{\partial_\su}
\right)^\nn\,.
\end{align}

As it was already said, the {\Tfs} associated to the set \(I\) are
non-zero only inside the lattice \(\HK(\kk_I,\mm_I)\) where \(\kk_I\) and \(\mm_I\)
denote the number of elements of \(I\) with gradings \(+1\) and \(-1\):
\begin{align}
  \kk_I=&\mathrm{Card}\{\ivp \in I | {\sg  \ivp}=1\}
&  \mm_I=&\mathrm{Card}\{\ivp \in I | {\sg  \ivp}=-1\}\,.
\end{align}
In particular, we see that \(\forall s\geq 1,
\rTf[][1][][\emptyset]=0\), so that the definition \eqref{eq:DefWIfun}
gives \(\Wt[][][\emptyset] = 
\gQf[][\emptyset]\).
Therefore, the {\Tfs} \(\rTf[][1][][I]\) can be reconstructed from
\eqref{eq:TQsugrOp1} by choosing a {\nesting} {\ppath} %
{\idest} an
ordering of the elements of \(I=\{\ivp_1,\ivp_2,\cdots,\ivp_{|I|}\}\)
: 
\begin{align}
\label{eq:WDressing}
   \framedline{\Wt[][][I]=}{\OOp[\ivp_{|I|}][\Il{|I|-1}]\cdot
   \OOp[\ivp_{|I|-1}][\Il{|I|-2}] \cdots \OOp[\ivp_{1}][\Il{0}]
   \gQf[][\emptyset]}\\
 \where \Il { \nn}\equiv&\{\ivp_1,\ivp_2,\cdots,\ivp_\nn\}\hspace{3cm}\And \Il { 0}=\emptyset.
\end{align}
This expression provides a ``dressing procedure'', in the sense that
it gives an expression of \(\Wt[][][I]=\sum \rTf[][1][][I] z^s\)
(hence it also gives an expression of \(\rTf[][1][][I]\)) in terms of the {\Qfs}
\(\gQf[][I']\) associated to subsets \(I'\subset I\) (which were defined
through the  ``undressing procedure''). To recover the {\Tfs}
associated to an arbitrary {\rp} \(\lambda\), one should simply
use the Cherednik-Bazhanov-Reshetikhin formula \eqref{eq:CBR} (we will
do it in section \ref{sec:wronsk-expr-qq}).

Let us now see how this ``dressing procedure'' allows to recover the
Bethe equations and the spectrum of our {\cds}, under the
assumption that {\Qfs} are polynomial functions of the variable \(\su\).

\subsubsection{{\Qfu}{\Qfu}-relations}
\label{sec:qq-relations}

\index{Q-functions@{\Qfs}!QQ-relations|(}

In the dressing procedure above, the choice of the {\nesting} {\ppath} is
arbitrary, whereas 
the expression of \(\Wt[][][I]\) should not depend on the
{\ppath}. We will show that in order to make \(\Wt[][][I]\) independent of
the {\ppath}, a  consistency condition called ``{\Qfu}{\Qfu}-relation'' has to
hold. For that purpose, let us write \eqref{eq:WDressing} for the two
{\nesting} paths
\((\ivp_1,\ivp_2,\cdots,\ivp_{\nn},\jvp,\kvp)\)  and
\((\ivp_1,\ivp_2,\cdots,\ivp_{\nn},\kvp,\jvp)\)  of the
set \(I,\jvp,\kvp\) (where \(I=\{\ivp_1,\ivp_2,\cdots,\ivp_{\nn}\}\)). For
these paths, the relation \eqref{eq:WDressing} gives
\begin{align}
  \Wt[][][I,\jvp,\kvp]=&
\OOp[\kvp][I,\jvp] \cdot
  \OOp[\jvp][I] \cdot \Wt[][][I]
=
\OOp[\jvp][I,\kvp] \cdot
  \OOp[\kvp][I] \cdot \Wt[][][I]\,,
\end{align}
which gives the 
consistency constraint: %
\begin{align}
\label{eq:qq-diff-fnct}
\OOp[\kvp][I,\jvp] \cdot
  \OOp[\jvp][I] =&\OOp[\jvp][I,\kvp] \cdot
  \OOp[\kvp][I]\,.
\end{align}

Let us show how to write a constraint on {\Qfs} from the equation
\eqref{eq:qq-diff-fnct}. For simplicity %
let us start with the case
\({\sg  \jvp}={\sg  \kvp}=1\). Then \eqref{eq:qq-diff-fnct} implies
\begin{align}
\label{eq:qq-diff-fnct2}
  \OOp[\kvp][I]^{-1} \cdot
\OOp[\jvp][I,\kvp]^{-1} =& 
\OOp[\jvp][I]^{-1} \cdot \OOp[\kvp][I,\jvp]^{-1}  \,.
\end{align}
If we plug the expression \eqref{eq:TQsugrOpD} into
\eqref{eq:qq-diff-fnct2}, we get
\begin{align}
\frac{\gQf[][I]}{\gQf[][I,\kvp]}
      \left(
1-x_\kvp~z~\frac{\gQf[\su-1][I]}{\gQf[\su][I]} e^{\partial_\su}
\right)
\frac{\gQf[][I,\kvp]}{\gQf[][I,\jvp,\kvp]}
      \left(
1-x_\jvp~z~\frac{\gQf[\su-1][I,\kvp]}{\gQf[\su][I,\kvp]} e^{\partial_\su}
\right)
-\jvp\leftrightarrow\kvp=&0\,,
\end{align}
where \(f(\jvp,\kvp)-\jvp\leftrightarrow\kvp\) denotes
\(f(\jvp,\kvp)-f(\kvp,\jvp)\). %
The term of degree \(0\) in \(z\) is
\(\frac{\gQf[][I]}{\gQf[][I,\jvp,\kvp]} -\jvp\leftrightarrow\kvp\),
which is trivially zero. The term of degree \(2\) in \(z\) is
\(z^2 x_{\jvp} x_{\kvp}
\frac{\gQf[\su-1][I]%
}{%
  \gQf[\su+1][I,{\jvp},\kvp]} -\jvp\leftrightarrow\kvp\), which is also
trivially zero. Therefore, \eqref{eq:qq-diff-fnct} reduces to %
\begin{align}
\left(x_{\jvp}\frac{\gQf[][I]\gQf[\su-1][I,\kvp]}{\gQf[][I,{\jvp},\kvp]\gQf[][I,\kvp]}
+x_{\kvp}\frac{\gQf[\su-1][I]\gQf[\su+1][I,\kvp]}{\gQf[\su+1][I,{\jvp},\kvp]\gQf[][I,\kvp]}
-\jvp\leftrightarrow\kvp
\right) z ~\eu=&0\,.
\end{align}
After division by \(z ~\eu\), and multiplication by
\({\gQf[][I,{\jvp},\kvp]\gQf[\su+1][I,{\jvp},\kvp]\gQf[][I,\kvp]}\gQf[][I,\jvp]\),
this equation can be written in terms of \(2\times 2\) determinants:
\begin{multline}
\gQf[\su+1][I,{\jvp},\kvp] \gQf[][I]
\begin{vmatrix}
  x_{\jvp} \gQf[][I,\jvp] & x_{\kvp} \gQf[][I,\kvp] \\ \gQf[\su-1][I,\jvp] & \gQf[\su-1][I,\kvp]
\end{vmatrix}\\=\gQf[][I,{\jvp},\kvp] \gQf[\su-1][I]
\begin{vmatrix}
  x_{\jvp} \gQf[\su+1][I,\jvp] & x_{\kvp} \gQf[\su+1][I,\kvp] \\ \gQf[\su][I,\jvp] & \gQf[\su][I,\kvp]
\end{vmatrix}\,.
\end{multline}
This is equivalent to 
\begin{gather}
  \mathcal{A}(\su+1)=\mathcal{A}(\su)\,\\
\where \mathcal{A}(\su)= \begin{vmatrix}
  x_{\jvp} \gQf[][I,\jvp] & x_{\kvp} \gQf[][I,\kvp] \\ \gQf[\su-1][I,\jvp] & \gQf[\su-1][I,\kvp]
\end{vmatrix} / \left(\gQf[][I,{\jvp},\kvp] \gQf[\su-1][I]\right)\,.
\label{eq:qqdperegq}
\end{gather}
Due to the polynomiality of {\Qfs}, \(\mathcal{A}\) is then a
constant. The value of this constant can be viewed as a normalization,
because the equations (\ref{eq:BT1},~\ref{eq:BT2}) are invariant under
the transformation
\begin{align}
 \gQf[][I]\rightsquigarrow& c_I \gQf[][I]\,,&
  \rTf[][][][I]\rightsquigarrow& c_I \rTf[][][][I]\,,
\label{eq:cIfreedom}
\end{align}
where \(c_I\) is independent of \(a\), \(s\), and \(\su\) (it is only a
function of \(I\)). For instance, the freedom \eqref{eq:cIfreedom} can be
used to enforce that the coefficient of highest degree in \(\gQf[][I]\)
is equal to one. With this choice, \eqref{eq:qqdperegq} becomes the
following ``{\Qfu}{\Qfu}-relation'':
  \begin{multline}
\gQf[][I,{\jvp},\kvp] \gQf[\su-1][I]=\begin{vmatrix}
  x_{\jvp} \gQf[][I,\jvp] & x_{\kvp} \gQf[][I,\kvp] \\ \gQf[\su-1][I,\jvp] & \gQf[\su-1][I,\kvp]
\end{vmatrix} / \left(x_{\jvp}-x_{\kvp}\right)\,.
\label{eq:qq++}\\
\If \qquad \sg {\jvp} = \sg {\kvp} =1\,.
\end{multline}

If \(\sg {\jvp}\) or \(\sg {\kvp}\) is equal to \(-1\), then one can repeat this %
chain of arguments %
for all possible values of \(\sg {\jvp}\) and \(\sg {\kvp}\).
This way, we get a generalization of \eqref{eq:qq++}
to arbitrary grading which reads (with the notations \eqref{eq:DeltaSugr}):
\begin{align} 
\framedline{\gQf[][I\Delta {\jvp}\Delta \kvp] \gQf[\su-1][I]=}{\begin{vmatrix}
  x_{\jvp} \gQf[][I\Delta \jvp] & x_{\kvp} \gQf[][I\Delta \kvp] \\ \gQf[\su-1][I\Delta \jvp] & \gQf[\su-1][I\Delta \kvp]
\end{vmatrix} / \left(x_{\jvp}-x_{\kvp}\right)}\,\,.
\label{eq:QQsu}
\end{align}
This is very natural because we can see from \eqref{eq:TQWs} that the
effect of gradings can be encoded into the notation \(I\Delta {\jvp}\), which
adds the element \({\jvp}\) if \(\sg {\jvp}=1\) or removes it if \(\sg {\jvp} =-1\).

In this section, %
{\Qfu}{\Qfu}-relations were derived as a %
consistency condition for \eqref{eq:WDressing}. They were obtained by
asking that the generating
series \(\Wt[][][I]\) is the same for two specific different {\nesting}
paths (where only the last two indices are interchanged). 

Actually, the {\Qfu}{\Qfu}-relations even imply that the generating
series \(\Wt[][][I]\) is the same for two arbitrary {\nesting}
paths \(({\ivp}_1,{\ivp}_2,\cdots,{\ivp}_\nn)\) and \(({\ivp}_{\sigma(1)},{\ivp}_{\sigma(2)},\cdots,{\ivp}_{\sigma(\nn)})\).
Indeed, the permutation \(\sigma\) can be written
as a product of transpositions of the form \(\tau_{[\nn,\nn+1]}\)
(defined in \eqref{eq:DefTAUij0}).
As the arguments above ensure that \(\Wt[][][I]\) is invariant under the
change of {\ppath} associated to each \(\tau_{[\nn,\nn+1]}\), they imply
that \(\Wt[][][I]\) is indeed the same for two arbitrary {\nesting} paths.

\index{Q-functions@{\Qfs}!QQ-relations|)}

\subsubsection{Bethe equations}
\label{sec:bethe-equations}

\index{Bethe equations |(}

Now %
we can show how to find the {\Qfs} which have to be plugged into
the {\ops} 
\(\OOp[\ivp][I]\) in 
 the
{\rhs} of \eqref{eq:WDressing}. We will find a set of equations on
these {\Qfs} (called the
Bethe equations)
\cite{Lai1974,PhysRevB.12.3795,Kulish:1983rd,Babelon:1981un}
 such that, if we find a solution of these equations,
we will be able to plug it into the {\rhs} of \eqref{eq:WDressing} and
to write a consistent ({\idest} polynomial) expression for the {\Tfs}. In
particular we will see that for \(\Kr=2\), we find the same Bethe
equation as 
the
equation \eqref{eq:QuantBetheN}, which was obtained in the
introductive section \ref{sec:integr-et-} for the XXX\(_{1/2}\)
Heisenberg {\cds}. %
 To show this, we will assume that
{\Qfs} are polynomials. %
 Let us then 
  denote by \(\su_I^{\rlb[\nrt]}\)
their roots, which we will call ``Bethe roots'': %

\index{Bethe roots}
\index{u(s@\ensuremath{\su} (spectral parameter)!uIn@\ensuremath{\su_I^{\rlb[\nrt]}}|see{Bethe roots}}
\begin{align}
\label{eq:QfromRoots}
  \gQ[][I]=&\alpha_I \prod_{\nrt=1}^{d_I}\left(\su-\su_I^{\rlb[\nrt]}\right)\,.
\end{align}
Then, 
the QQ-relation \eqref{eq:QQsu} can be written at positions
\(\su=\su_{I\Delta {\jvp}}^{\rlb[\nrt]}\) and \(\su=\su_{I\Delta
  {\jvp}}^{\rlb[\nrt]}+1\). In both of these positions one term is zero in the {\rhs},
and we get
\begin{empheq}[left=\empheqlbrace]{align}
 \label{eq:toBethe1}
\left.\left(x_{\jvp}-x_{\kvp}\right)  \gQf[\su][I\Delta {\jvp}\Delta
 \kvp] \gQf[\su-1][I]
+x_{\kvp}  \gQf[\su-1][I\Delta \jvp] \gQf[\su][I\Delta \kvp] \right|_{\su = \su_{I\Delta {\jvp}}^{\rlb[\nrt]}} =&0\,,\\
\left. \left(x_{\jvp}-x_{\kvp}\right)  \gQf[\su+1][I\Delta {\jvp}\Delta
 \kvp] \gQf[\su][I]
-x_{\jvp} \gQf[\su+1][I\Delta \jvp] \gQf[\su][I\Delta \kvp]\right|_{\su = \su_{I\Delta {\jvp}}^{\rlb[\nrt]}} =&0\,.
\label{eq:toBethe2}
\end{empheq}
We can then take a linear combination of \eqref{eq:toBethe1} and
\eqref{eq:toBethe2} in such a way that the coefficient of \(\gQf[\su_{I\Delta
  {\jvp}}^{\rlb[\nrt]}][I\Delta \kvp]\) cancels. This gives 
\begin{multline}
x_{\jvp} \gQf[\su+1][I\Delta \jvp]  \left(x_{\jvp}-x_{\kvp}\right)  \gQf[\su][I\Delta {\jvp}\Delta
 \kvp] \gQf[\su-1][I]\\
\left.  + x_{\kvp} \gQf[\su-1][I\Delta \jvp]  \left(x_{\jvp}-x_{\kvp}\right)  \gQf[\su+1][I\Delta {\jvp}\Delta
 \kvp] \gQf[\su][I]\right|_{\su=\su_{I\Delta {\jvp}}^{\rlb[\nrt]}}=0\,.
\end{multline}
An equivalent way to write it is 
\begin{align}
\label{eq:BetheEqQQ}
\standardline
{\frac{\gQf[\su_{I\Delta {\jvp}}^{\rlb[\nrt]}+1][I\Delta \jvp]  %
  \gQf[\su_{I\Delta {\jvp}}^{\rlb[\nrt]}][I\Delta {\jvp}\Delta
 \kvp] \gQf[\su_{I\Delta {\jvp}}^{\rlb[\nrt]}-1][I]}
{\gQf[\su_{I\Delta {\jvp}}^{\rlb[\nrt]}-1][I\Delta \jvp]  %
  \gQf[\su_{I\Delta {\jvp}}^{\rlb[\nrt]}+1][I\Delta {\jvp}\Delta
 \kvp] \gQf[\su_{I\Delta {\jvp}}^{\rlb[\nrt]}][I]}
=}{-\frac{x_{\kvp}}{x_{\jvp}}}\,\,.
\end{align}

In this equation, we want to choose \(I\), \(\jvp\) and \(\kvp\) such that
the {\Qfs} involved in \eqref{eq:BetheEqQQ} lie along a given {\nesting}
{\ppath}, so as to plug them into the expression \eqref{eq:WDressing} of
the generating series of the {\Tfs}. If both \(\jvp\) and \(\kvp\) have grading \(\sg \jvp=\sg \kvp
=1\), then from the definition \eqref{eq:DeltaSugr} we see that \(I\),
\(I\Delta \jvp=I\cup\{\jvp\}\), and \(I\Delta \jvp\Delta
\kvp=I\cup\{\jvp,\kvp\}\) can be chosen along the same {\nesting}
{\ppath}. Similarly, if \(\jvp\) and \(\kvp\) have grading \(\sg \jvp=\sg \kvp
=-1\), then \(I\), \(I\Delta \jvp=I\setminus\{\jvp\}\), and \(I\Delta \jvp\Delta
\kvp=I\setminus\{\jvp,\kvp\}\) can be chosen along the same {\nesting}
{\ppath}. By contrast, if \(\sg \jvp = - \sg \kvp\), then we see that \(I\)
and \(I\Delta \jvp \Delta \kvp\) have the same {\nesting} {\level} (for
instance, if \(\sg \jvp=1\) and \(\sg \kvp=-1\), then \(|I\Delta \jvp
\Delta \kvp|=|I\cup \{\jvp\} \setminus \{\kvp\}|=|I|\)), and therefore
they {\cannnot} lie on the same {\nesting} {\ppath}. In that case we should use
the relations 
\begin{align}
\frac{\gQf[\su_{I}^{\rlb[\nrt]}+1][I\Delta \jvp]
  \gQf[\su_{I}^{\rlb[\nrt]}][I\Delta \kvp]}
{\gQf[\su_{I}^{\rlb[\nrt]}][I\Delta \jvp]
 \gQf[\su_{I}^{\rlb[\nrt]}+1][I\Delta \kvp] } =&\frac
{x_{\kvp}}{x_{\jvp}}\,,
\quad 
\And 
\frac{\gQf[\su_{I\Delta\jvp\Delta\kvp}^{\rlb[\nrt]}][I\Delta \jvp]
  \gQf[\su_{I\Delta\jvp\Delta\kvp}^{\rlb[\nrt]}-1][I\Delta \kvp]}
{\gQf[\su_{I\Delta\jvp\Delta\kvp}^{\rlb[\nrt]}-1][I\Delta \jvp]
 \gQf[\su_{I\Delta\jvp\Delta\kvp}^{\rlb[\nrt]}][I\Delta \kvp] } =&\frac
{x_{\kvp}}{x_{\jvp}}\,,
\label{eq:Bethebf}
\end{align}
which arise by setting \(\su=\su_{I}^{\rlb[\nrt]}+1\) (resp \(\su=\su_{I\Delta\jvp\Delta\kvp}^{\rlb[\nrt]}\)) in \eqref{eq:QQsu}.

Finally, the Bethe equations are the set of \(\Kr+\Mr-1\) equations
corresponding to an arbitrary {\nesting} {\ppath} \(({\ivp}_1,{\ivp}_2,\cdots
{\ivp}_{\Kr+\Mr})\):

\Pv{\begin{subequations}
\label{eq:SuBetheSplit}
  \begin{equation*}
    \forall \mlvl \in \ninter 1 {\Kr+\Mr-1},\hspace{7cm}~
  \end{equation*}
  \begin{empheq}[left={\empheqlbrace}]{align}
    &
    \label{eq:SuBethebb}
    \frac{\gQf[\su_{I_\mlvl}^{\rlb[\nrt]}+1][I_\mlvl] %
      \gQf[\su_{I_\mlvl}^{\rlb[\nrt]}][I_{\mlvl+1}]
      \gQf[\su_{I_\mlvl}^{\rlb[\nrt]}-1][I_{\mlvl-1}]}
    {\gQf[\su_{I_\mlvl}^{\rlb[\nrt]}-1][I_\mlvl] %
      \gQf[\su_{I_\mlvl}^{\rlb[\nrt]}+1][I_{\mlvl+1}]
      \gQf[\su_{I_\mlvl}^{\rlb[\nrt]}] [I_{\mlvl-1}]}
    =-\frac{x_{{\ivp}_{\mlvl+1}}}{x_{{\ivp}_{\mlvl}}}\,,\\
    &&\llap{\If \sg{{\ivp}_\mlvl}= \sg {{\ivp}_{\mlvl+1}}=}+1
    \nonumber\\[.5cm]
    &
    \label{eq:SuBetheff}
    \frac{\gQf[\su_{I_\mlvl}^{\rlb[\nrt]}+1][I_\mlvl] %
      \gQf[\su_{I_\mlvl}^{\rlb[\nrt]}][I_{\mlvl-1}]
      \gQf[\su_{I_\mlvl}^{\rlb[\nrt]}-1][I_{\mlvl+1}]}
    {\gQf[\su_{I_\mlvl}^{\rlb[\nrt]}-1][I_\mlvl] %
      \gQf[\su_{I_\mlvl}^{\rlb[\nrt]}+1][I_{\mlvl-1}]
      \gQf[\su_{I_\mlvl}^{\rlb[\nrt]}] [I_{\mlvl+1}]}
    =-\frac{x_{{\ivp}_{\mlvl}}}{x_{{\ivp}_{\mlvl+1}}}\,,\\
    &&\llap{\If \sg{{\ivp}_\mlvl}= \sg {{\ivp}_{\mlvl+1}}=}-1
    \nonumber\\[.5cm]
    &
    \label{eq:SuBethebf}
    \frac{\gQf[\su_{I_\mlvl}^{\rlb[\nrt]}][I_{\mlvl-1}]
      \gQf[\su_{I_\mlvl}^{\rlb[\nrt]}-1][I_{\mlvl+1}]}
    {\gQf[\su_{I_\mlvl}^{\rlb[\nrt]}-1][I_{\mlvl-1}]
      \gQf[\su_{I_\mlvl}^{\rlb[\nrt]}][I_{\mlvl+1}] } =\frac
    {x_{\ivp_{\mlvl}}}{x_{\ivp_{\mlvl+1}}}\,,\\
    &&\llap{\If \sg{{\ivp}_\mlvl}=+1 \And \sg {{\ivp}_{\mlvl+1}}=}-1
    \nonumber\\[.5cm]
    &
    \label{eq:SuBethefb}
    \frac{\gQf[\su_{\Il\mlvl}^{\rlb[\nrt]}+1][\Il{\mlvl-1}]
      \gQf[\su_{\Il\mlvl}^{\rlb[\nrt]}][\Il{\mlvl+1}]}
    {\gQf[\su_{\Il\mlvl}^{\rlb[\nrt]}][\Il{\mlvl-1}]
      \gQf[\su_{\Il\mlvl}^{\rlb[\nrt]}+1][\Il{\mlvl+1}] } =\frac
    {x_{\ivp_{\mlvl+1}}}{x_{\ivp_{\mlvl}}}\,,\\
    &&\llap{\If \sg{{\ivp}_\mlvl}=-1 \And \sg {{\ivp}_{\mlvl+1}}=}+1 \nonumber
  \end{empheq}
  \begin{equation}
     \where \Il{\mlvl}\equiv\{\ivp_1,\ivp_2,\cdots,\ivp_\mlvl\}\,.
  \end{equation}
\end{subequations}
}

There, the cases \eqref{eq:SuBethebb} and \eqref{eq:SuBetheff} are
obtained from \eqref{eq:BetheEqQQ}. On the other hand,
\eqref{eq:SuBethebf} is obtained from \eqref{eq:Bethebf} by choosing
\({I=\Il{\mlvl-1}\cup\{\spi_{\mlvl+1}\}}\), and \(\jvp=\spi_{\mlvl+1}\),
\(\kvp=\spi_{\mlvl}\) (so that \(\Il{\mlvl-1}=I\Delta\jvp\),
\(\Il{\mlvl}=I\Delta\jvp\Delta\kvp\) and \(\Il{\mlvl+1}=I\Delta\kvp\)),
whereas 
\eqref{eq:SuBethefb} is obtained from \eqref{eq:Bethebf} by choosing
\(I=\Il{\mlvl}\), and \(\jvp=\spi_{\mlvl}\),
\(\kvp=\spi_{\mlvl+1}\) (so that \(\Il{\mlvl-1}=I\Delta\jvp\),
\(\Il{\mlvl}=I\) and \(\Il{\mlvl+1}=I\Delta\kvp\)).

The Bethe equations \eqref{eq:SuBetheSplit} 
should be viewed as a set of equations on the Bethe roots
\(\su_I^{\rlb[\nrt]}\), which fix the {\Qfs} on a {\nesting} {\ppath} (see
\eqref{eq:QfromRoots},  where the degree of freedom \(\alpha_I\) is
non-physical as can be seen from \eqref{eq:cIfreedom}).
Given a solution of this set of equations, we can write corresponding
polynomial expressions (from \eqref{eq:QfromRoots}) for the
{\Qfs} \(\gQ[][I_\mlvl]\) 
along a given {\nesting} {\ppath}, and 
plug them
into the {\ops} 
\(\OOp[\ivp][I]\) in the {\rhs} of
\eqref{eq:WDressing} in order to express the {\Tfs}.

Of course, there exist other ways to prove the Bethe equations:
interestingly one of them is to ask that every possible
pole vanishes when \(\rTf[][1][1]\) is
expressed from \eqref{eq:WDressing} (see next section).

\paragraph{Case of the XXX\(_{1/2}\) Heisenberg {\cds}}
\label{sec:case-xxx_12-heis}

Let us now show that \eqref{eq:SuBetheSplit}
allows to recover the Bethe equation
\eqref{eq:QuantBetheN} of the introductory section
\ref{sec:integr-et-}:
the {\Ham} \eqref{eq:hamilHeisenbergintro} corresponds to the
{\GL{2}} {\cds}, with all inhomogeneities set to \(\theta_\spi=0\).
Then \(\gQf[][\{1,2\}]=\su^\lcds\). If we assume that
\(\gQf[][\emptyset]\) is a constant (independent of \(\su\)), then we can
write \eqref{eq:SuBetheSplit} for the {\nesting} {\ppath} \(({\ivp}_1,{\ivp}_2)=(1,2)\):
\begin{gather}
\label{eq:BetheToToXXX7}
  \frac{\gQf[\su^{\rlb[\nrt]}+1][\{1\}]}
{\gQf[\su^{\rlb[\nrt]}-1][\{1\}]
}
{    \left(\frac{\su^{\rlb[\nrt]}}{\su^{\rlb[\nrt]}+1}\right)^\lcds }
=-\frac{x_{2}}{x_{1}}\,,\\
\where \qquad \su^{\rlb[\nrt]} \equiv \su_{\{1\}}^{\rlb[\nrt]}\,,
\end{gather}
or equivalently 
\begin{align}
  {    \left(\frac{\su^{\rlb[\nrt]}}{\su^{\rlb[\nrt]}+1}\right)^\lcds }
=&\frac{x_2}{x_1} \prod_{\mrt\neq \nrt}\frac {\su^{\rlb[\nrt]}-\su^{\rlb[\mrt]}-1}
{\su^{\rlb[\nrt]}-\su^{\rlb[\mrt]}+1}\,.
\label{eq:BetheToXXX}
\end{align}
In \eqref{eq:BetheToToXXX7} there is a minus sign in the
{\rhs}, whereas the left-hand side reads \(\prod_{\mrt}\frac
{\su^{\rlb[\nrt]}-\su^{\rlb[\mrt]}-1}{\su^{\rlb[\nrt]}-\su^{\rlb[\mrt]}+1}\). To get
\eqref{eq:BetheToXXX}, the factor \(\frac
{\su^{\rlb[\nrt]}-\su^{\rlb[\nrt]}-1}{\su^{\rlb[\nrt]}-\su^{\rlb[\nrt]}+1}\) is removed
from the product, and that exactly absorbs the minus sign.

The {\Ham} \eqref{eq:hamilHeisenbergintro} is obtained in the
limit \(\g\to 1\), where \(\frac {x_2}{x_1}\) is set to 1. Then if we
change the variables as \(\su^{\rlb[\nrt]}\equiv \frac{e^{\bi p_\nrt}}{1-e^{\bi
    p_\nrt}}\), the {\lhs} becomes exactly
\(e^{\bi ~\lcds~p_\nrt}\) while the {\rhs} becomes exactly
\(\prod_{\mrt\neq \nrt} {\Sscal}(p_\nrt,p_\mrt)\). Therefore \eqref{eq:BetheToXXX}
exactly gives the Bethe equation \eqref{eq:QuantBetheN}, found in the
introductory section. In the next section we will also see that this
formalism allows to recover the correct expression of the energy.

Assuming that the Bäcklund flow exists and is polynomial, we derived Bethe
equations \eqref{eq:BetheEqQQ} which generalize the equation
\eqref{eq:QuantBetheN} found for the XXX\(_{1/2}\) {\cds} in the 
introductory section. 

\index{Bethe equations |)}

\subsubsection{Energy spectrum}
\label{sec:energy-spectrum}

Under the assumptions of the previous sections ({\idest} if the Bäcklund flow
exists and is polynomial), one can even recover the spectrum of the {\Ham}.

If the {\Tfs} are the eigenvalues of the {\Toprs} 
defined in section \ref{sec:chaines-de-spin} for {\csds},
then the energy of a state is given by \(\frac 2 {\Kr+\Mr} \lcds - 2 
\left.\partial_{\su} \log \rTf[][1][1]\right|_{{\su}=0}\). This
expression is simply the
eigenvalue of the {\Ham} \(\Hami= \frac 2 {\Kr+\Mr} \lcds - 2 
\left.\partial_{\su} \log \rT[][1][1]\right|_{\su=0}\), where the {\Topr} \(\rT[][1][1]
  = 
  \gT[][\raisebox{-.07cm}{\includegraphics[scale=.2]{figYdiag_1}}]\)
  corresponds to the fundamental {\rp} (it was also denoted \(\Top(\su)\) in the section \ref{sec:equation-de-yang}).

Therefore,  we need to
express \(\rT[][1][1]\) in order to recover the spectrum of the
{\cds}. From \eqref{eq:DefWIfun} we see that \({\rT[][1][1]\equiv
\rT[][1][1][\ove]}\) is the coefficient of \(z^1\) in \(\Wt[][][\ove]\). If we
express \(\Wt[][][\ove]\) from \eqref{eq:WDressing},
we see that \(\rT[][1][1]\) is obtained by keeping the term in \(z^1\) in
one of the {\ops} \(\mathcal{O}_I\), and keeping the term of degree
\(z^0\) (which is equal to \(\frac{\gQf[][I,\jvp]}{\gQf[][I]}\)) for all
the other {\ops} \(\mathcal{O}_I\).
That gives
\begin{gather}
\label{eq:qtoT11}
 \rTf[][1][1]=\gQf[][\ove] \sum_{\kk=1}^{\Kr+\Mr}
\sg {\ivp_\kk} x_{\ivp_\kk} \frac{\gQf[\su-\sg {\ivp_\kk}][\Il {{\kk}-1}]\gQf[\su+\sg {\ivp_\kk}][\Il {\kk}]}{\gQf[][\Il {{\kk}-1}]\gQf[][\Il {\kk}]}\,.\\
\where \Il {{\kk}}\equiv \{\ivp_1,\ivp_2,\cdots,\ivp_\kk\}\,, \qquad\textrm{such
  that}\qquad \Il{\Kr+\Mr}=\ninter 1 {\Kr+\Mr}\,.
\end{gather}

In the case of the {\cds} with inhomogeneities \(\theta_\spi=0\), we
get \(\gQf[][\ove]=\su^\lcds\). Therefore, if the length of the
{\cds} is \(\lcds\geq 2\), then \(\gQf[0][\ove]=0\)
and \({\left.\partial_\su \gQf[][\ove]\right|_{\su
  =0}=0}\).
To write the energy of a state, we want to write \(\left.\frac{\partial_{\su}
    \gT[][{ }]}{\gT[][{ }]}\right|_{{\su}=0}\), and
in \eqref{eq:qtoT11}, we can see that the terms \(\kk\leq \Kr+\Mr-1\)
contain the prefactor \(\gQf[][\ove]\), hence these terms do not
contribute\footnote{Rigorously, this
   argument assumes that the %
   {\Qfs} on the denominator
   have no zero at \(\su=0\) when \(\kk<\Kr+\Mr\). As a consequence the
   argument holds provided the roots of the {\Qfs} \(\gQf[][\Il 1]\),
   \(\gQf[][\Il 2]\), \(\cdots\), \(\gQf[][\Il \Kr+\Mr-1]\) are all non-zero.
   In fact, \(\rTf[][1][1]\) is a rational function of all these roots,
   and by continuity, the expression \eqref{eq:left.p-log-tfus} will
   hold even if some \(\gQf[][\Il {\jvp}]\) has a zero at \(\su=0\).
} to \(\left.\partial_{\su}
    \gT[][{ }]\right|_{{\su}=0}\) and to \(\left.\gT[][{
    }]\right|_{{\su}=0}\). By contrast, the last term \(\kk=\Kr+\Mr\)
  contains \(\gQf[][\ove]\) both in the numerator and the denominator,
  so that this term reduces to 
\begin{gather*}
\sg {\ivp_{\Kr+\Mr}} x_{\ivp_{\Kr+\Mr}} \frac{\gQf[\su-\sg {\ivp_{\Kr+\Mr}}][\Il {{{\Kr+\Mr}}-1}]\gQf[\su+\sg {\ivp_{\Kr+\Mr}}][\ove]}{\gQf[][\Il {{{\Kr+\Mr}}-1}]}
\end{gather*}
which contributes to  \(\left.\partial_{\su}
    \gT[][{ }]\right|_{{\su}=0}\) and to \(\left.\gT[][{
    }]\right|_{{\su}=0}\).
This term gives
 \begin{gather}
\label{eq:left.p-log-tfus}
\left.\partial_{\su} \log \Tfu({\su})\right|_{{\su}=0}=\sg {{\ivp}_{\Kr+\Mr}}\lcds + \left.\frac{\partial_\su\gQf[][\Il
       {\Kr+\Mr-1}]}{\gQf[][\Il
       {\Kr+\Mr-1}]}\right|_{\su= - \sg {{\ivp}_{\Kr+\Mr}}} %
   - \left.\frac{\partial_\su\gQf[][\Il
       {\Kr+\Mr-1}]}{\gQf[][\Il
       {\Kr+\Mr-1}]}\right|_{\su=0}\,.
 \end{gather}

For the Heisenberg XXX\(_{1/2}\) {\cds}, we have \(\Kr=2\), \(\Mr=0\),
and the energy %
\(E=\lcds-2\left.\partial_{\su} \log \Tfu({\su})\right|_{{\su}=0}\) is equal
 to
\begin{align}
  E=&-\lcds+2 \sum_{\nrt} \left(\left.\partial_{\su} \log
    (\su-\su^{\rlb[\nrt]})\right|_{{\su}=0} - \left.\partial_{\su} \log
    (\su-\su^{\rlb[\nrt]})\right|_{{\su}=-1}\right)\\
=&-\lcds-2\sum_{\nrt} \frac 1
{\su^{\rlb[\nrt]}(\su^{\rlb[\nrt]}+1)}=-\lcds+4-4 \cos(p_\nrt)\,,
\label{eq:EHeisenFromBacklund}
\end{align}
where the last line, which is obtained by the change of
variables\footnote{ This change of variables 
 was already used to express the Bethe equations
in terms of the momenta \(p_\nrt\) of the introductory section
\ref{sec:integr-et-}.
} \(\su^{\rlb[\nrt]}\equiv \frac{e^{\bi p_\nrt}}{1-e^{\bi
    p_\nrt}}\),
coincides with the energy \eqref{eq:BetheEnergyN} obtained in section
\ref{sec:integr-et-}.

\subsubsection{Bethe equations for {\GL \Kr} with $\Kr>2$}
\label{sec:bethe-equations-when}

\index{Bethe equations |(}

The equations \eqref{eq:SuBetheSplit} (which reduce to
\eqref{eq:BetheEqQQ} in the case of {\GL \Kr}) are often
called ``{\nested} Bethe {\anz}'' equations \cite{PhysRevB.12.3795},
and it is also possible %
to derive them
by the methods of the introductory section
\ref{sec:integr-et-},
though it is  %
much more complicated than for the Heisenberg {\cds}. Instead of this
study, let us simply mention in what sense they are a generalization
of \eqref{eq:QuantBetheN}.

To do this, we should remember that for \(\Kr=2\), we have seen that 
each
excited state %
can be 
{\lbd} by a
different set of roots\footnote{These roots \(\su^{\rlb[\nrt]}\) were
  indeed associated (via a change of variables) to the 
  momenta of the spin waves of section \ref{sec:integr-et-}. } for
the {\Qfs}. This will be 
interpreted in section \ref{sec:expr-diff-Qop} as the fact that
each eigenstate is associated to a different eigenvalue of
the {\Qoprs}. These roots will be called Bethe roots, and for \(\Kr=2\)
they are identical (up to a simple change of variable) to  the
momenta parameterizing the states in \eqref{eq:AnsatzNPart}.

For \(\Kr>2\), the excited states should therefore be {\lbd} by
the \(\Kr-1\) sets \linebreak \(\{\su_{\{1\}}^{\rlb[{\nrt}]}|1\leq {\nrt}\leq d_{\{1\}}\}\),
\(\{\su_{\{1,2\}}^{\rlb[{\nrt}]}|{1\leq {\nrt}\leq d_{\{1,2\}}}\}\), \(\cdots\),
  \(\{\su_{\{1,2,\cdots \Kr-1 \}}^{\rlb[{\nrt}]}|{1\leq {\nrt}\leq
    d_{\{1,2,\cdots,\Kr-1\}}\}}\), where \(d_I\) denotes the degree of
  the polynomial \(\gQf[][I]\).
 They are the Bethe roots which define
  the {\Qfs}\footnote{There are only \(\Kr-1\) sets of Bethe roots,
    because the {\Qf} \(\gQf[][\ove]=\prod_{\spi=1}^\lcds \su_\spi\) is
    known, and the function \(\gQf[][\emptyset]\) is supposed to be a
    constant, which will be proven in the next section} of the {\nesting}
  {\ppath} \((1,2,\cdots,\Kr)\), and we will denote them as
  \begin{gather}
\label{eq:notaroot}
    \su^{\rlb[][\mlvl][\nrt]}\equiv\su_{\{1,2,\cdots \mlvl
      \}}^{\rlb[\nrt]}\,,\qquad\If \nrt\leq {\dg[\mlvl]}\,,\qquad \where {\dg[\mlvl]}\equiv d_{\{1,2,\cdots,\mlvl\}}\,.
  \end{gather}
In \eqref{eq:notaroot}, the integer \(\mlvl\) will be called the ``{\level}''
of the root.
  
The {\Ham} \eqref{eq:HaK1} arises in the limit where \({\g}=1\) and
where \(\forall \spi,\su_\spi=0\). In this limit, %
we can write the equation 
\eqref{eq:SuBetheSplit} at the {\nesting} {\level} \(\mlvl\) ({\ie} we set
\({I=(1,2,\cdots,\mlvl-1)}\), \(\jvp=\mlvl\) and \(\kvp=\mlvl+1\)) to
get
\begin{multline}
 \left(%
   \prod_{\substack{\lrt\neq\nrt\\
1\leq \lrt \leq {\dg[\mlvl]}%
}} 
\frac{\su^{\rlb[][\mlvl][\nrt]}-\su^{\rlb[][\mlvl][\lrt]}+1}{\su^{\rlb[][\mlvl][\nrt]}-\su^{\rlb[][\mlvl][\lrt]}-1}
 \right)
 \left(\prod_{\substack{
1\leq \lrt \leq {\dg[\mlvl+1]}%
}} 
\frac{\su^{\rlb[][\mlvl][\nrt]}-\su^{\rlb[][\mlvl+1][\lrt]}}{\su^{\rlb[][\mlvl][\nrt]}-\su^{\rlb[][\mlvl+1][\lrt]}+1}
 \right)\\
\times
 \left(\prod_{\substack{
1\leq \lrt \leq {\dg[\mlvl-1]}%
}} 
\frac{\su^{\rlb[][\mlvl][\nrt]}-\su^{\rlb[][\mlvl-1][\lrt]}-1}{\su^{\rlb[][\mlvl][\nrt]}-\su^{\rlb[][\mlvl-1][\lrt]}}
 \right)
 =%
 1
\end{multline}
if \(2\leq\mlvl\leq \Kr-1\). We also get
\begin{gather}
 \left(%
   \prod_{\substack{\lrt\neq\nrt\\
1\leq \lrt \leq {\dg[1]}%
}} 
\frac{\su^{\rlb[][1][\nrt]}-\su^{\rlb[][1][\lrt]}+1}{\su^{\rlb[][1][\nrt]}-\su^{\rlb[][1][\lrt]}-1}
 \right)
 \left(\prod_{\substack{
1\leq \lrt \leq {\dg[2]}%
}} 
\frac{\su^{\rlb[][1][\nrt]}-\su^{\rlb[][2][\lrt]}}{\su^{\rlb[][1][\nrt]}-\su^{\rlb[][2][\lrt]}+1}
 \right)
 =%
 1
\end{gather}
for \(\mlvl=1\), and finally 
\begin{multline}
 \left(%
   \prod_{\substack{\lrt\neq\nrt\\
1\leq \lrt \leq {\dg[\Kr-1]}%
}} 
\frac{\su^{\rlb[][\Kr-1][\nrt]}-\su^{\rlb[\Kr-1,\lrt]}+1}{\su^{\rlb[][\Kr-1][\nrt]}-\su^{\rlb[\Kr-1,\lrt]}-1}
 \right)
 \left(\prod_{\substack{
1\leq \lrt \leq {\dg[\Kr-2]}%
}} 
\frac{\su^{\rlb[][\Kr-1][\nrt]}-\su^{\rlb[][\Kr-2][\lrt]}-1}{\su^{\rlb[][\Kr-1][\nrt]}-\su^{\rlb[][\Kr-2][\lrt]}}
 \right)\\
 =%
 \left(\frac {\su^{\rlb[][\Kr-1][\nrt]}+1}{\su^{\rlb[][\Kr-1][\nrt]}} \right)^\lcds
\end{multline}
for \(\mlvl=\Kr-1\). These expressions are obtained exactly like in
\eqref{eq:BetheToXXX}, and in particular the minus sign in
\eqref{eq:SuBethebb} is absorbed into the condition \(\lrt\neq\nrt\) in
the first product of each equality.

These three equations can be rewritten as
\begin{align}
\label{eq:BethBigerRank}
  \framedline{
    \begin{array}{c}
\forall \mlvl \in \ninter 1 {\Kr-1}\\
\forall \nrt \in \ninter 1 {{\dg[\mlvl]}}
\end{array}
,\qquad
}{e^{\bi~ \lcds~ p^{\rlb[\mlvl]}\left(\su^{\rlb[][\mlvl][\nrt]}\right)}
=
\prod_{\substack{\klvl \in \ninter 1 {\Kr-1}\\
\lrt \in \ninter 1 {{\dg[\klvl]}}\\
(\klvl,\lrt)\neq (\mlvl,\nrt)
}} \Sscal^{\rlb[][\mlvl][\klvl]}(\su^{\rlb[][\mlvl][\nrt]}-\su^{\rlb[][\klvl][\lrt]})}\,\,,
\end{align}
where the product on the {\rhs} runs over all the \((\klvl,\lrt)\)
such that \(\klvl\neq \mlvl\) or \(\lrt\neq \nrt\), {\idest} over all the other Bethe
roots except the root \(\su^{\rlb[][\mlvl][\nrt]}\). In \eqref{eq:BethBigerRank},
we define
\begin{align}
\label{eq:pmRank}
  p^{\rlb[\mlvl]}\left(\su^{\rlb[][\mlvl][\nrt]}\right)=&\left\{
  \begin{array}{ccc}
    0&\If&\mlvl<\Kr-1\\
    \bi \log\left(\frac {\su^{\rlb[][\mlvl][\nrt]}+1}{\su^{\rlb[][\mlvl][\nrt]}}\right)&\If&\mlvl=\Kr-1
  \end{array}\right.\,,\\
\Sscal^{\rlb[\mlvl],\rlb[\klvl]}(\su-\sv)=&
\left\{
  \begin{array}{ccc}
    \frac{\su-\sv-1}{\su-\sv+1}&\If&\klvl=\mlvl\\
    \frac{\su-\sv+1}{\su-\sv}&\If&\klvl=\mlvl+1\\
    \frac{\su-\sv}{\su-\sv-1}&\If&\klvl=\mlvl-1\\
    1&\rlap{\oth}
  \end{array}\right.\,.
\label{eq:SmkRank}
\end{align}

The equation \eqref{eq:pmRank} shows that the roots of {\level} \(\mlvl <
\Kr-1\) have no momentum, 
which is consistent with the analysis of
section \ref{sec:energy-spectrum} where we 
see, (in \eqref{eq:left.p-log-tfus}) that only the polynomial
\(\gQf[][\Il {\Kr-1}]\) contributes to the energy, which means that
only the roots \(\su^{\rlb[][\Kr-1][\nrt]}\) are massive and carry an
energy.
The equation \eqref{eq:SmkRank}
shows that the interactions of ``particles'' ({\idest} the roots) depend on
their {\level}, and that a root of {\level} \(\mlvl\) ``interacts'' only with
roots of {\level} \(\mlvl+1\) or \(\mlvl-1\).

The form of equation \eqref{eq:BethBigerRank} is very general, and it
describes the {\ing} theories with various different types of
particles ({\lbd} here by the {\level} \(\mlvl\)): %
every excited state is {\lbd} by 
a set of variables \(({\su}^{\rlb[][\mlvl][\nrt]})\) satisfying
\eqref{eq:BethBigerRank}, and its energy can be extracted from these
variables (see the next paragraph).%

Hence if we manage to prove that these {\Qfs} exist and are polynomial
(which we will do in the next section), then we get the spectrum of
the \(\GL \Kr\) {\cds}, generalizing the results of section
\ref{sec:integr-et-}. 

\index{Bethe equations |)}

\paragraph{Energy spectrum}

The eigenstates of the {\cds}'s {\Ham} correspond to solutions of
\eqref{eq:SuBetheSplit}, and we can also compute their energy as in
\eqref{eq:left.p-log-tfus}. 
For a \(\GLKM\) {\cds}, this energy is in general equal to \(\frac 2 {\Kr+\Mr} \lcds - 2 
\left.\partial_{\su} \log \rT[][1][1]\right|_{{\su}=0}\), and it can be
computed as in \eqref{eq:EHeisenFromBacklund}, to get
\begin{gather}
  E=\left(
\frac 2 {\Kr+\Mr}-2 \sg {{\ivp}_{\Kr+\Mr}}\right) \lcds%
-2\sum_{\nrt\in\ninter 1 {\dg[\Kr+\Mr-1]}} \frac {\sg{{\ivp}_{\Kr+\Mr}}}
{\su^{\rlb[][\Kr+\Mr-1][\nrt]}(\su^{\rlb[][\Kr+\Mr-1][\nrt]}+\sg{{\ivp}_{\Kr+\Mr}})}\,
\end{gather}
where we should note that \(\sg{{\ivp}_{\Kr+\Mr}}=\pm 1\) denotes the {\GLKM}
grading introduced in section \ref{sec:superspins}, which should not
be confused with the momenta \(p_\nrt\) of the ``particles''.

\section{Differential expression of {\Qoprs}}
\label{sec:expr-diff-Qop}

In the previous section, we saw that if the ``undressing'' and
``dressing'' procedures apply and give polynomial {\Tfs} associated to
subgroups {\GL{\kk\ensuremath{|}\mm}} of {\GL{\Kr\ensuremath{|}\Mr}, then
  (under the extra assumption that \(\gQ[][\emptyset]\) is independent
  of \(\su\)), we recover the spectrum of the Heisenberg {\cds} and
  generalize it to higher-rank groups.

In this section, we will introduce original results of this {\PhD}
\cite{Kazakov:2010iu}, and we will explicitly construct 
the whole Bäcklund flow associated to
a {\cds}.
This will allow us to prove the assumptions of section
\ref{sec:transf-de-backl} ``from scratch''.

 To do that, we will 
define some {\Qoprs} and {\Toprs} at all levels of {\nesting}.
they will turn out to have 
a very simple expression in terms of differential {\ops} or
equivalently in terms of diagrammatic expressions. These expressions
will allow to show that they obey {\Bafra}.

The construction of 
``{\Qoprs}'' which we give in this section
 is quite
different from the constructions introduced in the literature for
several models\footnote{See for instance
  \cite{Baxter:1972hz,1992JPhA...25.5243P,1997CMaPh.190..247B,1999CMaPh.200..297B,2001NuPhB.604..580H,2002NuPhB.622..475B,2002cond.mat..7177F,2003math......6242K,2005NuPhB.709..578K,2005JPhA...38.6641K,2006JPhA...39..S11B,2007CMaPh.272..263B,2007JSMTE..01....5B,2008JPhA...41I5206K,2008NuPhB.805..451B,2009JPhA...42g5204D,2010JPhA...43O5208B,2010JSMTE..11..002B,2011NuPhB.850..148B,Staudacher:2010jz,2011NuPhB.850..175F,2011arXiv1112.3600F,2012arXiv1205.1471T}.},
and in particular, it defines the {\Qoprs} directly as {\ops} and
shows their polynomiality for these {\GLKM} {\csds}.

\subsection{%
Derivation of the simplest {\Qoprs}, when %
\texorpdfstring{\ensuremath{\lcds=1}}{\Lnothing=1}
}
\label{sec:first-qoperators}

To start with, let us see, in the case of {\GL \Kr} with one single
spin ({\idest} \(\lcds=1\)), how to write 
{\ops} satisfying the equations of section
\ref{sec:transf-de-backl}, and in particular the TQ-relation.
In terms of generating series, 
this
TQ-relation \eqref{eq:TQs} reads (for first {\level} of {\nesting} of the
{\GL \Kr} {\cds}) \index{T-operators@{\Toprs}!TQ-relation}\index{Q-operators@{\Qoprs}}
\begin{gather}
 \label{eq:TQw1}
    \Wt[][][\oj]\gQ[][\ove]=
    \Wt[][][\ove]\gQ[][\oj] - x_{\jvp} z
    \Wt[][\su+1][\ove]\gQ[\su-1][\oj]\,,\\
\where {\Wt[][][I]\equiv}{\sum_{s=0}^\infty z^s
  \rT[][1][][I]}\,,\qquad\qquad \gQ[][I]\equiv \rT[][0][0][I]
\\
 \ove\equiv\ninter 1 {\Kr}\,,\qquad \And \oj=\ninter 1
{\Kr}\setminus \{\jvp\}\,.
\end{gather}
\index{jbar@\ensuremath{\oj\equiv \ninter 1 {\Kr+\Mr} \setminus \{\jvp\}}}
\index{0ove@\ensuremath{\ove \equiv \ninter 1 {\Kr+\Mr}}}
For a {\cds} with length \(\lcds=1\),
we can insert  the explicit expressions of 
\(\Wt[][][\ove]\) and \(\gQ[][\ove]\) into this TQ-relation
\eqref{eq:TQw1}.
These expressions of \(\Wt[][][\ove]\) and \(\gQ[][\ove]\)
read\footnote{We remind here that \(\ove=\ninter 1 \Kr\), and that the
  {\ops} \(
\rT\) defined in 
section \ref{sec:chaines-de-spin} are now also denoted as \(\rT[][][][\ove]\).
}
\begin{gather}
\Wt[][][\ove]= \left(\su+ \frac
   {\g~z}{1-\g~z}\right) w(z),\qquad \And \gQ[][\ove]=\gQ =\su\,,
 \label{eq:Twgen}
\end{gather}
and allow to view \eqref{eq:TQw1}
as an equation on the function
\(\Wt[][][\oj]= \gQ[][\oj] + z
\rT[][1][1][\oj] + z^2 \rT[][1][2][\oj]+\cdots\).

To find a solution to this equation, it can be interesting to
remember, from  section \ref{sec:transf-de-backl}, what properties we
would like from this solution:
\begin{itemize}
\item In order to obtain Bethe equations, %
we want
  to
  find a solution where \(\Wt[][][\oj]\) is a
polynomial in \(\su\).
Moreover, one can expect that at each step of the ``undressing''
procedure, the {\Top}- and {\Qoprs} are simpler than at the previous
step, in the sense that they have smaller degree. Hence, we will look
for a solution  where \(\Wt[][][\oj]\) is a
polynomial
of degree not bigger than 1 (which is the degree
of \(\Wt[][][\ove]\)).
\item Moreover, we would like {\Qoprs} to commute with {\Toprs}.
For 1 spin, \eqref{eq:Twgen} ensures that in the basis
where \(\g\) is diagonal, all {\Toprs} are diagonal. 
Therefore, in order to commute with {\Toprs}, we expect that 
\(\gQ[][\oj]\) is also diagonal in the basis where  \(\g\) is diagonal.
\end{itemize}
We therefore expect that, in the basis where \(\g\) is diagonal, 
\begin{align}
  \gQ[][\oj]=&\mathrm{diag}(\alpha_1 \su +\beta _1,\alpha_2 \su
  +\beta _2,\cdots,\alpha_\Kr \su +\beta _\Kr)
\label{eq:QnestAnal}
\end{align}

To go further, let us notice that the {\lhs} 
of \eqref{eq:TQw1} is 
equal to  \(\su ~ \Wt[][][\oj]\), which is
a multiple of
 \(\su\). Therefore the {\rhs} %
 has to be zero when \(\su=0\), %
 which gives the following constraint on \(\gQ[][\oj]\):
 \begin{align}
   \frac{\g~z}{1-\g~z}\gQ[0][\oj]-\frac{x_{\jvp}~z}{1-\g~z}\gQ[-1][\oj]=&
   0\,.
\label{eq:ConstrNest00Qo}
 \end{align}
Now, plugging \eqref{eq:QnestAnal} into \eqref{eq:ConstrNest00Qo} gives 
 \begin{align}
   \frac{x_\kvp~z}{1-x_\kvp~z}(\beta_\kvp)-\frac{x_{\jvp}~z}{1-x_\kvp~z}(\beta_\kvp-\alpha_\kvp)=&
   0\,,
 \end{align}
which is solved by 
\begin{empheq}[left=\empheqlbrace]{align}
  \beta_{\kvp}=&\frac{\alpha_{\kvp}}{1-x_{\kvp}/x_{\jvp}}&\If&{\kvp}\neq {\jvp}\label{eq:Qnest01spin000}\\
  \alpha_{\kvp}=&0&\If&{\kvp}={\jvp}\,.\label{eq:Qnest01spin001}
\end{empheq}
Up to a normalization, we have then shown that for 1 spin, the {\Qopr}
\(\gQ[][\oj]\) is given by
\begin{align}
\label{eq:Qnest01spin}
  \gQ[][\oj]=&\left(
    \begin{array}{l}
      \su + 1/(1-x_1/x_{\jvp})\\
      \qquad \su + 1/(1-x_2/x_{\jvp})\\
      \qquad \qquad \quad \ddots\\
      \qquad \qquad \qquad \su + 1/(1-x_{{\jvp}-1}/x_{\jvp})\\
      \qquad \qquad \qquad \qquad \quad 1\\
      \qquad \qquad \qquad \qquad \qquad \su + 1/(1-x_{{\jvp}+1}/x_{\jvp})\\
      \qquad \qquad \qquad \qquad \qquad \qquad \quad \ddots\\
      \qquad \qquad \qquad \qquad \qquad \qquad \qquad\su + 1/(1-x_{\Kr}/x_{\jvp})\\
    \end{array}\right)\,.
\end{align}

From this point, writing the solution of the TQ-relation
\eqref{eq:TQw1} with the required analyticity properties is just a
matter of plugging the expression of \(\gQ[][\oj]\) into
\eqref{eq:TQw1}, and deducing \Wt[][][\oj].
But before we come to this point, let us %
notice that except for one eigenvalue where the behavior is singular, the
  {\op} \eqref{eq:Qnest01spin} can be viewed as
  \(\gQ[][\oj] = \su + 1 + \frac{\g/x_{\jvp}}{1-\g/x_{\jvp}}\). If we
  change the normalization by an (infinite) factor  \(w(1/x_{\jvp})\), we see
  that
  \(\gQ[][\oj] = \left[(\su + 1 + \hD)~~ w(1/x_{\jvp})\right]\).

  In the next sections we will see how to make this claim more
  rigorous, with respect to the singularities in this expression. We
  will also see that the generalization to more spins is very simply
  obtained by 
replacing 
 \((\su + 1 + \hD)\) 
with
\(\DL[\su_{\spi}+1+\hD]\).

\subsection{General expression of the Bäcklund flow}
\label{sec:gener-expr-backl}

In section \ref{sec:first-qoperators}, 
we have found the expression of the {\Qoprs} for a {\GL \Kr} spin
chain with one single spin, at the 
first {\nesting} {\level}. Their form involves the limit of 
\(\left[(\su + 1 + \hD)~~ w(z)\right]\) at \(z\to 1/x_{\jvp}\), which is a
singular limit. Let us now generalize this to
\(\lcds\in\bN\) (but still at 
the first {\level} of {\nesting}), and we will see that
the {\MID} \eqref{eq:MIDww} which
we just derived allows to write an explicit polynomial solution to the
TQ-relation \eqref{eq:TQw1} (which is simply the equation 
\eqref{eq:TQs}, written at the first
{\level} of {\nesting} in terms of the generating series). The \(\nn=2\)
case of the {\MID} \eqref{eq:MIDww} reads 
\begin{gather}
\label{eq:toTQOPN0}
  \left(t-z\right)\Wt[t,z][\su+1]\cdot\gQ[][\ove]= t ~
  \Wt[t][\su+1] \cdot \Wt[z][][\ove] %
- z ~ \Wt[t]\cdot \Wt[z][\su+1][\ove]\,,
\end{gather}
where \(\Wt\) and \(\Wt[][][\ove]\) denote the same object,
but \(\ove\) is added to draw attention to the
similarity with equation \eqref{eq:TQw1}. In order to deduce
\eqref{eq:TQw1} from \eqref{eq:toTQOPN0}, it would be very natural to
define \(\Wt[][][\oj]\equiv \lim_{t\to 1/x_\jvp}
\left(1-\frac z t\right) \Wt[t,z][\su+1]\) which would give
\(\gQ[][\oj]=\Wt[0][][\oj]=
\lim_{t\to 1/x_\jvp}
\Wt[t][\su+1]\). Under this definition, \eqref{eq:toTQOPN0} would
provide an explicit polynomial and operatorial solution of
the TQ-relation \eqref{eq:TQw1}.

Unfortunately \(\left(1-\frac z t\right) \Wt[t,z][\su+1]\) is diverging
at \(t\to 1/x_\jvp\), so that in order to identify the object
\(\Wt[][][\oj]\), we should first remove this
singularity. To investigate this singularity, it is convenient to
write \(\Wt[t,z]\) as a sum of \(\hD\)-diagrams. Such an expression is obtained
in appendix (see \eqref{eq:diag2Wtlots}), and though the exact expression
is not crucial for the present argument, %
it is instructive to see this expression, written {\below} 
when \(\lcds=2\): %
\begin{gather}
 { \Wt[t,z]}%
=%
\left(
\MyTwoNodes[{}][{}][\forget][{jj1/ii1,jj2/ii2}]
+\sum_{\kk=1}^2
\left(\raisebox{-13pt}{\MyTwoNodes[{jj1/ii1}][{}][\forget][{jj2/ii2}][{jj1.south/\kk}]}
+ \raisebox{-13pt}{\MyTwoNodes[{jj2/ii2}][{}][\forget][{jj1/ii1}][{jj2.south/\kk}]}
\right)
+\sum_{1\leq \kk,\kk'\leq 2}
\raisebox{-13pt}{\MyTwoNodes[{jj1/ii1,jj2/ii2}][{}][\forget][{}][{jj1.south/\kk,jj2.south/\kk'}]}
+\sum_{\kk=1}^2
\raisebox{-13pt}{\MyTwoNodes[{jj2/ii1}][{jj1/ii2}][\forget][{}][{jj1.south/\kk,jj2.south/\kk}]}
\right)%
w(z)w(t)\,,
\label{eq:diag2WtlotsM}
\end{gather}
where a double vertical line \(\MyLine[{}][{}][\forget][{jj1/ii1}]\) at
position \(\spi\) denotes the {\op} \(\su_\spi \bI\), whereas the
lines
\(\raisebox{-13pt}{\MyLine[{jj1/ii1}][{}][\forget][][{jj1.south/1}]}\),
\(\raisebox{-13pt}{\MyLine[{}][{jj1/ii1}][\forget][][{jj1.south/1}]}\),
\(\raisebox{-13pt}{\MyLine[{jj1/ii1}][{}][\forget][][{jj1.south/2}]}\),
and 
\(\raisebox{-13pt}{\MyLine[{}][{jj1/ii1}][\forget][][{jj1.south/2}]}\)
denote respectively the {\ops} \(\frac {\g~t}{1-\g~t}\), \(\frac
{1}{1-\g~t}\), \(\frac {\g~z}{1-\g~z}\) and \(\frac {1}{1-\g~z}\).

We see that when \(t\to \frac 1{x_{\jvp}}\), there are two sources of
singularities:
\begin{itemize}
\item The factor \(w(t)=\prod_{\jvp=1}^\Kr \frac 1
  {1-x_\jvp t}\) has a pole of order one at \(t\to 1/x_{\jvp}\). The pole is
  of order one because we assume that all the eigenvalues \(x_\jvp\) of
  the twist \(\g\) are distinct.
\item every time a {\cd} acts on \(w(t)\) it multiplies \(w(t)\) by
  a factor \(\frac{\g~t}{1-\g~t}\). 
 The {\op}  \(\frac{\g~t}{1-\g~t}\) has
  one eigenvalue (associated to the value \(x_{\jvp}\) of \(\g\)) equal to 
\(\frac{x_{\jvp}~t}{1-x_{\jvp}~t}\). This eigenvalue has a simple pole at \(t\to 1/x_{\jvp}\).
\end{itemize}
In order to have a well-defined limit at \(t\to 1/x_{\jvp}\), we can then
multiply \eqref{eq:toTQOPN0} by \(\frac{(1-\g~t)^{\otimes \lcds}}{w(t)}\), to
get
\begin{multline}
\label{eq:fract-zwt1}
\frac{t-z}{w(t)}(1-\g~t)^{\otimes \lcds}\cdot\Wt[t,z][\su+1]\cdot
\gQ[][\ove]= \frac{t}{w(t)} (1-\g~t)^{\otimes
  \lcds}\cdot\Wt[t][\su+1] \cdot \Wt[z][][\ove] \\
- \frac{z}{t}\frac{t}{w(t)}(1-\g~t)^{\otimes
  \lcds}\cdot \Wt[t]\cdot \Wt[z][\su+1][\ove]\,.
\end{multline}
This now involves the {\op} \(\frac{1}{w(t)}(1-\g~t)^{\otimes
  \lcds}\cdot\Wt[t,z][\su+1]\), which
 is a polynomial in the variables \(t\) 
 and \(\su\). Hence the limit at \(t\to 1/x_{\jvp}\) is well defined. Let us then
 define
 \begin{align}
\label{eq:WN01DefDL}
   \Wt[][][\oj]&\equiv \left(1-z~x_{\jvp}\right)\lim_{t\to 1/x_\jvp}
\frac{1}{w(t)}(1-\g~t)^{\otimes
  \lcds}\cdot \Wt[t,z][\su+1]\,,\\
&=\left(1-z~x_{\jvp}\right)\lim_{t\to 1/x_\jvp}
\frac{1}{w(t)}(1-\g~t)^{\otimes
  \lcds}\cdot
\left[\DLt[{\su}_\spi+1+\hD] w(t) w(z)\right]
\,.\nonumber
 \end{align}
With this definition, the limit of \eqref{eq:fract-zwt1} when \(t\to
1/x_{\jvp}\) is 
\begin{gather}
  \Wt[][][\oj]\cdot
\gQ[][\ove]=   \Wt[0][][\oj] \cdot \Wt[z][][\ove] %
- x_{\jvp} z  \Wt[0][\su-1][\oj]\cdot \Wt[z][\su+1][\ove]\,,%
\end{gather}
which exactly coincides with the TQ-relation \eqref{eq:TQw1}.

Before we generalize this expression to arbitrary {\nesting} levels, and
to the {\sugrs} {\GL{\Kr\ensuremath{|}\Mr}}, let us elaborate on
the definition \eqref{eq:WN01DefDL}, and see what it teaches about the
Bäcklund flow. First for zero spin (\(\lcds=0\)), %
we get
\begin{gather}
\If \lcds=0\qquad\qquad\Then   \Wt[][][\oj]\equiv \lim_{t\to
  1/x_\jvp}
\left(1-\frac z t\right) w(z)=\prod_{\substack{1\leq \kvp\leq
    \Kr\\\kvp\neq \jvp}} \frac 1 {1-x_\kvp z}%
=
w_{\oj}(z)\,\\
  \where\qquad w_{I}(z)\equiv \sum_{s\geq 0}\chas [1](\g[I])
  z^s =
  \prod_{{\jvp}\in I} \frac 1 {1-x_\jvp z}\,.%
\end{gather}
As expected, when \(\lcds=0\) we recover the characters:  \(\gT[][][I]=\cha \lambda 
(\g[I])\). 
For \(\lcds\geq 1\), it is proven in appendix \ref{sec:coder-eigenv}
that the definition \eqref{eq:WN01DefDL} is equivalent to
\begin{align}
\label{eq:WN01DefDLwNest}
     \Wt[][][\oj]=&
\lim_{t\to 1/x_\jvp}
\frac 1 {w(t)} (1-\g~t)^{\otimes
  \lcds}\cdot\left[\DLt[{\su}_\spi+1+\hD] w(t) w_{\oj}(z)\right]
\,.
\end{align}
The difference with \eqref{eq:WN01DefDL} is that the factor \(1-z~x_\jvp\)
was moved to the right of the \(\DL\), and multiplied with \(w(z)\) to get
\(w_{\oj}(z)\). If we expand \eqref{eq:WN01DefDLwNest} in powers of
\(z\), then we get 
\begin{align}
\label{eq:DefTFirstNestS}
  \rT[][1][][\oj]=&\lim_{t\to 1/x_\jvp}
\frac 1 {w(t)} (1-\g~t)^{\otimes
  \lcds}\cdot\left[\DLt[{\su}_\spi+1+\hD] w(t) \chas [1](\g [\oj])\right]
\,.
\end{align}

At this point, we can generalize this expression to arbitrary levels
of {\nesting}, and to arbitrary {\yn} diagrams, in the
{\GL{\Kr\ensuremath{|}\Mr}} case: let us define the ``{\nested}
{\Toprs}'' as\footnote{In what follows, we will show that they do
  indeed define a Bäcklund flow, which will justify the denomination
  of ``{\nested} 
{\Toprs}''.}
\index{T-operators@{\Toprs}}
\begin{equation}
\label{eq:DefNestTop}
\fdisp{
  \begin{aligned}
  \gT[][][I]=& \lim_{\substack{\forall \ivp\in \Ib\\t_{\ivp}\to1/x_{\ivp}}}
\prod_{\ivp \in \Ib}\left(\frac {(1-\g~t_\ivp)^{\otimes
  \lcds}} {\left(w(t_\ivp)\right)^{\sg \ivp}} \right)%
\\&~~~~\cdot
\left[\DLt[{\su}_\spi+\kk_{\Ib}-\mm_{\Ib}+\hD] \left(\prod_{\ivp \in
      \Ib}\left(w(t_\ivp)\right)^{\sg \ivp}\right) \cha \lambda (\g [I])\right]
  \end{aligned}
}  
\end{equation}
\begin{gather}
  \where \su_\spi=\su-\spi\,,\qquad
\Ib=\{1,2,\cdots,\Kr+\Mr\}\setminus
I\,
,\\
   {\kk_{\Ib}}=\mathrm{Card}\{\ivp \in \Ib | {\sg  \ivp}=1\}\,,\qquad
  {\mm_{\Ib}}=\mathrm{Card}\{\ivp \in \Ib | {\sg  \ivp}=-1\}\,.
\end{gather}
Compared to \eqref{eq:DefTFirstNestS}, the definition
\eqref{eq:DefNestTop} contains the following generalizations:
first, it says that for an arbitrary {\rp}, \(\chas[1]\) has to be
replaced with \(\cha \lambda\). Second it says that for an arbitrary
{\level} of {\nesting}, we should put several functions \(w(t_\ivp)\) on the
right of the {\cdrs}. We should put one such factor for each
\(\ivp\in \Ib\), {\idest} for each eigenvalue which is ``removed'' in the
character \(\cha \lambda (\g [I])\). Last, it says that for {\sugrs},
the definition should remain the same up to a few signs, which are
chosen in such a way that the equations of section
\ref{sec:transf-de-backl} are preserved (as we will show in the next
subsections).

This {\op} \(\gT[][][I]\) is well defined
because it is the limit (when \(t_{\ivp}\to1/x_{\ivp}\)) of a polynomial
function of \(t_{\ivp}\). With this definition, \(\gT[][][I]\) is also a
polynomial function of \(\su\), and 
we will see that 
it obeys the commutation relation
\begin{align}
 \standardline{ \forall \su,\sv, \lambda, \mu, I, J
 }{\quad 
\comm[0.45cm][-.05cm]{\lT[\su][][I]}
{\lT[\sv][\mu][J]}
   =0}\,.
\label{eq:TcommLambdaMuIJ}
\end{align}

Like in section \ref{sec:transf-de-backl}, we will denote
\index{Q-operators@{\Qoprs}}
\begin{align}
\label{eq:DefQNest}
  \framedline{\gQ[][I]\equiv}{\rT[][0][0][I]}
\\=&\lim_{\substack{\forall \ivp\in \Ib\\t_{\ivp}\to1/x_{\ivp}}}
\label{eq:DefQNestDer}
\BI_I
\cdot
\left[\DLt[{\su}_\spi+\kk_{\Ib}-\mm_{\Ib}+\hD] %
\PI_I
\right]\,,\\
\where& \fdisp{\BI_I \equiv %
  \frac {\prod_{\ivp \in \Ib} (1-\g~t_\ivp)^{\otimes
   \lcds}} {\PI_I} %
}
\quad
\And\quad\fdisp{\PI_I\equiv \prod_{\ivp\in \Ib} w(t_{\ivp})^{\sg \ivp}}
\,.
\label{eq:DefPIi}
\end{align}

We will also show that with this explicit definition of the {\Top}- and
{\Qoprs}, %
 the TQ-relation
\eqref{eq:TQs} is satisfied, and that for each \(I\), \(\gT[][][I]\) obeys the CBR
determinant formula. We will also show how the ``dressing'' procedure
discussed in section \ref{sec:dressing-procedure} gives a simple
determinant expression of \(\gT[][][I]\). Finally, this expression will
allow to show that these {\op}s obey \Bafra, and they also satisfy the 
polynomiality conditions 
out of which the spectrum
can be deduced (as it was done in section \ref{sec:energy-spectrum}).

\subsubsection{Proof of the TQ-relation}
\label{sec:tq-relation}

The first thing which can easily be shown is that the TQ-relation is
satisfied by the {\ops} defined above.
At the {\level} of eigenvalues, this relation reduces to the
{\Tfu}{\Qfu}-relation which was %
written in \eqref{eq:TQs}.

Let us show that the {\Toprs} defined above by \eqref{eq:DefNestTop}
obey the TQ-relation %
\index{T-operators@{\Toprs}!TQ-relation}
\begin{align}
\label{eq:TQop}
    \rT[\su][1][][I\Delta\jvp]\cdot \gQ[][I] -
  \gQ[][I\Delta\jvp]\cdot\rT[][1][][I] =& x_{\jvp} \rT[\su+1][1][s-1][I\Delta\jvp]\cdot
  \gQ[\su-1][I]\,.
\end{align}
At the {\level} of generating series, this relation reads
\begin{align}
\label{eq:TQopWtoProve}
  \Wt[][][I\Delta\jvp]\cdot \gQ[][I] -
  \gQ[][I\Delta\jvp]\cdot
\Wt[][][I] =& x_{\jvp}~z~\Wt[][\su+1][I\Delta\jvp] 
\cdot
  \gQ[\su-1][I]\,.
\end{align}

On the other hand, a particular case of the {\MID} \eqref{eq:MIDPi}
(proved in section \ref{sec:mid}) is
\begin{multline}
\label{eq:MIDforTQ0}
  \lefteqn{(z-t_{\jvp})  \left[
\DLt[\su_{\spi}+1+\hD] w(z)w(t_{\jvp})\PI_{I\Delta \jvp}\right]
\cdot \left[
\DLt \PI_{I\Delta \jvp}\right]}
\\
=~z~\left[ \DLt[\su_{\spi}+1+\hD] w(z)\PI_{I\Delta \jvp}\right]
\cdot \left[\DLt w(t_{\jvp})\PI_{I\Delta \jvp}\right]
\\
\hspace{3cm}
-t_{\jvp} \left[ \DLt w(z)\PI_{I\Delta \jvp}\right]\cdot
\left[\DLt[\su_{\spi}+1+\hD] w(t_{\jvp})\PI_{I\Delta \jvp}\right]\,.
\end{multline}
From the definition \eqref{eq:DefPIi} of \(\PI_I\) and \(\BI_I\), we
  can notice
that \(w(t_{\jvp})\PI_{I\Delta \jvp}= \PI_{I}\), and multiply by
\(\BI_{{I\Delta {\jvp}}}\cdot \BI_{I}=\BI_{{I}}\cdot \BI_{I\Delta {\jvp}}\), to get
\begin{multline}
\label{eq:MIDforTQmB}
  \lefteqn{(z-t_{\jvp}) \BI_{I}\cdot \left[
\DLt[\su_{\spi}+1+\hD] w(z)\PI_{I}\right]
\cdot \BI_{I\Delta {\jvp}} \cdot\left[
\DLt \PI_{I\Delta {\jvp}}\right]}
\\
=~z~\BI_{I\Delta {\jvp}}\cdot\left[ \DLt[\su_{\spi}+1+\hD] w(z)\PI_{I\Delta {\jvp}}\right]
\cdot \BI_{I}\cdot\left[\DLt \PI_{I}\right]
\\
\hspace{3cm}
-t_{\jvp} ~\BI_{I\Delta {\jvp}}\cdot\left[ \DLt w(z)\PI_{I\Delta {\jvp}}\right]\cdot
\BI_{I}\cdot\left[\DLt[\su_{\spi}+1+\hD] \PI_{I}\right]\,.
\end{multline}
To get this expression we used the fact that for any
\(\PI=\prod_{\kk=1}^\nn \left(w(z_\kk)\right)^{a_\kk}\), and any
\(\su_\spi\)'s, we have the commutation relation
\begin{align}
\label{eq:commDPIfg}
  \comm[1cm][-.27cm]{(f({\g}))^{\otimes\lcds}}{\left[\DLt \PI\right]}=&0\,.
\end{align}
This relation was already shown in section \ref{sec:cons-numb-part} if
\(\PI\) is the character \(\cha \lambda\), and by linearity, it holds for
an arbitrary sum of characters. It was shown in section \ref{sec:mid}
(in the second proof of the \MID) that \(\PI=\prod_{\kk=1}^\nn
\left(w(z_\kk)\right)^{a_\kk}\) is a linear combination of characters
\(\cha \lambda\), 
hence the relation \eqref{eq:commDPIfg}. This relation allowed to
commute the \(\BI\)'s through other factors in order to derive
\eqref{eq:MIDforTQmB} from \eqref{eq:MIDforTQ0}.

We can now see that, in each term of \eqref{eq:MIDforTQmB}, the factor
to the right is a {\Qopr} (in the limit \(\forall \ivp,~ t_{\ivp}\to
1/x_{\ivp}\)). The other factors contain \(w(z)=\sudet\frac 1{1-gz}\)
(see appendix \ref{sec:elements-de-theorie}). To
produce the correct function \(\Wt[][][I]\), we actually need to replace 
\(w(z)\) with \(w_{I}(z)\) or \(w_{I\Delta \jvp}(z)\) defined by
\eqref{eq:GenSerIs}. To this end we 
multiply \eqref{eq:MIDforTQmB} by \(\frac{w_{I\Delta\jvp}(z)}{w(z)}=\frac 1 {w_{\overline{I\Delta\jvp}}(z)}\), to
get (in
the limit \(\forall \ivp,~ t_{\ivp}\to 1/x_{\ivp}\)):
\begin{gather}
\label{eq:MIDforTQlim}
  {\lim_{\forall \ivp,~ t_{\ivp}\to 1/x_{\ivp}}
  -t_{\jvp}\frac{1-z/t_{\jvp}}{w_{\Ijb}(z)} 
\BI_{I}\cdot \left[
\DLt[\su_{\spi}+1+\hD] w(z)\PI_{I}\right]
\cdot 
\gQ[\sv+1][I\Delta {\jvp}]
}\qquad\qquad\qquad
 \nonumber
\\
=~z~\lim_{\forall \ivp,~ t_{\ivp}\to 1/x_{\ivp}}
\frac 1 {w_{\Ijb}(z)} \BI_{I\Delta {\jvp}}\cdot\left[ \DLt[\su_{\spi}+1+\hD]
  w(z)\PI_{I\Delta {\jvp}}\right] \cdot \gQ[\sv][I]
 \nonumber
\\
\qquad\qquad\qquad
-
\lim_{\forall \ivp,~ t_{\ivp}\to 1/x_{\ivp}}
t_{\jvp}
\frac 1 {w_{\Ijb}(z)} 
\BI_{I\Delta {\jvp}}\cdot\left[ \DLt w(z)\PI_{I\Delta {\jvp}}\right]\cdot
\gQ[\sv+1][I]\\
\where \sv\equiv\su-{\kk}_\Ib+{\mm}_\Ib = \su-{\kk}_\Ijb+{\mm}_\Ijb-1\,.
\end{gather}

In the 
{\lhs}, in the limit \(\forall \ivp,~
t_{\ivp}\to 1/x_{\ivp}\), the factor \(\frac {1-z/t_{\jvp}} {w_{\Ijb}(z)}\)
becomes \linebreak \(\frac {1-z~x_{\jvp}} {w_{\Ijb}(z)} = \frac 1 {w_{\Ib}(z)}\).
As it is shown in the appendix \ref{sec:coder-eigenv}, %
this factor
can be
``commuted'' to the right of the {\cdrs} due to
the presence of the factor \(\BI_I\). %
In the {\rhs} of \eqref{eq:MIDforTQlim}, the
same argument allows to commute the factor \(\frac{1%
}{w_{\Ijb}(z)}
\) to the right of the
{\cdrs}. If we remember that
\(\frac{w(z)}{w_{\Ijb}(z)}={w_{I\Delta \jvp}(z)}\), we get
\begin{align}
\label{eq:MIDforTQlimfin}
  -1/x_\jvp \lefteqn{\lim_{\forall \ivp,~ t_{\ivp}\to 1/x_{\ivp}}
\BI_{I}\cdot \left[
\DLt[\su_{\spi}+1+\hD] w_{I}(z)\PI_{I}\right]
\cdot 
\gQ[\sv+1][I\Delta {\jvp}]
}
\nonumber~
& \\
&~=~z~\lim_{\forall \ivp,~ t_{\ivp}\to 1/x_{\ivp}}
 \BI_{I\Delta {\jvp}}\cdot\left[ \DLt[\su_{\spi}+1+\hD]
  w_{I\Delta {\jvp}}(z)\PI_{I\Delta {\jvp}}\right] \cdot \gQ[\sv][I]
\nonumber\\
&~ ~~~-1/x_{\jvp}
\lim_{\forall \ivp,~ t_{\ivp}\to 1/x_{\ivp}}
\BI_{I\Delta {\jvp}}\cdot\left[ \DLt w_{I\Delta {\jvp}}(z)\PI_{I\Delta {\jvp}}\right]\cdot
\gQ[\sv+1][I]\,.%
\end{align}
This equation \eqref{eq:MIDforTQlimfin} can at last be written as
\begin{multline}
\label{eq:TQfromMID}
   -1/x_\jvp 
\Wt[][\sv+1][I]\cdot 
\gQ[\sv+1][I\Delta {\jvp}]
 \\
= z
\Wt[][\sv+2][I\Delta {\jvp}]\cdot \gQ[\sv][I]
-1/x_{\jvp}~
\Wt[][\sv+1][I\Delta {\jvp}]\cdot
\gQ[\sv+1][I]\,,
\end{multline}
which is the TQ-relation \eqref{eq:TQopWtoProve} at point
\(\su=\sv+1\).

We have then proven the TQ-relation, at all levels of {\nesting}, and the
proof relied mainly on  the {\MID} \eqref{eq:MIDPi}.

\subsubsection{Hirota equation}
\label{sec:hirota-equation}

It is also possible to show that at all {\nesting} levels \(I\), the
{\Toprs} defined in \eqref{eq:DefNestTop} satisfy
the CBR formula
\begin{align}
\label{eq:NestedCBR}
  \lT[][][I]
  =&\frac{\Det{\rT[\su+1-\ii][1][\lambda_\jj+\ii-\jj][I]}{1\leq\ii,\jj\leq
    |\lambda|}}{\prod_{{\coordk}=1}^{|\lambda|-1}
\gQ[\su-{\coordk}][I]}\,.
\end{align}
\index{CBR formula}

It is not straightforward to prove it with the methods of section
\ref{sec:hirota-equat-cher}, because the denominator is more
complicated than just \(\prod_{\spi=1}^\lcds \su_\spi\), so that the
expansion around \(\su_\spi\to \infty\) would be less direct than in section
\ref{sec:hirota-equat-cher}. 

On the other hand, it was shown in section \ref{sec:mid} that the CBR
formula \eqref{eq:CBR} is equivalent to the determinant expression
\eqref{eq:MIDdet}, which is itself equivalent to the bilinear relation
\eqref{eq:MIDww}. By the same argument, the ``{\nested} CBR formula''
\eqref{eq:NestedCBR} is equivalent to a determinant relation and to
the bilinear relation
\begin{align}
  \label{eq:MIDNest}
\lefteqn{(z_1-z_\nn)  \Wt[z_1,\cdots , z_\nn][\su+1][I] \cdot
\Wt[z_2,\cdots,z_{\nn-1}][][I]}\nonumber\phantom{\where\quad }%
& \\
&\qquad=~ z_1 \Wt[z_1,\cdots,z_{\nn-1}][\su+1][I] \cdot
\Wt[z_2,\cdots,z_\nn][][I]\nonumber\\
&\qquad\qquad~~-z_\nn \Wt[z_1,\cdots,z_{\nn-1}][][I] \cdot
\Wt[z_2,\cdots,z_\nn][\su+1][I]\\
\where\quad &\Wt[z_\lL,\cdots,z_\mm][][I]\equiv
\lim_{\forall \ivp,~ t_{\ivp}\to 1/x_{\ivp}}
\BI_I\cdot
\left[\vphantom{\DLt} \right.\nonumber\\&\left.\qquad\qquad\qquad
\DLt
 w_I(z_\lL)w_I(z_{\lL+1})w_I(z_{\lL+2})\cdots w_I(z_\mm)
\PI_I
\right]\,,
\end{align}
where \(\PI_I\) and \(\BI_I\) are defined by \eqref{eq:DefPIi}.

In order to prove the relation \eqref{eq:MIDNest}, one simply has to
remember that \(w_I(z)=\frac{w(z)}{w_\Ib(z)}\), and that the appendix
\ref{sec:coder-eigenv} allows to move the factor \(\frac{1}{w_\Ib(z)}\)
to the left of all {\cdrs}. This allows to write
\begin{gather}
\label{eq:NestedWFunction2}
  \Wt[z_\lL,\cdots,z_\mm][][I] = \prod_{\kk=\lL}^{\mm}\frac 1 {w_{\Ib}(z_\kk)}
\lim_{\forall \ivp,~ t_{\ivp}\to 1/x_{\ivp}}%
\BI_I\cdot
\left[%
\DLt
\left(\prod_{\kk=\lL}^{\mm} w(z)\right)
\PI_I
\right]\,.
\end{gather}
Due to this remark, %
the relation \eqref{eq:MIDNest} is easy to prove from
\eqref{eq:MIDPiDef} by choosing \(\PI=\PI_I \left(\prod_{\kk=2}^{\nn-1}
  w(z)\right)\), multiplying by \(\BI_I\) (which commutes with the other
{\ops}) and taking the limit \(t_{\ivp}\to 1/x_{\ivp}\).

This proves that the ``{\nested} CBR formula'' \eqref{eq:NestedCBR}
holds, and as a consequence, the following ``{\nested} Hirota equation''
also holds:
\begin{align}
  \label{eq:NestedHirota}
\framedline{\rT[\su+1][][][I] \rT[][][][I]}{ = \rT[\su+1][a+1][][I]\rT[][a-1][][I]+\rT[\su+1][][s-1][I]\rT[][][s+1][I]}\,\,.
\end{align}
\index{Hirota equation}

Moreover, %
there is a general commutation relation
\begin{align}
  \comm[0.65cm][-.135cm]{[\su-\theta_{\spi}+\hD] \PI~~}{~~\DLt[\sv-\theta_{\spi}+\hD] \PI'}=&0\,,
\end{align}
which is valid when \(\PI\) and \(\PI'\), are of the form
\(\prod_{\kk=1}^\nn \left(w(z_\kk)\right)^{a_\kk}\), as in
\eqref{eq:MIDPiDef}.
This relation is obtained from \eqref{eq:TcommLambdaMu2}, using the
fact that \(\PI\) and \(\PI'\) are linear combinations of characters, as
shown in section \ref{sec:mid}
(in the second proof of the \MID). 
Due to the form 
\eqref{eq:NestedWFunction2} of the generating series of the  {\Toprs}, 
this commutation relation implies 
\begin{gather}
  { \forall \su,\sv,s,s',I,J,}{\quad   
\comm[0.45cm][-.05cm]{\rT[][1][][I]}{\rT[\sv][1][s'][J]}=0\,.}
\end{gather}
Finally, the ``{\nested} CBR formula'' \eqref{eq:NestedCBR} allows to deduce the
commutation relation
\begin{align}
 \framedline{ \forall \su,\sv, \lambda, \mu, I, J
 }{\quad 
\comm[0.45cm][-.05cm]{\lT[\su][][I]}
{\lT[\sv][\mu][J]}
   =0}\,\,.
\label{eq:TcommLambdaMuIJ2}
\end{align}

\subsubsection{QQ-relations and {\Wronskian} expressions}
\label{sec:wronsk-expr-qq}

\index{Q-operators@{\Qoprs}!QQ-relations}

Let us now prove that the {\Toprs} which %
we defined in equation \eqref{eq:DefNestTop} 
correspond indeed to the Bäcklund flow of section
\ref{sec:bethe-equat-energy}. First, one can easily prove 
that the {\Qoprs} defined in \eqref{eq:DefQNestDer} obey the 
following
QQ-relation %
\begin{align}
  \gQ[][I\Delta {\jvp}\Delta \kvp] \gQ[\su-1][I]=\begin{vmatrix}
  x_{\jvp} \gQ[][I\Delta \jvp] & x_{\kvp} \gQ[][I\Delta \kvp] \\ \gQ[\su-1][I\Delta \jvp] & \gQ[\su-1][I\Delta \kvp]
\end{vmatrix} / \left(x_{\jvp}-x_{\kvp}\right)\,.
\label{eq:QQop}
\end{align}
It is proven %
directly\footnote{
An other proof of this QQ-relation could be obtained by repeating the
arguments of section \ref{sec:bethe-equat-energy}, which allow to
obtain \eqref{eq:QQop} from \eqref{eq:TQop}.
But with this method, it is not straightforward to derive the
denominator \(\left(x_{\jvp}-x_{\kvp}\right)\).
}
 from the {\MID}, like in the
proof of \eqref{eq:TQop} (see \cite{Kazakov:2010iu}).

As in section \ref{sec:jacobiident-bilin}, it is then straightforward
to show that \eqref{eq:QQop} is equivalent to the determinant
expression
\begin{gather}
  \gQ[] [I\Delta {\jvp_1}\Delta {\jvp_2}\Delta\cdots \Delta {\jvp_\nn}]
=\frac{\Det {x_{\jvp_\kk}^{1-\lL}\gQ[\su-\lL+1][I\Delta {\jvp_\kk}]}{1\leq\kk,\lL\leq\nn}
}{
\dVdM(x_{\jvp_1},x_{\jvp_2},\cdots,x_{\jvp_\nn})
\prod_{{\kk}=1}^{\nn-1}
  \gQ[\su-{\kk}][I]
}
\label{eq:QQopDet}\,,\\
\where \dVdM(x_{\jvp_1},x_{\jvp_2},\cdots,x_{\jvp_\nn}) \equiv \Det {x_{\jvp_\kk}^{1-\lL}}{1\leq\kk,\lL\leq\nn}
\end{gather}
which holds for arbitrary \(I\subset\{1,2,\cdots,\Kr+\Mr\}\) and
\(\jvp_1, \jvp_2,\cdots \jvp_\nn\) such that the \(\jvp_\kk\) are
distinct and obey
\begin{align}
  \forall \kk\in\ninter 1 \nn,& \qquad
\left\{
  \begin{array}{lcr}
    \sg{\jvp_\kk}=1&\And&\jvp_\kk\notin
      I\\
\Or\\
    \sg{\jvp_\kk}=-1&\And&\jvp_\kk\in
      I
  \end{array}
\right.\,.
\end{align}

In the case of the {\GL \Kr} group, this allows to write every {\Qopr}
as a determinant, in terms of the \(\Kr+1\) {\ops}  \(\gQ[][\emptyset]\), \(\gQ[][{1}]\),
\(\gQ[][{2}]\), \(\cdots\), \(\gQ[][{\Kr}]\) as
\begin{align}
  \gQ[][I]
=&\frac{\Det {x_{\jvp}^{1-\kk}\gQ[\su-\kk+1][\jvp]}{
\substack{\jvp\in I
\\
1\leq \kk\leq |I|
}}
}{
\dVdM\left(\left(x_{\jvp}\right)_{\jvp\in I}\right)
\prod_{{\kk}=1}^{|I|-1}
  \gQ[\su-{\kk}][\emptyset]
}\,.
\label{eq:QQopDetGLK}
\end{align}
We will even see (in section \ref{sec:degree-toprs}) that in this
expression, the {\op} \(\gQ[][\emptyset]\) 
in the denominator is a \(\su\)-independent {\op} which commutes with
all {\Top}- and {\Qoprs}. (In other words, \(\gQ[][\emptyset]=1\) up to
a normalization).

For the {\GL{\Kr\ensuremath{|}\Mr}} groups, a very similar expression
is written as:
\begin{gather}
    \gQ[][I]
=\frac{\Det {x_{\jvp}^{1-\kk}\gQ[\su-\kk+1][\jvp]}{
\substack{\jvp\in %
J
\\
1\leq \kk\leq |%
J|
}}
}{
\dVdM\left(\left(x_{F\Delta \jvp}\right)_{\jvp\in %
J}\right)
\prod_{{\kk}=1}^{|%
J|-1}
  \gQ[\su-{\kk}][F]
}
\label{eq:QQopDetGLKM}\,,\\
\where\quad F\equiv\left\{\jvp\in\ninter 1 {\Kr+\Mr}\middle| \sg \jvp
=-1\right\}\\
\And \quad %
J\equiv(F\cup I)\setminus(F\cap I)\,.
\end{gather}
This result, %
is very natural in the notations above (it just follows from 
\eqref{eq:QQopDet}), and it 
was called ``bosonization
trick'' in \cite{Gromov:2010km}, because it allows to manipulate the
{\GL{\Kr\ensuremath{|}\Mr}} QQ-relations using the same expressions as
in the {\GL \Kr} case. An important difference with
\eqref{eq:QQopDetGLK} is nevertheless that the {\op}  \(\gQ[][F]\)
in the denominator is a non trivial polynomial in the variable \(\su\).

\paragraph{Determinant expression of {\GL \Kr} {\Toprs}}
\label{sec:determ-expr-gl}

Next one can easily show that with these definitions 
(\ref{eq:DefNestTop},\ref{eq:DefQNest}) of {\Top}- and {\Qoprs},
the expression \eqref{eq:WDressing}
(which defines the ``dressing procedure'' derived from the
TQ-relation)
of the generating series 
of {\Tfs} %
 holds at the {\level} of
{\ops}. Hence, %
we get
an expression of {\Toprs} for symmetric
representations
in terms of {\Qoprs}.
 By plugging the expression \eqref{eq:QQopDetGLK} %
of {\Qoprs} %
we can get a simple expression of these {\Toprs}.
In the {\GL \Kr} case, that gives the expression
\begin{gather}%
\label{eq:ToSGLKfromNest}
    \rT[][1]= %
    {\gQ[\su-\Kr][\emptyset]}%
    \cdot \frac{\Det
      {x_{\jvp}^{1-\kk+s~\theta(1-\kk)}~%
        {\gQ[\su-\kk+1+s~\theta(1-\kk)][\jvp]}
      }{ 1\leq \jvp,\kk\leq \Kr }} {\VdM[\Kr][1][x] \gQ[\su+s-1][\emptyset]
      \prod_{a=2}^{\Kr}\gQ[\su-a][\emptyset] }\,,
\\
    \where \theta(\nn)=\left\{
      \begin{array}{ll}
        1&\If \nn\geq 0\\
        0&\mathrm{otherwise}
      \end{array}
    \right.\,.
  \end{gather}
This expression can be used to express the {\Toprs} associated to
arbitrary {\yn} diagrams, by means of the CBR formula
\eqref{eq:CBR}. This gives the following {\Wronskian} expression, for the
{\GL \Kr} {\Toprs}:
\begin{gather}
\label{eq:WronskianLambda}
\fdisp{\gT=
    {\gQ[\su-\Kr][\emptyset]}
    \cdot \frac{\Det
      {x_{\jvp}^{1-\kk+\lambda_\kk}~
        {\gQ[\su-\kk+1+\lambda_\kk][\jvp]}
      }{ 1\leq \jvp,\kk\leq \Kr }} {\VdM[\Kr][][x] 
\prod_{\kk=1}^\Kr \gQ[\su-\kk+\lambda_\kk][\emptyset]
 }}\,.
\end{gather}

Of course, the same expression is obtained as easily at an arbitrary
{\nesting} {\level} and it reads
\begin{gather}
\label{eq:NestedWronskianLambda}
\gT[][][I]=
    {\gQ[\su-|I|][\emptyset]}
    \cdot \frac{\Det
      {x_{\jvp}^{1-\kk+\lambda_\kk}~
        {\gQ[\su-\kk+1+\lambda_\kk][\jvp]}
      }{
\substack{\jvp\in I
\\
1\leq \kk\leq |I|
}
}} {
\dVdM\left(\left(x_{\jvp}\right)_{\jvp\in I}\right)
\prod_{\kk=1}^{|I|} \gQ[\su-\kk+\lambda_\kk][\emptyset]
 }\,,
\end{gather}
which holds if \(|\lambda|\leq |I|\). By contrast, the {\Topr} is zero
if \(|\lambda|> |I|\), because \(\cha \lambda (\g [I])\) is zero in
\eqref{eq:DefNestTop}.
Here we can notice that the Bäcklund transforms, which decrease
the size of the set \(I\), simply amount to keeping the minor
\eqref{eq:NestedWronskianLambda} of the determinant \eqref{eq:WronskianLambda}.

In order to generalize this to {\sugrs}, it is instructive to show
how this expression simplifies for rectangular representations:

\paragraph{Rectangular representations}
\label{sec:rect-repr}

First, one can
 notice
that \eqref{eq:ToSGLKfromNest} is equivalent to 
\begin{gather}%
\label{eq:ToSGLKfromNestexpanded}
    \rT[][1]=
    \sum_{\jvp=1}^\Kr\frac{x_\jvp^{s+{\Kr}%
        -1}\gQ[\su+s][\{\jvp\}] \gQ[\su-1][\oj]}
    {\gQ[\su+s-1][\emptyset]\prod_{\substack{1\leq\kvp\leq {\Kr}\\
\kvp\neq \jvp}}
\left(x_\jvp-x_\kvp\right)
}\,.
  \end{gather}
This can be seen by expanding the determinant
\eqref{eq:ToSGLKfromNest} with respect to the first column, and noticing
that the minors which remain are exactly {\Qoprs}, because they are of
the form \eqref{eq:QQopDetGLK}. 
For an arbitrary rectangular {\rp} \(\las \equiv
(\underbrace{s,s,\ldots,s,s}_{\ensuremath{a\textrm{ times}}},0,0,\ldots)\),
the determinant \eqref{eq:WronskianLambda} can also be expanded with
respect to the \(a\) first columns (in these columns, the {\Qoprs} are
shifted by \(\lambda_\kk=s\), whereas \(\lambda_\kk=0\) for the other columns).
This expansion
gives 
\begin{gather}%
\label{eq:TasGLKfromNestexpanded}
\fdisp{
    \rT=
\sum_{\substack{B\subset \ninter 1 \Kr
\\ |B|=a}}
\frac{\gQ[\su+s][B]\gQ[\su-a][\overline{B}] \prod_{\ivp\in B} x_\ivp^{s+{\Kr}-a}
    }
{\gQ[\su+s-a][\emptyset] \prod_{\substack{\ivp\in B%
}}
\prod_{\jvp\in \overline B}
\left(x_\ivp-x_\jvp\right)}
}
\,,
  \end{gather}
where the sum runs over all  subsets \(B\subset \ninter 1 {\Kr%
}\) %
of size \(|B|=a\).

We see that this expression is a sum of terms of the form
\(\gQ[\su+s][I]\gQ[\su-a][\Ib]\) (times a simpler factor). At \(a=0\),
only \(I=\ove\) is allowed and 
we recover \(\rT[][0]=\gQ[][\ove]\). At \(a=1\), \(I\) has the form
\(\{\jvp\}\) (see \eqref{eq:ToSGLKfromNestexpanded}), and each time \(a\)
increases by \(1\), %
the size of \(I\) is increased by one.

Of course the expression \eqref{eq:TasGLKfromNestexpanded} is less general than
\eqref{eq:NestedWronskianLambda} (because it only applies to
rectangular representations), but we will see that it is much
easier to manipulate, and to generalize. %
Moreover, in the next chapters of this manuscript, we will never have
to deal with non-rectangular representations.

\paragraph{Expression of {\Toprs} for {\sugrs}}
\label{sec:expr-toprs-super}

If we reproduce the proof of \eqref{eq:ToSGLKfromNest} ({\idest} we plug
\eqref{eq:QQopDet} into \eqref{eq:WDressing}) for the {\sugr}
{\GL{\Kr\ensuremath{|}\Mr}}, 
then we obtain
\begin{multline}
    \rT[][1]=
    \sum_{\substack{1\leq \jvp\leq \Kr+\Mr\\
\sg\jvp=1}}^\Kr\frac{x_\jvp^{s+{\Kr} -{\Mr} -1} \gQ[\su+s][\{\jvp\}]
      \gQ[\su-1][\oj]} 
    {\gQ[\su+s-1][\emptyset]}
{\prod_{\substack{1\leq\kvp\leq \Kr+\Mr\\
\kvp\neq \jvp}}
\left(x_\jvp-x_\kvp\right)^{-\sg{\kvp}}
}\\
\If s>\Mr-\Kr\,.
\end{multline}
This super-symmetric generalization of
\eqref{eq:ToSGLKfromNestexpanded} holds only if \(s>\Mr-\Kr\), whereas
if \(0\leq s \leq \Mr-\Kr\), one gets a slightly different expression:
\begin{multline}
    \rT[][1]=(-1)^s
    \sum_{\substack{F\subset \ninter 1 {\Kr+\Mr}\\
|F|=s\\
\forall \jvp\in F,~\sg \jvp=-1}}
\frac{\gQ[\su-1][F]\gQ[\su+s][\overline F]
\prod_{\ivp\in F}
 x_\ivp^{1-s+{\Mr}-{\Kr}}
}
{\gQ[\su+s-1][\emptyset] \prod_{\ivp\in F}\prod_{\jvp\in \overline F}
\left(x_\ivp-x_\jvp\right)^{\sg{\jvp}}}%
\\
\If s\leq \Mr-\Kr+1\,,
\end{multline}
and although it is not obvious, 
the QQ-relations imply that 
these two expressions are %
equivalent when \(s=\Mr-\Kr+1\).

If we use the CBR formula \eqref{eq:CBR} to express the {\Toprs} for
arbitrary representations, then it is not easy to recast the outcome
into the form of a simple determinant like
\eqref{eq:NestedWronskianLambda}. However, in the case of rectangular
representations it is possible to write
expressions analogous to \eqref{eq:TasGLKfromNestexpanded}:
\begin{gather}
\label{eq:super-rectangular-Toprs}
\fdisp{
  \rT=\left\{
    \begin{split}
      \sum_{\substack{B\subset \ninter 1 \Kr+\Mr
\\ |B|=a\\
\forall \jvp\in B, \sg \jvp=1
}}
\frac{\gQ[\su+s][B]\gQ[\su-a][\overline{B}] \prod_{\ivp\in B} x_\ivp^{s+{\Kr}-{\Mr}-a}
    }
{\gQ[\su+s-a][\emptyset] \prod_{\substack{\ivp\in B%
}}
\prod_{\jvp\in \overline B}
\left(x_\ivp-x_\jvp\right)^{\sg \jvp}}\\
\If s-a\geq \Mr-\Kr\,,\\
\\
   (-1)^{a~s} \sum_{\substack{F\subset \ninter 1 {\Kr+\Mr}\\
|F|=s\\
\forall \jvp\in F,~\sg \jvp=-1}}
\frac{\gQ[\su+s][\overline F] \gQ[\su-a][F] \prod_{\ivp\in F}
 x_\ivp^{a-s+{\Mr}-{\Kr}}}
{\gQ[\su+s-a][\emptyset] \prod_{\ivp\in F}\prod_{\jvp\in \overline F}
\left(x_\ivp-x_\jvp\right)^{\sg{\jvp}}}%
\\
\If s-a\leq \Mr-\Kr\,.
    \end{split}
\right.
}
\end{gather}
Once again, we see that this expression is a sum of terms of the form
\(\gQ[\su+s][I]\gQ[\su-a][\Ib]\) (times a simpler factor).
If we first describe the domain \(s\geq a+\Mr-\Kr\), we find that at \(a=0\), 
only \(I=\ove\) is allowed and 
we recover \(\rT[][0]=\gQ[][\ove]\). At \(a=1\), \(I\) has the form
\(\{\jvp\}\) where \(\sg \jvp=+1\), and each time \(a\)
increases by \(1\), %
\(I\) can have one more element \(\jvp_\kk\), which must have the grading
\(\sg{\jvp_\kk}=+1\). When \(a=\Kr\), \(I\) contains all the elements with
grading \(\sg{\jvp_\kk}=+1\), and we reach a boundary of the lattice. At
the boundary we have \(\rT[][\Kr][]=\gQ[\su+s][B]\gQ[\su-\Kr][F]\) (up to a
factor containing \(\gQ[][\emptyset]\) and the eigenvalues of twist),
where \(B=F\equiv\left\{\jvp\in\ninter 1 {\Kr+\Mr}\middle| \sg \jvp
=+1\right\}\) and \(F=F\equiv\left\{\jvp\in\ninter 1 {\Kr+\Mr}\middle| \sg \jvp
=-1\right\}\).
Next we can describe the domain where \(a\geq s-\Mr+\Kr\). First we can
notice that the boundary at \(s=\Mr\) has almost the same expression
\(\rT[][][\Mr]=\gQ[\su+\Kr][B]\gQ[\su-a][F]\)
as the previous boundary at \(a=\Kr\). Then if we decrease \(s\) up to
zero, we still obtain sums of terms of the form
\(\gQ[\su+s][I]\gQ[\su-a][\Ib]\), where \(I\) has one more element at each
step, and this element needs to have the grading
\(\sg{\jvp_\kk}=-1\).

 Therefore, we see that the structure of the determinant expression
 \eqref{eq:super-rectangular-Toprs} is essentially the same as
 \eqref{eq:TasGLKfromNestexpanded}, 
with a specificity that the indices with grading \(\sg{\jvp_\kk}=+1\)
are in some sense associated to the domain  \(s\geq
 a+\Mr-\Kr\),  whereas the indices with
 grading \(\sg{\jvp_\kk}=-1\) are associated to the domain  \(a\geq s-\Mr+\Kr\).

Exactly like for the {\GL{\Kr}} case, the expression
\eqref{eq:super-rectangular-Toprs} can be generalized to arbitrary
{\nesting} levels. 

This concludes the ``dressing'' process, and shows that the
{\Qoprs} obtained by the ``undressing procedure'' of fig
\ref{fig:Backlund}, explicitly constructed by equation
\eqref{eq:DefNestTop}, allow to reconstruct all {\Toprs}. %

\paragraph{Bäcklund flow and spectrum of the {\cds}}
\label{sec:baackl-flow-spectr}

We have obtained simple determinant expressions for the {\Toprs} that
we have defined in \eqref{eq:DefNestTop}, which take the particularly
simple form \eqref{eq:TasGLKfromNestexpanded} and
\eqref{eq:super-rectangular-Toprs} for rectangular representations. These
expressions can also be written at an arbitrary {\nesting} {\level} (see for
instance \eqref{eq:NestedWronskianLambda}). 

 From this point, some identities on determinants
 \cite{springerlink:10.1007/s002200050165} (called Plücker identities,
 which generalize the {\jacobi} identity \eqref{eq:{\jacobi}}) show that
 these determinant expressions imply that these {\Toprs} obey {\Bafra}.
This means that we have
an explicit, operatorial expression of a Bäcklund flow which satisfies
all the expected analyticity properties (indeed, we have shown that it
is polynomial, and 
as shown in the next section~\ref{sec:degree-toprs}, the polynomial
\(\gQ[][\emptyset]\) is a constant). 
As a consequence, this construction gives a derivation of
the spectrum of the model.

\subsection{Degree of the {\Toprs}}
\label{sec:degree-toprs}

By construction, the {\Toprs} of the Bäcklund flow defined by
\eqref{eq:DefNestTop} are polynomial functions of the variable \(\su\),
and we will now show that their degree depends on the eigenspace, and
is explicitly given by the {\op} \(\sum_{\jvp\in I}
\Mpo_\jvp\)
(where \(\Mpo_\jvp\) is the number of particles of type \(\jvp\)).
This means that for states %
\(\ket\psi\in E_{\Mp_1,\Mp_2,\cdots,\Mp_{\Kr+\Mr}}\) (where the set
\(E_{\Mp_1,\Mp_2,\cdots,\Mp_{\Kr+\Mr}}\) was defined in \eqref{eq:DefEMMMM}), 
the state \(\gT[][][I]\ket\psi\) is a polynomial function of \(\su\) which
has degree \(\sum_{\jvp\in I} \Mp_\jvp\).

\begin{proof}
  The most direct way to show this result is to write explicitly the
  expression \eqref{eq:DefNestTop} in terms of \(\hD\)-diagrams. 

  Using this method, we can first find the degree of the polynomials \linebreak
  \(\gQ[][I]=\lim~\BI_I
\cdot \left[\DLt[{\su}_\spi+\kk_{\Ib}-\mm_{\Ib}+\hD] \PI_I \right]
\) defined in \eqref{eq:DefQNestDer}. The expression of 
\(\left[\DLt[{\su}_\spi+\kk_{\Ib}-\mm_{\Ib}+\hD] \PI_I \right]\)
 in
  terms of \(\hD\)-diagrams is 
given in appendix \ref{sec:diagr-expr-co}. It is given by equation
\eqref{eq:diag2Wtlots} when \(\lcds=2\) and a general rule is given in
appendix \ref{sec:diagr-expr-co} to write it for arbitrary \(\lcds\).
It gives a sum of \(\hD\) diagrams multiplied by \(\PI_I\). This factor
\(\PI_I\) disappears when
\(\left[\DLt[{\su}_\spi+\kk_{\Ib}-\mm_{\Ib}+\hD] \PI_I \right]\) is
multiplied by \(\BI_I =
  \frac {\prod_{\ivp \in \Ib} (1-\g~t_\ivp)^{\otimes
   \lcds}} {\PI_I}\). But this multiplication also introduces the
factor \(\prod_{\ivp \in \Ib} (1-\g~t_\ivp)^{\otimes
   \lcds}\), which means that each line of every \(\hD\)-diagram should
 be multiplied by \(\prod_{\ivp \in \Ib} (1-\g~t_\ivp)\).

The consequence is that \(\BI_I
\cdot \left[\DLt[{\su}_\spi+\kk_{\Ib}-\mm_{\Ib}+\hD] \PI_I \right]\) is
a sum of \(\hD\)-diagrams, where each 
  double vertical line \(\MyLine[{}][{}][\forget][{jj1/ii1}]\)
  correspond to the {\op} \(\su_\spi \prod_{\jvp\in\Ib}(1-{\g}~t_\jvp)\),
  whereas the line
  \(\raisebox{-13pt}{\MyLine[{jj1/ii1}][{}][\forget][][{jj1.south/\kvp}]}\)
  denotes the {\op} \({\g}~t_\kvp
  \prod_{\jvp\in\Ib\setminus{\kvp}}(1-{\g}~t_\jvp)\) and the line
  \(\raisebox{-13pt}{\MyLine[{jj1/ii1}][{}][\forget][][{jj1.south/\kvp}]}\) 
  denotes the {\op} \(
  \prod_{\jvp\in\Ib\setminus{\kvp}}(1-{\g}~t_\jvp)\).

  Therefore, every dependence in \(\su\) comes from {\ops}
  \(\mathcal{O}^{(0)}_\spi=\su_\spi 
  \prod_{\jvp\in\Ib}(1-{\g}~t_\jvp)\). If we denote by
  \(\left({\vv}_\jvp\right)_{1\leq\jvp\leq \Kr+\Mr}\) the eigenvectors
  of \({\g}\), we see that in the limit \(t_{\ivp}\to 1/x_{\ivp}\), 
  \begin{equation*}
    \forall \jvp\in\Ib,\quad \mathcal{O}^{(0)}_\spi\ket{{\vv}_\jvp}=0\,.
  \end{equation*}
 Hence, the maximal possible degree in \(\su\) of
 \(\gQ[][I]\ket{{\vv}_{\jvp_1},{\vv}_{\jvp_2},\cdots,{\vv}_{\jvp_\lcds}}\)
 is the number of indices \(\spi\in \ninter 1 {\lcds}\) such that
 \(\mathcal{O}^{(0)}_\spi\) does not give \(0\), {\idest} the number of \(\spi\in
 \ninter 1 {\lcds}\) such that \(\jvp_\spi\in I\).

 This proves that for states belonging to the set
 \(E_{\Mp_1,\Mp_2,\cdots,\Mp_{\Kr+\Mr}}\), the degree of \(\gQ[][I]\)
 is at most \(\sum_{\jvp\in I} \Mp_\jvp\) .

In order to conclude the proof, one can check\footnote{This check is
  important, because the above arguments would not forbid
  \(\gQ[][\emptyset]\) to be equal to zero (or to have some
  eigenvalues equal 
to zero).} that \(\gQ[][\emptyset]\) has degree \(0\)
and is given explicitly by
 \begin{gather}
   \gQ[][\emptyset]=\prod_{\kk=1}^{\Kr+\Mr}
\Mpo_\kk !
\prod_{\substack{\jj=1\\ \jj\neq\kk}}^{\Kr+\Mr}(1-x_\kk/x_\jj)^{\Mpo_\kk}\,.
 \end{gather}
In the {\rhs}, each \(\Mpo_\kk\) is an {\op}, hence this identity means
that for a state \(\ket \psi \in %
E_{\Mp_1,\Mp_2,\cdots,\Mp_{\Kr+\Mr}}\), we have 
\(   \gQ[][\emptyset]\ket \psi=\prod_{\kk=1}^{\Kr+\Mr}
\Mp_\kk !
\prod_{\substack{\jj=1\\
    \jj\neq\kk}}^{\Kr+\Mr}(1-x_\kk/x_\jj)^{\Mp_\kk}\ket \psi\).

For the {\GL \Kr} group, %
if one of the {\ops} \(\gQ[][\{\jvp\}]\) had a degree smaller
than \(\Mpo_\jvp\), then we could write \(\gQ[][\ove]\) as the {\Wronskian}
determinant \eqref{eq:QQopDetGLK}
and
obtain that \(\gQ[][\ove]\) would be a polynomial of degree smaller than \(\sum_{\jvp=1}^{\Kr}
\Mpo_\jvp=\lcds\). That is impossible because
\(\gQ[][\ove]=\prod_{\spi=1}^\lcds \su_\spi\). Therefore each
polynomial \(\gQ[][\{\jvp\}]\) has degree \(\Mpo_\jvp\), and when we write
the {\Wronskian} expression \eqref{eq:NestedWronskianLambda} we deduce
that \(\gT[][][I]\) has degree \(\sum_{\jvp\in I}
\Mpo_\jvp\).

For {\sugrs}, the same conclusion is obtained by a slightly more
complicated argument, because we {\cannnot} use the relation
\eqref{eq:QQopDetGLK}. If we denote by \(d(\PP)\) the degree of the
polynomial \(\PP\), then the QQ-relation implies that 
\begin{gather}
  d(\gQn[][I,\jvp,\kvp])=d(\gQ[][I,\jvp])+d(\gQ[][I,\kvp])-d(\gQ[][I])\\
  \where I,\jvp,\kvp\equiv I\cup\{\jvp\}\cup\{\kvp\}\,,
\end{gather}
even for {\sugrs}. Then we can deduce recursively that 
\(d(\gQn[][I])=\sum_{\jvp\in I}d(\gQn[][\{\jvp\}])\), so that if any 
\(\gQ[][\{\jvp\}]\) had a degree smaller
than \(\Mpo_\jvp\), then \(d(\gQ[][\ove])\) would be smaller than
\(\sum_\jvp \Mpo_\jvp=\lcds\). This allows to conclude about the degree
of all {\Qoprs}. Finally the expression
\eqref{eq:ToSGLKfromNestexpanded} gives the degree of the {\Toprs}
associated to symmetric representations, from which the degree is
obtained for every {\rp}, using the CBR formula \eqref{eq:NestedCBR}.
\end{proof}

\section{Relation to the classical integrability}
\label{sec:quantum-classical}

As a %
conclusion to this chapter, it is interesting to note that, as
explained in \cite{Alexandrov:2011aa} this
construction actually corresponds to a general property of the rational
{\taufs} of the MKP hierarchy.
This hierarchy arises for instance in the study of some specific partial
differential equations, which are called ``{\ing}'' in the sense
that they can be solved exactly. 
In this section we will not introduce completely the classical
integrability\footnote{The name ``classical integrability'' emphasizes
the fact that it solves differential equations on functions, and not
on quantum {\ops}.} and the MKP hierarchy, but the reader can find in
\cite{Alexandrov:2011aa} (and in the references therein), an
introduction to the subject, which emphasizes the %
tools %
mentioned in
this section.

The aim of this section is not to cover in great details this MKP
hierarchy, but rather to explain that the construction of the previous
section finds a natural interpretation in this context. This shows
that the {\MID} out of which %
our
construction of the Bäcklund flow was
proposed is not just a surprising identity very specific to this
model. It is rather a meaningful identity arising in lots of
different contexts, %
which would allow
to generalize the
construction to other {\ing} {\csds} (non
polynomial ones, for instance), or to other {\ing}
systems if we prove that they are related to the MKP hierarchy.

In this section, we will restrict for simplicity to the group \(\GL
\Kr\) as opposed to the {\sugrs} of the previous sections.

\subsection{The MKP hierarchy and the CBR formula}
\label{sec:mkp-hierarchy-cbr}

The KP and MKP hierarchy describe some sets of functions (called
{\taufs}) which obey a given equation (see {\below}). Several
constructions are known for these functions, and each of these
functions allows to construct a solution of several
{\ing} differential equations.

More explicitly the {\taufs} of the {\KPhi} are specific functions of an infinite
sequence \(\tk=(t_1,t_2,\cdots)\) of variables called ``times''. %
A function of \(\tk\) is called a {\tauf} of the {\KPhi} if it satisfies
\begin{gather}
\label{eq:KPint}
\fbox{\ensuremath{\displaystyle {
\oint_{{\mathcal C}} e^{\xi ({\tk}-{\tk'},z)} %
 \tau%
 ({\tk}-[z^{-1}])\tau%
 ({\tk'}+[z^{-1}])\mathrm{d}z =0}}}\,\,,\\
 \where \tau(\tk+[z^{-1}])\equiv
 \tau\left(t_1+\frac{z^{-1}}{1},t_2+\frac{z^{-2}}{2},t_3+\frac{z^{-3}}{3}+\cdots\right),\\
 \And \xi(\tk,z) \equiv  \sum_{{\kk}\geq 1} t_{\kk} z^{\kk}\,.
\end{gather}
In \eqref{eq:KPint}, the integration over the complex variable \(z\) is
performed on %
a contour \({\mathcal C}\) which encloses the singularities of \( \tau%
 ({\tk}-[z^{-1}])\tau%
 ({\tk'}+[z^{-1}])\) but not
the singularities of \(e^{\xi
  ({\tk}-{\tk'},z)}\).

A very %
simple %
example of %
{\tauf} is the function
\begin{gather}
  \tau(\tk)=e^{\sum_{\jvp=1}^\Kr\xi(\tk,x_\jvp)}\,.
\end{gather}
In this case the factor \( \tau%
 ({\tk}-[z^{-1}])\tau%
 ({\tk'}+[z^{-1}])\) in \eqref{eq:KPint} is equal to
\( \frac{\tau%
 ({\tk})}{\tau  ({[z^{-1}]})}
\tau
 ({\tk'}) \tau
 ({[z^{-1}]})\)
which is independent of \(z\) and has
 no singularity. Then the contour \(\mathcal{C}\) does not enclose any
 singularity, and \eqref{eq:KPint} holds.

The modified {\KPhi} (or {\MnM}{\KPhi}) is obtained by adding one more
time \(\su\). Then a 
function \(\tau(\su,\tk)\) is a {\tauf} of the MKP hierarchy if
\begin{gather}
\label{eq:MKPint}
\fbox{\ensuremath{\displaystyle {\forall \nn \in \bN,\quad
\oint_{{\mathcal C}} e^{\xi ({\tk}-{\tk'},z)}  z^{\nn}
\tau(\su+\nn, {\tk}-[z^{-1}])
\tau(\su, {\tk'}+[z^{-1}])\mathrm{d}z =0}}}\,\,,
\end{gather}
where the contour \(\mathcal{C}\) encircles all the singularities of
\(\tau(\su+\nn, {\tk}-[z^{-1}]) 
\tau(\su, {\tk'}+[z^{-1}])\),  but does not encircle any
singularity of \(e^{\xi ({\tk}-{\tk'},z)}  z^{\nn}\).

Once again, a very simple {\tauf} of the MKP hierarchy is given by the
function 
\begin{gather}
\label{eq:tautforchar}
  \tau(\su,\tk)=e^{\sum_{\jvp=1}^\Kr\xi(\tk,x_\jvp)}\,.
\end{gather}

\paragraph{Expression in terms of {\Ydag}s}
\label{sec:expr-terms-ydag}

These {\taufs} are functions of the ``times'', but %
they can be %
transformed into
functions \(\uptau(\su,\lambda)\),
which are functions of arbitrary {\Ydag}s \(\lambda\):
\begin{gather}
\label{eq:ttolambda}
\tau(\su,\tk)=\sum_{\lambda}s_{\lambda}(\tk)\uptau(\su,\lambda)\,\\
\label{eq:ttolambdaDefS}
\where s_{\lambda}({\tk})=
\Det{h_{\lambda_\ii -\ii +\jj}({\tk})}{1\leq \ii,\jj\leq |\lambda|}\,,
\end{gather}
where \(s_{\lambda}({\tk})\) denote the Schur polynomials, which
can expressed in terms of the symmetric Schur polynomials
\(h_\ii=s_{\las [1][\ii]}\)
defined by 
\begin{align}
\label{eq:ttolambdaDefH}
  e^{\xi ({\tk} , z)}=\sum_{{\kk}\geq 0}h_{\kk} ({\tk})z^{\kk}\,.
\end{align}

One can show that it is also possible to go the other way round and
express \(\uptau(\su,\lambda)\) as 
\begin{gather}
\label{eq:tauttolamb}
\left. \phantom{\int}
\uptau(\su,\lambda)=s_{\lambda}(\tilde \partial )\tau(\su,\tk)\right
|_{\tk=0},\\
\where \tilde \partial =(\partial_{t_1} , \frac{1}{2}\partial_{t_2},
\frac{1}{3}\partial_{t_3}, \ldots \, ~)\,.
\end{gather}

For instance, for the {\tauf}\footnote{
Here, \(\g\in\GL\Kr\) denotes a matrix with eigenvalues \(x_\jvp\).}
\(\tau(\su,\tk)=e^{\sum_{\jvp=1}^\Kr\xi(\tk,x_\jvp)}=e^{\sum_{\kk\geq
    1}t_{\kk} \mathrm{tr}(\g^\kk)}\), the object %
\(s_{\lambda}(\tilde \partial )\tau(\su,\tk)\) is equal to %
\(s_{\lambda}(\tilde \tk)\tau(\su,\tk)\) where 
\begin{align}
  \tilde \tk=&\left(\mathrm{tr}(\g), \frac{\mathrm{tr}(\g^2)}{2}, \frac{\mathrm{tr}(\g^3)}{3},\cdots\right)\,.
\end{align}

Therefore the {\tauf} \(\tau(\su,\tk)=e^{\sum_{\jvp=1}^\Kr\xi(\tk,x_\jvp)}\)
is associated to \(\uptau(\su,\lambda)=s_{\lambda}(\tilde \tk
)\). To understand this object, let us first %
write \eqref{eq:ttolambdaDefH} to %
obtain \(h_{\kk} ({\tilde \tk})\):
\begin{gather}
\sum_{{\kk}\geq 0}h_{\kk} ({\tilde \tk})z^{\kk}=  e^{\xi ({\tilde \tk}
  , z)} = e^{\sum_{\kk\geq 1} z^\kk
  \frac{\mathrm{tr}(\g^\kk)}{\kk}}%
=e^{-\mathrm{tr}~\mathrm{log}\left(1-{\g} z\right)}=w(z)
\end{gather}
where \(w(z)=\det \frac 1 {1- {\g}~ z}=\sum_{s=0}^{\infty} z^s \chs s(\g)\)
is the generating series of the symmetric characters (see
\eqref{eq:DefWz}). This immediately implies that \(h_{s} ({\tilde
  \tk})=\chs s(\g)\) is the character of \(\g\) in the symmetric
{\rp} \(\las[1]\). Then the equation \eqref{eq:ttolambdaDefS}
ensures that for an arbitrary {\Ydag} \(\lambda\), \(s_{\lambda}(\tilde \tk )\) is
the character \(\cha \lambda(\g)\). This shows that the {\tauf}
\(\tau(\su,\tk)=e^{\sum_{\jvp=1}^\Kr\xi(\tk,x_\jvp)}\) 
is associated to \(\uptau(\su,\lambda)=\cha {\lambda}(\g)\).

\paragraph{Bilinear identities for {\taufs}}
\label{sec:bilin-ident-taufssss}

We have defined {\taufs} as functions satisfying the relation
\eqref{eq:MKPint}, and we have shown that the simplest example of
{\taufs} \eqref{eq:tautforchar} was tightly related to characters. We will now
see that the relation \eqref{eq:MKPint} is tightly related to
bilinear identities, and we will later see that for well-chosen
{\taufs}, these bilinear identities correspond to 
the relations obtained in section \ref{sec:expr-diff-Qop} for
characters and for {\Toprs}. 

Let us choose \({\nn}=1\) and \(\tk'=\tk-[{z_1}^{-1}]-[{z_2}^{-1}]\), and write
\eqref{eq:MKPint}. The factor \(e^{\xi ({\tk}-{\tk'},z)}  z^{\nn}\) is
then equal to
\begin{align}
\label{eq:exifor3lin}
  e^{\xi ([{z_1}^{-1}]+[{z_2}^{-1}],z)}z=&z~e^{\sum_{\kk\geq 1}
    \frac{\left(z/z_1\right)^\kk}{\kk}+\frac{\left(z/z_1\right)^\kk}{\kk}}\\=&
  z~e^{-\mathrm{log}(1-z/z_1)-\mathrm{log}(1-z/z_2)} = \frac{z}{
(1-\frac{z}{z_1})(1-\frac{z}{z_2})}\,.
\end{align}

The prescription for the contour \(\mathcal{C}\) in \eqref{eq:MKPint} is
that it should encircle all the singularities of \(\tau(\su+\nn, {\tk}-[z^{-1}]) 
\tau(\su, {\tk'}+[z^{-1}])\), but not the singularities of \(e^{\xi
  ({\tk}-{\tk'},z)}  z^{\nn}\). Moreover, \(\tau(\su+\nn, {\tk}-[z^{-1}]) 
\tau(\su, {\tk'}+[z^{-1}])\) is regular at \(z\to \infty\) where it
converges to \(\tau(\su+\nn, {\tk}) \tau(\su, {\tk'})\), and that
implies that the singularities of \(\tau(\su+\nn, {\tk}) \tau(\su,
{\tk'})\) lie only in a bounded domain of the complex plane.

Let us consider a contour \(\mathcal{C}_{\infty}\) which encloses all
the singularities of \linebreak \(e^{\xi ({\tk}-{\tk'},z)}  z^{\nn}
\tau(\su+\nn, {\tk}-[z^{-1}])
\tau(\su, {\tk'}+[z^{-1}])\). If we choose for instance a circle with a
large radius we can compute 
\begin{align}
\label{eq:MKPto3infty}
  \oint_{{\mathcal C}_\infty} e^{\xi ({\tk}-{\tk'},z)}  z^{\nn}
\tau(\su+\nn, {\tk}-[z^{-1}])
\lefteqn{\tau(\su, {\tk'}+[z^{-1}])\mathrm{d}z} \\=& 2 \bi \pi z_1z_2\tau(\su+\nn, {\tk})
\tau(\su, {\tk'}) \\=& 2 \bi \pi z_1 z_2 \tau(\su+1, {\tk})
\tau(\su, {\tk}-[{z_1}^{-1}]-[{z_2}^{-1}])
\end{align}
because the integrand is equivalent to \(z_1z_2\frac{\tau(\su+\nn, {\tk})
\tau(\su, {\tk'})}{z}\), in virtue of \eqref{eq:exifor3lin}.

The difference between the contour \({\mathcal C}_\infty\) and the
contour \({\mathcal C}\) of \eqref{eq:MKPint} is only the two
singularities of \(e^{\xi ({\tk}-{\tk'},z)}  z^{\nn}\), at positions
\(\su=z_1\) and \(\su=z_2\). Thus, the difference between these contours
is
\begin{multline}
\label{eq:MKPto3diff}
\left[  \oint_{{\mathcal C}_\infty} 
  -\oint_{{\mathcal C}} \right]
\left(  e^{\xi ({\tk}-{\tk'},z)}  z^{\nn}
\tau(\su+\nn, {\tk}-[z^{-1}])
\tau(\su, {\tk'}+[z^{-1}])\right)\mathrm{d}z \\=2 \bi \pi
\frac{z_1z_2}{z_1-z_2}\left(z_1 \tau(\su+\nn, {\tk}-[{z_1}^{-1}])
\tau(\su, {\tk'}+[{z_1}^{-1}])\right.\\ \left.-z_2 \tau(\su+\nn,
{\tk}-[{z_2}^{-1}]) 
\tau(\su, {\tk'}+[{z_2}^{-1}])\right)
\\=2 \bi \pi
\frac{z_1z_2}{z_1-z_2}\left(z_1 \tau(\su+1, {\tk}-[{z_1}^{-1}])
\tau(\su, {\tk}-[{z_2}^{-1}])\right.\\ \left.-z_2 \tau(\su+1,
{\tk}-[{z_2}^{-1}]) 
\tau(\su, {\tk}-[{z_1}^{-1}])\right)
\end{multline}
Finally, \eqref{eq:MKPint}, \eqref{eq:MKPto3infty} and
\eqref{eq:MKPto3diff} allow to conclude that
\begin{multline}
\label{eq:3termMPK}
  (z_1-z_2)\tau(\su+1, {\tk})
\tau(\su, {\tk}-[{z_1}^{-1}]-[{z_2}^{-1}])\\=
z_1 \tau(\su+1, {\tk}-[{z_1}^{-1}])
\tau(\su, {\tk}-[{z_2}^{-1}])\\
-z_2 \tau(\su+1, {\tk}-[{z_2}^{-1}])
\tau(\su, {\tk}-[{z_1}^{-1}])
\end{multline}

This equation is a 3-term consequence of \eqref{eq:MKPint}, and we
will now show that the {\MID} is nothing but this equation
\eqref{eq:3termMPK}, written for well-chosen {\taufs}.

\paragraph{{\taufs} for {\csds}}
\label{sec:taufssss-spin-chains}

In the previous section, we defined the {\Toprs} as %
\(\lT  =\left[\DLt
\cha \lambda(\g)\right]\), and we would like to identify them with %
some \(\uptau(\su,\lambda)\). %
If we remember that \(\uptau(\su,\lambda)=\cha \lambda(\g)\) was
associated to \(\tau(\su,\tk)=e^{\sum_{\kk\geq
    1}t_{\kk} \mathrm{tr}(\g^\kk)}\), we see that we have to define
\begin{gather}
\label{eq:DeftT}
\fdisp{  \tT = \left[\DLt
e^{\sum_{\kk\geq
    1}t_{\kk} \mathrm{tr}(\g^\kk)}
\right]}\,.
\end{gather}

Then the \(\tT\) are linear combinations of the previous {\Toprs}, written
as 
\begin{align}
  \tT=\sum_{\lambda}s_{\lambda}(\tk)\lT\,,
\end{align}
and therefore they commute with each other.

Let us now show that the {\MID} \eqref{eq:MIDPi} is the statement that the function 
\(%
\tT\) obeys the relation \eqref{eq:3termMPK}.
\begin{proof}
  First, let us see what \(\tT[][\tk+[z^{-1}]]\) means:
  \begin{gather}
\label{eq:tTtpzm1}
    \begin{aligned}
      \tT[][\tk+[z^{-1}]]=&\left[\DLt e^{\sum_{\kk\geq
            1}\left(t_{\kk}+\frac{z^{-\kk}}{\kk}\right)
          \mathrm{tr}(\g^\kk)}
      \right]\\
      =& \left[\DLt e^{-\mathrm{tr}~\mathrm{log}\left(1-{\g}/z\right)}
        e^{\sum_{\kk\geq 1} t_{\kk} \mathrm{tr}(\g^\kk)}
      \right]\\
      =& \left[\DLt w(1/z) \PI \right]\,
    \end{aligned}\\
\where \Pi = e^{\sum_{\kk\geq
    1}t_{\kk} \mathrm{tr}(\g^\kk)}\,.
  \end{gather}

Therefore, if we replace \(\tk\rightsquigarrow
\tk+[{z_1}^{-1}]+[{z_2}^{-1}]\) in \eqref{eq:3termMPK}, it reads
\begin{multline}
\label{eq:3termMPK2Cod}
  (\frac 1 {z_1}-\frac 1 {z_2})
\left[\DLt[{\su}_\spi+1+\hD] 
  w(1/z_1) w(1/z_2) \PI \right]\cdot
\left[\DLt \PI \right]\\
= \frac 1 {z_2} \left[\DLt[{\su}_\spi+1+\hD] 
  w(1/z_2)\PI \right]\cdot
\left[\DLt  w(1/z_1) \PI \right]\\
-\frac 1 {z_2} \left[\DLt[{\su}_\spi+1+\hD] 
  w(1/z_1)\PI \right]\cdot
\left[\DLt  w(1/z_2) \PI \right]
\end{multline}
This is exactly the equation \eqref{eq:MIDPi}, and the condition \(\Pi=\prod_{\kk=1}^\nn  
\left(w(z_\kk)\right)\) (resp \(\Pi=\prod_{\kk=1}^\nn  
\left(w(z_\kk)\right)^{a_\kk}\) in \eqref{eq:MIDPiDef}) corresponds to
the case when \(\tk=0+[z_1]+[z_2]+\cdots+[z_\nn]\) (resp \(\tk =\sum_{\kk=1}^\nn {a_\kk}~
 [{z_\kk}]\).
The more general case (where \(\tk\) is arbitrary) can also be written as a
limit of \(\tk =\sum_{\kk=1}^\nn {a_\kk}~ [{z_\kk}]\) when \(\nn\) tends
to \(\infty\).
\end{proof}

We saw in this section that the {\MID} is the same as an important
identity satisfied by the {\taufs} of the MKP hierarchy. We will not
detail it here, but we showed in \cite{Alexandrov:2011aa} that the
{\MID} (or more specifically the equivalent CBR formula
\eqref{eq:CBR}) allows to show that \(\tT\) is indeed a {\tauf} of the MKP
hierarchy. 

In the next sections we will see that {\another} property of the
{\Toprs}, namely the existence of the polynomial Bäcklund flow defined in
section \ref{sec:expr-diff-Qop}, is also %
related to
the properties
of specific {\taufs} of this MKP hierarchy (the rational solutions).

\subsection{The rational solution of the MKP hierarchy}
\label{sec:rati-solut-mkp}

The general polynomial solution of the {\KPhi} was constructed
by Krichever in \cite{Krichever78} (see also
\cite{Krichever83,DMKM88}), and can be directly extended to the MKP
hierarchy.
In this section, we will not reproduce in details this
construction, but simply give the most relevant expressions which
allow to compare with the %
Bäcklund flow introduced in the section \ref{sec:expr-diff-Qop}.

The general polynomial solution of the MKP hierarchy is %
given by the determinant %
\begin{gather}
   \tau({\su},\tk)=\frac{\Det{A_{\coordi}(\su-{\coordj},\tk)}{1\leq {\coordi},{\coordj}\leq
       {\NN}}}{\prod_{{\coordi}=1}^{\NN} p_{\coordi}^{\su}}\\
\where
A_{\coordi}(\su,\tk)=\left.\sum_{{\mm}=0}^{d_{\coordi}} a_{{\coordi},{\mm}} \partial_z^{\mm}\left( z^\su
    e^{\xi(\tk,z)}\right)\right| _{z=p_{\coordi}}
\end{gather}
and it is {\lbd} by an integer \({\NN}\geq 0\), by \({\NN}\) numbers
\(\{p_{\coordi}\}\), by \({\NN}\) numbers \(d_{\coordi}\) (which are the degree of the
polynomials \(\frac{A_{\coordi}(\su,\tk)}{p_{\coordi}^\su}\)), and by the coefficients
\(\{a_{{\coordi},{\mm}}\}\).

If we expand at large \(\su\), we see that 
 \begin{gather}
   A_{\coordi}(\su,\tk)=a_{{\coordi},d_{\coordi}}\su^{d_{\coordi}}p_{\coordi}^\su e^{\xi(\tk,p_{\coordi})}+\mathcal{O}\left(
     \su^{d_{\coordi}-1}p_{\coordi}^\su \right)\\
 \And \tau({\su},\tk)=\su^{\sum_{{\coordi}}d_{\coordi}}e^{\sum_{{\coordi}=1}^{\NN}
   \xi(\tk,p_{\coordi})}\Det{a_{{\coordi},d_{\coordi}} p_{\coordi}^{-{\coordj}}}{1\leq {\coordi},{\coordj}
 \leq {\NN}}+\mathcal{O}\left(
     \su^{(\sum_{{\coordi}}d_{\coordi})1}\right)
 \end{gather}

By comparison, \(\tT\) has the large \(\su\) asymptotic behavior
\begin{gather}
\tT \sim \su^{\lcds}
e^{\sum_{\jvp=1}^{\Kr}\xi(\tk,x_\jvp)}
\end{gather}
as can be seen in \eqref{eq:DeftT}, from which we deduce that \(\tT\)
corresponds to \({\NN}=\Kr\), \(p_{\coordi}=x_{\coordi}\) and \(\sum_{{\coordi}}d_{\coordi}=\lcds\).

From there, one can see that \(A_{\jvp}(\su,\tk+[z^{-1}])\) has a pole when
\(z\to
p_{\jvp}=x_{\jvp}\).
That allows to show that the residue of \(\tau(\su,\tk+[z^{-1}])\) at
\(z\to x_{\coordi}\) gives rise to the smaller determinant 
\begin{equation*}
\frac{\Det{A_{\kk}(\su-{\coordl},\tk)}{
\substack{{\kk}\in\oj\\1\leq {\coordl}\leq {\NN}-1}}
}{\prod_{{\coordi}\in \oj} p_{\coordi}^{\su}}\,.
\end{equation*}
Hence, this residue is a minor of the original determinant, exactly
like the Bäcklund transform reduces a determinant (eg
\eqref{eq:WronskianLambda}) into its minor (eg
\eqref{eq:NestedWronskianLambda}). This explains why the Bäcklund
flow was defined in \eqref{eq:DefNestTop} by taking a very singular
limit \(t_{\ivp}\to1/x_{\ivp}\), which amounts to taking a residue.
%
%
%%%%%%
%
%
%
%
A more detailed dictionary between this rational solution of MKP and
the Bäcklund flow constructed above is given in
\cite{Alexandrov:2011aa}, %
 but we can already see that the Bäcklund
transform comes from the limit \(z\to x_{\jvp}\) of
\(\tT[\su][{\tk+[z^{-1}]}])\). But as we saw in \eqref{eq:tTtpzm1},
\(\tk\rightsquigarrow\tk+[z^{-1}]\) is 
equivalent to a multiplication by \(w(1/z)\) on the right of all
{\cdrs}.
Therefore, we recover the prescription \eqref{eq:DefNestTop} which
says (for {\GL \Kr}) that every successive Bäcklund transform
inserts a \(w(t_{\jvp})\) on the right of the {\cdrs}, and prescribes
to take the limit
\(t_{\jvp}\to 1/x_{\jvp}\).

Moreover, as $\tk$ is related (through \eqref{eq:ttolambda})  to the
choice representation for the auxiliary space\footnote{Another way to
  phrase the same remark is by arguing that in view of 
\eqref{eq:TfromD}, 
  one should identify what stands to the rigth of the coredivatives
 in \eqref{eq:DefNestTop}
  with a character of some representation.}, these expressions
suggest that Q-operators correspond to a specific (actually
inifinite-dimensional) choice of representation in the auxiliary
space. This approach is frequently used in the litterature (see for instance \cite{Derkachov:2006fw,2009JPhA...42g5204D,Derkachov:2010qe,2010JSMTE..11..002B,2011NuPhB.850..175F})
and can certainly be shown to be equivalent to the present construction.

%%% Local Variables: ***
%%% mode:latex ***
%%% eval: (find-file "english.tex") ***
%%% TeX-master: "english.tex" ***
%%% End: ***

%% file: TBA.tex
As we saw in the introductory chapter
\ref{sec:gener-other-integr}, field theories (as opposed to the spin
chains studied in the previous chapter) may only be described by the
Bethe {\anz} in a regime where 
the spatial dimension is large enough. Therefore this Bethe {\anz}
is called the asymptotic Bethe {\anz}. By
contrast this chapter will 
be devoted to the exact computation of finite-size effects in these
theories.

A first step in the study of finite size effects was achieved by
Lüscher \cite{Luscher:1985dn,Luscher:1986pf}, who gave (order by
order) the first corrections to the asymptotic Bethe {\anz}.

But there are also several models for which an exact computation of
finite size effects 
can be obtained, in the sense that a set of (usually integral)
equations can be written, which
give the exact spectrum of the theory for arbitrary value of the size
{\LF}. One approach to get these equations is to define an {\ing}
discretization {\idest} to write a field theory as the limit of a
{\cds}. This approach was introduced by Destri and de Vega
\cite{Destri:1987ze}, and the corresponding equations are often called
the ``DdV'' equations. This approach was successfully applied to
models such as the Sine-Gordon model and Toda theories, but there
still exist several field theories which are believed to be {\ing} at infinite
size but for which we do not know any {\ing} discretization.

Another method was introduced by A. Zamolodchikov
\cite{Zamolodchikov:1989cf}, and is called the thermodynamic Bethe
{\anz} (TBA). This method
can be used for many relativistic sigma-models, and will be introduced
here in the example of the {\PCF}. 
This thermodynamic Bethe {\anz} seems very general, but one drawback is
that unlike the lattice discretization, it often leads to an infinite
set of integral equations. It was understood in
\cite{2009JHEP...12..060G} that in the particular case of the \(\SU 2
\times \SU 2\) {\PCM}, this infinite set of equations can be
recast into a single non-linear integral equation.

This chapter will introduce an important original result of this
{\thesis}: the 
existence of a
general procedure, based on the
{\Qfs} (expected to be the eigenvalues of {\Qoprs} constructed as in the chapter
\ref{part:qoperatorsspin}) which allows to recast the {\TBAE} of
many theories into a finite set of non-linear integral equations (we
will call such a set a ``FiNLIE'').\index{FiNLIE}

In the present chapter, we will 
illustrate this method on the example of the
{\PCF}, (as in the article
\cite{2010arXiv1007.1770K}). As we will see in the next chapter
\ref{cha:dualite-adscft}, we
can also apply
this method %
to the AdS/CFT
spectrum.  %

The section \ref{sec:example-princ-chir} will motivate the {\TBA} on the example
of the {\PCM}. It gives rise to a set of equations which
describes the spectrum of this field theory. The derivation of these
equations for numerous {\ing} models  is well presented in the literature, and the section
\ref{sec:example-princ-chir} does not aim at giving the most rigorous
proof of this construction. It is rather designed to introduce the
key concepts and hypotheses underlying the {\TBA}, and to show the
equations that arise from this procedure.
These equations
will be the starting point of original works of this {\PhD}, presented
in the next sections. 

The section \ref{sec:solution-generale-de} gives the typical solution
of the Hirota equation in several cases corresponding to different
{\ing} models (to a large extent this was already known in the
literature  before this {\PhD} (see in
particular \cite{springerlink:10.1007/s002200050165}), but some of
the results presented here
are original results \cite{Gromov:2010km} of this PhD). As explained in this section,
this general solution is a key ingredient to write FiNLIEs. Finally
the procedure allowing to write FiNLIEs is illustrated in the case of
the {\PCF}, introducing original results of this {\PhD}, written in
the article \cite{2010arXiv1007.1770K}.

\section{Example of the {\PCF}}
\label{sec:example-princ-chir}

\subsection{The asymptotic Bethe {\anz}}
\label{sec:asympt-bethe-anz}

In the asymptotic limit (when the spacial dimension is large enough),
\index{asym@asymptotic limit (\ensuremath{\LF\to\infty})}
the solution of the {\PCF} was obtained by Wiegmann
and Polyakov in
\cite{Wiegmann1984217,Polyakov1983121,Polyakov1984223}. 
Let us briefly introduce the model and the main arguments and
results of this approach (though without proof).

The {\PCF} is a two-dimensional relativistic field
theory characterized by the action 
\begin{align}
  \label{eq:ActPCM}
  \Sact=&\frac {-1} {2\alpha_0}~\int\!\!\!\int
  \mathrm{d}{x}~\mathrm{d}{t}~\mathrm{tr}\left(\left[\partial^\mu
      h\right]\cdot \left[\partial_\mu h^{-1}\right]\right)
= \frac {-1} {2\alpha_0}~\int\!\!\!\int
  \mathrm{d}{x}~\mathrm{d}{t}~\mathrm{tr}\left(h^{-1}\partial_\mu h\right)^2
\,
\end{align}
where the integration variable \(x\) is associated to a periodic space
dimension of size {\LF}, and the variable \(t\in \bR\) is associated to
the time. The field \(h(x,t)\) takes values in \(\SU \Np\), and the index
\(\mu\) refers to the direction \(x\) or \(t\). This action is
invariant under Lorentz transformations on the one hand, and on the
other hand under the
transformations 
\(h\rightsquigarrow h\cdot {\g}\) and \(h\rightsquigarrow  {\g}\cdot h\)
where \(\g \in \SU\Np\)
(these two transformations are called \(\SU{\Np}_\Rg\) and \(\SU
\Np_\Lf\) respectively). Therefore, this field theory will be called
the \(\SU\Np\times \SU\Np\) {\PCM}.

The integrability of this model (in the sense that when {\LF} is large, the
wave function is described by the Bethe {\anz}, and the spectrum is
obtained from Bethe equations)
can be motivated by
writing an infinite set of conserved charges \cite{Polyakov1977224},
and since the problem is relativistic, the momenta in
\eqref{eq:FieldBetheAnsatz} are parameterized 
by rapidities \(\phi_{\ii}\) as follows:
\begin{align}
  \label{eq:relat-rapid}
  p_\ii=&{\mmass}_\ii ~\mathrm{sinh}(\phi_\ii)\,.
\end{align}
They are also associated to energies
\begin{align}
  \label{eq:relat-ener}
  E_\ii=&{\mmass}_\ii ~\mathrm{cosh}(\phi_\ii)\,.
\end{align}

In the case of the Heisenberg {\cds}, one can note that the function
\(\Sscal(p,p')\) (in \eqref{eq:Smat}) can  
be expressed as \(\Sscal(p,p')=\Sscal(\su-\su')\) where \(\su \equiv \frac{e^{\bi p}}{e^{\bi
    p}-1}\) and \(\su' \equiv \frac{e^{\bi p'}}{e^{\bi
    p'}-1}\). This means
that the spectral parameter \(\su\) is an additive parameterization of
the momenta.
For the 
{\PCF}, 
the Lorentz invariance imposes that 
such an additive parameterization of the momenta is given by
the
rapidity \(\phi\)
and we get  \(\Smat(p_1,p_2)=\Smat(\phi)\) where \(\phi\equiv\phi_1-\phi_2\).
To have a notation similar to the section \ref{part:qoperatorsspin},
we will actually denote by the letter \(\us\) the quantity
\begin{align}
  \us=\frac{{\Np}}{2\pi}\phi.
\end{align}
\index{u(r@\ensuremath{\us} (rapidity)}}
In what follows, this quantity \(\us\) is actually what we will denote by the word ``rapidity''.

In the equation \eqref{eq:relat-rapid}, one can show that there are
massive particles of mass \({\mmass_1}\), given by  
\({\mmass_1}= \frac{\Lambda}{\alpha_0} e^{-\frac{4\pi
  }{{\Np} \alpha_{0}^{2}}}\) (where \(\Lambda\) is a cut-off). One can also show
(see the explanations {\below} and
\cite{Polyakov1983121,Wiegmann1984217} for more details) that
these particles give rise to 
 \(\Np-1\) different types of
bound states (configurations of multiple particles, 
 {\lbd} by \(a\in\ninter 1
{{\Np}-1}\))  with respective masses
\begin{align}
  \label{eq:PCFmasses}
  {\mmass}_a=&{\mmass}_1\frac{\sin\frac{\pi a}{{\Np}}}{\sin\frac{\pi }{{\Np}}}&\where&1\leq
  a\leq {\Np}-1\,.
\end{align}
In what follows this mass \(\mmass_1\) will usually be set to \(1\) by rescaling the
length \(\LF\) into a dimensionless parameter\footnote{In the Bethe
  equation, the mass and the length \(\LF\) only appear  in the expression \(e^{\bi~\LF~p}=
  e^{\bi~\LF~ \mmass_a~\mathrm{sinh}\phi}\), which means that the
  eigenstates only depend on the product \(\LF~\mmass_1\). 
}  \(\LF\rightsquigarrow \LF ~ \mmass_1\).
These massive particles carry spins for both \(\SU{\Np}_\Rg\) and \(\SU
\Np_\Lf\) {\idest} the wave function transforms ``covariantly'' under the
symmetry group \(\SU{\Np}_\Rg\times\SU \Np_\Lf\).
For the massive particles 
of type \(a=1\), the wave-function
transforms 
as
the bifundamental 
{\rp} 
(\(\left(\bC^\Np\right)_\Lf\otimes \left(\bC^\Np\right)_\Rg\))
under the symmetry group 
\(\SU{\Np}_\Rg\times\SU \Np_\Lf\). 

Then the matrix \(\Smat(p_1,p_2)=\Smat(\us)\) is
constrained by the relation 
\eqref{eq:YBfield}, by a unitarity condition \(\Smat(\us)\cdot \Smat(-\us)=1\)
(analogous to the constraint that for two particles, the two conditions
\eqref{eq:QuantCond2} can be recast into \(e^{\bi \LF p_\jj}=
\Sscal(p_\jj,p_\kk)\)), and by a crossing condition (see
\eqref{eq:PCFCrossing0}) which describes how the 
{\Sma} transforms when particles are replaced with
anti-particles. As shown in \cite{Zamolodchikov1979253}, this allows
to fix the {\Sma} uniquely (up to a scalar factor \(\CDD\)) and to get for the
fundamental massive particles (of type \(a=1\) in \eqref{eq:PCFmasses})
\begin{gather}
  \label{eq:SMatForPCF}
  \Smat_{\ii,\jj}(\us)=\CDD(\us)\cdot \Sscal_0(\us) \frac{\hat
    {\Rop}_{\Lf}(\us)}{\us-{\bi}} \otimes \Sscal_0(\us)  \frac{\hat
    {\Rop}_{\Rg}(\us)}{\us-{\bi}}\,,\\
  \where\quad
  \Sscal_0(\us)=       
  \frac{\Gamma \left( \bi\frac \us \Np \right)       
    \Gamma \left(-\bi \frac {\us+\bi} \Np \right)}       
  {\Gamma \left(-\bi\frac \us \Np \right)       
    \Gamma \left( \bi \frac {\us-\bi} \Np \right)}, \qquad   
  \CDD(\us)=\frac{\sinh (\pi\frac{ \us+\bi}{\Np})}{\sinh
    (\pi\frac{ \us-\bi}{\Np})}\,,
\label{eq:wher-sscal_0-fracg}\\
  \And \hat
  {\Rop}_{\Lf}(\us)\otimes \hat {\Rop}_{\Rg}(\us)=\left(\us~\bI+\bi~
    \perm_{\ii_{\Lf},\jj_{\Lf}}\right)\cdot \left(\us~\bI+\bi~
    \perm_{\ii_{\Rg},\jj_{\Rg}}\right)\,.
\label{eq:and-hat-rop_lf}
\end{gather}
This {\Sma} acts on the spaces corresponding to the particles \(\ii\)
and \(\jj\), like the {\op} \(\Rop(\su)\) of {\csds} (see
\eqref{eq:DefR}), and the main difference 
is that the ``physical space'' associated to each particle is
\(\left(\bC^\Np\right)_\Lf\otimes \left(\bC^\Np\right)_\Rg\), which
contains two copies of \(\bC^\Np\). The {\op} 
\(\perm_{\ii_{\Rg},\jj_{\Rg}}\) 
is the
permutation {\ops} (as defined in \eqref{eq:DefPerm}) acting on the
spaces \(\left(\bC^\Np\right)_\Rg\) associated to the particles \(\ii\)
and \(\jj\), while \(\perm_{\ii_{\Lf},\jj_{\Lf}}\) acts the same way on
the spaces \(\left(\bC^\Np\right)_\Lf\).

The ``crossing equation'' \cite{Zamolodchikov1979253} is the constraint 
\begin{align}
\label{eq:PCFCrossing0}
\frac{ \Sscal_0(\us+\bi\frac  \Np 2)^2 \CDD(\us+\bi \frac \Np 2)}
{ \Sscal_0(\us-\bi\frac  \Np 2)^2 \CDD(\us-\bi \frac \Np 2)}
  =&\left(\frac{\us+\bi \frac \Np 2-\bi}{\us+\bi \frac \Np 2}~~ \frac{\us-\bi
    \frac \Np 2+\bi} {\us-\bi \frac \Np 2}\right)^2
\end{align}
on the scalar part of the {\Sma}. We will actually use {\another}
(slightly stronger) equation {\below}
\begin{align}
\label{eq:PCFCrossing}
  \prod_{\nn=-\frac{\Np-1}2}^{\frac{\Np-1}{2}} \Sscal_0(\us+\bi \nn)^2
  \CDD(\us+\bi \nn)=&\left(\frac{\us-\bi\frac{{\Np}-1}2}{\us+\bi\frac{{\Np}-1}2}\right)^2\,
\end{align}
which implies the previous one \eqref{eq:PCFCrossing0}.
One can see that this equation actually comes from 
\begin{align}
  \prod_{\nn=-\frac{\Np-1}2}^{\frac{\Np-1}{2}} \Sscal_0(\us+\bi
  \nn)=&- \frac{\us-\bi\frac{{\Np}-1}2}{\us+\bi\frac{{\Np}-1}2}\,.
\end{align}
In (\ref{eq:PCFCrossing0},\ref{eq:PCFCrossing}), we see that when
\({\Np}\geq 3\), the non-trivial factor \(\CDD\) is a {\zerm}\footnote{If
  \({\Np}=2\), then the \(\CDD\) factor is equal to \(-1\) which does not contain
  any information. This is because there is no bound state, {\idest} no pole
  of the {\Sma} inside the physical strip.} (in the sense that
\(\prod_{\nn=-\frac{\Np-1}2}^{\frac{\Np-1}{2}} \CDD(\us+\bi \nn)=1\)), and 
is not imposed by the unitarity and crossing symmetry. This factor is
actually chosen to have a minimal number (and multiplicity) of poles in the
{\Sma}. Indeed one expects that the only poles of the {\Sma} (inside
the physical strip 
\(-\frac \Np 2 \leq \Im(\us) \leq \frac \Np 2\))
\index{strip!physical strip}
should be simple poles and correspond to bound states. In 
\eqref{eq:SMatForPCF}, the pole at \(\us=\pm\bi\)
indicates the existence of a bound state, made of 
two fundamental particles with rapidities \(\phi_1=\phi_0-\bi
\frac{\pi}{{\Np}}\) and \(\phi_2=\phi_0+\bi \frac{\pi}{{\Np}}\) (so that
\(\us=\frac {{\Np}}{2\pi}\left(\phi_1-\phi_2\right)=-\bi\)). This bound
state has an energy \({\mmass_1}
~\left(\mathrm{cosh}\left(\phi_1\right)+\mathrm{cosh}\left(\phi_2\right)\right)={\mmass_1}
\frac{\mathrm{sin}\left(\frac {2
      \pi}{{\Np}}\right)}{\mathrm{sin}\left(\frac {\pi}{{\Np}}\right)}
\mathrm{cosh}\left(\phi_0\right)\) and a momentum \({\mmass_1}
~\left(\mathrm{sinh}\left(\phi_1\right)+\mathrm{sinh}\left(\phi_2\right)\right)={\mmass_1}
\frac{\mathrm{sin}\left(\frac {2
      \pi}{{\Np}}\right)}{\mathrm{sin}\left(\frac {\pi}{{\Np}}\right)}
\mathrm{sinh}\left(\phi_0\right)\). It can therefore be viewed as a
single particle  of rapidity \(\phi_0\), and mass  \({\mmass_1}\frac{\sin\frac{2
    \pi}{{\Np}}}{\sin\frac{\pi }{{\Np}}}\). It is also possible to compute the {\Sma}
describing the interaction of this bound state with the fundamental
particles, and then the pole structure allows to find other bound
states (corresponding to bound states of three particles) and to
recursively identify the spectrum \eqref{eq:PCFmasses}. 

As 
we see, these bound states with mass \(\mmass_a\) can be viewed as
being 
made of several fundamental particles with
specific rapidities (such as \(\phi_1=\phi_0-\bi \frac{\pi}{{\Np}}\) and
\(\phi_2=\phi_0+\bi \frac{\pi}{{\Np}}\)).
Therefore we will restrict (for massive particles) to the fundamental
particles with mass \(\mmass_1\).

\paragraph{Bethe equations}
\label{sec:bethe-equations-1}

To summarize the Bethe equations obtained in this approach, let us
first remind how the various excited states are parameterized. In
the introductory section \ref{sec:gener-other-integr} 
it
appeared that 
in the Bethe {\anz},
the excited states are {\lbd} by rapidities of
``particles'' (denoting different excitations). In the present case, these ``particles'' are of several types:
\begin{itemize}
\item Massive particles with the mass \({\mmass}_1=1\) (after rescaling
  \(\LF\)). We will denote by 
  \(\theta_1,\theta_2,\cdots,\theta_{\dg[0]}\) the rapidities of these
  particles. These particles interact through the {\Sma} \eqref{eq:SMatForPCF}.

\item ``{\SU {\Np}} Magnons'' corresponding to the spin waves carried by
  the set of the \({\SU {\Np}}_\Lf\) and \({\SU {\Np}}_\Rg\) spins 
  of 
  the massive
  particles. As we saw in chapter \ref{part:qoperatorsspin} for an \(\SU
  {\Np}\) {\cds}, 
  the  rapidities of these ``magnons''
  are
  the roots of the \(\Np-1\) polynomials \(\gQf[][\Il\kk]\) along a {\nesting} {\ppath} and they
  obey the Bethe equations \eqref{eq:SuBetheSplit}, (or 
  \eqref{eq:BethBigerRank} in terms of Bethe roots). 

  Here, by contrast, we have both \({\SU {\Np}}_\Lf\) and \({\SU {\Np}}_\Rg\)
  spins. Therefore, we have two sets of polynomial {\Qfs}
  corresponding to the \({\SU {\Np}}_\Lf\) and \({\SU {\Np}}_\Rg\) spins. 
  \linebreak With the notations of chapter \ref{part:qoperatorsspin}, these
  polynomials are \linebreak denoted as
  \(\ffQ[][\{1\}][(\Rg)]\), \(\ffQ[][\{1,2\}][(\Rg)]\),
  \(\cdots\), \(\ffQ[][\{1,2,\cdots,{\Np}-1\}][(\Rg)]\) for the \({\SU
    {\Np}}_\Rg\) spins and \linebreak
  \(\ffQ[][\{1\}][(\Lf)]\), \(\ffQ[][\{1,2\}][(\Lf)]\), \(\cdots\),
  \(\ffQ[][\{1,2,\cdots,{\Np}-1\}][(\Lf)]\) for the \({\SU {\Np}}_\Lf\) spins.
\end{itemize}

The rapidities of these particles can be conveniently encoded into the
polynomials 
\index{Q-functions@{\Qfs}}
\begin{align}
  \label{eq:QPCFDef1}
  \fQ[][\mlvl] \equiv&\ffQ[-\frac {\mlvl} 2 -\bi
  \us][\{1,2,\cdots,{\Np}-\mlvl\}][(\Rg)]&\If&1\leq \mlvl\leq {\Np}-1\\ 
  \fQ[][0]\equiv&\varphi(\us)\equiv\prod \left(\us-\theta_{\ii}\right)\\
  \fQ[][\mlvl] \equiv&\ffQ[\frac {\mlvl} 2 -\bi \us][\{1,2,\cdots,{\Np}+\mlvl\}][(\Lf)]&\If&1-{\Np}\leq \mlvl\leq -1\\
  \fQ[][\mlvl]\equiv&1&\If&\mlvl=\pm {\Np}\,.
  \label{eq:QPCFDefl}
\end{align}
Their roots \(\us^{\rlb[][\mlvl][\nrt]}\) are defined as
\begin{align}
  \fQ[][\mlvl]=&\prod_{\nrt=1}^{\dg[\mlvl]}\left(\us-\us^{\rlb[][\mlvl][\nrt]}\right)\,.
\end{align}
\index{Bethe roots}
\index{u(r@\ensuremath{\us} (rapidity)!umn@\ensuremath{\us^{\rlb[][\mlvl][\nrt]}}|see{Bethe roots}}
where \(\mlvl\) denotes the different type of ``particles'',
\(\dg[\mlvl]\) denotes the number of particle of type \(\mlvl\), 
and \(\left\{\us^{\rlb[][\mlvl][\nrt]}\middle|1\leq \nrt \leq
  \dg[\mlvl]\right\}\) is the set of the rapidities of all the
particles of type \(\mlvl\).

The polynomial \(\fQ[][0]\), which describes the massive particles,
will be of special importance, and it will 
also be denoted\footnote{One should not confuse the symbol \(\phi\)
  in (\ref{eq:relat-rapid},~\ref{eq:relat-ener}) with the symbol \(\varphi\) (which denotes the
  polynomial \(\fQ[][0]\)).} as \(\varphi\).

One can notice that, compared to the {\Qfs} of chapter
\ref{part:qoperatorsspin}, the change of variables above contains a
``rotation'' \(\su\leadsto -\bi \us\). This is physically quite natural
because in chapter
\ref{part:qoperatorsspin} we had a change of variables
\(\su^{\rlb[\nrt]}\equiv \frac{e^{\bi p_\nrt}}{1-e^{\bi p_\nrt}}\) (for the
roots of the polynomial \(\ffQ[][\{1,2,\cdots,{\Np}-1\}][(\Rg)]\)). This
relation implied that \(p_\nrt\in\bR\Leftrightarrow
\Re(\su^{\rlb[\nrt]})=-\frac 1 2\). That is why we change the variables as
\(\fQ[][1] \equiv \ffQ[-\frac {1} 2 -\bi 
\us][\{1,2,\cdots,{\Np}-1\}][(\Rg)]\). For other {\Qfs}, we will see
that the change of variables (\ref{eq:QPCFDef1}-\ref{eq:QPCFDefl})
allows to have real {\Qfs}, 
and that will allow to consistently write the {\TBA}.

Then the Bethe equations take the same form as
\eqref{eq:BethBigerRank}, up to the change of variables
(\ref{eq:QPCFDef1}-\ref{eq:QPCFDefl}), and up to the specific behavior of
the massive particles, involving the {\Sma} \eqref{eq:SMatForPCF}. We
will also set the twist to one \(\g=\bI\) (as compared to chapter \ref{part:qoperatorsspin}).
Explicitly, these Bethe equations read \index{Bethe equations}
\begin{align}
  \label{eq:BethePCM}
  \lefteqn{  \forall \mlvl \in \ninter {-{\Np}+1} {{\Np}-1} \setminus \{0\} ,\quad
    \forall \nrt \in \ninter 1 {\dg[\mlvl]},}\nonumber\phantom{e^{\bi
      ~\LF~\sht}}\\
  -1=&\frac{\fQ[\us^{\rlb[][\mlvl][\nrt]}-\bi/2][\mlvl-1]\fQ[\us^{\rlb[][\mlvl][\nrt]}+\bi][\mlvl]\fQ[\us^{\rlb[][\mlvl][\nrt]}-\bi/2][\mlvl+1]}{\fQ[\us^{\rlb[][\mlvl][\nrt]}+\bi/2][\mlvl-1]\fQ[\us^{\rlb[][\mlvl][\nrt]}-\bi][\mlvl]\fQ[\us^{\rlb[][\mlvl][\nrt]}+\bi/2][\mlvl+1]}\\[.5cm]
  \lefteqn{
    \forall \nrt \in \ninter 1 {\dg[0]},}\nonumber\phantom{e^{\bi
      ~\LF~\sht}}\\
  \label{eq:BethePCM2}
  e^{\bi
    ~\LF~\sht}=&\frac{-1}{\Sscal(\theta_\nrt)}\frac{\fQ[\theta_\nrt-\bi/2][1]\fQ[\theta_\nrt-\bi/2][-1]}
  {\fQ[\theta_\nrt+\bi/2][1]\fQ[\theta_\nrt+\bi/2][-1]}\\
  \where~&\Sscal(\us)\equiv \prod_{\krt=1}^{\dg[0]} \Sscal_0(\us-\theta_\krt)^2
  \CDD(\us-\theta_\krt)\,.
\end{align}
The Bethe equation \eqref{eq:BethePCM} describes the ``magnons'' (as in
chapter \ref{part:qoperatorsspin}) and is sometimes called the
``auxiliary Bethe equation'', as opposed to the Bethe equation
\eqref{eq:BethePCM2} which describes the massive particles.

This can also be written (in the spirit of \eqref{eq:BethBigerRank}) as
\begin{align}
  \label{eq:BethPCMS}
  \framedline{
    \begin{array}{c}
      \forall \mlvl \in \ninter {1-{\Np}} {{\Np}-1}\\
      \forall \nrt \in \ninter 1 {\dg},
    \end{array}
    \quad
  }{e^{\bi~ \lcds~ p^{\rlb[\mlvl]}\left(\us^{\rlb[][\mlvl][\nrt]}\right)}
    = 
    \prod_{\substack{\klvl \in \ninter {1-{\Np}} {{\Np}-1}\\
        \lrt \in \ninter 1 {\dg[\klvl]}\\
        (\klvl,\lrt)\neq (\mlvl,\nrt)
      }} \Sscal^{\rlb[\mlvl],\rlb[\klvl]}(\us^{\rlb[][\mlvl][\nrt]}-\us^{\rlb[][\klvl][\lrt]})}\,\,,
\end{align}
where the product on the {\rhs} runs over all the \((\klvl,\lrt)\)
such that \(\klvl\neq \mlvl\) or \(\lrt\neq \nrt\), {\idest} over all the other Bethe
roots except the root \(\us^{\rlb[][\mlvl][\nrt]}\). In \eqref{eq:BethPCMS},
we  define
\begin{align}
  \label{eq:PCFBethepDef}
  p^{\rlb[\mlvl]}\left(\us\right)=&\left\{
    \begin{array}{cccc}
      0&\If&\mlvl\neq 0&\textrm{({\ie} for magnons)}\\
      \sht&\If&\mlvl=0&\textrm{({\ie} for massive particles)}
    \end{array}\right.\,,\\
  \Sscal^{\rlb[\mlvl],\rlb[\klvl]}(\us-\vs)=&
  \left\{
    \begin{array}{ccc}
      \frac{\us-\vs+\bi}{\us-\vs-\bi}&\If&\klvl=\mlvl\neq 0\\
      \frac{1}{\Sscal_0(\us-\vs)^2\CDD(\us-\vs)}&\If&\klvl=\mlvl =0\\
      \frac{\us-\vs-\frac\bi 2}{\us-\vs+\frac\bi 2}&\If&\klvl=\mlvl\pm 1\\
      1&\rlap{\oth}
    \end{array}\right.\,.
  \label{eq:PCFBetheSDef}
\end{align}

These constraints give equations on the rapidities \(\theta_\nrt\) of
the particles. Each excited state is associated to a solution of these
equations, and the corresponding energy is
\begin{gather} 
\label{eq:e=-sum_nrtchtth-}
  E= \sum_{\nrt}\cht[\theta_\nrt]\,.
\end{gather}

\subsection{Thermodynamic Bethe {\anz}}
\label{sec:{\anz}-de-bethe}

The {\TBA} is based on a ``double Wick rotation'' trick which goes as
follows:
in \eqref{eq:ActPCM}, the space is periodic \(x \in [0,\LF]\) while the time
\(t\in ℝ\) is not bounded. It means that \((x,t)\) belongs to a
cylinder of radius \(\LF\). This cylinder can be viewed as a torus where
one dimension has size \(\LF\), and the other one has size \(\RF\to
\infty\).
On this torus, we can write the partition function \(Z\) as an Euclidean
{\ppath} integral, 
which is
dominated by the vacuum when \(\RF\to
\infty\) ({\idest} at zero temperature):
\begin{align}
  \label{eq:zpathint}
  Z\approx&e^{-\RF ~E_0(\LF)}&\RF\to \infty\,,
\end{align}
where \(E_0\) is the vacuum energy, 
and
the symbol \(\approx\) denotes a logarithmic equivalent.

In this Euclidean {\ppath} integral, the roles of space and time are
symmetric, and they can be exchanged (the corresponding transformation
is called a ``Matsubara transform''). This means that \(Z\)
can as well be computed from the same {\PCF} with a space period
\(\RF\to \infty\) and an Euclidean time period \(\LF\). Back to the  Minkowski
signature, it means that the time has an imaginary period \(\LF\), which is
equivalent to the existence of an inverse temperature \(\beta=\LF\).
Therefore we see that a model with finite size is mapped to a ``mirror''
model with an infinite size but a finite temperature.  
\(E_0\)
is
then
 extracted from the free energy 
in the mirror model:
\begin{align}
  E_0({\LF})=&f({\LF})  \,.
\end{align}
In this mirror
model, the space period is \(\RF\to\infty\) so that the
the Bethe equations given in
the previous section can be used to compute the free energy
\(f({\LF})\).
Although the Bethe equations that we can write in this mirror
  model take 
  exactly the same form as (\ref{eq:BethePCM}-\ref{eq:BethePCM2}), 
  the roots \(\us^{\rlb[][\mlvl][\nrt]}\) entering these equations are
  ``virtual particles'' which do not have the same physical meaning as
  the original ones.
 Moreover,
 the finite temperature \(\beta=\LF\) gives 
a large number of (virtual) particles, and the Bethe equations have to be
written with such a large number of ``particles'' ({\ie} of
excitations). 

\subsubsection{The string hypothesis}
\label{sec:string-hypothesis}

Let us investigate the properties of the Bethe equations
\eqref{eq:BethePCM} for the magnons, in the mirror model where
the temperature is finite 
and 
thus each of the polynomials \(\fQ[][\mlvl]\) has a very large
degree\footnote{We remind that the degree \(\dg[\mlvl]\) of \(\fQ\) is
  equal to the number of particles of type \(\mlvl\).
}. We
will motivate the statement that, in the ``mirror'' theory with a
finite temperature, the roots will be grouped (in the complex plane)
in a very specific way. This statement will be called the ``string
hypothesis'', as its derivation {\below} is not completely rigorous.

Let us assume that the Bethe roots 
contributing to the free energy
are
symmetric with respect to complex-conjugacy, {\idest} that there is no
spontaneous breaking of the symmetry
of the Bethe equations (\ref{eq:BethePCM},~\ref{eq:BethePCM2})
 under complex-conjugation, then
we can see that if a given root \(\us^{\rlb[][\mlvl][\nrt]}\) (where \(\mlvl\neq
0\)) has a positive imaginary part, then 
\begin{align}
\label{eq:fracfq-nrt-bi2mlvl}
  \frac{\fQ[\us^{\rlb[][\mlvl][\nrt]}-\bi/2][\mlvl-1]
    \fQ[\us^{\rlb[][\mlvl][\nrt]}-\bi/2][\mlvl+1]}
  {\fQ[\us^{\rlb[][\mlvl][\nrt]}+\bi/2][\mlvl-1]
    \fQ[\us^{\rlb[][\mlvl][\nrt]}+\bi/2][\mlvl+1]} &\to 0
\end{align}
because\footnote{
\label{fn:1}
To understand the limit
  \eqref{eq:fracfq-nrt-bi2mlvl}, one can think that if
  \(\us^{\rlb[][\mlvl][\nrt]}=\xs+\bi\ys\), then the ratio 
\( \frac{\fQ[\us^{\rlb[][\mlvl][\nrt]}-\bi/2][\mlvl-1]
    \fQ[\us^{\rlb[][\mlvl][\nrt]}-\bi/2][\mlvl+1]}
  {\fQ[\us^{\rlb[][\mlvl][\nrt]}+\bi/2][\mlvl-1]
    \fQ[\us^{\rlb[][\mlvl][\nrt]}+\bi/2][\mlvl+1]}\) is roughly speaking equal
  to
\(\left(\frac
  {\xs+\bi\ys-\bi/2}{\xs+\bi\ys+\bi/2}\right)^\Mp\), where
\(\Mp\) denotes the degree of the polynomial
\(\fQ[][\mlvl-1] 
    \fQ[][\mlvl+1]\) (hence \(\Mp\) is very large).
But if \(\ys>0\), then \(|\xs+\bi\ys+\bi/2| > |\xs+\bi\ys-\bi/2|\) and
we obtain \eqref{eq:fracfq-nrt-bi2mlvl} in the limit \(\Mp\to\infty\).
} the polynomials \(\fQ[\us^{\rlb[][\mlvl][\nrt]}+\bi/2][\mlvl-1]\) and
\(\fQ[\us^{\rlb[][\mlvl][\nrt]}+\bi/2][\mlvl+1]\) are real and their degree becomes
infinite when the number of particles is infinite.

\begin{figure}
\fbox{\begin{minipage}{.95 \textwidth}
  \begin{center}
   \includegraphics{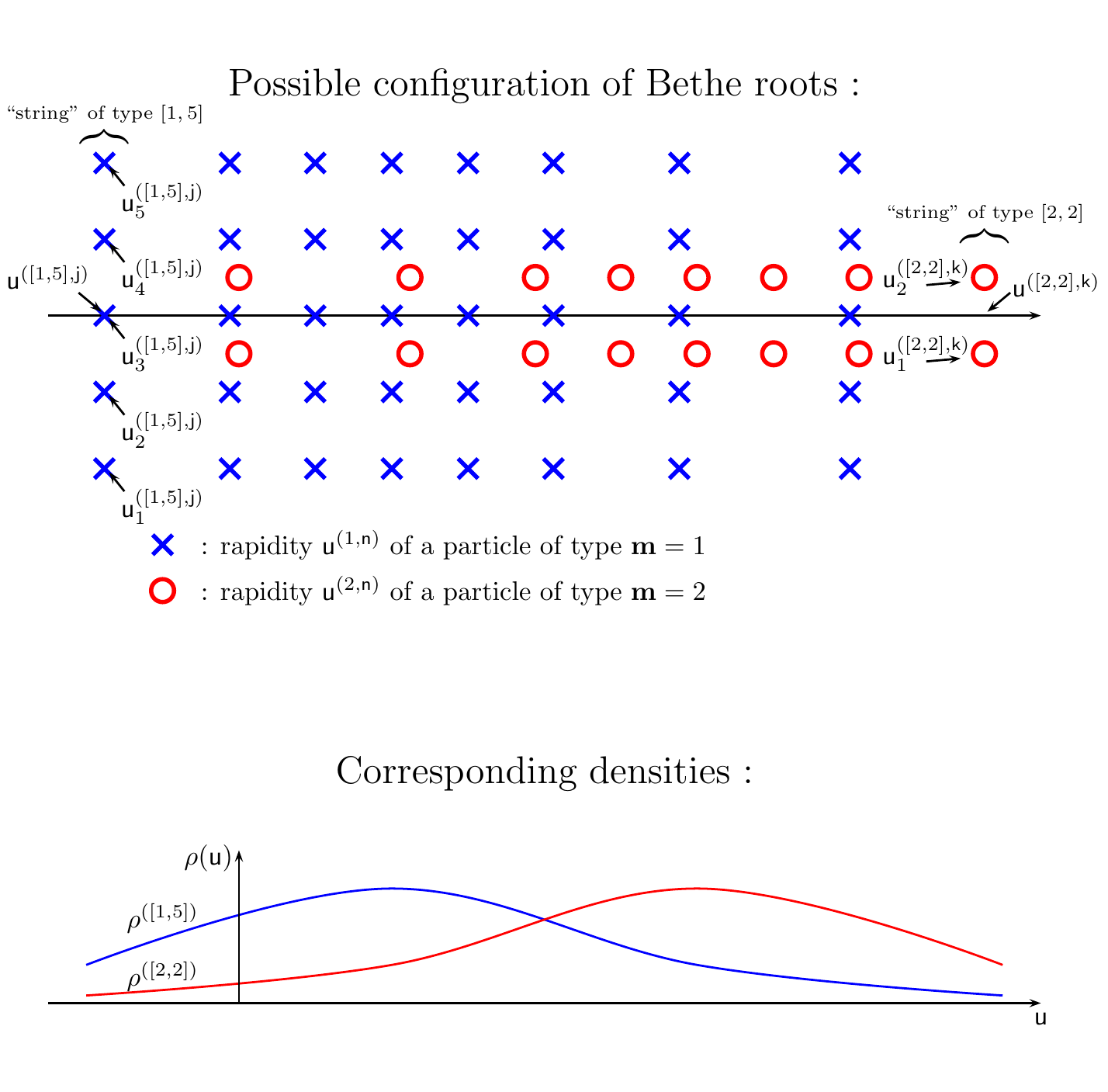}
   \caption{The ``string hypothesis'' and the
     introduction of densities.}
   \label{fig:densstring}
  \end{center}

This illustration of the ``string hypothesis'' shows a simplified
configuration of Bethe roots, where only two types of bound states
appear, namely the strings of type \(\strb[1][5]\) (made of five
particles of type \(\mlvl=1\)) and the strings {\lbd} by
\(\strb[2][2]\), which are made of two particles of type
\(\mlvl=2\). The rapidities of the particles inside a string are
{\lbd} as in \eqref{eq:BoundString}, and one real rapidity
\(\us^{\rlb[][\strb][\nrt]}\) describes each string. The position of each cross
(resp circle) on the plane denotes the rapidity \(\us^{\rlb[][1][\nrt]}\) (resp
\(\us^{\rlb[][2][\nrt]}\)) of a ``particle'' of type \(\mlvl=1\) (resp \(\mlvl=2\)).

Such a configuration can be described by densities (lower half of
the figure) for the rapidities \(\us^{\rlb[][\strb][\nrt]}\). 
For instance, the density \(\rho^{\rlb[{\strb[2][2]}]}\) stands for the
frequency at which the pattern of two red circles on top of each
other is repeated.
\end{minipage}}\end{figure}

Due to the Bethe equation \eqref{eq:BethePCM}, we therefore expect that in
this limit we have
\(\frac{\fQ[\us^{\rlb[][\mlvl][\nrt]}+\bi][\mlvl]}{\fQ[\us^{\rlb[][\mlvl][\nrt]}-\bi][\mlvl]}
\to \infty\), which means that \(\fQ[\us^{\rlb[][\mlvl][\nrt]}-\bi][\mlvl]\to 0\),
{\idest} that there exists\footnote{
To motivate the existence of a Bethe root with rapidity
\(\us^{\rlb[][\mlvl][\lrt]}=\us^{\rlb[][\mlvl][\nrt]}-\bi\), we assume that the
convergence \(\fQ[\us^{\rlb[][\mlvl][\nrt]}-\bi][\mlvl]\to 0\) does not
reduce to the same argument as in footnote \ref{fn:1}.

This can be motivated for instance by saying that the degree of \(\fQ\) is
smaller that the degree of \(\fQ[][\mlvl-1]\) (resp \(\fQ[][\mlvl+1]\)) if
\(\mlvl\) is positive (resp negative), as it was seen in section \ref{sec:degree-toprs}.
} {\another} Bethe root
\(\us^{\rlb[][\mlvl][\lrt]}\) such that
\(\us^{\rlb[][\mlvl][\lrt]}=\us^{\rlb[][\mlvl][\nrt]}-\bi\). If
\(\us^{\rlb[][\mlvl][\nrt]}-\bi\) also has a positive imaginary part, then we
can apply the same argument to deduce that there is {\another} root with rapidity \(\us^{\rlb[][\mlvl][\nrt]}-2 \bi\).
Iteratively, we deduce that if a
Bethe root does not lie on the real axis, then it belongs to a set of
roots separated by a distance \(\bi\).
As we assume that the
configuration of roots is symmetric with respect to complex-conjugacy,
these sets of roots have the form
\begin{gather}
  \label{eq:BoundString}
  \us^{\rlb[][{\strb[\mlvl][\kk]}][\nrt]}_a=\us^{\rlb[][{\strb[\mlvl][\kk]}][\nrt]}+ \bi
  \left(a-\frac{\kk+1} 2\right)\,
  \qquad \where a \in \ninter 1 \kk\,,
\end{gather}
where \(\kk\) is the number of roots in this ``string''. This
{\lbg} of the roots is illustrated in figure \ref{fig:densstring},
which shows a simplified configuration of Bethe roots where only two
types of strings are present.

Here \(\us^{\rlb[][{\strb[\mlvl][\kk]}][\nrt]}\in\bR\) is the rapidity of ``the
center of the string'', and the roots
\(\us^{\rlb[][{\strb[\mlvl][\kk]}][\nrt]}_a\) belonging to this ``string'' stand
above and {\below} it on the complex plane. This allows to formally
introduce a new type of particles {\lbd} by \(\strb[\mlvl][\kk]\), and
which corresponds to the strings of \(\kk\) elementary particles of type \(\mlvl\).

 This means that
when \(\mlvl\neq 0\), the polynomial
\(\fQ[][\mlvl]\) can be written in terms of the real numbers
\(\us^{\rlb[][{\strb[\mlvl][\kk]}][\nrt]}\), which correspond to the centers of
the strings:
\begin{align}
  \label{eq:StringHyp}
  \fQ[][\mlvl]=&\prod_{\kk=1}^\infty \prod_{\nrt=1}^{\dg[{\strb[\mlvl][\kk]}]}
  \prod_{a=1}^{\kk} \left(\us-\us^{\rlb[][{\strb[\mlvl][\kk]}][\nrt]}-\bi
    \left(a-\frac{\kk+1} 2\right)\right)\,,
\end{align}
where we have denoted by \(\dg[{\strb[\mlvl][\kk]}]\) the number of such sets of
roots (of \(\fQ[][\mlvl]\)) which have size \(\kk\).

For the polynomial \(\fQ[][0]\)
({\idest} for massive particles) there also exist
bound-states, discussed in the previous section, and which give rise
to the spectrum \eqref{eq:PCFmasses}. Hence for particles of type
\(\mlvl=0\), we also 
write \eqref{eq:StringHyp}, with the important difference that
\(\kk\leq {\Np}-1\), {\idest} that these bound states {\cannnot} contain more that
\({\Np}-1\) roots.

The relation \eqref{eq:StringHyp} is the so-called ``string
hypothesis'', and it will guide us to find the configurations of roots
contributing to the free energy of the finite-temperature {\PCF}.

Using this ``string hypothesis'', 
we should write
the Bethe equations
(\ref{eq:BethePCM},\ref{eq:BethePCM2}) for roots \(\us^{\rlb[][\mlvl][\nrt]}\)
which belong to a given string ({\idest}
\(\us^{\rlb[][\mlvl][\nrt]}=\us^{\rlb[][{\strb[\mlvl][\kk]}][\nrt]}_a\)). In fact, the equation
\eqref{eq:BethePCM} is not always completely well defined because the
numerator contains \(\fQ[\us^{\rlb[][\mlvl][\nrt]}+\bi][\mlvl]\) which is zero if
\(a<\kk\) (because \eqref{eq:BoundString} ensures that \(\fQ[][\mlvl]\)
has {\another} zero at \(\us^{\rlb[][\mlvl][\nrt]}+\bi\)), and (by the same
argument) the denominator is zero if \(a>1\). Therefore, we should write
\eqref{eq:BethePCM} for each \(a\in\ninter 1 \kk\) and multiply the
resulting equations.

The most concise way to do this is with the notations of
(\ref{eq:PCFBethepDef},\ref{eq:PCFBetheSDef}). With these notations,
one gets the following Bethe equation on the ``strings'':
\begin{gather}
  \label{eq:BetheString}
  \fbox{
    \ensuremath{\displaystyle{
        \begin{aligned}
          \lefteqn{\forall \mlvl \in \ninter {1-{\Np}} {{\Np}-1},\qquad
            \forall \kk \geq 1,\qquad
            \forall \nrt \in \ninter 1 {\dg[\strb]},
          }\qquad\qquad\\
          &
          e^{\bi~ \lcds~ p^{\rlb[\strb]}\left(\us^{\rlb[][\strb][\nrt]}\right)}
          = 
          \prod_{\substack{
              \jlvl\in \ninter {1-{\Np}} {{\Np}-1}\\
              \lL\geq 0\\
              \irt \in \ninter 1 {\dg[{\strb[\jlvl][\lL]}]}\\
              (\strb,\nrt)\neq(\strb[\jlvl][\lL],\irt)
            }} \Sscal^{\rlb[{\strb[]}],\rlb[{\strb[\jlvl][\lL]}]}(\us^{\rlb[][\strb][\nrt]}-\us^{\rlb[][{\strb[\jlvl][\lL]}][\irt]})
        \end{aligned}
      }}
  }\,,
  \\\where
  \Sscal^{\rlb[{\strb[]}],\rlb[{\strb[\jlvl][\lL]}]}(\us-\vs)\equiv
  \prod_{s=-\frac{\kk-1}2}^{\frac{\kk-1}2}\prod_{s'=-\frac{\lL-1}2}^{\frac{\lL-1}2}
  \Sscal^{\rlb[\mlvl],\rlb[\jlvl]}(\us-\vs+\bi(s-s'))\,,\\
  \And
  p^{\rlb[\strb]}\left(\us\right)\equiv\sum_{s=-\frac{\kk-1}2}^{\frac{\kk-1}2}p^{\rlb[\mlvl]}(\us+\bi
  s)
  =\left\{
    \begin{aligned}
      &0&&\If \mlvl\neq 0\\
      \frac{\sin\frac{\pi \kk}{{\Np}}}{\sin\frac{\pi }{{\Np}}}& \sht[\us]&~~&\If \mlvl=0
    \end{aligned}
  \right.\,.
\end{gather}
In \eqref{eq:BetheString}, it is implicit that if \(\mlvl=0\), then
\(\kk\) should be chosen as 
\(\kk\leq {\Np}-1\).  Moreover,  we see that the
product
in the {\rhs} of \eqref{eq:BetheString}
runs over all the 
\((\strb[\jlvl][\lL],\irt)\) 
such that \(\jlvl\neq \mlvl\) or
\(\lL\neq\kk\) or  \(\irt\neq \nrt\).

After taking the log, the equation \eqref{eq:BetheString} reads
\begin{gather}
  {
    \ensuremath{\displaystyle{
        \begin{aligned}
          \lefteqn{\forall \mlvl \in \ninter {1-{\Np}} {{\Np}-1},\qquad
            \forall \kk \geq 1,\qquad
            \forall \nrt \in \ninter 1 {\dg[\strb]},
          }%
          \\
          &
          \dkp %
          = 
          - \lcds~ p^{\rlb[\strb]}\left(\us^{\rlb[][\strb][\nrt]}\right)
          +
          \sum_{\substack{
              \jlvl\in \ninter {1-\Kr} {\Kr-1}\\
              \lL\geq 0\\
              \irt \in \ninter 1 {\dg[{\strb[\jlvl][\lL]}]}\\
              (\strb,\nrt)\neq(\strb[\jlvl][\lL],\irt)
            }} {F}^{\rlb[{\strb[]}],\rlb[{\strb[\jlvl][\lL]}]}(\us^{\rlb[][\strb][\nrt]}-\us^{\rlb[][{\strb[\jlvl][\lL]}][\irt]})
        \end{aligned}
      }}
  }\,,
  \\\where
  {F}^{\rlb[{\strb[]}],\rlb[{\strb[\jlvl][\lL]}]}(\us-\vs)\equiv
  \frac 1 \bi \mathrm{log}\left(
    \Sscal^{\rlb[\strb],\rlb[{\strb[\jlvl][\lL]}]}(\us-\vs)\right)\,.
\end{gather}
Here \(\dkp\) denotes an arbitrary multiple of \(2 \pi\).

This means that the  rapidities \(\us^{\rlb[][\strb][\nrt]}\) 
of the centers of the strings
\eqref{eq:BoundString} are real numbers,  solution
of an equation \(\dkp=f(\us^{\rlb[][\strb][\nrt]})\), where the function \(f\) is a
fixed real function (if the rapidities of the other roots are
fixed). All the rapidities \(\us^{\rlb[][\strb][\nrt]}\) belong to the set
\(\left\{\us\middle|\frac{f(\us)}{2\pi}\in \bZ\right\}\), 
but there can exist some values of \(\us\) such that \(\dkp=f(\us)\) which
are not the rapidities \(\us^{\rlb[][\strb][\nrt]}\) of 
strings 
\eqref{eq:BoundString} of roots. Every such value of \(\us\) is called
 a hole.

 In finite temperature there are many particles (and maybe also
many holes), and we can introduce a density \(\rho\) of particles and a
density \(\bar \rho\) of holes.
 They have to obey \(\rho+\bar \rho
= |\partial_\us \frac f{2\pi}|\). This relation says that the number
\(\int_{\us}^{\us'} (\rho+\bar \rho) \mathrm{d}\us\) of holes
and particles 
in the interval \([\us,\us']\) is equal to the number
\(|\frac{f(\us')-f(\us)}{2\pi}|\), which is the number of times
that \(\frac{f}{2\pi}\) takes an integer value in the interval
\([\us,\us']\) (if \(f\) is monotonous in the interval \([\us,\us']\)).

Therefore, we introduce densities of holes and particles
for each type \(\strb\) of particles
 (more precisely they are 
  the densities \(\rho^{\rlb[\strb]}\) of the centers 
\(\us^{\rlb[][{\strb[\mlvl][\kk]}][\nrt]}\in\bR\) of the strings,
as in figure \ref{fig:densstring}).
As explained above, they have to obey
the relation
\begin{gather}
  \label{eq:BetheRho}
  \rho^{\rlb[\strb]}+\bar \rho^{\rlb[\strb]} = \left|\vphantom{\int}\right. \frac \LF {2 \pi} \partial_\us p^{\rlb[\strb]}\left(\us\right)-\sum_{\substack{
      \jlvl\in \ninter {1-\Kr} {\Kr-1}\\
      \lL\geq 0
    }} \int_{\vs\in\bR}\Ker^{\rlb[{\strb[]}],\rlb[{\strb[\jlvl][\lL]}]}(\us-\vs)
  \rho^{\rlb[{\strb[\jlvl][\lL]}]}(\vs) \mathrm{d}\vs \left. \vphantom{\int} \right|\,\\
  \where \Ker^{\rlb[{\strb[]}],\rlb[{\strb[\jlvl][\lL]}]}(\us)\equiv
  \frac{1}{2\bi\pi} \partial_\us \mathrm{log}\left(\Sscal^{\rlb[\strb],\rlb[{\strb[\jlvl][\lL]}]}(\us-\vs)\right)\,.
\end{gather}

In the {\rhs}, the integral \(\int_{\vs\in\bR}\Ker^{\rlb[{\strb[]}],\rlb[{\strb[\jlvl][\lL]}]}(\us-\vs)
\rho^{\rlb[{\strb[\jlvl][\lL]}]}(\vs)\mathrm{d}\vs\) can also be written as
\(\Ker^{\rlb[{\strb[]}],\rlb[{\strb[\jlvl][\lL]}]} \st
\rho^{\rlb[{\strb[\jlvl][\lL]}]}\) where
\begin{align}
\label{eq:DefConv}
  (f_1 \st f_2)(\us)\equiv&\int_{\vs\in\bR}f_1(\us-\vs) f_2(\vs) \mathrm{d}\vs
\end{align}
denotes the usual convolution.
\index{* (convolution)@\ensuremath{\st}  (convolution)}

In general the sign inside the absolute value (in \eqref{eq:BetheRho})
is not completely obvious. But it is at least clear that when \(\us\) is
large enough this sign is positive, (because \(\frac \LF {2
  \pi} \partial_\us p^{\rlb[\strb]}\left(\us\right)\) is very large).
We will actually assume that the densities 
\(\rho\) and \(\bar \rho\) are analytic, and this imposes that 
this sign is always plus. 
Therefore, we will actually drop the absolute value in \eqref{eq:BetheRho}.

Let us now introduce a new {\lbg} for the densities ({\idest} for the
different types of particles):
\begin{align}
  \rho^{a,s}\equiv&\rho^{\rlb[{\strb[a][s]}]}&\If&s>0\\
  \rho^{a,0}\equiv&\rho^{\rlb[{\strb[0][a]}]}&\If&s=0\\
  \rho^{a,s}\equiv&\rho^{\rlb[{\strb[-a][-s]}]}&\If&s<0\,.
\end{align} 
In this notation, the densities are {\lbd} by two integers \((a,s)\in
\ninter 1 {{\Np}-1} \times \bZ\). We will use the same rule to label the
densities \(\bar \rho\) and the kernels \(\Ker\) (for instance if \(s>0\)
and \(s'>0\), then \(\bar \rho^{a,s}\equiv \bar \rho^{\rlb[{\strb[a][s]}]}\)
and \(\Ker^{a,s,a',s'}\equiv
  \Ker^{\rlb[{\strb[a][s]}],\rlb[{\strb[a'][s']}]}\)). This choice of {\lbg} may
not seam natural, but we will see that in these new variables, the
equations on the densities will turn out to be quite simple and
universal.

\subsubsection{Minimization of the free energy }
\label{sec:minim-free-energy}

In order to compute the {\ppath} integral \eqref{eq:zpathint}, we should
find the configuration of roots having the lowest free energy in the
mirror model. 
This free energy should contain two terms: one term \(%
E=
\int_{\us\in\bR}
\sum_{a=1}^{{\Np}-1} \frac{\sin\frac{\pi a}{{\Np}}}{\sin\frac{\pi }{{\Np}}}
\cht[\us] \rho^{(a,0)}(\us) \mathrm{d}\us
\)
corresponding to the energy, and an entropic term.

The entropic term \cite{Zamolodchikov:1989cf}
corresponds to the fact that many configurations are described by the
same densities of roots. This counterintuitive fact arises because
the densities only contain informations about the number \({\NN}_h=\bar \rho
(\us)\delta_\us\) of holes and the number \({\NN}_r=\rho (\us)\delta_\us\)
of roots in a given interval \([\us,\us+\delta_\us]\). But the solutions
of \(\dkp=f(\us^{\rlb[][\strb][\nrt]})\) can be reshuffled between roots and
holes in 
\begin{equation*}
  \frac{\left({\NN}_r+{\NN}_h\right)!}{\left({\NN}_r\right)!\left({\NN}_h\right)!}
\end{equation*}
different ways, without consequence on the densities \(\rho\) and \(\bar
\rho\).
As the entropy is the logarithm of the number of configurations,
one can show \cite{Zamolodchikov:1989cf} that
the entropy is equal to
\begin{equation*}
  \int_{\us\in\bR} \left(\rho+\bar \rho\right)\mathrm{log}\left(\rho+\bar
    \rho\right) - \rho~\mathrm{log}\left(
    \rho\right) - \bar \rho~\mathrm{log}\left(\bar
    \rho\right)~\mathrm{d}\us\,.
\end{equation*}

Therefore, the free energy 
is given by
\begin{multline}
  \label{eq:freeEn}
  f(\LF)=\Min_{\rho^{a,s},\bar \rho^{a,s}} \int_{\us\in\bR}\left(
    \sum_{a=1}^{{\Np}-1} \frac{\sin\frac{\pi a}{{\Np}}}{\sin\frac{\pi }{{\Np}}}\right.
  \cht[\us] \rho^{a,0}(\us) \\
  -\frac 1 \beta \sum_{\substack{a\in \ninter 1 {{\Np}-1}\\s\in\bZ}}
  \rho^{a,s} \mathrm{log}\left(1+\frac{\bar
      \rho^{a,s}}{\rho^{a,s}}\right) + \bar \rho^{a,s}
  \mathrm{log}\left(1+\frac{\rho^{a,s}}{ \bar
      \rho^{a,s}}\right)\left.\vphantom{\sum_{a=1}^{{\Np}-1}}\right)\mathrm{d}\us\,,
\end{multline}
where \(\beta=\LF\).

The minimum in \eqref{eq:freeEn} is a minimum among all the possible
densities which satisfy the Bethe equation \eqref{eq:BetheRho}. That
means that if we vary \(\rho^{a,s}\) by an amount \(\delta\rho^{a,s}\), then \(\bar
\rho^{a,s}\) has to vary by the amount 
\begin{equation}
  \delta\bar\rho^{a,s} =-\delta \rho^{a,s} -\sum_{\substack{
      a'\in\ninter 1 {{\Np}-1}\\
      s'\in\bZ }} 
  \Ker^{a,s,a',s'}\st 
  \delta \rho^{a',s'}
  \,.
\end{equation}

With this constraint, the minimization condition reads
\begin{align}
  0=& \LF \frac{\delta f}{\delta \rho}=\LF
  \sum_{a=1}^{{\Np}-1} \frac{\sin\frac{\pi a}{{\Np}}}{\sin\frac{\pi }{{\Np}}}
  \cht[\us]\delta_{s,0} 
  - \mathrm{log}\left(1+\frac{\bar
      \rho^{a,s}}{\rho^{a,s}}\right)
  - \mathrm{log}\left(1+\frac{\rho^{a,s}}{ \bar
      \rho^{a,s}}\right)  \frac{\delta \bar \rho}{\delta \rho}\\
  =&\LF
  \sum_{a=1}^{{\Np}-1} \frac{\sin\frac{\pi a}{{\Np}}}{\sin\frac{\pi }{{\Np}}}
  \cht[\us]\delta_{s,0}
  -\mathrm{log}\left(\frac{\bar
      \rho^{a,s}}{\rho^{a,s}}\right) +
  \sum_{\substack{
      a'\in\ninter 1 {{\Np}-1}\\
      s'\in\bZ }} 
  \Ker^{a,s,a',s'}\st \mathrm{log}\left(1+\frac{\rho^{a,s}}{ \bar
      \rho^{a,s}}\right)
  \label{eq:PCMTBArrb}
\end{align} 

This equation \eqref{eq:PCMTBArrb} is the ``TBA equation'' , which is
an equation on the ratio \(\frac{\rho^{a,s}}{ \bar
  \rho^{a,s}}\). 

We will see that if we denote this ratio by \(\Y\), then a simpler
``{\Ysys} equation'' arises.

\subsubsection{TBA equations and {\Ysys} equation}
\label{sec:tba-equations-ysys}

The TBA equation \eqref{eq:PCMTBArrb} is an equation on two types of
densities \(\rho\) and \(\bar \rho\). It can be rewritten in terms of the
quantities \index{Y-functions@{\Yfs}}
\Pv{\begin{subequations}
    \begin{empheq}[left={\Y=\empheqlbrace}]{align}
      \frac{\rho^{a,s}}{\bar \rho^{a,s}}&&\If&s=0\,,
      \\
      \frac{\bar \rho^{a,s}}{\rho^{a,s}}&&\If&s\neq 0\,.
    \end{empheq}
  \end{subequations}}

For instance, when \(s\neq 0\), the TBA equation \eqref{eq:PCMTBArrb} is
rewritten as
\begin{align}
  \label{eq:TBAeqRewriteY}
  \mathrm{log}\left(\Y\right)=&
  \sum_{\substack{
      a'\in\ninter 1 {{\Np}-1}\\
      s'\in\bZ }} 
  \Ker^{a,s,a',s'}\st \mathrm{log}\left(1+
    \left(\Y[a'][s']\right)^{\pm 1}
  \right)&\If s\neq&0\,,
\end{align}
where the sign \(\pm 1\) is equal to \(1\) if \(s'=0\) and (resp \(-1\) if
\(s'\neq 0\)).

Then one can compute the quantity
\(\mathrm{log}\left(\frac{\Y[][]\left(\us+\frac\bi 2\right)\Y[][]\left(\us-\frac\bi 2\right)}{\Y[][s+1]\Y[][s-1]}\right)\)
:
\begin{gather}
  \mathrm{log}\left( \frac{\Y[][]\left(\us+\frac\bi 2\right)\Y[][]\left(\us-\frac\bi 2\right)}{\Y[][s+1]\Y[][s-1]} \right)=
  \sum_{\substack{
      a'\in\ninter 1 {{\Np}-1}\\
      s'\in\bZ }} 
  \Ker^{a,s,a',s'}
  \st \mathrm{log}\left(1+
    \left(\Y[a'][s']\right)^{\pm 1}
  \right)\,,
\end{gather}
\begin{multline}
  \where
  \tilde\Ker^{a,s,a',s'}(\us)\equiv 
  \Ker^{a,s,a',s'}\left(\us+\frac\bi 2\right) + 
  \Ker^{a,s,a',s'}\left(\us-\frac\bi 2\right) \\
  - \Ker^{a,s+1,a',s'}\left(\us\right) 
  -  \Ker^{a,s-1,a',s'}\left(\us\right)\,.
\end{multline}

But one can show 
(see equation (30) in \cite{Gromov:2009bc}) that this combination
\(\tilde \Ker\) is remarkably simple:
\begin{align}
  \If a
  =a'\And |s|>1 \Then
  &&&
  \tilde \Ker^{a,s,a',s'} (\us) =
  \delta_{s,s'+1} \delta({\us}) + \delta_{s,s'-1} \delta({\us})
  \,\\
  \If a
  =a'\pm 1 \And |s|>1 \Then
  &&&
  \tilde \Ker^{a,s,a',s'} (\us) =
  { 
    -\delta_{s,s'} \delta({\us})}\,
\end{align}
whereas if \(|a-a'|>1\), then \(\Ker^{a,s,a',s'}=0\) for all \(s,s'\).

From there, we get that if \(|s|>1\), then
\begin{align}
  \label{eq:tbaYsysEq}
  \framedline{  \frac{\Y[][]\left(\us+\frac\bi 2\right)
      \Y[][]\left(\us-\frac\bi 2\right)} {\Y[][s+1]\Y[][s-1]}=} {
    \frac{1+1/\Y[][s+1]}{1+1/\Y[a+1]}\frac{1+1/\Y[][s-1]}{1+1/\Y[a-1]}}\,\,. 
\end{align}
This equation \eqref{eq:tbaYsysEq} is called the {\YsE}.

In this equation, the factor \({1+1/\Y[a-1]}\) (resp \({1+1/\Y[a+1]}\))
should be absent if \(a=1\) (resp \(a=\Np-1\)), because the sum over
\((a',s')\) (in \eqref{eq:TBAeqRewriteY}) does not contain \(s'=0\) or
\(s'=\Np\). The condition that this term is absent can also be
written as a boundary condition 
\begin{gather}
  \fdisp{  \Y[0]=\Y[\Np]=\infty}\,.
\end{gather}

When \(|s|\leq 1\), the derivation of the {\YsE}
\eqref{eq:tbaYsysEq} from the TBA equation \eqref{eq:PCMTBArrb} is
more technical. The case \(\Np=2\) is  performed in
\cite{2009JHEP...12..060G}, where it is also conjectured that it holds
as well for \(\Np>2\).

It is a bit more technical, but the TBA equation \eqref{eq:PCMTBArrb}
also implies that \eqref{eq:tbaYsysEq} holds even when \(|s|\leq 1\) (see for
instance \cite{2009JHEP...12..060G}).

Moreover, one can show \cite{2009JHEP...12..060G} that the free energy
\(f(\LF)\) \eqref{eq:freeEn} 
of the mirror theory (or the vacuum energy \(E_0(\LF)\) of the original model)
 can be expressed in terms of these {\Yfs} as
\begin{align}
  \label{eq:EzfromY}
  \framedline{E_0(\LF)=}{f(\LF)=-\frac 1 {\Np} \sum_{a=1}^{{\Np}-1}\frac{\sin\frac{\pi
      a}{{\Np}}}{\sin\frac{\pi }{{\Np}}} \int_{\us\in\bR} \cht[\us]
  \mathrm{log}\left( 1+\Y[][0](\us)\right)\mathrm{d}\us}\,.
\end{align}

As we saw, the {\TBA} gives rise to the TBA-equation
\eqref{eq:PCMTBArrb}, which is an equation on the densities of roots
and holes for the configurations of Bethe roots which minimize the
free energy. Solving this equation allows to compute the free energy in
the mirror theory with temperature, {\idest} to completely find the vacuum
energy (of the initial model) at any finite size.
 This equation can be rewritten in terms of ratios of
densities, celled the {\Yfs} 
 (see for instance \eqref{eq:TBAeqRewriteY}), and the resulting
 equations imply the {\Ysys} equation 
\eqref{eq:tbaYsysEq}. This equation is very universal, and the same
equation describes the finite-size effects of several other {\ing}
models.

\begin{figure}
  \centering
  \includegraphics{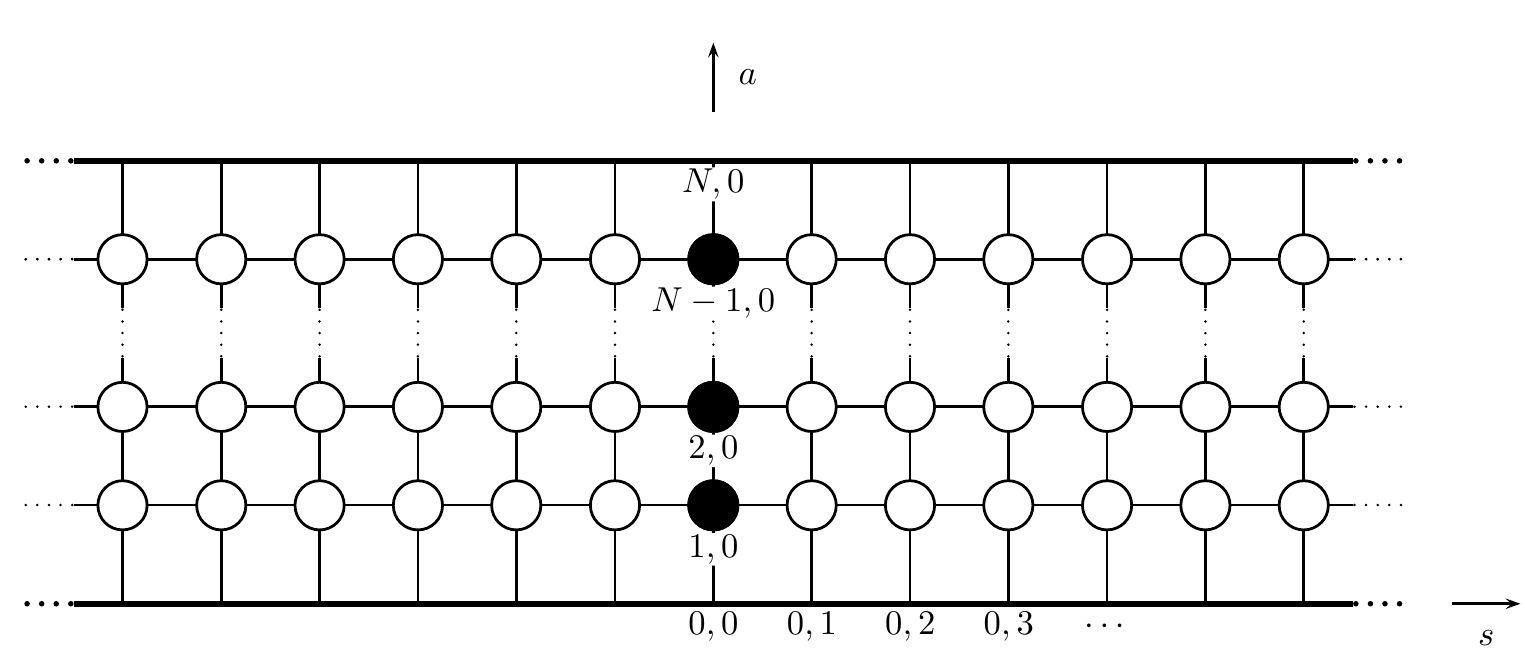}
  \caption{The \((a,s)\)-lattice for the {\PCM}}
  \label{fig:aslatt-pcm}
\end{figure}

A first point which characterizes this \(\SU{\Np}\times\SU \Np\)
{\PCM} is that  \((a,s)\), which labels the different types of particles, takes values in \(\ninter  1 {\Np-1} \times
\bZ\), and that in \eqref{eq:tbaYsysEq}, \(\Y[a-1]\) (resp \(\Y[a+1]\))
should be set to \(\infty\) if \(a=1\) (resp if \(a=\Np-1\)).
The lattice of these authorized values of \((a,s)\) is depicted in
figure \ref{fig:aslatt-pcm},
where the circles denote the values of \((a,s)\) corresponding to a
{\Yf}. The black disks in the middle stand for the functions
\(\Y[][0]\) which correspond to the massive particles.
Apart from
this property, the \(\SU{\Np}\times\SU \Np\) {\PCM} is characterized
by  its large \(\us\) behavior, namely 
\begin{align}
  \label{eq:YsysZeMo0}
  \mathrm{log}\left(\Y\right) + \LF
  \frac{\sin\frac{\pi a}{{\Np}}}{\sin\frac{\pi }{{\Np}}}
  \cht[\us]\delta_{s,0}\xrightarrow[\us \to \infty]{} c_{a,s}\,.
\end{align}
where \(c_{a,s}\) is a 
\(\us\)-independent number.  This
condition can be read from the {\TBAE} \eqref{eq:PCMTBArrb}, but it is
not a 
direct consequence of the {\YsE} \eqref{eq:tbaYsysEq}. This means that the TBA
equation \eqref{eq:PCMTBArrb} is slightly stronger than the {\Ysys} equation
\eqref{eq:tbaYsysEq}. 
Nevertheless we will see that 
it is sufficient to know the {\Ysys} equation \eqref{eq:tbaYsysEq} on
the one hand, and the 
asymptotic behavior \eqref{eq:YsysZeMo0} on the other hand.

\subsubsection{Excited states}
\label{sec:excited-states}

The construction given above allows to find the energy of the vacuum
at any finite size, but it does not apply to excited states.

What was proven for a few models, (and can be conjectured for many
other models) is that each excited state corresponds to a different
solution of the {\Ysys} equation, with different analyticity
properties (in particular regarding the existence of zeroes and poles
of the {\Yfs}) \cite{Bazhanov:1996aq,Dorey:1996re,Dorey:1998pt}. 

This means that the functions \(\Y\) are multi-valued functions of
\(\LF\), living on a specific Riemann surface. Each sheet of this Riemann
surface corresponds to one excited state, and there exists a {\ppath}
connecting this state to the vacuum (which means that the excited
states are obtained by an analytic continuation from the
vacuum). This analytic continuation preserves the {\Ysys} equation
\eqref{eq:YSysEq}, 
the asymptotic behavior \eqref{eq:YsysZeMo0},
and the form of the expression \eqref{eq:EzfromY}
of the energy. 
For this last equation \eqref{eq:EzfromY}, an ambiguity appears for
excited states because the integrand has singularities, and the
integration contour has to be specified\footnote{For the vacuum, we
  will see that the {\Yfs} are analytic (they do not have poles),
  whereas for excited states, they have poles giving rise to the
  ambiguity in the expression \eqref{eq:EzfromY}
of the energy.}.

A rigorous analysis of the analytic continuation from one sheet to
the other seems out of reach, but we will assume that 
for arbitrary excited states, 
there exists a
choice of contour such that the integral \eqref{eq:EzfromY}
gives the energy of this state. Under some natural hypotheses, we can
find this contour for several models, and in the case of the {\PCM},
that allows to go through a couple of non-trivial checks.

\section{General solution of Hirota equation}
\label{sec:solution-generale-de}

As we saw in the previous section, the {\TBA} gives rise to the
{\Ysys} equation 
\begin{align}
  \label{eq:YSysEq}
  \framedline{  \Y[][][+]   \Y[][][-]=}{\frac{1+\Y[][s+1]}{1+1/\Y[a+1]} \frac{1+\Y[][s-1]}{1+1/\Y[a-1]}}
\end{align}
where we have a set of functions \(\Y\equiv \Y(\us)\) 
of the
spectral parameter \(\us\), {\lbd} by two integers \((a,s)\).
In \eqref{eq:YSysEq}, the dependence in the spectral parameter \(\us\) is written as a
superscript:
\begin{align}
  \framedline{f^{\pm}\equiv}{f\left(\us\pm \frac \bi 2\right)}\,\,.
\end{align}
\index{00pmsup@\ensuremath{f^{\pm}\equiv f\left(\us\pm \frac \bi 2\right)}}

This {\Ysys} equation  \eqref{eq:YSysEq} is a very general equation,
which arises from the {\TBA} for a large variety of models. We will
see in this section that its general solution is exactly given by the
construction of the chapter \ref{part:qoperatorsspin}. This remark
will be the key point in order to recast these {\TBAE} into a FiNLIE.
It is motivated by noticing 
that under the change of variables
\begin{align}
  \label{eq:VarsYT}
  \framedline{  \Y=}{\frac{\T[][s+1]}{\T[a+1]}\frac{\T[][s-1]}{\T[a-1]}}\,\,,
\end{align}
\index{T-functions@{\Tfs}}
the {\Ysys} equation is equivalent to the following Hirota equation
\index{Hirota equation}
\begin{align}
  \label{eq:YHirota}
  \framedline{\T[][][+]\T[][][-]=}{\T[a+1]\T[a-1]+\T[][s+1]\T[][s-1]}\,\,,
\end{align}
where the 
\(\T\) are 
functions of the spectral
parameter \(\us\), {\lbd} by two integers \((a,s)\). 
The sense in which they are equivalent, as well the proofs of this
equivalence, will be given in \ref{sec:equiv-hirota-equat}, while the
present section only announces the key informations.

Up to the change of variables 
\begin{align}
  \label{eq:TchangeShift}
  {\T}(\us)=&\rTf[-\bi \us+\frac{a-s}{2}-\frac{\Np}{4}]
\end{align}
this equation is identical to the Hirota equation \eqref{eq:Hirota}
which we derived for {\csds}. For models with a known {\ing}
lattice regularization, this result is not surprising because the
field theory can be written as the limit of a {\cds}. But the
{\TBA} tells us that 
this Hirota equation even describes
models without any known {\ing}
lattice regularization, and allows to study them very efficiently.

As shown in chapter \ref{part:qoperatorsspin}, this Hirota equation is
equivalent (for typical solutions on the half plane \(a\geq 0\)) to the CBR determinant
formula\footnote{
The expression \eqref{eq:YCBR} for the CBR formula is identical to the
expression  \eqref{eq:CBR} obtained in chapter
\ref{part:qoperatorsspin}, up to the change of variables
\eqref{eq:TchangeShift}.
}
\begin{gather}
  \label{eq:YCBR}
  \T 
  =\frac{
    \Det{
      \T[1][s+\ii-\jj]\left(
        \us+\frac \bi 2 \left(a+1-\ii-\jj\right)
      \right)
    }{1\leq\ii,\jj\leq
      a}}{\prod_{{\coordk}=1}^{a-1}
    \T[0][0]\left(\us-\bi\frac s 2+\frac \bi 2 (a-2\coordk)\right)}
  \,.
\end{gather}

In this section we will write the general solution of
the Hirota equation \eqref{eq:YHirota} for several boundary
conditions, hence the solution 
of the {\YsE} \eqref{eq:YSysEq}. 
Before we delve into this, let us note the nature of the relation
between \eqref{eq:YHirota} and \eqref{eq:YSysEq}:
one can note that \(\Y\), written as a function of \(\T\) (see
\eqref{eq:VarsYT}), is invariant under the ``gauge'' transformation 
\begin{gather}
  \label{eq:GaugeFreedom}
  \fbox{\ensuremath{\displaystyle {   \T \rightsquigarrow \ga[a+s][1] \ga[a-s][2]
        \ga[-a+s][3] \ga[-a-s][4] \T}}}\,,\\
  \where \fbox{\ensuremath{\displaystyle { {f}^{[+\nn]}\equiv {f}\left(\us+\nn \frac \bi 2\right)}}}\,,
\end{gather}
\index{00pmsup@\ensuremath{f^{\pm}\equiv f\left(\us\pm \frac \bi 2\right)}!sh@\ensuremath{{f}^{[+\nn]}\equiv {f}\left(\us+\nn \frac \bi 2\right)}}
for four arbitrary  
functions 
\(\gan[][1]\),  \(\gan[][2]\),
\(\gan[][3]\) and 
\(\gan[][4]\) of the spectral parameter \(\us\). These functions will be
called ``gauge-functions''.
\index{gauge transformation}
The Hirota equation \eqref{eq:YHirota} is also invariant under this
gauge transformation.
We will see that every solution of the {\Ysys} equation corresponds to
a set of solutions of the Hirota equation, which are obtained from
each other by gauge transformations. The complete proof of this
statement is more technical than one would naively expect, and it is
given in subsection \ref{sec:equiv-hirota-equat}. Before that, we will
comment on the labels \((a,s)\) in the Hirota and {\Ysys} equations
(\ref{eq:YSysEq}, ~ \ref{eq:VarsYT},~\ref{eq:YHirota}).

\subsection{Examples of \texorpdfstring{$(a,s)$}{(a,s)} lattice}
\label{sec:a-s-lattice}

In section \ref{sec:example-princ-chir} we have derived that, for
the {\PCF}, the indices \(a\) and \(s\) {\lbg} the various {\Yfs}
belong to the set \(\ninter {1}{{\Np}-1}\times \bZ\). We also saw that
the {\Ysys} equation is satisfied at \(a=0\) and \(a={\Np}\) if we
define \(\Y[][0]=\Y[][{\Np}]=\infty\).

If we try to write that in terms of {\Tfs}, we see that \(\T\) needs
to be well defined when \((a,s)\in \ninter 0 {\Np} \times \bZ\), in
order to be able to compute the ratio \({
  \Y=}{\frac{\T[][s+1]}{\T[a+1]}\frac{\T[][s-1]}{\T[a-1]}}\). Moreover
the requirement \(\Y[][0]=\Y[][{\Np}]=\infty\) translates into
\(\T[-1]=0\) and \(\T[{\Np}+1]=0\). This is very natural and
corresponds very well to the analysis of chapter
\ref{part:qoperatorsspin}, where we saw that the {\Tfs} are zero
outside a given domain of the (a,s)-plane. We also saw that this
domain depends of the symmetry group of the model (see the 
\figpref{fig:BosHook} for {\GL \Kr} and the \figpref{fig:FatHook} for 
{\GL{\Kr\ensuremath{|}\Mr}).

  \subsubsection{The lattice $\St({\Np})$ of the {\PCM}}
  \label{sec:lattice-skn-pcm}

  Let us denote by \(\Yl{\St({\Np})}\) and \(\Tl{\St({\Np})}\) the lattices
  \Pv{\begin{empheq}[box=\fbox]{gather}
    {\Yl{\St({\Np})}\equiv\left\{(a,s)\middle| a\in \ninter 1
        {\Np-1} \And s\in \bZ\right\}}\\
    {\Tl{\St({\Np})}\equiv\left\{(a,s)\middle| a\in \ninter 0
        {\Np} \And s\in \bZ\right\}}\,.
  \end{empheq}}
  
  We will say that \(\Y\) is a solution of the {\YsE} on \(\St({\Np})\) if
  \(\Y\) is defined for all \((a,s)\in \Yl{\St({\Np})}\), and if it obeys
  the equation \eqref{eq:YSysEq} for all \((a,s)\in \Yl{\St({\Np})}\),
  with the prescription \(\Y[0]=\infty=\Y[{\Np}]\).
  This prescription can be summarized by rewriting the {\YsE} as 
  \begin{align}
    \label{eq:YSeBndInf}
    \frac{  \Y[][][+] \Y[][][-]}{
      \left(\Y[a+1]\right)^{1-\delta_{a,{\Np}-1}}
      \left(\Y[a-1]\right)^{1-\delta_{a,1}}}
    =& 
    \frac{1+\Y[][s+1]}{\left(1+\Y[a+1]\right)^{1-\delta_{a,{\Np}-1}}}
    \frac{1+\Y[][s-1]}{\left(1+\Y[a-1]\right)^{1-\delta_{a,1}}}\,.
  \end{align}
As we have seen, the {\Yfs} of the {\PCM} are an example of functions
which obey the {\YsE} on \(\St({\Np})\).

  We will also say that \(\T\) is a solution of the Hirota equation on
  \(\St({\Np})\) if \(\T\) is defined for all \((a,s)\in \Tl{\St({\Np})}\), and if
  it obeys the equation \eqref{eq:YHirota} for all \((a,s)\in
  \bZ\times\bZ\), with the prescription \(\T=0\) for all \((a,s)\not\in
  \Tl{\St({\Np})}\).

  With these definitions, we will see (in section
  \ref{sec:equiv-hirota-equat}) in what sense the {\YsE} on \(\St({\Np})\)
  is equivalent to the Hirota equation on \(\St({\Np})\).

  We can see a {\rp} of these lattices in 
  \figpref{fig:aslatt-pcm}: the set \(\Yl{\St({\Np})}\) of authorized values
  (of \((a,s)\)) for the {\Yfs} is denoted by circles (and disks),
  whereas the set \(\Tl{\St({\Np})}\) contains all the nodes of the grid
  in the background.

  This lattice corresponds to the symmetry group \(\SU\Np\times
  \SU\Np\) of the {\PCM}.

  \subsubsection{The lattices $\Sh({\Np})$ and $\HK(\Kr,\Mr)$ of {\GL
      \Kr} and {\GLKM} {\csds}}
  \label{sec:lattice-shn-glkr}

  For {\GL \Kr} and {\GLKM} {\csds}, we have seen in the
  previous chapter that the {\Tfs} live on the ``{\fat}-{\hook}'' lattice of the
  figures \ref{fig:BosHook} and \ref{fig:FatHook}. We will therefore
  denote by \(\Tl{\HK(\Kr,\Mr)}\) the lattice
  \begin{gather}
    \Tl{\HK(\Kr,\Mr)}\equiv\left\{(a,s)\in\bN\times\bZ \middle|
      \begin{array}{ll}
        s\geq 0 &\And 0\leq a\leq \Kr\\
        &\Or\\
        a\geq 0 &\And 0\leq s\leq \Mr\\
        &\Or\\
        a=0
      \end{array}
    \right\}\,.
  \end{gather}

  In the case of {\GL \Kr} (or {\SU \Np}), we can choose to emphasize
  the inclusion \linebreak \({{\SU \Np}\subset \SU\Np\times \SU\Np}\)
  by choosing a lattice included in {\St(\Np)}. Then we can define a
  lattice
  \begin{gather}
    \Tl{\Sh(\Np)}\equiv\left\{(a,s)\in\bN\times\bZ \middle|
      \begin{array}{ll}
        s\geq 0 &\And 0\leq a\leq \Np\\
        &\Or\\
        a= 0 &\Or a=\Np
      \end{array}
    \right\}\,.
  \end{gather}

  We will say that \(\T\) is a solution of the Hirota equation on
  \(\HK(\Kr,\Mr)\) (resp \(\Sh(\Np)\)) if \(\T\) is defined for all
  \((a,s)\in \Tl{\HK(\Kr,\Mr)}\) (resp \(\Tl{\Sh(\Np)}\)), and if it
  obeys the equation \eqref{eq:YHirota} for all \((a,s)\in
  \bZ\times\bZ\), with the prescription \(\T=0\) for all \((a,s)\not\in
  \Tl{\HK(\Kr,\Mr)}\) (resp \(\Tl{\Sh(\Np)}\)).

  The two lattices
  \(\Sh(\Kr)\) and \(\HK(\Kr,0)\) are equivalent in the
following sense: 
 if
  \(\T\) is a solution of Hirota equation at every \((a,s)\in\ninter 0
  {\Kr} \times \bN\) such that \(\T[-1]=0\), \(\T[][-1]=0\) when \(a \in\ninter 1
  {\Kr-1}\), and \(\T[\Kr+1][s]=0\) when \(s>1\), 
then \(\T\)
can be continued into a
  solution of Hirota on \(\Sh(\Kr)\) or into a solution of Hirota on
  \(\HK(\Kr,0)\) (both are possible),
by defining \(\T[0]\) and \(\T[\Np]\) (or \(\T[][0]\)) recursively.
 For instance, for \(\Sh(\Kr)\),
  this is done by a simple recurrence defining
  \(\T[\Kr][s]=\frac{\T[\Kr][s+1][+]\T[\Kr][s+1][-]}{\T[\Kr][s+2]}\)
  for \(s\leq -1\).
  These two lattices are depicted in figure
  \ref{fig:ShvsHK}.

  \begin{figure}
    \centering
    \subfigure[The lattice \(\HK(\Kr,0)\)]
    {\includegraphics{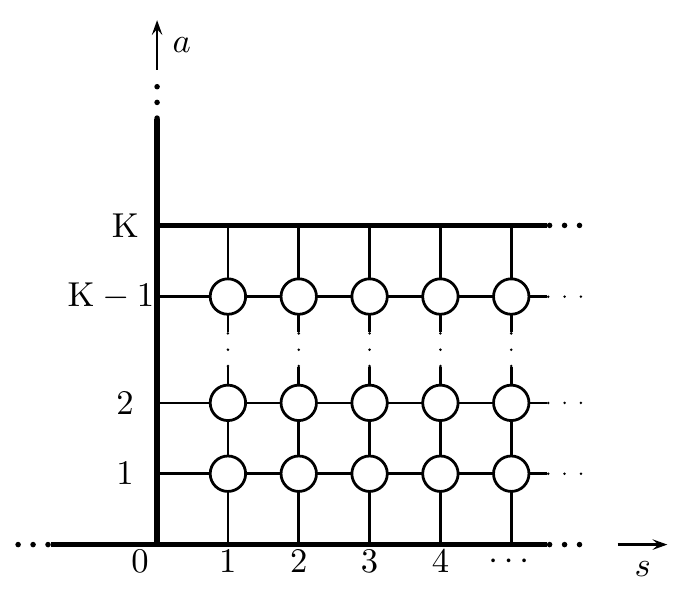}}
    \quad
    \subfigure[The lattice \(\Sh(\Kr)\)]
    {\label{fig:LatSh} \includegraphics{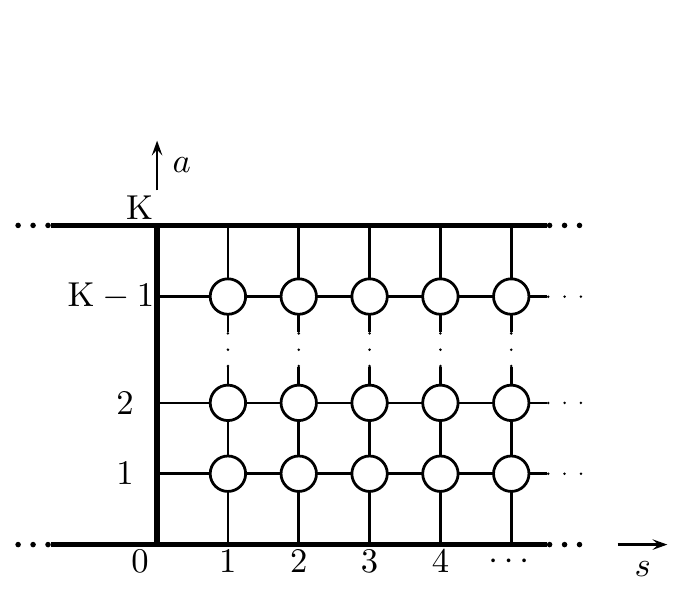}}

    \caption{The lattices \(\HK(\Kr,0)\) and  \(\Sh(\Kr)\) corresponding to the group {\GL\Kr}}
    \label{fig:ShvsHK}
  \end{figure}

  After the change of variables
  \({\Y=}{\frac{\T[][s+1]}{\T[a+1]}\frac{\T[][s-1]}{\T[a-1]}}\), we see
  that the authorized values for {\Yfs} are
  \begin{gather}
    \fdisp{\Yl{\HK(\Kr,\Mr)}\equiv\left\{(a,s)\in \bN\times\bN\middle|
        \begin{array}{rl}
          s\geq 1 &\And 1\leq a\leq \Kr-1\\
          &\Or\\
          a\geq 1 &\And 1\leq s\leq \Mr-1\\
          &\Or\\
          s=\Mr\geq1 &\And a=\Kr\geq1
        \end{array}
      \right\}}\,.
  \end{gather}

  We see that the if \(\Kr\geq1\) and \(\Mr\geq1\), then at
  \((a,s)=(\Kr,\Mr)\),
  \({\Y[\Kr][\Mr]=}{\frac{\T[\Kr][\Mr+1]}{\T[\Kr+1][\Mr]}\frac{\T[\Kr][\Mr-1]}{\T[\Kr-1][\Mr]}}\)
  is well defined. On the other hand, we get \(\Y[\Kr][s]=\infty\) if
  \(s>\Mr\) and \(\Y[a][\Mr]=0\) if \(a>\Kr\). If we plug this into the {\YsE}
  \eqref{eq:YSysEq} at \((a,s)=(\Kr,\Mr)\), then the {\rhs} is
  indeterminate, because it contains \(\frac \infty{1+1/0}\). Therefore we
  will say that \(\Y\) is a solution of the {\YsE} on \(
  \HK(\Kr,\Mr)\) if \(\Y\)
  is defined for all \((a,s)\in \Yl{\HK(\Kr,\Mr)}\), and if it obeys the
  equation \eqref{eq:YSysEq} for all \((a,s)\in \Yl{\HK(\Kr,\Mr)}\) except
  \((a,s)=(\Kr,\Mr)\). In this {\YsE}, we use 
  the prescriptions \(\Y[0]=\infty\), 
  \(\Y[][0]=0\), \(\Y[\Kr]=\infty\) if \(s>\Mr\) and  \(\Y[][\Mr]=0\) if
  \(a>\Kr\).

  This lattice \(\HK(\Kr,\Mr)\) is depicted in figure
  \ref{fig:HK-YvsT}. Physically it appears for the {\csds} of
  chapter \ref{part:qoperatorsspin}, but also for several fields
  theories, such as the Gross-Neveu model.

  \begin{figure}
    \centering
    \includegraphics{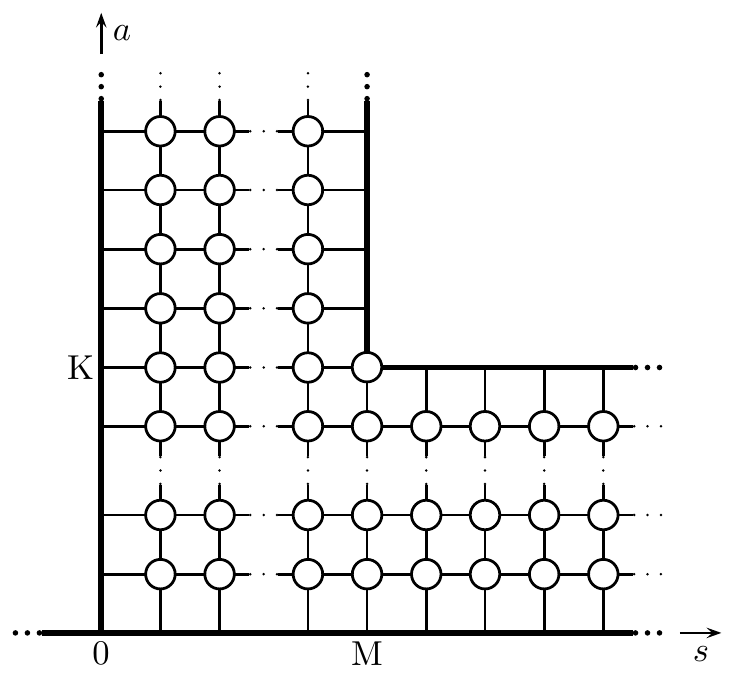}
    \caption{The ``{\fat}-{\hook}'' \(\HK(\Kr,\Mr)\): the lattice
      \(\Yl{\HK(\Kr,\Mr)}\) for {\Yfs} is denoted by circles whereas the
      lattice \(\Tl{\HK(\Kr,\Mr)}\) contains all the nodes of the grid
      in the background.}
    \label{fig:HK-YvsT}
  \end{figure}

  \subsubsection{The ``$\Tk$-{\hook}s'', such as in the case of  {\ADF}}
  \label{sec:t-hookfadscft}

  \begin{figure}
    \centering
    \includegraphics{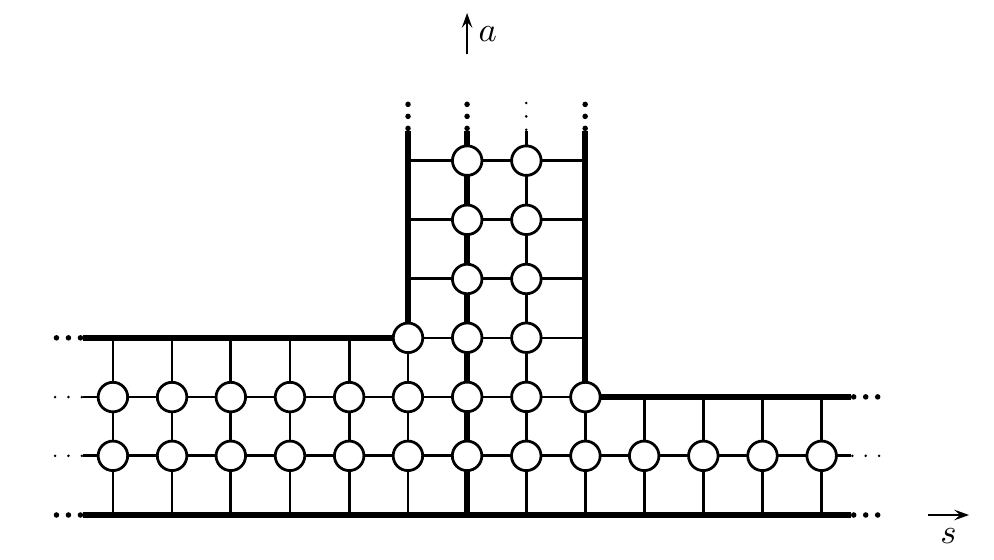}
    \caption{The ``\(\Tk\)-{\hook}'' \(\Tk(2,3|2+1)\): the lattice
      \(\Yl{\Tk(2,3|2+1)}\) for {\Yfs} is denoted by circles whereas the
      lattice \(\Tl{\Tk(2,3|2+1)}\) contains all the nodes of the grid
      in the background.}
    \label{fig:Tk-YvsT}
  \end{figure}

  We will see in the next chapter that {\another} shape of lattice occurs
  in the study of the AdS/CFT spectrum. This lattice is called a
  \(\Tk\)-shaped {\fat} {\hook}, or a ``\(\Tk\)-{\hook}'' in
  \cite{2009PhRvL.103m1601G,Gromov:2009bc}, and \cite{Volin:2010cq,Volin:2010xz}, and it is defined as
  \begin{gather}
        \Tl{\Tk(\Kr,\Kr'|\Mr+\Mr')}\equiv\left\{(a,s)\in\bN\times\bZ \middle|
      \begin{aligned}
        s\geq 0 &\And 0\leq a\leq \Kr\\
        &\Or\\
        s\leq 0 &\And 0\leq a\leq \Kr'\\
        &\Or\\
        a\geq 0 &\And -\Mr'\leq s\leq \Mr
      \end{aligned}
    \right\}\,,
   \end{gather}
   \begin{gather}
    \Yl{\Tk(\Kr,\Kr'|\Mr+\Mr')}\equiv\left\{(a,s)\in\bN\times\bZ \middle|
      \begin{aligned}
        s> 0 &\And 1\leq a\leq \Kr-1\\
        &\Or\\
        s< 0 &\And 1\leq a\leq \Kr'-1\\
        &\Or\\
        a > 0 &\And -\Mr'-1\leq s\leq \Mr-1\\
        &\Or\\
        a=\Kr\geq 1&\And s=\Mr\\& \qquad\If \Mr\geq -\Mr' + 1 \Or \Kr\leq\Kr'\\
        &\Or\\
        a=\Kr'\geq 1&\And s=-\Mr'\\&\qquad \If \Mr'\geq -\Mr + 1 \Or \Kr'\leq\Kr
      \end{aligned}
    \right\}\,,
   \end{gather}
   \begin{gather}
\where \Kr, \Kr', \Mr, \Mr' \geq 0\,.
  \end{gather}

This lattice is depicted in figure \ref{fig:Tk-YvsT}, and we will 
  say that \(\T\) is a solution of the Hirota equation on
  \(\Tk(\Kr,\Kr'|\Mr+\Mr')\) if \(\T\) is defined for all
  \((a,s)\in \Tl{\Tk(\Kr,\Kr'|\Mr+\Mr')}\), and if it
  obeys the equation \eqref{eq:YHirota} for all \((a,s)\in
  \bZ\times\bZ\), with the prescription \(\T=0\) for all \((a,s)\not\in
  \Tl{\Tk(\Kr,\Kr'|\Mr+\Mr')}\).

  Like in the case of the lattice \(\HK(\Kr,\Mr)\), the {\YsE} is
  ill-defined at the corners \((a,s)=(\Kr,\Mr)\) and
  \((a,s)=(\Kr',-\Mr')\).  Therefore we   will say that \(\Y\) is a
  solution of the {\YsE} on \(\Tk(\Kr,\Kr'|\Mr+\Mr')\) 
  if \(\Y\) is defined for all \((a,s)\in \Yl{\Tk(\Kr,\Kr'|\Mr+\Mr')}\), and if
  it obeys the equation \eqref{eq:YSysEq} for all \((a,s)\in
  \Yl{\Tk(\Kr,\Kr'|\Mr+\Mr')}\) except when \((a,s)=(\Kr,\Mr)\) and
  when \((a,s)=(\Kr',-\Mr')\). In this {\YsE}, we should use the
  prescriptions  \(\Y[0]=\infty\), \(\Y[\Kr]=\infty\) if \(s>\Mr\),
  \(\Y[\Kr']=\infty\) if \(s<-\Mr\), \(\Y[][\Mr]=0\) if \(a>\Kr\) and 
  \(\Y[][-\Mr']=0\) if \(a>\Kr'\). These prescriptions are exactly what
  comes from writing a ratio of {\Tfs} which are equal to zero outside
  the lattice \(\Tk(\Kr,\Kr'|\Mr+\Mr')\), and they make the {\YsE}
  singular at positions \((a,s)=(\Kr,\Mr)\) and
  \((a,s)=(\Kr',-\Mr')\).

  \subsubsection{The {\Wronskian} gauge}
  \label{sec:{\Wronskian}-gauge}

  Let us now define a physical choice of gauge, which will fix part of
  the gauge freedom of equation \eqref{eq:GaugeFreedom}. To this end, we
  should remind that the Hirota equation \eqref{eq:YHirota} is
  equivalent to the Hirota equation of chapter \ref{part:qoperatorsspin}
  up to the change of variables \eqref{eq:TchangeShift}. But in chapter
  \ref{part:qoperatorsspin}, 
  the \((a,s)\)-lattice had a clear interpretation in terms of
  {\rp}, which allowed to say that for instance
  \begin{gather}
    \label{eq:PhysWronBoun}
    \rT[][0]=\rT[][][0]=\rT[][0][0]\,.
  \end{gather}
  Additionally, we know that for {\GL\Kr},
  \(\chas[\Kr]({\g})=\ensuremath{\mathrm{det}}({\g})^s\), so that one can show
  (see the relation \eqref{eq:detshift} in appendix
  \ref{sec:ident-involv-co}), that
  \begin{gather}
    \label{eq:PhysWronGaug0}
    \rT[][\Kr]=\rT[\su+s][0][0] \ensuremath{\mathrm{det}}({\g})^s \,.
  \end{gather}
  In the present section, we do not introduce any twisted boundary
  condition. This means that we take the limit \(\g\to \bI\) of expressions
  like \eqref{eq:TfromD} (this limit \(\g\to \bI\) is taken after
  acting by {\cd}s). Therefore, we will disregard the factor
  \({\mathrm{det}}({\g})^s\) in \eqref{eq:PhysWronGaug0}.

  After the change of variables \eqref{eq:TchangeShift}, the constraints 
  \eqref{eq:PhysWronBoun} and  \eqref{eq:PhysWronGaug0} become
  \begin{align}
    \label{eq:PhyWronGauge}
    \T[0]
    =&\Ts[0][0][-s]
    &
    \T[][0]
    =&\Ts[0][0][+a]
    &
    \T[\Kr][s]=&\Ts[0][0][+\Kr+s]
  \end{align}

  In the {\TBA} approach, the gauge freedom \eqref{eq:GaugeFreedom} allows to
  restrict to {\Tfs} obeying the constraints \eqref{eq:PhyWronGauge}.
  \begin{proof}
    Let \(\T\) be an arbitrary solution of the Hirota equation on the
    lattice \(\Sh(\Kr)\). On the boundaries of the lattice, the Hirota
    equation implies that there exist six functions \(f_1\), \(f_2\),
    \(f_3\), \(f_4\), \(f_5\) and \(f_6\) of the spectral parameter \(\us\), such
    that 
    \begin{align}
      \label{eq:toPhyWronGauge}
      \T[0]
      =&f_1^{[+s]}f_2^{[-s]}
      &
      \T[][0]
      =&f_3^{[+a]}f_4^{[-a]}
      &
      \T[\Kr][s]=&f_5^{[+s]}f_6^{[-s]}\,
    \end{align}
    because one term is zero in the Hirota equation on the boundary of
    the lattice (for instance, at \(a=0\), we get
    \(\T[0][][+]\T[0][][-]=\T[0][s+1]\T[0][s-1]\), which implies that 
    \(\T[0]  = f_1^{[+s]}f_2^{[-s]} \)).

    If we set
    \begin{align}
      \label{eq:totototoWronGa}
      \gan[][1]=&\frac 1 {f_1 \gan[][3]},&
      \gan[][2]=&\frac{\ga[-2\Kr][3]f_4^{[-2\Kr]}}{f_6^{[-\Kr]}},&\gan[][4]=&\frac
      1 {f_4
        \gan[][3]},
    \end{align}
    then the function \(\tilT=\ga[a+s][1] \ga[a-s][2]
    \ga[-a+s][3] \ga[-a-s][4] \T\) gives the same {\Yfs} and obeys the
    additional gauge constraints \eqref{eq:PhyWronGauge}.
  \end{proof}

  In the case of {\sugrs}, we can also define an analogous gauge
  condition. The first difference is that we do not have an expression
  like \eqref{eq:PhysWronGaug0}, but on the other hand, the expression 
  \eqref{eq:super-rectangular-Toprs} gives
  \begin{align}
    \rT[][\Kr]\propto&\frac{\gQ[\su+s][B]\gQ[\su-\Kr][F]}
    {\gQ[\su+s-\Kr][\emptyset] }&\If&s\geq\Mr,\\
    \rT[][][\Mr]\propto&\frac{\gQ[\su+\Mr][B]\gQ[\su-a][F]}
    {\gQ[\su+\Mr-a][\emptyset] }&\If&a\geq\Kr,
  \end{align}
  \begin{gather}
    \where  B=\left\{\jvp\in \ninter 1 {\Kr+\Mr} ~\middle|~ \sg \jvp = +1\right\}\\
    \And F=\left\{\jvp\in \ninter 1 {\Kr+\Mr} ~\middle|~ \sg \jvp = -1\right\}
  \end{gather}
  where the symbol \(\propto\) means that the equality is true up to
  a function of the
  twist, analogous to the factor \(\ensuremath{\mathrm{det}}({\g})^s\) in
  \eqref{eq:PhysWronGaug0}. If we perform the change of variables
  \eqref{eq:TchangeShift}, keeping in mind that
  \(\gQ[][\emptyset]\) is \(\su\)-independent, then we get
  \begin{align}
    \label{eq:PhyWronGaugeHK}
    \T[0]=&\Ts[0][0][-s]  & \T[][0] =&\Ts[0][0][+a]
    & \T[\Kr+{\nn}][\Mr]=&\T[\Kr][\Mr+{\nn}]&\textrm{ for any }{\nn}\geq&0
  \end{align}

  This condition generalizes \eqref{eq:PhyWronGauge} to the lattice
  \(\HK(\Kr,\Mr)\). Like previously, the gauge freedom
  \eqref{eq:GaugeFreedom} is sufficient to restrict to {\Tfs} obeying
  the constraints \eqref{eq:PhyWronGaugeHK} (the proof is as easy as
  above, and is left to the reader).

  In what follows, we will often view the choice of a gauge obeying
  \eqref{eq:PhyWronGauge} (or \eqref{eq:PhyWronGaugeHK}) as a more
  physical choice, because if a lattice regularization turned out to
  exist and to give a meaning to the {\Tfs} (identifying the \((a,s)\)
  labels to rectangular {\yn} diagrams), then the {\Tfs} which would
  come out would satisfy the constraint \eqref{eq:PhyWronGaugeHK}.

  We will call ``{\Wronskian} gauge'' the gauges which obey these
  conditions. One should note that this requirement fixes only three out
  of four degrees of gauge-freedom ({\idest} the function \(\gan[][3]\) in
  \eqref{eq:totototoWronGa} is not fixed by this argument).

  For the lattice \(\Tk(\Kr,\Kr'|\Mr+\Mr')\) (which corresponds for instance to AdS/CFT), the
  gauge condition \eqref{eq:PhyWronGaugeHK} can be generalized as
  \begin{align}
    \label{eq:PhyWronGaugeTK}
    \T[0]=&\Ts[0][0][-s]   & \T[\Kr'+{\nn}][-\Mr']=&\T[\Kr'][-\Mr'-{\nn}]
    & \T[\Kr+{\nn}][\Mr]=&\T[\Kr][\Mr+{\nn}]&\textrm{ for any }{\nn}\geq&0
  \end{align}

As we will see, the ``{\Wronskian} gauges'' are gauges where
{\Wronskian} determinant expressions (similar to
\eqref{eq:NestedWronskianLambda} in the previous chapter) appear in the most natural way (hence the name of
{\Wronskian} gauge).
For instance, we used 
this gauge 
in the article
\cite{Gromov:2010km}, which gave the typical solution of Hirota
equation on the lattice  \(\Tk(2,2|2+2)\) in the form of a {\Wronskian}
determinant.

  \subsection{Equivalence of Hirota equation and {\Ysys} equation}
  \label{sec:equiv-hirota-equat}

  To see in what sense 
  the Hirota equation and the {\Ysys} equation
  are equivalent, 
  the first statement that we 
will
  prove is:
  \begin{statmt}
    \label{sta:HirotaGivesYsys}
    If \(\T\)  obeys the Hirota equation \eqref{eq:YHirota}, then
    \(\Y=\frac{\T[][s+1]}{\T[a+1]}\frac{\T[][s-1]}{\T[a-1]}\)  obeys the
    {\Ysys} equation \eqref{eq:YSysEq}
  \end{statmt}
  \begin{proof}
    Let \(\T\) be a set of functions which obeys the Hirota equation
    \eqref{eq:YHirota}.
    Then if we define
    \(\Y=\frac{\T[][s+1]}{\T[a+1]}\frac{\T[][s-1]}{\T[a-1]}\), we can notice
    that due to the Hirota equation, we have
    \begin{gather}
      \label{eq:opYefrac}  1+\Y=\frac{\T[][][+]\T[][][-]}{\T[a-1]\T[a+1]}\,,\qquad
      1+1/\Y=\frac{\T[][][+]\T[][][-]}{\T[][s-1]\T[][s+1]}\,.
    \end{gather}

    We can then compute 
    \begin{align}
      \lefteqn{\frac{1+\Y[][s+1]}{1+1/\Y[a+1]}
        \frac{1+\Y[][s-1]}{1+1/\Y[a-1]}}\quad
      \\=&
      \frac{\T[][s+1][+]\T[][s+1][-]         \T[a+1][s-1]\T[a+1][s+1]}
      {\T[a-1][s+1]\T[a+1][s+1]         \T[a+1][][+]\T[a+1][][-]}
      \frac{\T[][s-1][+]\T[][s-1][-]         \T[a-1][s-1]\T[a-1][s+1]}
      {\T[a-1][s-1]\T[a+1][s-1]         \T[a-1][][+]\T[a-1][][-]}
      \\=&
      \frac{\T[][s+1][+]\T[][s+1][-]         
      }
      {
        \T[a+1][][+]\T[a+1][][-]}
      \frac{\T[][s-1][+]\T[][s-1][-]         
      }
      {
        \T[a-1][][+]\T[a-1][][-]}
      =
      \Y[][][+]   \Y[][][-]\,,
    \end{align}
    which means that \(\Y\) obeys the {\Ysys} equation \eqref{eq:YSysEq}.
  \end{proof}

  The next statement which we want to prove is the following:
  \begin{statmt}
    \label{sta:TfromY}
    If \(\Y\) is a typical solution of the {\Ysys} equation
    \eqref{eq:YSysEq}, then there exists a typical solution \(\T\) of the
    Hirota equation  \eqref{eq:YHirota} such that \(\Y=\frac{\T[][s+1]}{\T[a+1]}\frac{\T[][s-1]}{\T[a-1]}\).
  \end{statmt}

As in the section \ref{sec:cbr-determ-form}, a typical solution of Hirota
equation (resp the {\YsE}), is a solution \(\T\) (resp \(\Y\)) such that
every small 
perturbation of a set of initial data (see {\below})
is associated to {\another} solution of Hirota
equation (resp the {\YsE}), which converges to \(\T\) (resp \(\Y\))
when the perturbation tends to zero.

For the lattice \(\St(\Np)\), for instance, one possible
\index{Typical (solution)}
    choice of initial values is the set \linebreak of functions 
    \(\left\{\Y[a][0]\middle| 1\leq a\leq {\Np}-1 \And 0\leq s\leq 1
    \right\}\), for the {\YsE}, and \linebreak
    \(\left\{\T[a][0]\middle| 0\leq a\leq {\Np}
      \And 0\leq s\leq 1
    \right\}\) for the Hirota equation.
Then a typical solution of the {\YsE} on \(\St(\Np)\) is a solution
\(\Y\) of the {\YsE} such that, for every small perturbation 
    \(\left[\Y[][0]\right]_\epsilon=  \Y[][0]+\mathcal{O}(\epsilon)\)
    (and \(\left[\Y[][1]\right]_\epsilon=
    \Y[][1]+\mathcal{O}(\epsilon)\)) of 
the initial values,  there
    exists a solution \(\left[\Y\right]_\epsilon\) such that
    \(\Y=\lim_{\epsilon\to 0} \left[\Y\right]_\epsilon\).
Similarly a typical solution of the Hirota equation on \(\St(\Np)\) is a
solution 
\(\T\) of the Hirota equation such that, for every small perturbation 
    \(\left[\T[][0]\right]_\epsilon=  \T[][0]+\mathcal{O}(\epsilon)\)
    (and \(\left[\T[][1]\right]_\epsilon=
    \T[][1]+\mathcal{O}(\epsilon)\)) of 
the initial values,  there
    exists a solution \(\left[\T\right]_\epsilon\) such that
    \(\T=\lim_{\epsilon\to 0} \left[\T\right]_\epsilon\).

An important properties of these typical solutions is that they are
completely characterized by the initial values (the various {\Tfu}- or
{\Yfs} are expressed recursively in terms of the initial values).

    From the thermodynamic Bethe {\anz} point of view, we will see (in
    the next sections) that the solution of {\Ysys} is 
    typical when the size \(\Lf\) of the spatial dimension is large. If we assume that the {\Yfs}
    are analytic in the parameter \(\LF\), it is quite natural to
    expect that at any size \(\LF\), the solutions are still typical.

    Moreover, for the vacuum (the case when we can derive the {\Ysys}
    equation), the functions \(\Y\) are all positive (due to their
    definition in section \ref{sec:example-princ-chir} as ratios of
    densities), and we can directly show that all the {\Yfs} are fixed
    uniquely by the initial values (because the iteration procedure
    {\cannnot} involve a denominator which is identically zero), and that
    the solution of the {\YsE}s is typical.

  \begin{proof}[Proof of \staref{sta:TfromY}]
    The \staref{sta:TfromY} is slightly trickier than the
    previous \staref{sta:HirotaGivesYsys}. We will prove it on a
    case-by-case basis (though we will see that the proof is
    identical for the lattices considered, and it will be detailed
    mainly  for
    the (a,s)-lattice 
    \(\St(\Np)\) of the {\PCM} (starting with the simplest case of
    \(\St(2)\)). The idea of the proof will be that, starting from a
    solution of the {\YsE}, we will find initial values for the {\Tfs}
    such that the initial values for the {\Yfs} are reproduced (when
    we write the ratio of these initial {\Tfs}). From these initial
    values we can (using the typicality condition) define all the
    other {\Tfs} by recurrence so that the Hirota equation is
    satisfied. Then the ratio
    \(\frac{\T[][s+1]}{\T[a+1]}\frac{\T[][s-1]}{\T[a-1]}\) obeys the
    {\YsE} and reproduces the initial values of the {\Yfs}, therefore
    it coincides with the {\Yfs} for all \((a,s)\).

    \paragraph{\(\St(2)\) case}

    Let us start with the simplest lattice \(\St(2)\) (corresponding to
    the  \(\SU 2\times \SU 2\) {\PCM}). In this case \(a=1\) is the only
    value of \(a\) for which \(\Y\) is 
    defined.
    Assuming the typicality condition stated above, we have shown that
    an arbitrary solution of the {\Ysys} equation is 
    uniquely fixed by the initial conditions \(\Y[1][0]\) and
    \(\Y[1][1]\) and 
    that the typical
    solution of the Hirota equation is fixed uniquely by the initial
    conditions \(\T[0][0]\), \(\T[0][1]\), \(\T[1][0]\), \(\T[1][1]\),
    \(\T[2][0]\) and \(\T[2][1]\). We can note that for this {\SU 2} {\PCM},
    the {\Yfs} are characterized by two initial conditions, whereas the
    {\Tfs} are characterized by six initial conditions. This is absolutely
    consistent with the fact that as compared to the {\Yfs}, the {\Tfs}
    have an extra gauge-freedom characterized by four functions (see
    \eqref{eq:GaugeFreedom}).

    If we know a solution \(\Y\) of the {\Ysys} equation, then we can
    write initial conditions for the {\Tfs}.
    They can be chosen for instance as 
    \begin{gather}
      \T[0][0]=\T[0][1]=\T[2][0]=\T[2][1]=1
    \end{gather}
    and \(\T[1][0]\) and \(\T[1][1]\) are solutions of 
    \begin{gather}
      \label{eq:invPMequivTY}
      \T[1][0][+] \T[1][0][-]=1+\Y[1][0],\qquad
      \T[1][1][+] \T[1][1][-]=1+\Y[1][1]\,.
    \end{gather}

    We will discuss how to solve these equations, but for the moment it is
    enough to say that solutions of \eqref{eq:invPMequivTY} do exist. For
    instance, one can choose \(\T[1][0](\us)= 1\) for \(\Im({\us})\in[0,1[\) and define
    \(\T[1][0](\us)\) by recurrence as \(\frac{1+\Y[1][0][-]}{\T[1][0][--]}\)
    if \(\Im({\us})\geq 1\) and as \(\frac{1+\Y[1][0][+]}{\T[1][0][++]}\)
    if \(\Im({\us})< 0\).

    Then the (typical) solution of Hirota equation characterized by these
    initial values gives 
    rise to a solution of the {\Ysys} which is characterized by the
    initial values \(\Y[1][0]\) and \(\Y[1][1]\). This proves that there
    exists a solution of the Hirota equation associated to this typical
    solution of the {\Ysys} equation \eqref{eq:YSysEq}.

    \paragraph{Case of \(\St({\Np})\)}
    \label{sec:case-sun}

    For the more general lattice \(\St(\Np)\) (corresponding to
    the  \(\SU \Np\times \SU \Np\) {\PCM}), we can follow exactly the
    same argument as for \(\St (2)\). For the {\Ysys} equation, one possible
    choice of initial values is the set of functions 
    \(\left\{\Y[a][0]\middle| 1\leq a\leq {\Np}-1 
      \And 0\leq s\leq 1
    \right\}\). For the Hirota equation, a possible choice of initial values
    is the set of functions
    \(\left\{\T[a][0]\middle| 0\leq a\leq {\Np}
      \And 0\leq s\leq 1
    \right\}\).

    Given a solution of the {\Ysys} equation, we can define 
    initial values for the Hirota equation which obey for instance
    \begin{gather}
      \label{eq:TfrYBnd}
      \T[0][0]=\T[0][1]=\T[{\Np}][0]=\T[{\Np}][1]=1,\\
      \And    \left(
        \begin{aligned}
          \mathrm{log} & \left( 1+\Y[1]\right)\\
          \mathrm{log} & \left( 1+\Y[2]\right)\\
          \vdots\\
          \mathrm{log} & \left( 1+\Y[{\Np}-1]\right)
        \end{aligned}
      \right)=
      \MM\cdot
      \left(
        \begin{aligned}
          \mathrm{log} & \left( \T[1]\right)\\
          \mathrm{log} & \left( \T[2]\right)\\
          \vdots\\
          \mathrm{log} & \left( \T[{\Np}-1]\right)
        \end{aligned}
      \right),\qquad
      \label{eq:TfrYeq}
      \textrm{for } 
      0\leq s\leq 1\,,\\
\label{eq:DefMMdiff}
      \where \MM=
      \left(
        \begin{array}{ccccc}
          {\DD}+{\DD}^{-1}&-1\\[.3cm]
          -1&{\DD}+{\DD}^{-1}&-1&\\[.3cm]
          &\ddots&\ddots&\ddots&\\[.3cm]
          &&-1&{\DD}+{\DD}^{-1}&-1\\[.3cm]
          &&&-1&{\DD}+{\DD}^{-1}
        \end{array}
      \right)\,,
      \\\And {\DD}\equiv e^{\frac{\bi}{2} \partial_\us }\,.
    \end{gather}
    The equation \eqref{eq:TfrYeq} is simply the requirement that
    \(\frac{\T[][][+]\T[][][-]}{\T[a-1]\T[a+1]}=1+\Y\) should hold.
    In order to
    show how to find \(\T\) such that \eqref{eq:TfrYeq} holds, 
    let us 
    show how, for \({\Np}-1\) functions
    \(y_1=\mathrm{log}\left( 1+\Y[1]\right)\), \(y_2=\mathrm{log}\left(
      1+\Y[1]\right)\), \(\cdots\), \(y_{{\Np}-1}=\mathrm{log}\left(
      1+\Y[\Np-1]\right)\), we can find \(\Np-1\) functions \(x_0\), \(x_1\),
    \(\cdots\), \(x_{\Np-1}\) such that \({\MM}\cdot (x_{\coordj})_{1\leq {\coordj} \leq \Np-1}
    = (y_{\coordj})_{1\leq {\coordj} \leq \Np-1}\). First one can notice that if a
    solution exists, then the first component of \({\MM}\cdot (x_{\coordj})-(y_{\coordj})\)
    is \(-x_2+x_1^++x_1^--y_1\). If it is zero we get
    \(x_2=x_1^++x_1^--y_1\). Then the second component gives
    \(x_3=x_1^{[+2]}+x_1^{[+0]}+x_1^{[-2]}-y_1^+-y_1^-\), etc. This
    gives
    \begin{gather}
      {\MM}\cdot (x_{\coordj})_{1\leq {\coordj} \leq \Np-1}
    = (y_{\coordj})_{1\leq {\coordj} \leq \Np-1}\qquad\Rightarrow \qquad \forall \kk,~~
    x_\kk=\FS[\kk%
    ]~x_1+\sum_{\lL=1}^{\kk-1} \FS[\kk-\lL]~ y_\lL\\
\where
      \FS[\coordi]\equiv
      {\DD}^{1-\coordi}+{\DD}^{3-\coordi}+\cdots +{\DD}^{\coordi-1}
      =
      \sum_{s=-{\frac{\coordi-1}{2}}}^{\frac{\coordi-1}{2}} {\DD}^{2s}\,.
  \end{gather}
  Finally, the last component of the equation \({\MM}\cdot (x_{\coordj})_{1\leq {\coordj} \leq \Np-1}
    = (y_{\coordj})_{1\leq {\coordj} \leq \Np-1}\) gives the
    constraint
    \begin{gather}
\label{eq:InvM}      \FS[\Np]~x_1+\sum_{\lL=1}^{\Np-1} \FS[\Np-\lL]~ y_\lL =0\,.
    \end{gather}
 One
easily sees that this equation always has a solution \(x_1\), 
which can for instance, 
    be defined by recurrence\footnote{
Let us give a possible way to solve the equation \(\FS[{\Np}] x_1
    =f(\us)\) : 
if we %
notice that
    \({\DD}^{-{\Np}+1}\left(\FS[{\Np}]-{\DD}^{-1} \FS[{{\Np}-1}]\right)=1\), we can define a solution of %
    \({\FS[{\Np}] x_1
      =f(\us)}\) as \(x_1= 0\) for
    \({\Im({\us})\in[0,{\Np}-1[}\) and %
then by recurrence as %
    \(x_1= 
    f\left(\us-\frac{{\Np}-1}{2}\right)- {\DD}^{-{\Np}} \FS[{{\Np}-1}] x_1
    \) if \(\Im({\us})\geq {\Np}-1\), and as %
    \(x_1 = 
    f\left(\us+\frac{{\Np}-1}{2}\right)- {\DD}^{{\Np}} \FS[{{\Np}-1}] x_1
    \) if \(\Im({\us})<0\).
}. Out of this solution \(x_1\),
we obtain a solution of the equation  \({\MM}\cdot (x_{\coordj})_{1\leq {\coordj} \leq \Np-1}
    = (y_{\coordj})_{1\leq {\coordj} \leq \Np-1}\).

    This shows %
    that we can find initial values \(\left\{\T[a][0]\middle| 0\leq a\leq {\Np}
      \And 0\leq s\leq 1
    \right\}\) which obey \eqref{eq:TfrYBnd} and
    \eqref{eq:TfrYeq}. Then a typical solution of Hirota equation can \linebreak be built
    from these initial values, and after the change of variables
    \(\Y=\frac{\T[][s+1]}{\T[a+1]}\frac{\T[][s-1]}{\T[a-1]}\), it
    gives a solution of the {\Ysys}
    equation which reproduces the initial values \linebreak \(\left\{\Y[a][0]\middle| 1\leq a\leq {\Np}-1 %
      \And 0\leq s\leq 1
    \right\}\). Therefore, it reproduces all the {\Yfs}.

    \paragraph{Other lattices}
    \label{sec:other-lattices}

    For other shapes of the \((a,s)\)-lattice, it is possible to proceed the
    same way to prove %
    this Statement \ref{sta:TfromY}.
    The
    set of initial
    conditions may be quite different, but the method is absolutely the
    same.

    For instance for \(\HK(\Kr,\Mr)\),
    the argument is identical to the
    construction above, \linebreak %
    if we use the initial values \(\left\{\Y[a][1]\middle| 1\leq a\leq \Kr+\Mr-1
    \right\}\) for {\Yfs} and \linebreak \(\{\T[0][0],\T[1][0]\}\cup
    \left\{\T[a][1]\middle| 0\leq a\leq \Kr+\Mr \right\}\) for {\Tfs}. An
    analogous of \eqref{eq:TfrYBnd} can be the choice
    \(\T[0][0]=\T[1][0]=\T[0][1]=\T[\Kr+\Mr][1]=1\), and the other initial {\Tfs}
    are expressed from the equation \eqref{eq:TfrYeq} at \(s=1\).

    For the lattice \(\Tk(2,2|4)\) of the AdS/CFT correspondence,
    the same argument can be used with the initial values
    \(\left\{\Y[1]\middle| -3\leq s\leq 3
    \right\}\) for {\Yfs} and \linebreak \(\{\T[0][0],\T[0][1]\}\cup
    \left\{\T[1]\middle| -4\leq s\leq 4 \right\}\) for {\Tfs}. An
    analogous of the gauge condition \eqref{eq:TfrYBnd} can be the choice
    \(\T[0][0]=\T[0][1]=\T[1][-4]=\T[1][4]=1\), and the other {\Tfs}
    are expressed like in \eqref{eq:TfrYeq}, by using the relation \(1+1/\Y=\frac{\T[][][+]\T[][][-]}{\T[][s-1]\T[][s+1]}\).
  \end{proof}

  We have now proven that every solution of the {\Ysys} equation
  corresponds to (at least) one solution of the Hirota equation. 

  We can now prove a concluding result which states that the gauge
  transformations \eqref{eq:GaugeFreedom} are the only additional
  freedom in the Hirota equation as compared to the {\Ysys}
  equation. More precisely the statement is 
  \begin{statmt}
    If \(\T\) and \(\tilT\) are two typical solutions of the Hirota
    equation \eqref{eq:YHirota} such that
    \(\frac{\T[][s+1]}{\T[a+1]}\frac{\T[][s-1]}{\T[a-1]} =
    \frac{\tilT[][s+1]}{\tilT[a+1]}\frac{\tilT[][s-1]}{\tilT[a-1]}\),
    then there exist four functions \(\gan[][1]\),  \(\gan[][2]\), \(\gan[][3]\) and
    \(\gan[][4]\) such that \(\tilT = \ga[a+s][1] \ga[a-s][2]
    \ga[-a+s][3] \ga[-a-s][4] \T\).
\label{sta:gaugetransfo}
  \end{statmt}
  \begin{proof}
    Let us write this proof
    for %
    all the lattices \(\HK(\Kr,\Mr)\) and \(\Tk(\Kr,\Kr'|\Mr+\Mr')\) at once. For all
    these lattices, the CBR determinant relation \eqref{eq:YCBR}
    holds\footnote{This CBR relation holds for all lattices such that
      \(\T=0\) is \(a<0\) and \(\T[1]\neq 0\)
      (at least for positive \(s\)), under a connexity condition  which holds
      for instance for \(\HK(\Kr,\Mr)\) and \(\Tk(\Kr,\Kr'|\Mr+\Mr')\).}.

    Then, we can use the relation
    \(\frac{\T[][][+]\T[][][-]}{\T[][s-1]\T[][s+1]}=1+1/\frac{\T[][s+1]}{\T[a+1]}\frac{\T[][s-1]}{\T[a-1]}\),
    which is just the Hirota equation (see \eqref{eq:opYefrac}), to
    rewrite the condition \(\frac{\T[][s+1]}{\T[a+1]}\frac{\T[][s-1]}{\T[a-1]} =
    \frac{\tilT[][s+1]}{\tilT[a+1]}\frac{\tilT[][s-1]}{\tilT[a-1]}\) as
    follows:
    \begin{gather}
      \frac{f_{(a,s)}^+f_{(a,s)}^-}{f_{(a,s+1)}f_{(a,s-1)}}=1,\qquad \where f_{(a,s)}=\frac{\tilT}{\T}\,.
    \end{gather}
    The general solution of this equation is \(f_{(a,s)}=h_a^{[+s]}\tilde
    h_a^{[-s]}\) where \(\{h_a\}\) and \(\{\tilde h_a\}\) are two sets of
    arbitrary functions {\lbd} by \(a\in \bN\).

    If we find four functions \(\gan[][1]\),  \(\gan[][2]\), \(\gan[][3]\) and
    \(\gan[][4]\) such that
    \begin{gather}
      \label{eq:YvsTfindGauge}
      h_0=\gan[][1]\gan[][3],\qquad    h_1=\gan[+][1]\gan[-][3],\qquad
      \tilde h_0=\gan[][2]\gan[][4],\qquad \tilde h_1=\gan[+][2]\gan[-][4],
    \end{gather}
    then we obtain the relation \(\tilT = \ga[a+s][1] \ga[a-s][2]
    \ga[-a+s][3] \ga[-a-s][4] \T\) when \({s=0}\) and when \({s=1}\).
    Due to the CBR formula, 
    these initial 
    data completely fix the \linebreak solutions \(\T\) and \(\tilT\) of the Hirota
    equation, which %
    allows to conclude that \linebreak \(\tilT = \ga[a+s][1] \ga[a-s][2]
    \ga[-a+s][3] \ga[-a-s][4] \T\) holds for arbitrary \(a\) and \(s\).

    Let us now show that a solution of \eqref{eq:YvsTfindGauge} does
    exist. For that, we can simply choose a
    solution \(\gan[][1]\) of the relation
    \(\frac{h_1}{h_0^-}=\frac{\gan[+][1]}{\gan[-][1]}\) (it can for
    instance be defined 
    by recurrence), and define
    \(\gan[][3]=\frac{h_0}{\gan[][1]}\).
    Then \(h_0=\gan[][1]\gan[][3]\) and
    \(h_1=\gan[+][1]\gan[-][3]\) are satisfied. If we also define 
    \(\gan[][2]\) as an arbitrary solution of 
    \(\frac{\gan[+][2]}{\gan[-][2]}=\frac{\tilde h_1}{\tilde h_0^-}\) and 
    \(\gan[][4]=\frac{\tilde h_0}{\gan[][2]}\), then
    \eqref{eq:YvsTfindGauge} is satisfied, which finishes the proof.
  \end{proof}

\paragraph{Conclusion}

In this section we have proven the statement that every solution of
the {\Ysys} equation corresponds to 
a set of solutions of the Hirota equation, which are obtained from
each other by gauge transformations. We have proven this statement in
the case of the lattices \(\St(\Np)\) and \(\Tk(\Kr,\Kr'|\Mr+\Mr')\),
and for their sublattices \(\HK(\Kr,\Mr)\) and
\(\Sh(\Kr)\simeq \HK(\Kr,0)\) as well, so that we covered a wide range
of possible thermodynamic Bethe Ansätze. 
These simple properties will be used extensively in the resolution of
the {\Ysys}s, for instance in section \ref{sec:application-au-champ}.

  \subsection{Typical solution of Hirota equation}
  \label{sec:gener-solut-hirota}

  In the previous sections, we have seen that the {\Ysys} equation
  describes the finite size effects of several {\ing} models, and we
  have seen in what sense this equation is equivalent to the Hirota
  equation \eqref{eq:YHirota}.

  We will now write the typical solution of the Hirota equation for
  different \((a,s)\) lattices. We will see that for the lattices
  \(\HK(\Kr,\Mr)\)
  of {\csds}, this solution will be the same as in
  section \ref{sec:expr-toprs-super}. More generally, we will see that
  for each of lattices introduced in section \ref{sec:a-s-lattice},
  the typical solution of the Hirota equation (which is an infinite set of
  equations on an infinite set of functions \(\T\)), is parameterized by
  a finite set of {\qfs}.

  Let us start with a statement \cite{springerlink:10.1007/s002200050165,Zabrodin:1996vm,1998TMP...116..782Z} for the lattice \(\St(\Np)\) of the
  \(\SU\Np\times \SU\Np\) {\PCM}: 
  \begin{statmt} \label{sec:statmtSolHirotPCM}
    Let \(\T\) be any typical solution of the Hirota equation on
    \(\St(\Np)\).

\index{Q-functions@{\Qfs}}
    Then there exist two sets of functions \(\gqs[1]\), \(\gqs[2]\),
    \(\cdots\), \(\gqs[{\Np}]\), \(\gps[1]\), \(\gps[2]\),
    \(\cdots\), \(\gps[{\Np}]\), and two additional functions \(\gqe\) and \(\gpe\)
    such that
    \begin{gather}
      \forall a\in \ninter 1 {\Np}, \qquad \forall s\in
      \bZ,%
      \qquad\qquad
      \fdisp{ \T=\sum_{\substack {I\subset \ninter 1 {\Np}
            \\ |I|=a}} \epsilon(I,\Ib) \gqfs[I][+s]\gpfs[\Ib][-s]}
      \label{eq:HirotaSolutPCM}
      \\
      \label{eq:qqPCMSolHir}
      \where
      \gqf[\{\iq_1,\iq_2,\cdots,\iq_\nn\}]{}=\frac{\Det{\gqss[\iq_\kk][{-1-\nn+2\lL}]}{1\leq
          \kk,\lL\leq \nn}}{\prod_{\kk=1}^{\nn-1}\gqes[{2\kk-\nn}]}\,,\qquad
      \qquad
      1\leq\iq_1<\iq_2<\cdots<\iq_\nn\leq {\Np}\,,
    \end{gather}
    \begin{gather}
    \And
      \gpf[\{\iq_1,\iq_2,\cdots,\iq_\nn\}]{}=\frac{\Det{\gpss[\iq_\kk][{-1-\nn+2\lL}]}{1\leq
          \kk,\lL\leq \nn}}{\prod_{\kk=1}^{\nn-1}\gpes[{2\kk-\nn}]}\,,
      \qquad
      \qquad
      1\leq\iq_1<\iq_2<\cdots<\iq_\nn\leq {\Np}\,.
      \label{eq:qqPCMSolHir2}
    \end{gather}
  \end{statmt}

  In the expression \eqref{eq:HirotaSolutPCM}, the sum runs over all
  subsets \(I\subset \ninter 1 {\Np}\)
having exactly \(a\) elements. For convenience
  we can %
write
  \(I=\{\iq_1,\iq_2,\cdots,\iq_a\}\) where
  \(\iq_\kk<\iq_{\kk+1}\), %
  \(\Ib\equiv\ninter 1 {\Np} \setminus
  I%
  =\{\jq_1,\jq_2,\cdots,\jq_{{\Np}-a}\}\) where
  \(\jq_\kk<\jq_{\kk+1}\). Then the sign \(\epsilon(I,\Ib)\) is defined as
  the signature of a permutation:
  \begin{gather}
    \fdisp{\epsilon(I,\Ib)=\epsilon(\sigma),\qquad
      \sigma(\nn)=\left\{
        \begin{array}{lc}
          \iq_\nn&\If \nn \in \ninter 1 {|I|}\\
          \jq_{\nn-|I|} & \If \nn \in \ninter {|I|+1} {\Np}
        \end{array}
      \right.
    }\,.\\
    \where I=\{\iq_1,\iq_2,\cdots,\iq_a\},\quad
    \Ib=\{\jq_1,\jq_2,\cdots,\jq_{{\Np}-a}\},\quad
    \iq_\kk<\iq_{\kk+1},\quad \jq_\kk<\jq_{\kk+1}\,.
  \end{gather}
  \index{e(s)@\ensuremath{\epsilon(\sigma)} (signature of a permutation)!eI@\ensuremath{\epsilon(I,\Ib)}}

  For instance, if \({\Np}=3\) and \(a=1\), then
  \(\epsilon(\{1\},\{2,3\})=+1=\epsilon(\{3\},\{1,2\})\) whereas
  \(\epsilon(\{2\},\{1,3\})=-1\). Hence 
  \(\T[1]=\gqss[%
  1%
  ][+s]\gpss[%
  2,3%
  ][-s]-\gqss[%
  2%
  ][+s]\gpss[%
  1,3%
  ][-s]+ \gqss[%
  3%
  ][+s]\gpss[%
  1,2%
  ][-s]\).

  \begin{proof}[Proof of the statement \ref{sec:statmtSolHirotPCM}]
    Like for the proofs of the previous section, we will use a set of
    initial values which characterize \(\T\). First, let us show that 
    the {\Tfs} 
    \(\T[0][0]\), \(\T[0][1]\), and
    \(\left\{\T[1][s]\middle|0\leq s\leq 2{\Np}-1\right\}\) are initial
    values, in the sense that %
    if we know these values, then we 
    can successively deduce all {\Tfs}: first all \(\T[0]\) can be
    expressed from \(\T[0][0]\) and \(\T[0][1]\). Then the Hirota
    equation %
    allows to express \(\left\{\T[2][s]\middle|1\leq
      s\leq 2{\Np}-2\right\}\), then \(\left\{\T[3][s]\middle|2\leq s\leq
      2{\Np}-3\right\}\) and iteratively up to
    \(\left\{\T[{\Np}][s]\middle|{\Np}-1\leq s\leq {\Np}\right\}\). As a
    consequence the {\Tfs} of \(\left\{\T\middle|0\leq s\leq 1\right\}\)
    are expressed from these initial conditions. Finally, a simple
    recurrence gives all the {\Tfs} out of \(\left\{\T\middle|0\leq s\leq 1\right\}\).

    The proof of \eqref{eq:HirotaSolutPCM} will %
    be obtained by finding functions
    \(\gqf[]{}\) such that \eqref{eq:HirotaSolutPCM} holds for these
    initial values and by deducing that it holds for all \((a,s)\).

    To this end, we can use the fact that if \eqref{eq:HirotaSolutPCM}
    holds, then the following determinant is zero:
    \begin{gather}
      \label{eq:DetZeroDefq}
      \forall  \iq\in\ninter 1 {\Np} \,,\qquad\qquad
      \BlockDet{\gqss[][2\lL]}{1\leq \lL\leq
        {\Np}+1}{\Ts[1][\kk+\lL-3][\lL-\kk+3]}{\substack{2\leq \kk\leq {\Np}+1\\
          1\leq \lL\leq
          {\Np}+1}}=0\,
    \end{gather}
    For instance when \({\Np}=2\), this determinant reads
    \begin{gather}
      \label{eq:SolvHirPCM2DetZero}
      \left|
        \begin{array}{ccc}
          \gqss[][+2]&\gqss[][+4]&\gqss[][+6]\\
          \Ts[1][0][+2]&\Ts[1][1][+3]&\Ts[1][2][+4]\\
          \Ts[1][1][+1]&\Ts[1][2][+2]&\Ts[1][3][+3]
        \end{array}
      \right|
      =0  \\
      \where \left\{
        \begin{aligned}
          (\Ts[1][0][+2], \Ts[1][1][+3], \Ts[1][2][+4])
          =&\gpss[2][+2](\gqss[1][+2], \gqss[1][+4],\gqss[1][+6])-
          \gpss[1][+2](\gqss[2][+2], \gqss[2][+4],\gqss[2][+6])  \\
          (\Ts[1][1][+1],\Ts[1][2][+2],\Ts[1][3][+3])
          =&\gpss[2][0](\gqss[1][+2], \gqss[1][+4],\gqss[1][+6])-
          \gpss[1][0](\gqss[2][+2], \gqss[2][+4],\gqss[2][+6])
        \end{aligned}\right.\,.
      \label{eq:SolvHirPCM2linesDef}
    \end{gather}
    The expression \eqref{eq:SolvHirPCM2linesDef} is directly read
    from \eqref{eq:HirotaSolutPCM}, and it implies that
    \eqref{eq:SolvHirPCM2DetZero} holds, because %
    for any \(\iq\in\{1,2\}\), the first line is a linear
    combination of the last lines. When \({\Np}>2\),
    \eqref{eq:HirotaSolutPCM} still implies that 
    the determinant
    \eqref{eq:DetZeroDefq} is %
    zero by the same argument.

    Let us now go in the opposite direction, and assume that we have
    {\Tfs}, but we do not know yet whether \eqref{eq:HirotaSolutPCM}
    holds.
    Then, we can view the equation
    \eqref{eq:DetZeroDefq} as a difference equation\footnote{By difference equation, we mean
``discrete differential equation'', also sometimes called ``recurrence equation''.} which defines
    \(\gqs\). This equation will be called the Baxter equation. In
    general ({\idest} under typicality condition on the {\Tfs}) this 
    equation has \({\Np}\) independent solutions, which we can choose to denote as
    \(\gqs[1]\), \(\gqs[2]\), \(\cdots\), \(\gqs[{\Np}]\).
    If we denote by \(\vv_\iq\) the vector
    \(\vv_\iq=(\gqss[\iq][+2],\gqss[\iq][+4],\cdots, \gqss[\iq][+2{\Np}+2])\), then
    the independence of these solutions means that
    the vectors \(\{\vv_\iq|1\leq
    \iq\leq {\Np}\}\) %
    span the \({\Np}\)-dimensional space
    \(\Vect{w_\kk|2\leq\kk\leq {\Np}+1}\) where
    \(w_\kk=(\Ts[1][\kk-2][4-\kk], \Ts[1][\kk-1][5-\kk],\cdots
    ,\Ts[1][\kk+{\Np}-2][{\Np}+4-\kk])\). 
    Therefore
    \(w_\kk\) is a linear
    combination of the vectors \(\vv_\iq\), and there exist functions
    \(\alpha_{\kk,\iq}(\us)\) such that
    \begin{gather}
      \label{eq:CombLinForSolHirPcm}
      w_\kk=\sum_{\iq=1}^{\Np}  \alpha_{\kk,\iq} \vv_\iq \\
      \ie ~~\forall \kk \in \ninter 2 {{\Np}+1},\quad \forall \lL \in \ninter  1
      {{\Np}+1},\qquad
      \Ts[1][\kk+\lL-3][\lL-\kk+3]= \sum_{\iq=1}^{\Np}  \alpha_{\kk,\iq} \gqss[][2\lL]\,.
    \end{gather}

    For instance if \({\Np}=2\), this equation reads %
    \begin{subequations}
      \begin{align}
        (\Ts[1][0][+2], \Ts[1][1][+3], \Ts[1][2][+4])
        =&\alpha_{2,1}(\gqss[1][+2], \gqss[1][+4],\gqss[1][+6])+
        \alpha_{2,2}(\gqss[2][+2], \gqss[2][+4],\gqss[2][+6])\,\\
        (\Ts[1][1][+1],\Ts[1][2][+2],\Ts[1][3][+3])
        =&\alpha_{3,1}(\gqss[1][+2], \gqss[1][+4],\gqss[1][+6])-
        \alpha_{3,2}(\gqss[2][+2], \gqss[2][+4],\gqss[2][+6])\,.
      \end{align}
    \end{subequations}
    Comparing the expressions that it gives for \(\T[1][1]\)
    and \(\T[1][2]\), we deduce that \(\alpha_{3,\iq}=\alpha_{2,\iq}^{[-2]}\)
    (assuming that \((\gqss[1][+4],\gqss[1][+6])\) and 
    \((\gqss[2][+4],\gqss[2][+6])\) are independent).

    For arbitrary \({\Np}\), the same argument gives
    \(\alpha_{\kk+1,\iq}=\alpha_{\kk,\iq}^{[-2]}=\alpha_{2,\iq}^{[4-2\kk]}\),
    if the vectors \(\tilde \vv_\iq=(\gqss[\iq][+4],\gqss[\iq][+6],\cdots,
    \gqss[\iq][+2{\Np}+2])\) are independent. Finally,
    if we define \(\gpf[\overline{\{\iq\}}]{}\equiv \alpha_{2,\iq}^{[-2]}\),
    then the relation 
    \eqref{eq:CombLinForSolHirPcm} says exactly that
    \eqref{eq:HirotaSolutPCM} holds when \(a=1\) and \(0\leq s\leq 2{\Np}-1\).

    Moreover we can find two functions \(\gqe\) and \(\gpf[\ove]{}\) such
    that for \(s\in\ninter 0 1\), \(\T[0]=\gqfs[\emptyset][+s]
    \gpfs[\ove][-s]\). For this it is sufficient to find \(\gqe\) such that
    \(\frac{\gqe[+]}{\gqe[-]}=\frac{\T[1][0]}{\T[0][0][-]}\), and to define 
    \(\gpf[\ove]{}\equiv \T[0][0] / \gqe\).

    Finally, the %
    determinant expression %
    \eqref{eq:qqPCMSolHir2} is
    equivalent (because of the {\jacobi} identity
    \eqref{eq:{\jacobi}}) to the bilinear identity
\index{Q-functions@{\Qfs}!QQ-relations}
    \begin{multline}
      \label{eq:tqtqrelation}
      \forall I\subset \ninter 1 {\Np}\,, \forall {\iq},{\jq} \in \Ib \textrm{ such
        that } {\iq} < {\jq},%
      \\
      \gpf[I]{}\gpf[I,{\iq},{\jq}]{}= \gpf[I,{\iq}]{-}\gpf[I,{\jq}]{+} -
      \gpf[I,{\iq}]{+}\gpf[I,{\jq}]{-}\,,\qquad\qquad \where I,{\iq},{\jq}\equiv I\cup \{{\iq},{\jq}\}
    \end{multline}
    The {\jacobi} identity also implies that this bilinear identity is
    equivalent to 
    \begin{gather}
      \gpf[\overline{\{\iq_1,\iq_2,\cdots,\iq_\nn\}}]{}=
      \frac{\Det{\gpfs[\overline{\{\iq_\kk\}}][{\nn+1-2\lL}]}{1\leq
          \kk,\lL\leq \nn}}{\prod_{\kk=1}^{\nn-1}\gpfs[\ove][{2\kk-\nn}]}\,.
    \end{gather}
    Therefore, we can express
    \(\gps[1]=\gpf[\overline{\{2,3,\cdots,{\Np}\}}]{}\),
    \(\gps[2]=\gpf[\overline{\{1,3,\cdots,{\Np}\}}]{}\), 
    \(\cdots\), \(\gps[{\Np}]=\gpf[\overline{\{1,2,\cdots,{\Np}-1\}}]{}\) in
    terms of the functions \(\gpf[\overline{\{\iq\}}]{}\) and
    \(\gpf[\ove]{}\) defined above.

    To summarize the above construction, we have seen that if \(\T\) is a
    typical solution %
    of the Hirota equation on the
    \((a,s)\)-lattice of the {\PCM} ({\idest} \(\T=0\) if \(a<0\) or
    \(a>{\Np}\)), then we can define \(\gqs[1]\), \(\gqs[2]\),
    \(\cdots\), \(\gqs[{\Np}]\), \(\gps[1]\), \(\gps[2]\),
    \(\cdots\), \(\gps[{\Np}]\), and \(\gqe\) and \(\gpe\)
    such that \eqref{eq:HirotaSolutPCM} holds when \(a=1\) and \(s\in\ninter 1
    {2{\Np}-1}\), and also when \(a=0\) and \(s\in \{0,1\}\). Moreover, there
    exists a %
    determinant identity, analogous to the {\jacobi} identity, which allows to
    prove \cite{springerlink:10.1007/s002200050165} that
    \eqref{eq:HirotaSolutPCM} is a solution of the Hirota equation. 
    Therefore the quantity \(\tilT\equiv\sum_{\substack {I\subset \ninter 1 {\Np}
        \\ |I|=a}} \epsilon(I,\Ib) \gqfs[I][+s]\gpfs[\Ib][-s]\) is a solution
    of the Hirota equation, which coincides with \(\T\) for a set of initial
    values. This implies that \(\tilT=\T\).
  \end{proof}

  In this derivation of \eqref{eq:HirotaSolutPCM}, we have assumed that
  the difference equation \eqref{eq:DetZeroDefq} has \({\Np}\) independent
  solutions, and that the vectors \(\tilde \vv_\iq=(\gqss[\iq][+4],\gqss[\iq][+6],\cdots,
  \gqss[\iq][+2{\Np}+2])\) are independent.

  There exist non-typical solutions of the Hirota equations for which
  these hypotheses do not hold, and for which the determinant expression
  \eqref{eq:HirotaSolutPCM} does not hold either. Though it is a bit
  more technical to prove, these hypotheses hold for typical solutions,
  hence the statement \ref{sec:statmtSolHirotPCM} is a statement about
  typical solutions.  As explained in the previous sections, this
  typicality condition should hold for the solution of Hirota equations
  obtained from the solutions of the {\Ysys} equations arising
  from the {\TBA}.

  In \eqref{eq:HirotaSolutPCM}, we see that two sets of functions,
  denoted by \(\qfu\) and \(\pfu\) play a completely symmetric role. In
  what follows, both of them will be called {\qfs}, although they are
  denoted by two different letters \(\qfu\) and \(\pfu\).

  \paragraph{Expression in terms of forms}
  \label{sec:expr-terms-forms}

Let us comment a little bit on the form of the expression
\eqref{eq:HirotaSolutPCM} proven above. We will see {\below} that the
expression \eqref{eq:HirotaSolutPCM} has the same structure as the
expressions for the rectangular representations of {\csds},
written in section \ref{sec:wronsk-expr-qq} of the previous chapter.

In particular they can be written as a determinant
\cite{springerlink:10.1007/s002200050165},
like for {\csds} where we saw that the expression
\eqref{eq:TasGLKfromNestexpanded} (which has the same form as
\eqref{eq:HirotaSolutPCM}) was identical to the determinant expression
\eqref{eq:WronskianLambda} (as it could be seen by expanding the
determinant with respect to the first lines).
More explicitly, in the present case the expression
\eqref{eq:HirotaSolutPCM} is identical to
\begin{gather}
\label{eq:DetSolPCM}
  \fdisp{ \T= \frac{
\BlockDet{\gqss[\jq][+s-1-a+2\ii]}{\substack{1\leq \jq\leq\Np\\
1\leq \ii \leq a
}}{\gpss[\jq]%
[-s-1-a-\Np+2\ii]
}
{\substack{1\leq \jq\leq\Np\\
a+1\leq \ii \leq \Np-a
}}
}{\prod_{\kk=1}^{a-1}\gqes[+s-a+2\kk] \prod_{\kk=1}^{\Np-a-1}\gpes[-s-\Np+a+2\kk]}}.
\end{gather}
For instance when \(a=1\), we see that if we expand the determinant
\eqref{eq:DetSolPCM} with respect to the first line, we get \(\Np\)
terms where each term is equal to \((-1)^\jq \gqss[\jq][+s]\) times the
corresponding \((\Np-1)\times(\Np-1)\) minor. Finally, using the
definition \eqref{eq:qqPCMSolHir2} of multi-indexed {\qfs}, we
recognize that the expression \eqref{eq:DetSolPCM} and
\eqref{eq:HirotaSolutPCM} are identical when \(a=1\). When \(a\neq 1\),
one can use the ``generalized Laplace expansion formula'', which says
how to expand a determinant with respect to several lines at once. If
we expand with respect to the \(a\) first lines at once, this formula
gives a sum of terms which are products of \(a\times a\) minors
(corresponding to the first lines) multiplied by
\((\Np-a)\times(\Np-a)\) minors (corresponding to the last lines), and
we obtain that the {\Wronskian} determinant \eqref{eq:DetSolPCM} is
identical to the expression \eqref{eq:HirotaSolutPCM}.

\index{exterior forms}
Another convenient way to rewrite this expression is by means of
``exterior forms'' \cite{Gromov:2011cx}. Let us define \(\Np\) objects
\(\xi_1\), \(\xi_2\), 
\(\cdots\), \(\xi_\Np\), and an antisymmetric (and linear) product
\(\wedge\)  such that \(\xi_1 \wedge \xi_2 \wedge \cdots \wedge
\xi_\Np=1\). Then the linearity and antisymmetry of the product
\(\wedge\) imply that for arbitrary coefficients
\index{exterior forms}
\(\left(c_{\ii,\jj}\right)_{\substack{1\leq\ii,\jj\leq \Np}}\), we
have
\begin{align}
\left(\sum_{\ii=1}^{\Np} c_{{\ii},1}~\xi_1\right)\wedge \left(\sum_{\ii=1}^{\Np}
  c_{{\ii},2}~\xi_2\right)\wedge \cdots \wedge \left(\sum_{\ii=1}^{\Np}
  c_{{\ii},\Np}~\xi_\Np\right) = \Det{c_{\ii,\jj}}{1\leq\ii,\jj\leq \Np}\,,
\end{align}
so that we can view this product \(\wedge\) as a formal %
antisymmetric multiplication of 
objects, designed to give rise to determinants.
In what follows,
we call \(1\)-form an arbitrary linear
\index{exterior forms!(n)-form}
combination of \(\xi_1\), \(\xi_2\),
\(\cdots\), and  \(\xi_\Np\). The product of \(\nn\) (1)-forms is an
(\(\nn\))-form, and by definition a (0)-form is a usual
scalar.

Let us now rewrite, in this language, the expression
\eqref{eq:DetSolPCM} of {\Tfs} (which is identical to the expression
\eqref{eq:HirotaSolutPCM}). To this end, let us define %
  the (1)-forms 
  \begin{gather}
\label{eq:-fdisp-gqf1equiv}
    {\gqf[(1)]{}\equiv \sum_{\iq=1}^\Np \gqs[\iq] \xi_\iq},\qquad\qquad
    {\gpf[(1)]{}\equiv \sum_{\iq=1}^\Np \gps[\iq] \xi_\iq}\,.
  \end{gather}
  
  If we also introduce the ({\nn})-forms 
  \begin{gather}
\label{eq:Defqnform}
    {\gqf[({\nn})]{}\equiv 
      \frac{\gqfs[(1)][-{\nn}+1]\wedge \gqfs[(1)][-{\nn}+3]\wedge
        \gqfs[(1)][-{\nn}+5]\wedge \cdots\wedge \gqfs[(1)][{\nn}-1]}
      {\gqes[-{\nn}+2]\gqes[-{\nn}+4]\cdots \gqes[{\nn}-2]}},\qquad\qquad {\nn}>1
    \,,\\
\label{eq:Defpnform}
    {\gpf[({\nn})]{}\equiv 
      \frac{\gpfs[(1)][-{\nn}+1]\wedge \gpfs[(1)][-{\nn}+3]\wedge
        \gpfs[(1)][-{\nn}+5]\wedge \cdots\wedge \gpfs[(1)][{\nn}-1]}
      {\gpes[-{\nn}+2]\gpes[-{\nn}+4]\cdots \gpes[{\nn}-2]}},\qquad\qquad {\nn}>1\,,
    \\
    \gqf[(0)]{}\equiv \gqe,\qquad\qquad\gpf[(0)]{}\equiv \gpe
    \,,
\label{eq:gqf0-gqe-qquadqq}
  \end{gather}
  then we see that
the formula \eqref{eq:DetSolPCM} can be rewritten as
  \begin{gather}
    \label{eq:formHirotaPCM}
    \fdisp{\T=\gqfs[(a)][+s]\wedge \gpfs[(\Np-a)][-s]}\,.
  \end{gather}

Moreover, if we wish to compare this expression with the initial
notation (\ref{eq:HirotaSolutPCM}-\eqref{eq:qqPCMSolHir2}),
we see that (if \(\iq_1<\iq_2<\cdots<\iq_\nn\)) the function
  \(\gqf[\{\iq_1,\iq_2,\cdots,\iq_\nn\}]{}\) is the coefficient of 
  \(\xi_{\iq_1} \wedge \xi_{\iq_2} \wedge \cdots \wedge \xi_{\iq_\nn}\) in 
  \(\gqf[({\nn})]{}\), and we see that the expression
  \eqref{eq:formHirotaPCM} and \eqref{eq:HirotaSolutPCM} are
  identical.

The notation \(\gqf[\{\iq_1,\iq_2,\cdots,\iq_\nn\}]{}\) (with curly
brackets around the indices) suggests that,
as in chapter \ref{part:qoperatorsspin}, we have 
  \begin{align}
\forall \sigma \in \Sgrp \nn,
\qquad
\qquad
\gqs[\iq_{\sigma(1)},\iq_{\sigma(2)},\cdots,\iq_{\sigma(\nn)}]{}=\gqs[\iq_1,\iq_2,\cdots,\iq_\nn]{}\,.
  \end{align}
On the other hand, as we see that  \(\gqf[\{\iq_1,\iq_2,\cdots,\iq_\nn\}]{}\) is the coefficient of 
  \(\xi_{\iq_1} \wedge \xi_{\iq_2} \wedge \cdots \wedge
  \xi_{\iq_\nn}\) (which is antisymmetric), it is actually more
  natural to use a notation (denoted without curly brackets) where
\begin{align}
\forall \sigma \in \Sgrp \nn,
\qquad
\qquad
\gqf[\iq_{\sigma(1)},\iq_{\sigma(2)},\cdots,\iq_{\sigma(\nn)}]{}=\epsilon(\sigma) ~
\gqf[\iq_1,\iq_2,\cdots,\iq_\nn]{}\,,
\label{eq:Defqmulti1}\\
\And \gqf[\iq_1,\iq_2,\cdots,\iq_\nn]{} =
\gqs[\iq_1,\iq_2,\cdots,\iq_\nn]{} \,, \qquad\qquad\If
\iq_1<\iq_2<\cdots<\iq_\nn\,.
\label{eq:Defqmulti2}
  \end{align}
With this notation we see that \(\gqf[\iq_1,\iq_2,\cdots,\iq_\nn]{}\) is
the coefficient of \(\xi_{\iq_1} \wedge \xi_{\iq_2} \wedge \cdots \wedge \xi_{\iq_\nn}\) in 
  \(\gqf[({\nn})]{}\). For the simplicity of notations, we will try to
  use, as much as possible, this antisymmetric definition, denoted
  without curly brackets.

We can also note that the definition \eqref{eq:Defqnform} is
equivalent\footnote{This equivalence has exactly the same
    proof as in section \ref{sec:jacobiident-bilin} where the {\jacobi}
    identity shows the equivalence between a determinant expression
    and a bilinear identity.
} to the following qq-relation \cite{0022-3719-16-34-009,1992PhRvB..4614624B,Bazhanov:2001xm,2000JPhA...33.8267P,Dorey:2006an}: \index{Q-functions@{\Qfs}!QQ-relations}
\begin{gather}
\label{eq:qqrelation}
  \gqf[%
  \cdots,%
  \bjq,\bkq]{}
  \gqf[%
  \cdots]{} =
  \gqf[%
  \cdots,%
  \bjq]{-}
  \gqf[%
  \cdots,%
  \bkq]{+} -
  \gqf[%
  \cdots,%
  \bjq]{+}
  \gqf[%
  \cdots,%
  \bkq]{-}\,.
\end{gather}
where ``\(_{\cdots}\)'' stands for an arbitrary set of indices, and
where the {\Qfs} \(\gqf[\iq_1,\iq_2,\cdots,\iq_\nn]{}\), defined in 
(\ref{eq:Defqmulti1}-\ref{eq:Defqmulti2}), are the coordinates of the
form \(\gqf[({\nn})]{}\) defined by \eqref{eq:Defqnform}.

\paragraph{Generalization to other lattices}
\label{sec:gener-other-latt}

  Let us now generalize this result to other lattices %
  starting with the lattice \(\Sh(\Kr)\) (corresponding for instance
  to an \(\SU\Kr\)-symmetric {\cds}):
  \begin{statmt}
\label{sec:statmtSolHirSh}    Let \(\T\) be any typical solution of the Hirota equation on
    \(\Sh(\Kr)\) such that the ``{\Wronskian} gauge'' condition
    \eqref{eq:PhyWronGauge} is satisfied. Then there exist \(\Kr\)
    functions \(\gqf[1]{}\), \(\gqf[2]{}\),   \(\cdots\), \(\gqf[\Kr]{}\) such
    that 
    \begin{gather}
      \label{eq:HirotaSolSpChn}
      \fdisp{\T=
        (-1)^{(\Kr-1)(\Kr-a)}
        \gqfs[(a)][+s]\wedge \gqfs[(\Kr-a)][-s-\Kr]}\,,\\
      \where 
      {\gqf[(1)]{}\equiv \sum_{\iq=1}^\Kr \gqf[\iq]{} \xi_\iq}\,,\qquad\qquad
      \And \gqf[(0)]{}\equiv 1\,,\\
      \And
      \gqf[({\nn})]{}\equiv 
      \gqfs[(1)][-{\nn}+1]\wedge \gqfs[(1)][-{\nn}+3]\wedge
      \gqfs[(1)][-{\nn}+5]\wedge \cdots\wedge \gqfs[(1)][{\nn}-1]
      \qquad \If {\nn}>1\,.
    \end{gather}
  \end{statmt}
  We can notice that the main difference with the equation
  \eqref{eq:HirotaSolutPCM} is that for the lattice \(\St(\Np)\), ({\idest} for
  the symmetry group \(\SU\Np\times \SU\Np\)), we had two sets of \(\Np\)
  functions, whereas here we only have one set of \(\Kr\) functions (for
  the group \(\SU \Kr\)). Another remark is that the gauge condition
  \eqref{eq:PhyWronGauge} allows to set \(\gqf[(0)]{}=1\).

  This solution of Hirota equation was first introduced in
  \cite{1997CMaPh.190..247B} for \(\Kr=2\), and generalized in
  \cite{springerlink:10.1007/s002200050165} to arbitrary rank \(\Kr\).
  \begin{proof}
To prove this statement, we will first show that (when \(s\geq -1\)),
\(\T\) is given by the same expression \eqref{eq:HirotaSolutPCM} (or
\eqref{eq:DetSolPCM} in terms of a {\Wronskian} determinant, or
\eqref{eq:formHirotaPCM} in terms of exterior forms) as in the
previous statement. Then we will translate the additional constraint
\(\T[][-1]=0\) (which makes the difference between the lattice \(\Sh(\Np)\)
and \(\St(\Np)\)) into a constraint on the {\qfs}, which will allow to
prove that the functions \(\gpf[\iq]{}\) and \(\gqf[\iq]{}\) coincide up to a shift and
to a rescaling.

    The construction used in the proof of the
    \stapref{sec:statmtSolHirotPCM} allows to find two sets of
    functions \(\gqf[1]{}\), \(\gqf[2]{}\), 
    \(\cdots\), \(\gqf[{\Np}]{}\), \(\gpf[1]{}\), \(\gpf[2]{}\), 
    \(\cdots\), \(\gpf[{\Np}]{}\), and two additional functions \(\gqe\) and
    \(\gpe\) 
    such that the equation \eqref{eq:HirotaSolutPCM} (or equivalently
    \eqref{eq:formHirotaPCM}) holds (at least for initial values). In this construction,
    the two 
    functions \(\gqe\) and \(\gpf[\ove]{}\) simply have to obey
    \begin{equation}
      \forall s\in\ninter 0
      1,\qquad\T[0]=\gqfs[\emptyset][+s] \gpfs[\ove][-s]\,.
    \end{equation}
    In view of the {\Wronskian} gauge condition \eqref{eq:PhyWronGauge}, we can choose
    \(\gqf[\emptyset]{}=1\) and \(\gpf[\ove]{}=\T[0][0]\). We can also notice
    that a particular case of \eqref{eq:HirotaSolutPCM} is 
    \(\T[\Kr]=\gqf[\ove]{[+s]}\gpf[\emptyset]{[-s]}\). Then the gauge condition
    \(\T[\Kr][s+1] = \T[\Kr][][+]\) gives
    \(\gpf[\emptyset]{+}=\gpf[\emptyset]{-}\), which means that
    \(\gpf[\emptyset]{}\) is \(\bi\)-periodic.

    Compared to the \stapref{sec:statmtSolHirotPCM}, {\another} important
    difference is that now, the {\Tfs} should obey the boundary condition
    \begin{equation}
      \label{eq:Tm1eZ}
      \forall a\in\ninter 1 {\Kr-1},\qquad  \T[][-1]=0\,.
    \end{equation}

    In the proof of \staref{sec:statmtSolHirotPCM}, the equation %
    \eqref{eq:HirotaSolutPCM} (or equivalently
    \eqref{eq:formHirotaPCM}) holds by construction for initial values,
    and by recurrence it holds for every \((a,s)\). In the present case,
    the same recurrence proves that \eqref{eq:HirotaSolutPCM} holds when
    \(s\geq -1\). If \(s\) is smaller, then the recurrence {\cannnot} be
    performed because it would involve a division by zero.
    But this result is exactly enough to rewrite \eqref{eq:Tm1eZ}
    in terms of the expression \eqref{eq:formHirotaPCM}, and obtain %
    the vanishing of 
    the following determinants: %
    \begin{equation}
      \label{eq:Tm1DeZ}  \forall a\in\ninter 1 {\Kr-1},\qquad  \T[][-1]=
      \BlockDet{\gqfs[\iq][-a-2+2\jj]}{\substack{
          1\leq \iq\leq \Kr\\
          1\leq\jj\leq a}}
      {\gpfs[\iq][-\Kr-a+2\jj]}
      {\substack{
          1\leq \iq\leq \Kr\\
          a+1\leq\jj\leq \Kr}}=0\,.
    \end{equation}
    At \(a=\Kr-1\), it implies that there exist \(\Kr-1\) functions \(c_\kk\)
    such that
    \begin{gather}
      \label{eq:linCombqTildQ}
      \gpf[(1)]{}=\sum_{\kk=0}^{\Kr-2}c_\kk~\gqf[(1)]{[-\Kr+2\kk]}\,,\\
      \where {\gpf[(1)]{}\equiv \sum_{\iq=1}^\Kr \gpf[\iq]{}
        \xi_\iq}\,,\qquad \And {\gqf[(1)]{}\equiv \sum_{\iq=1}^\Kr \gqf[\iq]{}
        \xi_\iq}\,.
    \end{gather}

    Then we can write the condition \eqref{eq:Tm1DeZ} at \(a=\Kr-2\). 
    The determinant has its two last lines made of functions
    \(\gpf[\iq]{}\), and in particular, the last line is
    \(\gpf[(1)]{[+2]}\). We have just shown that \( \gpf[(1)]{[+2]} = 
    \sum_{\kk=0}^{\Kr-2}c_\kk^{[+2]}\gqf[(1)]{[-\Kr+2\kk+2]}
    \), which allows to replace the last line of the determinant by this
    sum.
    The terms \(c_0^{[+2]} \gqf[(1)]{[-\Kr+2]}\),  \(c_1^{[+2]}
    \gqf[(1)]{[-\Kr+4]}\), \(\cdots\), \(c_{\Kr-3}^{[+2]} \gqf[(1)]{[+\Kr-4]}\) give a
    determinant equal to zero (where two lines are equal), and the only
    remaining term is \(c_{\Kr-2}^{[+2]} \gqf[(1)]{[-\Kr+2]}\). That gives
    \begin{align}
      - c_{\Kr-2}^{[+2]} \BlockDet{\gqss[][-{\Kr}+2\jj]}{\substack{
          1\leq \iq\leq \Kr\\
          1\leq\jj\leq \Kr-1}}
      {\gps[]}
      {\substack{
          1\leq \iq\leq \Kr}}=&0\,,
    \end{align}
    where we can notice that the determinant is equal to
    \(\T[\Kr-1][0]\neq 0\). Therefore we obtain \(c_{\Kr-2}=0\). 

    Next we can write the condition \eqref{eq:Tm1DeZ} at \(a=\Kr-3\). We can
    plug the expression \eqref{eq:linCombqTildQ} into the last line of the
    determinant, and we obtain \(c_{\Kr-3}=0\). We can the repeat the
    argument for \(a=\Kr-5\), \(a=\Kr-7\), etc up to \(a=1\). That gives
    \(c_{\Kr-2}=c_{\Kr-5}=\cdots=c_{1}=0\). Hence we obtain that there
    exists a function \(c_0(\us)\) such that 
    \begin{align}
      \gpf[(1)]{}=c_0~\gqf[(1)]{[-\Kr]}\,.
    \end{align}

    Let us now find an expression for \(c_0\). To this end, let us write down
    \begin{align}
      \T[\Kr-1][0]=&\BlockDet{\gqfs[\iq][-{\Kr}+2\jj]}{\substack{
          1\leq \iq\leq \Kr\\
          1\leq\jj\leq \Kr-1}}
      {\gpf[\iq]{}}
      {\substack{
          1\leq \iq\leq \Kr}}\\=&(-1)^{\Kr-1}c_0 \Det {\gqfs[\iq][-{\Kr}-2+2\jj]}{\substack{
          1\leq \iq,\jj\leq \Kr}}=\frac {(-1)^{\Kr-1}c_0}{\gpe[-]}\T[\Kr-1][0][-]\,.
    \end{align}
    As compared to the gauge condition \eqref{eq:PhyWronGauge}, this gives
    \begin{align}
      c_0=(-1)^{\Kr-1} \gpe[-]\,.
    \end{align}

    Finally, we should rescale the functions \(\gqf[\iq]{}\) in order to
    obtain the expression \eqref{eq:HirotaSolSpChn}. Let us define 
    \begin{gather}
      \gqbf[\iq]{}
      =f~%
      {\gqf[\iq]{}}%
      \,,\qquad \where f\equiv
      \left(\gpes[+\Kr+1]\right)^{1/\Kr}\,.
    \end{gather}
    This definition is such that 
    \Pv{\begin{align}
        \T[\Kr]=&\gqf[\ove]{[+s]}\gpf[\emptyset]{[-s]}=
        \gpf[\emptyset]{[-s]} \Det {\gqfs[\iq][-{\Kr}-1+2\jj+s]}{\substack{
            1\leq \iq,\jj\leq \Kr}}\\
        =&\Det {%
          {\gqbf[\iq]%
          {[-{\Kr}-1+2\jj+s]}}}{\substack{
            1\leq \iq,\jj\leq \Kr}} \frac{\gpf[\emptyset]{[-s]}} {
          \prod_{\jj=1}^\Kr f^{[-\Kr-1+2\jj+s]}
        }\\
        =&\Det {%
          {\gqbf[\iq]%
          {[-{\Kr}-1+2\jj+s]}}}{\substack{
            1\leq \iq,\jj\leq \Kr}} = %
        {\gqbf[(\Kr)]%
        {[+s]}}\,,
      \end{align}
      \begin{gather}%
        \where 
        {%
          {\gqbf[(0)]{}}}\equiv 1,\quad \And
        {%
          {\gqbf[({\nn})]{}}\equiv 
          {%
            {\gqbf[(1)]%
            {[-{\nn}+1]}}\wedge %
            {\gqbf[(1)]%
            {[-{\nn}+3]}}\wedge
            \cdots\wedge %
            {\gqbf[(1)]%
            {[{\nn}-1]}}}
        },\quad \textrm{ for } {\nn}>1\,.
      \end{gather}}
    To see this, we used the identity \(\frac{\gpf[\emptyset]{[-s]}} {
      \prod_{\jj=1}^\Kr f^{[-\Kr-1+2\jj+s]} }=1\), which arises because 
    \(\gpf[\emptyset]{}\) is \(\bi\)-periodic.

    We can also notice that (for \(s\geq -1\))
    \begin{align}
      \T[\Kr-1]=&\BlockDet {\gqfs[\iq][-{\Kr}+2\jj+s]}{\substack{
          1\leq \iq\leq \Kr\\1\leq \jj\leq \Kr-1}}
      {\gpfs[\iq][-s]}{\substack{
          1\leq \iq\leq \Kr}}
      =c_0^{[-s]}
      \BlockDet {\gqfs[\iq][-{\Kr}+2\jj+s]}{\substack{
          1\leq \iq\leq \Kr\\1\leq \jj\leq \Kr-1}}
      {\gqfs[\iq][-\Kr-s]}{\substack{
          1\leq \iq\leq \Kr}}
      \\
      =&(-1)^{\Kr-1} \frac{\gpes[-s-1]}{\left(f^{[\Kr+s]}\right)^\Kr} \BlockDet {
        {\gqbfs[\iq]%
          [-{\Kr}+2\jj+s]%
      }}{\substack{
          1\leq \iq\leq \Kr\\1\leq \jj\leq \Kr-1}}
      {%
          \gqbfs[\iq]%
          [-\Kr-s]%
      }{\substack{
          1\leq \iq\leq \Kr}}
      \\ =&(-1)^{\Kr-1} %
        \gqbf[(\Kr-1)]%
      {[+s]}\wedge %
      {\gqbf[(1)]%
      {[-\Kr-s]}}
    \end{align}
    Therefore we have proven that 
\(\T=
        (-1)^{(\Kr-1)(\Kr-a)}
        \gqbfs[(a)][+s]\wedge \gqbfs[(\Kr-a)][-s-\Kr]\) 
holds %
    when
    \(a=\Kr\) and when \(a=\Kr-1\), which implies\footnote{Indeed, the
      set
    \(\left\{\T[\Kr][1]
    \right\}\cup \left\{\T[\Kr-1]\middle|  \leq s\leq \Kr-1
    \right\}\)
can be used as initial value for the Hirota equation on the lattice
\(\Sh(\Kr)\) under the {\Wronskian} gauge condition.
} that it holds on the whole
    lattice \(\Sh(\Kr)\). If we finally rename the {\qfs} as
    \({\gqbf[]{}}\rightsquigarrow {\gqf[]{}}\), this proves the
    relation \eqref{eq:HirotaSolSpChn}.
  \end{proof}

  We can note that the solution \eqref{eq:HirotaSolSpChn} is very
  similar to the expression \eqref{eq:TasGLKfromNestexpanded} obtained
  for {\csds}. This is not a surprise because these two expressions %
  are solution of
  the Hirota equation on the same lattice \(\Sh(\Kr)\). More precisely we
  can see that these expressions coincide up to the change of variables
  \begin{gather}
    \label{eq:qvsQ}
    \gqf[%
    \iq_1,\iq_2,\cdots,\iq_\nn%
    ]{}(\us)=
    \ffQ[-\bi \us+%
    \frac{\nn%
    }{2}][\{\iq_1,\iq_2,\cdots,\iq_\nn\}][{ }]
    \prod_{\kk=1}^\nn x_{\iq_\kk}^{-\bi {\us}-\frac \nn 2}
    \prod_{\lL=\kk+1}^\nn (x_\lL-x_\kk)\,.%
  \end{gather}

  The {\Qfs} of chapter \ref{part:qoperatorsspin} obeyed QQ-relations
  which involved the eigenvalues \(x_\jvp\) of the twist \(\g\) (see
  \eqref{eq:qq++}), whereas after the change of variables
  \eqref{eq:qvsQ}, the qq-relation does not contain a twist anymore (see 
  for instance \eqref{eq:tqtqrelation} and \eqref{eq:qqrelation}). 
  Moreover, we see that in \eqref{eq:qvsQ}, we have defined the left
  hand side as an antisymmetric function of the indices 
\(\iq_1,\iq_2,\cdots,\iq_\nn\), motivated by the exterior forms
  formalism.
  In the {\rhs}, we also see an antisymmetry which comes from
  the factor \(\prod_{\lL=\kk+1}^\nn (x_\lL-x_\kk)\). We see that the
  twist is responsible for this difference between the present
  chapter and the chapter \ref{part:qoperatorsspin}, where we
  explicitly constructed {\Qoprs} which are symmetric functions of
  their indices.

  Another difference between the {\Qfs} of chapter
  \ref{part:qoperatorsspin} and the present {\qfs} is that in general the
  {\qfs} are not polynomial. Indeed, they are constructed for an
  arbitrary solution of the Hirota equation, which may or may not be
  polynomial.

  Due to these differences, we use a small letter for the {\qfs} of this
  chapter, as opposed to the capital letter of the {\Qfs} of chapter
  \ref{part:qoperatorsspin}. The capital letter
  {\Pv{\ensuremath{\fQf}}} will also be used in 
  this section to denote polynomial functions, 
whose roots will be the rapidities of excitations (``particles'').

  For an arbitrary ``\(\Tk\)-{\hook}'' \(\Tk(\Kr,\Kr'|\Mr+\Mr')\), %
  the typical solution of Hirota equation can also be easily written
  in terms of {\qfs}.
  This solution is written in the statement {\below}, and it was first
  written for \((\Kr, \Kr'+\Mr'|\Mr)=(2,0|1)\) in
  \cite{2007JSMTE..01....5B,2008NuPhB.805..451B}, then generalized to
  an arbitrary lattice \(\HK(\Kr,\Mr)=\Tk(\Kr,0|\Mr)\) (corresponding to
  {\GLKM} {\csds}) in \cite{2010NuPhB.826..399T}.
  For ``\(\Tk\)-{\hook}'', the character solution ({\idest} the solution of
  Hirota equation without spectral parameter) was first written in
  \cite{Gromov:2010vb} for the AdS/CFT ``\(\Tk\)-{\hook}''
  \(\Tk(2,2|4)\). We then generalized it to the \(\us\)-dependent
  {\Wronskian} solution of the Hirota equation  in
  \cite{Gromov:2010km}, and the 
  generalization to an arbitrary ``\(\Tk\)-{\hook}'' was written in \cite{Tsuboi:2011iz}.
 The proof can in principle be obtained by the same
  method as the \stapref{sec:statmtSolHirSh}, {\idest} by first finding a
  {\Wronskian} expression in terms of too many functions\footnote{More
    precisely, we would a priori obtain \(2\Kr+2\) functions to describe
  the domain \(s\geq \Mr+a-\Kr\), plus \(2(\Mr+\Mr')+2\) functions to describe
  the domain where \(a\geq s+\Kr-\Mr\) and  \(a\geq -s\), plus \(2\Kr'+2\) functions to describe
  the domain \(s\leq \Kr'-\Mr'-a\).}, and then showing that they are not
independent of each other and they can be expressed in terms of only
\(\Kr+\Kr'+\Mr+\Mr'+1\) functions. However, this is more technical
than for the \staref{sec:statmtSolHirSh}, and is not 
  crucial for what follows, hence this solution will be given here
  without proof.
  \begin{statmt}
\label{sta:Thooksol}
     Let \(\T\) be any typical solution of the Hirota equation on
    \(\Tk(\Kr,\Kr'|\Mr+\Mr')\) such that the ``{\Wronskian} gauge'' condition
    \eqref{eq:PhyWronGaugeTK} is satisfied.

    Then there exists a set of functions \( \cqbf[1]{}\), \( \cqbf[2]{}\),
    \(\cdots\), \( \cqbf[\Kr+\Kr'+\Mr+\Mr']{}\)  and an additional function
     \(\cqbe\)
    such that 
    \Pv{\begin{empheq}[left={\T=\empheqlbrace}]{align}
\label{eq:TTHook1}
      &\sum_{\substack {I\subset\ninter 1 \Kr\\
        |I|=a}} \epsilon(I,\Ib)
    \cqfs[I][+s+\frac{\Kr-\Kr'-\Mr+\Mr'}2]\cqfs[\Ib][-s-\frac{\Kr-\Kr'-\Mr+\Mr'}2]\\[-1cm]
     &&\mathllap{\If s\geq \Mr+a-\Kr}\nonumber\\[1cm]
\label{eq:TTHook2}
&\upepsilon^{\uu}_{a,s}\sum_{\substack {\mathclap{F\subset\ninter {\Kr+1} {\Kr+\Mr+\Mr'}}\\
        |F|=\Mr-s\\
      I=\ninter 1 \Kr \cup F}}
  \epsilon(I,\Ib) \cqfs[I][+a
+\frac{-\Kr-\Kr'+\Mr+\Mr'}2
] \cqfs[\Ib][-a-\frac{-\Kr-\Kr'+\Mr+\Mr'}2
]\\[-1cm]
 &&\mathllap{\If a\geq s+\Kr-\Mr \And  a\geq -s} \mathrlap{+\Kr'-\Mr'}\nonumber\\[1cm]
\label{eq:TTHook3}
&\upepsilon^{\lL}_{a,s}\sum_{\substack {\mathclap{B\subset\ninter {\Kr+\Mr+\Mr'+1} {\Kr+\Kr'+\Mr+\Mr'}}\\
        |B|=\Kr'-a\\
      ~~I=\ninter 1 {\Kr+\Mr+\Mr'} \cup B~~}}
  \epsilon(I,\Ib) \cqfs[I][-s+\frac{-\Kr+\Kr'+\Mr-\Mr'}2]
  \cqf[\Ib]{\mathrlap{[+s-\frac{-\Kr+\Kr'+\Mr-\Mr'}2]}}\\[-1cm]
&&\mathllap{\If s\leq -a+\Kr'-\Mr'}\nonumber
    \end{empheq}}
  \begin{gather}
    \where \upepsilon^{\uu}_{a,s} \equiv(-1)^{(a+1+\Kr+\Kr'+\Mr+\Mr')(s+\Kr'+\Mr')},\\
    \upepsilon^{\lL}_{a,s} \equiv
(-1)^{a(\Kr+\Kr'+\Mr+\Mr')}
  \end{gather}
  \begin{gather}
\label{eq:and-gqfiequiv-gqbfj}
\And \cqf[I]{}\equiv \cqbf[J]{},\qquad \where 
J\equiv(F\cup I)\setminus(F\cap I)\,,\qquad
F=\ninter{\Kr+1}{\Kr+\Mr+\Mr'}\,,\\
\And
    \cqbf[\{\jq_1,\jq_2,\cdots,\jq_\nn\}]{}=
    \frac{\Det{\cqbfs[\jq_\kk][{-1-\nn+2\lL}]}{1\leq \kk,\lL\leq
        \nn}}{\prod_{\kk=1}^{\nn-1}\cqbes[{2\kk-\nn}]}\,,\qquad
\If \jq_1<\jq_2<\cdots<\jq_\nn\,.
\label{eq:and-gqbfjq_1-jq_2}
  \end{gather}

Moreover, the 
functions \( \cqbf[1]{}\), \( \cqbf[2]{}\),
    \(\cdots\), \( \cqbf[\Kr+\Kr'+\Mr+\Mr']{}\)  and 
     \(\cqbe\)
can be chosen such that
\begin{gather}
\label{eq:THookQemptyo}
      \cqbf[F]{}=
    \frac{\Det{\cqbfs[\Kr+\kk][{-1-\nn+2\lL}]}{1\leq \kk,\lL\leq
        \Mr+\Mr'}}{\prod_{\kk=1}^{\nn-1}\cqbes[{2\kk-\nn}]}=1.
\end{gather}

  \end{statmt}
The letter {\cqfu} (resp {\(\cqbfu\)}) denote here the {\Qfs} associated
to a {\sugr} (and to a lattice \(\Tk(\Kr,\Kr'|\Mr+\Mr')\) or
\(\HK(\Kr,\Mr)\), cf \figref{fig:HK-YvsT} and \figpref{fig:Tk-YvsT}),
whereas the letter {\qfu} (in the previous statements) denotes the {\qfs} 
associated to the group \(\SU \Kr\) or \(\SU \Np \times \SU \Np\). The
choice of a different letter is simply aimed at preventing confusions
in the chapter \ref{cha:dualite-adscft}, and does not have a deep
physical meaning.

We can see that the expressions (\ref{eq:TTHook1}-\ref{eq:TTHook3})
have the same structure as the equation
\eqref{eq:super-rectangular-Toprs}: namely the {\Tfs} are sums of
terms of the form \(\cqfs[I][+\alpha]\cqfs[\Ib][-\alpha]\) where
\(\alpha\) is equal to a constant plus \(a\) or \(|s|\) (depending on the domain).
In
the row \(a=0\) of the ``right band'' (which means the domain \(s\geq
\Mr+a-\Kr\)), the set \(I\) is equal to 
\(\emptyset\), and when \(a\) increases, \(I\) acquires new elements which
belong to \(\ninter 1 \Kr\). Then we arrive to the ``upper band'' ({\idest} the domain \(a\geq
s+\Kr-\Mr \And  a\geq -s\)). At the boundary (\(s=\Mr\)) the
{\Tfs} are the same as at the boundary \(a=\Kr\) of the right band,
and that is a consequence of the gauge constraint
\eqref{eq:PhyWronGaugeTK}. Then, when \(s\) decreases, \(I\) acquires new
elements which belong to the set \(\ninter {\Kr+1}
{\Kr+\Mr+\Mr'}\). Finally, in the last domain \(s\leq -a+\Kr'-\Mr'\) (the
``left band''), the set \(I\) acquires the elements of \(\ninter
{\Kr+\Mr+\Mr'+1} {\Kr+\Kr'+\Mr+\Mr'}\).
Therefore the set of indices \(\ninter 1 \Kr\) is associated to the
right band, while \(\ninter {\Kr+1}
{\Kr+\Mr+\Mr'}\) is associated to the upper band and \(\ninter
{\Kr+\Mr+\Mr'+1} {\Kr+\Kr'+\Mr+\Mr'}\) is associated to the left band.
This structure is summarized in figure \ref{fig:HkBnd} (in the
notation of exterior forms).

Like in the particular case of the lattice \(\HK(\Kr,\Mr)\), which was
studied in chapter \ref{part:qoperatorsspin} for {\csds}, we will
see (see (\ref{eq:cQcQbb},\ref{eq:cQcQbf})) that the
{\cqfu}{\cqfu}-relations are 
modified by the grading of the indices. On the other hand the
functions {\cqbfu} obey {\cqbfu}{\cqbfu}-relations which are
independent of the grading. This constructions comes from the
``bozonisation trick'' of equation \eqref{eq:QQopDetGLKM} in section
\ref{sec:wronsk-expr-qq} (see also \cite{Gromov:2010km}). More
precisely, the formula 
\eqref{eq:and-gqbfjq_1-jq_2} ensures the following
{\cqbfu}{\cqbfu}-relation, which is not grading-dependent.
\begin{gather}
\label{eq:cqbcqbrelation}
  \cqbf[\iq_1,\iq_2,\cdots,\iq_\nn,~\bjq,\bkq]{}
  \cqbf[\iq_1,\iq_2,\cdots,\iq_\nn]{} =
  \cqbf[\iq_1,\iq_2,\cdots,\iq_\nn,~\bjq]{-}
  \cqbf[\iq_1,\iq_2,\cdots,\iq_\nn,~\bkq]{+} -
  \cqbf[\iq_1,\iq_2,\cdots,\iq_\nn,~\bjq]{+}
  \cqbf[\iq_1,\iq_2,\cdots,\iq_\nn,~\bkq]{-}\,, \\[.5cm]
\where 
\cqbf[\iq_{\sigma(1)},\iq_{\sigma(2)},\cdots,\iq_{\sigma(\nn)}]{}=\epsilon(\sigma) ~
\cqbf[\iq_1,\iq_2,\cdots,\iq_\nn]{}\,,\qquad\qquad \textrm{ for
  arbitrary }\sigma \in \Sgrp \nn\\
\And \cqbf[\iq_1,\iq_2,\cdots,\iq_\nn]{} =
\cqbs[\iq_1,\iq_2,\cdots,\iq_\nn]{} \,, \qquad\qquad\If \iq_1<\iq_2<\cdots<\iq_\nn\,.
\end{gather} \index{Q-functions@{\Qfs}!QQ-relations}
After the change of {\lbg} \eqref{eq:and-gqfiequiv-gqbfj}, we
obtain the following  {\cqfu}{\cqfu}-relations \cite{Tsuboi:1998ne,2007JSMTE..01....5B,2008NuPhB.805..451B,Kazakov:2007fy,Zabrodin:2007rq,2003CzJPh..53.1041G,Gromov:2007ky,Gromov:2010km}:
\Pv{\begin{empheq}[left={\empheqlbrace}]{align}
  \cqf[\iq_1,\iq_2,\cdots,\iq_\nn,~\bjq,\bkq]{}
  \cqf[\iq_1,\iq_2,\cdots,\iq_\nn]{} = &\sg \bjq\left(
  \cqf[\iq_1,\iq_2,\cdots,\iq_\nn,~\bjq]{-}
  \cqf[\iq_1,\iq_2,\cdots,\iq_\nn,~\bkq]{+}\right.\nonumber
\\&\qquad\qquad\left. -
  \cqf[\iq_1,\iq_2,\cdots,\iq_\nn,~\bjq]{+}
  \cqf[\iq_1,\iq_2,\cdots,\iq_\nn,~\bkq]{-}\right)\,, \label{eq:cQcQbb}\\
&&\mathllap{\If \sg \bjq = \sg \bkq\,,}\nonumber\\[.5cm]
  \cqf[\iq_1,\iq_2,\cdots,\iq_\nn,~\bjq]{}
  \cqf[\iq_1,\iq_2,\cdots,\iq_\nn,~\bkq]{} = &\sg \bjq\left(
  \cqf[\iq_1,\iq_2,\cdots,\iq_\nn,~\bjq,\bkq]{-}
  \cqf[\iq_1,\iq_2,\cdots,\iq_\nn]{+}\right.\nonumber
\\&\qquad\qquad\left. -
  \cqf[\iq_1,\iq_2,\cdots,\iq_\nn,~\bjq,\bkq]{+}
  \cqf[\iq_1,\iq_2,\cdots,\iq_\nn]{-}\right)\,,
\label{eq:cQcQbf}
 \\
&&\mathllap{\If \sg \bjq \neq \sg \bkq\,,}\nonumber
\end{empheq}
}
\begin{gather}
\where 
\cqf[\iq_{\sigma(1)},\iq_{\sigma(2)},\cdots,\iq_{\sigma(\nn)}]{}=\epsilon(\sigma) ~
\cqf[\iq_1,\iq_2,\cdots,\iq_\nn]{}\,,\qquad\qquad \textrm{ for
  arbitrary }\sigma \in \Sgrp \nn\\
\And \cqf[\iq_1,\iq_2,\cdots,\iq_\nn]{} =
\cqs[\iq_1,\iq_2,\cdots,\iq_\nn]{} \,, \qquad\qquad\If \iq_1<\iq_2<\cdots<\iq_\nn\,.
\end{gather}
Here, the grading is \((-1)^{\gr
{\jq}}=-1\) for the indices \(\jq\) associated to upper band, {\idest} the
indices in \(\ninter {\Kr+1} {\Kr+\Mr+\Mr'}\) (and \((-1)^{\gr
{\jq}}=+1\) for all the other indices).

\paragraph{Expression in terms of forms}
\label{sec:expr-terms-forms-1}

\begin{figure}
  \centering
  \includegraphics{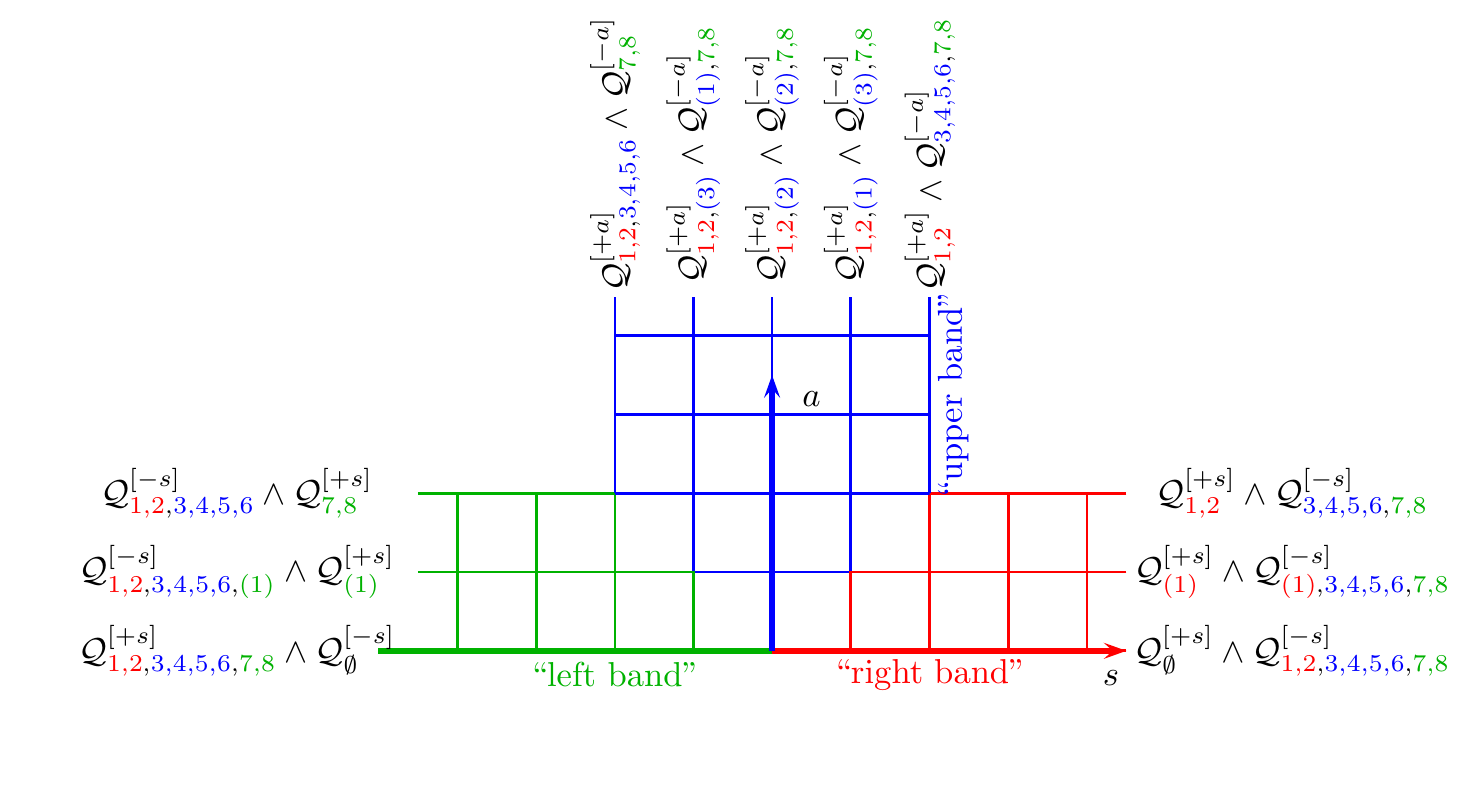}
  \caption{The typical solution of Hirota equation for the
    \(\Tk\)-{\hook} \(\Tk(2,2|2+2)\) of AdS/CFT (expressions are written up to a sign)}
  \label{fig:HkBnd}
\end{figure}

In terms of from, the expressions (\ref{eq:TTHook1}-\ref{eq:TTHook3})
are conveniently rewritten as\footnote{The article \cite{Forms} will
  include a deeper study of these expressions written in terms of forms.}
\Pv{\begin{empheq}[left={\T=\empheqlbrace}]{align}
\label{eq:TTHook1f}
      &\cqfs[(a)][+s+\frac{\Kr-\Kr'-\Mr+\Mr'}2]\wedge\cqfs[(\Kr-a),\Kr+1,\Kr+2,\cdots,\Kr+\Mr+\Mr'+\Kr'][-s-\frac{\Kr-\Kr'-\Mr+\Mr'}2]
\\
     &&\mathllap{\If s\geq \Mr+a-\Kr}\nonumber\\[.5cm]
\label{eq:TTHook2f}
&\upepsilon^{\uu}_{a,s}~
\cqfs[1,\cdots,\Kr,(\Mr-s)][+a
+\frac{-\Kr-\Kr'+\Mr+\Mr'}2
]\wedge \cqfs[(\Mr'+s),
\Kr+\Mr+\Mr'+1,\cdots,\Kr+\Mr+\Mr'+\Kr'
][-a-\frac{-\Kr-\Kr'+\Mr+\Mr'}2
]
\\
 &&\mathllap{\If a\geq s+\Kr-\Mr \And  a\geq -s}
 \mathrlap{+\Kr'-\Mr'}\nonumber\\[.5cm]
\label{eq:TTHook3f}
&\upepsilon^{\lL}_{a,s}~
\cqfs[1,\cdots,\Kr+\Mr+\Mr',(\Kr'-a)][-s+\frac{-\Kr+\Kr'+\Mr-\Mr'}2]
  \cqf[(a)]{\mathrlap{[+s-\frac{-\Kr+\Kr'+\Mr-\Mr'}2]}}
\\
&&\mathllap{\If s\leq -a+\Kr'-\Mr'}\nonumber
    \end{empheq}}
In the case of the \(\Tk\)-{\hook} \(\Tk(2,2|2+2)\) of AdS/CFT, these
expressions are summarized in figure \ref{fig:HkBnd}.
The differential forms which appear in
(\ref{eq:TTHook1f}-\ref{eq:TTHook3f}) are defined by
\begin{gather}
  \cqf[a,\cdots,b,(\nn)]{}\equiv  \cqf[(\nn),a,\cdots,b]{}\equiv \prod_{\iq=a}^{b}
  \left(\xi_\iq \partial_{\xi_\iq}\right)
  \cqf[(\nn+1+b-a)]{},\qquad\qquad \If a<b\\
\label{eq:cQnform}
\where \cqf[(\nn)]{}\equiv \sum_{\iq_1<\iq_2<\cdots<\iq_\nn} \cqf[
\iq_1,\iq_2,\cdots,\iq_\nn]{}~~\xi_{\iq_1}\wedge\xi_{\iq_2}\wedge\cdots\wedge\xi_{\iq_\nn}\,,
\end{gather}
\begin{gather}
\And \left(\xi_\jq \partial_{\xi_\jq}\right)
\xi_{\iq_1}\wedge\xi_{\iq_2}\wedge\cdots\wedge\xi_{\iq_\nn} = \left\{
  \begin{aligned}
    &\xi_{\iq_1}\wedge\xi_{\iq_2}\wedge\cdots\wedge\xi_{\iq_\nn}
    &&\If\jq\in\{\iq_1,\iq_2,\cdots,\iq_\nn\}\\
&0&&\oth\!\!.
  \end{aligned}
\right.
\end{gather}
With these definitions, we see that the form
\(\cqf[a,\cdots,b,(\nn)]{}\) contains all the {\cQfs} {\lbd} by a set of
indices %
which contains
\(\ninter a b\), and \(\nn\)
additional other indices. Moreover, the change of variable
\eqref{eq:and-gqfiequiv-gqbfj} between \(\cqf[]{}\) and \(\cqbf[]{}\)
allows to compute the form \(\cqf[(\nn)]{}\) in \eqref{eq:cQnform} as a
partial Hodge dual\footnote{
Hodge duality is simply a formalism to rewrite (in terms of forms),
the change of variable \eqref{eq:and-gqfiequiv-gqbfj} between
\(\cqf[]{}\) and \(\cqbf[]{}\). It says that up to a sign, \(\cqf[(\nn)]{}\) can be
  formally viewed as
\(\sum_\nn \cqf[(\nn)]{}\equiv \pm 
\left(\sum_{\substack{%
\sg\jq=-1}}
\left(\xi_\jq+\partial_{\xi_\jq}\right)\right)
\sum_ {\nn'}\cqbf[(\nn')]{}
\), where a sum runs over all indices \(\jq\) such that \(\sg \jq = -1\).
} of the form \(\cqbf[(\nn)]{}\) defined by 
\begin{gather}
\cqbf[(0)]{}\equiv \cqbf[\emptyset]{},\qquad\qquad \qquad\qquad 
{\cqbf[(1)]{}\equiv \sum_{\iq=1}^{\Kr+\Mr+\Mr'+\Kr'} \cqbf[\iq]{}
  \xi_\iq},\\
\And
  {\cqbf[({\nn})]{}\equiv 
      \frac{\cqbfs[(1)][-{\nn}+1]\wedge \cqbfs[(1)][-{\nn}+3]\wedge
        \cqbfs[(1)][-{\nn}+5]\wedge \cdots\wedge \cqbfs[(1)][{\nn}-1]}
      {\cqbes[-{\nn}+2]\cqbes[-{\nn}+4]\cdots
        \cqbes[{\nn}-2]}},\qquad\qquad \when {\nn}>1
    \,.%
\end{gather}

\paragraph{Freedom in the choice of the {\qfs}}
\label{sec:freedom-choice-qfs}

In section \ref{sec:equiv-hirota-equat}, we have seen that if two
different (typical) solutions of the Hirota equation give rise to the
same {\Yfs}, then they are equal up to a gauge transformation.

In the same spirit, let us now conclude the present section by a
statement describing the 
case when two different sets of {\qfs} give rise to the same {\Tfs}:
in such a case, 
the two sets of {\qfs} have to
obey the same difference equation
\eqref{eq:DetZeroDefq}, which means that they are obtained from each other
by linear transformation. Going carefully through the construction
given in the proof of the \stapref{sec:statmtSolHirotPCM} we find the
following statement, written for the lattice {\St(\Np)} of the
\(\SU\Np\times \SU\Np\) {\PCM}.

\begin{statmt}
\label{sta:freedom-choice-qfs-1}
  Let \(\T\) be a typical solution of the Hirota equation on
  \(\St(\Np)\), and let \(\gqs[1]\), \(\gqs[2]\),
    \(\cdots\), \(\gqs[{\Np}]\), \(\gps[1]\), \(\gps[2]\),
    \(\cdots\), \(\gps[{\Np}]\), and \(\gqe\) and \(\gpe\) be such that
    \eqref{eq:HirotaSolutPCM} holds, {\idest} such that, in the notations
    (\ref{eq:-fdisp-gqf1equiv}-\ref{eq:gqf0-gqe-qquadqq}), we have
    \(\T=\gqfs[(a)][+s]\wedge \gpfs[(\Np-a)][-s]\). Let \(\gqbs[1]\),
\(\gqbs[2]\), 
    \(\cdots\), \(\gqbs[{\Np}]\), \(\gpbs[1]\), \(\gpbs[2]\),
    \(\cdots\), \(\gpbs[{\Np}]\), and \(\gqbf[\emptyset]{}\) and
    \(\gpbf[\emptyset]{}\) be {\another} set of {\qfs} such that
    \eqref{eq:HirotaSolutPCM} holds, {\idest} such that in these
    notations
    \(\T=\gqbfs[(a)][+s]\wedge \gpbfs[(\Np-a)][-s]\).

    Then, there exists an \({\bi}\)-periodic  matrix \(\HH({\us})\in\SL\Np\), and two
    \({\bi}\)-periodic functions \(C(\us)\) and \(F(\us)\) such that
\begin{align}
\label{eq:gpbs=f-cnp-2h_ij}
\gpbs[]=&F\,C^{\Np-2}{\HH_{{\coordi}}}^{\coordj} \gps[\jq]\,, & \gqbs[]=&F^{-1}C^{\Np-2}{\HH_{{\coordi}}}^{\coordj} \gqs[\jq]\,,\\
\gpbe=&F\,C^{\Np}\gpe\,,&\gqbe=&F^{-1}C^{\Np}\gqe\,.
\label{eq:gpbe=f-cnpgpe-gqbe=f}
\end{align}
\end{statmt}
The converse is also true: one can easily check that this
transformation does indeed leave the {\Tfs} invariant, and out of an
arbitrary set of {\qfs} this transformation produces {\another} set of {\qfs} giving rise to
the same {\Tfs}.

Here, the statement is written only for the lattice \(\St(\Np)\) of the
\(\SU\Np\times \SU\Np\) {\PCM}, but the same statement\footnote{
In the case of an arbitrary \(\Tk\)-{\hook}, a remarkable difference
is that if we write a statement which generalizes the
\staref{sta:freedom-choice-qfs-1}, then \(\HH\) becomes a block-matrix,
where several blocks of coefficients are forced to be zero.
} can be written
for an arbitrary \(\Tk\)-{\hook}. Actually, even in the case of the
\(\Tk\)-{\hook} \(\Tk(2,2|2+2)\) of {\ADF}, we will usually restrict
to either the right band or the upper band, where the
\staref{sta:freedom-choice-qfs-1} written above for \(\St(\Np)\) will be
sufficient.

\subsection{Gauge conditions and {\Wronskian} expressions of the {\Tfs}}
\label{sec:gauge-cond-wronsk}

As we have seen in the previous sections, the expression of the {\Tfs}
in terms of {\qfs} is quite simple in the ``{\Wronskian}'' gauges obeying
the constraints \eqref{eq:PhyWronGaugeTK}. We also said that an
arbitrary solution of the Hirota equation can be transformed into a
solution which obeys these conditions, and this transformation will
essentially fix three out of fours degrees of gauge freedom.

This subsection will elaborate on the meaning of these statements at
the {\level} of {\qfs}.

\paragraph{Relaxing the gauge constraints}

Let us repeat {\below} the constraints defining the {\Wronskian} gauges for
the lattice\footnote{The other lattices
 \(\Sh(\Kr)\) and  \(\HK(\Kr,\Mr)\) for which the {\Wronskian} gauge was
 defined in section \ref{sec:a-s-lattice} can be viewed as sublattices
 of \(\Tk(\Kr,\Kr'|\Mr+\Mr')\).
} \(\Tk(\Kr,\Kr'|\Mr+\Mr')\) :
\begin{subequations}
  \begin{gather}
\label{eq:GaWroBo}
    \T[0]=\Ts[0][0][-s]\\
\T[\Kr'+{\nn}][-\Mr']=\T[\Kr'][-\Mr'-{\nn}]
    \qquad\qquad \T[\Kr+{\nn}][\Mr]=\T[\Kr][\Mr+{\nn}]\qquad\qquad\textrm{ for any }{\nn}\geq0\label{eq:GaWro}
  \end{gather}
\end{subequations}

We can see that the {\Wronskian} solution
(\ref{eq:TTHook1}-\ref{eq:TTHook3}) of Hirota equation  directly
implies the condition 
\eqref{eq:GaWro}. Thus, this condition \eqref{eq:GaWro} is  necessary
in order to write the {\Wronskian} solution of Hirota equation. 
 On the other hand, {\Wronskian} expression 
(\ref{eq:TTHook1}-\ref{eq:TTHook3}) of the {\Tfs} does
not directly imply that \(\T[0]=\Ts[0][0][-s]\). As a consequence, we
see that although the condition \eqref{eq:GaWroBo} is physically
meaningful, it is not necessary in order to write the {\Wronskian}
expressions of {\Tfs}. 

For instance, for the lattice \(\Sh(\Np)\), we see that
the condition
\eqref{eq:GaWroBo} is reflected
 in the \stapref{sec:statmtSolHirSh}  by the 
relation \(\gqf[(0)]{}\equiv\gqf[\emptyset]{}= 1\).  If we relax this
gauge constraint,
we obtain that the general solution of the Hirota equation on
    \(\Sh(\Kr)\) such that 
    \begin{align*}
          \T[][0]%
    =&\T[0][-a]%
    &\And&&
    \T[\Kr][s]=&\T[0][-\Kr-s]
    \end{align*}
is given by 
\begin{gather}
      {\T=
        (-1)^{(\Kr-1)(\Kr-a)}
        \gqfs[(a)][+s]\wedge \gqfs[(\Kr-a)][-s-\Kr]}\,,\\
      \where 
      {\gqf[(1)]{}\equiv \sum_{\iq=1}^\Kr \gqs[\iq] \xi_\iq}\,, \qquad\qquad
       \And \gqf[(0)]{}\equiv \gqe\,,
      \\
      \And
      \gqf[({\nn})]{}\equiv \frac{
      \gqfs[(1)][-{\nn}+1]\wedge \gqfs[(1)][-{\nn}+3]\wedge
      \gqfs[(1)][-{\nn}+5]\wedge \cdots\wedge \gqfs[(1)][{\nn}-1]}
{\gqes[-{\nn}+2]\gqes[-{\nn}+4]\cdots \gqes[{\nn}-2]}
      \qquad \If {\nn}>1\,.
\label{eq:and-gqfnnequiv-frac}
    \end{gather}
At the {\level} of these expressions, the only difference, compared to  %
the case
when the gauge constraint \eqref{eq:GaWroBo} is enforced, is the
presence of a denominator in \eqref{eq:and-gqfnnequiv-frac}.

Similarly, for the case of the \(\Tk\)-{\hook} of
\stapref{sta:Thooksol}, the same {\Wronskian} solution holds if we do not
impose the gauge condition \eqref{eq:GaWroBo}, but we only impose 
\eqref{eq:GaWro}. The only difference is that in this case, the
condition \eqref{eq:THookQemptyo} does not hold any more.

\paragraph{Gauge transformations and {\qfs}}
\label{sec:gauge-transf-qfs-1}

Let us now investigate the form of the gauge transformations
\( \T \rightsquigarrow \ga[a+s][1] \ga[a-s][2]
        \ga[-a+s][3] \ga[-a-s][4] \T\)
preserving the gauge constraint \eqref{eq:GaWro}.
First, in order to preserve the gauge constraint \eqref{eq:GaWro}, we require that the condition
\(\frac{\T[\Kr+1][\Mr]}{\T[\Kr][\Mr+1]}=1\) still holds after the gauge
transformation. This imposes   
\begin{equation*}
  \frac{\ga[+\Kr-\Mr+1][2]}{\ga[+\Mr-\Kr+1][3]}
~\frac{
        \ga[+\Mr-\Kr-1][3]}{\ga[+\Kr-\Mr-1][2]
        }=1\,.
\end{equation*}
Without loss of generality\footnote{In principle, one can only
  conclude that
\(\alpha \equiv \ga[+\Kr-\Mr][2]=\ga[+\Mr-\Kr][3]\) is an \(\bi\)-periodic
function. But then we can redefine the functions \(\gan[][2]\) and
\(\gan[][3]\) as 
\(\gan[][2]\rightsquigarrow \gan[][2] / \sqrt{ \alpha^{[+\Mr-\Kr]} } \)
and \(\gan[][3]\rightsquigarrow \gan[][3]  \sqrt{ \alpha^{[-\Mr+\Kr]} }
\).
Due to the \(\bi\)-periodicity of \(\alpha\), this transformation leaves
the product \(\ga[a+s][1] \ga[a-s][2]
        \ga[-a+s][3] \ga[-a-s][4]\), which means that we describe the
        same gauge transformation, but in terms of ``rescaled'' gauge
        functions 
\(\gan[][2]\) and
\(\gan[][3]\) which now obey 
\(\ga[+\Kr-\Mr][2]=\ga[+\Mr-\Kr][3]\).
 }, we can deduce that
\(\ga[+\Kr-\Mr][2]=\ga[+\Mr-\Kr][3]\). If we proceed the same way to
require that the gauge condition 
\(\frac{\T[\Kr'+1][-\Mr']}{\T[\Kr'][-\Mr'-1]}=1\) is preserved by the
gauge transformation, then we see that \(\ga[+\Kr'-\Mr'][1]=\ga[-\Kr'+\Mr'][4]\).

Hence we see that the general gauge transformation preserving the
gauge condition \eqref{eq:GaWro} takes the form
\begin{gather}
  \label{eq:GaugeWronFree}
  \T \rightsquigarrow \ga[
+a+s%
][1] \ga[+a-s
][2]
        \ga[-a+s-2\Mr+2\Kr][2] \ga[-a-s+2\Kr'-2\Mr'][1]  \T\,%
.
\end{gather}
If we compare this with the expression
(\ref{eq:TTHook1}-\ref{eq:TTHook3}) of the {\Tfs}, we immediately see
that this transformation corresponds exactly, at the {\level} of {\qfs},
to  
\begin{gather}
\label{eq:qtransfgaug}
\forall I,\qquad\qquad
  \cqf[I]{}\rightsquigarrow f_1^{[+\kk_I-\mm_I]} f_2^{[-\kk_I+\mm_I]}  \cqf[I]{}\\
\where ~~  \kk_I=\mathrm{Card}\{\iq \in I | {\sg  \iq}=1\}\qquad\qquad
\And~~  \mm_I=\mathrm{Card}\{\iq \in I | {\sg  \iq}=-1\}\,,
\end{gather}
where we should set \(f_1= \ga[-\frac{\Kr-\Kr'-\Mr+\Mr'}2][1]\)  and
\(f_2=\ga[-\frac{\Kr-\Kr'-\Mr+\Mr'}2][1]\) in order to reproduce
\eqref{eq:GaugeWronFree}. One can show that this transformation
\eqref{eq:qtransfgaug} preserves the determinant expressions such as
(\eqref{eq:and-gqfiequiv-gqbfj}-\eqref{eq:and-gqbfjq_1-jq_2}), because
it also preserves the underlying qq-relation.

We see that if we only impose the gauge constraint \eqref{eq:GaWro},
then the remaining gauge freedom takes the form 
\eqref{eq:qtransfgaug},
expressed in terms of two independent
functions \(f_1\) and \(f_2\).
 Of course, if we also add the constraint \eqref{eq:GaWroBo}, then the
 functions \(f_1\) and \(f_2\) are not independent anymore, and we have
 \(f_2=1/f_1\). We see that in this case, %
 there is only one degree of gauge freedom left.

  \subsection{Writing FiNLIEs}
  \label{sec:writing-finlies}

  These results about the typical solutions of the {\Ysys} equation and
  of the Hirota equation allow to build FiNLIEs for several different
  models. This means that for several {\ing} models, where we know
  that the {\TBA} gives rise to the {\Ysys} equation \eqref{eq:YSysEq},
  we will derive a finite set of non-linear integral equations (FiNLIE).
  \index{FiNLIE}

  Let us sketch the major steps of this procedure, which we will detail
  in section \ref{sec:application-au-champ} for the case of the {\SU {\Np}}
  {\PCM}. 

  \begin{enumerate}
  \item First, one has to find the {\Ysys} equation or the TBA
    equations. Though the form \eqref{eq:YSysEq} of the {\Ysys}
    equation is quite universal, each {\ing} model is characterized
    by a different \((a,s)\)-lattice and by a different asymptotic
    behavior at large 
    \(\us\). This asymptotic behavior is an additional constraint, to be put ``on top
    of'' the {\Ysys} equation (we will see that it corresponds to
    a ``{\zerm}'' of     the {\Ysys} 
    equation). It can either be read from the TBA
    equations, or ``guessed'' from the asymptotic limit (the limit when
    the size of the space is large).
    \label{{\idest}:1}

  \item Then one has to understand the ``asymptotic limit'', which is
    the limit of an infinite size model. In this limit, one should find
    a solution of Hirota equation which gives rise to the correct
    asymptotic Bethe equations. %
    \label{{\idest}:2}%
    This step is important, not only because it allows to %
    reproduce the initial asymptotic Bethe equations (which were the
    starting point to write the {\TBA}), but rather because it %
    also allows to understand the behavior of the {\Yfs}
    in this limit.

  \item For an arbitrary finite size \(\Lf\),
    the {\TBA} gives rise to an infinite set of {\Yfs}, which obey the
    {\YsE}. These 
{\Yfs}
 can
    be written in terms of (an infinite set of) {\Tfs}, which are themselves written in terms
    of (a finite set of) {\qfs}. Finding the  solution of the {\Ysys}
    equation therefore 
    reduces to identifying these {\qfs}. %
    \label{{\idest}:3} \\
    In order to write the {\Tfs}, we would a priori have to know
    these {\qfs} on the whole complex plane. By contrast the
    {\TBAE} only involve {\Yfs} on the real axis. Therefore we
    replace an infinite number of functions on the real axis by a finite
    number of functions on the complex plane. To make this interesting
    it will be necessary to find a convenient parameterization of these
    {\qfs}, and this will be done in two steps:
    \begin{itemize}
    \item A first step is to identify some domains in the complex
      plane (called analyticity strips), where the 
{\Yfs} (resp the  {\Tfu}- and {\qfs}) are analytic. %
These analyticity strips are fixed by the {\TBAE} (more precisely
their zero-modes), or they can also be read from 
      the large \(\us\) asymptotic %
      behavior, and sometimes from some additional
symmetries
      of the {\Ysys}.

    \item The next step is to show, from %
      the
      existence of these analyticity strips, %
      that a parameterization of the {\qfs} in 
      terms of a finite 
      number of functions on the real axis can be found. %

      This step will express the general solution of
      Hirota equation with given analyticity strips in terms of a
      finite number of functions on the real axis, but these functions
      are still to be fixed in the next steps.
    \end{itemize}

  \item Finally, one has to write non-trivial equations putting enough constraints
    on the {\qfs}, so that only one solution of the {\Ysys} is kept. To
    do this, one should manage to write equations containing the
    zero-modes of the {\TBAE} (or equivalently the large \(\us\)
    asymptotic). Non-trivial analyticity constraints or symmetries of
    the model can also be necessary at this point.
    \label{{\idest}:4}
\\
   These equations will usually take the form of closed equations on
   (the functions parameterizing) the {\qfs}, which can for instance
   be solved numerically by a fix-point approach. These equations will
   be called a FiNLIE (finite set of non-linear integral equations).

  \item Last, but not least, %
    we can
    express the energy
    associated to a given solution of the {\YsE}. 
    \label{{\idest}:5}
    This allows to answer the initial question of finding the
    finite-size spectrum of a given {\ing} model.
  \end{enumerate}

  \section{FiNLIE for the {\PCF}}
  \label{sec:application-au-champ}

Following the general procedure of the above section
\ref{sec:writing-finlies}, we will
  now see how to proceed explicitly and to write a FiNLIE in the
  case of the {\PCM} introduced in the section
  \ref{sec:example-princ-chir}. The number of the subsections will
  reflect the steps listed in section \ref{sec:writing-finlies}.

  \subsection{{\Ysys} equation}
  \label{sec:ysys-equation}
 
  As it was already discussed, the finite size effects of the {\PCM} are
  encoded into an infinite set of {\Yfs}, which obey the {\Ysys}
  equation \eqref{eq:YSysEq}. They obey the following, large \(\us\)
  asymptotic behavior (see \eqref{eq:YsysZeMo0}):
  \begin{align}
    \label{eq:YsysZeMo}
    \mathrm{log}\left(\Y\right) + \LF
    \frac{\sin\frac{\pi a}{{\Np}}}{\sin\frac{\pi }{{\Np}}}
    \cht[\us]\delta_{s,0}\xrightarrow[\us \to \infty]{} c_{a,s}\,,
  \end{align}
  where \(c_{a,s}\) is an arbitrary \(\us\)-independent number. This
  condition can be read from \eqref{eq:PCMTBArrb}, and we will see that
  it can be viewed as the insertion of a {\zerm} of the {\Ysys}
  equation \eqref{eq:YSysEq}.

  We see that in this limit the {\Yfs} \(\Y\) are non-zero constants,
  except for \(\Y[][0]\) which is roughly equal to \(e^{-\LF \frac{\sin\frac{\pi a}{{\Np}}}{\sin\frac{\pi }{{\Np}}}
    \cht[\us]}\ll 1\). The only place where the quantity \(\LF\) appears
  in the {\Ysys} is via this asymptotic behavior, and it only changes
  the speed at which \(\Y[][0]\xrightarrow[\us\to\infty]
  {}0\). Therefore, we expect that
the asymptotic behavior \eqref{eq:YsysZeMo} is independent of \(\LF\),
({\idest} that the constants \(c_{a,s}\) are independent of \(\LF\)), and that
the limit \(\LF\to\infty\) (where we also have \(e^{-\LF \frac{\sin\frac{\pi a}{{\Np}}}{\sin\frac{\pi }{{\Np}}}
    \cht[\us]}\ll 1\)) is essentially the same as the limit
  \(\us\to\infty\).

Hence, we also have
  \begin{align}
    \label{eq:YsysZeMoL}
    \mathrm{log}\left(\Y\right) + \LF
    \frac{\sin\frac{\pi a}{{\Np}}}{\sin\frac{\pi }{{\Np}}}
    \cht[\us]\delta_{s,0}\xrightarrow[\LF %
    \to\infty]{}c_{a,s}\,.
  \end{align}

  \subsection{Asymptotic limit}
  \label{sec:asymptotic-limit}

  The asymptotic limit is the limit when \(\LF\) is very large. In that
  case, the equation \eqref{eq:YsysZeMo} ensures that
  \begin{align}
    \forall a \in& \ninter 1 {{\Np}-1}\,,&\Y[][0]{}\xrightarrow[\LF\to\infty]{}&0
  \end{align}

  If we really set \(\Y[][0]=0\), then we get a quite peculiar solution of
  the {\YsE} \eqref{eq:YSysEq} %
  in the sense
  that when \(s=0\), the {\lhs} is zero and the denominator in the
  {\rhs} is infinite. But since this {\YsE} is
  degenerate at \(s=0\), one can see that the two sets of functions
  \(\left\{\Y\middle| s>0\right\}\) and \(\left\{\Y\middle| s<0\right\}\)
  are completely independent (each of them has to obey independent
  {\Ysys} equations). That means that if we really set \(\Y[][0]=0\), we
  describe a quite degenerate, non-typical solution.

  Therefore we will keep \(\Y[][0]\) small, but not exactly zero.
  We will show that then, there is a (typical) solution of the
  {\YsE} \eqref{eq:YSysEq} which allows to recover the 
  asymptotic Bethe {\anz}  of section \ref{sec:asympt-bethe-anz}.
  We will first study the degenerate leading order (corresponding to 
  \(\Y[][0]=0\)  for all \(a\in\ninter 1 {{\Np}-1}\)). We will then add
  exponentially small terms which will make this solution consistent and
  make Bethe equations arise.

  \subsubsection{Splitting $\St({\Np})$ into two half strips $\Sh({\Np})$}
  \label{sec:splitting-stn-into}

  In the approximation \(\Y[][0]=0\)  for all \(a\in\ninter 1 {{\Np}-1}\), 
  the general solution of the {\Ysys} equation on the lattice
  \(\St({\Np})\) (see \figpref{fig:aslatt-pcm}) is given by 
  \begin{equation}
    \label{eq:Ysplit}
    \Y(\us)=
    \left\{
      \begin{array}{lcr}
        \Y[][][(\Rg)](\us)&\If&s>0\\
        0&\If&s=0\\
        \Y[][-s][(\Lf)](\us)&\If&s<0
      \end{array}
    \right.\,,
  \end{equation}
  where \(\Y[][][(\Rg)]\) and \(\Y[][][(\Lf)]\) are two arbitrary
  (independent) solutions of the {\Ysys} equation on the lattice
  \(\Sh({\Np})\) of \figpref{fig:LatSh}.

  At the {\level} of {\Tfs}, we can define two sets \(\{\T[][][(\Rg)]\}\) and
  \(\{\T[][][(\Lf)]\}\) of {\Tfs} which obey the Hirota equation on the
  lattice \(\Sh({\Np})\), such that
  \begin{align}
    \Y[][][(\Rg)]=&\frac{\T[][s+1][(\Rg)]}{\T[a+1][][(\Rg)]}\frac{\T[][s-1][(\Rg)]}{\T[a-1][][(\Rg)]}\,, &
    \Y[][][(\Lf)]=&\frac{\T[][s+1][(\Lf)]}{\T[a+1][][(\Lf)]}\frac{\T[][s-1][(\Lf)]}{\T[a-1][][(\Lf)]}.
  \end{align}

  To make this solution less degenerate, we 
  will find gauge functions
  \(\gan[][1]\),  \(\gan[][2]\), \(\gan[][3]\) and \(\gan[][4]\)
  and glue \(\T[][][(\Rg)]\)
  with \(\T[][][(\Lf)]\) as follows:
\Pv{  \begin{subequations}
    \label{eq:Tsplit}
    \begin{empheq}[left={\T=\empheqlbrace}]{align}
      \label{eq:Tright}
      &\T[][][(\Rg)]&\If&s\geq 0\,,\\
      &\ga[a+s][1] \ga[a-s][2] \ga[-a+s][3] \ga[-a-s][4]
      \T[][-s][(\Lf)]&\If&s\leq 0\,,
      \label{eq:Tleft}
    \end{empheq}
  \end{subequations}}
  This is consistent if the two expressions coincide at \(s=0\), {\idest} if
  \begin{equation}
    \ga[a][1] \ga[a][2] \ga[-a][3] \ga[-a][4]=\frac{\T[][0][(\Rg)]}{\T[][0][(\Lf)]}
  \end{equation}
  This equation always has a solution because \(\T[][0][(\Rg)]\) and
  \(\T[][0][(\Lf)]\) are of the form \(f^{[+a]} \tilde f^{[-a]}\) (this
  can be seen from the relation
  \(\frac{\T[][0][+]\T[][0][-]}{\T[a-1][0]\T[a+1][0]}=1+\Y[][0]\simeq
  1\)).
  The expression \eqref{eq:Tright} ensures that \(\T\) obeys the Hirota
  equation when \(s>0\), while \eqref{eq:Tleft} ensures that \(\T\) obeys
  the Hirota equation when \(s<0\). To get an approximate solution of
  Hirota equation at \(s=0\), which obeys \(\Y[][0]\ll 1\), we will choose
  the gauge functions 
  \(\gan[][1]\),  \(\gan[][2]\), \(\gan[][3]\) and \(\gan[][4]\) such
  that 
  \(\ga[a-1][1] \ga[a+1][2] \ga[-a-1][3] \ga[-a+1][4]\ll 1\). This will
  ensure that
  \begin{align}
    \T[][-1]=\ga[a-1][1] \ga[a+1][2] \ga[-a-1][3] \ga[-a+1][4]
    \T[][1][(\Lf)]\ll 1
  \end{align}
  is exponentially small (typically like \(e^{-\Lf \mathrm{cosh}(\us)}\)).
  This will imply that at \(s=0\),
  \begin{align}
    \T[][0][+]\T[][0][-]&=(\T[][0][(\Rg)])^+(\T[][0][(\Rg)])^-=\T[a-1][0][(\Rg)]\T[a+1][0][(\Rg)]
    \\ &\simeq \T[a-1][0][(\Rg)]\T[a+1][0][(\Rg)]+\T[][-1] \T[][1][(\Rg)] = 
    \T[a-1]\T[a+1]+\T[][-1]\T[][+1]\,,
  \end{align}
  which means that the Hirota equation is satisfied (to the leading
  order) at \(s=0\).

  At the {\level} of {\Yfs}, the prescription \eqref{eq:Tsplit}
  implies that 
  \eqref{eq:Ysplit} should be replaced with
  \begin{equation}
    \label{eq:Ysplit2}
    \Y(\us)=
    \left\{
      \begin{array}{lcr}
        \Y[][][(\Rg)](\us)&\If&s>0\\
        \frac{\T[][1][(\Rg)]}{\T[a-1][0][(\Rg)]\T[a+1][0][(\Rg)]}
        \ga[a-1][1] \ga[a+1][2] \ga[-a-1][3] \ga[-a+1][4]
        \T[][1][(\Lf)]
        \ll 1
        &\If&s=0\\ 
        \Y[][-s][(\Lf)](\us)&\If&s<0
      \end{array}
    \right.\,,
  \end{equation}
  which is a solution of the {\Ysys} equation to the leading order.

  \subsubsection{Explicit expression of the {\Tfs}}
  \label{sec:expl-expr-tfs}

  In the asymptotic limit, we have already seen that the 
  Bethe equation \eqref{eq:BethePCM} (from the asymptotic Bethe
  {\anz}) was the same as the Bethe equation for a spin 
  chain with inhomogeneities \(\theta_\ii\) (which are the rapidities of
  the massive particles).

  At the {\level} of {\Tfs}, that means that the {\Tfs} \(\T[][][(\Rg)]\) and
  \(\T[][][(\Rg)]\), (which obey the Hirota equation on \(\Sh({\Np})\simeq
  \HK({\Np},0)\)), correspond to two \(\SU \Np\) {\csds}. 
  These {\Tfs} can then be expressed for instance from the {\Wronskian}
  expression
  (\ref{eq:WronskianLambda},\ref{eq:TasGLKfromNestexpanded}) derived
  in the chapter \ref{part:qoperatorsspin} for {\csds}.

  After the change of
  variables \eqref{eq:TchangeShift} %
  these {\Tfs} can %
  be expressed %
  in terms of a set of {\qfs}, as in 
  \eqref{eq:HirotaSolSpChn}:
  \begin{align}
\label{eq:TRTL0}
    \T[][][(\Rg)]=&
    (-1)^{(\Np-1)(\Np-a)}
    {\gqf[(a)]{(\Rg)}} \left(\us+\bi\frac s 2\right)\wedge
    \gqf[(\Np-a)]{(\Rg)}\left(\us-\bi\frac{s+\Np}2\right)\\
    \T[][][(\Lf)]=&
    (-1)^{(\Np-1)(\Np-a)}
    {\gqf[(a)]{(\Lf)}} \left(\us+\bi\frac s 2\right)\wedge
    \gqf[(\Np-a)]{(\Lf)}\left(\us-\bi\frac{s+\Np}2\right)%
  \end{align}
  \begin{gather}
        \where  
      {\gqf[(1)]{(\RL)}\equiv \sum_{\iq=1}^\Np \gqs[\iq][(\RL)] \xi_\iq},\qquad\qquad
      \And \gqf[(0)]{(\RL)}\equiv 1\,,\\
\And 
        {%
          {\gqf[({\nn})]{(\RL)}}\equiv 
          {%
            {\left.\gqf[(1)]{(\RL)}\right.}^{[-{\nn}+1]}\wedge %
            {\left.\gqf[(1)]{(\RL)}\right.}^{[-{\nn}+3]}\wedge
            \cdots\wedge %
            {\left.\gqf[(1)]{(\RL)}\right.}^{[+{\nn}-1]}}
        },\quad \textrm{ for } {\nn}>1\,.
  \end{gather}

This form for the {\Tfs} arises from a typical solution of the
  Hirota equation (due to \stapref{sec:statmtSolHirSh}), where the
  Hirota equation itself comes from the {\Ysys} equation. On the
  other hand, we know that if the {\qfs} are polynomial, this solution
  corresponds to the transfer matrices of {\csds}, defined in the
  chapter \ref{part:qoperatorsspin}, up to the change of variable
  \eqref{eq:qvsQ}. Moreover, if these {\qfs} are polynomial, their
  zeroes obey the Bethe equations
  (\ref{eq:BethBigerRank}-\ref{eq:SmkRank}), which are identical to
  \eqref{eq:BethePCM} up to the change of variables
  (\ref{eq:QPCFDef1}-\ref{eq:QPCFDefl}).

This motivates the identification of {\qfs}, in the asymptotic limit,
as
\begin{align}
  \gqf[%
\{
  1,2,\cdots,\Np-\mlvl%
\}
  ]{(\Rg)} = &
\fQ[  \us+\bi\frac
  \Np 4][\mlvl]
&
  \gqf[%
\{
  1,2,\cdots,\Np-\mlvl%
\}
  ]{(\Lf)} = &
\fQ[  \us+\bi\frac
  \Np 4][-\mlvl][(\Lf)]\,,&\when &\Lf\to\infty
\label{eq:TRTLN}
\end{align}

This identifications says that in the asymptotic limit
(\(\Lf\to\infty\)), the {\Yfs} (which are ratios of the densities of
holes and particles in the mirror model) are simply obtained as the
ratios of the polynomial {\Tfs} constructed in chapter
\ref{part:qoperatorsspin} as the eigenvalues of the {\Toprs}.

In other words, the zeroes of the {\qfs} are identified with the
rapidities of the ``particles'' (the excitations) described in section
\ref{sec:asympt-bethe-anz}, and this identification is motivated by
noticing that the zeroes of the {\qfs} obey the right Bethe
equations \eqref{eq:BethePCM}. In the upcoming subsection
\ref{sec:bethe-equations-2}, we will further motivate the
identification \eqref{eq:TRTLN}, by showing that it also gives rise to
the Bethe equation \eqref{eq:BethePCM2}.

In terms of the Hasse diagram (see \figpref{fig:Hasse}), 
the identification \eqref{eq:TRTLN} tells us the asymptotic limit of
the {\qfs} along a given {\nesting}
{\ppath}. Using the qq-relation \eqref{eq:qqrelation}, this allows to
deduce the other {\qfs}.

  \subsubsection{Middle nodes equation}
  \label{sec:middle-node-equation}

In this subsection, we will find some relations which describe
  in more details how the two solutions \(\T[][][(\Rg)]\)
  and \(\T[][][(\Lf)]\) can be glued together, and which will allow to
  recover the Bethe equation \eqref{eq:BethePCM2} in the next subsection.
  To this end, we will
  write a ``middle nodes equation'', which holds not only in the asymptotic
  limit, but is also at any finite size. %

One should be aware that the results which will be found in the present
  section do not only hold in the asymptotic limit (\(\Lf\to\infty\)),
  but also for an arbitrary finite size \(\LF\).

In the asymptotic limit, %
we wish
how the {\Tfs}
\(\T[][][(\Rg)]\)  and \(\T[][][(\Lf)]\) are glued together, by
investigating the {\Ysys} at \(s=0\). In view of the description
\eqref{eq:Tsplit} of this gluing procedure, we will usually denote
\(\T[][]=\T[][][(\Rg)]\) (especially if \(s\geq 1\)), whereas
\(\T[][][(\Lf)]\) denotes a different gauge (which is suitable when
\(s\leq 0\)). 

When the size \(\LF\) is finite, we will also use two different gauges
\(\T[][][(\Rg)]\)  and \(\T[][][(\Lf)]\), and we will usually write
\(\T[][]=\T[][][(\Rg)]\). We will see in the next sections how these
gauges arise at finite size \(\LF\). The reason why they are interesting
is that since the {\Yfs} are gauge-invariant, they can be expressed
arbitrarily in terms of either \(\T[][][(\Rg)]\) or \(\T[][][(\Lf)]\).

Let us write the 
  {\Ysys} equation at \(s=0\), in the form \eqref{eq:YSeBndInf} (which
  takes into account the boundary condition
  \(\Y[0]=\infty=\Y[{\Np}]\)). If we express 
 each {\Yf} as a
  ratio of {\Tfs}, %
we get
\begin{subequations}
  \label{eq:YmidDelt}
  \begin{align}
    \frac{\Y[][0][+]\Y[][0][-]}
    {\left(\Y[a+1][0]\right)^{1-\delta_{a,{\Np}-1}}
      \left(\Y[a-1][0]\right)^{1-\delta_{a,1}} }=&
    \frac{1+\Y[][1]}{\left(1+\Y[a+1][0]\right)^{1-\delta_{a,{\Np}-1}}}
    \frac{1+\Y[][-1]}{\left(1+\Y[a-1][0]\right)^{1-\delta_{a,1}}}\\
    =& \frac{\frac{\T[a][1][+]\T[a][1][-]}{\T[a+1][1]\T[a-1][1]}}
    {\left(\frac{\T[a+1][0][+]\T[a+1][0][-]}{\T[a+2][0]\T[a][0]}\right)^{1-\delta_{a,{\Np}-1}}}
    \frac{
      \frac{\T[a][1][(\Lf)](\us+\bi/2)\T[a][1][(\Lf)](\us-\bi/2)}{\T[a+1][1][(\Lf)]\T[a-1][1][(\Lf)]}
    }
    {\left(\frac{\T[a-1][0][+]\T[a-1][0][-]}{\T[a][0]\T[a-2][0]}\right)^{1-\delta_{a,1}}
    }\,.
  \end{align}
\end{subequations}
where \(\T\) denotes here \(\T[][][(\Rg)]\). We %
have replaced each factor \(1+\Y\) by a ratio of {\Tfs} in a given
gauge (usually chosen as \(\T[][][(\Rg)]\) except for \(1+\Y[][-1]\) which
is expressed through \(\T[][][(\Lf)]\)).

The equation \eqref{eq:YmidDelt} has to be written for all \(a\in
\ninter 1 {\Np-1}\). Then the {\lhs} is \(\frac{\Y[1][0][+] \Y[1][0][-]}{
  \Y[2][0]}\) (resp  \(\frac{\Y[2][0][+] \Y[2][0][-]}{
  \Y[3][0]\Y[1][0]}\), resp \(\cdots\)) when \(a=1\) (resp \(a=2\), resp
\(\cdots\)). Hence its logarithm is
\begin{align*}
  \left(
    \begin{array}{c}
 \mathrm{log}
       \frac{\Y[1][0][+] \Y[1][0][-]}{
   \Y[2][0]}\\
 \mathrm{log} \frac{\Y[2][0][+] \Y[2][0][-]}{
   \Y[3][0]\Y[1][0]}\\
\vdots\\
\mathrm{log}
\frac{\Y[\Np-2][0][+] \Y[\Np-2][0][-]}{
  \Y[\Np-1][0]\Y[\Np-3][0]}\\
 \mathrm{log}
 \frac{\Y[\Np-1][0][+] \Y[\Np-1][0][-]}{
   \Y[\Np-2][0]}
     \end{array}
\right)=       \left(
        \begin{array}{cccc}
          e^{\frac \bi 2 \partial_\us}+e^{-\frac \bi 2 \partial_\us}&-1\\[.3cm]
          -1&&\ddots\\[.3cm]
          &\ddots&&-1\\[.3cm]
          &&-1&e^{\frac \bi 2 \partial_\us}+e^{-\frac \bi 2 \partial_\us}
        \end{array}
      \right)
  \cdot 
      \left(
        \begin{aligned}
          \mathrm{log} & \left( \Y[1][0]\right)\\
          \mathrm{log} & \left( \Y[2][0]\right)\\
          \vdots\\
          \mathrm{log} & \left( \Y[{\Np}-2][0]\right)\\
          \mathrm{log} & \left( \Y[{\Np}-1][0]\right)
        \end{aligned}
      \right),%
\end{align*}

This {\lhs} of equation \eqref{eq:YmidDelt} will then be denoted as
\(\Y[][0][\cdot\MM]\) which is a short  
notation for \(\mathrm{exp}(\MM\cdot \mathrm{log}~ \Y[][0])\), where
\(\MM\) is the matrix defined in equation \eqref{eq:DefMMdiff}.
Using this notation the equation \eqref{eq:YmidDelt} reads
\begin{align}
\label{eq:YMidEquThroughM}
  \Y[][0][\cdot\MM]=\left(\frac{\T[a][1] \T[a][1][(\Lf)]}{\T[a-1][0]
      \T[a+1][0]}\right)^{\cdot \MM} \times \left(
    \frac{\T[\Np][0][+]\T[\Np][0][-]}{\T[\Np][1]\T[\Np][1][(\Lf)]}\right)^{\delta_{a,{\Np}-1}}
   \left( \frac{\T[0][0][+]\T[0][0][-]}{\T[0][1]\T[0][1][(\Lf)]}\right)^{\delta_{a,1}}   \,,
\end{align}
where the last factors have to be added to get the correct expression
at \(a=1\) and at \(a=\Np-1\).

From this point, it would be natural to multiply by \(\MM^{-1}\) and to
deduce \(\Y[][0]\). One actually has to do something slightly less
direct, because %
\(\MM^{-1}\) is not completely well
defined. Indeed we saw in section \ref{sec:equiv-hirota-equat} that to
invert \({\MM}\), one has to solve an equation of the form \(\FS[\Np]
x=f\) (see \eqref{eq:InvM}), and the solution to this equation is not
unique.
One possible way to invert \({\MM}\) is by introducing a matrix \(\tilde \MM\) with coefficients \(\tilde
     \mm_{\coordi,\coordj}\) defined by 
    \begin{gather}
      \label{eq:DefTilMinvLapl}
      \tilde \mm_{\coordi,\coordj} =
      \left\{
        \begin{array}{ccc}
          \FS[\coordj] ~ \FS[{{\Np}-\coordi}]&\If&\coordi>\coordj\\{}
          \FS[\coordi] ~ \FS[{{\Np}-\coordj}]&\If&\coordi\leq\coordj
        \end{array}
      \right.\\
      \where \FS[\coordi]\equiv
      {\DD}^{1-\coordi}+{\DD}^{3-\coordi}+\cdots +{\DD}^{\coordi-1}
      =
      \sum_{s=-{\frac{\coordi-1}{2}}}^{\frac{\coordi-1}{2}}
      {\DD}^{2s}\,,\qquad\where \DD=e^{\frac \bi 2 \partial_\us}\,.
    \end{gather}
     One can easily check (for instance with {\Matm}), that this matrix
    \(\tilde \MM\) is the adjugate matrix (or
    co-matrix)\footnote{Here, we call adjugate matrix
      of \(\MM\) 
      the transpose of the matrix whose elements are the co-factors
      ({\idest} the minors) of \(\MM\).} of \(\MM\), and it obeys
    \begin{gather}
      \label{eq:propmtildinvlap}
      \tilde \MM \cdot \MM=\FS[{\Np}]\bI\,
    \end{gather}
    where {\bI} denotes the identity matrix of size \(({\Np}-1)\times({\Np}-1)\).
This equation \eqref{eq:propmtildinvlap} means that
 \begin{gather}
  \left( \left(f_a\right)^{\cdot \MM}\right)^{\cdot \tilde \MM}=
f_a^{\FS[{\Np}]},\\
\where f_a^{\FS[{\Np}]}\equiv \mathrm{exp}\left(\FS[{\Np}]~\mathrm{log}~f_a\right)
=
\prod_{\kk=-\frac{\Np-1}{2}}^{\frac{\Np-1}{2}} f_a^{[+2 \kk]}\,.
\end{gather}

With these notations, the equation \eqref{eq:YMidEquThroughM} becomes
\begin{subequations}
\label{eq:MidStillFS}
  \begin{align}
    \Y[][0][{{\FS[{\Np}]}}]=&\left(\frac{\T[a][1]
        \T[a][1][(\Lf)]}{\T[a-1][0] \T[a+1][0]}\right)^{\FS[{\Np}]}
    \left(\frac{\T[\Np][0][+]\T[\Np][0][-]}{\T[\Np][1]\T[\Np][1][(\Lf)]}\right)^{\tilde
      {\mm}_{a,\Np-1}} \left(
      \frac{\T[0][0][+]\T[0][0][-]}{\T[0][1]\T[0][1][(\Lf)]}\right)^{\tilde
      {\mm}_{a,1}}\\
    =&\left(\frac{\T[a][1] \T[a][1][(\Lf)]}{\T[a-1][0]
        \T[a+1][0]}\right)^{\FS[{\Np}]}
    \left(\frac{\T[\Np][0][+]\T[\Np][0][-]}{\T[\Np][1]\T[\Np][1][(\Lf)]}\right)^{
      \FS[a]} \left(
      \frac{\T[0][0][+]\T[0][0][-]}{\T[0][1]\T[0][1][(\Lf)]}\right)^{\FS[\Np-a]}\,.
  \end{align}
\end{subequations}
In order to obtain \(\Y[][0][{{\FS[{\Np}]}}]\), we have to invert the
transformation \(f\mapsto f^{\FS[\Np]}\). By doing a Fourier transform,
we see that the kernel \(\Ker_\Np\) with Fourier transform 
\begin{gather}
  \label{eq:KerNpDef}        
  \widetilde{\Ker_\Np}(\omega)=\frac 1 {\sum_{{\jj}=-\frac {\Np-1} 2}^{\frac        
      {{\Np}-1} 2} e^{2{\bi} \pi {\jj} \omega}}
\end{gather}
obeys the equation
\begin{gather}
\label{eq:PropDefKN}
  \left(f^{\st \Ker_\Np}\right)^{\FS[\Np]}=f,\qquad \where f^{\st
    \Ker_\Np}\equiv \mathrm{exp}\left(\Ker_\Np\st \mathrm{log}~f_a\right)\,.
\end{gather}
After inverse Fourier transform from \eqref{eq:KerNpDef}, we obtain
the following explicit expression of the kernel \(\Ker_\Np\):
\begin{equation}        
\label{eq:ker_np-pi-leftfr}
\Ker_{\Np}(\us)=\frac{\mathrm{tan}\left(\frac{1}{2} \pi        
    \left(\frac{1}{{\Np}}-\frac{2 {\bi}        
        \us}{{\Np}}\right)\right)+\mathrm{tan}\left(\frac{1}{2} \pi        
    \left(\frac{1}{{\Np}}+\frac{2 {\bi} \us}{{\Np}}\right)\right)}{2 {\Np}}\,.
\end{equation} 

Let us note at this point that the property \eqref{eq:PropDefKN} means
that a particular solution of the equation \(f^{\FS[\Np]}={\gnothing}\) is given by \(f={\gnothing}^{\st
  \Ker_\Np}\).
This solution is not unique, and the general solution of this equation
is \(f={\gnothing}^{\st
  \Ker_\Np}~z\), where \(z\) obeys \(z^{\FS[\Np]}=1\). Such a function \(z\)
is called a ``{\zerm}''.

If we did not take the zero-modes into account, 
we
could naively expect %
(from \eqref{eq:MidStillFS}) that
  \begin{align*}
    \Y[][0]=&\left(\frac{\T[a][1]
        \T[a][1][(\Lf)]}{\T[a-1][0] \T[a+1][0]}\right)
    \left(\left(\frac{\T[\Np][0][+]\T[\Np][0][-]}{\T[\Np][1]\T[\Np][1][(\Lf)]}\right)^{
      \FS[a]}\left(
      \frac{\T[0][0][+]\T[0][0][-]}{\T[0][1]\T[0][1][(\Lf)]}\right)^{
      \FS[{\Np}-a]}\right)^{\st \Ker_\Np} \,.
  \end{align*}
With this expression, we would actually obtain an incorrect behavior
at large \(\us\), in the sense that \(\mathrm{log}\left(\Y[][0]\right)\)
would have a finite limit  at \(\us\to\infty\). %
In order to
reproduce \eqref{eq:YsysZeMo}, we should %
therefore add an extra factor ({\zerm}) and
write
  \begin{gather}
\label{eq:MidNodEq}
    \fdisp{\Y[][0]\bumpeq
    e^{-E_a}
    \left(\frac{\T[a][1]
        \T[a][1][(\Lf)]}{\T[a-1][0] \T[a+1][0]}\right)
    \left(\left(\frac{\T[\Np][0][+]\T[\Np][0][-]}{\T[\Np][1]\T[\Np][1][(\Lf)]}\right)^{
      \FS[a]}\left(
      \frac{\T[0][0][+]\T[0][0][-]}{\T[0][1]\T[0][1][(\Lf)]}\right)^{
      \FS[{\Np}-a]}\right)^{\st \Ker_\Np} }\,,\\
\where
E_a\equiv \LF
    \frac{\sin\frac{\pi a}{{\Np}}}{\sin\frac{\pi }{{\Np}}}
    \cht[\us]\,, 
\label{eq:EaDef}
\\
\And f_1\bumpeq f_2 \qquad \textrm{ denotes } \qquad (f_1)^{\FS[{\Np}]}=(f_2)^{\FS[{\Np}]}\,.
  \end{gather}
\index{0796@\ensuremath{\bumpeq}\addcontentsline{toc}{chapter}{Index and notations}
\addcontentsline{lot}{chapter}{Index and notations}
\label{sec:index}
}
Here the symbol \(\bumpeq\) is used to denote an equality which holds up to a
{\zerm} of \(\FS[{\Np}]\), ({\idest} \(f_1\bumpeq f_2\) means that \(f_1=z~f_2\)
where \(z\) obeys \({z}^{\FS[{\Np}]}=1\)).
 In \eqref{eq:MidNodEq}, this {\zerm} 
has to converge to a constant when \(\us\to\infty\), and since it
  is \(\bi\Np\)-periodic (because \(\frac {z^{[+\Np]}}{z^{[-\Np]}} =\frac
  {\left(z^{\FS[{\Np}]}\right)^+}{\left(z^{\FS[{\Np}]}\right)^-}=1\)),
  the Liouville theorem states that it is characterized by its pole
  structure in the strip \(\Ast {\Np}\).
   For instance it can contain
\(\CDD\) factors (defined in
\eqref{eq:wher-sscal_0-fracg}).  We will
see in the next section how to fix this {\zerm}.

This equation \eqref{eq:MidNodEq} will be called the ``middle nodes equation'', and it will
be very useful because it encodes the large \(\us\) asymptotic behavior
\eqref{eq:YsysZeMo} into an equation on the {\Tfs}.

  \subsubsection{Bethe equations}
  \label{sec:bethe-equations-2}

Let us now see how this ``middle nodes equation'' can be used to
understand better the large \(\LF\) limit of the {\Ysys}, and to obtain
the Bethe equations \eqref{eq:BethePCM} and
\eqref{eq:BethePCM2}. To this end, we will use the expressions
(\ref{eq:TRTL0}-\ref{eq:TRTLN}) of the functions \(\T[][][(\Rg)]\) and
\(\T[][][(\Lf)]\).

A first remark is that these expressions correspond to polynomial spin
chains (as in chapter \ref{part:qoperatorsspin}) and therefore they
have to obey the equation \eqref{eq:BethePCM}.

Next, let us express the {\Tfs} on the {\rhs} of the ``middle
node equation'' \eqref{eq:MidNodEq}. 
First, we obtain (from (\ref{eq:TRTL0}-\ref{eq:TRTLN})
\begin{gather}
\label{eq:TA=phiNaive1}
  \T[0][][(\Rg)]=\gqf[(\Np)]{(\Rg)}\left(\us-\bi\frac{s+\Np}2\right)=
\vf[{[-s-\Np/2]}]
  \,,\qquad \T[0][][(\Lf)]=\vf[{[-s-\Np/2]}]
  \,,\\
  \T[\Np][][(\Rg)]=  \label{eq:TA=phiNaive2}
\gqf[(\Np)]{(\Rg)}\left(\us+\bi\frac{s}2\right)=
\vf[{[+s+\Np/2]}]
  \,,\qquad \T[\Np][][(\Lf)]=\vf[{[+s+\Np/2]}]%
  \,,\\
  \T[][0][(\Rg)]=\gqf[(\Np)]{(\Rg)}\left(\us+\bi\frac{a-\Np}2\right)=
\vf[{[+a-\Np/2]}]
  \,,\qquad \T[][0][(\Lf)]=
\vf[{[+a-\Np/2]}]
  \,,
\label{eq:TA=phiNaive3}
\end{gather}
which (unlike the equations of the previous section) are only valid in
the \(\LF\to\infty\) limit. %
In these expressions, \(\vf\) denotes the polynomial \(\fQ[][0]\). %

We will show in the next section that these expressions only hold
inside specific strips of the complex plane. These strips will be such
that 
for instance, the function \(1+\Y[1][0]=
\frac{\T[1][0][+]\T[1][0][-]}{\T[2][0]\T[0][0]}\) will
actually\footnote{As we will see, this zero arises because the
  numerator has a zero due to \eqref{eq:TA=phiNaive3}, whereas the
  denominator lies in a domain of the complex plane where 
\eqref{eq:TA=phiNaive1} does not hold, hence the denominator does not
have a zero at the same position.
} have a zero at every
\(\us= \theta_\jrt + \bi\frac \Np 4\), where \(\theta_\jrt\) denotes an arbitrary
root of \(\vf=\fQ[][0]\). For the moment, let us just show that if \(1+\Y[1][0]=
\frac{\T[1][0][+]\T[1][0][-]}{\T[2][0]\T[0][0]}\) has a zero at
\(\us= \theta_\jrt + \bi\frac \Np 4\), then we recover the Bethe equation
\eqref{eq:BethePCM}.

To this end, let us see what comes out if the expressions
(\ref{eq:TA=phiNaive1}-\ref{eq:TA=phiNaive3}) are plugged into 
\eqref{eq:MidNodEq}. First the factor \(    \left(\left(
\frac{\T[0][0][+]\T[0][0][-]}{\T[0][1]\T[0][1][(\Lf)]}\right)^{\FS[{\Np}-a]}\right)^{\st
\Ker_\Np} \) 
becomes\footnote{Let us remind here that where \(\T\) is
written (as opposed to \(\T[][][(\Lf)]\)), it implicitly denotes
\(\T[][][(\Rg)]\) which is expressed from
(\ref{eq:TA=phiNaive1}-\ref{eq:TA=phiNaive1}).}   
\begin{gather}
\label{eq:leftl-fract00+t00-t0}
   \left(\left(
\frac{\T[0][0][+]\T[0][0][-]}{\T[0][1]\T[0][1][(\Lf)]}\right)^{\FS[{\Np}-a]}\right)^{\!\!\!\st %
\Ker_\Np%
}\bumpeq
   \left(\left(
\frac{\vf[{[1-{\Np}/2]}]
} {\vf[{[-1-{\Np}/2]}]%
} \right)^{\st \Ker_\Np}\right)^{\!\!\! %
\FS[{\Np}-a]%
}%
\bumpeq \left((-1)^{\frac {\dg[0]}\Np} %
\frac{{\Sb}^{[+{\Np}/2]}\vf[{[+{\Np}/2]}]}
{\vf[{[+{\Np}/2-2]}]}\right)^{\!\!\!\FS[{\Np}-a]}\,,\\
\where \Sb(\us) =\prod_{\nrt=1}^{\dg[0]} \Sscal_0(\us-\theta_\nrt)\,.
\end{gather}
Here \(\Sb\) is the product of \(\prod \Sscal_0(\us-\theta_\nrt)\), which runs over all
the roots \(\theta_\nrt\) of \(\varphi\), and the symbol \(\bumpeq\)
stresses that %
the
equality %
holds only up to a
{\zerm} of \(\FS[{\Np}]\).
 This equality is obtained from the crossing
relation \eqref{eq:PCFCrossing}, which gives \(\left( 
\frac{\Sb^{[+{\Np}/2]}\vf[{[+{\Np}/2]}]}
{\vf[{[+{\Np}/2-2]}]}\right)^{\FS[{\Np}]}=\frac{\vf[{[1-{\Np}/2]}]}
{\vf[{[-1-{\Np}/2]}]}\).
The expression \eqref{eq:leftl-fract00+t00-t0} can still be
``simplified'' a little, as:
\begin{align}
   \left(\left(
\frac{\T[0][0][+]\T[0][0][-]}{\T[0][1]\T[0][1][(\Lf)]}\right)^{\FS[{\Np}-a]}\right)^{\st
\Ker_\Np}\bumpeq&-\frac{\vf[{[3{\Np}/2-a-1]}]}{\vf[{[-{\Np}/2+a-1]}]}
\left((-1)^{\frac {\dg[0]}\Np}
\Sb^{[+{\Np}/2]}\right)^{\FS[{\Np}-a]}\\
\bumpeq&-\frac{\vf[{[3{\Np}/2-a-1]}]}{\vf[{[-{\Np}/2+a-1]}]} \frac{\vf[{[-{\Np}/2-a+1]}]}{\vf[{[3{\Np}/2-a-1]}]}
\left(\frac {(-1)^{\frac {\dg[0]}\Np}} {\Sb^{[-{\Np}/2]}}\right)^{\FS[a]}\,.
\end{align}
Performing the same computation for the factor \(
    \left(\left(\frac{\T[\Np][0][+]\T[\Np][0][-]}{\T[\Np][1]\T[\Np][1][(\Lf)]}\right)^{
      \FS[a]}\right)^{\st \Ker_\Np}\), one obtains:
\begin{align}
\label{eq:y0bumpeq}
  \Y[][0]\bumpeq e^{-E_a} \left(\frac{\T[a][1]
        \T[a][1][(\Lf)]}{\T[a-1][0] \T[a+1][0]}\right)
\frac{\vf[{[-{\Np}/2-a+1]}]}{\vf[{[-{\Np}/2+a-1]}]}
\frac{\vf[{[-{\Np}/2-a+1]}]}{\vf[{[-{\Np}/2+a+1]}]} 
\left(
\frac 1 {\left(\Sb^{[-{\Np}/2]}\right)^2}
\right)^{\FS[a]}\,.
\end{align}

\paragraph{Pole structure of \(\Y[][0]\)}
\label{sec:pole-structure-y0}

The {\zerm} in this expression can be found by investigating the
pole structure of \(\Y[][0]\): due to the expression
\(1+\Y[][0]=\frac{\T[a][0][+] \T[a][0][-]}{\T[a+1][0]\T[a-1][0]}\), we
see that \(1+\Y[][0]\) has no pole except maybe when
\(\T[a+1][0]\T[a-1][0]\) cancels, {\idest} at positions
\(\theta_\jrt+\bi/2\left(-a\pm1+{\Np}/2\right)\). Actually, both
the numerator and the denominator have zeroes at a position which
tends to \(\theta_\jrt+\bi/2\left(-a\pm1+{\Np}/2\right)\) when
\(\LF\to\infty\). But if they do not coincide
exactly, then they give rise to a pole of \(1+\Y[][0]\) (and of \(\Y[][0]\)
as well). Therefore we see that at most, \(\Y[][0]\) has simple poles at
positions \(\theta_\jrt+\bi/2\left(-a\pm1+{\Np}/2\right)\).

This allows to find the correct {\zerm} in \eqref{eq:y0bumpeq}.
Explicitly, we obtain
\begin{gather}
\label{eq:y0-=-e}
  \Y[][0] =  e^{-E_a} \left(\frac{\T[a][1]
        \T[a][1][(\Lf)]}{\T[a-1][0] \T[a+1][0]}\right)
\frac{\vf[{[-{\Np}/2-a+1]}]}{\vf[{[-{\Np}/2+a-1]}]}
\frac{\vf[{[-{\Np}/2-a+1]}]}{\vf[{[-{\Np}/2+a+1]}]} 
\left(
\frac 1 {\left(\Sb^2\CDb\right)^{[-{\Np}/2]}}
\right)^{\FS[a]}\,,\\
\where \CDb(\us)=\prod_\nrt \CDD(\us-\theta_\nrt)\,,
\end{gather}
where we see that in the denominator, the product 
\(\T[a-1][0] \T[a+1][0] \vf[{[-{\Np}/2+a-1]}]\vf[{[-{\Np}/2+a+1]}]\)
 has double zeroes at positions
\(\theta_\jrt+\bi/2\left(-a\pm1+{\Np}/2\right)\). Due to the presence of
the factor \(\CDb\), the product \(\Sb^2\CDb\) has simple poles at
positions  \(\theta_\jrt\pm \bi\), hence \(\left(
\frac 1 {\left(\Sb^2\CDb\right)^{[-{\Np}/2]}}
\right)^{\FS[a]}\) has simple zeroes at the positions
\(\theta_\jrt+\bi/2\left(-a\pm1+{\Np}/2\right)\). Thus we see that the
presence of the factor \(\CDb\) is necessary for the consistency of the
pole structure of \(\Y[a][0]\).

It is unfortunately not straightforward to show that there {\cannnot} 
be any other {\zerm} in the expression \eqref{eq:y0-=-e}
of the function \(\Y[][0]\). In order to exclude the possibility of
other zero-modes, one actually has to require (in addition to the pole
structure described above) a minimality of the number of zeroes of
\(\Y[][0]\).

While the 
condition that \(\Y[][0]\) has simple poles at
positions 
\(\theta_\jrt+\bi/2\left(-a\pm1+{\Np}/2\right)\) %
is a clear
consequence  of %
our analyticity conditions
namely the behavior of the
{\Tfs} at \(\LF\to\infty\), the minimality of the number of zeroes of
\(\Y[][0]\) is (by contrast) rather a naturality condition, which can
be viewed as an 
additional analyticity condition on our solution of the {\Ysys}.

\paragraph{Bethe equation}
\label{sec:bethe-equation}

As indicated above, we will
now show how to recover the Bethe equations under the assumption that 
\(%
  1+\Y[1][0]\) has  zeroes at positions \(
{\us= \theta_\jrt + \bi\frac \Np 4}\). 
Therefore, %
we would like to express the {\Tfs}\footnote{Let us remind here that where \(\T\) is
written (as opposed to \(\T[][][(\Lf)]\)), it implicitly denotes
\(\T[][][(\Rg)]\).}    of the ratio \(\left(\frac{\T[a][1]
        \T[a][1][(\Lf)]}{\T[a-1][0] \T[a+1][0]}\right)\) which remains
    on the {\rhs} of \eqref{eq:y0-=-e}.  To do this, we can
    use the relation \eqref{eq:qtoT11} which gives (after the changes
    of variables \eqref{eq:TchangeShift} and (\ref{eq:QPCFDef1}-\ref{eq:QPCFDefl}))
    \begin{align}
     \T[1][1][(\Rg)]\left(\us+\bi \frac {\Np} 4\right)= \rT[-\bi
     \us][1][1]=
\varphi(\us)\sum_{\mlvl=0}^{\Np+1}\frac{\fQ[\us+\bi+\bi \frac\mlvl
  2][\mlvl]%
}{\fQ[\us+\bi \frac\mlvl 2][\mlvl]%
} \frac{\fQ[\us-\bi+\bi \frac{\mlvl+1} 2][\mlvl+1]%
}{\fQ[\us+\bi \frac{\mlvl+1} 2][\mlvl+1]%
}\,.
    \end{align}
Therefore we see that in \(\T[1][1][(\Rg)]\left(\theta_\nrt+\bi \frac {\Np}
  4\right)\), the factor \(\varphi(\theta_\nrt)\) in front of the sum
vanishes. Therefore, only the term \(\mlvl=0\) survives, because the
denominator also contains \(\varphi(\us)\equiv \fQ[][0]\). Hence, we get 
    \begin{gather}
     \T[1][1][(\Rg)]\left(\theta_\nrt+\bi \frac {\Np} 4\right)=
\frac{\varphi\left(\theta_\nrt+\bi \right)  \fQ[\theta_\nrt-\frac \bi 2][+1]}
{\fQ[\theta_\nrt-\frac \bi 2][+1]}\,,
\\
      \T[1][1][(\Lf)]\left(\theta_\nrt+\bi \frac {\Np} 4\right)=
\frac{\varphi\left(\theta_\nrt+\bi\right)  \fQ[\theta_\nrt-\frac \bi 2][-1]}
{\fQ[\theta_\nrt-\frac \bi 2][-1]}\,.
\end{gather}

Inserting this expression into \eqref{eq:y0-=-e}, 
the equation \(\Y[1][0](\theta_\nrt+ \bi\frac \Np 4)=-1\) gives 
the Bethe
equation \eqref{eq:BethePCM2}
\begin{gather}
\label{eq:fdisp-1-=}
  \fdisp{
  -1 = \left. e^{-\bi \LF\,p_1} 
\frac{\fQ[\us-\bi/2][1]\fQ[\us-\bi/2][-1]}
  {\fQ[\us+\bi/2][1]\fQ[\us+\bi/2][-1]}
\frac 1 {\left(\Sb^2\CDb\right)}
\right|_{\us=\theta_\nrt}}\,,\\
\where~ \LF\,p_a\equiv \LF
    \frac{\sin\frac{\pi a}{{\Np}}}{\sin\frac{\pi }{{\Np}}}
    \sht[\us]=-\bi E_a(\us + \bi\frac\Np4)\,.
\end{gather}

This showed that the Bethe equation 
 arises
naturally from the {\Ysys}, 
if the {\Tfs} take polynomial values corresponding to two different
\(\SU\Np\) chains.
This motivates the identification \eqref{eq:TRTLN} which says
  that in the asymptotic limit, the {\Yfs} 
(which are ratios of the densities of
holes and particles in the mirror model) are the ratios of the
polynomial {\Tfs} constructed in chapter \ref{sec:integr-et-} for {\csds}.

 These expressions will be a starting point to
understand the properties of {\Tfs} when the size \(\Lf\) is finite. We
will also see that the analysis of the next sections  will
explain the fact that \(\Y[1][0](\theta_\nrt+ \bi\frac \Np 4)=-1\),
which was assumed in the above argument.

\subsection{Parameterization of the {\qfs}}
\label{sec:parameterization-qfs}

As explained in section \ref{sec:writing-finlies}, an important step which
we now have to perform in order to obtain a FiNLIE is to parameterize
the {\qfs} in a way which encodes their analyticity properties. To do
this, we will first investigate the analyticity properties of the {\Tfu}-
and {\Yfs}. Then, we will deduce the analyticity properties of the
{\qfs}, and we will encode these properties into a convenient
parameterization of these {\qfs}.

\subsubsection{Analyticity strips for the {\Yfs}}
\label{sec:analyt-strips-yfs}

One can see from \eqref{eq:YsysZeMoL} that the limit of the {\Yfs}
when \(\LF\to\infty\) is not analytic in the whole complex, and hence
the {\Tfs} (and the {\qfs}) which are the building blocks of the
{\Yfs} {\cannnot} be analytic on the whole complex plane.
Indeed, we have
\begin{align}
  \If |\Im({\us})|< \frac \Np 4,&\qquad
  \Y[a][0]\xrightarrow[]{\LF\to\infty}0,\\
  \If |\Im({\us})|\in \left]\frac \Np 4,3\frac\Np4\right[,&\qquad |\Y[a][0]|\xrightarrow[]{\LF\to\infty}\infty,
\end{align}
because \(\Y[][0]\sim e^{\LF
    \frac{\sin\frac{\pi a}{{\Np}}}{\sin\frac{\pi }{{\Np}}}
    \cht[\us]}\).

Therefore we will call ``analyticity strip'' of \(\Y[a][0]\) the domain
\(\Ast{\Np/2}\) defined by
\begin{gather}
  \Ast \nn \equiv \left\{z\in\bC\middle| ~~|\Im(z)|<\frac \nn 2\right\}\,.
\end{gather}
\index{A0@\ensuremath{\Ast a}}
The correct statement defining this strip is that \(\Y[a][0]\) is
meromorphic 
on \(\Ast{\Np/2}\) %
and its 
limit %
at \(\LF \to \infty\) is well defined
and meromorphic on \(\Ast{\Np/2}\). %

We will denote this statement as \(\Y[a][0]\in \Af[\mm] {\Np/2}\), where 
\index{A1@\ensuremath{\Af [\mm] \nn}}
\(\Af [\mm]\nn\) denotes the set of meromorphic functions which have a
meromorphic limit on \(\Ast \nn\) when \(\LF\to\infty\).

For \(s\neq0\), let us show that
\begin{gather}
\label{eq:yinafmm-s+np2-}
  \Y\in\Af[\mm] {|s|+\Np/2}\,.
\end{gather}
That means that on this domain \(\Y[a][s]\)
tends to the asymptotic solution of section
\ref{sec:asymptotic-limit} when \(\LF\to \infty\).

First, we see that the {\Yfs} obtained from the {\TBAE} are analytic
on the real axis, 
  as it can be read from \eqref{eq:PCMTBArrb} (or
  \eqref{eq:TBAeqRewriteY} in terms of {\Yfs}). Moreover, we have seen
  that the {\TBAE} imply the {\YsE}, which can be rewritten as 
  \begin{gather}
\label{eq:y=-e-lf}
    \Y\bumpeq
e^{-\delta_{s,0}~ E_a}
\left(\frac{1+\Y[][s+1]}{1+1/\Y[a+1]}
  \frac{1+\Y[][s-1]}{1+1/\Y[a-1]}\right)^{\st \Ker_2}\,,\\
\where
E_a\equiv \LF
    \frac{\sin\frac{\pi a}{{\Np}}}{\sin\frac{\pi }{{\Np}}}
    \cht[\us]
\,\qquad\qquad\And f^{\st
    \Ker_\Np}\equiv \mathrm{exp}\left(\Ker_\Np\st \mathrm{log}~f_a\right)\,.
  \end{gather}
In principle, this expression contains a {\zerm} \(z\) (denoted by the
symbol \(\bumpeq\)) such that
\(z^+z^-=1\). This {\zerm} is \(2\bi\)-periodic (because
\(\frac{z^{[+2]}}{z^{[-2]}} = \frac{z~z^{[+2]}}{z^{[-2]} z} =1\)) and
bounded at infinity. Therefore the Liouville theorem shows that it is
completely characterized by its 
singularities inside \(\Ast 2\) (which can be deduced from its zeroes
and its singularities in \(\Ast 1\) if we use the relation \(z^+z^-=1\)). 
In our construction we assume the poles structure of all {\Yfs}
smoothly converges to their asymptotic pole structure, and therefore
we can expect that this {\zerm} is a meromorphic function with a
smooth limit when \(\LF\to\infty\), hence this {\zerm} will be assumed
to preserve the analyticity strips.

We will show iteratively, by using the expression \eqref{eq:y=-e-lf},
that the analyticity strips are given by \eqref{eq:yinafmm-s+np2-}.
To this end, we will disregard the %
{\zerm} which could be
hidden in the symbol \(\bumpeq\), as motivated above. A more rigorous
version of
the argument can also be written using the {\TBAE}, and %
it would give
the same result, without having to disregard any {\zerm}.

To start with, we can use the definition %
of the convolution \eqref{eq:DefConv} %
to
write
\begin{align}
    \Y(\us+\alpha \bi/2)=
e^{-\delta_{s,0}~ E_a(\us+\alpha \bi/2)}
\left(\frac{1+\Y[][s+1]}{1+1/\Y[a+1]}
  \frac{1+\Y[][s-1]}{1+1/\Y[a-1]}\right)^{\st \Ker_2^{[+\alpha]}}\,,
\end{align}
which holds when \(|\alpha|<1\). As the right-hand side gives an
analytic expression of \(\Y(\us)\) ion \(\Ast 1\), 
we can
conclude that
\( \Y\in\Af[\mm] {1}\).  At \(|\alpha|\geq1\), the pole of the kernel
\(\Ker_2\) (defined in \eqref{eq:ker_np-pi-leftfr}) at
position \(\pm \bi/2\) prevents us from going through this
argument, and we should understand the consequences of this pole.

To this end
let us comment on the analytic structure of
\(\Ker_\Np\st f\) when the function \(f\) is analytic
on a strip  \(\Ast \nn\), where \(\nn>\Np-1\). Due to the presence of a pole
in \(\Ker_\Np\), the 
function
\begin{equation}
h_1(\us)\equiv \int_{\vs\in\bR}\Ker_\Np(\us-\vs) f(\vs)\mathrm{d}\vs
\end{equation}
has two cuts where it suddenly jumps, when \(\Im({\us})=\pm \frac {\Np-1}2\),
because \(\Ker_\Np\) has a pole at position \(\pm\bi\frac {\Np-1}2\),
with
residue \(\pm \frac 1 {2\bi \pi}\). The %
Cauchy theorem then shows that (for \(\us\in\bR\)),
\begin{equation}
 \lim_{\substack{\epsilon\to 0\\\epsilon>0}} ~~~h_1\left(\us \pm \bi\frac
   {\Np-1}2+\bi\epsilon\right) -  h_1\left(\us \pm \bi\frac
   {\Np-1}2-\bi\epsilon\right) = \mp f\left(\us%
 \right)\,.
\end{equation}
On the other hand, we can define\footnote{This function \(h_2\) can
  equivalently be defined as 
\begin{gather}
h_2(\us)\equiv \int_{\vs\in\bR}\Ker_\Np(\vs) f(\us-\vs)\mathrm{d}\vs
\end{gather}
} 
\begin{gather}
h_2(\xs+\bi \ys)\equiv \int_{\vs\in\bR}\Ker_\Np(\xs-\vs) f(\vs+\bi
\ys)\mathrm{d}\vs,\qquad\qquad \xs,\ys\in\bR
\end{gather}
This function \(h_2\) coincides with \(h_1\) if \(\ys<\bi \frac {\Np-1}2\),
as it can be seen by a simple contour manipulation. But unlike \(h_1\),
the function \(h_2\) has no jump at \(\bi \frac {\Np-1}2\).

In what follows,
we will not explicitly choose
 between the definitions \(h_1\) and \(h_2\), which are equivalent on
 the real axis. We will simply remark that as shown above, we can
 write (if \(f\) is analytic on \(\Ast \nn\), and \(\us\) is real)
\begin{gather}
\label{eq:Redistribue-conv-shifts}
  \left(\Ker_\Np\st f\right)^{[a+b]}=\Ker_\Np^{[+a]}\st
  f^{[+b]},\qquad\qquad |a|<\Np-1,|b|<\nn,\\
\where 
  \left(\Ker_\Np\st f\right)^{[a]}%
  \equiv e^{\bi \frac a
    2 \partial _\us}\left(\Ker_\Np\st f\right)
\equiv \sum_{\nn=0}^\infty\frac{\left(\bi \frac a
    2\right)^\nn \partial_\us^\nn}{{\nn}!}\left(\Ker_\Np\st f\right)\,.
\end{gather}
Here, we have a non-ambiguous definition of the convolution when
\(\us\) is real , and we formally define the shifted convolution
\(\left(\Ker_\Np\st f\right)^{[a]}\) as an analytic continuation. The
above discussion about \(h_1\) and \(h_2\) teaches us that the shift can be
distributed between the functions \(f\) and \(\Ker_\Np\) (see
\eqref{eq:Redistribue-conv-shifts}) as long as we do not meet poles.

The same analysis can actually also be performed if \(f\) contains
poles, but then an extra term should be 
added to the equation
\(  \left(\Ker_\Np\st f\right)^{[b]}=\Ker_\Np\st
  f^{[+b]}\) in order to take into account %
  the contribution of shifting the integration
  contour across the pole\footnote{
For instance, if \(f\) has a pole at position \(\bi/2\) with residue
\({\RR}_0\), then we see that 
\(\Ker_Np\st \left(f^{[+1+\epsilon]}\right)-\Ker_\Np\st \left(f^{[+1-\epsilon]}\right)\) is equal (when
\(\epsilon\to 0\)) to \(-2\bi\pi \frac{\Ker_\Np (0)}{\us} {\RR}_0 \). Hence
in order to define \(  \left(\Ker_\Np\st f\right)^{[b]}\) as the
analytic continuation of \(\left(\Ker_\Np\st f\right)^{[b]}\), we
should define \(  \left(\Ker_\Np\st f\right)^{[b]}=\Ker_\Np\st
  f^{[+b]} -2\bi\pi \frac{\Ker_\Np (0)}{\us-\bi/2} {\RR}_0     \) when \(b>1\).
}. This can be done case by case, and at the
  price of an integration by part, it also works if \(f\) has a
  logarithmic pole (in order to define \(f^{\st \Ker_\Np}\)).

  Keeping these statements in mind, we can proceed to prove
  \eqref{eq:yinafmm-s+np2-}. First, as we said, the expression
  \eqref{eq:y=-e-lf} (where the {\rhs} is analytic at least on the real axis)
  gives immediately \( \Y\in\Af[\mm] {1}\), which is true for arbitrary
  \((a,s)\). This result teaches that the ratio
  \(\frac{1+\Y[][s+1]}{1+1/\Y[a+1]}
  \frac{1+\Y[][s-1]}{1+1/\Y[a-1]}\)
  on the {\rhs} of \eqref{eq:y=-e-lf} is not
  only analytic around the real axis, but even on \(\Ast 1\). Therefore
  \(\left(\frac{1+\Y[][s+1]}{1+1/\Y[a+1]} 
  \frac{1+\Y[][s-1]}{1+1/\Y[a-1]}\right)^{\st \Ker_2}\) is analytic on
\(\Ast 2\). If \(s\neq 0\) or if \(\Np\geq 4\), \(e^{-\delta_{s,0}~ E_a}\) is also
analytic on \(\Ast 2\). By this statement we mean \(e^{-\delta_{s,0}~
  E_a}\in \Af [\mm]2\) as defined above, and this is true only\footnote{By contrast the statement ``for fixed \(\LF\), 
\(e^{-\delta_{s,0}~ E_a}\) is an analytic function of the variable
\(\us\)'' is always true.} if \(s\neq 0\) or
if \(\Np\geq 4\). 
Hence we deduce that \(\Y \in \Af[\mm] 2\) (if \(s\neq 0\) or
if \(\Np\geq 4\)).

In order to prove \eqref{eq:yinafmm-s+np2-}, we can therefore use an
iterative procedure, where at the step \(\nn\), we show that for several
values of \((a,s)\), we have \(\frac{1+\Y[][s+1]}{1+1/\Y[a+1]} 
  \frac{1+\Y[][s-1]}{1+1/\Y[a-1]}\in \Ast \nn\) and we deduce \(\Y\in
  \Ast {\nn+1}\). 
After the step\footnote{Here \(\lceil \cdots\rceil\)
  denotes the ceiling function.}  \(\nn%
=
\lceil {\Np}/2 \rceil\), the
iterations do not teach anything for \(s=0\),
because
\(e^{-E_a}\) is only analytic on the strip \(\Ast
{{\Np}/2}\). Therefore we only get \(\Y[a][0]\in \Af[\mm] {\Np/2}\).
Similarly, after the iteration \(\nn=1+\lceil {\Np}/2 \rceil\),
we {\cannnot} gain analyticity for \(\Y[][\pm1]\), because the ratio
\(\frac{1+\Y[][2]}{1+1/\Y[a+1][1]} 
  \frac{1+\Y[][0]}{1+1/\Y[a-1][1]}\) is only analytic on \(\Ast
  {\Np/2}\). Therefore we only get \(\Y[a][\pm1]\in \Af[\mm]
  {1+\Np/2}\). In the same way, for each value of \(|s|\), the iteration
  stops when the {\Yfs} on the {\rhs} stop gaining analyticity, and we
  get
  exactly
  \begin{equation*}
      \Y\in\Af[\mm] {|s|+\Np/2}\,.%
  \end{equation*}

\subsubsection{Analyticity strips for the {\Tfs}}
\label{sec:analyt-strips-tfs}

The analyticity strips identified in the previous section for the
{\Yfs} can now be used to find analyticity strips for {\Tfs}. The
analyticity properties of {\Tfs} depend on the gauge, because they 
can easily be spoilt by %
a gauge transformation
\eqref{eq:GaugeFreedom} having poor analyticity properties.

One can show that there exist gauges where the {\Tfs} have the
analyticity strip 
\begin{gather}
\label{eq:tinafmm-s+1+nn2-}
\begin{aligned}
  \T\in&\Af[\mm] {s+1+{\Np}/2}&&\If a \in \ninter 1 {\Np-1}\\
  \T\in&\Af[\mm] {s+{\Np}/2}&&\oth
\end{aligned}
\end{gather}
To show this one can simply use the proof of the \stapref{sta:TfromY},
which shows how to find {\Tfs} out 
of the set of {\Yfs} \(\left\{\Y[a][0]\middle| 1\leq a\leq {\Np}-1 
      \And 0\leq s\leq 1
    \right\}\). In this proof an expression of 
    \(\left\{\T[a][0]\middle| 0\leq a\leq {\Np}
      \And 0\leq s\leq 1
    \right\}\) is obtained in a specific gauge, %
    through the
    inversion of the {\op} \(f\mapsto f^{\FS[\Np]}\). This inversion
    can for instance\footnote{The inverse of \(f\mapsto
      f^{\FS[\Np]}\) is not unique, and choosing one specific inverse
      corresponds to a
      choice of gauge. If we use the kernel \(\Ker_\Np\) and we do not
    add any {\zerm}, then we get meromorphic {\Tfs}.} be done by
  means of the kernel \(\Ker_\Np\), and in 
    that case, we obtain {\Tfs} such that the analyticity condition
    \eqref{eq:tinafmm-s+1+nn2-} holds for \(0\leq s\leq 1\).
To conclude that this analyticity condition holds for all \(a\) and \(s\),
one can for instance do a recurrence over \(s\), and proceed quite
similarly\footnote{This recurrence can be sketched as follows: if
  we know the analyticity strip for \(\T[][s-1]\) and \(\T\), we can use
  the Hirota equation \eqref{eq:YHirota} to obtain
  \(\T[][s+1]\in\Af[\mm] {s+{\Np}/2}\) if \(a \in \ninter 1 {\Np-1}\) and 
  \(\T[][s+1]\in\Af[\mm] {s-1+{\Np}/2}\) if \(a \in \{0,\Np\}\). We can
  then deduce \(\T\) slightly outside its analyticity strip by writing 
  \(\T[][s+1]{+}={1+\Y[][s+1]}\T[a+1][s+1]\T[a-1][s+1]/\T[][s+1]\) which
  allows to deduce the wider strips written in
  \eqref{eq:tinafmm-s+1+nn2-}.
} to the %
proof of \eqref{eq:yinafmm-s+np2-}.

Moreover, we can for instance choose a gauge where the limit of \(\T\) is
\(\T[][][(\Rg)]\), which does not have any pole. %
We will assume that there exists (at least) one such gauge with
\begin{align}
\label{eq:tinaf-s+1+nn2-}
  \T\in&\Af {s+1+{\Np}/2}&&\If a \in \ninter 1 {\Np-1}\,;&&\And&&
  \T\in&\Af {s+{\Np}/2}&&\If a\in\{0,\Np\}\,.
\end{align}
Here 
\index{A2@\ensuremath{\Af \nn}}
\(\Af \nn\) denotes the set of holomorphic\footnote{One should be
  careful to distinguish the symbols \(\Af[\mm]\nn\), introduced to
  describe meromorphic functions, and the symbol \(\Af \nn\), introduced to
  describe holomorphic functions.
} functions which have an
holomorphic limit on \(\Ast \nn\) when \(\LF\to\infty\).

The analyticity constraint \eqref{eq:tinaf-s+1+nn2-} shows that the
analyticity properties of the {\Tfs} are simpler than the properties of
{\Yfs}. In particular, we see that the {\Tfs} do not have any poles.

Moreover, it is not difficult to show that  we can choose a gauge
where we have (like in the asymptotic limit) 
\begin{align}
\label{eq:t0s=t0-0-s}
  \T[0][s]=&\T[0][
0][{[-s]}],&\T[\Np][s]=&\T[\Np][0][{[+s]}].
\end{align}
In that case, we
 see that \(\T[0][0]=\T[0][s][{[+s]}]\) is analytic when \(\Im(\us)\in[-{\Np}/4-s 
 ,{\Np}/4  ]\), and this statement holds for arbitrary \(s\geq 0\).
Therefore \(\T[0][0]\) is
analytic as long as \(\Im({\us})<\frac \Np 4\). Proceeding the same way 
for \(\T[\Np][s]\) we obtain
\begin{align}
\label{eq:t0textrm-analyt-when}
  \T[0]&&\textrm{is analytic when} &&\Im({\us})<&\frac \Np 4+\frac s 2\\
  \T[\Np]&&\textrm{is analytic when} &&\Im({\us})>&-\frac \Np 4-\frac s
  2\,.
\label{eq:tnpt-analyt-when}
\end{align}

It is noteworthy that the analyticity domain for the functions 
\(\T[0][0]= \gpfs[\ove][-{\Np}]\) and \(\T[0][0]= \gqf[\emptyset]{}\) are
half-planes, and not just strips (like it is the case for the {\Tfu}-
and {\Yfs}). This property of {\qfs} is general in the sense that it
holds for 
{\qfs} with an arbitrary number of indices, and remarkably it also
holds for other models than the {\PCM}.

In the next sections, we will see how to find a solution of the Hirota
equation under these analyticity conditions, and we will see that the
solution obtained this way passes several nontrivial consistency checks.

\paragraph{Remark}
\label{sec:remark}

The analyticity strips defined above (in \eqref{eq:tinafmm-s+1+nn2-})
give non-zero analyticity 
strips when \(s\geq 0\) (or at least when \(s\geq -\Np/4\)), whereas 
for
\(s<-{\Np}/2\), the
analyticity strip has size zero. This corresponds to one possible
choice of gauge, but one can also choose a gauge having similar
properties for the ``left band'', {\idest} such that 
\eqref{eq:tinaf-s+1+nn2-} is replaced with
\begin{align}
\label{eq:tinaf--s+1+np2if}
  \tilT\in&\Af {-s+1+{\Np}/2}&&\If a \in \ninter 1 {\Np-1}\,;&&\And&&
  \tilT\in&\Af {-s+{\Np}/2}&&\If a\in\{0,\Np\}\,.
\end{align}

\subsubsection{Analyticity strips for the {\qfs}}
\label{sec:analyt-strips-qfs}

As we have seen, the {\qfs} are defined as a set of independent
solutions of the difference equation 
    \begin{gather}
      \label{eq:DetZeroDefqbis}
      0=\BlockDet{\gqss[][2\lL]}{1\leq \lL\leq
        {\Np}+1}{\Ts[1][\kk+\lL-3+s][\lL-\kk+3-s]}{\substack{2\leq \kk\leq {\Np}+1\\
          1\leq \lL\leq
          {\Np}+1}}\,,\qquad\qquad 1\leq \iq\leq {\Np}
    \end{gather}
If \(s=0\), this equation coincides with \eqref{eq:DetZeroDefq}, whereas
 if
\(s\neq 0\) %
it
follows from
\eqref{eq:HirotaSolutPCM} by the same argument as \eqref{eq:DetZeroDefq}.

In this equation, we can see that each coefficients
\(\Ts[1][\kk+\lL-3+s][\lL-\kk+3-s]\) entering the determinant is
analytic in the domain \(\Im({\us})\in]-\frac {\Np} 4-\lL-\frac 1 2, \frac {\Np}
4+\kk+s-\frac 5 2[\)
(where \(s\) can be chosen arbitrarily large).

This allows to choose the {\qfs} such that
\begin{gather}
\label{eq:textrmf-each-iqinn}
  \textrm{For each }\iq\in\ninter 1 \Np,
\qquad
\gqs[]%
\textrm{ is analytic
    when} %
\quad
  \Im({\us})>%
  -1/2-{\Np}/4\,,\\
\label{eq:textrmf-each-iqinn-1}
  \textrm{For each }\iq\in\ninter 1 \Np, \qquad%
  \gps[]%
  \textrm{ is analytic when} \quad %
  \Im({\us})<%
  \mathrlap{
  1/2+{\Np}/4\,,}\phantom{-1/2-{\Np}/4\,,}\\
  \And%
\quad
  \gqf[\emptyset]{}=1 ,\qquad\qquad\gpf[\emptyset]{}=1,
\label{eq:andgqf-}
\end{gather}
where the last constraint \eqref{eq:andgqf-} corresponds to the gauge
constraint \eqref{eq:t0s=t0-0-s}. %

The {\qfs} with multiple indices can be computed through the {\Wronskian}
determinant (\ref{eq:qqPCMSolHir}-\ref{eq:qqPCMSolHir2}) to get   %
\begin{align}
\label{eq:textrmfor-each-ii}
  &\textrm{For each }I \subset\ninter 1 \Np,\qquad%
  \gqf[I]{}%
  \textrm{ is analytic
    when} \qquad%
  \Im({\us})>%
  -1-{\Np}/4+%
  {|I|}/2\,,\\
  &\textrm{For each }I \subset \ninter 1 \Np, \qquad  %
  \gpf[I]{}%
  \textrm{ is
    analytic when} \qquad%
  \Im({\us})<%
  1+{\Np}/4-{|I|}/2\,,
\label{eq:textrmfor-each-ii-1}
\end{align}
which implies in turn that
\begin{align}
\label{eq:t=gqfs-gpfsnp-a}
  \T=\gqfs[(a)][+s]\wedge \gpfs[(\Np-a)][-s]\in \Af [{[-a+{\Np}/2]}]{s+2}\,,
\end{align}
where \(\Af[{[a]}]s\) denotes the set of holomorphic functions which have an
holomorphic limit  when \(\LF\to\infty\), for \(\us\in \Ast[{[a]}] s\equiv \left\{\us\in\bC\middle|\Im({\us})\in[\frac {-s-a}2,\frac{s-a}2]\right\}\).
We see that when we compute the {\Tfs} from the {\Wronskian} determinant
expression \(\T=\gqfs[(a)][+s]\wedge \gpfs[(\Np-a)][-s]\), we automatically obtain %
the analyticity strip \eqref{eq:t=gqfs-gpfsnp-a}, but this analyticity
strip is smaller than
in \eqref{eq:tinaf-s+1+nn2-}. The wider analyticity strip
\eqref{eq:tinaf-s+1+nn2-} means that in fact, there are
situations where the coefficients of the determinant are not analytic,
but some cancellations inside the determinant allow the determinant to
be analytic. Such situations will also be described in the chapter \ref{cha:dualite-adscft}.

\subsubsection{Cauchy representations of analytic functions}
\label{sec:cauchy-repr-analyt}

In the above subsections we have found analyticity constraints on the
set of {\qfs}. Let us now show, in a quite general context, that the
information we have about analytic functions often allows to
parameterize them in a very simple way. 

To demonstrate this, we will give a theorem which solves a very simple
``Riemann Hilbert'' problem. This type of problems (called Riemann
Hilbert problems) are situations where we know some analyticity
properties of a function on the complex plane and its behavior at
\(z\to\infty\), and they allow to uniquely fix this function.

\begin{statmt}
\label{sta:Cauchy}
  Let \(F(z)\) be an holomorphic function of \(z\) in the domain
  \(\Im(z)\geq 0\), and let \(G(z)\) be an holomorphic function of \(z\) in
  the domain 
  \(\Im(z)\leq 0\).

  If \(F(z)\) and \(G(z)\) go to zero at infinity, at least as a power
  law ({\idest} if there exists \(\epsilon>0\) such that \(F(z)
  z^{\epsilon}\xrightarrow [|z|\to\infty]{Im(z)\geq 0} 0\) and \(G(z)
  z^{\epsilon}\xrightarrow [|z|\to\infty]{Im(z)\leq 0} 0\)), then
we have the equality
\Pv{
  \begin{subequations}
\label{eq:begin-displ-frac-0}
    \begin{empheq}[left={ \displaystyle \frac 1
        {2\bi\pi}\int_{\vs\in\bR}\frac{F(\vs)-G(\vs)}{\vs-\us}\mathrm{d}\vs
        =\empheqlbrace},box=\fbox]{align}
      &F(\us) &\If&~~\Im(\us)>0
\label{eq:begin-displ-frac}
\\
      &G(\us) &\If&~~\Im(\us)<0
\label{eq:begin-displ-frac-1}
    \end{empheq}
  \end{subequations}
  \begin{equation}
    \label{eq:princpertinte}
    \frac{F(\us)+G(\us)}2=\frac 1
        {2\bi\pi}\pint_{\vs\in\bR}\frac{F(\vs)-G(\vs)}{\vs-\us}\mathrm{d}\vs,\qquad\qquad
        \If\quad \us\in\bR\,,
  \end{equation}
}
where the symbol \(\pint\) in \eqref{eq:princpertinte} denotes a
principal part integration.
\end{statmt}

\begin{proof}
  Let us prove (for instance) the equality
  \eqref{eq:begin-displ-frac}. First, we notice that the condition
  that \(F(z)\) and \(G(z)\) go to zero at infinity, at 
  least as a power law, ensures that the integral is convergent and is
  equal to \(\lim_{{\RR}\to\infty} 
\frac 1
{2\bi\pi}\int_{-{\RR}}^{\RR}\frac{F(\vs)-G(\vs)}{\vs-\us}\mathrm{d}\vs\). 

 Next
  we can compute the integral
  \(I_{\RR}=\int_{-{\RR}}^{\RR}\frac{F(\vs)}{\vs-\us}\mathrm{d}\vs +
  \int_{\mathcal{C}_{\RR}} \frac{F(\vs)}{\vs-\us}\mathrm{d}\vs\),
  where \(\mathcal{C}_{\RR}\) is an half circle of radius \({\RR}\) which closes
  the contour in the direction \(\Im(\sv)>0\), (in other words %
  \({\mathcal{C}_{\RR}=\left\{{\RR} e^{\bi \theta}\middle|
  \theta\in[0,\pi]\right\}}\)).
In the case \(\Im({\us})>0\), the integrand has one single singularity
at \(\vs=\us\). Hence as soon as \({\RR}>|\us|\), the integral \(I_{\RR}\)
is equal to \(I_{\RR}={2\bi\pi} F(\us)\).

Then we compute the integral   \(I'_{\RR}=\int_{-{\RR}}^{\RR}\frac{G(\vs)}{\vs-\us}\mathrm{d}\vs +
  \int_{\mathcal{C}'_{\RR}} \frac{G(\vs)}{\vs-\us}\mathrm{d}\vs\),   where
  \(\mathcal{C}'_{\RR}\) is the half circle \(\left\{{\RR} e^{-\bi \theta}\middle|
  \theta\in[0,\pi]\right\}%
\) %
of radius \({\RR}\) which closes 
  the contour in the direction \(\Im(\sv)<0\). %
In the case \(\Im({\us})>0\), the integrand does not have any singularity,
and  \(I'_{\RR}=0\).

Finally, one easily checks that \(\lim_{{\RR}\to\infty}
\int_{\mathcal{C}_{\RR}} \frac{F(\vs)}{\vs-\us}\mathrm{d}\vs = 
\lim_{{\RR}\to\infty}
\int_{\mathcal{C}'_{\RR}} \frac{G(\vs)}{\vs-\us}\mathrm{d}\vs =0\). Hence
\(\frac 1
        {2\bi\pi}\int_{\vs\in\bR}\frac{F(\vs)-G(\vs)}{\vs-\us}\mathrm{d}\vs\)
        is the limit of \(\frac 1 {2\bi\pi} \left(I_{\RR}-I'_{\RR}\right)\) when
        \({\RR}\to\infty\), and we obtain the result \eqref{eq:begin-displ-frac}.

The proof of \eqref{eq:begin-displ-frac-1} is absolutely identical,
and one just has to write the correct sign in the residue theorem.

Finally, we can use \eqref{eq:begin-displ-frac-0} to
write  
\(%
\frac{F(\us+\bi \epsilon )+G(\us-\bi \epsilon)}2 =
\frac 1 {2\bi\pi}\int_{\vs\in\bR} \frac{(\vs-\us)^2}{(\vs-\us)^2+\epsilon^2}  \frac{F(\vs)-G(\vs)}{\vs-\us}\mathrm{d}\vs
\)  when \(\us\in\bR\). Then, if we notice that the ratio
\(\frac{(\vs-\us)^2}{(\vs-\us)^2+\epsilon^2}\) is an even function of
\(\us-\vs\), which tends to one if \(|\us-\vs|\gg \epsilon\) and to zero
if \(|\us-\vs|\ll \epsilon\). Therefore the limit of \(\frac 1 {2\bi\pi}\int_{\vs\in\bR} \frac{(\vs-\us)^2}{(\vs-\us)^2+\epsilon^2}  \frac{F(\vs)-G(\vs)}{\vs-\us}\mathrm{d}\vs
\) when \(\epsilon\to0\) is 
exactly the principal value integral which gives \eqref{eq:princpertinte}.
\end{proof}

\paragraph{Cauchy kernel}
The expressions in the {\lhs} of \eqref{eq:begin-displ-frac-0}
can also be written as \(\CK\st\left(F-G\right)\), where \(\CK\) denotes
the Cauchy kernel
\begin{align}
  \CK(\us)\equiv \frac 1
        {2\bi\pi} \frac {-1}{\us}\,.
\end{align}
Therefore, expressions like \(\CK\st\left(F-G\right)\) (which this
theorem gives) are called ``Cauchy representations'' of complex
functions.

The next subsection will show that this theorem allows to find
convenient parameterizations of the {\qfs}. We will also see in
chapter \ref{cha:dualite-adscft} that the same theorem allows to find
non-trivial 
equations giving rise to a FiNLIE for the AdS/CFT {\Ysys}.

\subsubsection{Parameterization of the {\qfs}}
\label{sec:parameterization-qfs-1}

\paragraph{Reality condition}
\label{sec:reality-condition}

In the asymptotic limit (\(\Lf\to\infty\)), the {\qfs} were related to 
the real polynomials \(\fQ[][\mlvl]\) by the change of variables
\eqref{eq:TRTLN}.
Let us therefore introduce the following shifted {\qfs}:
\begin{align}
\label{eq:wqfiequiv-gqfsi-nn2}
  \wqf[I]{}\equiv& \gqfs[I][-{\Np}/2] &   \wpf[I]{}\equiv&  \gpfs[I][+{\Np}/2]\,.
\end{align}
In the asymptotic limit, we have \(\gpf[I]{}=\gqfs[I][-{\Np}]\), which
ensures that \(\wqf[I]{}=\wpf[I]{}\). The relation  \eqref{eq:TRTLN}
also ensures that in the asymptotic limit
  \(\wqf[
\{
  1,2,\cdots,\Np-\mlvl
\}
  ]{%
  }(\us) =
\fQ[][\mlvl][{ }]
\), where we have dropped the labels \((\RL)\) for simplicity.

Outside the asymptotic limit, {\idest} when the size \(\LF\) is finite, we
{\cannnot} a priori assume that these functions will still be real.
 But in
order to obtain a real energy from equation \eqref{eq:EzfromY}, we can
expect that the equation \(\overline{
\Y%
}=\Y[{\Np-a}]%
\) (which holds in the asymptotic limit), will 
hold even at finite size. 
In this relation \(\overline {\Y}\) denotes the complex-conjugate of the
function \(\Y\), defined by
\begin{gather}
\label{eq:DefBarCompConj}
  \overline F (\us)\equiv \overline{F(\overline\us)}\,.
\end{gather}
\index{0bar@\ensuremath{\overline F \equiv u\mapsto
    \overline{F(\overline\us)} }}
 From this hypothesis, we can %
 deduce that  %
 one can choose a gauge (for the {\Tfs})
such that
\begin{align}
\label{eq:overl-tus=tnp-abar}  \overline{
\T(\us)}=(-1)^{\frac{\Np(\Np-1)}{2}}\T[{\Np-a}](\bar \us)\,.
\end{align}
\begin{proof}
 Let us assume that \(\overline{\Y(\us)}=\Y[{\Np-a}](\bar \us)\), and let us 
  denote
by \(\T\)
 a solution of the Hirota equation %
  such that
  \(\Y=\frac{\T[][s+1]}{\T[a+1]}\frac{\T[][s-1]}{\T[a-1]}\). %
  Let \(\tilT(\us)\equiv 
\overline{\T[{\Np-a}](\bar \us)}
\), which is {\another} solution of the Hirota equation. Then we notice
that
\(\frac{\tilT[][s+1]}{\tilT[a+1]}\frac{\tilT[][s-1]}{\tilT[a-1]} = 
\overline{\Y[{\Np-a}](\bar
  \us)}=\Y=\frac{\T[][s+1]}{\T[a+1]}\frac{\T[][s-1]}{\T[a-1]}\), and we
deduce (from \stapref{sta:gaugetransfo}) that there exist
four gauge functions \(\gan[][1]\),  \(\gan[][2]\), \(\gan[][3]\) and
    \(\gan[][4]\) such that \(\tilT = \ga[a+s][1] \ga[a-s][2]
    \ga[-a+s][3] \ga[-a-s][4] \T\).
This shows that the transformation 
\begin{gather}
  \T\rightsquigarrow \bi^{\frac{\Np(\Np-1)}{2}}\sqrt {\ga[a+s][1]}\sqrt { \ga[a-s][2]}\sqrt {
    \ga[-a+s][3]}\sqrt { \ga[-a-s][4]} \T = \sqrt{\T\tilT}\,,
\end{gather}
is a gauge transformation into a gauge where the condition
\eqref{eq:overl-tus=tnp-abar} is satisfied. One should also note that
the above construction does not spoil the analyticity strip of the {\Tfs}.
\end{proof}

Finally, (potentially at the price of a transformation of the form
(\ref{eq:gpbs=f-cnp-2h_ij}-\ref{eq:gpbe=f-cnpgpe-gqbe=f})), we can
ensure that
  \begin{align}
\label{eq:forall-usinbc-forall}
\overline{\wqf[\emptyset]{}(\us)}&= \wpf[\emptyset]{}(\overline \us)\,,
&&\And&
\forall \us&\in\bC, &\forall \iq&\in\ninter{1}\Np& \overline{\wqs{}(\us)}&= \wps{}(\overline \us)\,.
\end{align}
In the article \cite{2010arXiv1007.1770K}, we have therefore denoted
these functions as \(q\) and \(\overline q\).

The relation \eqref{eq:forall-usinbc-forall} is proven by the same
argument as \eqref{eq:overl-tus=tnp-abar}: we start from arbitrary
{\qfs} producing the {\Tfs} which obey
\eqref{eq:overl-tus=tnp-abar}.
Then we notice that the {\qfs}
\begin{align}
\wqbf[\emptyset]{}=&\overline{\wpf[\emptyset]{}},&
\wpbf[\emptyset]{}=&\overline{\wqf[\emptyset]{}},&
  \wqbs{}=&\overline \wps{},%
&\wpbs{}=&\overline \wqs{},
\end{align}
reproduce the same {\Tfs}.
Indeed, the {\Wronskian} determinant expression gives
\begin{gather}
  \wqbf[(\nn)]{}=(-1)^{\frac{\nn(\nn-1)}2}\overline{\wpf[(\nn)]{}}\,,
\qquad\qquad\qquad\And
\wpbf[(\nn)]{}=(-1)^{\frac{\nn(\nn-1)}2}\overline{\wqf[(\nn)]{}}\,,\\%
\hence  \wqbf[(a)]{[+s]}\wedge \wpbf[(\Np-a)]{[-s]} =
(-1)^{\frac{\Np(\Np-1)}{2}} ~~\overline{\,\wqf[(\Np-a)]{[+s]}\wedge \wpf[(a)]{[-s]}\,}\,.
\end{gather}
Therefore, the \stapref{sta:freedom-choice-qfs-1}
shows that they are related by a transformation of the form
(\ref{eq:gpbs=f-cnp-2h_ij}-\ref{eq:gpbe=f-cnpgpe-gqbe=f}).
Up to the factors\footnote{We can for instance get rid of these factors by
  restricting to the case 
where \(\wqf[\emptyset]{}=\wpf[\emptyset]{}=1\).
} \(F\) and \(C\), this means that
\(\overline{\wqs[\iq]}={\HH_{{\coordi}}}^{\coordj} \wps[\jq]\), and that
\(\overline{\wps[\iq]}={\HH_{{\coordi}}}^{\coordj} \wqs[\jq]\). The
consistency of these two relations imposes \(\overline{\HH}\cdot \HH
=1\), which allows to decompose \(\HH\) as \(\HH=\left.\overline{A}\right.^{\:-1} A\),%
and to redefine \(\wqs[\jq]\rightsquigarrow
{A_{{\coordi}}}^{\coordj}\wqs[\jq]\) and \(\wps[\jq]\rightsquigarrow
{A_{{\coordi}}}^{\coordj}\wps[\jq]\), so as to ensure 
the relation \eqref{eq:forall-usinbc-forall}.

In addition, it is easy to see that we can simultaneously constrain
the gauge to obtain %
\begin{gather}
  \wqf[\emptyset]{}=1
\end{gather}

\paragraph{Parameterization of the {\qfs}}
\label{sec:parameterization-qfs-2}

Let us now write down how to parameterize the {\qfs}. As stated above,
we can choose a gauge where 
\begin{gather}
  \wqf[\emptyset]{}=1\,.
\end{gather}
Roughly speaking, the reality condition 
\eqref{eq:forall-usinbc-forall}
 fixes two out of four
degrees of gauge freedom, whereas the above condition
\(\wqf[\emptyset]{}=1\) fixes one more degree of gauge freedom.

As discussed in section \ref{sec:ysys-equation}, the limit
\(\us\to\infty\) should be essentially the same when \(\LF\) is finite as
in the limit \(\LF\to\infty\). This allows to deduce that there exist
polynomials \(\Pf[\iq]\) (where \(\iq\in\ninter 1 \Np\)) such that
\(\wqs(\us)-\Pf(\us)\xrightarrow[]{\us\to\infty}0\) and
\(\wps(\us)-\Pf(\us)\xrightarrow[]{\us\to\infty}0\). These polynomials
have the same degree as the polynomial
\(\wqs[][\LF=\infty](\us)\) which describes the asymptotic limit.
Then, we can use the 
\stapref{sta:Cauchy} %
to write 
\Pv{
  \begin{subequations}
\label{eq:paramwpq}
    \begin{empheq}[left={ 
\Pf(\us) + 
\displaystyle \frac 1
        {2\bi\pi}\int_{\vs\in\bR}\frac{\rf[](\vs)}{\vs-\us}\mathrm{d}\vs
        =\empheqlbrace},box=\fbox]{align}
      &\wqs(\us) &\If&\Im(\us)>0
\\
      &\wps(\us) &\If&\Im(\us)<0
    \end{empheq}
    \begin{gather}
      \where \rf[]\equiv \wqs(\us) - \wps(\us)\,.
    \end{gather}
  \end{subequations}
}
To obtain this equation, we used the \staref{sta:Cauchy} for the
functions \(F=\wqs-\Pf\) and \(G=\wps-\Pf\), and this argument relies on the
analyticity properties
(\ref{eq:textrmf-each-iqinn}-\ref{eq:textrmf-each-iqinn-1}).

The expression \eqref{eq:paramwpq} allows to parameterize all the
{\Yfu}- {\Tfu}- and {\qfs} in terms of the functions \(\rf\) and the
polynomials \(\Pf\).
We can also notice that the ``jump density''\footnote{The denomination
``jump density'' will be used to emphasize the fact that in
\eqref{eq:paramwpq}, we define a function which is analytic on the
whole complex plane except on the real axis, and which is obtained by
``gluing'' the function \(\wqs(\us)\) (for \(\Im(\us)>0\)) with the
function \(\wps(\us)\) (for \(\Im(\us)<0\)). This ``gluing'' gives rise to
a function which ``jumps'' by the amount \(\rf\) on the real axis.

One should not try to interpret physically the word density in ``jump
density'', because the function \(\rf[]\equiv \wqs - \wps\) is
imaginary, and even after division by \(\bi\) it does not necessarily
have a constant sign.
} \(\rf\) is exponentially
small in the \(\LF\to\infty\) limit, which means that it describes the finite
size corrections. Moreover \(\rf(\us)\equiv \wqs(\us) - \wps(\us)\) is also exponentially small in
the limit \(\us\to\infty\), which will be convenient for {\NuMr}
computation (it allows to approximate it to a good accuracy by a
function with finite support).

A last comment about the function \(\rf(\us)\equiv \wqs(\us) -
\wps(\us)\) is that it is 
holomorphic when \(|\Im(\us)|<1/2\) (see
(\ref{eq:textrmfor-each-ii}-\ref{eq:textrmfor-each-ii-1})). This means
that if we know this 
function slightly shifted from the real axis, we can write for
instance (if \(|\alpha|<1\))
\Pv{    \begin{empheq}[left={
\Pf(\us+\alpha \frac \bi 2) + 
\displaystyle \frac 1
        {2\bi\pi}\int_{\vs\in\bR}\frac{\rf[](\vs
+\alpha \frac \bi 2)}{\vs
-\us}\mathrm{d}\vs
        =\empheqlbrace}]{align}
      &\wqs(\us+\alpha \frac \bi 2) &\If&\Im(\us)>0
\label{eq:paramwq}
\\
      &\wps(\us+\alpha \frac \bi 2) &\If&\Im(\us)<0
\label{eq:paramwp}
    \end{empheq}}
This is obtained by analytic continuation from \eqref{eq:paramwpq},
and it allows to express the functions \(\wqs\) on their
whole analyticity domain.

\paragraph{Case of the {\U 1} sector}
\label{sec:case-Uo-1}

Let us now study a specific class of states ({\idest} a specific class of
solutions of the {\Ysys}) called the ``{\U 1} sector''. It is the set
\index{sector!U(1) sector}
of states such that in the asymptotic limit all the polynomials
\(\fQ[][\mlvl][{ }]\) are constant polynomials ({\idest} they are equal to
one up to a normalization), except the polynomial
\(\fQ[][0][{ }]\). This means that these states have no spin-wave
excitations, and one can show that for instance, the vacuum and the
first excited state (defining the mass gap) belong to this sector.
 For the states in this sector, the functions
\(\T[][][(\Rg)]\) and \(\T[][][(\Lf)]\) are equal. This property is
clear in the \(\LF\to\infty\) limit (because \(\fQ=\fQ[][-\mlvl]\)), and
we will construct solutions of the {\Ysys} which obey this property at
finite size, and converge to the asymptotic solution when
\(\LF\to\infty\). 

One can then %
express explicitly the polynomials {\Pf} (which converge to 
\(\wqs[][\LF=\infty](\us)\) when \(\LF\to\infty\))
 because in the
asymptotic limit we have
\begin{align}
\forall  & \mlvl\geq 1,&
\Det{\wqss[\iq][{-1-\Np+\mlvl+2\jj}]}{1\leq
          \iq,\jj\leq \Np-\mlvl}
&
=
\wqf[
\{
  1,2,\cdots,\Np-\mlvl
\}
  ]{%
  } =
\fQ[][\mlvl][{ }]=1\,.
\end{align}
At \(\mlvl=\Np-1\), this equation gives \(\wqs[1][\LF=\infty]=1\), then at
\(\mlvl=\Np-2\) it gives \(\wqs[2][+]-\wqs[2][-]=1\), which can for
instance be solved by \(\wqs[2][\LF=\infty](\us)=-\bi \us\).
Another solution would be \(\wqs[2][\LF=\infty](\us)=-\bi \us+c\), for
an arbitrary constant \(c\), but we will disregard this term because it can be absorbed by transformations
of the form
(\ref{eq:gpbs=f-cnp-2h_ij}-\ref{eq:gpbe=f-cnpgpe-gqbe=f}). For the same
reason, we will disregard the factor 
\(-\bi\) in
\(\wqs[2][\LF=\infty]\). %

By proceeding iteratively, we obtain
\begin{gather}
\label{eq:fdispforall-iq-leq}
\fdisp{\forall \iq \leq \Np-1,\qquad \Pf(\us)=
\frac{\us^{\iq-1}} {(\iq-1)!} }\,,%
\end{gather}
whereas \(\Pf(\us)\) is a polynomial of degree \(\Np-1+\Mp\), where
\(\Mp\equiv\dg[0]\) 
is the degree of \(\varphi\equiv\fQ[][0]\). Out of the \(\Np+\Mp\) coefficients of this
polynomial, only \(\Mp\) coefficients are relevant because \(\Np-1\) of
them can be removed by transformations of the form
(\ref{eq:gpbs=f-cnp-2h_ij}-\ref{eq:gpbe=f-cnpgpe-gqbe=f}), and one of them
is an overall multiplicative factor (which can be absorbed into a
gauge transformation).
Moreover, we have (still in the asymptotic limit)  \(\vf=\wqf[\ove]{}=\Det
{\wqss[\iq][{-1-\Np+2\jj}]}{1\leq
          \iq,\jj\leq \Np}= \Det
{\Pf[][{[-1-\Np+2\jj]}]}{1\leq
          \iq,\jj\leq \Np}
\) (if \(\Im(\us)\geq \Np/4\)). Using the expression
\eqref{eq:fdispforall-iq-leq} of \(\Pf[1]\), \(\Pf[2]\), \(\cdots\),
\(\Pf[\Np-1]\), we obtain
\begin{gather}
\label{eq:vf=b-1np-22left}
  \vf=\bi^{\frac{(\Np-1)(\Np-2)}2}\left( e^{\frac \bi
  2 \partial_\us}-e^{-\frac \bi
  2 \partial_\us}\right)^{{\Np}-1} \Pf [\Np]\,,
\end{gather}
in the limit \(\LF\to\infty\).

The above arguments show the expression \eqref{eq:fdispforall-iq-leq}
in the asymptotic limit \(\LF\to\infty\). At finite size, this limit
fixes only the large \(\us\) behavior, {\idest} the leading coefficient of
\(\Pf\). But then we can argue, exactly like in the \(\LF\to\infty\), that
the other coefficients can be dropped using a
transformation of the form
(\ref{eq:gpbs=f-cnp-2h_ij}-\ref{eq:gpbe=f-cnpgpe-gqbe=f}).

Moreover, we still have a free degree of gauge freedom which takes the
form
\begin{gather}
  \wqf[I]{}\rightsquigarrow h^{\FS[\nn]}\wqf[I]{},\qquad\qquad\where
  \nn=|I|\,,\\
  \wpf[I]{}\rightsquigarrow {\overline
    h}^{\FS[\nn]}\wpf[I]{},\qquad\qquad\where \overline {h}(\us)\equiv
  \overline{h(\overline \us)}\,,
\end{gather}
where \(h(\us)\) is an holomorphic function when \(\Im(\us)>-1/2\), which
goes to one when \(\LF\to\infty\) or \(\us\to\infty\).
This degree of freedom is sufficient to impose for instance
\begin{equation}
  \fdisp{\wqs[1]{}=1}\,\qquad\qquad \ie~ \rf[1]=0\,.
\end{equation}

In what follows we will restrict to the study of these ``{\U 1}
sector'' states, as we did in \cite{2010arXiv1007.1770K}.
We see that for these states, the {\qfs} (and hence the {\Tfu}- and
{\Yfs}) are parameterized by a set of \(\Np-1\) functions \(\rf[2]\),
\(\rf[3]\), \(\cdots\), \(\rf[\Np]\), and by \(\Mp\) relevant
coefficients of the 
polynomial \(\Pf[\Np]\).

\subsection{Set of equations}
\label{sec:set-equations}

Now that we have parameterized all the relevant functions, we will
write equations which allow to fix them uniquely. For simplicity, we
will explain this procedure in the \(\U 1\) sector, and we will
construct the solution of the {\Ysys} obeying the conditions written in
the previous sections (such as the reality condition
\(\overline{
\Y(\us)}=\Y[{\Np-a}](\bar \us)\), the  symmetry condition
\(\Y[][-s]=\Y\), the analyticity strips), and
converging (at \(\LF\to\infty\)) to the asymptotic solution written in
section \ref{sec:asymptotic-limit}. Hence we will identify these
solutions to the finite-size description of these \(\U 1\) sector states.

To write these equations, we will
use mainly the ``middle nodes equation'' of section
\ref{sec:middle-node-equation}, which describes the behavior of the
{\Ysys} at \(s=0\). In particular, we will rewrite this equation in a form where
no {\zerm} needs to be added.

This ``middle nodes equation'' is necessarily a key ingredient to
solve the {\Ysys}, because it is the place where the size \(\LF\)
appears in an equation. We will rewrite this equation into an equation
on the densities \(\rf\), in such a way that this equation can be solved
iteratively.

\subsubsection{Equation on the densities $\rf$}
\label{sec:middle-nodes-equat}

Let us now insert the parameterization \eqref{eq:paramwpq} of the
{\qfs} into the middle 
nodes equation \eqref{eq:MidNodEq}. First, we can notice that our
parameterization ensures that
\begin{gather}
\label{eq:tnp=wqfs-qquadqq-t0=}
  \T[\Np]=\wqfs[\ove][+s+\Np/2],\qquad\qquad \T[0]=\wpfs[\ove][-s-\Np/2],\\
\begin{aligned}
\hence 
\quad  \left(\frac{\T[\Np][0][+]\T[\Np][0][-]}{\T[\Np][1]\T[\Np][1][(\Lf)]}\right)^{
    \FS[a]}
    \left(\frac{\T[0][0][+]\T[0][0][-]}{\T[0][1]\T[0][1][(\Lf)]}\right)^{
    \FS[{\Np}-a]} =& \left(\frac{ \wqfs[\ove][{\Np}/2-1]}{ \wqfs[\ove][{\Np}/2+1]}\right)^{
    \FS[a]}\left( \frac{ \wpfs[\ove][-{\Np}/2+1]}{\wpfs[\ove][-{\Np}/2-1]}\right)^{
    \FS[{\Np}-a]}\\
  =&\frac{ \wqfs[\ove][{\Np}/2-a]}{ \wqfs[\ove][{\Np}/2+a]}
\frac{ \wpfs[\ove][\Np/2-a]}{ \wpfs[\ove][-3\Np/2+a]}
\end{aligned}
\end{gather}

Therefore the middle nodes equation \eqref{eq:MidNodEq} simplifies to 
 \begin{gather}
\Y[][0]\bumpeq
    e^{- E_a}
      \frac{{\left.{\T[a][1]}\right.}^2}{\T[a-1][0] \T[a+1][0]}%
\left(\frac{ \wqfs[\ove][{\Np}/2-a]}{ \wqfs[\ove][{\Np}/2+a]}
\frac{ \wpfs[\ove][\Np/2-a]}{ \wpfs[\ove][-3\Np/2+a]}\right)^{\st
\Ker_\Np}\,,\\
\where f_1\bumpeq f_2 \qquad \textrm{ denotes } \qquad (f_1)^{\FS[{\Np}]}=(f_2)^{\FS[{\Np}]}\,,
  \end{gather}
where we have also used the fact that (for the {\U 1} sector states),
the functions 
\(\T[][][(\Rg)]\) and \(\T[][][(\Lf)]\) are equal.
Using the definition of
\(\Y[][0]=\frac{\T[a][1]\T[a][-1]}{\T[a-1][0] \T[a+1][0]}\), we deduce
\begin{gather}
\T[][-1]\bumpeq
    e^{- E_a}
      \T[a][1]
\left(\frac{ \wqfs[\ove][{\Np}/2-a]}{ \wqfs[\ove][{\Np}/2+a]}
\frac{ \wpfs[\ove][\Np/2-a]}{ \wpfs[\ove][-3\Np/2+a]}\right)^{\st \Ker_\Np}\,.
  \end{gather}
On the other hand, this {\Tf} is also equal to the determinant
\begin{gather}
  \T[][-1]=
  \left|
    \begin{array}{ccccc}
      \wqss[1][-a+{\Np}/2]&\wqss[2][-a+{\Np}/2]&\wqss[3][-a+{\Np}/2] & \cdots &
      \wqss[\Np][-a+{\Np}/2]\\ 
      \wqss[1][-a+2+{\Np}/2]&\wqss[2][-a+2+{\Np}/2]&\wqss[3][-a+2+{\Np}/2] &
      \cdots & \wqss[\Np][-a+2+{\Np}/2]\\ 
      \vdots&&&&\vdots\\
      \wqss[1][+a-2+{\Np}/2]&\wqss[2][+a-2+{\Np}/2]&\wqss[3][+a-2+{\Np}/2] &
      \cdots & \wqss[\Np][+a-2+{\Np}/2]\\ 
      \wpss[1][+a+2-3{\Np}/2]&\wpss[2][+a+2-3{\Np}/2]&\wpss[3][+a+2-3{\Np}/2] & \cdots &
      \wpss[\Np][+a+2-3{\Np}/2]\\ 
      \wpss[1][+a+4-3{\Np}/2]&\wpss[2][+a+4-3{\Np}/2]&\wpss[3][+a+4-3{\Np}/2] & \cdots &
      \wpss[\Np][+a+4-3{\Np}/2]\\ 
      \vdots&&&&\vdots\\
      \wpss[1][-a+{\Np}/2]&\wpss[2][-a+{\Np}/2]&\wpss[3][-a+{\Np}/2] & \cdots &
      \wpss[\Np][-a+{\Np}/2]\\ 
    \end{array}
  \right|\,.
\end{gather}
In this determinant (which is just an explicit rewriting of the
expression \eqref{eq:formHirotaPCM} after the change of variables
\eqref{eq:wqfiequiv-gqfsi-nn2}), we can notice that the first line and
the last line are the functions \(\wqs\) and \(\wps\) with the same shift,
and they differ only by the jump density \(\rf\). Therefore we can
subtract or add these two lines, so as to replace the coefficients of
the first line by \(\wqss[][-a+{\Np}/2]-\wpss[][-a+{\Np}/2]\), and the coefficients of the last
line by \(\frac{\wqss[][-a+{\Np}/2]+\wpss[][-a+{\Np}/2]}2\). If we write this
argument for \(\T[][-1][{[a-{\Np}/2]}]\) instead of \(\T[][-1]\) (in order to
obtain \(\wqs[]-\wps[]\) instead of \(\wqss[][-a+{\Np}/2]-\wpss[][-a+{\Np}/2]\)),
we get
  \begin{gather}
\label{eq:t-1-leftus+bi}
  \T[][-1][{[%
        a-{\Np}/2%
]}]
    =\sum_{\jq=2}^{\Np}
  d_{a,\jq} \rf[\jq]\\[.5cm]
\where d_{a,\jq}\equiv (-1)^\jq 
  \left|
     \begin{array}{cccccc}
      \wqss[1][+2]&\cdots&\wqss[\jq-1][+2]&\wqss[\jq+1][+2] &
      \cdots & \wqss[\Np][+2]\\ 
      \wqss[1][+4]&\cdots&\wqss[\jq-1][+4]&\wqss[\jq+1][+4] &
      \cdots & \wqss[\Np][+4]\\ 
      \vdots&&&&&\vdots\\
      \wqss[1][+2a-2]&\cdots&\wqss[\jq-1][+2a-2]&\wqss[\jq+1][+2a-2] &
      \cdots & \wqss[\Np][+2a-2]\\ 
      \wpss[1][+2a-2{\Np}+2]&\cdots&\wpss[\jq-1][+2a-2{\Np}+2]&\wpss[\jq+1][+2a-2{\Np}+2] & \cdots &
      \wpss[\Np][+2a-2{\Np}+2]\\ 
      \wpss[1][+2a-2{\Np}+4]&\cdots&\wpss[\jq-1][+2a-2{\Np}+4]&\wpss[\jq+1][+2a-2{\Np}+4] & \cdots &
      \wpss[\Np][+2a-2{\Np}+4]\\ 
      \vdots&&&&&\vdots\\
      \wpss[1][-2]&\cdots&\wpss[\jq-1][-2]&\wpss[\jq+1][-2] & \cdots &
      \wpss[\Np][-2]\\ 
      \Pf[1]&\cdots&\Pf[\jq-1]&\Pf[\jq+1]&\cdots&\Pf[{\Np}]
     \end{array}
\label{eq:fkq}
  \right|\,,
\end{gather}
where we also used the fact that \(\Pf=
\frac{\wqs+\wps}2\).

Therefore, if we denote by
\(\left(c_{\iq,a}\right)_{\substack{2\leq\iq\leq\Np\\
1\leq a\leq \Np-1}}\)
the adjugate matrix\footnote{The ``adjugate matrix'' is the transpose
  of the matrix of the cofactors. In other words, it is defined by
\(c_{\iq,a}
=(-1)^{\iq+a} \Det{d_{b,\jq}}{\substack{
b\in\ninter 1 {\Np-1}\setminus \{a\}
\\
\jq\in\ninter 2 \Np \setminus \{\iq\}
}}\). %
With this definition of the adjugate matrix, the
inverse of a matrix \(\MM\) is its adjugate matrix divided by its
determinant \(\det \MM\).
} of \(\left(d_{a,\jq}\right)_{\substack{1\leq a
\leq\Np-1\\2\leq\jq\leq  {\Np}}}\), we
obtain
\begin{gather}
\label{eq:fdisprf=s-c_iq-a}
  \fdisp{\rf=\frac{\displaystyle {\sum_{a=1}^{\Np-1}
  c_{\iq,a} \T[][-1][{[a-{\Np}/2]}]}}{\Det{d_{a,\jq}}{\substack{1\leq a
\leq\Np-1\\2\leq\jq\leq  {\Np}}}}}\,.
\end{gather}
In these equations, the coefficients \(c_{\iq,a}\) can be written in
terms of the jump densities \(\rf\) (and of the  \(\Mp\) coefficients of the
polynomial \(\Pf[\Np]\)), because when \(\us\) is real, all the functions
\(\wqs\) which enter the determinant \(d_{a,\jq}\) have a positive shift, %
while the functions \(\wps\) have a negative shift ({\idest} they appear under
the form \(\wpss[][-s]\) where \(s>0\)), hence they are expressed via \eqref{eq:paramwpq}.

This expression \eqref{eq:fdisprf=s-c_iq-a} is interesting because at
large \(\Lf\), the coefficients \(d_{b,\jq}\) have a well-defined limit,
and the determinant in the denominator as well. On the other hand
\(\T[][-1]\) is very small, and one easily obtains for instance the
leading order expression of \(\rf\) out of the leading order expression
of \(\T[][-1]\). At finite size the equation is less simple because the
coefficients \(d_{b,\jq}\) are functions of the \(\rf\), but we will
see that
this equation \eqref{eq:fdisprf=s-c_iq-a} is suitable for an iterative
{\NuMr} resolution.

\paragraph{Middle nodes equation and \(\CDD\) factor}
\label{sec:middle-node-equation-1}

In order to efficiently use this equation 
\eqref{eq:fdisprf=s-c_iq-a}
to write an equation on the
jump densities \(\rf\), we should express (when \(\us\in\bR\)) the quantity
\begin{gather}
\label{eq:t-1a-nn2bumpeq}
\T[][-1][{[a-{\Np}/2]}]\bumpeq
    e^{- E_a^{[+a-{\Np}/2]}}
      \T[a][1][{[a-{\Np}/2]}]
\left(\frac{ \wqfs[\ove][{\Np}/2-a]}{ \wqfs[\ove][{\Np}/2+a]}
\frac{ \wpfs[\ove][\Np/2-a]}{ \wpfs[\ove][-3\Np/2+a]}\right)^{\st \Ker_\Np^{[+a-{\Np}/2]}}\,,
  \end{gather}
in terms of the densities \(\rf\). The factor \( \T[a][1][{[a-{\Np}/2]}]\)
does not pose any problem, because \( \T[a][1]\) is taken inside the
strip \(\Ast %
{2}\) where it is easily expressed from the
parameterization \eqref{eq:paramwpq} of the {\qfs}. 
For the other factors, let us remind that since \(\wqf[\ove]{}(\us)\)
(resp \(\wpf[\ove]{}(\us)\)) is expressed as a {\Wronskian} determinant of
the functions \(\wqs\) and \(\wps\), the parameterization of {\qfs} only
allows to compute \(\wqf[\ove]{}(\us)\)
(resp \(\wpf[\ove]{}(\us)\)) on the domain where
 \(\Im({\us})\geq \frac {{\Np}-1} 2\) (resp \(\Im({\us})\leq -\frac {{\Np}-1} 2\)).
Therefore, if \(a\) is smaller than \(\frac {\Np} 2 -1\), we {\cannnot} express
\(\wqfs[\ove][{\Np}/2+a]\) from this parameterization. But what we can do is
to redistribute\footnote{As we already said, redistributing the shifts
in a convolution amounts to moving an integration contour, and this
is allowed if the functions are analytic enough (otherwise the
contribution of some poles may occur). Here, this shift of contour is
allowed because we know from \eqref{eq:tnpt-analyt-when} and
\eqref{eq:tnp=wqfs-qquadqq-t0=} that
\(\wqf[\ove]{}(\us)\) is analytic as soon as \(\Im({\us})>0\).
} the shifts between \(\wqfs[\ove][{\Np}/2+a]\) and the kernel
\(\Ker_\Np\). 

For instance we can write
\begin{gather}
  \left(\frac{ 1}{ \wqfs[\ove][{\Np}/2+a]}
\frac{1}{ \wpfs[\ove][-3\Np/2+a]}\right)^{\st \Ker_\Np^{[+a-{\Np}/2]}}
=
  \left(\frac{ 1}{ \wqfs[\ove][{\Np}+a-1]}
\right)^{\st \Ker_\Np^{[+a-{\Np}+1]}}
  \left(\frac{1}{\wpfs[\ove][+a-2{\Np}-1]}
\right)^{\st \Ker_\Np^{[+a-1]}}\,.
\end{gather}

For the numerator, one could try to proceed the same way, but it would
fail because we would have to shift the kernel \(\Ker_\Np\) by more that
\(\pm (\Np-1)\). Instead of that, we can use the relation
\begin{gather}
 \left( \frac{\wpf[\ove]{+}}{\wpf[\ove]{-}}\right)^{\FS[\Np]} =
 \frac{\wpf[\ove]{[+\Np]}}{\wpf[\ove]{[-\Np]}}\,,\qquad\qquad \ie~~
\wpf[\ove]{\st \Ker_\Np} \bumpeq \left(\wpf[\ove]{[-2\Np]}\right)^{\st
  \Ker_\Np} \frac{\wpf[\ove]{[+1-\Np]}}{\wpf[\ove]{[-1-\Np]}}
\end{gather}
Using this relation, the numerator \(\left(%
    \wqfs[\ove][{\Np}/2-a]%
  \wpfs[\ove][\Np/2-a]%
\right)^{\st \Ker_\Np^{[+a-{\Np}/2]}}\) can be rewritten as 
\begin{multline}
  \left(    \wqfs[\ove][{\Np}/2-a]
  \wpfs[\ove][\Np/2-a]
\right)^{\st \Ker_\Np^{[+a-{\Np}/2]}} \\ \bumpeq
\frac{\wqfs[\ove][+{\Np}-1]}{\wqfs[\ove][+{\Np}+1]} 
\frac{\wpfs[\ove][-{\Np}+1]}{\wpfs[\ove][-{\Np}-1]} 
  \left(    \wpfs[\ove][-2\Np-a+1]
\right)^{\st \Ker_\Np^{[+a-1]}} 
  \left(    \wqfs[\ove][+2\Np+\Np-a+1]
\right)^{\st \Ker_\Np^{[-\Np+a+1]}} 
\end{multline}
Now the {\rhs} only involves {\qfs} in the domain where they
are easily expressed from the densities \(\rf\).

If we insert these expressions into \eqref{eq:t-1a-nn2bumpeq}, then we
obtain
\begin{multline}
\label{eq:t-1a-nn2bumpeq-1}
  \T[][-1][{[a-{\Np}/2]}]\bumpeq\\
    e^{- E_a^{[+a-{\Np}/2]}}
      \T[a][1][{[a-{\Np}/2]}]
\frac{\wqfs[\ove][+{\Np}-1]}{\wqfs[\ove][+{\Np}+1]} 
\frac{\wpfs[\ove][-{\Np}+1]}{\wpfs[\ove][-{\Np}-1]} 
  \left(   \frac{ \wpfs[\ove][-2\Np-a+1]}{\wpfs[\ove][+a-2{\Np}-1]}
\right)^{\st \Ker_\Np^{[+a-1]}} 
  \left( \frac   {\wqfs[\ove][+2\Np+\Np-a+1]}{\wqfs[\ove][{\Np}+a-1]}
\right)^{\st \Ker_\Np^{[-\Np+a+1]}} 
\,,
\end{multline}
for \(a\in\ninter 1 {\Np-1}\).

An important and interesting statement is then 
\Pv{\begin{empheq}[box=\fbox]{multline}
\label{eq:middle-node-equation-2}
  \T[][-1][{[a-{\Np}/2]}]=
    e^{- E_a^{[+a-{\Np}/2]}}
      \T[a][1][{[a-{\Np}/2]}]
\frac{\wqfs[\ove][+{\Np}-1]}{\wqfs[\ove][+{\Np}+1]} 
\frac{\wpfs[\ove][-{\Np}+1]}{\wpfs[\ove][-{\Np}-1]} \\\times 
  \left(   \frac{ \wpfs[\ove][-2\Np-a+1]}{\wpfs[\ove][+a-2{\Np}-1]}
\right)^{\st \Ker_\Np^{[+a-1]}} 
  \left( \frac   {\wqfs[\ove][+2\Np+\Np-a+1]}{\wqfs[\ove][{\Np}+a-1]}
\right)^{\st \Ker_\Np^{[-\Np+a+1]}} 
\,,
\end{empheq}}
which says that there is no {\zerm} in the equation
\eqref{eq:t-1a-nn2bumpeq-1}. To be more exact there can still be a
phase \(e^{2\bi\kk\pi/\Np}\), but the operation \(f\rightsquigarrow f^{\st
    \Ker_\Np}\equiv \mathrm{exp}\left(\Ker_\Np\st
    \mathrm{log}~f_a\right)\) itself is defined up to a phase
  \(e^{2\bi\kk\pi/\Np}\), corresponding to the choice of the branch of
  the logarithm.

Exactly like for the {\zerm} \(\CDD\) in the asymptotic expression
\eqref{eq:y0-=-e} of \(\Y[][0]\), there is no complete proof of this
statement, which should rather be viewed as a condition that we impose
on the solution of the {\Ysys}.
It can be easily motivated by the fact that it reproduces the correct
pole structure and analyticity strip for the {\Yfs}. In order to
really prove that the {\zerm} in \eqref{eq:middle-node-equation-2}
is exact, one should use a minimality principle for the number of
zeroes of the \(\Y[][0]\) and it is not absolutely clear, whether or not
this minimality principle follows from the present construction (and
in particular from the regularity of the {\Tfs}).

\paragraph{Relation to \(\CDD\)}
\label{sec:relation-cdd}

In the asymptotic limit (\(\LF\to\infty\)), we can use the expressions
(\ref{eq:TA=phiNaive1}-\ref{eq:TA=phiNaive3})
of \(\T\), to  say that inside
the analyticity strips, we have 
\begin{gather}
  \wqf[\ove]{}%
  =\T[\Np][][{[-\Np/2]}]=\varphi,\qquad\qquad\qquad   \wpf[\ove]{}%
  =\T[0][][{[+\Np/2]}]=\varphi.
\end{gather}
When we plug this into the equation \eqref{eq:middle-node-equation-2},
we can compute \(\Y[][0]=\frac{\T[][1]\T[][-1]}{\T[a+1][]\T[a-1][]}\)
by explicitly computing the convolutions which appear in
\eqref{eq:middle-node-equation-2}. Interestingly, that gives exactly
\eqref{eq:y0-=-e}, and we find that the factor \(\CDD\) is therefore
reproduced by the equation \eqref{eq:middle-node-equation-2}.
The factors \(\Sb\) and \(\CDb\) which appear in \eqref{eq:y0-=-e} are
well defined only in the asymptotic limit, and we see that they now
appear from the convolution of (the logarithm of)  polynomial 
{\Tfs} by
the kernel \(\Ker_\Np\).
At finite size, the {\Tfs} are not polynomial anymore, but the 
convolution is still well defined for
arbitrary size \(\LF\).

In section \ref{sec:bethe-equations-2} we used the expression of
\(\Y[][0]\) to find Bethe equations under the  hypothesis that 
\(\Y[1][0](\theta_\nrt+ \bi\frac \Np 4)=-1\), for any Bethe 
root \(\theta_\nrt\).
In the next section, we will see that at finite size, the
generalization of the Bethe equations is obtained by a quite different
method, but let us nevertheless elaborate on the hypothesis that in the
asymptotic limit,
\(\Y[1][0](\theta_\nrt+ \bi\frac \Np 4)=-1\).
 In terms of {\Tfs}, we know that 
\begin{gather}
  1+\Y[1][0] \left(\us+ \bi\frac \Np 4\right)=
\frac{\T[1][0](\us+ \bi\frac \Np 4+\frac \bi 2)
  \T[1][0](\us+ \bi\frac \Np 4- \frac \bi 2)}{\T[2][0](\us+ \bi\frac \Np 4)\T[0][0](\us+ \bi\frac \Np 4)}\,.
\end{gather}
The analyticity
strips are such that in the numerator \( \T[1][0](\us+ \bi\frac \Np 4-
\frac \bi 2)\xrightarrow{\LF\to\infty} \vf(\us)\) when
\(\us\in\bR\). That is because the argument \(\us+ \bi\frac \Np 4-
\frac \bi 2\) is inside the analyticity strip \(\Ast {1+\Np/2}\) where
the asymptotic expression \eqref{eq:TA=phiNaive3} is valid.
On the other hand in the denominator, the argument of  \(\T[0][0](\us+
\bi\frac \Np 4)\) stands at the very boundary of the analyticity strip (see
\eqref{eq:t0textrm-analyt-when}), and we {\cannnot} say that it converges
to \(\vf(\us)\).
This is why it makes sense to expect that, in the asymptotic limit, \(  1+\Y[1][0]
(\us+ \bi\frac \Np 4)\) has a zero at every root \(\theta_\nrt\) of the
polynomial \(\vf\).

\paragraph{Equations on the densities}
\label{sec:equations-densities}

The middle nodes equation, as rewritten in
\eqref{eq:middle-node-equation-2}, can be inserted into the equation
\eqref{eq:fdisprf=s-c_iq-a} for the densities.
That gives
\begin{multline}
\label{eq:fdisprf=fr-sum_-1}
    {\rf=\frac{1}{\Det{d_{a,\jq}}{\substack{1\leq a
\leq\Np-1\\2\leq\jq\leq  {\Np}}}}}
\sum_{a=1}^{\Np-1}
  c_{\iq,a} ~
    e^{- E_a^{[+a-{\Np}/2]}}
      \T[a][1][{[a-{\Np}/2]}]
\frac{\wqfs[\ove][+{\Np}-1]}{\wqfs[\ove][+{\Np}+1]} 
\frac{\wpfs[\ove][-{\Np}+1]}{\wpfs[\ove][-{\Np}-1]} \\\times 
  \left(   \frac{ \wpfs[\ove][-2\Np-a+1]}{\wpfs[\ove][+a-2{\Np}-1]}
\right)^{\st \Ker_\Np^{[+a-1]}} 
  \left( \frac   {\wqfs[\ove][+2\Np+\Np-a+1]}{\wqfs[\ove][{\Np}+a-1]}
\right)^{\st \Ker_\Np^{[-\Np+a+1]}} 
\,.
\end{multline}

In this equation, all the functions in the {\rhs} are
parameterized in terms of the densities \(\rf\) (and of the polynomial
\(\Pf[\Np]\)). This means that this equation is a closed
equation\footnote{More precisely, \eqref{eq:fdisprf=fr-sum_-1} is a
  set of \(\Np-1\) coupled equations on the \(\Np-1\) densities
  \(\rf\). Indeed, we should write the equation
  \eqref{eq:fdisprf=fr-sum_-1} for each \(\iq\in\ninter 2 {\Np}\).}
 on
the densities \(\rf\). This 
equation \eqref{eq:fdisprf=fr-sum_-1}, which rewrites the middle
  nodes equation \eqref{eq:MidNodEq} in terms of the densities \(\rf\),
is convenient because
the {\rhs} is made of \( e^{- E_a^{[+a-{\Np}/2]}}\) (which is
very small in the asymptotic limit), multiplied by several functions
having a smooth limit when \(\LF\to\infty\). As we will see, this
makes the equation suitable for an iterative resolution.

\subsubsection{Equation on the polynomial $\Pf [\Np]$}
\label{sec:equat-polyn-pf}

As explained in section \ref{sec:parameterization-qfs-2}, the {\qfs}
are defined by the set of \(\Np-1\) functions \(\rf[2]\),
\(\rf[3]\), \(\cdots\), \(\rf[\Np]\), and by the \(\Mp\) relevant
coefficients of the 
polynomial \(\Pf[\Np]\).

We have just obtained the equation \eqref{eq:fdisprf=fr-sum_-1}, which
is a closed equation on the densities \(\rf\). %
Actually, the same equation also fixes the %
polynomial \(\Pf[\Np]\), if we require that the {\qfs} (and hence the
densities \(\equiv \wqs(\us) - \wps(\us)\))
 remain
analytic.
Indeed, %
this equation 
involves a division
by the determinant \(\Det{c_{\iq,a}}{\substack{2\leq\iq\leq\Np\\
1\leq a\leq \Np-1}}\). This determinant is a function of \(\us\) and
we will see that for excited states, it has some zeroes\footnote{More precisely, we prove this
statement in a vicinity of \(\LF=\infty\), and we numerically checked it
at finite size for several states when the size \(\LF\) is smaller.}.
If this is the case, then the coefficients of the
polynomial \(\Pf[\Np]\) should be fitted in such a way that these
zeroes do not give rise to any pole in the functions \(\rf\) in the
{\lhs} of \eqref{eq:fdisprf=fr-sum_-1}.

Let us simply illustrate this on the case \(\Np=3\). We will see that
for arbitrary \(\LF\) it gives an equation on the polynomial \(\Pf
[\Np]\), and that when \(\LF\) tends to \(\infty\), this equation
reduces to the Bethe equation \eqref{eq:fdisp-1-=}.

\paragraph{Finite size Bethe equations in the \(\Np=3\) case}
\label{sec:finite-size-bethe}

Let us consider a state in the {\U 1} sector
of the \(\SU3\times \SU3\) {\PCM},
 which has real asymptotic Bethe
roots \(\theta_\nrt\).

For such a state, the linear system \eqref{eq:t-1-leftus+bi} can be written   
as   
\begin{gather}
  \left(   
    \begin{array}{cc}   
      A&B\\   
      \overline{A}&\overline{B}   
    \end{array}   
  \right)   
  \left(   
    \begin{array}{c}   
      \rf[2]\\   
      \rf[3]   
    \end{array}   
  \right)=   
  \left(   
    \begin{array}{c}   
      \T[1][-1][{[-1/2]}]\\
      \T[2][-1][{[+1/2]}]
    \end{array}   
  \right)\,,%
\qquad\qquad
\where \overline{A}(\bar \us)\equiv \overline {A(\us)}.
\end{gather}
where\footnote{The expressions   \(A=\wpss[3][-2]-\Pf[3]\) and
  \(B=\wpss[2][-2]-\Pf[2]\) are obtained directly form \eqref{eq:fkq},
  if we remember 
that \(\wps[1]{}=\wqs[1]{}=\Pf[1]=1\), and that the functions \(\wqs\) and
\(\wps\) are the complex conjugate of each other, in the sense of
equation 
\eqref{eq:forall-usinbc-forall}.%
}   
\(A=\wpss[3][-2]-\Pf[3]\) and \(B=\Pf[2]-\wpss[2][-2]\).

By inverting the matrix \(  \left(   
    \begin{array}{cc}   
      A&B\\   
      \overline{A}&\overline{B}   
    \end{array}   
  \right)\), some singularity could occur at the zeroes of its determinant
  \(A\overline{B}-\overline{A}B\), {\idest} when the determinant is   
  zero. If we want every \(\rf\) to be regular, we need  the   
  numerator 
 to vanish (in \eqref{eq:fdisprf=fr-sum_-1}) at the same positions as
 the zeroes of the denominator, to cancel this pole. This gives the
 following finite size    
  Bethe equation:   

  \begin{multline}
\label{eq:finitBE}   
\textrm{For every zero \(\tth_\jrt\) of \(A\overline{B}-\overline{A}B\),}\\
    \quad       \left\{   
      \begin{array}{rcl}   
\T[1][-1](\tth_{\jrt}-\bi/4)\overline A(\tth_{\jrt})&=&\T[2][-1](\tth_{\jrt}+\bi/4) A(\tth_{\jrt})\\   
\T[1][-1](\tth_{\jrt}-\bi/4)\overline B(\tth_{\jrt})&=&\T[2][-1](\tth_{\jrt}+\bi/4) B(\tth_{\jrt})   
\end{array}   
\right.   
\end{multline}
One can notice that at such  \(\tth_{\jrt}\)   
the two conditions in the {\rhs}  
 are equivalent ({\idest} we have only one constraint for each \(\tth_\jrt\)).

The \(\tth_{\jrt}\) are a finite size analogue of the Bethe roots \(\theta_\jrt\). %
In particular %
we see
that at large \(\LF\), the roots of   
\(A\overline{B}-\overline{A}B\) are precisely the Bethe roots. Indeed,    
at large \(\LF\), \(B\simeq \bi\) and \(A\simeq \Pf[3][{[-2]}]-\Pf[3]\), giving   
\(A\overline{B}-\overline{A}B\simeq -\bi(\Pf[3][{[-2]}]-\Pf[3]+Pf[3][{[+2]}]-\Pf[3])\simeq   
-\varphi\).
This means that the roots \(\tth_\jrt\) coincide (in the asymptotic
limit) with the roots \(\theta_\jrt\) of \(\vf\).
 Moreover,
the complex-conjugacy relation \eqref{eq:overl-tus=tnp-abar}, together
with the limit   \(B\simeq \bi\),  
 implies that the second  relation  in the {\rhs} of \eqref{eq:finitBE}  
  reduces to the reality condition   
\(\frac{\overline{\T[1][-1](\tth_{\jrt}-{\bi}/4)}}{~~\T[1][-1](\tth_{\jrt}-{\bi}/4)~~}=-1\). Using the   
leading-order large \(\LF\) expression \eqref{eq:y0-=-e} of
\(Y_{a,0}\) in terms of
\(\Sb\) and \(\CDb\), we   
get at large \(\LF\)   
\begin{eqnarray}   
  \T[1][-1](\us-\bi/4)&\simeq&\varphi^{[-2]}\frac{\varphi+2\varphi^{[-2]}}{2\varphi {\bSact}^{[-2]}}   
  e^{-\LF\cosh(\frac{2\pi}3 (\us-\bi/4))}  \\ 
\textrm{where }{\bSact}(\us)&=&\prod_jS_0^2(\us-\theta_{\jrt})\chi_{_{CDD}}(\us-\theta_{\jrt}) 
\end{eqnarray}   
Using the fact that \(\varphi(\tth_{\irt})=0\) at each \(\tth_{\irt}\) (in the
asymptotic limit), and   
dividing by the complex conjugate, the large \(\LF\) regularity   
requirement becomes   
\begin{eqnarray}   
\left.  \frac{\left(\varphi^{[-2]}\right)^2}{\varphi   
    {\bSact}^{[-2]}}\frac{\varphi}{\left(\varphi^{[+2]}\right)^2{\bSact}^{[+2]}}
\right|_{\us=\tth_{\jrt}}e^{\bi \LF \sinh(\frac{2\pi}3 \tth_{\jrt})}&=&-1   
\end{eqnarray}   
Using the crossing relation, the {\lhs} becomes simply 
\({\bSact}(\tth_{\jrt})e^{\bi \LF \sinh(\frac{2\pi}3 \tth_{\jrt})}\), 
so that the  
finite size regularity condition stated above is equivalent at large   
\(\LF\) to the asymptotic Bethe equations \eqref{eq:fdisp-1-=}.   

For \(\Np>3\), one can also write finite size Bethe equations
corresponding to the absence of poles for the {\qfs}, and check that
they reproduce the asymptotic Bethe equations in the limit
\(\LF\to\infty\)  (see \cite{2010arXiv1007.1770K}).

\paragraph{Iterative solution of the FiNLIE}
\label{sec:iteart-solut-finl}

Let us now explain how to solve iteratively
the %
finite set of non-linear integral 
equations \eqref{eq:fdisprf=fr-sum_-1}, to obtain the finite-size
spectrum of the {\PCM}.
We should start by finding the asymptotic Bethe roots \(\theta_\jrt\)
(hence the 
polynomial \(\vf\)), and deducing a polynomial \(\Pf[\Np]\) which obeys
\eqref{eq:vf=b-1np-22left}. We also start with \(\rf=0\) (to match the
asymptotic limit). Then, to solve the equation 
\eqref{eq:fdisprf=fr-sum_-1}, we
proceed iteratively: at each step, we compute the {\rhs} of
\eqref{eq:middle-node-equation-2} and the coefficients \(d_{a,\jq}\) of
\eqref{eq:t-1-leftus+bi}, 
 using the previous values of
\(\Pf[\Np]\) and of the densities \(\rf\)
(which parameterize the \(\T\) and the {\qfs}). Then we find the
position of 
the zeroes of the numerator of \eqref{eq:fdisprf=fr-sum_-1}, and
update the polynomial \(\Pf[\Np]\) such 
that the denominator 
\(\Det{c_{\iq,a}}{\substack{2\leq\iq\leq\Np\\
1\leq a\leq \Np-1}}\) has zeroes at the same position (which ensures
that the {\rhs} of \eqref{eq:fdisprf=fr-sum_-1} has no pole). Finally we update
the densities according to \eqref{eq:fdisprf=fr-sum_-1}, and from
these new densities we can start a new iteration.
In its spirit this algorithm is a fix point algorithm, and if it
converges, then the densities to which it converges are solutions of
the equations
\eqref{eq:fdisprf=fr-sum_-1}.

If \(\LF\) is large enough, then one can prove that this algorithm
converges\footnote{The convergence of the algorithm when \(\LF\) is
  large is obtained %
from
  the presence (in the equation \eqref{eq:fdisprf=fr-sum_-1})of the
  factor \(e^{- 
    E_a^{[+a-{\Np}/2]}}\) which is very small. This factor
  means that we are looking for the fixed points of a function which
  is a contraction mapping, hence the convergence of the algorithm.
}, whereas when \(\LF\) is finite, we can find {\NuMr} evidence of a
convergence, and find a {\NuMr} approximate solution.

\subsection{Computation of the energy}
\label{sec:expression-energy}

In the previous sections, we have written the analyticity constraints
on the {\Tfs} which allow to completely constrain the {\qfs} and
reduce the {\Ysys} to a closed set of a finite number of equations.

In particular, we can see that the vacuum %
is a solution where the {\Yfs} do
not have any zero or pole, 
whereas the excited states correspond to solutions which have several
zeroes and poles. This was shown in the asymptotic limit in section
\ref{sec:asymptotic-limit}, where the poles of the {\Yfs} %
were given by the zeroes of the polynomial \(\vf\equiv
\fQ[][0]\). For the vacuum, \(\vf=1\) has no zero, whereas it has zeroes
for all the excited states. At finite size, the zeroes of the {\Tfs}
are modified, but it is a common belief (supported by our numerics)
that for the vacuum, the {\Yfs} still have no pole when \(\LF\) is
finite.

In the expression \eqref{eq:EzfromY} of the energy, the existence of
poles or zeroes of the function \(1+\Y[][0](\us)\) requires to specify
the integration contour for excited states.

In the section \ref{sec:expr-energy-excit} we will propose an
expression for the energy of the excited states in the {\U 1}
sector. The guideline to write such an expression is that we will
require the energy to be given by the same expression
\eqref{eq:EzfromY} as in the case of vacuum, except that the
integration contour may be changed. We will require that this gives a
real energy, which converges (when \(\LF\to\infty\)) to the 
asymptotic energy \eqref{eq:e=-sum_nrtchtth-}.

Then the section \ref{sec:{\NuMr}-results} will discuss the {\NuMr}
results and their consistency with known analytic results.

\subsubsection{Expression of the energy of excited states}
\label{sec:expr-energy-excit}

In order to generalize the expression \eqref{eq:EzfromY} of the energy
to excited 
states, let us first describe (at least when \(\LF\) is large enough) the
singularities of the integrand  \linebreak
\({\cht[\us]
  \mathrm{log}\left( 1+\Y[][0](\us)\right)}\).

To this end, let us denote by \(\theta_\jrt^{(a)}\) the zeroes of
\(\T[][0][{[-a+{\Np}/2]}]\). In the asymptotic limit,
\(\T[][0][{[-a+{\Np}/2]}] = \vf\) and these zeroes coincide with the
roots \(\theta_\jrt\) of \(\vf\). At finite size, by contrast, the zeroes
\(\theta_\jrt^{(a)}\) of \(\T[][0][{[-a+{\Np}/2]}]\) are distinct in
the sense that in
general, the set
\(\left\{\theta_\jrt^{(a)}\middle|1\leq\jrt\leq\Mp\right\}\) %
depends on
\(a\).

This means that the function
\(1+\Y[][0](\us)=\frac{\T[][0][+]\T[][0][-]}{\T[a-1][0]\T[a+1][0]}\) has
zeroes at positions \linebreak \({\theta_\jrt^{(a)}+\bi/2\left(-a+{\Np}/2\pm1\right)}\)
and poles at position
\(\theta_\jrt^{(a\pm1)}+\bi/2\left(-a\pm1+{\Np}/2\right)\). When \({\LF}\) is
large, the sets
\(\left\{\theta_\jrt^{(a)}\middle|1\leq\jrt\leq\Mp\right\}\) almost
coincide with %
the roots of \(\vf\), which means that
in
\(1+\Y[][0](\us)=\frac{\T[][0][+]\T[][0][-]}{\T[a-1][0]\T[a+1][0]}\),
each pole 
 almost coincides with a zero.
We also noticed numerically (by
iterating the algorithm given in section \ref{sec:set-equations}) that
even when \(\LF\) is small, the 
distance between the zeroes and the poles of
\(1+\Y[][0](\us)=\frac{\T[][0][+]\T[][0][-]}{\T[a-1][0]\T[a+1][0]}\)
remains quite small (compared to \(\bi/2\)).

These zeroes and poles of \(1+\Y[][0](\us)\) give rise to logarithmic
singularities (branch points) in the integrand \(\cht[\us]
  \mathrm{log}\left( 1+\Y[][0](\us)\right)\) of %
the expression \eqref{eq:EzfromY} of the energy. %
In order
to
understand the impact of these singularities when the contour is
modified, we can for instance rewrite the integral through an
integration by parts:
\begin{align}
\label{eq:EippPCM}
E=%
\sum_{a=1}^{{\Np}-1}\frac{\sin\frac{\pi
      a}{{\Np}}}{\sin\frac{\pi }{{\Np}}} 
\int_{\us\in\bR} \sht[\us]
\frac {\partial_\us\Y[][0]}{1+\Y[][0]}
\frac{\mathrm{d}\us}{2\pi}\,.
\end{align}
Then we see that if we denote by
\(\zs_0=\theta_\jrt^{(a\pm1)}+\bi/2\left(-a\pm1+{\Np}/2\right)\) the position
of a pole of \(\Y\), then \(\frac
{\partial_\us\Y[][0]}{1+\Y[][0]}\sim -\frac 1 {\us-\zs_0}\) in the
vicinity of \(\us=\zs_0\), hence the integrand has a pole with residue 
\(\frac {-1}{2\pi} \sht[\zs_0]\). Similarly, if we denote by
\(\zs_0=\theta_\jrt^{(a)}+\bi/2\left(-a\pm1+{\Np}/2\right)\) the position
of a zero of \(\Y\), then \(\frac
{\partial_\us\Y[][0]}{1+\Y[][0]}\sim \frac 1 {\us-\zs_0}\) in the
vicinity of \(\us=\zs_0\), hence the integrand has a pole with residue 
\(\frac {1}{2\pi} \sht[\zs_0]\).

Having noticed this, we can try to find the correct integration
contour, which we write {\below} in the case when 
\(\Np\) is odd.

\paragraph{Case when \(\Np\) is odd}
\label{sec:case-when-np}

If \(\Np\) is odd, then the function \(1+\Y[\frac{{\Np}-1}2][0]\) has zeroes at position 
\({\theta_\jrt^{(\frac{{\Np}-1}2)}+\bi /4 \pm \bi /2}\) and it also has poles
at position 
\({\theta_\jrt^{(\frac{{\Np}-1}2 \pm 1)}+\bi /4 \pm \bi /2}\). In particular,
we will denote
\begin{align}
 \bth_\jrt\equiv \theta_\jrt^{(\frac{{\Np}-1}2)}\,.
\end{align}
These \(\bth_\jrt\) are %
the zeroes of \(\T[\frac{{\Np}-1}2][0][{[+1/2]}]\)
and in
the asymptotic limit, they converge to the Bethe roots. In finite
size, we will call them the ``finite size Bethe root''\footnote{One
  should note that in this construction, several different objects
  tend to the Bethe roots in the limit \(\LF\to\infty\).
In the previous section, we defined the \(\tth_\jrt\), which obey finite
size Bethe equations, and tend to the \(\theta_\jrt\) when  \(\LF\) tends
to \(\infty\). Here, we define {\another} finite size version of the
asymptotic Bethe roots, which enters the expression of the energy.
} as they
generalize the asymptotic Bethe roots. We can also notice that due to
the complex-conjugacy relation \eqref{eq:overl-tus=tnp-abar}, we have
\begin{align}
 \theta_\jrt^{(\frac{{\Np}+1}2)} = \overline{\bth_\jrt}\,.
\end{align}

Let us then imagine %
a contour which encircles the zeroes
\(\left\{\bth_\jrt-\bi/4\middle|1\leq\jrt\leq\Mp\right\}\) of
\(1+\Y[\frac{{\Np}-1}2][0]\) and the zeroes \(\left\{\overline{\bth_\jrt}+\bi/4\middle|1\leq\jrt\leq\Mp\right\}\) of
\(1+\Y[\frac{{\Np}+1}2][0]\). %
This choice, which we will motivate {\below}, gives the following
expression of the energy
\begin{multline}
 \label{eq:oddNU1energy}        
E(\LF) = -\frac{1}{{\Np} } \sum_{a = 1}^{\Np-1} \frac{\sin(\frac{a        
    \pi}{\Np})}{\sin(\frac{\pi}{\Np})} \int_{\us\in\bR}       
\cosh\left(\frac{2 \pi}{\Np} \us \right) \log \left( 1 + \Y[][0](\us)        
\right) \mathrm{d}\us\\        
+\bi \sum_{\jrt} \frac {
\mathrm{cos}\frac \pi{2 \Np}}       
{\mathrm{sin}\frac \pi \Np}\left[\sinh \left(\frac{2 \pi }{\Np} \left(\bth_{\jrt}        
      -{\bi}/4\right)\right)-\sinh \left(\frac{2\pi}{\Np}         
    \left(\bar\bth_{\jrt}        
      +{\bi}/4\right)\right)\right]\,.
\end{multline}
In this expression, we have added the contribution of these
singularities to the integral on the real axis. For instance the
zero of \(1+\Y[\frac{{\Np}-1}2][0]\) at position \(\bth_\jrt-\bi/4\)  stands
{\below} the real axis, hence if the contour is deformed from the real
axis to enclose this singularity, the deformation of the contour
should enclose it counter-clockwise (see figure
\ref{fig:contour3}). Hence this singularity 
contributes as
\(\bi \frac {
{
\mathrm{sin}\left(\frac {\Np-1} 2 \frac
  \pi  \Np\right)
}}
{\mathrm{sin}\frac \pi \Np}
\sinh \left(\frac{2 \pi }{\Np} \left(\bth_{\jrt}        
      -\bi/4\right)\right)\), as it can be seen from the integration by
  part \eqref{eq:EippPCM}. Using the same argument for %
  the
zero of \(1+\Y[\frac{{\Np}+1}2][0]\) at position
\(\overline{\bth_\jrt}+\bi/4\), we obtain\footnote{To obtain 
\eqref{eq:oddNU1energy}
we also used the simplification \({\mathrm{sin}\left(\frac {\Np-1} 2 \frac
  \pi  \Np\right)}
=\mathrm{cos}\frac \pi{2 \Np}
\).
} the expression \eqref{eq:oddNU1energy} above. %

\colorprint{\FigContour{}\FigEnergiesPCMO{}\FigEnergiesPCMT{}}{}
Moreover, one can see that if \(\bth_\jrt\) has a positive imaginary
part, then the following simple contour integration reproduces exactly 
the expression \eqref{eq:oddNU1energy}: (cf figure \ref{fig:contour3})
\begin{gather}
\label{eq:EnergyOdd}   
E(\LF) = -\frac{1}{\Np} \sum_{a = 1}^{\Np-1} \frac{\sin(\frac{a        
    \pi}{\Np})}{\sin(\frac{\pi}{\Np})} \int_{\us\in \bR+\bi(\frac a 2
  -\frac \Np 4)}  
\cosh\left(\frac{2 \pi}{\Np} \us\right) \log \left( 1 +        
  \Y[][0](\us) \right) \mathrm{d}\us
\,.  
\end{gather}
This case when \(\bth_\jrt\) has a positive imaginary part corresponds
for instance to the first excited state, and more generally to the
states having Bethe roots with even momentum number. In this
statement, we call ``momentum number'' the integer \(\kappa\) such that
\index{momentum number}
in the Bethe equation \eqref{eq:BethePCM2}, we have  \(\dkp=\bi~ \lcds~
\sht + \mathrm{log}\left( \Sscal(\theta_\nrt) \right)\).  This can be
shown at least when \(\Lf\) is large enough, by the arguments which we
will use in section \ref{sec:comp-with-known} to reproduce
the ``Lüscher
correction'',  %
 and we have no {\NuMr}
evidence that the situation is different at smaller \(\LF\).

The contour manipulation showing the equivalence of the expressions
\eqref{eq:EnergyOdd} and \eqref{eq:oddNU1energy} (when \(\bth_\jrt\)
has positive imaginary part) is illustrated in figure 
\ref{fig:contour3}, in the case \(\Np=3\).
\colorprint{}{\FigContour{}}

For states where some \(\bth_\jrt\) have negative imaginary part, the
contour \eqref{eq:EnergyOdd} is not satisfactory anymore, and should
be slightly deformed in the vicinity of the roots
\(\bth_\jrt\). With this deformation, 
we will still get
the expression \eqref{eq:oddNU1energy}.

Let us now motivate the particular choice of the expression \eqref{eq:oddNU1energy} for
the energy:
\begin{itemize}
\item First of all, this expression is real. That is why the contour
  has to take into account the singularities \(\bth_\jrt\) and
  \(\overline{\bth_\jrt}\) on the same footing.
\item Next, we see that in the asymptotic limit, \(\bth_\jrt\) becomes
  real. Then the term \(\sinh \left(\frac{2 \pi }{\Np} \left(\bth_{\jrt}        
      -{\bi}/4\right)\right)-\sinh \left(\frac{2\pi}{\Np}         
    \left(\bar\bth_{\jrt}        
      +{\bi}/4\right)\right)\) becomes equal to \(\cht\), and the sum (the
  second term in \eqref{eq:oddNU1energy}) becomes equal to
\(\sum_{\jrt}\cht[\theta_\jrt]\), as expected from
\eqref{eq:e=-sum_nrtchtth-}. In this asymptotic limit, the integral
term in \eqref{eq:oddNU1energy} is exponentially small, and we recover
the asymptotic expression of the energy.
\item The integration contour is natural enough in the sense that it
  remains inside the analyticity strip, and that it is not
  self-intersecting. This non-intersection condition was used in the
  discussion above   to fix the natural sign of the contributions of the
  singularity. This   condition can also be used to exclude contours
  which wind several 
  times around the same singularity.
\item At least for \(\Np=3\), the total number of non-intersecting
  contours %
  is small: for each singularity of a {\Yf}, if the
  singularity lies inside the analyticity strip, then the contour goes
  either above or {\below} this singularity. As the singularities of
  \(1+\Y[1][0]\) (resp \(1+\Y[1][0]\)) are only at position
  \(\bth_\jrt-\bi/4\) and \(\overline{\bth_\jrt}-\bi/4\)  (resp
  \(\bth_\jrt+\bi/4\) and \(\overline{\bth_\jrt}+\bi/4\)) if we
  restrict to the interior of the analyticity strip, then one quickly
  sees that \eqref{eq:oddNU1energy} is the only choice which obeys the
  above conditions and which reproduces the correct energy in the
  \(\LF\to\infty\) limit.

  If \(\Np>3\), \eqref{eq:oddNU1energy} can be viewed as a natural
  generalization of the \(\Np=3\) case.
\end{itemize}

\paragraph{Case when \(\Np\) is even}
\label{sec:case-when-npe}

If \(\Np\) is even, then a contour can also be proposed which obeys the
same naturality conditions. We proposed such a contour in %
\cite{2010arXiv1007.1770K}. As we will discuss in the next section, we
actually did not yet manage to perform serious {\NuMr} and analytic
checks of this expression when \(\Np>3\), and it should be viewed as one
possible conjecture.

\subsubsection{Numerical results}
\label{sec:{\NuMr}-results}

In order to check the consistency of the above construction, and to
show the efficiency of the FiNLIE, we iterated numerically the
algorithm given above to solve the {\Ysys}. As the functions \(\rf\)
decrease exponentially at large \(\us\), we could approximate them by
functions with a finite support ({\idest} we introduced a cutoff for the
variables \(\rf\)). In practice these functions were internally defined
by a polynomial interpolation from their values on a finite set of
points (about 500 points) belonging to this finite support.
The convolutions which appear in these FiNLIE are linear {\ops},
and could be expressed through matrices. We could write the exact
convolution of an interpolation function by the kernel \(\Ker_\Np\) or
the Cauchy kernel (which allows to compute the {\qfs} out of the
densities {\rf}) as the multiplication\footnote{As {\rf} is defined by
its values at a finite number of positions, it is internally viewed as
a vector. If functions are viewed as vectors, then the convolution is
a linear {\op} which maps one vector to {\another} one, hence it is
described by a matrix-multiplication.} by a matrix whose coefficients
are known analytically. This allowed to iterate the FiNLIE algorithm
at a reasonable speed.

When \(\LF\) is large enough, one can prove that the algorithm above
does converge to a solution, because we find the fix point of a
complicated function by iteratively defining \(x_{{\nn}+1}=F(x_{\nn})\). When
\(\LF\) is large enough, \(F\) is a contraction mapping in some 
vicinity of \(\rf=0\), and the sequence \(x_{\nn}\) is therefore converging.
Numerically, when \(\LF\) is large, we could indeed immediately notice
that the algorithm converges to a 
solution which is very close to the asymptotic limit. 
Then, when the length \(\LF\) decreases, the algorithm looks worse and  
worse converging, and the densities become more and more peaked around
the endpoints of the distribution. These endpoints are not artifacts
from the cutoffs, but come simply from the fact that when \(\LF\)
  is small, \(e^{-\Lf \mathrm{cosh}(\us)}\) is almost equal to one in a
  wide range of \(\us\) (as long as \(\mathrm{cosh}(\us)\leq 1/\LF\)) and
  then it quickly becomes very small when \(\mathrm{cosh}(\us)\gg 1/\LF\).
  It turned out that most of the non-trivial behavior of the Y-,
  {\Tfu}- and {\qfs} occurs precisely in a small vicinity of \(\us\simeq
  \mathrm{argcosh}(1/\LF)\). By   
choosing a small enough interpolation step\footnote{
The algorithm searches for the fix point of a function \(F\) by
defining \(x_{{\nn}+1}=F(x_{\nn})\), and saying that if \(x_{\nn}\) converges at
\({\nn}\to\infty\), then its limit is a solution of \(x=F(x)\). Then a trick
which often improves the convergence is to define \(x_{{\nn}+1}=\alpha
F(x_{\nn})+(1-\alpha) x_{\nn}\), where \(\alpha\in]0,1]\). If this sequence
converges, it also converges to a fix point of \(F\).

In the {\NuMr} resolution of the FiNLIE, it was necessary to use this
trick to get a satisfactory convergence when \(\Lf\) is small.}, it was nevertheless
possible to make the algorithm reasonably convergent for
several excited states in a range of length \(10^{-3}\leq\LF\leq 100\),
when \(\Np=3\).
These results were written in \cite{2010arXiv1007.1770K}, and they are
presented in the figure \ref{fig:PCMen}. %
They
can be improved with respect to these results, and one can reach
much smaller length \(\LF\). These results should be soon available in
the version 2 of \cite{2010arXiv1007.1770K}.

\colorprint{}{\FigEnergiesPCMO}

Unfortunately, at \({\Np}\geq 4 \) the calculations become heavier and 
(with the size of interpolation steps we can afford)  our algorithm becomes 
unstable already   for \(\LF\) of order \(\sim 1 \) (which means we  
{\cannnot} really check the conformal limit for instance). At the moment we  
{\cannnot} say whether this has a physical meaning (like some symmetry  
breaking down, or some new type of singularity appearing) or whether  
it is just a {\NuMr} artifact, due to a poor {\NuMr} accuracy, or to  
the choice of the equations. For instance, 
the  
function that we iterate %
may stop
to be a contraction mapping but still 
have a fix point. 
One could for instance expect that for \(\Np\geq 4\), if we 
rewrite slightly this
function it could become a contraction again, and extracting its
fix point would be possible by iterations. 

\subsubsection{Comparison with known limits}
\label{sec:comp-with-known}

Once we have iterated and numerically solved the FiNLIE, we can check
that it matches the known features of the {\PCM}. 

\paragraph{Conformal limit}
\label{sec:conformal-limit}

The conformal limit is the limit where the length \(\LF\) is very
small. In this limit, the action can be linearized giving rise to a
2-dimensional conformal theory with \(\Np^2-1\) massless bosons.

As explained in \cite{2010arXiv1007.1770K}, this linearization %
shows that when \(\LF\ll 1\), the energy behaves like 
\begin{align}
\label{eq:esimfr-pilfl-fracnp2}
  E\sim\frac{2 \pi}{\LF}\left(-\frac{\Np^2-1} {12} + \sum _\jrt |\nn_\jrt|\right)
\end{align}
where \(\nn_\jrt\) denotes the momentum number associated to the Bethe
root \(\theta_\jrt\). Our numerics are compatible with this result, as
they show that when all particles have zero momentum, \(E \frac{\LF}{2
  \pi}\) converges to \(-2/3\), at a logarithmic speed. These numerics are
also compatible with the fact that the states \(\theta_1\) and
\(\theta_2\) have energies which behave like  \(\frac{2
  \pi}{\LF}\left(-1/3+1\right)\) and \(\frac{2
  \pi}{\LF}\left(-1/3+2\right)\) respectively (see figure
  \ref{fig:PCMen}).

In \cite{2010arXiv1007.1770K}, we also computed the first correction
to \eqref{eq:esimfr-pilfl-fracnp2}, and 
found a logarithmic speed of convergence which matches quite well our
{\NuMr} results.

\paragraph{Asymptotic limit}
\label{sec:asymptotic-limit-2}

We already discussed the fact that by construction, the FiNLIE
reproduces the known asymptotic limit (\(\LF\to\infty\)) of the {\PCM}.
In fact one can perform deeper consistency checks of the large \(\LF\)
behavior of the FiNLIE. More precisely there exists a general
procedure, initiated by Lüscher \cite{Luscher:1985dn,Luscher:1986pf}
(see also \cite{Klassen:1990ub}). This procedure allows to find the
first corrections to the asymptotic energy \eqref{eq:e=-sum_nrtchtth-}
when the size \(\LF\) is large, but finite. Following
\cite{Klassen:1990ub}, it is is easy to show that when \(\Np=3\),  these
Lüscher 
corrections predict that the mass gap (the difference between the
energy of the state denoted \(\theta_0\) on figure \ref{fig:PCMen} and
the energy of the vacuum) is given by 
\begin{equation} 
\label{eq:MGLuscher} 
  E^{\textrm{mass gap}}_{\LF\to\infty}\simeq 1 - \left(\frac{32 e^{-\sqrt 3 \LF/2} \pi^3}{ \Gamma\left( \frac  
  1 3\right)^6}\right)\,,
\end{equation} 
where the term \(\left(\frac{32 e^{-\sqrt 3 \LF/2} \pi^3}{ \Gamma\left( \frac  
  1 3\right)^6}\right)\) is a so- called \(\mu\)-term.

Let us now show that this result can also be obtained analytically
from our FiNLIE and from the prescription \eqref{eq:oddNU1energy} for
the energy. First we should note that when \({\Np}=3\) it suffices to
compute \(\Y[1][0]\) in order to obtain the energy, because
\(Y_{2,0}=\overline{\Y[1][0]}\).
    
As we saw in the previous sections, \(\Y[1][0]\) is given by
\eqref{eq:y0-=-e}, which reads
\begin{align}
\label{eq:y0Luscher}
  \Y[1][0] =&  e^{-\LF  \cht[\us]} %
    \frac{\left(\T[1][1]\right)^2
}{\T[0][0] \T[2][0]}%
\frac{\vf[{[-3/2]}]}{\vf[{[+1/2]}]} 
\frac 1 {\left(\Sb^2\CDb\right)^{[-3/2]}}
\\=&
e^{-\LF  \cht[\us]} %
    {\left(\frac{  3 \us+5 \bi/4 }
{\us+\bi/4}\right)^2
}\,,%
\frac 1 {\left(\Sb^2\CDb\right)^{[-3/2]}}
\end{align}
where the last line is obtained by replacing the {\Tfs} with their
explicit value as it can be computed from section \ref{sec:expl-expr-tfs}.

At large \(\LF\), this expression allows to compute the leading order
of the integral term in \eqref{eq:oddNU1energy}. We see that this term
is of the order \(\mathcal{O}\left(e^{-\LF}\right)\), which is much
smaller than the \(\mu\)-term \(\frac{32 e^{-\sqrt 3 {\LF}/2} \pi^3}{ \Gamma\left( \frac  
  1 3\right)^6}\) which we want to reproduce. This suggests that the
second term in \eqref{eq:oddNU1energy} gives the leading correction to
the mass gap, as we will show now.

Finding the behavior of this term is a bit more tricky, as it  
involves the   
position of the Bethe root. This position can be estimated by  
computing the densities to the leading order, to deduce the first  
correction to \(\T[1][0]\), in order to solve the equation \(\T[1][0](\theta_0  
+{\bi}/4)=0\).  

For the mass gap, this root should be at the origin, up to exponential
corrections in \(\LF\). Moreover one can show\footnote{These large \(\LF\) 
  expressions are obtained by neglecting integral terms in the  
  determinant expression of \(\T[1][0]\).} that \(\T[1][0](0+\bi/4)\sim \frac \bi
6 \rf[2](0)+\bi \rf[3](0)=\mathcal{O}(e^{-{\LF} \sqrt{3}/2})\), while
\(\partial_\us\T[1][0](0+\bi/4)\sim \bi\), so that \(\T[1][0](\theta_0  
+\bi/4)=0\) gives \(\bth_0\sim - {\frac 1
  6}\rf[2](0)-\rf[3](0)\). Using the asymptotic  
expression for \(\rf[{\jq}]\)'s (which can be extracted by keeping only the
asymptotic expressions of \(\T[][-1]\) and of \(d_{a,\jq}\) in the formula 
\eqref{eq:fdisprf=s-c_iq-a}), one gets   
\(\bth_0   
 \sim \frac{\bi e^{-\sqrt{3} {\LF}/2}        
  \Gamma\left(-\frac{1}{3}\right)^2        
  \Gamma\left(\frac{2}{3}\right)^2}{\sqrt{3} \pi        
  \Gamma\left(\frac{1}{3}\right)^2}  
\), so that the second term in (\ref{eq:oddNU1energy}),        
which is \(\sinh \left(\frac{2 \pi }{3} \left(\bth_0        
      -{\bi}/4\right)\right)-\sinh \left(\frac{2\pi}{3}         
    \left(\bar\bth_0        
      +{\bi}/4\right)\right)\nonumber         
  \) can be computed at leading        
order.        
    
That gives 
\begin{equation} 
  E\simeq 1 - \left(\frac{32 e^{-\sqrt 3 \LF/2} \pi^3}{ \Gamma\left( \frac  
  1 3\right)^6}\right)\,,
\end{equation} 
which coincides exactly with the \(\mu\)-term of the Lüscher corrections,
and this a good non-trivial test of the expression
\eqref{eq:oddNU1energy} of the energy.

Moreover, it is in good agreement with our {\NuMr} results, which is a
consistency that the algorithm has no obvious mistake. The figure
\ref{fig:MGraph3} shows this consistency with the numerics, and we see that
for this lowest-lying excited state,
the Lüscher correction \eqref{eq:MGLuscher} gives a good
approximation of the energy up to lengths of order one, while the
expressions from the conformal limit give a good approximation when
the length is smaller than (and up to) of order one.

\colorprint{}{\FigEnergiesPCMT}

\section{Conclusion}
\label{sec:conclusion}

  For many {\ing} two-dimensional field theories (or ``sigma models''), the TBA  approach of
  Al. Zamolodchikov %
  gives rise to
  a very universal system of
  functional equations, the \(Y\)-system.  This {\Ysys} is equivalent  
  to the same Hirota equation which arises for {\csds} in chapter 
  \ref{part:qoperatorsspin}. %
The
  Hirota equation is associated to variables \(a\) and \(s\) belonging to
  a %
  lattice %
which is fixed by the symmetry algebra of the model.
  In general, the {\YsE} is not sufficient to characterize and to
  solve a model, and some additional information, namely the
  analyticity properties with respect to the spectral parameter \(\us\)
  must be specified.

  As we saw, the typical solution of the {\Ysys} for a large variety of
  lattices is parameterized by a finite set of {\qfs} (where the
  number of {\qfs} is essentially equal to the rank of the symmetry
  group), and the resolution of the {\Ysys} reduces to finding the
  {\qfs} which reproduce the correct analyticity constraints on the
  {\Yfs}.

   We illustrated this procedure in the case of the \(\SU\Np\times
  \SU\Np\) {\PCM}, generalizing some results of
  \cite{2009JHEP...12..060G} by using the {\qfs}, which provide, as we
  showed, the typical solution to the {\Ysys}. Moreover, we showed how
  the additional analyticity constraints on the {\Yfs} are rewritten
  as natural constraints on the {\Tfu}- and {\qfs}. That allowed
for instance  to generalize to any finite size \(\LF\) the Bethe equations fixing the
  position of Bethe roots. We also saw that these analyticity
  conditions on the {\Tfs} allowed, when the size \(\LF\) is large, to recover
  the  asymptotic Bethe equations, including the factor \(\CDD\) in the
  phase of the {\Sma}.

  The {\NuMr} and analytical checks that we performed confirm the
  consistency of the finite set of equations that we obtain,  and of
  its iterative resolution, and in particular
  for \(\Np=3\), the Lüscher corrections provide a serious check that
  the contour proposed and motivated in section
  \ref{sec:expr-energy-excit} to define the energy is very consistent.

  At the present the numerics are still perfectible, and in particular
  it is left to understand why we have some convergence issues at
  length of order one when \(\Np\geq 4\). This point would certainly be
  an important step in order to understand, at the {\level} of the
  {\Ysys}, the large \(\Np\) behavior of the {\PCM}, which exhibits a
  well-studied phase transition.

 It would also be very
  enlighting to understand analytically how our FiNLIE behave in the
  conformal limit, and how they give rise to the analytic
  expressions  known from conformal field theory.

  In the next chapter we will see that in the example of the {\ADF}
  duality the same approach also allows to write a FiNLIE. This
  shows that this procedure based on {\qfs} applies to several
  different models. We will see that %
a lot of work has
  to be done on a
  case-by-case basis, even though several common features arise.

  %%% Local Variables: ***
  %%% mode:latex ***
  %%% eval: (find-file "english.tex") ***
  %%% TeX-master: "english.tex" ***
  %%% End: ***

%% file: AdsCFT.tex
In this chapter, we will see how the methods of the
previous chapter can be applied to the {\Ysys} of AdS/CFT.

This {\Ysys} was conjectured in %
\cite{2009PhRvL.103m1601G} and then understood in terms
of the {\TBA} approach
\cite{2009JPhA...42K5401B,Gromov:2009bc,Arutyunov:2009ur}, and it
is believed to describe  the exact scaling dimensions of the {\ops}
in the {\SYM} conformal field theory.
Its derivation is conceptually slightly different to the {\Ysys} of
(for instance) the {\PCM}, because {\SYM} is not %
  two-dimensional, %
and its
integrability, 
 comes out of a mapping
between some {\ops} (the 
single trace {\ops}) and the states of an {\ing} {\cds}.

This integrability was first noticed and understood in high-energy QCD
\cite{Lipatov:1993yb,Faddeev:1994zg}, and then in
{\SYM} \cite{Minahan:2002ve,Beisert:2003yb,Beisert:2003tq}.
Inspired by the considerable activity in the string side
 of the
duality
\cite{Gubser:2002tv,Frolov:2002av,Russo:2002sr,Minahan:2002rc,Frolov:2003qc,Beisert:2003xu,Arutyunov:2003za,Arutyunov:2003uj,Beisert:2003ea,Kruczenski:2004wg,Kazakov:2004qf},
where integrability was also noticed 
\cite{Mandal:2002fs,Bena:2003wd}, it was shown that integrability allowed to write Bethe
equations for {\SYM} \cite{Beisert:2004hm,Arutyunov:2004vx}.

In order to write these Bethe equations, one key step is to find the
{\Sma}. Like for the {\PCM} in chapter \ref{cha:ansatzs-de-bethe}, it
turns out that this {\Sma} is fixed, up to an overall phase, by the
symmetries of the model and consistency requirements
\cite{Staudacher:2004tk,Beisert:2005fw,Beisert:2005tm}. 
This overall phase is fixed by a crossing equation which was
identified in \cite{Janik:2006dc}.

The {\Ysys}, which was conjectured from these Bethe equations,
  was successfully tested in both the weak 
  coupling regime, (by
comparison with perturbative expansion in {\SYM}
\cite{Janik:2007wt,Heller:2008at,Bajnok:2008bm,Fiamberti:2008sm,Velizhanin:2008pc,Minahan:2009wg,Arutyunov:2010gb,Balog:2010xa}),
and in the strong coupling regime \cite{Gromov:2009tq}. On spectacular
prediction of the {\Ysys}, (latter checked against
perturbative expansion in {\SYM}) was the prediction of the first
subleading corrections to the dimension of the ``Konishi operator'' \cite{Gromov:2009zb,Frolov:2010wt}.

In this chapter, we will not see in great details how this {\Ysys} was
conjectured, but we will use it as the starting point of an analysis
in terms of  {\Qfs}. This analysis \cite{Gromov:2011cx} is an original
contribution of this {\PhD}, and it allows to recast the infinite set
of equations arising from the {\TBA} into a finite set of integral
equations (FiNLIE). 

The numbering of the sections follows the general road-map of section
\ref{sec:writing-finlies}, %
 where the steps needed in order
to write a FiNLIE for a given model are listed.

\section{The {\Ysys} for AdS/CFT}
\label{sec:ysys-adscft}

The {\Ysys} describing the energy spectrum of the AdS strings (or
equivalently the scaling dimensions of {\SYM} {\ops}) is reviewed
for instance in \cite{Gromov:2010kf}.
It holds in the ``planar limit'' of {\SYM}, which is the limit when
\(\Np_c\to\infty\) and \(\lambda\equiv \G_{YM}^2 \Np_c\) is finite, where
\(\G_{YM}\) is the coupling constant of {\SYM} and \(\Np_c\) is the rank
of the gauge group. The constant \(\lambda\) was introduced by %
't Hooft \cite{Hooft1974461} who noticed that in this limit, all
non-planar Feynman diagrams are suppressed. Therefore, \(\lambda\) is
called the
't Hooft coupling. %

In this planar limit, the spectrum of AdS/CFT %
is given by
the general {\YsE}
\eqref{eq:YSysEq} on the lattice \(\Tk(2,2|2+2)\), which reflects the
PSU\((2,2|4)\) symmetry of the model. The dispersion relation is more
subtle than for relativistic models:
  the massive particles (corresponding to the nodes at \(s=0\)
  in the {\Ysys})  carry an energy and a momentum, given by  %
\begin{gather}
\label{eq:DispRelAdSCFT}
E_a\equiv a+\frac{2\bi \G}{x^{[+a]}}-\frac{2\bi \G}{x^{[-a]}},\qquad\qquad
  e^{\bi p_a}\equiv \frac{x^{[+a]}}{x^{[-a]}}\\
\where x^{[\pm a]}\equiv x\left(\us\pm \frac \bi 2 a\right) \equiv
\frac 1 2 \left(\frac {\us\pm \frac \bi 2 a} \G +\bi \sqrt
  {4-\frac{\left(\us\pm \frac \bi 2 a\right)^2}{\G^2}}\right)\,,
\label{eq:DefXMir}
\end{gather}
where \(\G=\frac{\sqrt{\lambda}}{4\pi}\) is the coupling constant.
In this expression, the function \(\sqrt z\) denotes the holomorphic
function on \(\bC\setminus \bR^-\) which coincides with the usual square
root on \(\bR^+\) ({\idest} the square root on the complex plane has a cut on \(\bR^-\)).
We see that the function \(x(\us)\) defined in \eqref{eq:DefXMir} obeys
\begin{align}
\label{eq:xpooxeuog}
  x(\us)+\frac 1 {x(\us)}=\frac \us \G\,,
\end{align} 
 and that it has
cuts on \(]-\infty,-2\,\G]\cup [2\,\G,\infty[\) on 
the real axis\footnote{These cuts correspond to
  \(4-\frac{\us^2}{\G^2}<0\).}. This function \(x(\us)\) can also be
viewed as a double-valued function of \(\us\), and %
\eqref{eq:DefXMir} gives its expression on a specific Riemann
sheet. The function \(1/x(\us)\) corresponds to {\another} sheet of the
same double-valued function (because it %
  also %
 obeys \eqref{eq:xpooxeuog}),
and it still has a cut on \({\bZc_0\equiv]-\infty,-2\,\G]\cup
[2\,\G,\infty[}\).
\index{Zhukovsky cuts!Zc@\ensuremath{{\bZc}_\nn}}
 This
cut of square-root type, associated to branch points at 
position \(\pm 2\,\G\), will be called Zhukovsky cuts. %
Another
choice can for instance be 
\begin{align}
\label{eq:DefXPhys}
  \hx(\us)\equiv \frac 1 2 \left(\frac \us \G+\sqrt{\frac
      \us\G-2\,}\sqrt{\frac \us \G +2\,}\right)=\left\{
      \begin{aligned}
        x(\us)&&\If&\Im(\us)>0\\
        1/x(\us)&&\If&\Im(\us)<0
      \end{aligned}
      \right.\,.
\end{align}

The choice \eqref{eq:DefXPhys} has no cut on \([2\,\G,\infty[\), but has a
Zhukovsky cut \([-2\,\G,2\,\G]\). It does not have a cut on \(]-\infty,-2\,\G]\) either,
because the cuts from the two square roots compensate each other.
This definition \eqref{eq:DefXPhys} 
  \index{sheet (Riemann sheet)!magic sheet}
will be called the ``magic
branch'' %
of \(x(\us)\), %
and more generally the ``magic'' sheet of multi-valued-functions will be
a sheet\footnote{One should notice that in general, the cut structure
  does not fix uniquely the Riemann sheet. Indeed, \(\hx\) and \(1/\hx\)
  are two different determinations of \(x\), which have the same cut
  structure, but correspond to distinct Riemann sheets.

  In the literature, there exists {\another} choice of sheet which 
has ``short'' cuts, like the magic branch and which is called
``physical sheet''. For the function \(x(\us)\), it coincides with the
``magic sheet'', and in what follows this ``physical sheet'' will not
play an important role.
} where all cuts are ``short'' {\idest} of the form \([-2\,\G,2\,\G]\) or
\([-2\,\G+\frac \bi 2 a,2\,\G+\frac \bi 2 a]\). On the other hand the
definition \eqref{eq:DefXMir} will be called the ``mirror''
branch. This branch \cite{Arutyunov:2007tc} is the most frequent in
the study of the {\Ysys} 
and will be simply denoted as \(x(\us)\) (the same symbol as the
multi-valued function). %
  These different branches of \(x(\us)\) are
presented on figure \ref{fig:Xriemann}. %

\colorprint{}{\XFig}

The expression (\ref{eq:DispRelAdSCFT}-\ref{eq:DefXMir}) of the
dispersion relation is an expression in the mirror kinematics. One can
also write this dispersion in %
  {\another} branch (called ``physical''), %
where it reads
\begin{gather}
\label{eq:PhysDispRelAdSCFT}
\hat E_a\equiv a+\frac{2\bi \G}{\hx^{[+a]}}-\frac{2\bi \G}{\hx^{[-a]}},\qquad\qquad
  e^{\bi \hat p_a}\equiv \frac{\hx^{[+a]}}{\hx^{[-a]}}\,.
\end{gather}

Then, the AdS/CFT spectrum is obtained by solving the {\YsE} %
on the lattice \(\Tk(2,2|2+2)\), under some analyticity conditions which
include %
the asymptotic behavior 
\begin{align}
\label{eq:AsymAdSCFT}
  Y_{a,0}(\us)\simeq \left(\frac {x^{[-a]}} {x^{[+a]}}\right)^\LF\,,
\end{align}
where \(\LF\) is the length of an {\op}\footnote{%
    As mentioned in the
  introductory section (see for instance \cite{Minahan:2010js}) the
  {\SYM} {\op} %
  {\(\mathrm{tr}(XXYX)\)} can be mapped to 
the state \(\ket{↑↑↓↑}\) of an \(\SU 2\) {\cds} of size
\(\lcds=4\).
Then \(\LF=4\) is called the ``length'' of this {\op}. The conformal
dimension of this {\op} is then given by one solution of the
{\Ysys}, obeying the asymptotic behavior \eqref{eq:AsymAdSCFT} with \(\LF=4\).
}.
Then the anomalous dimension
\(\gamma\) of this {\op} is given by
\begin{align}
\label{eq:Eadscft}
\gamma=E-\Mp,\quad \where
  E=
\sum_{\jrt=1}^{\Mp}\hat E_1\left(\us^{\rlb[\jrt]}\right)+
\sum_{a=1}^\infty\int_{\us\in\bR}\frac{\mathrm{d}\us}{2\bi\pi}
  \frac{\partial E_a(\us)}{\partial \us} \mathrm{log}
  \left(
1+\Y[][0](\us)
  \right)\,,
\end{align}
where \(E_a(\us)\) (resp \(\hat E_a(\us)\)) are defined in
\eqref{eq:DispRelAdSCFT}  (resp \eqref{eq:PhysDispRelAdSCFT}),
and where \(\left\{\us^{\rlb[\jrt]}\middle| 1\leq\jrt\leq \Mp\right\}\)
is the set of the ``momentum-carrying'' Bethe roots (the analogue of
the roots \(\theta_\jrt\) of the polynomial \(\fQ[][0]\) in the
{\PCM}).
\index{Bethe roots}

An %
important remark is that the {\YsE} must hold on the ``mirror
sheet'' \cite{Arutyunov:2007tc,Gromov:2009bc}.  %
One {\cannnot} always continue analytically the {\YsE} to other sheets
than the mirror sheet, and 
  one can even show \cite{Gromov:2011cx} that on the ``magic'' sheet, the %
{\YsE} takes a quite different
form at \((a,s)=(1,1)\) for instance.

To give a complete set of equations, the {\YsE} has to be supplemented
with the asymptotic behavior \eqref{eq:AsymAdSCFT} and with
conditions at the corner (namely, the function \(\Y[1][1]\) has to be
the analytic continuation of \(1/\Y[2][2]\), %
  as we will see in
\eqref{eq:FermConstrYadscft}). For 
instance, the {\TBAE} \cite{Gromov:2009bc} contain these conditions,
though they are hard to understand from the integral TBA equations.
 In this
chapter we will see that in terms of the {\Tfs}, %
a
closed system of equations %
  is obtained %
 by imposing a few natural
analyticity 
constraints \cite{Gromov:2011cx}. %

\section{The asymptotic limit}
\label{sec:asymptotic-limit-1}

The asymptotic limit (\(\LF\to\infty\)) %
  is well presented in the {\Ysys} literature 
(see for
instance \cite{2009PhRvL.103m1601G} or the review
\cite{Gromov:2010kf}). 
In this {\PhD}, this limit was studied in \cite{Gromov:2010km} where
we first wrote the {\Wronskian} solution of 
the Hirota equation on a ``\(\Tk\)-{\hook}'' (the expressions
(\ref{eq:TTHook1}-\ref{eq:TTHook3})) and wrote the explicit
expressions of all {\cQfs} in the asymptotic limit. %

In this section, we will not repeat the results of this article, but
simply mention a few observations %
which arose in this study of the asymptotic limit. First of all let us
briefly mention that the expressions of {\cQfs} were obtained in this
paper %
from the knowledge of the {\cQfs} corresponding to a given {\nesting}
{\ppath}. %
The {{\cqfu}{\cqfu}-relations} were
used to express all {\cQfs} in terms of basis of nine\footnote{This set
of nine {\cQfs} is denoted as \(\mathcal{B}_2\) in \cite{Gromov:2010km}.
Using the gauge constraint \(\T[0]=\Ts[0][0][-s]\) they can be reduced
to a set of eight functions called \(\mathcal{B}_1\). As the definition of
the {\Wronskian} gauge fixes only three (out of four) gauge freedom, only
seven of these eight functions are really independent.
} {\cQfs} (using
\eqref{eq:QQopDetGLKM}), and these nine functions were expressed out
of the {\cQfs} lying on the {\nesting} {\ppath}, using the QQ-relations. They
are given by the formulae (5.6-5.9) (or (B.1-B.8) for the most general
states) and proven in appendix B of \cite{Gromov:2010km}.

We will not copy these formulae here, because they will not be used
directly in what follows. Instead, let us summarize a few
properties of the asymptotic solution, which were found in
\cite{Gromov:2010km}, and which correspond to symmetries of the
{\Ysys}. %

\paragraph{Super-determinant}
\label{sec:super-determinant}

 A first nontrivial property of the asymptotic solution is that the ratio 
\(\frac{\cqe[+]}{\cqe[-]}\frac{\cqf[\ove]{-}}{\cqf[\ove]{+}}\) is
equal to one.
  This relation  means that there exists a gauge where
\(\cqe{}=\cqf[\ove]{}=1\), {\idest} where \(\T[0]=1\), in addition to the gauge
constraints \eqref{eq:PhyWronGaugeTK}. In the absence of a non-trivial
symmetry, there would %
a priori not always exist such a gauge where \(\T[0]=1\).
Indeed, it is in
general possible to choose \(\cqe{}=1\) and that corresponds to the gauge
choice \(\T[0]=\Ts[0][0][-s]\) in \eqref{eq:PhyWronGaugeTK}, but then, the
remaining degree of gauge freedom\footnote{Let us remind here that the
gauge constraint \eqref{eq:PhyWronGaugeTK} fixes only three out of the
four degree of gauge freedom in \eqref{eq:GaugeFreedom}.} takes the
form
\(\T\rightsquigarrow \left({\gan[{\FS[s]}][]}\right)^{\FS[a]}\). There \({\gan[][]}\) is an
arbitrary function of \(\us\), and 
\begin{gather}
\label{eq:DefFS}
  {\gan[{\FS[s]}][]}\equiv\left\{
    \begin{aligned}
&\prod_{\nn=-\frac{s-1}2}^{\frac{s-1}2}
  \ga[2\nn][],&\If& s \geq 1\,,\\
&1&\If&s=0\,,\\
&1/\gan[{\FS[-s]}][]&\If&s<0\,.
    \end{aligned}\right.
\end{gather}
Since \(\T[0]\) is invariant under this transformation we see that
\(\T[0]=1\) {\cannnot} be enforced by a gauge transformation.
This means that the relation
\(\frac{\cqe[+]}{\cqe[-]}\frac{\cqf[\ove]{-}}{\cqf[\ove]{+}}=1\) is
not just %
obtained by a gauge
transformation, but it is the manifestation of some
symmetry. Interestingly enough, this relation can also be written in a
gauge-invariant way as \(\Y[1][-1]\Y[1][-2]=\Y[1][1]\Y[1][2]\). 
On this form, we see that 
this property is trivial for the states having the symmetry \(\Y[a][s]=\Y[a][-s]\), 
and what we noticed in the asymptotic limit is that this constraint
holds 
even for states which do not exhibit the symmetry
\(\Y[a][s]=\Y[a][-s]\).

In
\cite{Gromov:2010km}, we identified  the physical meaning of the equality
 \(\frac{\cqe[+]}{\cqe[-]}\frac{\cqf[\ove]{-}}{\cqf[\ove]{+}}=1\), as
the requirement that the group symmetry \(\PST\) %
  only involves matrices
with
super-determinant equal to one. %

More explicitly, there exists a limit called the classical limit (when
\(\G\to\infty\)), where the {\Tfs} are the characters of a monodromy
matrix \(\Omega\in\PST\), in some rectangular representations {\lbd} by
indices \((a,s)\in\Tl{\Tk(2,2|2+2)}\) (see \cite{Gromov:2010vb}). As it
belongs to \(\PST\), this matrix has its super-determinant equal to one.

This interpretation holds only in the 
\(\G\to\infty\) limit,
but in the more general case of finite \(\G\), 
 the {\cQfs} can be viewed as a
generalization of the eigenvalues of \(\Omega\), and the condition
\(\frac{\cqe[+]}{\cqe[-]}\frac{\cqf[\ove]{-}}{\cqf[\ove]{+}}=1\) can
be interpreted as the condition that the super-determinant of \(\Omega\)
is equal to
one. 

  We see that although the symmetry groups \(\PST\) and
  \(\U{2,2\ensuremath{|}4}\) are associated to the Hirota equation on
  the same (a,s)-lattice \(\Tk(2,2|2+2)\), the \(\PST\) case gives rise to
  additional constraints on the {\Tfs}. The same thing can be noticed in
the %
setup of the chapter
\ref{part:qoperatorsspin}: 
we have seen that a {\cds} with symmetry \(\GL \Kr\) or with
symmetry \(\SU \Kr\) gives rise to two {\Toprs} which obey the Hirota
equation on the same lattice. But the equation
\eqref{eq:PhysWronGaug0} shows that the relation
\(\rT[][\Kr]=\rT[\su+s][0][0]\) is true only for \(\SU \Kr\) {\csds}
(unless the twist takes a very specific value).

\paragraph{Structure of the right band}
\label{sec:structure-right-band}

Some other symmetries appear in the asymptotic limit, which we will
illustrate here with the 
``right band'' (when \(s\geq a\)) in the {\SL 2} sector, for
\index{sector!SL(2) sector}
the simplicity of notations.  %
  This %
so-called {\SL 2} sector \cite{Minahan:2010js}, %
denotes the states having only one type of Bethe roots: the momentum
carrying roots \(\us^{\rlb[\jrt]}\) which enter in the expression
\eqref{eq:Eadscft} of the 
energy. This sector 
is analogous to the {\U 1} sector of the {\PCM} studied in section
\ref{sec:application-au-champ}.

In the asymptotic limit, we obtain (see formula (5.6) and (5.12) in
\cite{Gromov:2010km})
\begin{gather}
\frac{\cqs[2]{}}{\cqs[1]{}}=-\bi\us+\frac 1 2
\frac{{\frac{B^{(+)}}{B^{(-)}}}+1}{{{\frac{B^{(+)}}{B^{(-)}}}-1}}\,,
\qquad\qquad
 \frac{\cqf[\overline{\{1\}}]{}} {\cqf[\overline{\{2\}}]{}}=-\bi\us-\frac 1 2
  \frac{{\frac{{\RR}^{(-)}}{{\RR}^{(+)}}}+1}{{{\frac{{\RR}^{(-)}}{{\RR}^{(+)}}}-1}}
=-\bi\us+\frac 1 2
\frac{{\frac{{\RR}^{(+)}}{{\RR}^{(-)}}}+1}{{{\frac{{\RR}^{(+)}}{{\RR}^{(-)}}}-1}}
\,,\\
\where B^{(\pm)}\equiv\prod_{\jrt=1}^{{\Mp}}\sqrt{\frac {\G}{\hx_\jrt^{\mp}}}\left(\frac {1}{x}-\hx_\jrt^{\mp}\right),\qquad
{\RR}^{(\pm)}\equiv\prod_{\jrt=1}^{{\Mp}}\sqrt{\frac
  {\G}{\hx_\jrt^{\mp}}}\left({x}-\hx_\jrt^{\mp}\right),\\
\And
\hx_\jrt^{\mp} \equiv \hx\left(\us^{\rlb[\jrt]}\mp \frac \bi 2\right)\,.
\end{gather}

In view of the expression \eqref{eq:TTHook1}, this
means that up to a gauge transformation, the right band of the 
\(\Tk\)-{\hook} of {\ADF} is given by
\begin{gather}
\forall s\geq 1,\qquad
  \T[1]=\left|
    \begin{array}{cc}
      1 & \gqf[]{[+s]} \\%
      1 & \gqbf[]{[-s]} \\%
    \end{array}
\right|\,,\qquad \where  \gqf[]{}\equiv
\frac{\cqs[2]{}}{\cqs[1]{}}\And \gqbf[]{}\equiv  \frac{\cqf[\overline{\{1\}}]{}} {\cqf[\overline{\{2\}}]{}}
\end{gather}

In this asymptotic solution, we can notice a couple of properties:
\begin{itemize}
\item We see that \(\T\) is real and that
  \(-\gqbf[]{}%
  \) is the
  complex-conjugate of \(\gqf[]{}%
  \) (which is defined on the mirror-sheet). This property
  will still hold at finite size, and it simply comes from the reality
  of the {\Yfs}~: by the same argument as in section
  \ref{sec:reality-condition}, the   reality of the {\Yfs} allows to
  choose a gauge where the {\Tfs} are real, and in turn this allows to
  choose {\cQfs} which are complex-conjugated to each other, up to a
  sign (see for instance (5.12) in
\cite{Gromov:2010km}). %
As the reality of {\Yfs} is not specific to the asymptotic limit, this
property will still hold for arbitrary (finite) \(\Lf\), as we will
discuss in  section~\ref{sec:analyticity-qfs}. %
\item As a function of \(\us\), we see that \(\T[1]\) is analytic on the whole
  complex plane, except on \({\bZc}_{s}\) and \({\bZc}_{-s}\), where we
  use the notation %
\begin{align}
\label{eq:DefbZcshift}
  {\bZc}_\nn\equiv \left\{\xs+\bi \frac {\nn} 2\middle| \xs\in]-\infty,-2\,\G]\cup [2\,\G,\infty[\right\}\,.
\end{align}
\index{Zhukovsky cuts!Zc@\ensuremath{{\bZc}_\nn}}
We see this analyticity 
property %
from the fact that (due to the function \(x(\us)\)), the
functions \(B^{(\pm)}\) and \({\RR}^{(\pm)}\) are analytic on the whole
complex plane except on \(\bZc_0\).

Moreover, when \(\us\) is large, we see that
\begin{gather}
  \T[1]\xrightarrow[\us\to\infty]{} \alpha~ s
\end{gather}
where \(\alpha\) is a constant, independent
of \(\us\) and \(s\), which we can absorb into a gauge transformation if we
wish. This means that 
\begin{gather}
  \Y[1]\xrightarrow[\us\to\infty]{} s^2-1
\end{gather}
\item Another fact which was noticed in \cite{Gromov:2010km}, is that
  the functions \(\frac{\cqs[2]{}}{\cqs[1]{}}\) and
  \(\frac{\cqf[\overline{\{1\}}]{}} {\cqf[\overline{\{2\}}]{}}\) are
  equal up to the replacement \(B^{(\pm)}\leftrightarrow {\RR}^{(\pm)}\) (or
  equivalently \(x\leftrightarrow 1/x\)).  Let us remind here that the
  {\YsE} holds in the mirror sheet, and that therefore the above
  expressions define \(\T[1]\) in the mirror sheet.

Let us   %
introduce 
\begin{gather}
{\hqf[]{}}\equiv
-\bi\us+\frac 1 2
\frac{{\frac{\hat B^{(+)}}{\hat B^{(-)}}}+1}{{{\frac{\hat
        B^{(+)}}{\hat B^{(-)}}}-1}}\,\qquad\qquad 
\And   \hT[1]=\left|
    \begin{array}{cc}
      1 & \hqf[]{[+s]} \\
      1 & \hqf[]{[-s]} \\
    \end{array}
\right|\,,\\
\where \hat B^{(\pm)}\equiv\prod_{\jrt=1}^{{\Mp}}\sqrt{\frac {\G}{\hat
    x_\jrt^{\mp}}}\left(\frac {1}{\hx}-\hx_\jrt^{\mp}\right)\,.
\end{gather}
We can then notice that %
when \(|\Im(\us)|<s/2\), we have
\(\hqf[]{}(\us+\bi\frac s 2)=\gqf[]{}(\us+\bi\frac s 2)\) and
\(\hqf[]{}(\us-\bi\frac s 2)=\gqbf[]{}(\us-\bi\frac s 2)\) as we can see
from the relation \eqref{eq:DefXPhys} between \(x\) and \(\hx\). Hence we
deduce that \(\hT[1]\) coincides with \(\T\) when \(|\Im(\us)|<s/2\). 

This function \(\hT[1]\) defines {\another} sheet for the function \(\T[1]\),
which only differs from the mirror sheet when \(|\Im(\us)|\geq s/2\). We
see that it is analytic on the whole complex plane except 
on \({\hbZc}_{s}\) and \({\hbZc}_{-s}\), where we
  use the notation %
\begin{align}
\label{eq:DefhbZcshift}
  {\hbZc}_\nn\equiv \left\{\xs+\bi \frac {\nn} 2\middle| \xs\in[-2\,\G,2\,\G]\right\}\,.
\end{align}
\index{Zhukovsky cuts!hZc@\ensuremath{{\hbZc}_\nn}}

This new choice of sheet exhibits the symmetry
\begin{gather}
\label{eq:ZfAsLim}
  \hT[1]=-\hT[1][-s]\,,
\end{gather}
which we interpret as a generalization of the \(\Zf\) symmetry of the
classical string theory on \(AdS_5\times {\Ssph}^5\) (see section
\ref{sec:zf-symmetry}). We will see that this symmetry is one of the
fundamental analyticity properties which leads to our FiNLIE.
\end{itemize}

As a last side remark about the right band, let us note that only the
expression of \(\T[1]\) is relevant, and it allows to express the
product \(\T[0]\T[2]=\T[1][][+] \T[1][][-] -\T[1][s+1] \T[1][s-1]\).
On the other hand, the individual expression of \(\T[0]\) versus \(\T[2]\)
is much less relevant than their product, because there
exist\footnote{
Indeed, 
\(a=0\) and \(a=2\), are boundaries of the (a,s)-lattice, which implies
that we necessarily
have \(\T[0]=f_1^{[+s]}f_2^{[-s]}\) and \(\T[2]=f_3^{[+s]}f_4^{[-s]}\)
(for some functions \(f_1\), \(\cdots\), \(f_4\)). Hence there exist a
gauge where \(\T[0]=1\) (obtained as \(
\left(f_1^{[+s]}f_2^{[-s]}\right)^{\FS[a-1]} \T\)) and {\another} gauge where \(\T[2]=1\) (obtained as \(
\left(f_3^{[+s]}f_4^{[-s]}\right)^{\FS[1-a]} \T\)), %
  and these two
  gauges transformations leave \(\T[1]\) invariant.}%
a
gauge where \(\T[0]=1\) and {\another} gauge where \(\T[2]=1\).
 Therefore, when
we discuss the right band of the \(\Tk\)-{\hook}, we will usually
focus on \(\T[1]\), and we may even omit to mention the existence of
\(\T[0]\) and \(\T[2]\).

The structure of the upper band and the left band can also be analyzed
in the same way. The left band is simply equal to the right band (up
to a gauge transformation), while the upper band has quite a 
degenerate structure in the asymptotic limit. This structure can be
read from \cite{Gromov:2010km}, and for instance it shows that there
exists a gauge where \(\T[][1] = G^{[+a]}+\overline{G}^{[-a]}\), where \(G\)
has essentially the same properties as the ratio
\(\frac{\cqs[2]{}}{\cqs[1]{}}\) of the right band: it is analytic on
the whole complex plane except \(\bZc_0\), and is an imaginary
polynomial when \(\us\to\infty\). The degree of this polynomial turns
out to be equal to \(\Mp-1\) where \(\Mp\) is the number of Bethe roots.

Moreover it is possible, as in the case of the {\PCM}, to derive the
Asymptotic Bethe equation \cite{Beisert:2004hm,Arutyunov:2004vx}
(including the crossing equation \cite{Janik:2006dc}) from this
asymptotic solution of the 
{\Ysys}. This was already done in the first paper
\cite{2009PhRvL.103m1601G} conjecturing the {\Ysys} of {\ADF}. This
derivation actually assumed that the {\zerm} (denoted as \(\phi\) in
\cite{2009PhRvL.103m1601G}) is a phase in the ``physical sheet''. This
condition can actually also be viewed as a consequence of the \(\Zf\)
symmetry which we will use in our construction.

\section{Parameterization of the {\Tfu}- and {\qfs}}
\label{sec:parameterization-qfs-3}

In this section, we will introduce the parameterization of the {\Tfu}-
and {\qfs} %
  which we will use to write the FiNLIE at any finite size \(\LF\). %
As in the chapter
\ref{sec:parameterization-qfs-1}, this parameterization will arise
by
understanding %
the analyticity strips of all the Y-, {\Tfu}-
and {\qfs}. Therefore we will start by quickly discussing this
analyticity, in the section \ref{sec:analyticity-strips}.

Then, we will deduce a parameterization of the {\qfs}, in the same
spirit as in section \ref{sec:parameterization-qfs}. The
parameterization of the {\qfs} will involve a polynomial
(corresponding to the \(\us\to\infty\) behavior, which will be extracted
from the asymptotic limit), and a Cauchy integral (given by the
\stapref{sta:Cauchy}).
In order to partially  fix these polynomials, we will even restrict to
states having two symmetric Bethe roots ({\idest} \(\Mp=2\) and
\(\us^{\rlb[1]}=-\us^{\rlb[2]}\)). 
 
\subsection{Analyticity strips}
\label{sec:analyticity-strips}

\subsubsection{Analyticity of the {\Yfs}}
\label{sec:analyticity-yfs}

Like in the example of the {\PCM} in chapter
\ref{cha:ansatzs-de-bethe}, one can find the analyticity strips of the
various {\Yfs} out of the {\TBAE}, or out of the {\YsE}, but if
we want to derive them from the {\YsE} then we need to
know additional constraints such as the asymptotic behavior
\eqref{eq:AsymAdSCFT}. The analyticity properties of the {\Yfs} were
decrypted in the papers
\cite{Cavaglia:2010nm,Balog:2011nm,Balog:2011cx}. In 
particular it was shown that if it is supplemented with these
analyticity properties, then the {\YsE} becomes equivalent to the {\TBAE}.

The most elementary analytic properties of the {\Yfs}, namely their
analyticity strips, can be obtained by the same method which we used in section
\ref{sec:analyt-strips-yfs} for the {\PCM}. In this section, there
were a few nodes 
(the middle nodes) where the analyticity was limited by the asymptotic
behavior, and the other {\Yfs} had increasingly big analyticity strips
when \(|s|\) increased, {\idest} when they were located further and
further %
{away} from the 
middle nodes.

In the case of {\ADF}, the asymptotic behavior
\eqref{eq:AsymAdSCFT} prevents \(\Y[][0]\) from being analytic on a
strip wider that \(\Ast a\), because \(\left(\frac {x^{[-a]}}
  {x^{[+a]}}\right)^\LF\) has four branch points at %
positions \(\pm 2\,\G+\bi\frac a 2\) and \(\pm 2\,\G-\bi\frac a 2\). Hence, on
the mirror sheet, this factor has cuts on \({\bZc}_a\) and
\({\bZc}_{-a}\), %
(where the notation \({\bZc}_a\) was defined in \eqref{eq:DefbZcshift}).

One should note that in the {\PCM}, the factor \(e^{-\LF
  \frac{\sin\frac{\pi a}{{\Np}}}{\sin\frac{\pi }{{\Np}}}
  \cht[\us]}\) was an analytic function for any finite \(\LF\), and only
its \(\LF\to\infty\) limit was not analytic. By contrast the factor \(\left(\frac {x^{[-a]}}
  {x^{[+a]}}\right)^\LF\) has branch points (and square root cuts) even when \(\LF\) is
finite. Therefore the meaning of ``analyticity strip'' is slightly
different from the case of the {\PCM}.
These analyticity strips  are now the biggest strip of the form \(\Ast \nn\), where the
{\Yfs} are meromorphic. Inside these strips, one shows that the {\Yfs}
have a well-defined limit when \(\LF\to\infty\).

In addition to the statement that \(\Y[][0]\) behaves like \(\left(\frac {x^{[-a]}}
  {x^{[+a]}}\right)^\LF\), {\another} analyticity constraint must be used
to fix these analyticity strips. This extra analyticity condition
substitutes to the {\YsE} at \((a,s)=(2,\pm2)\) (where this {\YsE} is
ill-defined, %
  as discussed in section \ref{sec:t-hookfadscft}). It reads %
\Pv{
  \begin{subequations}
\label{eq:FermConstrYadscft}
    \begin{empheq}[left={ \forall \us\in\bZc_{0}, \qquad
        \empheqlbrace}]{gather}
      \Y[1][1]\left(\us+\bi\epsilon\right)\xrightarrow[\epsilon\in
      \bR]{\epsilon \to
        0}1/\Y[2][2]\left(\us-\bi\epsilon\right)\,,\\
      \Y[1][-1]\left(\us+\bi\epsilon\right)\xrightarrow[\epsilon\in
      \bR]{\epsilon \to 0}1/\Y[2][-2]\left(\us-\bi\epsilon\right)\,.
    \end{empheq}
  \end{subequations}
}
This says that the {\Yfs} standing at position  \((a,s)\) such that
\(a=|s|\) ({\idest} \((a,s)\in\{(1,1),(2,2),(1,-1),(2,-2)\}\)) have a Zhukovsky
cut on the real axis , and that the analytic continuation of
\(\Y[1][1]\) (resp \(\Y[1][-1]\)) through this Zhukovsky is \(1/\Y[2][2]\)
(resp \(1/\Y[2][-2]\)). This imposes that the analyticity strip of
\(\Y[1][\pm1]\) and \(\Y[2][\pm2]\) reduces to \([-2\,\G,2\,\G]\).

Then we can use the same arguments as in section
\ref{sec:analyt-strips-yfs} to iteratively deduce the analyticity
strips under the conditions
(\ref{eq:AsymAdSCFT},~\ref{eq:FermConstrYadscft}). This way we obtain
\begin{gather}
\label{eq:AnalYADF}
  \fdisp {\Y\in\Af[\mm]{\left|a-|s|\right|}},\\
{\ie}~
\Y[][0]\in\Af[\mm] a%
,\qquad
\Y[][\pm1]\in \Af[\mm]{a-1}%
,\qquad
\Y[1]\in\Af [\mm] 
 {|s|-1}\,.
\end{gather}

\paragraph{Remark}
\label{sec:remark-2}

One should note that the analyticity of the convolution kernels 
which appear in the {\TBAE} \cite{Gromov:2009bc} is quite complicated,
and involves several Zhukovsky cuts. But 
the explicit expression of these kernels strongly suggest 
that the only possible %
branch points for the function \(\Y\) are
\(\left\{\pm 2 \G + \bi\frac {a+s} 2 +\bi\, \nn\middle | \nn\in\bZ\right\}\).
We will therefore assume that outside its analyticity strip, the
function \(\Y\) is still analytic except on
\begin{equation*}
  \bigcup_{\nn\in\bZ} \bZc_{a+s+2\nn}\,.
\end{equation*}
In particular, we see that with these choices of cuts which correspond to the
mirror sheet, \(\Y(\us)\) is analytic when \(|\Re(\us)|\leq 2\,\G\).

Moreover (as suggested by the form of the convolution kernels
appearing in the {\TBAE}), we expect that the %
branch points
are always of square root type. This assumption means that for any
closed contour \(\gamma\), \(\Y\left(\GmC[{\GmC}]\right)=\Y\), where the notation
\(F\left(\GmC\right)\) denotes the analytic
continuation of a function \(F\) following 
the contour \(\gamma\) from the point \(\us\) to the same point \(\us\).

  These two assumptions are very standard in this subject, and are
  used in all the {\Ysys} literature. Therefore, we will also use them in the
  present manuscript. %

\subsubsection{Analyticity of the {\Tfs}}
\label{sec:analyticity-tfs}

We can then deduce analyticity strips for the {\Tfs}. In the case of
the {\PCM}, we saw that there exists a gauge where the {\Tfs} of the
``right band'' have a larger and larger analyticity strip when \(s\) increases (see \eqref{eq:tinaf-s+1+nn2-}). We also saw that there exists
{\another} gauge where the {\Tfs} of the ``left band'' are analytic
inside a wider and wider strip as \(-s\) increases ({\idest} when we get
further away from the middle nodes %
{located} at \(s=0\)).

For {\ADF}, if we %
deduce analyticity strips for
{\Tfs} out of 
the analyticity strips of the {\Yfs}, then the same general pattern
appears:
there exist gauges where the {\Tfs} of the ``upper band'' (the domain
where \(a\geq |s|\), see \figpref{fig:HkBnd}) have the analyticity strip
\begin{gather}
\label{eq:AdFanalUp}
  \upT\in\Af {a-|s|+1}\,.
\end{gather}
There also exist other gauges where the {\Tfs} of the ``right band''
(the domain 
where \(s\geq a\), see fig \ref{fig:HkBnd}) have the analyticity strip
\begin{gather}
\label{eq:AdFanalRight}
  \riT\in\Af {s-a+1}\,.
\end{gather}
Finally, there is a third type of gauges where the {\Tfs} of the 
``left band''
(the domain 
where \(s\leq -a\), see fig \ref{fig:HkBnd}) have the analyticity strip
\begin{gather}
\label{eq:AdFanalLeft}
  \leT\in\Af {-s-a+1}\,.
\end{gather}
In these properties (\ref{eq:AdFanalUp}-~\ref{eq:AdFanalLeft}), the
symbols \(\upT\), \(\riT\) and \(\leT\) denote {\Tfs} which differ only by
the choice of the gauge.

\subsubsection{Analyticity strips for the {\qfs}}
\label{sec:analyticity-qfs}

One can easily see that the analyticity strips for the {\Tfs} would
most naturally arise from %
having {\qfs} analytic on half planes, as in chapter
\ref{cha:ansatzs-de-bethe}. More precisely, we will use two different
gauges for the upper band and the right band, and only a subset of the
\(2^8\) {\cQfs} of the general {\Wronskian} expression
(\ref{eq:TTHook1}-\ref{eq:TTHook3}) is analytic in 
each of these gauges. The {\Qfs} of this subset will be denoted by the
letter {\qfu}.

\paragraph{{\qfs} for the right band}
\label{sec:right-band-1}

More explicitly we %
will use the following notations for a gauge (to be specified in the
next paragraph) where the right band is
analytic:
\begin{gather}
 \forall s\geq 1,\qquad \riT[1] = 
\rqss[1][+s] \rpss[2][-s] - \rqss[2][+s] \rpss[1][-s]\,,
  \end{gather}
where the arrow under the {\qfs} denotes the fact that we write
expressions in the gauge \(\riT\) which obeys
\eqref{eq:AdFanalRight}. As compared to the {\cQfs} of the 
general {\Wronskian} expression
(\ref{eq:TTHook1}-\ref{eq:TTHook3}), we see that 
\(\rqs[1]\) (resp \(\rqs[2]\)) denotes \(\cqs[1]\) (resp \(\cqs[2]\)) in this
specific gauge, while \(\rps[1]\) (resp \(\rps[2]\)) denotes
\(\cqf[\overline{\{2\}}]{}\) (resp \(\cqf[\overline{\{1\}}]{}\)) in 
{this}
gauge. %
  They are analytic in the following half-planes %
\begin{align}
  \rqf[I]{}&&\textrm{is analytic when} &&\Im({\us})>\frac{|I|-1}2&&(\And I\subset\{1,2\})\\
  \rpf[I]{}&&\textrm{is analytic when} &&\Im({\us})<\frac{1-|I|}2&&(\And I\subset\{1,2\})\,.
\end{align}
These analyticity strips are designed to reproduce exactly the
analyticity strips of the functions \(\riT\) (see
\eqref{eq:AdFanalRight}). As in section
\ref{sec:application-au-champ}, they can be deduced from the
analyticity strips of the {\Tfs} using the Baxter equation
\eqref{eq:DetZeroDefq}.

Moreover, the same arguments as in section
\ref{sec:parameterization-qfs-1} allow to impose the reality of the
functions \(\riT\). That even allows to impose, at the {\level} of the
{\qfs}, the following complex-conjugacy conditions:
\begin{align}
\label{eq:CompConjRightGauge}
  \rps[1]=-\rbqs[1],\qquad\qquad\qquad \rps[2]=\rbqs[2]\,.
\end{align}
  In view of our definition \eqref{eq:DefBarCompConj} of the
complex-conjugate of a function, %
these relations mean for instance that for arbitrary \(\us\in\bC\), \(\rps[1](\us)\)
is the complex-conjugate of \(\rqs[1](\overline{\us})\) (see the
definition \eqref{eq:DefBarCompConj}).
For {\Tfs}, it gives
\begin{gather}
\label{eq:riTexpr}
 \forall s\geq 1,\qquad
  \riT[1]=\rqss[1][+s] \rbqs[2][{[-s]}]+
\rqss[2][+s] \rbqs[1][{[-s]}]\,.
\end{gather}

Moreover, as mentioned in section \ref{sec:structure-right-band}, the
expression of \(\riT[0]\) is not very relevant and can be arbitrarily changed by
a gauge transformation which leaves \(\riT[1]\) invariant. Therefore, we
can choose
\begin{gather}
\label{eq:rit0=}
  \riT[0]=1\,.
\end{gather}
In that case,
\(\riT[2]=\riT[1][][+]\riT[1][][-]-\riT[1][s+1]\riT[1][s-1]\) gives
\begin{gather}
\label{eq:rit2=}
\forall s\geq 2,\qquad  \riT[2]=
  \left(\rqs[1][+]\rqs[2][-]-\rqs[1][-]\rqs[2][+]\right)^{[+s]}
  \left(\rbqs[1][-]\rbqs[2][+]-
\rbqs[1][+]\rbqs[2][-]\right)^{[-s]}\,.
\end{gather}

\paragraph{{\qfs} for the upper band}
\label{sec:upper-band-1}

For the upper band, we can find a gauge where the {\Tfu} have
analyticity strips given by \eqref{eq:AdFanalUp}, and we will see that we
can even choose the %
{\qfs} to be analytic on half-planes. In this gauge, we have
\begin{gather}
\label{eq:TupGaugeFromq}
 \forall a \geq |s|,\qquad 
\upT = \upqf[(2-s)]{[+a]}\wedge \uppf[(2+s)]{[-a]}\,,
  \end{gather}
  where the brace symbol under
the {\qfs} emphasizes the choice of the gauge \(\upT\) which obeys
\eqref{eq:AdFanalUp}.
As compared to the {\cQfs} (or actually the forms built out of the
{\cQfs}) of the 
general {\Wronskian} expression
(\ref{eq:TTHook1f}-\ref{eq:TTHook3f}), we see that 
\(\upqf[(\nn)]{}\) can be viewed as %
\(\cqf[1,2,(\nn)]{}\), whereas
\(\uppf[(\nn)]{}\) corresponds to %
\(\cqf[(\nn),7,8]{}\). But we can equivalently view them as the exterior
forms which allow to rewrite the \stapref{sec:statmtSolHirotPCM} as the
equation \eqref{eq:formHirotaPCM}. They are then defined\footnote{
Here we should notice that the sign of the shifts in the spectral
parameter is changed compared to
(\ref{eq:Defqnform},\ref{eq:Defpnform}). This change is aimed at
reminding that in the original {\Wronskian} expression
(\ref{eq:TTHook1}-\ref{eq:TTHook3}) for the whole ``\(\Tk\)-{\hook}''
\(\Tk(2,2|2+2)\), these {\qfs}, associated to the upper band, are
associated to indices with grading \((-1)^{\gr {\jq}}=-1\). This means
that the QQ-relation has a sign (see \eqref{eq:cQcQbb}), which
differs from \eqref{eq:qqrelation}.

This overall sign, which has no deep meaning and can be absorbed into a
gauge transformation, reproduces the sign of \cite{Gromov:2011cx}.
} as
  \begin{gather}
    {\upqf[(1)]{}\equiv \sum_{\iq=1}^4 \upqf[\iq]{} \xi_\iq},\qquad\qquad
    {\uppf[(1)]{}\equiv \sum_{\iq=1}^4 \uppf[\iq]{} \xi_\iq}\,,\qquad\qquad%
    \upqf[(0)]{}\equiv \upqe,\qquad\qquad\uppf[(0)]{}\equiv \uppe
\\
\label{eq:DetUpq}
    {\upqf[({\nn})]{}\equiv 
      \frac{\upqfs[(1)][+{\nn}-1]\wedge \upqfs[(1)][+{\nn}-3]\wedge
        \upqfs[(1)][+{\nn}-5]\wedge \cdots\wedge \upqfs[(1)][-{\nn}+1]}
      {\upqes[-{\nn}+2]\upqes[-{\nn}+4]\cdots \upqes[{\nn}-2]}},\qquad\qquad {\nn}>1
    \,,
  \end{gather}
  \begin{gather}
\label{eq:DetUpp}
    {\uppf[({\nn})]{}\equiv 
      \frac{\uppfs[(1)][+{\nn}-1]\wedge \uppfs[(1)][+{\nn}-3]\wedge
        \uppfs[(1)][+{\nn}-5]\wedge \cdots\wedge \uppfs[(1)][-{\nn}+1]}
      {\uppes[-{\nn}+2]\uppes[-{\nn}+4]\cdots \uppes[{\nn}-2]}},\qquad\qquad {\nn}>1\,,
  \end{gather}
and their coordinates obey the QQ-relations (see chapter
\ref{cha:ansatzs-de-bethe}, and 
\cite{0022-3719-16-34-009,1992PhRvB..4614624B,Tsuboi:1998ne,Bazhanov:2001xm,2000JPhA...33.8267P,Dorey:2006an,2007JSMTE..01....5B,2008NuPhB.805..451B,Kazakov:2007fy,Zabrodin:2007rq,2003CzJPh..53.1041G,Gromov:2007ky,Gromov:2010km}
) \index{Q-functions@{\Qfs}!QQ-relations}
\begin{gather}
\label{eq:upqqrelation}
  \upqf[\cdots,\bjq,\bkq]{}
  \upqf[\cdots]{} =
  \upqf[\cdots,\bjq]{+}
  \upqf[\cdots,\bkq]{-} -
  \upqf[\cdots,\bjq]{-}
  \upqf[\cdots,\bkq]{+}\,,\\
\label{eq:uppprelation}
  \uppf[\cdots,\bjq,\bkq]{}
  \uppf[\cdots]{} =
  \uppf[\cdots,\bjq]{+}
  \uppf[\cdots,\bkq]{-} -
  \uppf[\cdots,\bjq]{-}
  \uppf[\cdots,\bkq]{+}\,,
\end{gather}
where ``\(_{\cdots}\)'' stands for an arbitrary set of indices.

Then, if we want the expression \eqref{eq:TupGaugeFromq}
 to reproduce the analyticity strips
\eqref{eq:AdFanalUp}, it is natural to guess that
\begin{align}
\label{eq:UpQanal}
  \upqf[(\nn)]{}&&\textrm{is analytic when} &&\Im({\us})>
-\frac 1 2 + \left|\frac {\nn-2} 2\right|\,,\\
  \uppf[(\nn)]{}&&\textrm{is analytic when} &&\Im({\us})<
\frac 1 2 - \left|\frac {\nn-2} 2\right|\,.
\label{eq:UpPanal}
\end{align}
Indeed, we
see that if (\ref{eq:UpQanal}-\ref{eq:UpPanal}) hold, then
\(\upqf[(s-2)]{[+a]}\) (resp \(\uppf[(2-s)]{[-a]}\)) is analytic when
\(\Im(\us)> 
-\frac {1+a-|s|} 2\) (resp \(\Im(\us)< 
\frac {1+a-|s|} 2\)), and hence \(\upT = \upqf[(s-2)]{[+a]}\wedge
\uppf[(2-s)]{[-a]} \in \Af {a-|s|+1}\).

The conditions (\ref{eq:UpQanal}-\ref{eq:UpPanal}) are
therefore a very natural guess %
  and one %
can actually prove that a
choice 
of {\qfs} obeying (\ref{eq:UpQanal}-\ref{eq:UpPanal}) does exist (see
appendix D.6 in \cite{Gromov:2011cx}).

Moreover, the same arguments as in section
\ref{sec:parameterization-qfs-1} allow to impose the reality of the
functions \(\upT\). That even allows to impose, at the {\level} of the
{\qfs}, the following complex-conjugacy conditions:
\begin{gather}
\label{eq:upp=barq}
\uppf[\ove]{}=%
{\upbqf[\emptyset]{}},%
\qquad
\uppf[234]{}=%
{\upbqf[2]{}},%
\qquad \uppf[134]{}=%
-{\upbqf[1]{}},%
\qquad
\uppf[124]{}=%
{\upbqf[4]{}},%
\qquad
\uppf[123]{}=%
-{\upbqf[3]{}},\\
\uppf[34]{}={\upbqf[12]{}},\quad
\uppf[23]{}=-{\upbqf[23]{}},\quad
\uppf[14]{}=-{\upbqf[14]{}},\quad
\uppf[24]{}={\upbqf[24]{}},\quad
\uppf[13]{}={\upbqf[13]{}},\quad
\uppf[12]{}={\upbqf[34]{}}.
\label{eq:upp=barq2}
\end{gather}
where the notation \(\overline {f}\) denote the function \(\us\mapsto
\overline{f(\bar\us)}\) (where \(f\) is an arbitrary holomorphic
function).
Let us also remind that in these expressions %
the function \(\upqf[\iq_1,\iq_2,\cdots,\iq_\nn]{}\) (resp
\(\uppf[\iq_1,\iq_2,\cdots,\iq_\nn]{}\))  denotes the
coefficient of \(\xi_{\iq_1} \wedge \xi_{\iq_2} \wedge \cdots \wedge \xi_{\iq_\nn}\) in 
  \(\upqf[({\nn})]{}\) (resp \(\uppf[({\nn})]{}\)).

These expressions (proven in appendix D.8 in \cite{Gromov:2011cx}), are
designed to produce real {\Tfs}: with the %
  relations %
\eqref{eq:upp=barq}, we obtain real expressions for \(\upT[][1]\) and
\(\upT[][2]\):
\begin{align}
\forall &a\geq 2,&  \upT[][2]=&\upqfs[\emptyset][+a]
{\upbqf[\emptyset]{[-a]}}\,, \\
\forall &a\geq 1,&\upT[][1]=&\upqfs[1][+a]{\upbqf[2]{[-a]}} +
\upqfs[2][+a]{\upbqf[1]{[-a]}} +
\upqfs[3][+a]{\upbqf[4]{[-a]}} +
\upqfs[4][+a]{\upbqf[3]{[-a]}}\,.
\label{eq:UpTa1=qbq}
\end{align}
On the other hand, the qq-relations allow to derive\footnote{For
  instance, 
one gets
\(\uppf[14]{}=\frac{\uppf[143]{+}\uppf[142]{-}-\uppf[143]{-}\uppf[142]{+}}{\uppf[1432]{}}=\frac{{\upbqf[1]{+}}{\upbqf[4]{-}}
-{\upbqf[1]{-}}{\upbqf[4]{+}}}{{\upbqf[\emptyset]{}}}=-{\upbqf[14]{}}\) from
the qq-relation (or the determinant expression
  \eqref{eq:DetUpp}). The
same argument allows to derive each expression in \eqref{eq:upp=barq2}.}
\eqref{eq:upp=barq2} from \eqref{eq:upp=barq}, and to get 
\begin{multline}
\forall a\geq 0,\quad \upT[][0]=\upqfs[12][+a]{\upbqf[12]{[-a]}} -
\upqfs[13][+a]{\upbqf[24]{[-a]}} -
\upqfs[14][+a]{\upbqf[23]{[-a]}} \\
-\upqfs[23][+a]{\upbqf[14]{[-a]}} -
\upqfs[24][+a]{\upbqf[13]{[-a]}} +
\upqfs[34][+a]{\upbqf[34]{[-a]}}\,. 
\end{multline}
In the same way, we can write expressions for the {\pfs} with three
 or four indices, by using the qq-relations. That also gives rise to the following
 real expressions for \(\upT[][-1]\) and \(\upT[][-2]\)
 \begin{align}
\forall &a\geq 1,&\upT[][-1]=& \upqf[{123}]{[+a]} {\upbqf[124]{[-a]}} +
\upqf[{124}]{[+a]} {\upbqf[123]{[-a]}} + \upqf[{134}]{[+a]}
{\upbqf[234]{[-a]}}+\upqf[{234}]{[+a]} {\upbqf[134]{[-a]}}
 \end{align}

These reality conditions say that \(\uppf[(\nn)]{}\) is
the complex conjugate of \(\upqf[(4-\nn)]{}\), up to a transformation of
the form (\ref{eq:gpbs=f-cnp-2h_ij}-\ref{eq:gpbe=f-cnpgpe-gqbe=f}),
where the matrix \(\HH\) only has a few nonzero coefficients, which are
equal to \(\pm 1\) (see \cite{Gromov:2011cx}).

Moreover, if we restrict to symmetric states where \(\Y[][s]=\Y[][-s]\)
(like in the {\SL 2} sector), then we know (from
\ref{sta:freedom-choice-qfs-1}) that there exists a gauge
transformation which transforms \(\upT\) into \(\upT[][-s]\).
One can actually show (see appendix D.8. of \cite{Gromov:2011cx}) that 
a transformation of the form
(\ref{eq:gpbs=f-cnp-2h_ij}-\ref{eq:gpbe=f-cnpgpe-gqbe=f}) on the
{\qfs} allows to also impose
\begin{subequations}
  \label{eq:Uqqup}
  \begin{gather}
    \upqf[123]{}=%
    U^2 \upqf[1]{}\,,\qquad\qquad%
    \upqf[124]{}=%
    U^2 \upqf[2]{}\,,\phantom{{%
        \qquad\qquad \upqf[\ove]{}=%
        \left(U^+U^-\right)^2 \upqf[\emptyset]{}\,,}}\\%
    \upqf[134]{}=%
    U^2 \upqf[3]{}\,,\qquad\qquad%
    \upqf[234]{}=%
    U^2 \upqf[4]{}\,,{%
      \qquad\qquad \upqf[\ove]{}=%
      \left(U^+U^-\right)^2 \upqf[\emptyset]{}\,,}\\
    \upT[][-s]=%
    \left(\left(U^{[+a]}\overline{U}^{[-a]}\right)^{\FS[-s]}\right)^2
    \upT\,,
\label{eq:UisGauge}
  \end{gather}
\end{subequations}
  where \(U\) is a function of \(\us\), which is analytic when \(\Im(\us)\geq
0\), and we see that it defines the relation between \(\upT[][-s]\) and
\(\upT\). In this statement, the non-trivial claim is that
the transformation of the form
(\ref{eq:gpbs=f-cnp-2h_ij}-\ref{eq:gpbe=f-cnpgpe-gqbe=f}) used to
ensure the relation \eqref{eq:Uqqup} does not spoil the
complex-conjugacy relations \eqref{eq:upp=barq}.

\paragraph{Gauge freedom}
\label{sec:geuge-freedom}
The requirements above do not fix completely the gauge {\upTft}, and
one can easily see that we still have one degree of gauge freedom for
the right band and two degrees of gauge freedom for the upper
band. These degrees of freedom take the form
\begin{gather}
\label{eq:rqs1r-gan1-rqs1}
  \rqs[1]{}\rightsquigarrow \gan[][1] \rqs[1]{},\qquad\qquad
  \rqs[2]{}\rightsquigarrow \gan[][1] \rqs[2]{},\qquad\qquad
\riT[1]\rightsquigarrow \ga[+s][1]\overline{\gan[][1]}^{[-s]} \riT[1],
 \\
 \upqf[(\nn)]{}\rightsquigarrow \ga[+\nn][2]\ga[-\nn][3]
 \upqf[(\nn)]{},\qquad\qquad \upT\rightsquigarrow 
\ga[+s-2][2]\ga[+2-s][3]
\overline{\gan[][2]}^{[+2-s]}\overline{\gan[][3]}^{[+s-2]}
 \upT.
\label{eq:upqfnnr-ga+nn2ga-nn3}
\end{gather}
where \(\gan[][1](\us)\), \(\gan[][2](\us)\) and \(\gan[][3](\us)\) are analytic when \(\Im(\us)>0\).

\paragraph{Remark}
\label{sec:remark-1}

Like in the section \ref{sec:parameterization-qfs} where we studied
the case of the \(\SU\Np\times
  \SU\Np\) {\PCM}, we see that the {\qfs} are analytic inside half
  planes. This result is a manifestation of the fact that the
  analyticity strips for the {\Yfs} grow with \(|s|\) (resp with \(a\)) in the
  ``right band'' and the ``left band'' (resp the ``upper band'') of
  the \(\Tk\)-{\hook}. %
There actually exist other Y-systems (for instance for the
amplitudes of {\ADF} \cite{Gaiotto:2010fk}) which do not obey these
properties, and 
it would be interesting to see how much the properties above would %
differ
for these Y-systems.

\subsection{Parameterization of the {\qfs}}
\label{sec:parameterization-}

Let us now specify more precisely the gauges \(\upT\) and \(\riT\)
considered above. In addition to the analyticity strips found above, 
we can impose the behavior at \(\us\to\infty\).
Indeed, one can notice that 
when \(\us\to\infty\), 
\begin{gather}
\label{eq:smallways}
  \left(\frac{x^{[-a]}}{x^{[+a]}}\right)^\LF\sim \left(\frac{\G/\us}{\us/\G}\right)^\LF\qquad\qquad
  \when |\Im(\us)|<\frac a 2 \And   |\us|\to\infty \,.
\end{gather}
We see that the factor \(\left(\frac{x^{[-a]}}{x^{[+a]}}\right)^\LF\), which
defines the asymptotic behavior of \(\Y[][0]\) is the only place where
\(\LF\) appears, and the limit \(\LF\to\infty\) makes this factor very
small (because \(\left|\frac{x^{[-a]}}{x^{[+a]}}\right|<1\)). As this
factor is already small in the limit \(|\us|\to\infty\), we see that the
limit \(\us\to\infty\) should be essentially independent on \(\LF\).

Hence we expect that the features of the {\Yfs} should be the same in the
\(\us\to \infty\) limit as in the \(\LF\to \infty\) case. 

Exactly like in the case of the {\PCM}, we can use this argument to
fix the polynomial behavior of the {\qfs}, which dominates at
\(\us\to\infty\), and to which an integral term will be added, exactly
like in \eqref{eq:paramwpq}. The polynomial terms can be extracted
from the asymptotic limit which was briefly discussed in section
\ref{sec:asymptotic-limit-1} (see \cite{Gromov:2010km} for more
details).

\paragraph{{\qfs} in the right band}
\label{sec:qfs-right-band}

Like in section \ref{sec:parameterization-qfs}, the large \(\us\)
behavior %
allows
the following parameterization\footnote{Let us note that, as in the
  case of the {\PCM}, 
fixing
  the large \(\us\) behavior leaves
some gauge freedom. Indeed the asymptotic behavior restricts the
asymptotic behavior of the function \(\gan[][1]\) in
\eqref{eq:rqs1r-gan1-rqs1}. It still leaves enough 
  freedom %
  to fix \(\rqs[1]{}=1\).
}:
\Pv{\begin{empheq}[box=\fbox]{gather}
\label{eq:RightBandParam}
  \rqs[1]{}=1,\qquad\qquad\qquad \rqs[2]{}=-\bi \us + \CK\st
  \rho\equiv -\bi \us + \frac 1
        {2\bi\pi}\int_{\vs\in\bR}\frac{\rho(\vs)}{\vs-\us}\mathrm{d}\vs\,,
      \end{empheq}}
      where \(\rho= \rqs[2]{}+\overline{\rqs[2]{}}\) is a real function on the
real axis. More precisely the parameterization of \(\rqs[2]{}\) should
be understood as
\Pv{    \begin{empheq}[left={ 
-\bi \us + \frac 1
        {2\bi\pi}\int_{\vs\in\bR}\frac{\rho(\vs)}{\vs-\us}\mathrm{d}\vs
        =\empheqlbrace}]{align}
      \rqs[2]{}(\us)& &\If&\Im(\us)>0
\label{eq:RGparamq2sup}
\\[.3cm]
      -\overline{\rqs[2]{}}(\us)& &\If&\Im(\us)<0.
\label{eq:RGparamq2inf}
    \end{empheq}}

\paragraph{%
    Polynomial behavior}%
\label{sec:polynomial-behavior}

One should note that the asymptotic limit (\(\LF\to\infty\)) only
specifies the leading order of \(\rqs[2]{}\) at \(\us\to\infty\). In
principle, we could very well have \(\rqs[2]{}=-\bi \us +\alpha+ \CK\st
  \rho\) where the number \(\alpha\) is a constant term. Then, we
  could impose  \(Re(\alpha)=0\), using
  transformations of the form
  (\ref{eq:gpbs=f-cnp-2h_ij}-\ref{eq:gpbe=f-cnpgpe-gqbe=f}).
  If we restrict to states having
  symmetric Bethe roots, we actually also have the symmetry
  \(\Y(\us)=\Y(-\us)\), and each 
  {\qf} is symmetric in the sense that \(\gqf[I]{}(-\us)=\pm
  \bgqf[I]{}(\us)\) (where the sign \(\pm\) depends on the set of indices
  \(I\)). This imposes %
  \(\alpha=0\), and it explains the %
  parameterization above.

\paragraph{%
    {\Tfs} in the right band%
}

As compared to the expression 
\eqref{eq:riTexpr} of \(\riT[1]\), this parameterization gives
\begin{align}
 \forall s\geq 1,\qquad
      \riT[1]=&\rqss[2][+s] + \rbqs[2][{[-s]}]%
= -\bi \us^{[+s]} +
  \CK^{[+s]}\st \rho +\bi \us^{[-s]} -   \CK^{[-s]}\st \rho
\\
=& s+\left(\CK^{[+s]}-\CK^{[-s]}\right)\st \rho
\end{align}
which holds if \(|\Im(\us)|<s/2\) (due to the condition on \(\Im(\us)\) in
\eqref{eq:RGparamq2sup}). Hence, we obtain
\begin{gather}
\label{eq:DefT1RR}
 \forall s>2\,|\Im(\us)|, \qquad
  \fdisp{  \riT[1]
= s+\CK_s \st \rho},\\
  \where \fdisp{\CK_s\equiv \CK^{[+s]} - \CK^{[-s]}}.
\end{gather}
Moreover, the equations (\ref{eq:rit0=},\ref{eq:rit2=}) give (with
this parameterization)
\begin{align}
\forall&\us,~~\forall s,&  \riT[0]=&1,\\
\forall& s>1+2\,|\Im(\us)|,&
\riT[2]=&
  \left(\rqss[2][+s+1]-\rqss[2][+s-1]\right)%
  \left(\rbqss[2][-s-1]-
\rbqss[2][-s+1]\right)%
\\ &&=&
  \left(
1+\CK^{[+s]}_1 \st \rho\right)
\left(
1+\CK^{[-s]}_1 \st \rho\right)\,.
\end{align}

\paragraph{{\qfs} in the upper band}
\label{sec:qfs-upper-band}

For the upper band, we will have to choose a slightly less explicit
parameterization of {\qfs}, in order to exactly reproduce the
analyticity strips \eqref{eq:AdFanalUp} (or
(\ref{eq:UpQanal}-\ref{eq:UpPanal}) at the {\level} of {\qfs}).

The simplest possible parameterization of the {\qfs} would be to
define the functions \(\upqf[(0)]{}\), \(\upqf[1]{}\), \(\upqf[2]{}\),
\(\upqf[3]{}\) and 
\(\upqf[4]{}\), and then the other {\qfs} ({\idest} the coordinates of the
forms \(\upqf[(\nn)]{}\)) would be computed through the {\Wronskian}
determinant \eqref{eq:DetUpq}. %
  In order %
to reproduce the analyticity domains given in
\eqref{eq:UpQanal}%
, the parameterization of
\(\upqf[(0)]{}\) would be analytic when \(\Im(\us)>1/2\), whereas the parameterization of
 \(\upqf[1]{}\), \(\upqf[2]{}\), \(\upqf[3]{}\) and
\(\upqf[4]{}\) would be analytic when \(\Im(\us)>0\). But then the
{\Wronskian} expression \eqref{eq:DetUpq} would only show that
\(\upqf[(2)]{}\) is analytic when \(\Im(\us)>1/2\), whereas we would
expect it to be analytic as soon as \(\Im(\us)>-1/2\). This phenomenon
is exactly like in section \ref{sec:analyticity-strips}, where the
{\Wronskian} expression of {\Tfs} sometimes had non-analytic coefficients
inside the determinant, but the determinant was nevertheless analytic
due to some cancellations of the various non-analyticities. In the
present case of {\ADF}, the analyticity strips have a much more
physical meaning than for the {\PCM} and it is really crucial to
  produce %
the correct analyticity strips. Therefore we 
have to find
a more subtle parameterization of the {\qfs} %
  than what was
  suggested above. %

Therefore, we will choose to express all the {\qfs} in terms of the
functions \(\upqf[1]{}\), \(\upqf[2]{}\), \(\upqf[12]{}\), \(\upqf[123]{}\)
and \(\upqf[124]{}\). We will show that if 
  the parameterization of these five functions obeys the
  analyticity constraints %
\eqref{eq:UpQanal},
then the other {\qfs} can be expressed using the qq-relations, and
they will be automatically analytic in the correct domain.

  We should first %
find the
large \(\us\) behavior of the functions \(\upqf[1]{}\),
\(\upqf[2]{}\), %
  from the asymptotic limit. %
In %
  this %
\(\LF\to\infty\) 
limit, \(\Y[][0]\) is small and the {\Ysys} splits
into two lattices \(\HK(2,2)\) (exactly like in section
\ref{sec:asymptotic-limit}, where we saw that in  the asymptotic
limit, the lattice \(\St(\Np)\) of the {\PCM} splits into two
sublattices \(\Sh(\Np)\)). One can see \cite{Gromov:2010km} that there
is a choice of {\qfs} where this splitting simply corresponds to 
\begin{gather}
   \upqf[I]{}\xrightarrow[\LF\to\infty]{}0 \qquad\qquad \textrm{if and
   only if}\quad 3\in I \Or 4\in I\,. 
\end{gather}
Then we see that \eqref{eq:UpTa1=qbq} becomes simply
\(\upT[][1]=\upqfs[1][+a]{\upbqf[2]{[-a]}} +
\upqfs[2][+a]{\upbqf[1]{[-a]}} \). Comparing with the
asymptotic limit (which imposes the behavior at large \(\us\)), we
obtain the parameterization
\begin{gather}
  \upqf[1]{}=1,\qquad \qquad    \upqf[2]{}=W,\\
\where   W(\us)=\Pf[0](\us) + \frac 1
        {2\bi\pi}\int_{\vs\in\bR}\frac{\tilde{\rho}_2(\vs)}{\vs-\us}\mathrm{d}\vs\,,\qquad \If\Im(\us)>0\,,
\label{eq:ParamDefW}
\end{gather}
which parameterizes \(\upqf[1]{}\) and \(\upqf[2]{}\) in terms of
a polynomial \(\Pf[0]\) of degree \(\Mp-1\) and 
 a
real function \(\tilde\rho_2\) on the real axis. We note that in order
to fix \(\upqf[1]{}=1\), we used one of the two gauge freedoms of
equation \eqref{eq:upqfnnr-ga+nn2ga-nn3}.

Next we can fix the function \(\upqf[12]{}\). %
We will see in the next sections that in %
  a very %
natural choice of
gauge, \(\T[][0]\)  will have a double zero at each Bethe root
\(\us^{\rlb[\jrt]}\). But the
above discussion ensures that in the asymptotic limit
\(\upT[][0]=\upqfs[12][+a]{\upbqf[12]{[-a]}}\). Therefore, we
can use one degree of gauge freedom to choose 
\begin{gather}
   \upqf[12]{}=\tbaQ
\end{gather}
where \(\tbaQ\) is a polynomial of degree \(\Mp\), which converges, in
the asymptotic limit, to the polynomial
\({\baQ}=\prod_{\jrt=1}^\Mp(\us-\us^{\rlb[\jrt]})\).

Finally, we should parameterize the functions \(\upqf[123]{}\) and
\(\upqf[124]{}\). To this end, we simply 
use the relation \eqref{eq:Uqqup}
to write
\begin{gather}
  \upqf[123]{}=U^2,\qquad\qquad \upqf[124]{}=U^2 W,\\
\where
\label{URhoU}
    U^2 =-\frac 1 {{\hx}^{%
        {\LF + \gamma} %
        -1}} ~ 
\frac 1 {
\vphantom{{\hx}^{%
    {\LF} %
  }}
2 \pi \bi}
 \int_{-\infty}^{+\infty} \frac
    {\rhou(\vs)}{\us-\vs}\mathrm{d}\vs\,,\qquad
\qquad\If \Im(\us)>0\\
\And \rhou(\us)=2\, \Re\left(U^2 \cdot {\hx}^{%
    {\LF + \gamma} %
    -1} \right)\,.
\end{gather}

This function \(U\) defines a gauge transformation between \(\upT[][-s]\)
and \(\upT\) (see \eqref{eq:UisGauge}) and its parameterization %
corresponds to the
asymptotic behavior \(U\sim {\us}^{(-{\LF}-\gamma)/2}\) of \(U\) when \(\us\) is
large, which will be motivated in the next sections.

Finally, our parameterization of the functions \(\upqf[I]{}\) reads
\Pv{\begin{empheq}[box=\fbox]{gather}
  \upqf[1]{}=1,\qquad \qquad    \upqf[2]{}=W,%
  \qquad \qquad
  \upqf[12]{}=\tbaQ,\\%
  \upqf[123]{}=U^2,\qquad \qquad    \upqf[124]{}=U^2~W.
\end{empheq}}
where the functions \(W\), \(\tbaQ\) and \(U\) are parameterized by two
densities \(\tilde{\rho}_2\) and \(\rhou\)(see \eqref{eq:ParamDefW} and
\eqref{URhoU}) and two polynomials \(\Pf[0]\) and \(\tbaQ\).

\paragraph{Other {\qfs} in the upper band}
\label{sec:qfs-upper-band-2}

This defines a basis of five {\qfs}. Let us show that 
all the other {\qfs}
on the Hasse diagram can be expressed in terms of these five functions
by means of the qq-relations:

First, we know that
\begin{gather}
  \upqf[\emptyset]{}=
  \frac{\upqf[1]{+}\upqf[2]{-}-\upqf[1]{-}\upqf[2]{+}}{\upqf[12]{}}=\frac{W^--W^+}{
   \vphantom{{\baQ}^{{\baQ}^{{\baQ}^{\baQ}}}}
\tbaQ}\,,
\end{gather}
where we see that \(\upqf[\emptyset]{}(\us)\) is analytic when
\(\Im(\us)>1/2\). This relation was obtain by choosing
\(\raisebox{.1cm}{\ensuremath{_{ {{\cdots}}}}}=%
      \emptyset%
\), 
\(\bjq=1\), and \(\bkq=2\) in
\eqref{eq:upqqrelation}. On the other hand, if we set 
\(\raisebox{.1cm}{\ensuremath{_{ {\cdots}}}}=1\) and
\(\bjq=2\), \(\bkq=3\), we get
\begin{gather}
    \upqf[1,2,3]{}
  \upqf[1]{} =
  \upqf[1,2]{+}
  \upqf[1,3]{-} -
  \upqf[1,2]{-}
  \upqf[1,3]{+}\,,\\
\hence \left(\frac{\upqf[1,3]{}}{\upqf[1,2]{}}\right)^+
-\left(\frac{\upqf[1,3]{}}{\upqf[1,2]{}}\right)^- = - \frac{     \upqf[1,2,3]{}
  \upqf[1]{}
}{\upqf[1,2]{+}\upqf[1,2]{-}}=-\frac{U^2}{{\left.\tbaQ\right.}^+{\left.\tbaQ\right.}^-}. 
\label{eq:upq13eq}
\end{gather}
This allows to write
\begin{gather}
\label{eq:upq13solved}
  \upqf[1,3]{} =  \tbaQ~~  \sum_{\kk=1}^\infty \left(\frac{U^2}{{\left.\tbaQ\right.}^+{\left.\tbaQ\right.}^-}\right)^{[+2\kk-1]}\,.
\end{gather}
At first sight, the equation \eqref{eq:upq13eq} only tells that %
\(\frac{\upqf[1,3]{}}{\upqf[1,2]{}}  -  \sum_{\kk=1}^\infty
\left(\frac{U^2}{{\left.\tbaQ\right.}^+{\left.\tbaQ\right.}^-}\right)^{[+2\kk-1]}\)
is an \(\bi\)-periodic
function. But we know its behavior at \(|\us|\to\infty\), where
\(\frac{\upqf[1,3]{}}{\upqf[1,2]{}}\) tend to zero,  and so does \(\sum_{\kk=1}^\infty
\left(\frac{U^2}{{\left.\tbaQ\right.}^+{\left.\tbaQ\right.}^-}\right)^{[+2\kk-1]}\).
That %
allows\footnote{%
    Indeed, the Liouville theorems implies that
    any \(\bi\)-periodic function which decreases to zero at infinity
    (and is analytic on the upper half-plane) is
  equal to zero on the whole complex plane.} %
to write the equation \eqref{eq:upq13solved}.
We can repeat the arguments to find \(\upqf[1,4]{}\), \(\upqf[2,3]{}\) and
\(\upqf[2,4]{}\). We obtain
\begin{gather}
  \left(\frac{\upqf[1,4]{}}{\upqf[1,2]{}}\right)^+
-\left(\frac{\upqf[1,4]{}}{\upqf[1,2]{}}\right)^- = - \frac{     \upqf[1,2,4]{}
  \upqf[1]{} }{\upqf[1,2]{+}\upqf[1,2]{-}} ~~\Longrightarrow ~~ 
  \upqf[1,4]{} = \tbaQ ~~  \sum_{\kk=1}^\infty \left(\frac{W~
      U^2}{{\left.\tbaQ\right.}^+{\left.\tbaQ\right.}^-}\right)^{[+2\kk-1]}\,,\\
  \left(\frac{\upqf[2,3]{}}{\upqf[1,2]{}}\right)^+
-\left(\frac{\upqf[2,3]{}}{\upqf[1,2]{}}\right)^- = - \frac{     \upqf[1,2,3]{}
  \upqf[2]{} }{\upqf[1,2]{+}\upqf[1,2]{-}} ~~\Longrightarrow ~~ 
  \upqf[2,3]{} = \tbaQ ~~  \sum_{\kk=1}^\infty \left(\frac{W~U^2}{{\left.\tbaQ\right.}^+{\left.\tbaQ\right.}^-}\right)^{[+2\kk-1]}\,,\\
  \left(\frac{\upqf[2,4]{}}{\upqf[1,2]{}}\right)^+
-\left(\frac{\upqf[2,4]{}}{\upqf[1,2]{}}\right)^- = - \frac{     \upqf[1,2,4]{}
  \upqf[2]{} }{\upqf[1,2]{+}\upqf[1,2]{-}} ~~\Longrightarrow ~~ 
  \upqf[2,4]{} = \tbaQ ~~  \sum_{\kk=1}^\infty \left(\frac{W^2U^2}{{\left.\tbaQ\right.}^+{\left.\tbaQ\right.}^-}\right)^{[+2\kk-1]}\,.
\end{gather}
From these expression, we can notice that \(\upqf[2,3]{}=\upqf[1,4]{}\),
and that all these {\qfs} with two indices are analytic functions of
\(\us\) when \(\Im(\us)>-1/2\). 

Next we can find \(\upqf[3]{}\) and \(\upqf[4]{}\) by the same
methods. This gives
\begin{gather}
  \left(\frac{\upqf[3]{}}{\upqf[1]{}}\right)^+
-\left(\frac{\upqf[3]{}}{\upqf[1]{}}\right)^- = - \frac{\upqf[1,3]{}
  \upqf[\emptyset]{} }{\upqf[1]{+}\upqf[1]{-}} \quad\Longrightarrow \quad 
  \upqf[3]{} = \sum_{\kk=1}^\infty \left(\upqf[1,3]{}\upqf[\emptyset]{}\right)^{[+2\kk-1]}\,,%
\\
  \left(\frac{\upqf[4]{}}{\upqf[1]{}}\right)^+
-\left(\frac{\upqf[4]{}}{\upqf[1]{}}\right)^- = - \frac{\upqf[1,4]{}
  \upqf[\emptyset]{} }{\upqf[1]{+}\upqf[1]{-}} \quad\Longrightarrow \quad 
  \upqf[4]{} = \sum_{\kk=1}^\infty
  \left(\upqf[1,4]{}\upqf[\emptyset]{}\right)^{[+2\kk-1]}\,. %
\end{gather}
Finally, the qq-relations allow to express \(\upqf[3,4]{}\) as follows
\begin{gather}
  \left(\frac{\upqf[3,4]{}}{\upqf[1,3]{}}\right)^+
-\left(\frac{\upqf[3,4]{}}{\upqf[1,3]{}}\right)^- = - \frac{\upqf[1,3,4]{}
  \upqf[3]{} }{\upqf[1,3]{+}\upqf[1,3]{-}} \quad\Longrightarrow \quad 
  \upqf[3,4]{} = \upqf[1,3]{} \sum_{\kk=1}^\infty \left(\frac{U^2
  \left.\upqf[3]{}\right.^2 }{\upqf[1,3]{+}\upqf[1,3]{-}}\right)\,.
\end{gather}

These expressions express all the {\uqfs} with zero, one or two
indices, and they show that with this parameterization, 
\(\upqf[(0)]{}\) (resp \(\upqf[(1)]{}\), resp \(\upqf[(2)]{}\)) is analytic
when \({\Im({\us})>1/2}\) (resp \(\Im({\us})>0\), resp \(\Im({\us})>-1/2\)).

One can write in the same way the functions with three or four
indices, but the expressions that we obtain simply reproduce the
equation \eqref{eq:Uqqup}. %
{Hence}
\(\upqf[(3)]{}\) (resp \(\upqf[(4)]{}\)) is analytic
when \(\Im({\us})>0\) (resp \(\Im({\us})>1/2\)).

This shows that with this parameterization, the analyticity strips
\eqref{eq:UpQanal} of all the {\uqfs} are directly imposed by our
choice of parameterization. If we remind that the {\upfs} are
essentially the complex-conjugate of the {\uqfs} (see
\eqref{eq:upp=barq}), we see that the analyticity strips \eqref{eq:UpPanal}  for the {\upfs}
are also imposed by this parameterization.

Therefore, this parameterization allows to express all the {\upTfs},
inside their analyticity strips \eqref{eq:AnalYADF}, in terms of 
three densities and two polynomials. This parameterization, defined
above, was obtain at the price of introducing two gauges denotes
\(\upT\) and \(\riT\).  In what follows, these two gauges will be called
``parameterization gauges''. More precisely, \(\riT\) is the parameterization gauge
associated to the ``right band'', whereas  \(\upT\) is the parameterization gauge
associated to the ``upper band band''.

\section{Set of equations}
\label{sec:set-equations-1}

Now that we have parameterized the various {\Tfs} in terms of three
densities and two polynomials, we will write down the equations which
allow to fix these densities (and these polynomials). For simplicity
we focus on states in the  {\SL 2} sector, and when we will fix these
polynomials, we will even restrict to states having two symmetric
Bethe roots ({\idest} \(\Mp=2\) and \(\us^{\rlb[1]}=-\us^{\rlb[2]}\)).

First, we will impose some constraints motivated by the symmetries of
the model. These constraints are on the one hand the existence of the
``Physical gauge'', and on the other hand the ``{\Zf}'' symmetry.
These constraints can be motivated from physical considerations, but
they can also be derived if we assume that the {\TBAE} hold.
On the other hand, the {\TBAE} can be derived from our construction,
which means that our constraints are equivalent to the {\YsE}, for
which they provide an alternative formulation, motivated by the
symmetries of the model.

Next we will see how to deduce equations on the densities, in order to
write an iterative algorithm.

\subsection{The ``Physical Gauge''}
\label{sec:physical-gauge-1}

In the previous chapter, we have defined a set of more
physical gauges called ``{\Wronskian} gauges''.
We will now define one such %
gauge, which differs slightly from the parameterization
gauges constructed above. This gauge obeys specific properties
(guessed in the subsection \ref{sec:physical-gauge})
which should make this gauge more physical if a 
physical construction (such as a lattice
regularization) turns out to exist in the case of {\ADF}. The
subsection \ref{sec:exist-phys-gauge} will show that the existence of
such a basis can be derived from the {\TBAE}, whereas the
subsection \ref{sec:relat-param-gaug} will show how express this
``Physical gauge'' in terms of the ``Parameterization gauges'' defined
in the previous sections.

\subsubsection{Properties of the physical gauge}
\label{sec:physical-gauge}

We will denote by a bold letter \(\bTft\) the {\Tfs} is this particular
``physical'' gauge  %
(whereas the slant letters \(\Tft\) will %
denote {\Tfs} in an unspecified gauge). The first natural conditions 
that we impose is that  %
the {\bTfs} obey the {\Wronskian} gauge condition
\eqref{eq:PhyWronGaugeTK} and the reality condition
\begin{align}
\overline{\bT(\us)} = \bT(\overline{\us})\,.
\end{align}

The {\Wronskian} gauge condition ensures that in this gauge, the {\bTfs}
are expressed through {\cQfs}, as in 
(\ref{eq:TTHook1}-~\ref{eq:TTHook3}). Then the condition
\(\frac{\cqe[+]}{\cqe[-]}\frac{\cqf[\ove]{-}}{\cqf[\ove]{+}}=1\)
(where \(\cqe=1\) and \(\cqf[\ove]{}=\T[0][s][{[-s]}]\)) imposes that 
\begin{align}
\label{eq:TphysBoun}
 \bT[0][0][+]&= \bT[0][0][-],&
 \bT[0][s]&= \bT[0][0][{[+s]}]= \bT[0][0][{[-s]}]\,.
\end{align}
This tells that
\( \bT[0][0]\) should be an \(\bi\)-periodic function in the mirror
sheet (indeed the mirror sheet is the sheet where the {\YsE} holds).
This
means that the function
\begin{equation}
\label{DefF}
\CF\equiv\sqrt{ \bT[0][0]}\,,
\end{equation}
which is an
\(\bi\)-periodic function on the mirror sheet, %
could a priori have a %
periodic
structure of
infinitely many
 \(\bZc\)-cuts.

In principle one could do a periodic gauge-transformation 
which sets \(\bT[0][0]\) to one (for instance \(\bT\rightsquigarrow 
\bT / \bT[0][0][{[+a+s]}]\)).
But such a gauge transformation would involve the multiplication by
\(\bT[0][0]\), which %
has a periodic structure of
Zhukovsky cuts (separated by \(\bi\)), so that this gauge transformation spoils
the analyticity of the {\bTfs}. If we do not do any such
transformation, then we see that \(\bT[0][s]\) a priori has a poor
analyticity (due to periodic cuts, it is at most analytic on \(\Ast
1\)).

Therefore we can exclude that in the physical gauge, the {\bTfs} obey
the analyticity constraint \eqref{eq:AdFanalRight} in the right band,
or the analyticity constraint \eqref{eq:AdFanalLeft} in the %
{left}
band. But as this gauge is supposed to have a physical origin, it
should obey some analyticity conditions, and therefore it must obey
the analyticity constraint \eqref{eq:AdFanalUp} in the upper band:
\begin{gather}
\label{eq:PhysTAdfAnal}
  \bT\in\Af {a-|s|+1}\,.  
\end{gather}
In particular, we obtain that \(\bT[0][0]\in\Af {1}\).
Then as we know that \(\CF\) is \(\bi\)-periodic in the mirror sheet, we can
deduce that 
\(\CF\) is analytic on the whole complex plane except on
\(\bigcup_{\nn\in\bZ}\bZc_{2\nn+1}\). %

\begin{figure}
\fbox{\begin{minipage}{.95\textwidth}
  \begin{center}
  \includegraphics{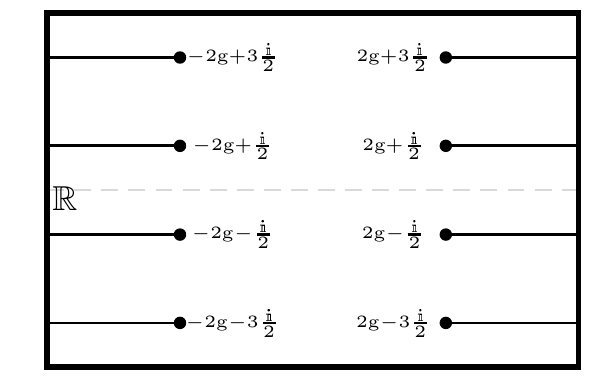}
   \caption{The cut structure of the function \(\CF\).}
   \label{fig:Fcuts}
  \end{center}

The cuts of the function \(\CF\) on the mirror sheet are shown as solid
lines. The periodicity condition \eqref{eq:TphysBoun} ensures that
this picture is periodic, whereas the analyticity strip of \(\bT[0][0]\)
fixes the position of the branch points, which are at position
\(\pm2\,\G+\bi\frac {2\nn+1}2\), where \(\nn\in\bZ\).
\end{minipage}}\end{figure}

The periodic cuts structure of the function \(\CF\) is illustrated in
figure \ref{fig:Fcuts}.

\colorprint{\XFig}{}

Moreover, we will restrict to excited states belonging to the
so-called {\SL 2} sector \cite{Minahan:2010js}, which denotes the states
having only one type of Bethe roots: the momentum carrying roots which
enter in the expression \eqref{eq:Eadscft} of the energy. This sector
is analogous to the {\U 1} sector of the {\PCM} studied in section
\ref{sec:application-au-champ}. For these states the {\Yfs} are
symmetric \(\Y=\Y[][-s]\). Then we expect that a physical construction
of the {\bTfs} would preserve this symmetry, hence the relation
\begin{gather}
\label{eq:bt=bTsymS-SS}
  \bT=\bT[][-s]\,.
\end{gather}

\subsubsection{Existence of the physical gauge}
\label{sec:exist-phys-gauge}

Now that we have written the physical properties that we expect from
this physical gauge, it is actually possible to show that such a gauge
does
exist. Indeed, we know that there exist several gauges
where the analyticity constraint \eqref{eq:AdFanalUp} is
satisfied. Therefore we will start from one such gauge %
  (for instance
\(\T=\upT\)), %
and  
we will show how
to construct, out of this solution of Hirota
equation, some {\Tfs} which obey the same analyticity properties
and which are additionally real (\(\T(\us)=\overline{\T(\overline
  {\us})}\)) and symmetric (\(\T=\T[][-s]\)).
Finally, we will need to use the TBA equations in order to show that
in this gauge, the {\Wronskian} gauge conditions
\eqref{eq:PhyWronGaugeTK} hold.

Let us then start with {\Tfs} \(\T\), which are %
such that the 
analyticity condition \eqref{eq:AdFanalUp} holds.
As
 the
{\Yfs} obey \(\Y=\Y[][-s]\), we know\footnote{The argument here uses
  the \staref{sta:gaugetransfo} (page
  \pageref{sta:gaugetransfo}). Indeed the ratio of the \(\T\) (resp of
  \(\T[][-s]\)) gives rise to \(\Y\) (resp to \(\Y[][-s]=\Y\)). 
  We deduce that \(\T\) and   \(\T[][-s]\) give rise to the same {\Yfs},
  hence they are equal up to a gauge transformation.
 } that there exists four gauge
functions \(\gan[][1]\),  \(\gan[][2]\),
\(\gan[][3]\) and 
\(\gan[][4]\) such that 
\(\T[a][-s]=\ga[a+s][1] \ga[a-s][2]
        \ga[-a+s][3] \ga[-a-s][4] \T\). From this we deduce that the
 gauge transformation
 \begin{gather*}
   \T\rightsquigarrow %
     {%
\sga[a+s][1] \sga[a-s][2]
        \sga[-a+s][3] \sga[-a-s][4]
}%
\T = \sqrt{\T \T[][-s]}
 \end{gather*}
allows to define a {\Tf} which still has the same analyticity strips,
but which obeys the symmetry condition \(\T=\T[][-s]\). 

The same argument can be used to %
  ensure that %
the {\Tfs} are also real. %
Since \(\Y\) is
real %
({\idest} \(\overline{\Y(\us)} =
\Y(\overline{\us})\)), we know that there exist four gauge functions
such that\footnote{Let us remind that here, \(\overline{\T}\) denotes
  the function
\(\us\mapsto \overline{{\T}(\overline\us)}\). Hence, when we say that \(\T\)
is real, it means \(\T=\overline{\T}\), which means that
if \(\us\) is real, then \(\T(\us)\) is real.
} \(\overline{\T}=\ga[a+s][1] \ga[a-s][2]
        \ga[-a+s][3] \ga[-a-s][4] \T\). Then we can use the gauge
        transformation %
\begin{gather*}
   \T\rightsquigarrow %
   {
\sga[a+s][1] \sga[a-s][2]
        \sga[-a+s][3] \sga[-a-s][4]}\T = \sqrt{\T \overline {\T}}
 \end{gather*}
which preserves the analyticity strips and
the condition \(\T=\T[][-s]\), %
and %
  makes the %
{\Tfs} real.

\paragraph{Continuity relation}
\label{sec:continuity-relation}

At this point we have shown how to obtain a gauge where the {\Tfs} are
analytic in the upper band, and symmetric with respect to complex
conjugacy and to the exchange \(s\leftrightarrow-s\). In order to 
obtain a 
{\Wronskian} gauge conditions
\eqref{eq:PhyWronGaugeTK}, we now need to use %
analyticity
conditions,  %
contained in the {\TBAE}, %
  which should be enforced in addition to
  the %
{\YsE}. In \cite{Cavaglia:2010nm}, a set of analyticity conditions is
written, and it is shown that these analyticity conditions are equivalent to
the {\TBAE} (in the sense that if both these analyticity conditions
and the {\YsE} hold, then we can derive the {\TBAE}, and that conversely the
{\TBAE} imply that both the {\YsE} and these analyticity conditions hold).
One of these analyticity conditions (namely the equation (1.7) in \cite{Cavaglia:2010nm}) reads as follows (in the present
notation):
\begin{gather}
\label{discY11Y22our}
\forall \nn\geq 1,\qquad
        \disc \left((\log
          \Y[1][1]\Y[2][2])^{[+2{\nn}]}\right)=-\sum_{a=1}^{{\nn}}\disc\left(\log(1+\Y[a][0][{[+2{\nn}-a]}])\right)\,\\
\where \disc F\equiv F^{[+0]}-F^{[-0]}\equiv \lim_{\epsilon\to 0} F^{[+\epsilon]}-F^{[-\epsilon]}\,.
\end{gather}
This equation is written at the {\level} of {\Yfs}, %
and at the {\level} of {\Tfs}, it reads
\begin{align}
\label{discY11Y22Tour}
\forall \nn\geq 1,\qquad
        \disc \left((\log
          \Y[1][1]\Y[2][2])^{[+2{\nn}]}\right)=&-\disc \left(\log \left(\prod_{a=1}^{{\nn}}
\frac{\T[][0][{[+2\nn-a+1]}]\T[][0][{[+2\nn-a-1]}]}{
\T[a+1][0][{[+2\nn-a]}]\T[a-1][0][{[+2\nn-a]}]}
\right)\right)\\
=& - \disc \left(\log \left(\frac{\T[1][0][{[+2{\nn}]}]\T[{\nn}][0][{[+{\nn}-1]}]}
{\T[{\nn}+1][0][{[+{\nn}]}]\T[0][0][{[+2{\nn}-1]}]}\right)\right)\,.
\end{align}

 Since we know that 
\(\disc \left(\log
  \left(\T[{\nn}][0][{[+{\nn}-1]}]/\T[{\nn}+1][0][{[+{\nn}]}]\right)\right)=0\),
(due to the analyticity strips \eqref{eq:UpPanal} of the {\Tfs}), we
obtain that the ratio 
\begin{align}
  \frac 1 {\Y[1][1] \Y[2][2]}  \frac{\T[1][0]}{\T[0][0][-]}
\end{align}
has no discontinuity over the cuts \(\bZc_{2\nn}\).
Moreover, we know from the remark at the end of the section
\ref{sec:analyticity-yfs} that the only possible non-analyticity of
this %
ratio are Zhukovsky cuts 
located precisely on \(\bZc_{2\nn}\) (where \(\nn\in\bZ\)). Hence we
obtain that
\begin{gather}
\label{eq:anal-to-get-physical-gauge}
\frac 1 {\Y[1][1] \Y[2][2]}  \frac{\T[1][0]}{\T[0][0][-]}
\quad\textrm{ is analytic when }\Im(\us)>0.
\end{gather}

\paragraph{Gauge transformation constructing the physical gauge}
\label{sec:gauge-transf-phys}

The analyticity condition \eqref{eq:anal-to-get-physical-gauge} allows
to define a function \(f\) such that 
\begin{gather}
f(\us)
\quad\textrm{ is analytic when }\Im(\us)>-1/2,\qquad\And\qquad
\left(\frac {f^-}{f^+}\right)^2 = \frac 1 {\Y[1][1] \Y[2][2]}
\frac{\T[1][0]}{\T[0][0][-]}\,.
\end{gather}
Using this function, we can define the {\bTfs} in the physical gauge as
\begin{gather}
\label{eq:ConstrPhysGaugeADF}
  \bT=f^{[a+s]}f^{[a-s]}\bar f^{[-a+s]}\bar f^{[-a-s]}\T\,.
\end{gather}
With this definition, \(\bT\) is analytic in the upper band (in the
sense of equation \eqref{eq:PhysTAdfAnal}), and it is symmetric with
respect to complex
conjugacy and to the exchange \(s\leftrightarrow-s\). Moreover, the
definition of the function \(f\) which appears in
\eqref{eq:ConstrPhysGaugeADF} is such that 
\begin{gather}
\label{eq:ratConstrPhysGaugeADF}
  \frac{\bT[3][2]\bT[0][1]}{\bT[2][3]\bT[0][0][-]}=\frac{\T[3][2]\T[0][1]}{\T[2][3]\T[0][0][-]}\left(\frac {f^+}{f^-}\right)^2=\frac{\T[3][2]\T[0][1]}{\T[2][3]\T[1][0]}{\Y[1][1] \Y[2][2]}=1\,.
\end{gather}
To conclude, one can use the reality of the {\bTfs}, and write the
complex conjugate of \eqref{eq:ratConstrPhysGaugeADF} to obtain \(
\frac{\bT[3][2]\bT[0][1]}{\bT[2][3]\bT[0][0][+]}=1\). Dividing it by
\eqref{eq:ratConstrPhysGaugeADF} gives
\(\bT[0][0][+]=\bT[0][0][-]\). Together with the relation
\(\bT[0]=\bT[0][-s]\) and with the Hirota equation at \(a=0\), this allows
to deduce that \eqref{eq:TphysBoun} holds in the physical gauge which
we have just constructed. Finally, if we insert this equality into
\eqref{eq:ratConstrPhysGaugeADF}, then we obtain
\(\bT[3][2]=\bT[2][3]\). This equality implies that the gauge condition
\eqref{eq:GaWro} holds (with \(\Kr=\Kr'=\Mr=\Mr'=2\) in the present case
of {\ADF}). This shows that the {\bTfs} of the physical gauge
constructed above do obey the ``{\Wronskian}'' gauge condition.
 
Hence we have explicitly constructed the ``physical gauge'' obeying
the required conditions. Moreover, one can show (see appendix E.1 of
\cite{Gromov:2011cx}) that this gauge is unique if we impose the
behavior of the {\Tfs} when \(\us\to\infty\) and the structure of its
zeroes.

\subsubsection{Relation to the ``parameterization gauges''}
\label{sec:relat-param-gaug}

In the argument used above to construct the physical basis, We can
choose to start from the gauge \(\T=\upT\). This allows to deduce the
following relation between the {\bTfs} and the {\upTfs}:
\begin{gather}
\label{eq:bT=gaugexupT}
  \bT=f^{[a+s]}f^{[a-s]}\bar f^{[-a+s]}\bar
  f^{[-a-s]}\left(U^{[+a]}\overline{U}^{[-a]}\right)^{\FS[-s]}
  \upT\,,\\
\where \qquad\left(\frac {f^-}{f^+}\right)^2 =\bB= \frac 1 {\Y[1][1] \Y[2][2]}
\frac{\upT[1][0]}{\upT[0][0][-]}\,.
\label{eq:DefBandf}
\end{gather}

Let us now find how to express the function \(f\) defined above. We
know from the TBA equations that the function \(\bB\) defined above 
is analytic on the upper-half plane. That allows to write a Cauchy
{\rp} of this function as
\Pv{\begin{gather}
\label{eq:bB=KCstarrhoB} \log \bB = \CK \st \rhob
=\frac 1
        {2\bi\pi}\int_{\vs\in\bR}\frac{\rhob(\vs)}{\vs-\us}\mathrm{d}\vs
,\qquad\qquad \If\Im(\us)>0\,,
\end{gather}
where \(\rhob%
=\mathrm{log}(\bB~\overline{\bB})\) is equal to
  \begin{subequations}
\label{eq:rhob=blabla}
    \begin{empheq}[left={
 \displaystyle
\rhob(\vs)
=\empheqlbrace}%
]{%
gather}
  \log \frac{\upT[1][0][2]}{\upT[0][0][-]\upT[0][0][+]\Y[1][1][2]
    \Y[2][2][2]}%
\label{eq:rhoBinside}
  \qquad%
  \If%
~~
  \us\in]-2\,\G,2\,\G[\\
\label{eq:rhoBoutside}
    \log \frac{\upT[1][0][2]}{\upT[0][0][{[-1+0]}]\upT[0][0][{[+1-0]}]}%
    \qquad%
    \If%
~~    %
\us\in ]-\infty,-2\,\G[\cup 
]2\,\G,\infty[\,.
    \end{empheq}
  \end{subequations}
}
The expression \( \log \bB = \CK \st \rhob\) is obtained from the \stapref{sta:Cauchy} with the
functions \(F=\log \bB\) and \(G=-\log \overline\bB\), using the behavior
of \(\bB\) at \(|\us|\to\infty\) : \(\bB\xrightarrow[\Im(\us)\geq 0]{|\us|\to\infty}1\).

\paragraph{Shifts and cuts structure in \eqref{eq:rhob=blabla}}
\label{sec:relat-param-gaug-1}

Let us elaborate on the shifts in equation \eqref{eq:rhob=blabla}:
first let us note that at position \(\us\in\bR\), the quantity
\(\bB~\overline{\bB}\) is defined by continuity as \(\lim_{\epsilon\to
  0^+}
\left.{\vphantom{\overline{\bB}}\bB}\right.^{[+\epsilon]}{{\overline{\bB}}}^{[-\epsilon]}\),
which we can also denote as
\(\left.{\vphantom{\overline{\bB}}\bB}\right.^{[+0]}{{\overline{\bB}}}^{[-0]}\)
. We
can also see that although the definition \eqref{eq:bB=KCstarrhoB} is
written for \(\Im(\us)>0\), it allows to write \(\bB\) on the real
axis as
follows (see \eqref{eq:princpertinte} where one can set \(F=\log \bB\) and \(G=-\log
\overline\bB\)):
\begin{gather}
  \label{eq:bB=KCstarrhoBpp} \log \bB = \sK \st \rhob + \frac 1 2 \rhob
=\frac 1
        {2\bi\pi} \pint_{\vs\in\bR}\frac{\rhob(\vs)}{\vs-\us}\mathrm{d}\vs
+ \frac 1 2 \rhob
,\qquad\qquad \If\Im(\us)=0\,.%
\end{gather}

Let us now elaborate on the expression \eqref{eq:rhoBinside}, which
expresses \(\bB~\overline{\bB}\) when \linebreak \({\us\in]-2\,\G,2\,\G[}\). This
expression %
arises from
the reality of the functions \(\Y[1][1]\), \(\Y[2][2]\), \(\upT[1][0]\) and
\(\upT[0][0]\) which appear in the definition \eqref{eq:DefBandf} of
\(\bB\). As we said, \(\bB~\overline{\bB}\) actually denotes the limit
\(\left.{\vphantom{\overline{\bB}}\bB}\right.^{[+0]}{{\overline{\bB}}}^{[-0]}\)
of
\(\left.{\vphantom{\overline{\bB}}\bB}\right.^{[+\epsilon]}{{\overline{\bB}}}^{[-\epsilon]}\)
when \(\epsilon\to0\). Therefore, in the expression \(\log \frac{\upT[1][0][2]}{\upT[0][0][-]\upT[0][0][+]\Y[1][1][2]
    \Y[2][2][2]}\), the factor \(\upT[0][0][-]\upT[0][0][+]\) arises from
  the limit of
  \(\upT[0][0][{[-1+\epsilon]}]\upT[0][0][{[+1-\epsilon]}]\). As the function
  \(\upT[0][0]\) is analytic only in \(\Ast 1\) (see \eqref{eq:AdFanalUp}),
  this prescription could be %
  important %
  to
  ensure that the argument of \(\upT[0][0]\) is inside the
  analyticity strip (which means that we can compute it from the
  densities introduced in section \ref{sec:parameterization-}). In the
  present case, as \(\us\in]-2\,\G,2\,\G[\), the function \(\upT[0][0]\)
  (which is defined in the mirror sheet) is regular at \(\us\pm \bi/2\),
  which means that 
  \(\upT[0][0][{[-1+\epsilon]}]\upT[0][0][{[+1-\epsilon]}]\) is regular
  at \(\epsilon=0\), and its limit is simply \(\upT[0][0][+]\upT[0][0][-]\).

To finish with, let us elaborate on the expression \eqref{eq:rhoBoutside}, which
expresses the limit
\(\left.{\vphantom{\overline{\bB}}\bB}\right.^{[+0]}{{\overline{\bB}}}^{[-0]}\)
when %
\(\us\in ]-\infty,-2\,\G]\cup 
[2\,\G,\infty[\). One should note that the functions \(\Y[1][1]\) and
\(\Y[2][2]\) have cuts on the real axis (see
\eqref{eq:FermConstrYadscft}) and as they are real functions, we have 
\begin{gather}
 \overline{ \Y[1][1][{[+0]}] } = \Y[1][1][{[-0]}]=\frac
 1{\Y[2][2][{[+0]}]},\qquad\qquad\qquad \where \us\in\bZc_0\,.
\end{gather}
This implies that \(\frac 1 {\Y[1][1][{[+0]}] \Y[1][1][{[+0]}]} \frac 1
{\Yb[1][1][{[-0]}] \Yb[1][1][{[-0]}]}=1\), which
explains that the expression \eqref{eq:rhoBoutside} does not
contain the functions \(\Y[1][1]\) and \(\Y[2][2]\). Moreover, one should
note that, unlike the equation \eqref{eq:rhoBinside},
we have to use the notation
\(\upT[0][0][{[-1+0]}]\upT[0][0][{[+1-0]}]\), which denotes the limit of
  \(\upT[0][0][{[-1+\epsilon]}]\upT[0][0][{[+1-\epsilon]}]\) when
  \(\epsilon\) is positive and tends to zero.
It was important to specify this prescription here, because the
function \(\upT[0][0]\) has a discontinuity at \(\us\pm\bi/2\) (when
\(\us\in\bZc_0\)).

\paragraph{Equation of the function \(f\).}
\label{sec:equation-function-f}

As we can see from \eqref{eq:DefBandf}, the function \(f\) which appears
in the gauge transformation \eqref{eq:bT=gaugexupT} between the gauge
\({\upTft}\) and the gauge \(\bTft\) has to obey the relation
\begin{gather}
\label{eq:2leftlog-f-}
   2\left({\log f^-}- {\log f^+}\right) =\log \bB = \CK \st \rhob\,.
\end{gather}

This relation is easily solved if we impose that \(f\) decreases to zero
at \(|\us|\to\infty\)  and is analytic in the upper-half plane. It gives
\begin{gather}
\label{eq:f=psi}
 \fdisp{ 2\log f=\Psi^+\st \rhob}\,,
\end{gather}
where \(\Psi\) denotes the convolution kernel
\begin{gather}
\label{eq:psi=}  \Psi(\us)=-\frac{\psi(-iu)}{2\pi}= \frac{\gamma}{2\pi}+ \sum_{{\nn}=0}^\infty\left({\CK}^{[2{\nn}]}-\frac 1{2 \pi ({\nn}+1)}\right)\,,
\end{gather}
where \(\gamma\) denotes Euler's constant
and \(\psi\) denotes the derivative of \(\log \Gamma\). Roughly speaking,
this means that the equation \eqref{eq:2leftlog-f-} is solved by 
\begin{gather*}
   2\log f=\sum_{{\nn}=0}^\infty {\CK}^{[2{\nn}]} \st \rhob\,,
\end{gather*}
up to a normalization (and this normalization should compensate the
fact that the sum diverges) .

In fact, the equation \eqref{eq:2leftlog-f-}
 only fixes \(\log f\) up to an additive constant. This constant is
 irrelevant because it only fixes the normalization of the {\bTfs} (in
 the physical gauge). Therefore we can freely choose a 
normalization
 such that the equation \eqref{eq:f=psi} holds.

\paragraph{{\qfs} for the physical gauge}
\label{sec:qfs-physical-gauge}

Having expressed the gauge transformation between the physical gauge
and the parameterization gauge for the upper band, we can deduce and
expression of the {\bTfs} in the upper band, in terms of {\qfs}. It
reads
\begin{subequations}
\label{eq:BoldT=qqqqqq}
  \begin{align}
    \forall &a\geq 2,
    &\bT[][2]=&\bqfs[\emptyset][+a]{\bqbf[\emptyset]{[-a]}}\,,
    \\
\forall &a\geq 1,&\bT[][1]=&\bqfs[1][+a]{\bqbf[2]{[-a]}} +
\bqfs[2][+a]{\bqbf[1]{[-a]}} +
\bqfs[3][+a]{\bqbf[4]{[-a]}} +
\bqfs[4][+a]{\bqbf[3]{[-a]}}\,,\\
\forall &a\geq 0,&\bT[][0]=&\bqfs[12][+a]{\bqbf[12]{[-a]}} -
\bqfs[13][+a]{\bqbf[24]{[-a]}} -
\bqfs[14][+a]{\bqbf[23]{[-a]}} \\
&&&~~-\bqfs[23][+a]{\bqbf[14]{[-a]}} -
\bqfs[24][+a]{\bqbf[13]{[-a]}} +
\bqfs[34][+a]{\bqbf[34]{[-a]}}\,. \\
\forall &a,s,&\bT[][-s]=&\bT\,.
\end{align}
\end{subequations}

The bold letter \(\bqf[{ }]{}\) denotes the {\qfs} in the physical
gauge. These \(\bqf[{ }]{}\)-functions are related to the {\uqfs} by
the relation (in terms of (\(\nn\))-forms)
\begin{gather}
\label{eq:Defbq}
  \bqf[(\nn)]{}\equiv U^{\FS[2-\nn]}f^{[+s-\nn]}f^{[-s+\nn]}\,,
\end{gather}
where \(U^{\FS[\nn]}\) is defined by \eqref{eq:DefFS}.
The definition \eqref{eq:Defbq} is designed
to reproduce the relation \eqref{eq:bT=gaugexupT}.

In other words, the \(\bqf[{ }]{}\)-functions
are
 defined by means of the
\(\bqf[{ }]{}\)\(\bqf[{ }]{}\)-relation, from the basis
\Pv{\begin{empheq}{gather}
  \bqf[1]{}=U\,f^+f^-,\qquad \qquad    \bqf[2]{}=U\,f^+f^-W,%
  \qquad \qquad
  \bqf[12]{}=f^2\tbaQ,\\%
  \bqf[123]{}=\bqf[1]{},\qquad \qquad    \bqf[124]{}=\bqf[2]{}.
\end{empheq}}

\paragraph{Right band}
\label{sec:right-band}

Up to here, we have written completely distinct gauges for the upper
band and the right band. In particular, the {\bTfs} defined above are
analytic only in the upper band, and we would also be interested in
relating it to the {\riTfs} which we defined for the right band. To
this end, we will introduce an intermediate gauge {\wT}, which is very
similar to the physical gauge \(\bT\), but is analytic in the right
band. Next, we will show how this gauge is related to the
parameterization gauge \(\riT\).

 Let us write the simplest gauge transformation which makes the {\Tfs} analytic in
 the right band: %
 \begin{equation}
\label{eq:DefwT}
\fdisp{ \wT=(-1)^{a(s+1)} \bT(\CF^{[a+s]})^{a-2}}\,.
\end{equation}
The factor \((\CF^{[a+s]})^{a-2}\) is necessary to obtain
\(\wT[0]\in\Af{s+1}\) (it even gives \(\wT[0]=1\)), and it also gives 
\(\wT[2]\in\Af {s-1}\), because \(\wT[2]=\bT[2]=\bT[s][2]\in\Af
{s-1}\) (when \(s\geq 2\)).  Actually, we will see that this gauge
even obeys the 
analyticity condition \(\wT[1]\in\Af {s}\), which shows that it
has the analyticity strips \eqref{eq:AdFanalRight}. The sign
\((-1)^{a(s+1)}\) has no consequence on the analyticity strips, but it
is actually necessary in order to have functions with a simple
asymptotic behavior when \(\us\to \infty\).

This claim that \(\wT\) (defined above) is analytic in the right band
can either be 
viewed as a fundamental hypothesis describing our solution of the
Hirota equation, and out of which we will write several non-trivial
equations 
(and we will eventually see that these equations imply the {\TBAE})
or we can adopt {\another} point of view and %
start from the known features of the
{\Ysys}, namely the {\TBAE}, and %
deduce this
analyticity property. %
Let us sketch the proof that if the {\TBAE} hold, then we obtain
\(\wT[1]\in\Af {|s|}\), and we will see that this proof %
is very similar to the proof of the
existence of the physical gauge. This proof relies on the
discontinuity relation (F.5) in \cite{Cavaglia:2010nm}. One can see
(like in section \ref{sec:exist-phys-gauge}, see also appendix C.2 in
\cite{Gromov:2011cx}) 
that this discontinuity relation %
means that the ratio
\begin{equation}\label{relatingBUC2}
   \bfC\equiv%
   \frac{\Y[1][1]}{\Y[2][2]}\frac{\upT[0][0][-]}{\upT[1][0]}\left(\frac{\upT[2][1]}{\riT[1][2]}\frac{\riT[1][1][-]}{\upT[1][1][-]}\right)^2
\end{equation}
is analytic when \(\Im(\us)>0\).

To understand better this statement, let us introduce the function \(h\)
such that 
\begin{gather}
\label{eq:GaugeHwTriT}
  \wT[1]=h^{[+s]}\bar h ^{[-s]} \riT[1]\,.
\end{gather}
This function exists because \(\riT\) and \(\wT\) are real {\Tfs} which
differ only by a gauge. 
Moreover, due to the definition \(\wT[0]=1=\riT[0]\), we obtain
\begin{gather}
\label{eq:wThupT}
\fdisp{  \wT[]=\left(h^{[+s]}\bar h ^{[-s]}\right)^{\FS[a]} \riT}\,.
\end{gather}
In order to prove that \(\wT\) is analytic in the right band, we will
prove  prove that \(h\) is analytic on the upper-half plane ({\idest}
when \(\Im(\us)>0\)).%
  To this end, %
we
will see that the function \(\bfC\) defined in \eqref{relatingBUC2} can
be rewritten in terms of \(h\). %
  To show this, let us %
write
\(\frac{\Y[1][1]}{\Y[2][2]}\) in terms of the {\bTfs}:
\begin{gather}
 \frac{\Y[1][1]}{\Y[2][2]}=
\frac{\bT[1][0](\bT[1][2])^2}{\bT[0][0][-](\bT[2][1])^2}
= 
\left[\frac{\bT[1][0](\bT[1][1][-])^2}{\bT[0][0][-](\bT[2][1])^2}\right]
\frac{(\bT[1][2])^2}{(\bT[1][1][-])^2},\\%
\where 
\frac{\bT[1][0](\bT[1][1][-])^2}{\bT[0][0][-](\bT[2][1])^2}=
\left(\frac{U }{U ^{[+2]}}
  \frac{f^{+}}{f^{[+3]}}\right)^2\frac{\upT[1][0](\upT[1][1][-])^2}{\upT[0][0][-](\upT[2][1])^2}\,,\qquad%
\And \frac{\bT[1][2]}{\bT[1][1][-]} = - \frac{h^{[+2]}}h
\frac{\riT[1][2]}{\riT[1][1][-]}\\
\hence \fdisp{\bfC
=\frac{\Y[1][1]}{\Y[2][2]}\frac{\upT[0][0][-]}{\upT[1][0]}\left(\frac{\upT[2][1]}{\riT[1][2]}\frac{\riT[1][1][-]}{\upT[1][1][-]}\right)^2
=
\left(\frac{U }{U ^{[+2]}}
     \frac{f^{+} h^{[+2]}}{f^{[+3]} h}\right)^2 } \,.
\label{eq:hence-fdispbfc-=frac}
\end{gather}
Therefore, we see that the analyticity of \(\bfC\) on the upper half
plane proves that \(h^{[+2]}/h\) is analytic on the upper-half
plane, %
  but it is not yet sufficient in order to prove that \(h\) is analytic on the upper-half
plane. Indeed, one can easily find a solution \(h_0\) of the equation \(\left(\frac{U }{U ^{[+2]}}
     \frac{f^{+} h_0^{[+2]}}{f^{[+3]} h_0}\right)^2 = \bfC
\) such that \(h_0\) is analytic on the upper-half plane, but then we
have \(h=h_0\,X\) where \(X\) is an arbitrary \(\bi\)-periodic 
function.

We can notice that the function \(\tilde
h=%
{h_0}{\sqrt{X\,\overline X}}\) also obeys \(\wT[1]=\tilde
h^{[+s]}\overline{\tilde h} \vphantom{\tilde h} ^{[-s]} \riT[1]\), and 
we will actually show that  \(\tilde
h\)
is analytic on the upper-half plane, which will allow to conclude by
renaming the functions as \(\tilde h \rightsquigarrow h\).
To this end we simply have to notice that \(\riT[2]\in \Af{s-1}\)
whereas \(\wT[2]=\bT[2]=\bT[s][2]\in \Af{s-1}\), so that their ratio is 
\begin{equation}
\frac {\wT[2]}{\bT[2]}=  h^{[+s+1]}  h^{[+s-1]}   \bar h^{[-s+1]}  \bar h^{[-s-1]} \in \Af{s-1}\,,
\end{equation}
which gives \(X\,\overline X \in \Af{s-2}\) for arbitrary \(s\). Hence
\(X\,\overline X\) is analytic on the whole complex plane, which allows
to conclude that \(\tilde h\) is analytic on the upper-half-plane. %

  The above argument proves the analyticity of the gauge \(\wT\) in
  the right strip, and in particular, it %
allows to conclude (from the
analyticity property 
\eqref{eq:AdFanalRight} of the {\riTfs} and from their relation
\eqref{eq:GaugeHwTriT} to the {\wTfs}) that
\begin{gather}
  \wT[1]\in\Af s\,.
\end{gather}

\subsection{The \texorpdfstring{$\Zf$}{Z4} symmetry}
\label{sec:zf-symmetry}

Let us now introduce {\another} fundamental analyticity condition on the
{\Yfu}- {\Tfu}- and {\qfs}, which we call the \(\Zf\) symmetry. As we
have already noticed in the asymptotic limit, this symmetry is %
related to a symmetry between \(s\) and \(-s\) in a very specific Riemann
sheet, which we will introduce.

\subsubsection{The ``magic'' sheet and the $\Zf$ symmetry.}
\label{sec:mirror-sheet-zf}

As we have seen already, the {\YsE} holds only on a very specific
Riemann sheet, called the mirror sheet, and where the {\Yfu}- {\Tfu}-
and {\qfs} have cuts at positions \(\bZc_\nn\equiv \left\{\xs+\bi \frac
  {\nn} 2\middle| \xs\in]-\infty,-2\,\G]\cup [2\,\G,\infty[\right\}\). On
the other hand, we have seen in the asymptotic limit (in section
\ref{sec:asymptotic-limit-1}) that there exists {\another} sheet, which
coincides with the mirror sheet inside the analyticity strip, but has
only ``short'' Zhukovsky cuts, at position \({\hbZc}_\nn\equiv
\left\{\xs+\bi \frac {\nn} 2\middle| \xs\in[-2\,\G,2\,\G]\right\}\). 

\paragraph{``magic'' {\Tfs}}
\label{sec:htfs}

We will now explain this \(\Zf\) symmetry in terms of {\Tfs}. This
symmetry will be the finite-size generalization of the relation
\eqref{eq:ZfAsLim}. %
With the present notations for the various gauges, it reads
\begin{gather}
\label{eq:Z4rightandphys}
 \fdisp{ \hrT[][-s]=(-1)^a\hrT}\,,\qquad\qquad\qquad  \fdisp{ \hbT[-a]=(-1)^s\hbT}\,,
\end{gather}
where the {\Tfs} with a ``hat'' symbol denote an analytic continuation in the
variable \(a\) (resp \(s\)) performed in a sheet with ``short'' cuts of
the form \(\hbZc_\nn\). For instance for the right band, it means that 
\begin{subequations}
    \label{eq:DefhriT}
  \begin{align}
    \label{eq:DefhriT0}
    \forall &s\in\bZ, & \hrT[0] = &1\,,\\
    \label{eq:DefhriT1}
    \forall &s\in\bZ, & \hrT[1] = &
    \hrqss[1][+s] \hrbqs[2][{[-s]}]+
    \hrqss[2][+s] \hrbqs[1][{[-s]}]\,,\\
    \label{eq:DefhriT2}
    \forall &s\in\bZ, & \hrT[2] = &
    \left(\hrqs[1][+]\hrqs[2][-]-\hrqs[1][-]\hrqs[2][+]\right)^{[+s]}
    \left(\hrbqs[1][-]\hrbqs[2][+]-
      \hrbqs[1][+]\hrbqs[2][-]\right)^{[-s]}
  \end{align}
\end{subequations}
where \(\hrqs\) denotes the analytic continuation of \(\rqs\) to a sheet
having only ``short'' cuts.
  This definition of \(\hrqs\) means that %
\begin{gather}
  \If\Im(\us)>0,\qquad\Then \hrqs(\us)\equiv \rqs(\us),\\
\And ~~ \hrqs(\us)~~\textrm{is analytic
  when}~~\us\in\bC\setminus\bigcup_{\nn\leq 0} \hbZc_{2 \nn}\\
\whereas \rqs(\us)~~\textrm{is analytic
  when}~~\us\in\bC\setminus\bigcup_{\nn\leq 0} \bZc_{2 \nn}\,.
\end{gather}
We can define \(\hrbqs\) by the same prescription (namely
that it coincides with \(\rbqs\) when \(\Im(\us)<0\) and that it
has only shorts Zhukovsky cuts), and we then notice that
\begin{equation}
\hrbqs = \overline{\hrqs}\,.
\end{equation}
The above definition \eqref{eq:DefhriT} of \(\hrT\) differs from the
expressions (\ref{eq:rit0=},\ref{eq:riTexpr},\ref{eq:rit2=}) of \(\riT\)
(without ``hat'') by two features:
\begin{itemize}
\item The expressions (\ref{eq:rit0=},\ref{eq:riTexpr},\ref{eq:rit2=})
  of \(\riT\) (without ``hat'') 
  were valid only when \(s\geq a\). Indeed, the solution of Hirota
  equation on a \(\Tk\)-{\hook} is given by three different {\Wronskian}
  expressions in the upper band, the right band and the left band (see
  \stapref{sta:Thooksol}). If we use the same {\Wronskian} expression in
  the left band, it means that for \(s\leq a\) the {\Tfs} with a
  ``hat'' do not corresponds to the {\Tfs} of the \(\Tk\)-{\hook}. In
  particular the Hirota equation at \(a=2\) ensures that for all \(s\),
  \(\hrT[3]=0\), which %
    shows that the upper band does not exist
    for the functions \(\hrT\). %

  This definition of \(\hrT\) having the same {\Wronskian} expression when
  \(s\leq a\) %
    as %
  when \(s\geq a\) means that we are doing an analytic
  continuation in the variable \(s\).
\item The {\qfs} \(\rqs\) are replaced by \(\hrqs\). This means that we
  are working in a Riemann sheet having only ``short'' cuts. We see
  that if \(s\geq a\), then \(\hrT\) coincides with \(\riT\) inside the
  analyticity strip \(\Ast {s+1-a}\), and they only differ outside this
  analyticity strip. That is why we say that \(\hrT\) is defined on the
  ``magic sheet'', which is the Riemann sheet which coincides with the
  mirror sheet inside the analyticity strip
but has only ``short'' cuts.
\end{itemize}

These new {\Tfs}, obtained by an analytic continuation in the
variable \(s\) performed in a sheet with ``short'' cuts, will
be called ``magic'' {\Tfs}. They can also be defined for the upper
band, and then the analytic continuation will be performed with
respect to the variable \(a\) (instead of \(s\)) and we obtain:
\begin{subequations}
  \begin{align}
    \forall &a\in\bZ,
    &\hbT[][2]=&\hbqfs[\emptyset][+a]{\hbqbf[\emptyset]{[-a]}}\,,
    \\
    \forall &a\in \bZ,&\hbT[][1]=&\hbqfs[1][+a]{\hbqbf[2]{[-a]}} +
    \hbqfs[2][+a]{\hbqbf[1]{[-a]}} + \hbqfs[3][+a]{\hbqbf[4]{[-a]}} +
    \hbqfs[4][+a]{\hbqbf[3]{[-a]}}\,,\\
    \forall &a\in \bZ,&\hbT[][0]=&\hbqfs[12][+a]{\hbqbf[12]{[-a]}} -
    \hbqfs[13][+a]{\hbqbf[24]{[-a]}} -
    \hbqfs[14][+a]{\hbqbf[23]{[-a]}} \\
    &&&~~-\hbqfs[23][+a]{\hbqbf[14]{[-a]}} -
    \hbqfs[24][+a]{\hbqbf[13]{[-a]}} +
    \hbqfs[34][+a]{\hbqbf[34]{[-a]}}\,. \\
    \forall &a,s,&\hbT[][-s]=&\hbT\,.
  \end{align}
\end{subequations}

As compared to the expression \eqref{eq:BoldT=qqqqqq}, we see that an
analytic continuation with respect to the variable \(a\) is performed,
and we define the functions \(\hbqf[I]{}(\us)\) which coincide with
\(\bqf[I]{}(\us)\) when \(\Im(\us)\geq -1/2+\left|\frac{|I|-2}2\right|\)
(where it is analytic), and which differs from \(\bqf[I]{}(\us)\) by
the fact %
that it has only ``short'' cuts.

\paragraph{\(\Zf\) symmetric gauges}
\label{sec:stateent-zf-symmetry}

With these definitions of
the ``magic'' {\Tfs} (denoted with a ``hat''), there are
several gauges which obey the \(\Zf\) symmetry. This allows to write, 
for the gauges introduced above, 
\begin{align}
\label{eq:ZfRi}
& \forall a\leq 3,~~ \forall s\in\bZ,\qquad\qquad { \hrT[][-s]=(-1)^a\hrT}\,,
\qquad\And~ { \hwT[][-s]=(-1)^a\hwT}\,,
\\ 
\label{eq:ZfUp}
& \forall a\in\bZ,~~ \forall s\in\ninter {-2}{2},\qquad\qquad
{ \hbT[-a]=(-1)^s\hbT}\,.
\end{align}

\subsubsection{Motivation from the strong coupling limit}
\label{sec:motiv-from-strong}

We already saw that in the asymptotic limit, this symmetry is (at
least for the right band) easily seen from the explicit expressions of
the {\riTfs}. There actually exists {\another} limit where this symmetry
gets all its meaning as a symmetry of the string theory on
\(AdS_5\times {\Ssph}^5\). 

This limit is the strong coupling limit, where the coupling \(\G\) is
very large. In the literature, one method to study this limit is
called the ``finite gap'' approach
\cite{Gromov:2009tq,Gromov:2010vb}. Applying this approach to the 
string theory on \(AdS_5\times {\Ssph}^5\), one can show that the
{\Tfs} %
 become characters,  explicitly
written in \cite{Gromov:2010km} (see eqs. (4.12-21) there),  in the
highest weight {\rp}
\(\las \) of \(\PST\):
\begin{equation}\label{eq:TT-Omega}
\T=\mathrm{trace}_{\las}\Omega(\us/\G)\equiv \chas \Omega(\us/\G)\,.
\end{equation}
Here \(\Omega(\us)\in \PST\) is the classical monodromy matrix, which
depends on the ratio \(\us/\G\). For these representations (like for the
characters used in chapter \ref{part:qoperatorsspin}), the characters
obey an \(\us\)-independent Hirota equation, {\idest}
\begin{gather}
\left( \chas  %
  \Omega \right)^2 =
{\chas %
  [a+1]}(\Omega)~
{\chas %
  [a-1]}(\Omega) + {\chas %
  [][s+1]}(\Omega)~
 {\chas %
  [][s+1]}(\Omega)\,.
\end{gather}
By comparison, it is the product
\(
{{\chas %
  }\left(\Omega(\frac{\us+\bi /2}\G)\right)~
{\chas %
  }\left(\Omega(\frac{\us-\bi/2}\G)\right)}\) %
  which would appear in
the {\lhs} of 
 the Hirota equation \eqref{eq:YHirota}. %
But as \(\G\to \infty\), the shift \(\frac {\bi /2}\G\) is
negligible and the expression \eqref{eq:TT-Omega} becomes a solution
of the Hirota equation in the strong coupling limit.

In other words, \(\Omega(\us/\G)\) varies only if \(\us\) varies by amounts of
order \(\G\). By comparison, the shift \(\pm\bi/2\) in the Hirota equation
can be neglected. One should also note 
that in \cite{Gromov:2010km}, this identification of \(\T=\chas \left(\Omega(\us/\G)\right)\)
was made  in the
mirror sheet and the expression \eqref{eq:TT-Omega} should be
considered in the mirror kinematics.
As a function of \(\us\), \(\Omega(\us/\G)\) 
has only one Zhukovsky cut on \(\bZc_0\) with an
essential singularity at the branch points 
\(\us=\pm 2\,\G\). This can be understood as the fact that an infinite number
of \(\bZc\)-cuts located on \(\bZc_\nn\) collide (because the shift \(\bi \nn /2\) can
be neglected when \(\G\to\infty\)) into a single cut \(\bZc_0\).

In this limit, the matrix \(\Omega\) has a physical definition
(as a string's monodromy matrix), and the \(\Zf\) symmetry of the
corresponding  coset sigma model \cite{Bena:2003wd} imposes
a constraint on \(\Omega\) \cite{Beisert:2005bm}, which can be written
in terms %
of
its eigenvalues
\((\mu_1,\mu_2,\cdots,\mu_8)\). This constraint reads
\begin{align}\label{eq:Z4x}
\mu_1(\us)&= 1/\mu_2(\GmC[\us]),& \mu_3(\us)&= 1/\mu_4(\GmC[\us]),&
\mu_5(\us)&= 1/\mu_6(\GmC[\us]),& \mu_7(\us)&= 1/\mu_8(\GmC[\us])\,,
  \end{align}
where we denote by \(F(\GmC[\us])\) the result of the analytic
continuation of a function \(F\) following 
a contour which encircles the branch point \(\us=2\,\G\), but which avoids
the other singularities arising in the finite gap
solution\footnote{
These singularities which we avoid are 
square root cuts of the functions \(\mu_\iq(\us)\), but these cuts  are
absent in the monodromy matrix}.

An explicit expression of the characters \(\chas\) of a matrix, as
functions of its eigenvalues, can be written explicitly (see \cite{Gromov:2010vb}),
and it allows to see that the property (\ref{eq:Z4x}) implies
  the following symmetry
  \begin{subequations}
\label{eq:TaT-a}
    \begin{empheq}[left={\empheqlbrace}]{align}
\label{eq:TaTright}
          \T(\us)&=(-1)^s \T[][-\widehat {s}](\GmC[\us]),    \qquad\qquad \If~
          |s|\geq a\,, \\
\label{eq:TaTleft}
          \T(\us)&=(-1)^a \T[-\widehat{a}](\GmC[\us]),    \qquad\qquad  \If
          ~ a\geq |s|\,,
    \end{empheq}
  \end{subequations}
where the functions \(\T[a][-\widehat {s}](\GmC[\us])\) (resp
\(\T[-\widehat{a}][s](\GmC[\us])\)) denote  the analytic
continuations of the functions \({\Tft}\) with respect to the argument
\(s\)  (resp \(a\)) from the values \(s>a\) (resp \(a>|s|\)).

One can expect that the symmetry \eqref{eq:TaT-a}, will hold even at
finite \(\G\), but if we want to generalize from the strong coupling to
the ``quantum case'' (when \(\G\) is finite), one difficulty is that we
have to identify the contour \(\gamma\). Indeed, at strong coupling, we
saw that several branch point ``collide'' into the position \(\pm 2\,\G\),
whereas in the quantum case, there are distinct branch points at the
position \(\pm 2 \g+\frac \bi 2 \nn\), and the position of the contour
with respect to each branch point has to be specified. From the
study of the asymptotic limit and from our study of the analytic
properties of the {\Tfu} and {\qfs}, we can propose one natural
generalization of the equation \eqref{eq:TaT-a} to the quantum case.

Let us first consider the {\Tfs} in the right band. They can be
expressed  as follows in terms of {\qfs}:
\begin{align}
 \forall s&\geq 1,&\T[1] = &
\gqss[1][+s] \gpss[2][-s] - \gqss[2][+s] \gpss[1][-s]\,,\\
 \forall s&\in\bZ,&\T[1][\widehat{s}] = &
\gqss[1][+s] \gpss[2][-s] - \gqss[2][+s] \gpss[1][-s]\,.
\end{align}
If we restrict to the {\qfs} (as opposed to the {\pfs}), we can see
that in \eqref{eq:TaTright}, the transformation
\(\T(\us)\rightsquigarrow \T[][-\widehat {s}](\us)\) can be rewritten as
\begin{equation*}
  \T(\us)\rightsquigarrow \T[][-\widehat {s}](\us)
  \qquad\Rightarrow\qquad
  \gqs\left(\us+\frac \bi 2 s\right) \rightsquigarrow \gqs\left(\us-\frac \bi 2 s\right).
\end{equation*}
To obtain the transformation of equation \eqref{eq:TaTright},
we should also apply the transformation  %
\(\T[][-\widehat {s}](\us)\rightsquigarrow
\T[][-\widehat {s}](\GmC[\us])\) which reads
\begin{equation*}
\T[][-\widehat {s}](\us)\rightsquigarrow \T[][-\widehat {s}](\GmC)
  \qquad\Rightarrow\qquad
  \gqs\left(\us-\frac \bi 2 s\right) \rightsquigarrow
  \gqs\left(\GmC[\us-\frac \bi 2 s] \right),
\end{equation*}
where \(\GmC[\us-\frac \bi 2 s]\) denotes the analytic continuation from
position \(\us-\frac \bi 2 s\), around the branch point \(\us=2\g\), and then
back to position \(\us-\frac \bi 2 s\). If we perform this continuation
clockwise, as in \colorprint{\figpref{fig:appA}}{\figref{fig:appA}},
then we notice that  
the transformation 
\begin{equation*}
  \T(\us)\rightsquigarrow \T[][-\widehat {s}](\GmC)
  \qquad\Rightarrow\qquad
  \gqs\left(\us+\frac \bi 2 s\right) \rightsquigarrow \gqs\left(\GmC[\us-\frac \bi 2 s]\right),
\end{equation*}
is the
continuation \(\us+\frac \bi 2 s\rightsquigarrow \GmC[\us-\frac \bi 2
s]\), which is nothing but the continuation from position \(\us+\frac
\bi 2 s\) into position \(\us-\frac \bi 2 s\) avoiding a short
Zhukovsky cut \(\hbZc_0\).
 The same observation holds for {\pfs} if the continuation
 \(\GmC[\us]\) is made counterclockwise\footnote{There is no apparent
   contradiction
 in this prescription which uses  opposite directions for the contour
 \(\gamma\)
in the continuation of the {\pfu}- and {\qfs}.
Indeed, we see in \eqref{eq:Z4x} that
\(\Omega\left(\frac{\GmC[{\GmC}]}{\G}\right)=\Omega\left(\frac{\us}{\G}\right)\), which means
that the continuation through the contours \(\gamma\) and \(\gamma^{-1}\)
are identical.}. Therefore (\ref{eq:TaT-a}) can be reformulated as
follows (for the right band): 
\begin{figure}%
  \centering%
  \begin{tikzpicture}[baseline]%
    \draw (-3,-2) rectangle (3,2);%
    \tikzstyle{boule} = [shape=circle,fill=black,scale=.4] %
    \definecolor{dgreen}{rgb}{0,0.5,0}%
    \tikzstyle{Bcross} = [color=dgreen,thick]%
    \draw[Bcross] (-1.5,0) ++ (.2,.2) -- ++ (-.4,-.4);%
    \draw[Bcross] (-1.5,0)  ++ (-.2,.2) -- ++ (.4,-.4);%
    \draw[Bcross] (1.5,0)  ++ (.2,.2) -- ++ (-.4,-.4);%
    \draw[Bcross] (1.5,0)  ++ (-.2,.2) -- ++ (.4,-.4);%
    \node at (-1.5,-.4) {\(-2\G\)};  \node at (1.5,-.4) {\(2\G\)};%
    \node[boule] (up) at (.5,1.) {};%
    \node[above right] at (up.north) {\(\us+\bi \frac{s}2\)};%
    \node[boule] (down) at (.5,-1.) {};%
    \node[below right] at (down.south) {\(\us-\bi \frac{s}2\)};%
    \draw[-stealth,color=red,thick] (up)--(down);%
    \draw[-stealth,color=red,thick] (down.north east) .. controls (1,1.5) and (4,0) ..  (down);%
  \end{tikzpicture} \!\!\!=\qquad \begin{tikzpicture}[baseline]%
    \draw (-3,-2) rectangle (3,2);%
    \tikzstyle{boule} = [shape=circle,fill=black,scale=.4] %
    \definecolor{dgreen}{rgb}{0,0.5,0}%
    \tikzstyle{Bcross} = [color=dgreen,thick]%
    \draw[Bcross] (-1.5,0) ++ (.2,.2) -- ++ (-.4,-.4);%
    \draw[Bcross] (-1.5,0)  ++ (-.2,.2) -- ++ (.4,-.4);%
    \draw[Bcross] (1.5,0)  ++ (.2,.2) -- ++ (-.4,-.4);%
    \draw[Bcross] (1.5,0)  -- (-1.5,0);%
    \draw[Bcross] (1.5,0)  ++ (-.2,.2) -- ++ (.4,-.4);%
    \node at (-1.5,-.4) {\(-2\G\)};  \node at (1.5,-.4) {\(2\G\)};%
    \node[boule] (up) at (.5,1.) {};%
    \node[above right] at (up.north) {\(\us+\bi \frac{s}2\)};%
    \node[boule] (down) at (.5,-1.) {};%
    \node[below right] at (down.south) {\(\us-\bi \frac{s}2\)};%
    \draw[-stealth,color=red,thick] (up) .. controls (3,1) and (3,-1)%
    .. (down);%
  \end{tikzpicture}%
  \caption{\label{fig:appA}Equivalent representations of the analytic
    continuation of the {\qfs} in \eqref{eq:TaT-a}.}%
\end{figure}
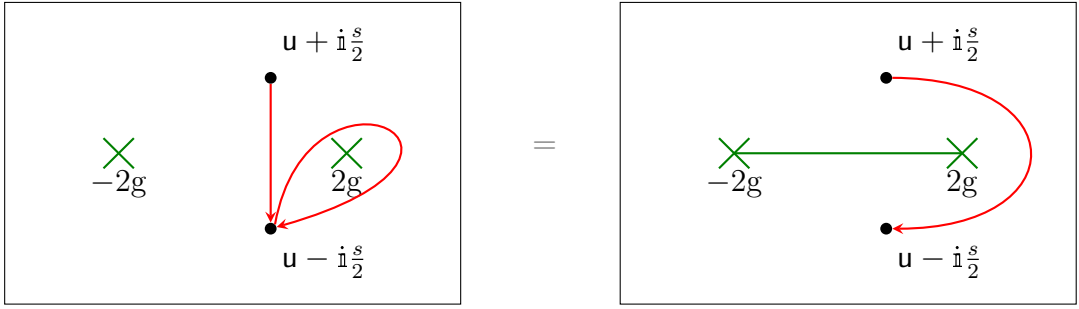 %
} %
\begin{equation}
\label{appA4}
  \T(\us)=(-1)^a \hT[][-s](\us)\,.
\end{equation}
where \(\hT[][-s]\) is defined by the analytic continuation
\(s\rightsquigarrow -s\) performed in sheet which only has ``shorts''
Zhukovsky cuts.

Generically, the {\qfs} have an infinite number of cuts
in the quantum case (when \(\G\) is finite), and then the above argument
does not explain what {\ppath} should be used (which branch points should, or
should not be encircled by the {\ppath} \(\gamma\)). But we will see that in
the right band, in the gauges \(\riTft\) and \(\Tft\), the {\hQfs} have
only one single Zhukovsky cut (on the real axis), and the above
argument is non-ambiguous.
Hence, we obtain that in the right band, the relation \eqref{appA4} is
the most natural generalization of the classical \(\Zf\) symmetry
\eqref{eq:TaTright}.

A similar analysis  for the upper band is more tricky.
However, one can show (see appendix (C.4) in \cite{Gromov:2011cx}),
that the \(\Zf\) symmetry \eqref{eq:ZfUp} for the upper band can be
derived from the \(\Zf\) symmetry \eqref{appA4} in the right band.

\subsubsection{Relation to the {\TBA}}
\label{sec:relation-tba}

Exactly like the statements about the physical gauge in the previous
section, the above statement of the {\Zf} symmetry can be derived
from the {\TBAE}. We presented this proof in \cite{Gromov:2011cx} and
we will not repeat it here in details. This proof relies on the
{\TBAE} \cite{Gromov:2009bc} for the functions \(\Y[1]\) (where \(s\geq
2\)), where we can write the {\Yfs} as \(\Y[1] =
1/\left(\frac{\T[1][][+]\T[1][][-]}{
\T[1][s+1]\T[1][s-1]
}-1\right)\) (see \eqref{eq:opYefrac}), which allows to notice the cancellations
between several {\Tfs} (the idea is the same as in section
\ref{sec:exist-phys-gauge}, but it is more technical). At the end of
the day we obtain \(\hrT[1][0]=0\). As compared to the expression
\eqref{eq:DefhriT1} of \(\hrT[1]\) (where we should note, from
\eqref{eq:CompConjRightGauge} and \eqref{eq:RightBandParam}, that
\(\hrqs[1]{}=1=-\hrps[1]\) and that \(\hrps[2]{}=\hrbqs[1]\)),it gives
\begin{gather}
  \hrT[1][0] =
    \hrqs[1] \hrbqs[2]+
    \hrqs[2] \hrbqs[1] = \hrbqs[2] - \hrqs[2] =0\,.
\end{gather}
This shows that the functions \(\hrbqs[2]\) and \(\hrqs[2]\) are equal,
which implies that 
\begin{gather}
  \hrT[1][s] = \hrqs[2][-s] - \hrqs[2][+s] = - \hrT[1][-s]\,.
\end{gather}
We will see in the next section how to prove that the function \(h\)
(in \eqref{eq:GaugeHwTriT}) gives rise to a function \(\hat h\) which
has only one cut \(\hbZc_0\) on the magic sheet, but this property allows to prove
that the \(\Zf\) symmetry holds in the gauge \(\wT\) as well, which gives the
equation \eqref{eq:ZfRi}. Finally, one can show that the condition
\(\hrT[1][0]=0\) (which is obtained from the {\TBAE}) allows to prove
that \(\hbT[0][1]=0\), which gives the equation \eqref{eq:ZfUp} (see
\cite{Gromov:2011cx} for more details).

\subsubsection{Relation to the analyticity of $\Y[1][1]$ and $\Y[2][2]$}
\label{sec:relat-analyt-y11}

Instead of giving here a detailed proof of this \(\Zf\) symmetry from
the {\TBAE}, let us show on a simple example (for the right band) the
relation between this symmetry and the analyticity conditions
\eqref{eq:FermConstrYadscft} on the function \(\Y[1][1]\) and
\(\Y[2][2]\).

The analyticity conditions \eqref{eq:FermConstrYadscft} are equivalent
the conditions\footnote{We use here the notation 
\(F\args[\pm 0\, \bi]\equiv F^{[\pm0]}\) to denote the limit of
\(F\args[\pm\bi\,\epsilon]\) when \(\epsilon\) tends to zero (but
\(\epsilon >0\)).

More generally, we will use the notation \(F\args[][\nn\pm 0\, \bi]\equiv
F^{[+\nn\pm0]}\) to denote the limit of 
\(F\args[][\nn\pm\bi\,\epsilon]\) when \(\epsilon\) tends to zero (but
\(\epsilon >0\)).
}
\Pv{  \begin{subequations}
\label{eq:MagRatYadscft}
    \begin{empheq}[left={ \forall \us\in ]-\infty,-2\,\G]\cup
[2\,\G,\infty[, \qquad
        \empheqlbrace}]{gather}
  \rmr\args[+ 0\, \bi]=1/\rmr\args[- 0\, \bi],
\label{eq:rmrprop}
\\
\label{eq:smrprop}
  \smr\args[+ 0\, \bi]=1/\smr\args[- 0\, \bi],
\end{empheq}
\end{subequations}
\begin{gather}
  \where \rmr \equiv \frac{1+1/\Y[2][2]}{1+\Y[1][1]}
=\frac{\T[2][2][+]\T[2][2][-]\T[0][1]}{\T[1][1][+]\T[1][1][{-}]\T[2][3]}
\,,\qquad\qquad
\smr \equiv \frac{1+\Y[2][2]}{1+1/\Y[1][1]}
=\frac{\T[2][2][+]\T[2][2][-]\T[1][0]}{\T[1][1][+]\T[1][1][{-}]\T[3][2]}
\,.
\end{gather}
}

When these (gauge-invariant) ratios are written in terms of {\Tfs},
we see that
\(\smr\) only involves the functions \(\T\) where \(a\geq |s|\)
({\idest} the {\Tfs} which lie in the upper band). We also see that
 \(\rmr\) only involves the functions \(\T\) where \(s\geq a\)
({\idest} the {\Tfs} which lie in the right band). This allows to write
the gauge invariant ratio \(\rmr\) in terms of the {\Tfs} in the gauge
\(\riTft\), 
which we parameterized in section
\ref{sec:parameterization-} when \(s\geq a\):
\begin{gather}
  \rmr 
=\frac{\riT[2][2][+]\riT[2][2][-]\riT[0][1]}{\riT[1][1][+]\riT[1][1][{-}]\riT[2][3]}
=\frac{  \left(\rqss[2][+2]-\rqs[2]\right)
  \left(-\rbqs[2]+\rbqss[2][-2]\right) }{  \left(\rqss[2][+2]+\rbqs[2]\right)
  \left(\rqs[2]+\rbqss[2][-2]\right) }\,.
\end{gather}
In order to understand the relation between \(\rmr^{[+0]}\) and
\(\rmr^{[-0]}\), we can notice that due to the analyticity domains given
in (\ref{eq:RGparamq2sup},\ref{eq:RGparamq2inf}), we have
\begin{gather}
\forall \us\in\bR,\qquad\qquad
\rqss[2][+2+0]=\rqss[2][+2-0],\qquad\qquad\And ~ \rbqss[2][-2+0]=\rbqss[2][-2-0]\,.
\end{gather}
Hence we see that the condition \eqref{eq:rmrprop}  reads
\begin{gather}
\forall \us\in \bZc_0\,,\qquad
  \frac{  \left(\rqss[2][+2]-\rqss[2][+0]\right)
  \left(-\rbqss[2][+0]+\rbqss[2][-2]\right) }{
  \left(\rqss[2][+2]+\rbqss[2][+0]\right) 
  \left(\rqss[2][+0]+\rbqss[2][-2]\right) } = 
  \frac{
  \left(\rqss[2][+2]+\rbqss[2][-0]\right) 
  \left(\rqss[2][-0]+\rbqss[2][-2]\right) }
{  \left(\rqss[2][+2]-\rqss[2][-0]\right)
  \left(-\rbqss[2][-0]+\rbqss[2][-2]\right) }
\end{gather}
The simplest way to ensure this property is to have
\begin{subequations}
\begin{align}
\label{eq:MgR1}
\forall&\us\in\bZc_0\,,&  \rqss[2][-0]=&-\rbqss[2][+0]\,,\\%
\label{eq:MgR2}
\forall&\us\in\bZc_0\,,&    \rbqss[2][-0]=&-\rqss[2][+0]\,.
\end{align}
\end{subequations}

We will now show that
the condition \eqref{eq:MgR2}\footnote{The present argument is a
  motivation (not a proof) for the \(\Zf\) symmetry, hence the equation
  is simply guessed \eqref{eq:MgR2}. A more rigorous proof, sketch in the
  previous section, can be found in \cite{Gromov:2011cx}.}
is exactly  the \(\Zf\) property \linebreak\(\hrT[][-s]=(-1)^a\hrT\).
Then we will show how to obtain \eqref{eq:MgR1} from \eqref{eq:MgR2}.
That will show that the \(\Zf\) symmetry implies the
relation \eqref{eq:rmrprop}. The same program can be followed for the
ratio \(\rmr\) (see \cite{Gromov:2011cx}), but we will not repeat it
here.

\paragraph{\(\Zf\) symmetry from \eqref{eq:MgR2}}
\label{sec:zf-symmetry-from}
As we see from the definition
(\ref{eq:RGparamq2sup},\ref{eq:RGparamq2sup}) (and from the 
\stapref{sta:Cauchy}), %
the jump density \(\rho\) which parameterizes the function \(\rqs[2]\)
is equal to \(\rho= \rqss[2][+0]+\rbqss[2][-0]\). Therefore,
the condition \eqref{eq:MgR2} implies that 
\begin{align}
\label{eq:rho=0outside}
\forall&\us\in\bZc_0\,,&  
\rho(\us)=&0\,.
\end{align}
This means that the function \(G\equiv -\bi \us + \CK\st
  \rho\) has no jump on \(\bZc_0\), and is therefore analytic on
  \(\bC\setminus [-2\,\G,2\,\G]\). Hence wee see that \(G\) is a function which
  coincides with \(\rqs[2]\) when \(\Im(\us)>0\) and which has only short
  Zhukovsky cuts. In our notations, this means that
  \(G=\hrqs[2]\). Moreover we also see that \(G\) coincides with
  \(\rbqs[2]\) when \(\Im(\us)<0\), which gives  \(\hrbqs[2]=G=\hrqs[2]\). Hence we
  have
  \begin{align}
    \hT[0]=&1=\hT[0][-s]\\
    \hT[1]=&\hrqss[2][+s] + \hrbqs[2][{[-s]}]=\hrqss[2][+s] +
    \hrqss[2][-s] = - \hT[1][-s].\\
    \hT[2]=&  \left(\hrqss[2][+s+1]-\hrqss[2][+s-1]\right)
  \left(\hrbqss[2][-s-1]-
\hrbqss[2][-s+1]\right)\\ = &
\left(\hrqss[2][+s+1]-\hrqss[2][+s-1]\right)
  \left(\hrqss[2][-s-1]-
\hrqss[2][-s+1]\right)=\hT[1][1][{[+s]}] \hT[1][1][{[-s]}]= \hT[2][-s]%
\,.
\label{eq:=huT2}
  \end{align}
  This shows that the \(\Zf\) symmetry \(\hrT[][-s]=(-1)^a\hrT\) (see
  \eqref{eq:Z4rightandphys}) is implied by the property
  \eqref{eq:MgR2}. One easily shows that the converse is true, and
  that the equation \(\hrT[][-s]=(-1)^a\hrT\), which states the \(\Zf\)
  symmetry of the right band, is equivalent to the condition \eqref{eq:MgR2}.

\paragraph{Relation between \eqref{eq:MgR1}  and \eqref{eq:MgR2} }
\label{sec:relation-beetween-}

It is usual in this study on integrability to assume that the branch
points are quadratic, in the sense that for a 
contour \(\gamma\) which encircles a branch point of a function \(F\), we
have \(F\left(\GmC[{\GmC}]\right)=F(\us)\), where the notation \(F(\GmC)\),
introduced in section \ref{sec:motiv-from-strong}, denotes the
analytic continuation of \(F(\us)\) along the contour \(\gamma\).

\begin{figure}
 \centering
\fbox{\begin{minipage}{.95\textwidth}
   \begin{center}
   \includegraphics{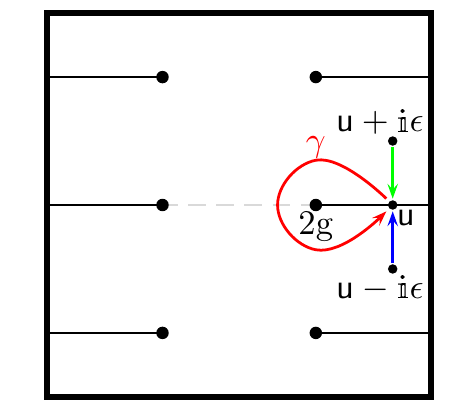}
   \caption{{\ppath} \(\gamma\) around the branch point \(2\,\G\).}
\label{fig:gammacut}
   \end{center}

The {\ppath} \(\gamma\) (in red) is a {\ppath} which starts from point
\(\us\in\bZc_0\), encircles the branch point at position \(2\,\G\), and comes back
to the point \(\us\). We see that if a function \(F\) has a long cut on
\(\bZc_0\), then \(F(\us+\bi\,0)\) denotes the limit of
\(F(\us+\bi\epsilon)\), obtained by continuation of the function \(F\) along
the green arrow. This means that the argument \(\us\) of \(F(\us+\bi 0)\)
sits exactly on the real axis, but that the value \(F(\us+\bi 0)\) is
obtained by continuation from the expression of \(F\) above the
cut. Then we see that after the continuation along the contour
\(\gamma\) (in red), we still obtain the function \(F\) at position
\(\us\in\bR\), but now \(\us\) is approached from {\below}, exactly
like a continuation from 
\(F(\us-\bi\epsilon)\) along the blue arrow which defines
F(\us-\bi\,0).
Hence, 
we have \(F(\GmC[\us+\bi\,0])=F(\us-\bi\,0)\).
\end{minipage}}
\end{figure}

For \(\us>2\,\G\) (for instance), we can define a contour \(\gamma\) which
encircles \(2\,\G\) in the 
counter-clockwise direction (see figure \ref{fig:gammacut}) 
and then for an arbitrary function \(F\) having a cut on \(\bZc_0\), we see that \(F(\GmC[
\us+\bi\,0])=F\args[-\bi \,0]\). Hence we see
 that the
condition \eqref{eq:MgR2} can be rewritten as 
\begin{gather}
  \rqs[2]\args[+\bi\,0] = - \rbqs[2]\left(\GmC[
\us+\bi\,0]\right)\,.
\end{gather}
Assuming that the cuts are of square root type, this allows to deduce
that
\begin{gather}
\label{eq:MgR1gam}
  \rqs[2]\left(\GmC[\us+\bi\,0]\right) = - \rbqs[2]\args[+\bi\,0]\,,
\end{gather}
which
exactly gives \(\rqss[2][-0]=-\rbqss[2][+0]\),
which is the identity \eqref{eq:MgR1}. This shows that how
\eqref{eq:MgR1}  follows from \eqref{eq:MgR2}, under the assumption
that the branch point is of square root type.

\paragraph{conclusion}
\label{sec:conclusion-1}
 
As we said the \(\Zf\) symmetry can be either motivated from physical
symmetries of the string theory on \(AdS_5\times {\Ssph}^5\), or
proven from the {\TBAE}. For simplicity, we did not repeat here the
proof of this symmetry from the {\TBAE} (see \cite{Gromov:2011cx}),
but we illustrated it on a simple example, showing that the
analyticity condition \eqref{eq:rmrprop} was implied by
the \(\Zf\) symmetry of the right band. The same program can be
performed for the upper band, to relate the condition
\eqref{eq:smrprop} to the \(\Zf\) symmetry of the upper band.
As the condition \eqref{eq:MagRatYadscft} is equivalent to the
condition \eqref{eq:FermConstrYadscft}, we will not view the condition
\eqref{eq:FermConstrYadscft} as a fundamental input, but as a
consequence of the \(\Zf\) symmetry.

\subsubsection{Consequences on the parameterization of the {\qfs}.}
\label{sec:cons-param-qfs}

In the previous subsection, we obtained the equation
\eqref{eq:rho=0outside}, which states that
\begin{align}
\label{eq:rho=0outsideb}
\hrbqs[2]=&\hrqs[2]&{\idest}&&\forall&\us\in\bZc_0\,,&  
\rho(\us)=&0\,.
\end{align}
  This means that the density \(\rho\) introduced in section
\ref{sec:parameterization-} actually only has the finite support \([-2\,\G,2\,\G]\). %

For the upper band, similar equations can be obtained. Indeed, we have
\(\hbT[0][1]=0\), and \(\huT[0][1]=0\). As compared to the equation
\eqref{eq:UpTa1=qbq}, this gives
\begin{gather}
\huT[0][1]= 0 = %
{\hubqf[2]{%
  }} +
\huqf[2]{}%
+
\huqf[3]{}%
{\hubqf[4]{%
  }} +
\huqf[4]{}%
{\hubqf[3]{%
  }} \,.
\end{gather}
We can write this expression on the real axis, but then we have to
specify on which side of the cut. For instance if we choose to have a
positive
imaginary part, then this equation reads
\begin{gather}
\huT[0][1][{[+0]}]= 0 = 
{\hubqf[2]{ [+0]
  }} +
\huqf[2]{ [+0]}
+
\huqf[3]{ [+0]}
{\hubqf[4]{ [+0]
  }} +
\huqf[4]{ [+0]}
{\hubqf[3]{ [+0]
  }}\,.
\end{gather}
If we remember that by definition the functions \(\huqf[I]{}\) only have
short Zhukovsky cuts, we get
\begin{align}
  &\forall \us\in\bZc_0,\qquad \hubqf[\iq]{ [+0]  } = \hubqf[\iq]{ [-0]
  } = \upbqf[\iq]{ [-0]
  },\\%
\hence~~&\forall \us\in\bZc_0,\qquad 
{\upbqf[2]{ [-0]
  }} +
\upqf[2]{ [+0]}
+
\upqf[3]{ [+0]}
{\upbqf[4]{ [-0]
  }} +
\upqf[4]{ [+0]}
{\upbqf[3]{ [-0]
  }} =0\,.
\label{eq:UpBdRhZr}
\end{align}
On this form, this equation involves {\qfs} which can be written in
terms of the parameterization written in section
\ref{sec:parameterization-} (because the functions 
\(\upqf[\iq]{}\) have positive shift and the functions 
\(\upbqf[\iq]{}\) have negative shift).

Moreover, we know that both in the \(\LF\to\infty\) limit and
in the \(\us\to\infty\) limit, the two last terms 
in
\eqref{eq:UpBdRhZr} %
tend to zero.
 In the \(\us\to\infty\) limit, this allows to deduce that (in view of
 the parameterization \eqref{eq:ParamDefW} of \({\upbqf[2]{ [-0]
  }} +
\upqf[2]{ [+0]}\)
 \begin{align}
   \overline{\Pf[0]}=- \Pf[0]\,,
 \end{align}
which means that the polynomial \(\Pf[0]\) is imaginary. Moreover in the
case of a state with two symmetric Bethe roots ({\idest} \(\Mp=2\) and
\(\us^{\rlb[1]}=-\us^{\rlb[2]}\)), the polynomial \(\Pf[0]\) has degree
one. Then (up to an irrelevant normalization which we set to one),
\(\Pf[0](\us)\) has to be equal to \(-\bi \us+\alpha\), where
\(\alpha\in\bi\,\bR\) is a constant term (which is a pure imaginary). But
for a symmetric configuration of roots, we know that \(\gqf[2]{}(-\us)=\pm
  \bgqf[2]{}(\us)\), which imposes \(\alpha=0\). Hence, the definition
  \eqref{eq:ParamDefW} of \(W\) becomes
  \begin{align}
\label{eq:WnewPoly}
    W(\us)=-\bi\us + \frac 1
        {2\bi\pi}\int_{\vs\in\bR}\frac{\tilde{\rho}_2(\vs)}{\vs-\us}\mathrm{d}\vs\,,\qquad \If\Im(\us)>0\,.
  \end{align}

The equation \eqref{eq:UpBdRhZr} can also be used to constrain the
density \(\tilde{\rho}_2\). Indeed, the first terms are \({\upbqf[2]{ [-0]
  }} +
\upqf[2]{ [+0]} = \tilde{\rho}_2
\), hence we obtain
\begin{gather}
  \forall \us\in\bZc_0,\qquad 
\tilde{\rho}_2
=-
\upqf[3]{ [+0]}
{\upbqf[4]{ [-0]
  }} -
\upqf[4]{ [+0]}
{\upbqf[3]{ [-0]
  }} \,.
\end{gather}
Hence we can for instance define a function \(\rho_2\equiv \tilde{\rho}_2
+
\upqf[3]{ [+0]}
{\upbqf[4]{ [-0]
  }} +
\upqf[4]{ [+0]}
{\upbqf[3]{ [-0]
  }}\), which is a real
function on \(\bR\), such that
\begin{subequations}
\label{eq:beging-rho_2=0}
  \begin{gather}
    \rho_2=0\qquad\qquad \when \us\in\bZc_0,\\
    W=-\bi \us + \CK\st\left( \rho_2 - \upqf[3]{ [+0]} {\upbqf[4]{
          [-0] }} - \upqf[4]{ [+0]} {\upbqf[3]{ [-0] }}\right)\qquad
    \when
    \Im(\us)>0.%
  \end{gather}
\end{subequations}
Then the expressions \eqref{eq:rho=0outsideb} and
\eqref{eq:beging-rho_2=0} are useful because they give rise to
densities with finite support, and because they allow to encode the
\(\Zf\) symmetry into the parameterization of the {\qfs}.

\colorprint{\FigNumAdsCFT \renewcommand\floatpagefraction{.5}}{}

\subsection{Set of equations and iterative algorithm}
\label{sec:set-equat-iter}

Let us now derive a finite set of equations which allows to write an
iterative algorithm leading to a solution to the {\YsE}
which obeys the analyticity constraints mentioned above. That will allow
to solve the {\Ysys} and to solve the initial spectral problem. As it
was said in section \ref{sec:parameterization-}, we will restrict for
simplicity to the states in the {\SL 2} sector, and  more
specifically to states having two symmetric Bethe roots ({\idest}
\(\Mp=2\) and \(\us^{\rlb[1]}=-\us^{\rlb[2]}\)).

\subsubsection{Equation on $\Y[1][1]$ and $\Y[2][2]$}
\label{sec:equation-y11-y22}

\paragraph{Equation on the product \(\Y[1][1] \Y[2][2]\)}
\label{sec:equation-product-y11}

We already noticed in section \ref{sec:physical-gauge-1} that the
existence of the physical gauge was related to the analyticity of the
function \(\bB(\us)\) (defined by \eqref{eq:DefBandf})  when
\(\Im(\us)>0\). We also noticed that this analyticity condition allows
to write a spectral {\rp} \eqref{eq:bB=KCstarrhoB} of \(\bB\)
in terms of a jump density \(\rhob\), where
\(\rhob=\mathrm{log}(\bB~\overline{\bB})\) has the piecewise expression
\eqref{eq:rhob=blabla}. In this piecewise expression, we could see
that reality conditions could make the functions \(\Y[1][1]\) and
\(\Y[2][2]\)  partially disappear in the expression of \(\rhob\).

Inspired by these observations, we can introduce a new function \(\tbB\)
which has a similar expression, but is specifically chosen in such a way
that \(\Y[1][1]\) and \(\Y[2][2]\) disappear from the expression of the
corresponding density. Let us define this function as
\begin{equation}\label{eq:DeftbB}
\log \tbB(\us)\equiv \frac{\log
  \bB(\us)}{\sqrt{4\G^2-{\us}^2}}\,,\qquad\qquad \where \bB = \frac 1 {\Y[1][1] \Y[2][2]}
\frac{\upT[1][0]}{\upT[0][0][-]}\,.
\end{equation}
Then, if we use the \stapref{sta:Cauchy} with the functions
\(F(\us)=\log \tbB(\us)\) and \(G(\us)=\log \tbBb(\us)\equiv\log
\overline{\tbB(\overline \us)}\), we obtain
\begin{gather}
 \log \tbB(\us)=
\CK \st \rhotB,\qquad\qquad \when \Im(\us)>0
\end{gather}
where \(\rhotB(\us)\equiv\log \tbB(\us) - \log
\overline{\tbB(\overline \us)}\) is equal to
\Pv{
  \begin{subequations}
\label{eq:rhotb=blabla}
    \begin{empheq}[left={
 \displaystyle
\rhotB(\us)%
 =\empheqlbrace}%
]{align}
\label{eq:begin-displ-rhotb}
\frac{1}{\sqrt{4 \G
    ^2-\us^2}}\log&\frac{\upT[0][0][+]}{\upT[0][0][-]}\,,& \If&
\us\in \hbZc_0\\
\frac{\bi}{\sqrt{\us-2\,\G}\sqrt{\us+2\,\G}}\log&\frac
{\upT[1][0][2]}{\upT[0][0][{[-1+0]}]\upT[0][0][{[+1-0]}]}\,,&\If&
\us\in \bZc_0
\label{eq:begin-displ-rhotb-1}
    \end{empheq}
  \end{subequations}
}
where \(\frac{\bi}{\sqrt{\us-2\,\G}\sqrt{\us+2\,\G}}\) appears %
because %
it
coincides on the real axis with the
continuation \(\left(\frac 1 {\sqrt{4\G^2-{\us}^2}}\right)^{[+0]}\) of
\(\frac 1 {\sqrt{4\G^2-{\us}^2}}\) (which is analytic on the
upper-half-plane) to the real axis. %

The expression \eqref{eq:begin-displ-rhotb} is obtained for
\(\us\in[-2\,\G,2\,\G]\) because \(\frac{1}{\sqrt{4 \G
    ^2-\us^2}}\) is real, hence \(\rhotB(\us) = \frac{\log
  \left(\bB(\us)/ \overline{\bB}(\us) \right)}{\sqrt{4 \G
    ^2-\us^2}}\), where 
\(\log
  \left(\bB(\us)/ \overline{\bB}(\us)
  \right)=\log\frac{\upT[0][0][+]}{\upT[0][0][-]}\) (due to the
  reality of the Y- and {\Tfs}), 
which gives \eqref{eq:begin-displ-rhotb}.
On the other hand, the expression \eqref{eq:begin-displ-rhotb-1} is
obtained for \(\us\in ]-\infty,-2\,\G]\cup [2\,\G,\infty[\)
 because \(
\left(\frac 1 {\sqrt{4\G^2-{\us}^2}}\right)^{[+0]} =
\frac{\bi}{\sqrt{\us-2\,\G}\sqrt{\us+2\,\G}}\) is a phase, hence
\(\rhotB(\us) =  \bi  \frac{\log
  \left(\bB(\us) \overline{\bB}(\us)
  \right)}{{\sqrt{\us-2\,\G}\sqrt{\us+2\,\G}}}\), where
\(\log
  \left(\bB(\us) \overline{\bB}(\us)
  \right)=\rhob=\log
  \frac{\upT[1][0][2]}{\upT[0][0][{[-1+0]}]\upT[0][0][{[+1-0]}]}\). This shows that
  the %
  expression
  \eqref{eq:DeftbB} of \(\tbB\) is indeed such that we
  have a density where \(\Y[1][1]\) and \(\Y[2][2]\) disappear completely.

These arguments allow to write the equation 
\begin{align}
\CK \st \rhotB=\log \tbB(\us) = \frac{\log
  \bB(\us)}{\sqrt{4\G^2-{\us}^2}}  =  \frac{1}{\sqrt{4\G^2-{\us}^2}}
\log\left( \frac 1 {\Y[1][1] \Y[2][2]}
\frac{\upT[1][0]}{\upT[0][0][-]} \right)
\end{align}
which gives 
\begin{gather}
\label{eq:Y11Y22Up}
\fdisp{  \log \left(\Y[1][1] \Y[2][2]\right) =   \log
\left(\frac{\upT[1][0]}{\upT[0][0][-]} \right) - 
{\sqrt{4\G^2-{\us}^2}}\,  \CK \st \rhotB}\qquad\qquad \when \Im(\us)>0\,.
\end{gather}

\paragraph{Remark}
\label{sec:remark-3}
One can also plug the %
expression \eqref{eq:rhotb=blabla} of the density \(\rhotB\)
into the equation, to write an equivalent form of
\eqref{eq:Y11Y22Up}. We wrote this expression (see equation (5.24) in
\cite{Gromov:2011cx}), which we do not repeat here to keep the
simplest possible notations. 

Interestingly, the form obtained by this substitution is (after a
shift of integration contour, allowed by understanding the structure
of the zeroes of the {\Tfs}, written in section
\ref{sec:bethe-equation-1}) directly related to the {\TBAE}. It is
one of the elements which allows to prove that the {\TBAE} are implied
by our analyticity conditions and by the resulting FiNLIE.

\paragraph{Equation on the ratio \(\Y[1][1] / \Y[2][2]\)}
\label{sec:equation-ratio-y11}

The same procedure applies to writing an equation on the ratio
\(\Y[1][1] / \Y[2][2]\). To this end, we should consider the function
\(\bfC\) defined in \eqref{eq:DefbfC}~:
\begin{equation}\label{eq:DefbfC}
   \bfC\equiv%
   \frac{\Y[1][1]}{\Y[2][2]}\frac{\upT[0][0][-]}{\upT[1][0]}\left(\frac{\upT[2][1]}{\riT[1][2]}\frac{\riT[1][1][-]}{\upT[1][1][-]}\right)^2
\end{equation}

This ratio is analytic when \(\Im(\us)>0\), and it therefore admits a
Cauchy {\rp} as
\begin{gather}
\label{eq:log-bfc-=}
  \log \bfC = \CK\st \rhoc,\qquad\qquad \when \Im(\us)>0,\\
\where \rhoc \equiv \log \bfC-\log \bfCb = \log
\frac{\upT[0][0][{[-1+0]}]}{\upT[0][0][{[+1-0]}]}\left(\frac{\riT[1][1][{[-1+0]}]}{\upT[1][1][{[-1+0]}]}
\frac{\upT[1][1][{[+1-0]}]}{\riT[1][1][{[+1-0]}]}\right)^2\,.
\label{eq:where-rhoc-equiv}
\end{gather}

By inserting the relation \eqref{eq:log-bfc-=} into the definition
\eqref{eq:DefbfC} of \(\bfC\), we can write 
\begin{gather}
\label{eq:equaforYoY}
\fdisp{\log   \frac{\Y[1][1]}{\Y[2][2]}
= \CK\st \rhoc 
-\log\left(
   \frac{\upT[0][0][-]}{\upT[1][0]}\left(\frac{\upT[2][1]}{\riT[1][2]}\frac{\riT[1][1][-]}{\upT[1][1][-]}\right)^2\right)}\qquad\when
\Im(\us)\in ]0,1[\,.
\end{gather}

\paragraph{Remark}

Like for the previous expression, it is possible to simplify slightly
this expression if we know the structure of the zeroes of the {\Tfs}. 
Indeed %
we can plug the
expression \eqref{eq:where-rhoc-equiv} of \(\rhoc\) into the equation
\eqref{eq:equaforYoY}. Then, we can redistribute the shifts by
moving the integration contours, and this steps requires a precise
knowledge of the zeroes of the {\Tfs} (to know what singularities are
involved when we move the contours). That gives rise to the
expression\footnote{Here \({\baQ}\) denotes the polynomial \(\prod
  (\us-\us^{\rlb[\jrt]})\), %
  where the product runs
  over the Bethe roots \(\us^{\rlb[\jrt]}\). At a finite size
  \(\LF\neq\infty\), these Bethe roots will be defined in the subsection
\ref{sec:bethe-equation-1}}
\begin{gather}
\log   \frac{\Y[1][1]}{\Y[2][2]}
= \log
\frac{\upT[1][0]}{{\baQ}^+{\baQ}^-}\left(\frac{\riT[1][2]}{\upT[2][1]}\right)^2-\sK_1\st\log\frac{\upT[0][0]}{{\baQ}^2}\left(\frac{\riT[1][1]}{\upT[1][1]}\right)^2\,,
\label{eq:log-fracy11y22-=}
\end{gather}
which turns out to be directly related to the
{\TBAE}.

 \subsubsection{Equation on $\CF$}
\label{sec:equation-cf}

The {\Wronskian} gauge condition \eqref{eq:PhyWronGaugeTK}, satisfied by
the {\bTfs}, implies that
\begin{align}
  \Y[1][1]\Y[2][2]=&\frac{\bT[1][0]\bT[2][3]}{\bT[0][1]\bT[3][2]}=
\frac{\bT[1][0]}{\left(\CF^+\right)^2}\,.
\end{align}
If we note that \(\bT[1][0]\) is regular on the real axis, and that
\(\frac{\Y[1][1][{[+0]}]\Y[2][2][{[+0]}]}{\Y[1][1][{[-0]}]\Y[2][2][{[-0]}]}
= \left(\Y[1][1][{[+0]}]\Y[2][2][{[+0]}]\right)^2\) for \(\us\in\bZc_0\) (see
\eqref{eq:FermConstrYadscft}), we can deduce that
\begin{align}
\label{eq:equatCF0}
  \forall
  \us&\in\bZc_0,&\left(\Y[1][1][{[+0]}]\Y[2][2][{[+0]}]\right)^2 = &
\left(\frac{\CF^{[+1-0]}}{\CF^{[+1+0]}}\right)^2\,.
\end{align}

As we know, the function \(\log \CF\) is \(\bi\)-periodic and has cuts on
\(\bigcup_{\nn\in\bZ}\bZc_{2\nn+1}\). The equation \eqref{eq:equatCF0}
tells us that on each of these cuts, \(\log \CF\) jumps by the amount 
\begin{align}
\label{eq:JumpF}
  \log \CF^{[+2\nn+1+0]} - \log \CF^{[+2\nn+1-0]}=&- \log
  \left(\Y[1][1][{[+0]}]\Y[2][2][{[+0]}]\right)\,&\where ~\us&\in\bZc_0\,.
\end{align}
In the {\rhs}, the expression of \(\Y[1][1][{[+0]}]\Y[2][2][{[+0]}]\)obtained in
\eqref{eq:Y11Y22Up} can also be plugged into this equation \eqref{eq:JumpF}.

The above equation \eqref{eq:JumpF} can be solved to express \(\CF\)
as a function of 
\(\Y[1][1][{[+0]}]\Y[2][2][{[+0]}]\). To this end we should anticipate
on the section \ref{sec:bethe-equation-1}, where we will show that
\(\CF\) has simple zeroes at the positions \(\us^{\rlb[\jrt]}\) of the Bethe
roots. This allows to fix \(\CF\) as
\begin{gather}
\label{eq:CF=CF0LAMBDA}
 \CF(\us)=\CF_0(\us) \Lambda_\CF\prod_{\jrt=1}^{\Mp}\sinh\left(\pi\left(\us-\us^{\rlb[\jrt]}\right)\right)\,,%
\end{gather}
where \(\CF_0\) is defined by the integral expression
\begin{gather}
  \CF_0(\us)=\exp\left[ \int_{{\vs\in{\bZc}_0}}\frac
    {1}{2\bi}\left(\tanh\pi(\us- \vs)+\sign(\vs) \rule{0pt}{2.3ex}
    \right)\log \left(\Y[1][1](\vs+\bi\,0)\Y[2][2](\vs+\bi\,0)
      \rule{0pt}{2.3ex} \right) \mathrm{d}\vs \right]\,,
\end{gather}
and the constant \(\Lambda_\CF\) is a normalization. This normalization
depends on the normalization of the gauge \(\bT\), which was already fixed by
the equation \eqref{eq:f=psi}. %
It can be
fixed %
by the constraint
\begin{gather}
  \CF=f\bar f\sqrt{\upT[0][0]}\,.
\end{gather}

\subsubsection{Equation on $h$}
\label{sec:equation-h}

Let us now obtain the equation on the gauge function \(h\) defined in
\eqref{eq:GaugeHwTriT}. In particular we will show that 
in the mirror sheet, this function gives rise to a function \(\hat h\)
which has a single, short, Zhukovsky cut on \(\hbZc_0\).

\paragraph{derivation of the equation}
\label{sec:derivation-equation}

To obtain this equation we first write the Hirota equation \eqref{eq:YHirota}
for the {\upTfs}~:
\begin{align}
\label{eq:HirforH1}
\upT[2][2][+]\upT[2][2][-]=&\upT[3][2]\upT[1][2]+\upT[2][1]\upT[2][3]\,,\\
\label{eq:HirforH2}
\upT[1][1][+]\upT[1][1][-]=&\upT[1][0]\upT[1][2]+\upT[2][1]\upT[0][1]\,.
\end{align}
But we know that \(\upT[0][1]=1\) (see \eqref{eq:rit0=}) and that
\(\huT[2][s]=\huT[1][1][{[+s]}] \huT[1][1][{[-s]}]\) (see
\eqref{eq:=huT2}). \linebreak We also have \(\upT[3][2]=
\wT[3][2] / \left(h^{[+4]}h^{[+2]}h^{[+0]}\bar h ^{[-0]}\bar h ^{[-2]}\bar h
  ^{[-4]}\right)\) (see \eqref{eq:wThupT}), where \linebreak
\(\wT[3][2]
 = -\bT[3][2] \CF^{+} = - \bT[2][3] \CF^{+} =- \wT[2][3]  \CF^{+}
=-\upT[2][3]  \CF^{+} h^{[+4]}h^{[+2]}\bar h ^{[-2]}\bar h^{[-4]}
\)
(see \eqref{eq:DefwT}), %
hence
\(\upT[3][2]=-\upT[2][3]  \CF^{+}/ \left(h^{[+0]}\bar h ^{[-0]}\right)\).
We also know from the {\Wronskian} gauge condition on the {\bTfs} that
\(\Y[1][1]\Y[2][2]=\bT[1][0]/\bT[0][1]=-\wT[1][0]/\CF^{+}=
- h^{[+0]}\bar h ^{[-0]} \upT[1][0]/\CF^{+}\). Finally,
we should note%
\footnote{
In this argument, the condition \(\us\in[-2\,\G,2\,\G]\) is crucial since
it allows to write
\(\upT[1][1][{[+1+0]}]=\upT[1][1][{[+1-0]}]=\upT[1][1][+]\). To write
this, we use the fact that 
\(\upT\) is defined on the mirror sheet, hence it has cuts only when \(|\Re(\us)|>2\,\G\).

In general the {\lhs} of the Hirota equation
\eqref{eq:HirforH2} should be written either as
\(\upT[1][1][{[+1+0]}]\upT[1][1][{[+1+0]}]\) or as
\(\upT[1][1][{[-1+0]}]\upT[1][1][{[-1+0]}]\), and there is always one of
the two factors which is outside the analyticity strip, giving rise to
difficulties if \(|\Re(\us)|>2\,\G\).

We see here that it is important to remember that the Hirota
equation holds specifically in the mirror sheet.
} that for \(\us\in[-2\,\G,2\,\G]\),
\(\upT[1][1][+]=\upT[1][1][{[+1-0]}]=\huT[1][1][{[+1-0]}]\) (because the
argument is inside the inside the analyticity strip), and similarly we have
\(\upT[2][2][\pm]=\huT[2][2][{[\pm1\mp0]}]\). With all these substitutions,
the Hirota
equations (\ref{eq:HirforH1},\ref{eq:HirforH2}) can be rewritten as
\begin{align}
\label{eq:HirforH1b}
\forall\us&\in[-2\,\G,2\,\G],&
\huT[1][1][{[+1+0]}]
\huT[1][1][{[-1-0]}]%
=&
-\upT[1][2]
\CF^{+}/ \left(h^{[+0]}\bar h ^{[-0]}\right)
+
\upT[2][1]
\,,\\
\label{eq:HirforH2b}
\forall\us&\in[-2\,\G,2\,\G],&
\huT[1][1][{[+1-0]}]
\huT[1][1][{[-1+0]}]=&-\upT[1][2]
\Y[1][1]\Y[2][2]\CF^{+}/\left(h^{[+0]}\bar h ^{[-0]}\right)
+\upT[2][1]\,.
\end{align}
If we subtract these two equations, then we obtain
\begin{align}
\label{eq:eqtoh}
  \forall\us&\in[-2\,\G,2\,\G],&
\frac{\huT[1][1][{[+1+0]}]
\huT[1][1][{[-1-0]}] - \huT[1][1][{[+1-0]}]
\huT[1][1][{[-1+0]}]}{\upT[1][2]}=&-\left(1-
\Y[1][1]\Y[2][2]\right) \CF^{+}/ \left(h^{[+0]}\bar h
^{[-0]}\right)\,.
\end{align}
Using the parameterization \eqref{eq:DefT1RR}, we also see that
\begin{align}
  \begin{lefteqn} {\huT[1][1][{[+1+0]}] \huT[1][1][{[-1-0]}] -
      \huT[1][1][{[+1-0]}] \huT[1][1][{[-1+0]}]}%
  \end{lefteqn}
\quad\nonumber\\
~~&=
{\left(1+\CK^{[+2]}\st \rho - \sK\st\rho -\rho/2\right)
  \left(1+\sK\st \rho -\rho/2 
 - \CK^{[-2]}\st\rho \right)}
\nonumber\\
&\qquad-
\left(1+\CK^{[+2]}\st \rho - \sK\st\rho +\rho/2\right)
  \left(1+\sK\st \rho +\rho/2 
 - \CK^{[-2]}\st\rho \right)%
\\
&=-\rho\,\left(2+\CK^{[+2]}\st \rho - \CK^{[-2]}\st\rho \right) =
-\rho \, {\upT[1][2]}\,.
\end{align}
Plugging this expression into the {\lhs} of
\eqref{eq:eqtoh}, we finally obtain
\begin{gather}
\label{eq:forh1}
 \forall\us\in[-2\,\G,2\,\G],\qquad\qquad
\fdisp{  h^{[+0]}\bar h^{[-0]} =\frac{\left(1-
\Y[1][1]\Y[2][2]\right) \CF^{+}}{\rho}}\,.
\end{gather}

\paragraph{Analyticity and \(\Zf\) symmetry}
\label{sec:analyt-zf-symm}

In this expression, we will now show that the {\rhs} has no branch
point at \(\pm2\,\G\). To this end, we will denote by \(\gamma\) a contour which
encircles \(2\,\G\) (or \(-2\,\G\)) but no other singularity, and we will use the
notations of section 
\ref{sec:zf-symmetry} ({\idest}  we denote by \(F(\GmC[\us])\) the result of the analytic
continuation of a function \(F\) following 
the contour \(\gamma\)). If we note that \(\rho=
\huqss[2][+0]-\huqss[2][-0]\),
we can write
\begin{align}
\forall\us\in[-2\,\G,2\,\G]\,,\qquad  \rho(\GmC)=&\huqs[2]\left(\GmC[\us+\bi\,0]\right)
  -\huqs[2]\left(\GmC[\us-\bi\,0]\right) \nonumber\\
=&\huqs[2]\left(\us-\bi\,0\right)
  -\huqs[2]\left(\us+\bi\,0\right)
  = -\rho(\us)\,.
\end{align}
Here we used the fact that, for instance, if \(\gamma\) a clockwise
contour around \(2\,\G\), then \(\huqs[2]\left(\GmC[\us+\bi\,0]\right) =
\huqs[2]\left(\us-\bi\,0\right)\). We also used the assumption that the
cuts are of quadratic type, so that this relation 
implies \(
\huqs[2]\left(\us+\bi\,0\right)=
\huqs[2]\left(\GmC[{\GmC[\us+\bi\,0]}]\right) =
\huqs[2]\left(\GmC[\us-\bi\,0]\right)\). By analytic continuation from
\([-2\,\G,2\,\G]\), this also implies that \linebreak \(\rho(\GmC)=-\rho(\us)\) for
arbitrary \(\us\).

The same arguments allow to write
\begin{gather}
  \left(1- \Y[1][1]\Y[2][2]\right)\left(\GmC[\us]\right) = 
  1- \frac 1
  {\Y[1][1](\us)\Y[2][2](\us)}%
\,,\\%
\And\qquad\CF\left(\GmC[\us+\bi/2]\right) = %
{\CF\left(\us+\bi/2\right) }{\Y[1][1](\us)\Y[2][2](\us)}\,.
\end{gather}
Hence we deduce that the ratio \(\frac{\left(1-
\Y[1][1]\Y[2][2]\right) \CF^{+}}{\rho}\) is regular on the real axis,
{\idest} that it is invariant under analytic continuation along the {\ppath}
\(\gamma\).

This regularity exactly allows to deduce that \(\hwT[1][0]=0\), which
means that the \(\Zf\) symmetry of the {\riTfs} is also satisfied by the {\wTfs}. This property also allows to deduce that \(\hat
{h}=\hat{\overline{h}}\), where we define \(\hat h\) (resp
\(\hat{\overline{h}}\)) as the function which coincides with \(h\) (resp
\(\bar h\)) when \(\Im(\us)>0\) (resp \(\Im(\us)<0\)) and which only has
short Zhukovsky cuts. Then the equation \eqref{eq:forh1} can be
written as 
\begin{gather}
\label{eq:forh2}
 \forall\us\in[-2\,\G,2\,\G],\qquad\qquad
\fdisp{  \hat h^{[+0]} \hat h^{[-0]} =\frac{\left(1-
\Y[1][1]\Y[2][2]\right) \CF^{+}}{\rho}}\,,
\end{gather}
where the function \(\hat h\) is real and is analytic on \(\bC\setminus
\hbZc_0\) (see also the appendix C.3 in \cite{Gromov:2011cx}).

\paragraph{Expression of \(h\)}
\label{sec:expression-h}

The solution of the equation \eqref{eq:forh2} which is analytic on \(\bC\setminus
\hbZc_0\) and has the correct behavior at \(\us\to\infty\) (see \cite{Gromov:2011cx})
is solved by the following convolution
\begin{gather}
\label{eq:fdisplog-hat-h=}
          \fdisp{\log \hat h=-\frac{\LF+2}{2}\log{\hx}+
        {\cal Z}\,{\hat \st}\,\log\left(\frac{\CF^{+}(1-\Y[1][1]\Y[2][2])}{\rho}\right)},
\end{gather}
where we introduce (for any function \(F\)), the convolution  \({\cal
  Z}\,{\hat \st}\, F\) defined by
\begin{gather}
  {\cal
  Z}\,{\hat \st}\, F(\us)=\int_{%
  -2\,\G}^{2\,\G} \frac {-1}{2\bi\pi}
\frac{\sqrt{4\G^2-\us^2}}{\sqrt{4\G^2-\vs^2}}\frac 1 {\us-\vs} F(\vs)\mathrm{d}\vs\,.
\end{gather}

\subsubsection{Equation on  $U$}
\label{sec:equation-UU}

In section \ref{sec:physical-gauge-1}, we have already defined a
function \(\bfC\) (defined by \eqref{relatingBUC2}) which is analytic on
the upper half plane. This allows to introduce a Cauchy {\rp}
of this function as
\begin{gather}
\label{eq:log-bfc-=2}
  \log \bfC = \CK\st \rhocd,\qquad\qquad \when \Im(\us)>0,\\
\end{gather}
where
\(\rhocd \equiv \log \bfC+\log \bfCb\). We see that this {\rp}
differs slightly from \eqref{eq:log-bfc-=}, and it is chosen because
\(\rhocd\) decreases more quickly when \(|\us|\to\infty\) (see
\cite{Gromov:2011cx}). 

As we know that \(\bfC
=
\left(\frac{U }{U ^{[+2]}}
     \frac{f^{+}\hat h^{[+2]}}{f^{[+3]}\hat h}\right)^2 \) (see
   \eqref{eq:hence-fdispbfc-=frac}), this {\rp} allows to
   write 
   \begin{gather}
\label{eq:fdisplog-UU-=}
     \fdisp{\log U = \log \Lambda + \log \frac{\hat h}{f^+} + \frac 1
       2 \Psi \st \rhocd}\,,
   \end{gather}
where \(\Psi\) denotes the convolution kernel defined by
\eqref{eq:psi=}. In this equation, \(\Lambda\) denotes a normalization
constant which can for instance be fixed \cite{Gromov:2011cx} from the
relation 
\begin{equation}
       \sqrt{\upT[0][0][+]\upT[0][0][-]}=
       {U\bar
         U}\upT[0][1]=
       {U\bar
         U}\frac{\rho_2}{1-\Y[1][1]\Y[2][2]}\,,\qquad\when \us\in[-2\,\G,2\,\G]\,,
\label{eq:Normalization}
\end{equation}
which is derived by the same arguments as \eqref{eq:forh2}.

\subsubsection{Equation on the densities $\rho$ and $\rho_2$}
\label{sec:equat-dens-rho}

In order to write an iterative algorithm in the same spirit as  in
chapter \ref{cha:ansatzs-de-bethe}, we should express the densities in
terms of which all the Y-, {\Tfu}- and {\qfs} are parameterized. We
already showed how to express the density \(U\) in terms of the
functions \(f\), \(h\), \(\Y[1][1]\), \(\Y[2][2]\) and \(\rho\). From the
equations of the previous sections, we know how to express all these
functions in terms of the three densities \(\rho\), \(\rho_2\) and \(\rhou\)
and the
polynomial \(\tilde {\baQ}\), which parameterize all our {\qfs}. Hence we have
already written the closed equation on the function \(U\).

Let us now see how to write equations for the functions \(\rho\) and
\(\rho_2\), even though it is less explicit. We have seen in the
previous sections how to express to product \(\Y[1][1]\Y[2][2]\) and the
ratio \(\frac{\Y[1][1]}{\Y[2][2]}\). This allows to compute both
\(\Y[1][1]\) and \(\Y[2][2]\), and even the ratio \(\rmr \equiv
\frac{1+1/\Y[2][2]}{1+\Y[1][1]}\). But on \([-2\,\G,2\,\G]\), we know that
this ratio should be equal to
\begin{gather}
  \rmr =\frac{(1+\sK^+_1\st
    \rho-\frac{\rho}{2})(1+\sK^-_1\st\rho-\frac{\rho}{2})}
{(1+\sK^+_1\st \rho+\frac{\rho}{2})(1+\sK^-_1\st\rho+\frac{\rho}{2})}
\qquad\when~\us\in \hbZc_0\,.
\label{eq:rmr-=frac1+sk+_1st-r}
\end{gather}
Analytically, it is not clear how to invert this equation and write
\(\rho\) and a function of \(\rmr\), but numerically it
allows to express quite easily \(\rho\) as a function of \(\rmr\).
For instance, we can analytically use \eqref{eq:rmr-=frac1+sk+_1st-r}
 to express \(\rho\) as a function of \(\rmr\) and \(\sK^+_1\st
    \rho\). This expression is used to iteratively find \(\rho\) as a
    function of \(\rmr\) (using a fixed-point algorithm).

Although it is numerically slightly more complicated, the same
procedure can be used to extract \(\rho_2\) from the ratio \(\smr \equiv
\frac{1+\Y[2][2]}{1+1/\Y[1][1]}\). Due to our parameterization, this
ratio takes almost the same form as \eqref{eq:rmr-=frac1+sk+_1st-r},
except that \(\rho\) is replaced with \(\rho_2\), and that several other
terms (involving \(\upqs[3]\) and \(\upqs[4]\))
appear. Numerically\footnote{Quite interestingly,
it is numerically much easier to invert the relation
\eqref{eq:rmr-=frac1+sk+_1st-r} and find \(\rho\) as a function of
\(\rmr\) than to invert the analogous expression for \(\smr\) and express
\(\rho_2\) as a function of \(\smr\) and \(\upqs[3]\) and \(\upqs[4]\). What
we numerically found is that we can invert the equation and find
\(\rho_2\) efficiently if we impose the condition that 
\(\upT[1][2](\us^{\rlb[\jrt]}\pm\bi/2)=0\) for all the Bethe roots
\(\us^{\rlb[\jrt]}\). We will derive this condition in section \ref{sec:bethe-equation-1}.
} this
equation allows to write \(\rho_2\) as a function of \(\smr\) and \(\upqs[3]\) and \(\upqs[4]\).

\subsubsection{Bethe equation}
\label{sec:bethe-equation-1}

As in chapter \ref{cha:ansatzs-de-bethe}, we would now like to fix the
coefficients of the polynomial \(\tilde {\baQ}\) contained in our
parameterization. We expect that an analyticity condition like the
absence of poles of the {\Tfs} could impose a Bethe equation on these
coefficients, as it was the case in section \ref{sec:equat-polyn-pf}
for the {\PCM}.

In order to find equations on these polynomials, we should first
investigate the properties of the zeroes and poles of the {\bTfs}.
As explained in section \ref{sec:analyticity-tfs}, the {\Tfs} have no
pole inside their 
analyticity strip, but they can have zeroes which give rise to poles
of the {\Yfs}. We will analyze these zeroes assuming that the gauges
\(\bTft\) and \(\wTft\) %
do not have poles inside
their analyticity strip.

Let us suppose that \(\bT[0][0]\) has some zeroes in its analyticity strip \(\Ast 1\). Since \(\CF=\sqrt{\bT[0][0]}\)
defines the gauge transformation between 
the gauges
\(\bTft\) and \(\wTft\) (see \eqref{eq:DefwT}), %
\(\bT[0][0]\) should not have zeroes with an odd multiplicity, since it
would give rise to branch points in \(\CF=\sqrt{\bT[0][0]}\), which
would spoil the analyticity of the \(\bTft\)- or {\wTfs}.
Hence we obtain that \(\bT[0][0]\) only has 
double\footnote{The function \(\bT[0][0]\) may also have zeroes with
  even multiplicity \(2\nn>2\). If this case arise, we say that it has
  \(\nn\) double zeroes which coincide.} zeroes  so that \(\CF\)
 has only simple zeroes.
We will denote these zeros as \({\us^{\rlb[\jrt]}}\) and assume that there are \({\Mp}\) such zeroes. 

\begin{figure}[t]
\centering
\includegraphics{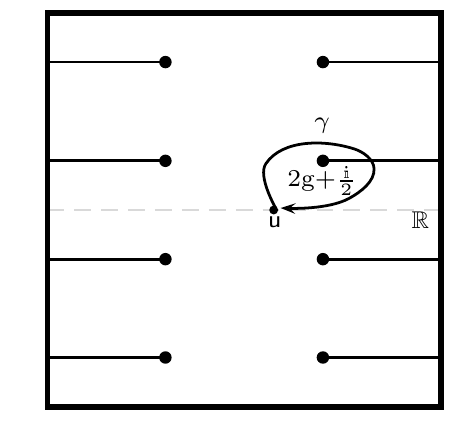}
\caption{\label{fig:contourbethe}Paths used for analytical
  continuation in the Bethe equation \eqref{eq:BethefromTBA}.}
\end{figure}

If we compare with the Bethe roots of the {\TBAE}, we can
see that these zeroes \({\us^{\rlb[\jrt]}}\) are exactly the Bethe roots,
and they should satisfy
\begin{equation}
\Y[1][0]\left(\GmC[\us^{\rlb[\jrt]}]\right)=-1\,,\label{eq:BethefromTBA}
\end{equation}
where \(\Y[1][0]\left(\GmC\right)\) denotes the analytic
continuation of the function \(\Y[1][0]\) following 
the contour \(\gamma\) defined on figure \ref{fig:contourbethe}. This
contour encircles one single branch point of the function \(\Y[1][0]\)
(at position \(2\,\G+\frac \bi 2\)), and then comes back to the point \(\us^{\rlb[\jrt]}\).

This condition can actually be rewritten as a regularity condition on
the {\Tfs}, as we will now see. Since \(\wT[1][2]=\bT[1][2]/\CF^+\), the absence of poles in \(\wT[1][2]\) is only possible if \(\bT[1][2]\) has zeroes at positions \({\us}^{\rlb[\jrt]}\pm {\bi}/2\).
Assuming that \(\bT[2][1]\) does not have zeroes at \({\us}^{\rlb[\jrt]}\pm
{\bi}/2\),  \(\Y[2][2]=\bT[2][1]/\bT[1][2]\) should have poles at
\({\us}^{\rlb[\jrt]}\pm {\bi}/2\). On the other hand, 
we can write the {\YsE} at \(a=1,s=1\), and continue it along  the
contour \(\gamma\). Using 
\eqref{eq:FermConstrYadscft}, we get
\begin{equation}
       \left(1+\Y[1][0]\left(\GmC\right)\right){\Y[2][2][-](\us)} =
       \left(\frac{\Y[1][1][+](1+1/\Y[2][1])}
         {(1+\Y[1][2])}\right)\left(\GmC\right)  
\end{equation}
The equation \eqref{eq:BethefromTBA} is then the condition that, in
the {\lhs}, the poles of \(\Y[2][2][-]\)  at 
the Bethe roots are canceled  with the zeroes of
\(\left(1+\Y[1][0]\left(\GmC\right)\right)\).  Therefore we see that
the
condition \eqref{eq:BethefromTBA}, which fixes the position of the Bethe
roots from the {\TBAE}, can as well be viewed as  the condition that the {\rhs} is regular at  \({\us}=\us^{\rlb[\jrt]}\).

More details about this Bethe equations can be found in
\cite{Gromov:2011cx}. In particular it is shown in appendix E.2  that the condition
\eqref{eq:BethefromTBA} can be rewritten as 
\begin{equation}
\left.   \left(\frac{\hat h^+}{\hat
       h^-}\right)^2+\,\frac{\Y[2][2][+]}{\Y[2][2][-]}\frac{\riT[1][2][+]}{\riT[1][2][-]}\frac{\hrT[1][1][{[-2]}]}{\hrT[1][1][{[+2]}]}\right|_{\us=\us^{\rlb[\jrt]}}=~0
\label{eq:BetheAn}
\end{equation}
This expression is numerically more convenient than the expression
\eqref{eq:BethefromTBA}, because it only involves functions with
argument inside the analyticity strip (where we do not have to do any
analytic continuation around a branch point). It means that this
equation allows to update the position \(\us^{\rlb[\jrt]}\) of the Bethe
roots, provided we know the densities \(\rho\), \(\rho_2\) and \(\rhou\) (and the
polynomial \(\tilde {\baQ}\)), which parameterize the {\qfs} as defined in
section \ref{sec:parameterization-}.

\paragraph{Expression of \(\tilde {\baQ}\)}
\label{sec:expression-tilde-q}

The above discussion only fixes the position of the Bethe roots, which
are the zeroes of (for instance) \(\bT[0][0]\). The position of these
Bethe roots is important because in enters several equations (see for
instance \eqref{eq:CF=CF0LAMBDA} and \eqref{eq:log-fracy11y22-=}), but
it can also be used to write a constraint on the polynomial \(\tilde
{\baQ}\). Indeed if we find the Bethe roots from the equation
\eqref{eq:BetheAn}, then we can write the constraint
\({\bT[0][0](\us^{\rlb[\jrt]})=\bT[1][0](\us^{\rlb[\jrt]}\pm\bi/2)=0}\) to fix the
coefficients of the polynomial \(\tilde {\baQ}\).

\subsubsection{Expression of the energy}
\label{sec:expression-energy-1}

The expression of the energy (or the anomalous dimension \(\gamma\) of
{\ops}), is given by \eqref{eq:Eadscft} in terms of the {\Yfs}. Once
we know the pole structure of the {\Yfs}, as identified in the
previous subsection, it is possible to show that \(E\) is exactly given
by the large \(\us\) behavior of the product \(\Y[1][1] \Y[2][2]\). More
precisely we have
\begin{equation}
\fdisp{\log\left(  \Y[1][1][{[+0]}] \Y[2][2][{[+0]}]\right)\sim \bi
E/\us}\,,\qquad \when \us\to\pm \infty,\qquad \us\in\bR\,.
\end{equation}
This is proven in \cite{Gromov:2011cx} using the {\TBAE}, but
interestingly enough, it can be rewritten (using \eqref{eq:equatCF0})
as 
\begin{equation}\label{eq:EfromT00}
      E=\frac{1}{2}\lim_{\us\to\infty} \us\partial_\us\log\bT[0][0]\,\qquad\qquad\us\in\bR\,.
\end{equation}
One can expect that if a physical construction of the {\Toprs} for {\ADF}
can be found, for instance from a lattice regularization, or from
string theory, then this expression will come naturally because the
{\bTfs} define a gauge which is expected to have a physical origin. This
expression is quite similar to the expression of the energy in chapter
\ref{part:qoperatorsspin}, which involves 
the derivative of the logarithm of a {\Topr}.

\paragraph{Remark and large \(\us\) behavior}
\label{sec:remark-large-us}

In the definition of the parameterization of the {\qfs} it was used
that the behavior of the {\qfs} when  \(\us\to\infty\)  can be extracted
from the asymptotic limit (\(\LF\to\infty\)). This assumption allowed
for instance to fix the degree of the polynomial in the definition of
\(W\) (see \eqref{eq:WnewPoly}) and of \(\rqs[2]{}\) (see
\eqref{eq:RightBandParam}). These polynomial asymptotic were in direct
relation to the limit of \(\Y[][1]\) (resp \(\Y[1]\)) when
\(|\us|\to\infty\) 
and \(|\Im(\us)|<\frac{a-1}2\), (resp \(|\Im(\us)|<\frac{s-1}2\)),
which %
can %
be extracted from the \(\LF\to\infty\) limit, as one
can check from the {\TBAE}. Roughly speaking, this was motivated by
equation \eqref{eq:smallways}, which shows that inside the analyticity
strip, \(\left(\frac{x^{[-a]}}{x^{[+a]}}\right)^\LF\) is small both in
the \(\LF\to\infty\) limit and in the \(\us\to\infty\) limit.

The same argument {\cannnot} easily be used for the functions \(\Y[1][1]\)
and \(\Y[2][2]\), because their analyticity strip has size
zero. Therefore the asymptotic behavior of these {\Yfs} is not exactly
the same as in the \(\LF\to\infty\) limit. One can actually show (see
the end 
of appendix B in \cite{Gromov:2011cx}) that several functions behave at \(\us\to\infty\) as a
power of \(\us\), where the exact power does not coincide with the
\(\LF\to\infty\) limit.

For instance we obtain \(f\sim \us^{\gamma/2}\), \(U\sim
\us^{-({\LF}+\gamma)/2}\), etc.

\subsubsection{Iterative algorithm}
\label{sec:iterative-algorithm}

From all these equation, one can write an iterative algorithm, exactly
like in section \ref{sec:application-au-champ} for the {\PCM}. In this
algorithm, we can  start from the asymptotic %
limit (\(\LF\to\infty\))
of the
densities \(\rho\) and  \(\rho_2\), %
of the Bethe roots
and of the polynomial \(\tilde {\baQ}\). We can
also start from \(U=0\), %
  because \(U\) is exponentially small when 
\(\LF\to\infty\). %
Then %
for each
iteration, we should use
these densities to compute the
{\qfs} and the functions \(\upT\) (when 
\(a\geq |s|\)) and \(\riT\) (when \(s\geq a\)) inside their analyticity
strips (see section \ref{sec:parameterization-}).
Hence, we can 
 compute
the functions \(\Y[1][1]\) and \(\Y[2][2]\) (see \eqref{eq:Y11Y22Up} and
\eqref{eq:log-fracy11y22-=}), as well as the functions \(f\) (see
\eqref{eq:f=psi}) and \(h\) (see \eqref{eq:fdisplog-hat-h=}).
The expressions of \(\Y[1][1]\) and \(\Y[2][2]\) can then be used to find
a new expression of the densities \(\rho\) and \(\rho_2\), whereas the
expression of \(h\) and \(f\) can be plugged into the equation
\eqref{eq:fdisplog-UU-=} for \(U\). Finally, we should update the value
of the Bethe roots (to use a more accurate position of these roots in
the next iteration), and the polynomial \(\tilde {\baQ}\) (see section
\ref{sec:bethe-equation-1}).

And the end of an iteration, we get new expressions for the densities,
the positions of the Bethe roots, and the polynomial \(\tilde {\baQ}\). These
expressions can be used as starting point of a next iteration. If the
algorithm converges, then it provides a solution to the equations
written above.

\section{Numerical results}
\label{sec:{\NuMr}-results-1}

We iterated this algorithm in the case of the {\op} called Konishi
{\op}, which has two symmetric Bethe roots ({\idest} \(\Mp=2\) and
\(\us^{\rlb[1]}=-\us^{\rlb[2]}\)) and has length \(\LF=2\). Our {\NuMr}
results, obtained for a coupling \(\G\) of order \(1\) (or smaller),
confirmed that our FiNLIE reproduces exactly the solution of the
{\TBAE}, with the main difference that the number of functions (and of
equations) is finite. Our {\NuMr} results, shown on 
\colorprint{\figpref{fig:Yfuns}}{\figref{fig:Yfuns}}
 show a very good agreement with previous results obtained
from the {\TBAE}, and confirms the equivalence of our equations with
the {\TBAE}.

This equivalence is also proven analytically in the appendices of \cite{Gromov:2011cx}.

\colorprint{}{\FigNumAdsCFT}

\section{Conclusion}
\label{sec:conclusion-2}

We have shown in this chapter that for the {\Ysys} of AdS/CFT, a
finite set of nonlinear integral equation allows to express the
exact anomalous dimension of a few simple {\ops} of the {\SYM}
conformal field theory. As we saw, this result was obtained by first
solving the {\YsE} in terms of a finite number of {\qfs}, and then by
imposing some analyticity properties which completely fix these
{\qfs}. We argued that these analyticity conditions look very
physical, and are certainly the hints that a physical construction of
the {\Tfs} is possible, and obeys natural regularity conditions. In
particular, we identified a quantum \(\Zf\) symmetry which generalizes
the \(\Zf\) symmetry of the classical string theory to the quantum
{\level}.

 At the moment we {\cannnot} derive these analyticity conditions from an
 explicit, physical construction of the {\Tfs} (or of
 the {\Toprs}) corresponding to this {\Ysys} from a lattice
 regularization or directly from string theory, and we can simply
 prove that these conditions are equivalent to the known {\TBAE}. 
 But it will be very
 interesting to see if such a construction can be obtain. In
 particular that would allow to derive more rigorously that the energy
 spectrum of the {\ADF} duality is expressed from the {\Yfs}
 of the {\Ysys}. This would be 
particularly interesting
since the
 known derivations 
 of these equations are still not completely well controlled (for
 instance regarding the analytic continuation from the vacuum to the 
 excited states).

In this work, we restricted for simplicity to excited states which
belong to the {\SL 2} sector and have a symmetric configuration of two
Bethe roots ({\idest} \(\Mp=2\) and
\({\us^{\rlb[1]}=-\us^{\rlb[2]}}\)). 
The condition that there are only two symmetric Bethe roots was only used to drop
a few coefficients in the polynomial behavior of the {\qfs} in the
limit \(\us\to\infty\), and it would probably not be very difficult to
relax this condition. 
For states outside the {\SL 2} sector, by contrast, more work would
certainly be needed. We already know the expression of the {\cQfs} in
the asymptotic limit
\cite{Gromov:2010km} for these states, but in order to write a FiNLIE
for these states, it will be necessary to know very well the position
of poles and zeroes of the Y- {\Tfu}- and {\qfs}, and to see whether they affect
the equations we have written. In principle the asymptotic limit
should already contain important informations about these poles, so
that it is certainly feasible to generalize our equations. We expect
that if we perform this generalization, then we will have equations on
five densities instead of three, because in general \(\Y[][-s]\) will
not be equal to \(\Y\).

The efficiency of this finite system of equations, both analytically and
numerically, remains to be studied. The {\NuMr} interest in computing
energies from the {\Ysys} has already led to several important
results, and it will be interesting to see whether our FiNLIE allows a
better accuracy or a generalization to more excited states. At the
analytic {\level}, it may be 
better suited than the usual {\Ysys} to derive analytic expansions in
the limit \(\G\to\infty\) or \(\G\to 0\). In the strong coupling limit, in
particular, our formula looks more directly related to the symmetries
of the classical string theory, and we can have reasonable hope that
it will allow to perform an analytic expansion of energy.

%%% Local Variables: ***
%%% mode:latex ***
%%% eval: (find-file "english.tex") ***
%%% TeX-master: "english.tex" ***
%%% End: ***

%% file: concl.tex
\chapter*{Conclusion and outlook}
\label{cha:conclusion-outlook}

 \addcontentsline{lot}{chapter}{Conclusion and outlook}
\addcontentsline{toc}{chapter}{Conclusion and outlook}
\label{app:ccl}

This manuscript presented the research performed during this PhD,
which was devoted to the study of several integrable models. We saw that
various models, which are qualitatively very different from each
other, involve the same equation (the Hirota equation), related to the
existence of {\Qfs}.

We clarified the existence of these {\Qfs}, by constructing them
explicitly for spin chains (where they are the eigenvalues of the
{\Qoprs}), and by proving their existence under a typicality condition
in the case of integrable quantum field theories. In both of these
cases, we showed that they are the building blocks of the {\Tfs} (in
the sense that the {\Tfs} are written as a Wronskian determinant of
these {\Qfs}). Interestingly enough, we also showed (in the case of
spin chains) the relation
between this construction and the general polynomial solution of the
MKP hierarchy.

It would be very interesting to deeper understand the relation
between these objects. In particular,
we may wonder whether
explicit expressions 
like
the ones we wrote in chapter \ref{part:qoperatorsspin} for polynomial
spin chains can be written for more complicated models such as
non-polynomial spins chains or field theory. In chapter
\ref{part:qoperatorsspin}, our construction is
written by means of an ad-hoc operator \(\hD\), specifically designed to
give rise to polynomial spin chains, but we saw in section
\ref{sec:quantum-classical} that the construction which we obtained
can as well be written in terms of \(\tau\)-functions, making the
origin of our construction clearer. This remark will certainly be very
inspiring in order to generalize our construction to other integrable
models.

We also saw that even without understanding explicitly their
construction for integrable field theory, the existence of these {\Qfs}
allowed to simplify noticeably the study of the finite size
corrections. In particular, the usual formulation of the {\Ysys} can
be replaced by a few analyticity conditions on these {\Qfs}. These
requirements can also be written as a finite set of nonlinear integral
equations (FiNLIE) which is expected to be more efficient, both for
its numerical resolution and for analytical expansions.

Interestingly, it was proven \cite{Caetano:2010zd} in the case of the
\(\SU 2\times \SU 2\) 
{\PCM} that the equation obtained by this method in
\cite{2009JHEP...12..060G} is equivalent to the ``DdV'' equations
obtained from a lattice regularization. As opposed to a lattice
regularization (which is not always known to exist) our method seems
to be quite generally applicable to numerous models (since we know how
to express the
solution of the Hirota equation in terms of a finite number of
functions). However, we saw that an important part of the analysis
(namely the study of the analytical properties) had
to performed on a case-by-case basis, even though some common features
emerge (such as the fact that the {\qfs} are analytic on
half-planes). In the cases where DdV equations are known, 
it would be very interesting to clarify the relation between this DdV 
equation and this study of the analyticity properties. In particular
this may lead to a better understanding of the existence of the {\Ysys}.

%%% Local Variables: 
%%% mode: latex
%%% eval: (find-file "english.tex")
%%% TeX-master: "english"
%%% End: 

%% file: groups.tex
\index{Representation|(}

The first section \ref{sec:notat-tens-prod} of this annex consists
of a few definitions, 
mainly introducing the tensor product. It is not necessary to linearly
read this subsection, and the reader can definitely postpone reading these
definitions until they are referred to at some point.

The next sections \ref{sec:repr-de-su2}, \ref{sec:diagrammes-de-{\yn}}
and \ref{sec:gener-glkens}, will introduce some representations of
the matrix groups such as {\GL\Kr}, {\SU\Kr} and 
{\GLKM}.
 These representations will be {\lbd} by
so-called ``{\yn} diagrams''.

To start with, let us recall that a {\rp} of a group \(G\) is
defined by a 
vector space \(V\) and a morphism \(\pge: G\to \GL{V}\), {\ie} a map
from \(G\) to {\GL{V}} such that 
\begin{align}
  \forall {\g},{\g}'\in G,&~\pg[{\g}\cdot {\g}'][{ }]=\pg[][{
  }]\cdot\pg[{\g}'][{ }]\,.
\label{eq:ReprMorpDef}
\end{align}
We will sometimes identify a {\rp} \((V,\pge)\) to the space
\(V\) alone, if the morphism \(\pge\) is unambiguously defined by the context.

\index{Representation!character}
The character of a {\rp} \((V,\pge)\) is the map
\begin{align}
\label{eq:charDef}
  \chi : &\left\{
    \begin{array}{ccc}
      G&\to&ℂ\\
      {\g}&\mapsto&\mathrm{tr}\left(
\pg[][{ }]
\right)
    \end{array}
\right.\,.
\end{align}

\index{Representation|)}

\section{Notations and tensor product}
\label{sec:notat-tens-prod}

Let us first briefly remind what is meant by a tensor product of
Hilbert spaces, and introduce the corresponding notations.

We will consider a set of Hilbert spaces
\(\Hilbl_{1}, \Hilbl_{2},\cdots \Hilbl_{\lcds}\), where each
\(\Hilbl_{\spi}\) has (finite) dimension \(d_{\spi}\) and is defined through an
orthonormal basis denoted as \(\nket{{1}}_\spi,\nket{{2}}_\spi,\cdots,\nket{{d_\spi}}_\spi\).

Then one can construct a bigger Hilbert space \(\Hilb\equiv
\bigotimes_{\spi=1}^{\lcds} \Hilbl_\spi \) (denoted by a
bigger \(\Hilb\) letter), defined through an orthonormal basis which is
the %
set of %
the vectors 
\begin{align}
  \nket{{\nn}_1,{\nn}_2,\cdots,{\nn}_{\lcds}}\equiv&\nket{{}{{\nn}_1}}_1\otimes
  \nket{{}{{\nn}_2}}_2\otimes \cdots \otimes
  \nket{{}{{\nn}_{\lcds}}}_{\lcds}\,,
\label{eq:tensketbasis}
\end{align}
where each \({\nn}_\spi\) belongs to \(\ninter 1 {d_\spi}\).

For an arbitrary linear {\op} \(\mathcal{O}\) on %
this space, it is convenient to introduce the coordinates %
\begin{align}
  \mathcal{O}_{~{\coordj}_1,{\coordj}_2,\cdots,{\coordj}_\lcds}^{{\coordi}_1,{\coordi}_2,\cdots,{\coordi}_\lcds}=&\nbra{{\coordi}_1,{\coordi}_2,\cdots,{\coordi}_\lcds}\mathcal{O}\nket{{\coordj}_1,{\coordj}_2,\cdots,{\coordj}_\lcds}\,.
\end{align}

For instance, if \(A\) and \(B\) are {\ops} on \(\Hilbl\) and \(\Hilbl'\), %
their
tensor product %
is an {\op} on \(\Hilbl\otimes\Hilbl'\) with
coordinates
\begin{align}
  \left(A\otimes
    B\right)_{~{\coordj}_1,{\coordj}_2}^{{\coordi}_1,{\coordi}_2}=&
A_{~{\coordj}_1}^{{\coordi}_1}
B_{~{\coordj}_2}^{{\coordi}_2}\,.
\end{align}

There is a specific set of {\ops} which will be of crucial
importance in what follows, and which exists if all the ``local''
spaces \(\Hilbl_\spi\) are isomorphic: the permutation {\ops} defined by
\begin{align}
\label{eq:Pgen}
  \perm_{(\sigma)}\,:\,&\nket{{\nn}_1,{\nn}_2,\cdots,{\nn}_{\lcds}}
  \mapsto
    \nket{{\nn}_{\sigma(1)},\cdots,{\nn}_{\sigma(\lcds)}}%
\\
  \ie\quad&
\left(\perm_\sigma\right)_{~{\coordj}_1,{\coordj}_2,\cdots,{\coordj}_\lcds}^{{\coordi}_1,{\coordi}_2,\cdots,{\coordi}_\lcds}=%
\prod_{\spk=1}^\lcds %
  \delta^{{{\coordi}_\spk}}%
_{\sigma({\coordj}_\spk)}
\end{align}
for \index{Pa (permutation {\op})@\perm (permutation {\op})}  any permutation \(\sigma\in\Sgrp\lcds\) (where
\(\Sgrp\lcds\) denotes the
\index{S (symmetric group)@\ensuremath{\mathcal{S}} (symmetric group)}
  set of all permutations of \(\ninter 1 \lcds\)).
In the particular case when \(\sigma\) is the identity permutation, one
gets the identity {\op} \(\bI\). On the other hand, if
\(\sigma=\tau_{[\spk,\spl]}%
\) is the transposition \(\spk\leftrightarrow\spl\) defined in \eqref{eq:DefTAUij0}, then one gets the permutation
{\op} \(\perm_{\spk,\spl}\) such that
\begin{align}
\label{eq:defPcoord}
  ({\perm}_{\spk,\spl})_{~{\coordj}_1,{\coordj}_2,\cdots,{\coordj}_\lcds}^{{\coordi}_1,{\coordi}_2,\cdots,{\coordi}_\lcds}=&
  \delta_{{\coordj}_{\spk}}^{{\coordi}_{\spl}}\delta_{{\coordj}_{\spl}}^{{\coordi}_{\spk}}\prod_{\spn\in\ninter 1 \lcds\setminus \{\spk,\spl\}}
\delta_{{\coordj}_{\spn}}^{{\coordi}_{\spn}}\,.
\end{align}
This {\op} satisfies for instance (for \(\lcds=2\))
\begin{align}
  \perm_{1,2}\left(A\otimes B\right)=&\left(B\otimes A\right)\perm_{1,2}\,,
\end{align}
and it will turn out to have a crucial role in what follows.

To finish this section,
let \(\mathcal{O}\) be
an {\op} %
on 
\(\bigotimes_{\spi=1}^{\lcds} \Hilbl_\spi\). %
One can define its partial
trace \( \mathrm{tr}_\lcds\mathcal{O} \) with respect to  \(\Hilbl_\lcds\)
(for instance): it is an 
{\op} on \(\bigotimes_{\spi=1}^{\lcds-1} \Hilbl_\spi\) defined by
\begin{align}
\label{eq:tracpart}
   (\mathrm{tr}_\lcds\mathcal{O})
   _{~{\coordj}_1,{\coordj}_2,\cdots,{\coordj}_{\lcds-1}}^{{\coordi}_1,{\coordi}_2,\cdots,{\coordi}_{\lcds-1}}=&
\sum_{\coordk=1}^{d_\lcds}
\mathcal{O}   _{~{\coordj}_1,{\coordj}_2,\cdots,{\coordj}_{\lcds-1},\coordk}^{{\coordi}_1,{\coordi}_2,\cdots,{\coordi}_{\lcds-1},\coordk}\,.
\end{align}
In particular it satisfies \(\mathrm{tr}_2(A\otimes B)= A
 ~\mathrm{tr}(B)\).

\index{Representation|(}

\section{Representations of \SU{2}}
\label{sec:repr-de-su2}

\index{Representation!of SU(2)}

The representations of {\SU 2} are the simplest example of
representations, and are very frequently encountered in quantum
mechanics, where they describe the spin of different objects.

The three following \(2\times 2\) matrices
\begin{align}
\label{eq:defJ}
  J^{({\coordl})}=&\frac{\sigma^{({\coordl})}}{2}&{\coordl}=&1,2,3\,,
\end{align}
(where  \(\sigma^{({\coordl})}\)
denote the {\pau} matrices defined by \eqref{eq:DefPauli})
form a linear basis of the space of
all traceless, {\her} %
matrices. This  allows us to write\footnote{One can note that for
  \(U\in \SU 2\), there are in general several different vectors
  \((\phi_1,\phi_2,\phi_3)\in ℝ^3\) such that 
\(  U=e^{ⅈ \sum_{{\coordl}=1}^3 \phi_{\coordl} J^{({\coordl})}}\).
For instance, we have \(\bI=e^{4 \pi J^{({1})}}=e^{4 \pi J^{({2})}}=e^{4 \pi J^{({3})}}\).
}
\begin{align}
  \SU{2}=&\{U~|~\exists (\phi_1,\phi_2,\phi_3)\in ℝ^3 :
  U=e^{ⅈ \sum_{{\coordl}=1}^3 \phi_{\coordl} J^{({\coordl})}}
\}\,.
\label{eq:su2generators}
\end{align}
Therefore, we say that
\(J^{({\coordl})}\) are the ``generators'' of {\SU 2}.
\index{Generators (of a matrix group)}
\index{Generators (of a matrix group)!of SU(2)}

The irreducible representations\footnote{
The definition of an irreducible {\rp} will be given in the
section \ref{sec:YoungTab}.} 
of 
\SU{2} are {\lbd} by a number
\(\js\in ℕ/2\), and \({\js}\) is usually called the spin of the
representations. The {\rp} with spin \({\js}\) 
is given by the vector space \(%
V_{\js}\)
 and the morphism
\(\pge_{\js}\) defined by
\begin{align}
\label{eq:DefReprSU2f}
  V_{\js}=&\Vect{\ket{{\js},{\js}},\ket{{\js},{\js}-1},\ket{{\js},{\js}-2},\cdots,\ket{{\js},-{\js}}}\\
\label{eq:DefReprSU2fun}
  \pge_{\js}~~:~~&e^{\bi \sum_{{\coordl}=1}^3 \phi_{\coordl}
    J^{({\coordl})}}\mapsto e^{\bi \sum_{{\coordl}=1}^3 \phi_{\coordl}
    J_{\js}^{({\coordl})}}\\
\label{eq:DefReprSU2Gen}
  \where &J_{\js}^{(3)}\ket{{\js},{\mm}}={\mm} \ket{{\js},{\mm}}\\
  \left( J_{\js}^{(1)}\pm ⅈ\right.
  &\left.J_{\js}^{(2)}\right)\ket{{\js},{\mm}}=%
\sqrt{\left({\js}\mp {\mm}\right)\left({\js}\pm {\mm}+1\right)}
\ket{{\js},{\mm}\pm 1}
\label{eq:DefReprSU2Sqrt}
\,,
\end{align}
where \(\ket{\js,\mm}\) denotes an orthonormal basis of a
\(2\js+1\)-dimensional Hilbert space.
In \eqref{eq:DefReprSU2Sqrt}, we use the convention
\(\ket{{\js},{\js}+1}=0=\ket{{\js},-{\js}-1}\).

\begin{proof}[Proof that
  (\ref{eq:DefReprSU2f}-~\ref{eq:DefReprSU2Sqrt}) is a morphism ]

First, one can check that the \linebreak equation \eqref{eq:DefReprSU2fun} does indeed define a
function \(\pge_{\js}\). What has to be checked is %
that if \linebreak
\({e^{\bi \sum_{{\coordl}=1}^3 \phi_{\coordl} J^{({\coordl})}}=e^{\bi
  \sum_{{\coordl}=1}^3 \psi_{\coordl} J^{({\coordl})}}}\), then \(e^{\bi
  \sum_{{\coordl}=1}^3 \phi_{\coordl} J_{\js}^{({\coordl})}}=e^{\bi 
  \sum_{{\coordl}=1}^3 \psi_{\coordl} J_{\js}^{({\coordl})}}\). One way
to check it is by noticing\footnote{The matrix coefficients of \(e^{\bi
  \sum_{{\coordl}=1}^3 \phi_{\coordl} J_{\js}^{({\coordl})}}\) are 
not a very simple function of \((\phi_1,\phi_2,\phi_3)\), and it is not
very easy to notice that they are a
polynomial function of the matrix coefficients of \(e^{\bi
  \sum_{{\coordl}=1}^3 \phi_{\coordl} J^{({\coordl})}}\). 

We will actually see in the next section that there exists {\another}
construction of the function \(\pge_{\js}\), making it easy to write
the explicit (polynomial) expression of
the coefficients of the matrix \(e^{\bi
  \sum_{{\coordl}=1}^3 \phi_{\coordl} J_{\js}^{({\coordl})}}\) as a
function of the coefficients of the matrix 
\(e^{\bi
  \sum_{{\coordl}=1}^3 \phi_{\coordl} J^{({\coordl})}}\).
} that the coefficients of the matrix \(e^{\bi
  \sum_{{\coordl}=1}^3 \phi_{\coordl} J_{\js}^{({\coordl})}}\) are a
polynomial function of the coefficients of the matrix \(e^{\bi
  \sum_{{\coordl}=1}^3 \phi_{\coordl} J^{({\coordl})}}\).

Next, one should check that 
the relation 
 \eqref{eq:ReprMorpDef} holds. To do this, one can notice that
  \(J_{\js}^{({\coordl})}\) obeys the same commutation
  relation\footnote{Here, the symbol \(\epsilon
  ^{{\coordl},{\coordm},{\coordn}}\) denotes
the antisymmetric function of \({\coordl}\), \({\coordm}\) and \({\coordn}\)
such that
\(\epsilon^{1,2,3}=1\).
}
  \([J_{\js}^{({\coordl})},J_{\js}^{({\coordm})}]=ⅈ~ \epsilon
  ^{{\coordl},{\coordm},{\coordn}}J_{\js}^{({\coordn})}\) as the
  {\ops} \(J^{({\coordl})}\) defined in \eqref{eq:defJ}. From this
  we can prove that \(\pge_\js\) is a morphism, {\idest} that
  \begin{align}
    \label{eq:ifSU2prod}
    \If e^{\bi \sum_{{\coordl}=1}^3 \phi_{\coordl} J^{({\coordl})}} \cdot
    e^{\bi\sum_{{\coordl}=1}^3 \phi'_{\coordl}
      J^{({\coordl})}} = e^{\bi \sum_{{\coordl}=1}^3 \psi_{\coordl} J^{({\coordl})}},\\
    \Then e^{\bi\sum_{{\coordl}=1}^3 \phi_{\coordl}
      J_{\js}^{({\coordl})}} \cdot e^{\bi\sum_{{\coordl}=1}^3
      \phi'_{\coordl} J_{\js}^{({\coordl})}} = e^{\bi\sum_{{\coordl}=1}^3
      \psi_{\coordl} J_{\js}^{({\coordl})}}\,.
    \label{eq:thenSU2prod}
  \end{align}
  This result is obtained by the Baker-Campbell-Hausdorff formula,
  which allows to express \(\psi\) as a function of \(\phi\) and
  \(\phi'\). This Baker-Campbell-Hausdorff formula reads
  \begin{align}
    e^X e^ Y%
    =& \mathrm{exp}\left( X + Y + \frac{1}{2}[X,Y] +
      \frac{1}{12}[X - Y,[X,Y]] +\cdots \right)\,,
  \end{align}
  and it only involves the commutation relations between the
  \(J^{({\coordl})}\). 
This Baker-Campbell-Hausdorff formula provides 
the explicit expression of a \(\psi\in\bR^3\)  such that
\eqref{eq:ifSU2prod} holds. But since the \(J_{\js}^{({\coordl})}\)
obey the 
  same commutation relation as \(J^{({\coordl})}\), we immediately see
  that the same  expression \(\psi\) obeys \eqref{eq:thenSU2prod} as well.
\end{proof}

Among these representations, the \({\js}=0\) case corresponds to \(V_0=ℂ\) and
\(\forall {\g}, \textrm{ } \linebreak[2] \pg[][0]=𝕀\). This
{\rp} is usually called the trivial {\rp}.

By contrast, the {\rp} with spin \(\js=1/2\) corresponds to 
\(V_{1/2}=ℂ^2\) and \(\forall {\g},~\pg[][1/2]={\g}\). It will be
called the fundamental {\rp} of {\SU{2}}.

For the unity of notations, let us also introduce the following
notation:
\begin{align}
  \pe[J^{({\coordl})}][\js]\equiv&J_{\js}^{({\coordl})}\,,
\end{align}
were the bold letter \(\pee\) denotes the transformation of the
generators, whereas the letter \(\pge\) denotes the transformation of a
group element.

\section{{\yn} diagrams and representations of \GL{{\Kr}}}
\label{sec:YoungTab}

We will now see that the representations defined above can be
generalized not only to {\SU \Kr}, but even\footnote{
In general it is obvious that every {\rp} of {\GL \Kr} gives
rise to a {\rp} of the subgroup \(\SU{{\Kr}}\subset
\GL{{\Kr}}\).
But it is %
a priori not trivial that these representations of
\(\SU{{\Kr}}\) also define representations of {\GL \Kr}.
} to {\GL \Kr}.

In this section, we will define some representations of \GL{{\Kr}}, 
which are indexed by {\yn} diagrams. Their restriction to unitary
matrices will give the irreducible representations of {\SU \Kr}. 
In this manuscript, the construction will be introduced with less
details than (for instance) in \cite{FH}, but it will be generalized
to super-groups, following for instance  \cite{Baha}.

\subsection{{\yn} diagrams}
\label{sec:diagrammes-de-{\yn}}

\index{Young diagrams}
The {\yn} diagrams are diagrams which can be identified to
non-increasing sequences \(\lambda_{\ii}\geq 0\) of integers where
there is an \(\nn\) such that \(\lambda_{\ii}=0\) for all \(\ii\geq \nn\).
The identification between diagrams (made out of ``boxes'') and these
sequences goes as follows:
\begin{align}
\label{eq:DefYoung}
\framedline{
  (5,3,2,2,0,0,0,\cdots)~~\leftrightarrow} {~~\raisebox{-.45cm}{\includegraphics[scale=.25]{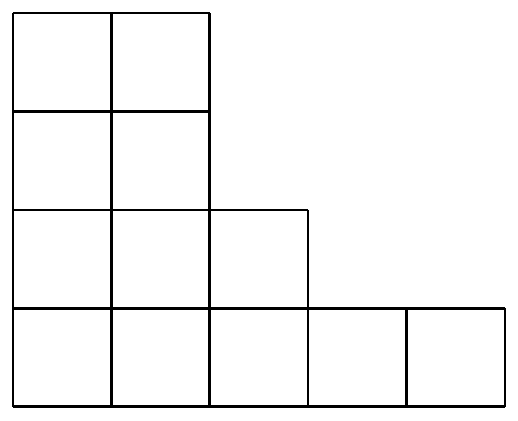}}}\,\,.
\end{align}
We see that \(\lambda_{\ii}\) becomes the number of boxes in the
\({\ii}^{\textrm{th}}\) row (counted from {\below}) of the diagram.

\index{Young tableaux}
A {\yn} tableau will denote a {\yn} diagram where positive integers
are written inside each box. We will see that each {\yn} diagram labels
a {\rp} of {\GL \Kr},
and that if we write integers in each box of the diagram (following a
given rule\footnote{This rule will be that the integers are increasing
in each column and non-decreasing in each row.}), then we obtain {\yn} tableaux which label a basis of this
{\rp}.

\subsection{Decomposition of the tensor {\rp} of \GL{{\Kr}}}
\label{sec:repr-tens-de}
\index{Representation!of GL({\Kr})}

A very natural {\rp} of 
\GL{{\Kr}} is given by the space \(V_{{\NN}}=\left(ℂ^{\Kr}\right)^{\otimes
  {\NN}}\) and the morphism 
\(\pge_{\NN}~~:~~{\Mnothing}\mapsto {\Mnothing}^{\otimes {\NN}}\). Then \(\pg[\Mnothing][{\NN}]\)
acts on the basis \eqref{eq:tensketbasis} as follows
\begin{align}
\label{eq:TensorRepr}
  \pg[\Mnothing][{\NN}] \nket{{\nn}_1,{\nn}_2,\cdots,{\nn}_{{\NN}}}=&\sum_{{\nn}'_1,{\nn}'_2,\cdots,{\nn}'_{\NN}
      \in {\ninter {1} {{\Kr}}}^{\NN}} \prod_{\spi=1}^{{\NN}}
    {\Mnothing}_{~{\nn}'_\spi}^{{\nn}_\spi} \nket{{\nn}'_1,{\nn}'_2,\cdots,{\nn}'_{{\NN}}}\,.
\end{align}
If \({\NN}=1\), this defines the  ``fundamental'' {\rp} of {\GL \Kr}, {\idest} the
{\rp} such that \(\pg[\Mnothing][{ }]=\Mnothing\).
This {\rp} will be denoted by the {\yn} diagram 
{\raisebox{-.1cm}{\includegraphics[scale=.3]{figYdiag_1}}}.
Then the tensor {\rp}  \eqref{eq:TensorRepr} will be denoted by
\(\underbrace{\raisebox{-.1cm}{\includegraphics[scale=.3]{figYdiag_1}}  ⊗ 
\raisebox{-.1cm}{\includegraphics[scale=.3]{figYdiag_1}} ⊗ 
\cdots ⊗ 
\raisebox{-.1cm}{\includegraphics[scale=.3]{figYdiag_1}}}_{\NN}\).

\paragraph{Reducibility of the tensor {\rp}}
\label{sec:reduc-tens-repr}

If \({\NN}\geq 2\), this {\rp} is reducible.
\index{Representation!irreducible}
This means that there exists at least one sub-space of  \(V_{\NN}\)
which is stable under \(\pg[\Mnothing][{\NN}]\), for every \({\Mnothing}\in {\GL
  \Kr}\).
For instance we will see that the set of all symmetric tensors is
stable under all \(\pg[\Mnothing][{\NN}]\). This set will be denoted by
\(V_{\underbrace{\includegraphics[scale=.2]{figYdiag2}}_{{\NN}}}\), which
is {\lbd} by the {\yn} tableau
\(({\NN},0,0,\cdots)=\underbrace{\includegraphics[scale=.3]{figYdiag2}}_{{\NN}}\)
(see \eqref{eq:DefYoung}). It
can be defined using the projector
\(\proj_{%
  {\includegraphics[scale=.2]{figYdiag2}}%
}\):
\begin{gather}
\label{eq:DefVSym}
V_{\underbrace{\includegraphics[scale=.2]{figYdiag2}}_{{\NN}}}\equiv
\mathrm{Im}(\proj_{\underbrace{\includegraphics[scale=.2]{figYdiag2}}_{{\NN}}})\\
\where 
 \proj_{\underbrace{\includegraphics[scale=.2]{figYdiag2}}_{{\NN}}}\nket{\nn_1,\nn_2,\nn_3,\cdots,\nn_{\NN}}=
\frac 1 {{\NN} !}\left(
 \sum_{\sigma\in \Sgrp {\NN}} \nket{{\nn}_{\sigma(1)},{\nn}_{\sigma(2)},\cdots,{\nn}_{\sigma({\NN})}}
\right)\,,\nonumber\\
 \label{eq:DefPsym}
\ie~
\proj_{\underbrace{\includegraphics[scale=.2]{figYdiag2}}_{{\NN}}}
=
\frac 1 {{\NN} !}
 \sum_{\sigma\in \Sgrp {\NN}} \perm_{\sigma}\,,
\end{gather}
where the permutation {\op} \(\perm_{\sigma}\) is defined by \eqref{eq:Pgen}.

An orthonormal basis of \(V_{%
  {\includegraphics[scale=.2]{figYdiag2}}%
}\) can be written as 
\begin{gather}
\label{eq:SymBasis}
  \mathcal{B}_{  {\includegraphics[scale=.2]{figYdiag2}}} \equiv
\left\{\left.
 \raisebox{-.35cm}{\includegraphics{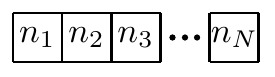}}
 ~~\right|
~~
1\leq {\nn}_1\leq {\nn}_2\leq\cdots\leq {\nn}_{\NN}\leq {\Kr} \right\}\\
\where   \raisebox{-.35cm}{\includegraphics{figYtabS}} \equiv \Normed\left(
 \proj_{\underbrace{\includegraphics[scale=.2]{figYdiag2}}_{{\NN}}}\nket{\nn_1,\nn_2,\nn_3,\cdots,\nn_{\NN}}\right)
\\
\where   \Normed\left(\ket\psi\right)\equiv
\frac{\ket\psi}{\sqrt{\braket\psi\psi}}\,.
\end{gather}
We see that the elements of this basis are {\yn} tableaux obtained by
filling the corresponding
{\Ydag} with ordered numbers belonging to \(\ninter 1 \Kr\).

The very definition of the {\ops} \(\perm_{\sigma}\) implies that
\begin{align}
\label{eq:ForPermCom}
  \left(\mathcal{O}_1\otimes\mathcal{O}_2\otimes \cdots \otimes \mathcal{O}_{\NN}\right)\cdot
  \perm_\sigma=& \perm_\sigma\cdot\left(
    \mathcal{O}_{\sigma(1)}\otimes\mathcal{O}_{\sigma(2)}\otimes
    \cdots \otimes \mathcal{O}_{\sigma({\NN})}\right)\,,
\end{align}
for any set of {\ops} \((\mathcal{O}_1,\mathcal{O}_2, \cdots,
\mathcal{O}_{\NN})\) on \(\Hilbl_1\) (resp \(\Hilbl_2\) resp \(\cdots\)). In
particular we see that  \({\Mnothing}^{\otimes {\NN}}\) commutes with any
\({\perm}_{\sigma}\), hence it also commutes with 
\(\proj_{%
  {\includegraphics[scale=.2]{figYdiag2}}
}\).
As a consequence,
\(V_{%
  {\includegraphics[scale=.2]{figYdiag2}}%
} =
\mathrm{Im}(\proj_{%
  {\includegraphics[scale=.2]{figYdiag2}}%
})\)
is stable under all \(\pg[\Mnothing][{\NN}]\), for all \(\Mnothing\in \GL\Kr\).

 This
allows to introduce a {\rp} of {\GL\Kr}, {\lbd} by the {\yn}
diagram
\(({\NN},0,0,\cdots)=\underbrace{\includegraphics[scale=.3]{figYdiag2}}_{{\NN}}\),
defined by the space 
\(V_{
  {\includegraphics[scale=.2]{figYdiag2}}%
}\)
and the morphism \(\pge_{\NN}\) {\below}:
\begin{align}
  \pg[\Mnothing][{\includegraphics[scale=.2]{figYdiag2}}] :  \left\{
  \begin{array}{ccl}
     V_{%
  {\includegraphics[scale=.2]{figYdiag2}}%
}&\to&  V_{%
  {\includegraphics[scale=.2]{figYdiag2}}%
}\\
    \ket{\Psi}&\mapsto &{\Mnothing}^{⊗ {\NN}}\ket{\Psi}
  \end{array}\right.\,.
\end{align}

Hence we have shown that when  \({\NN}\geq 2\), the {\rp}
\({\raisebox{-.1cm}{\includegraphics[scale=.3]{figYdiag_1}}  ⊗ 
\raisebox{-.1cm}{\includegraphics[scale=.3]{figYdiag_1}} ⊗ 
\cdots ⊗ 
\raisebox{-.1cm}{\includegraphics[scale=.3]{figYdiag_1}}}%
\) 
is reducible, which means that it contains at least one stable
subspace. This stable subspace also defines {\another} {\rp}
of {\GL \Kr}, which has a smaller dimension.

\paragraph{Decomposability of the tensor {\rp}}
\label{sec:decomp-tens-repr}

In addition to being reducible, 
the {\rp} \(V_{\NN}\) is actually decomposable, which means that
it can be written as a direct sum of representations corresponding to
stable subspaces. For instance, when \({\NN}=2\), we will show that one
can write
\begin{align}
\label{eq:V2Irred}
\raisebox{-.1cm}{\includegraphics[scale=.3]{figYdiag_1}}  ⊗ 
\raisebox{-.1cm}{\includegraphics[scale=.3]{figYdiag_1}}
  =&\raisebox{-.1cm}{\includegraphics[scale=.3]{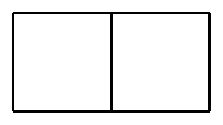}}⊕
  \raisebox{-.2cm}{\includegraphics[scale=.3,angle=90]{figYdiag3}}\,, 
\end{align}
where the {\rp}
\(\raisebox{-.2cm}{\includegraphics[scale=.3,angle=90]{figYdiag3}}\)
is defined from the space of antisymmetric tensors. For arbitrary \({\NN}\),
we introduce the {\rp} %
\begin{align}
\label{eq:DefVantisym}
V_{\left.\raisebox{-.55cm}{\includegraphics[scale=.2,angle=90]{figYdiag2}}\right\}{\NN}}\equiv&
~\mathrm{Im}%
  \big(
  \proj_{\left.\raisebox{-.55cm}{\includegraphics[scale=.2,angle=90]{figYdiag2}}\right\}{\NN}}
\big)&\where
\proj_{\left.\raisebox{-.55cm}{\includegraphics[scale=.2,angle=90]{figYdiag2}}\right\}{\NN}}\equiv&\frac
1 {{{\NN}!}}\sum_{\sigma
  \in \Sgrp {\NN}} \epsilon(\sigma)~ \perm_\sigma\,.
\end{align}
Here, \(\epsilon(\sigma)\equiv
\prod_{\ii<\jj}\frac{\sigma(\ii)-\sigma(\jj)}{\ii-\jj}\) is the
signature of the permutation \(\sigma\).

An orthonormal basis of this space can be written as 
\begin{align}
  \mathcal{B}_{\left.\raisebox{-.55cm}{\includegraphics[scale=.2,angle=90]{figYdiag2}}\right\}{\NN}}\equiv&
\left\{\left.
\raisebox{-1.2cm}{\includegraphics{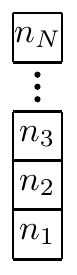}}
~~\right|~~
1\leq {\nn}_1< {\nn}_2<\cdots< {\nn}_{\NN}\leq {\Kr}\right\}\\
\where & \raisebox{-1.2cm}{\includegraphics{figYtabA}} \equiv
\sqrt{{\NN}!}~~\proj_{\left.\raisebox{-.55cm}{\includegraphics[scale=.2,angle=90]{figYdiag2}}\right\}{\NN}}
\ket{\nn_1,\nn_2,\nn_3,\cdots,\nn_{\NN}}\,.
\end{align}

Since \(\Mnothing^{\otimes {\NN}}\) commutes with all \(\perm_\sigma\) (see
\eqref{eq:ForPermCom}), this space is invariant under the action of
\(\pg[\Mnothing][{\NN}]\) for every \({\Mnothing\in\GL\Kr}\), and it therefore defines a
{\rp} of {\GL\Kr}. The {\yn} diagram
\({\NN}\left\{\raisebox{-.825cm}{\includegraphics[scale=.3,angle=90]{figYdiag2}}\right.
= (\underbrace{1,1,1,\cdots,1}_{\NN},0,0,0,\cdots)\) will now denote this {\rp}.

The equality \eqref{eq:V2Irred} states that when \({\NN}=2\),
the spaces \(V_{\includegraphics[scale=.15,angle=0]{figYdiag3}}\) and
\(V_{\includegraphics[scale=.15,angle=90]{figYdiag3}}\) are not only
stable under \(\pg[\Mnothing][{\NN}]\) for all \(\Mnothing\) (and therefore they define
{\rp}s), but even that they are linearly independent and span
the whole space 
\(V_{2}=\left(ℂ^{\Kr}\right)^{\otimes 2}\). 

\begin{proof}[Proof of \eqref{eq:V2Irred}]
  The only thing which was not proven above, is that the spaces \(V_{\includegraphics[scale=.15,angle=0]{figYdiag3}}\) and
\(V_{\includegraphics[scale=.15,angle=90]{figYdiag3}}\) are linearly independent and span
the whole space 
\(V_{2}=\left(ℂ^{\Kr}\right)^{\otimes 2}\). First their linear independence is easily shown from the equality 
\(\proj_{\includegraphics[scale=.15,angle=0]{figYdiag3}}\circ
\proj_{\includegraphics[scale=.15,angle=90]{figYdiag3}} =0=
\proj_{\includegraphics[scale=.15,angle=90]{figYdiag3}} \circ
\proj_{\includegraphics[scale=.15,angle=0]{figYdiag3}}\). Indeed, if
\(\proj_{\includegraphics[scale=.15,angle=0]{figYdiag3}}\ket{\phi} +
\proj_{\includegraphics[scale=.15,angle=90]{figYdiag3}}\ket{\psi}=0\),
then we get 
\({\proj_{\includegraphics[scale=.15,angle=0]{figYdiag3}}\left( \proj_{\includegraphics[scale=.15,angle=0]{figYdiag3}}\ket{\phi} +
\proj_{\includegraphics[scale=.15,angle=90]{figYdiag3}}\ket{\psi}\right)
= \proj_{\includegraphics[scale=.15,angle=0]{figYdiag3}}\ket{\phi}=0}\).
Moreover, the dimension of these spaces is \(\frac{{\Kr}({\Kr}+1)}2\)
and 
\(\frac{{\Kr}({\Kr}-1)}2\), as we can see from 
\eqref{eq:SymBasis} and \eqref{eq:DefVantisym}. It follows that %
their direct sum is the whole space
\(V_{2}=\left(ℂ^{\Kr}\right)^{\otimes 2}\).
\end{proof}

If \({\NN}>2\), the relation \eqref{eq:V2Irred}
can be generalized, and the {\rhs} will be a sum over the {\yn}
diagrams with \({\NN}\) boxes, which label different {\rp}s.
 For instance, when  \({\NN}=3\), we will show that
\begin{align}
\raisebox{-.1cm}{\includegraphics[scale=.3]{figYdiag_1}}  ⊗ 
\raisebox{-.1cm}{\includegraphics[scale=.3]{figYdiag_1}}⊗ 
\raisebox{-.1cm}{\includegraphics[scale=.3]{figYdiag_1}}
 =&\raisebox{-.1cm}{\includegraphics[scale=.3]{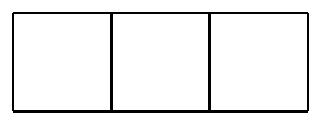}}⊕
 2~ \raisebox{-.2cm}{\includegraphics[scale=.3]{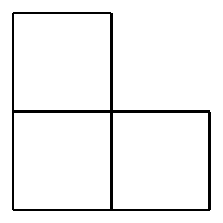}}⊕
  \raisebox{-.3cm}{\includegraphics[scale=.3]{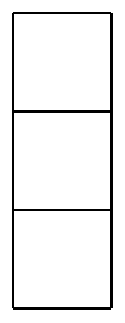}}\,.
\label{eq:Tens3Dec}
\end{align}
In the {\rhs}, the {\rp}s 
\(\raisebox{-.1cm}{\includegraphics[scale=.3]{figYdiag_3}}\) and \(
\raisebox{-.3cm}{\includegraphics[scale=.3]{figYdiag_111}}\) 
were already introduced in (\ref{eq:DefVSym},\ref{eq:DefVantisym}).
On the other hand the {\rp} 
\(\raisebox{-.2cm}{\includegraphics[scale=.3]{figYdiag_21}}\) 
is obtained by combining symmetrizations (as in \eqref{eq:DefPsym})
and antisymmetrizations (as in \eqref{eq:DefVantisym}). Let us define
the spaces
\begin{align}
\label{eq:DefVhook}
  V_{\includegraphics[scale=.15]{figYdiag_21}}\equiv&\mathrm{Im}\left(
c_{\includegraphics[scale=.15]{figYdiag_21}}
\right)\equiv\mathrm{Im}\left(
\left(1-{\perm}_{1,3}\right)\left(1+{\perm}_{1,2}\right)
\right)\\
\et   \tilde V_{\includegraphics[scale=.15]{figYdiag_21}}\equiv&\mathrm{Im}\left(
\tilde c_{\includegraphics[scale=.15]{figYdiag_21}}
\right)\equiv\mathrm{Im}\left(
\left(1+{\perm}_{1,3}\right)\left(1-{\perm}_{1,2}\right)
\right)\,.
\end{align}
These two spaces \(V_{\includegraphics[scale=.15]{figYdiag_21}}\) and
\(\tilde V_{\includegraphics[scale=.15]{figYdiag_21}}\) give rise
to two isomorphic\footnote{A funny way to see that they are isomorphic is
  as follows:

First notice that \(\mathrm{Im}\left(
\left(1+{\perm}_{1,3}\right)\left(1-{\perm}_{1,2}\right)
\right)\) and \(\mathrm{Im}\left(
\left(1+{\perm}_{1,2}\right)\left(1-{\perm}_{1,3}\right)
\right)\) are identical up to a relabelling of \(\Hilb_2\) and
\(\Hilb_3\). Then the Yang-Baxter equation \eqref{eq:RRR=RRR} tells us that 
\(\left(1+{\perm}_{1,2}\right)\left(1-{\perm}_{1,3}\right)\) and
\(\left(1-{\perm}_{1,3}\right)\left(1+{\perm}_{1,2}\right)\) 
are equal up to the multiplication, to the right and to the left, by
invertible matrices which commute with \(\pg[\Mnothing][3]\). This implies
that these representations are isomorphic indeed. 
} {\rp}s of {\GL\Kr}, which will both be denoted by \(\raisebox{-.2cm}{\includegraphics[scale=.3]{figYdiag_21}}\) in
\eqref{eq:Tens3Dec}.
The two {\ops} 
 \(c_{\includegraphics[scale=.15]{figYdiag_21}}\) and \(\tilde
 c_{\includegraphics[scale=.15]{figYdiag_21}}\) generalize the
 projection {\ops} introduced in  \eqref{eq:DefPsym} and
 \eqref{eq:DefVantisym} for the symmetric and antisymmetric
 {\rp}.
One important difference is that
\(c_{\includegraphics[scale=.15]{figYdiag_21}}\) and \(\tilde 
 c_{\includegraphics[scale=.15]{figYdiag_21}}\) are nor projectors.

 \begin{proof}[Proof of \eqref{eq:Tens3Dec}]
   First let us show that the four subspaces corresponding to these
   representations are linearly independent:
 \begin{gather}
   \IF
   \Sigma\equiv\proj_{\includegraphics[scale=.15]{figYdiag_111}}\ket{\Psi_1} + 
   \proj_{\includegraphics[scale=.15,angle=90]{figYdiag_111}}\ket{\Psi_2}
   +c_{\includegraphics[scale=.15]{figYdiag_21}}\ket{\Psi_3}
   +\tilde
   c_{\includegraphics[scale=.15]{figYdiag_21}}\ket{\Psi_4}=0\\
   \Then 
   \left.\begin{array}{l}
     0=\proj_{\includegraphics[scale=.15]{figYdiag_111}} \Sigma =
     \proj_{\includegraphics[scale=.15]{figYdiag_111}}\ket{\Psi_1}
     \\
     0=\proj_{\includegraphics[scale=.15,angle=90]{figYdiag_111}} \Sigma  =
     \proj_{\includegraphics[scale=.15,angle=90]{figYdiag_111}}\ket{\Psi_2}
   \end{array}\right\}~~\hence ~~\tilde \Sigma\equiv c_{\includegraphics[scale=.15]{figYdiag_21}}\ket{\Psi_3}
   +\tilde
   c_{\includegraphics[scale=.15]{figYdiag_21}}\ket{\Psi_4}=0\nonumber\\
   \et  0=(1-{\perm}_{1,3})\tilde \Sigma=2
   c_{\includegraphics[scale=.15]{figYdiag_21}}\ket{\Psi_3}\,, 
 \end{gather}
 which shows that they are independent indeed.

Finally, we will conclude by finding the dimension of
\(V_{\includegraphics[scale=.15]{figYdiag_21}}\). To do this, we
just have to notice that the vectors in the set
\begin{align}
\label{eq:basH}
\mathcal{B}_{\includegraphics[scale=.15]{figYdiag_21}}\equiv&\left\{\left.
  \raisebox{-.5cm}{\includegraphics{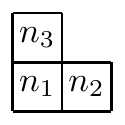}}
 ~~\right|
~~{\nn}_1\leq {\nn}_2~\et {\nn}_1<{\nn}_3\right\}\\
\where &
\raisebox{-.5cm}{\includegraphics{figYtabH}}
\equiv 
\Normed\left( c_{\includegraphics[scale=.15]{figYdiag_21}}
  \ket{{\nn}_1,{\nn}_2,{\nn}_3}\right)
\end{align}
are linearly independent\footnote{
The linear independence of these vectors can for instance be shown by
projecting on the space
\(\Vect{\left.\ket{x,y,z}~~\right|~~x\leq y ~\et x<z}\).
}, so that the dimension of 
\(\raisebox{-.2cm}{\includegraphics[scale=.3]{figYdiag_21}}\) is at
least \(\frac{({\Kr}-1){\Kr}({\Kr}+1)}{3}\).
If we note that the dimensions of
\(\raisebox{-.1cm}{\includegraphics[scale=.3]{figYdiag_3}}\) and
\(\raisebox{-.3cm}{\includegraphics[scale=.3]{figYdiag_111}}\) are
respectively \(\left(\begin{array}{cc} 
 {\Kr}+2\\3
   \end{array}
 \right)\) and \(\left(\begin{array}{cc}
 {\Kr}\\3
   \end{array}
 \right)\), we conclude that  \(\raisebox{-.1cm}{\includegraphics[scale=.3]{figYdiag_3}}⊕
 2~ \raisebox{-.2cm}{\includegraphics[scale=.3]{figYdiag_21}}⊕
  \raisebox{-.3cm}{\includegraphics[scale=.3]{figYdiag_111}}\) 
has dimension at least \({\Kr}^3\). This dimension argument implies that
\eqref{eq:Tens3Dec} holds and that \eqref{eq:basH} defines a basis of 
  \(V_{\includegraphics[scale=.15]{figYdiag_21}}\).
 \end{proof}

For arbitrary \({\NN}\), the representations associated to {\yn} diagrams
with \({\NN}\) boxes are constructed like above, by describing some invariant
subspaces of \({\raisebox{-.1cm}{\includegraphics[scale=.3]{figYdiag_1}}  ⊗ 
\raisebox{-.1cm}{\includegraphics[scale=.3]{figYdiag_1}}⊗ \cdots 
⊗ 
\raisebox{-.1cm}{\includegraphics[scale=.3]{figYdiag_1}}}\). 
For instance, when \({\NN}=4\), we get
\begin{align}
\label{eq:SplitV4}
  \raisebox{-.1cm}{\includegraphics[scale=.3]{figYdiag_1}}  ⊗ 
\raisebox{-.1cm}{\includegraphics[scale=.3]{figYdiag_1}}⊗ 
\raisebox{-.1cm}{\includegraphics[scale=.3]{figYdiag_1}}⊗ 
\raisebox{-.1cm}{\includegraphics[scale=.3]{figYdiag_1}}
 =&
\raisebox{-.1cm}{\includegraphics[scale=.3]{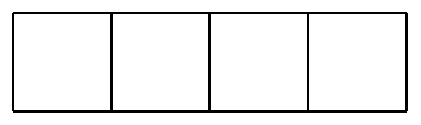}} 
⊗3 \raisebox{-.25cm}{\includegraphics[scale=.3,angle=0]{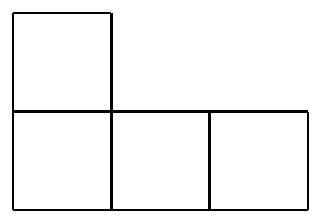}}
⊗2 \raisebox{-.25cm}{\includegraphics[scale=.3,angle=0]{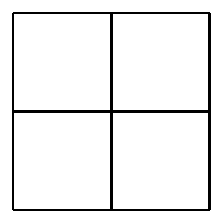}}
⊗3 \raisebox{-.35cm}{\includegraphics[scale=.3,angle=0]{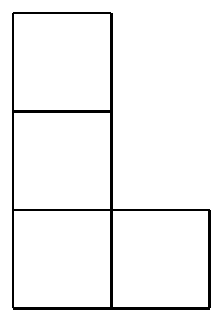}}
⊗
\raisebox{-.5cm}{\includegraphics[scale=.3,angle=90]{figYdiag4}}
\,.
\end{align}

For a general
{\yn} diagram, the associated {\rp} is
 the image of a ``{\yn}
symmetrizer'' generalizing the {\op}
\(c_{\includegraphics[scale=.15]{figYdiag_21}}\) of
\eqref{eq:DefVhook}. For instance for the representations in equation  \eqref{eq:SplitV4},
 one
possible expression of the {\yn} symmetrizers is
\begin{align}
  c_{\includegraphics[scale=.15]{figYdiag_31}}=&\left(
1-\perm_{1,4}
\right)\cdot\left(
1+\perm_{1,2}+\perm_{2,3}+\perm_{1,3}+\perm_{(1,2,3)}+\perm_{(3,2,1)}
\right),\\
  c_{\includegraphics[scale=.15]{figYdiag_22}}=&\left(
1-\perm_{1,3}\right)\cdot\left(1-\perm_{2,4}
\right)\cdot\left(
1+\perm_{1,2}\right)\cdot\left(1+\perm_{3,4}
\right),
\label{eq:sym22}
\\
  c_{\includegraphics[scale=.15]{figYdiag_211}}=&\left(
1-\perm_{1,3}-\perm_{3,4}-\perm_{1,4}+\perm_{(1,3,4)}+\perm_{(4,3,1)}
\right)\cdot\left(
1+\perm_{1,2}
\right),
\end{align}
where \(\perm_{(4,3,1)}\) (for instance) is the permutation {\op}
\eqref{eq:Pgen} associated to the permutation \(\sigma\) such that
\(\sigma(4)=3\), \(\sigma(3)=1\), \(\sigma(1)=4\) and \(\sigma(2)=2\). %
The expressions of these symmetrizers is obtained by first writing a
{\yn} tableau with the numbers \(1\), \(2\), \(\cdots\), \(\NN\). For
instance, let us demonstrate how to get \eqref{eq:sym22}, by writing the tableau
\(\raisebox{-6pt}{\includegraphics[scale=.9]{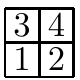}}\). Then for
each column, we 
should write the {\op} which antisymmetrizes with respect to the
corresponding indices: that gives \(\left(
1-\perm_{1,3}\right)\cdot\left(1-\perm_{2,4}
\right)\). We should also write for each line the {\op} which
symmetrizes with respect to these indices to get \(\left(
1+\perm_{1,2}\right)\cdot\left(1+\perm_{3,4}
\right)\). Finally we multiply them, and we see that \eqref{eq:sym22}
arises from the tableau
\(\raisebox{-6pt}{\includegraphics[scale=.9]{figYtab22}}\).

The construction above associates a {\rp} to each {\yn}
diagram. It can be proven\footnote{The converse is not true, and there
exist some irreducible representations which are not described by
{\yn} diagrams. 
The correct statement is that all the irreducible, polynomial
representations of {\GL\Kr} are described by {\yn} diagrams.
} that this {\rp} is actually 
irreducible. In the literature \cite{FH}, the approach
suggested above  is justified using the ``Schur-Weyl'' duality, which
relies on the fact that the {\rp}s of the symmetric group
\(\Sgrp {\NN}\) are also {\lbd} by {\yn} diagrams.

\subsection{Finite rank groups}
\label{sec:groupes-de-rang}

In the previous section, we saw how to construct irreducible {\rp}s of
\GL{{\Kr}}, and we wrote a basis for each of them (in terms of {\yn} tableaux).
One thing that can be noted is then that the  antisymmetric {\rp}
\eqref{eq:DefVantisym} has dimension \(\left(\begin{array}{cc} 
 {\Kr}\\N
   \end{array}
 \right)\).
If \(\NN={\Kr}\), we get a 1-dimensional {\rp} given by 
\(\pg[\Mnothing]=\det({\Mnothing})\). But if \({\NN}>{\Kr}\), the dimension would be
zero, {\idest} the antisymmetric {\rp} is not defined. This means that the
projector
\(\proj_{\left.\raisebox{-.55cm}{\includegraphics[scale=.2,angle=90]{figYdiag2}}\right\}{\NN}}\)
is zero as soon as \({\NN}>\Kr\).

More generally, a {\yn} diagram \(\lambda\) gives rise to a
{\rp} of \(\GL \Kr\) only if \(|\lambda|\leq\Kr\), where
\(|\lambda|\) is the number of rows in the {\yn} diagram \(\lambda\), as
defined in \eqref{eq:aDef}. Indeed, if \(|\lambda|>\Kr\), then when we
write the {\yn} symmetrizer, the factor corresponding to first column
is
\(\proj_{\left.\raisebox{-.55cm}{\includegraphics[scale=.2,angle=90]{figYdiag2}}\right\}|\lambda|}\),
which is zero.

Furthermore, the representations that we have defined are {\rp}s of
\GL{{\Kr}},
thus they also define {\rp}s of \SL{{\Kr}} and \SU{{\Kr}},
which can actually be shown to be 
irreducible.

\paragraph{Relation to the representations of {\SU 2}}
\label{sec:relation-with-teh}

Finally, let us show that the representations of {\SU 2} built in
section \ref{sec:repr-de-su2} are also described by {\yn} diagrams.
The {\rp} with spin \(0\) is the trivial {\rp}
associated to the empty {\yn} diagram\footnote{As explained in
  section \ref{sec:repr-de-su2}, this {\rp} corresponds to
  the vector space \(\bR\), and to the morphism \(\pge : {\g}\mapsto \bI\)}. Let us denote this
{\rp} by \((0)\). Then the condition 
\(\det({\Mnothing})=1\) (which is automatic for \({\Mnothing}\in SU(2)\)) shows that
\(\raisebox{-.2cm}{\includegraphics[scale=.3,angle=90]{figYdiag3}} =
(0)\). Indeed we have seen that this {\rp} obeys 
\(\pg[\Mnothing]=\det({\Mnothing})\).
One can also show that \(\raisebox{-.2cm}{\includegraphics[scale=.3]{figYdiag_21}} =
\raisebox{-.1cm}{\includegraphics[scale=.3]{figYdiag_1}}\), and this result
generalizes to bigger diagrams.
As the only possible {\yn} diagrams ({\idest} the diagrams obeying
\(|\lambda|\leq \Kr=2\)) have at most two rows ({\idest}
\(\lambda=(\lambda_1,\lambda_2,0,0,\cdots)\)), we obtain %
that each
{\rp} associated to one of these {\yn} diagrams is isomorphic
to a symmetric {\rp} \eqref{eq:DefVSym} (corresponding to
\(\lambda'=(
\lambda_1-\lambda_2,0,0,\cdots)\)). We will show {\below} that the {\rp}
with spin \({\js}\) can be identified with the symmetric {\rp}
\eqref{eq:DefVSym} associated to the {\Ydag}
\(\las[1][2\js]=(2\js,0,0,\cdots)\).
In this setup, the equality \eqref{eq:V2Irred} corresponds to the
well-known fact that the product of two spins \(1/2\) is written as
the sum of a spin \(0\) and a spin
\(1\). In the same way, \eqref{eq:Tens3Dec} implies that the product
of three spins \(1/2\) is made of two spins \(1/2\) and one spin
\(3/2\).

Let us now prove the identification between the {\rp}s of {\SU2} defined in
section \ref{sec:repr-de-su2} and the symmetric representations
\eqref{eq:DefVSym}. To do this, let us define the generators 
associated to {\rp}s of \(GL(\Kr)\). In the fundamental {\rp}
\index{Generators (of a matrix group)!of GL({\Kr})}
we can define the {\ops} \(e_{\coordi,\coordj}\)
\begin{align}
  e_{\coordi,\coordj}\nket{{\coordk}}=&\delta_{{\coordj},{\coordk}} \nket{{\coordi}}\,.
\end{align}
It means that \(e_{\coordi,\coordj}\) can be viewed as a matrix with
all coefficients equal to zero except the coefficient at position
\((\coordi,\coordj)\).

They are such that 
\begin{align}
  \GL{{\Kr}}=&\{{\Mnothing}~|~\exists \left(\phi_{\coordi,{\coordj}}\right)_{1\leq
    \coordi,{\coordj}\leq \Kr}\in \bC^{\Kr\times\Kr} :
  {\Mnothing}=\mathrm{exp}\left({\sum_{1\leq
    \coordi,{\coordj}\leq \Kr} \phi_{\coordi,{\coordj}} e_{\coordi,{\coordj}}}\right)
\}\,.
\label{eq:GLKgenerators}
\end{align}
This relation can for instance  be shown by means of a Jordan
decomposition, and it means that the {\ops} \(e_{\coordi,\coordj}\)
are the generators of {\GL{{\Kr}}} in the fundamental
{\rp}. They obey the commutation relation 
\begin{align}
  [e_{\coordi,\coordj},e_{\coordk,\coordl}]
=&
\delta_{\coordj,\coordk} e_{\coordi,\coordl} - \delta_{\coordl,\coordi} e_{\coordk,\coordj}\,.
\end{align}

By writing the action \(\pg[\Mnothing]\) of a group element \(\Mnothing=\exp
\left({\epsilon~ e_{\coordi,\coordj}}\right)\) in the {\rp}
\(\lambda\), one can easily find the expression
\(\pe[e_{\coordi,\coordj}]\) of the corresponding generator, which is
such that
\begin{align}
\label{eq:GenRep}
  \pg[\exp
\left({\epsilon~ e_{\coordi,\coordj}}\right)]=&\exp
\left({\epsilon~ \pe[e_{\coordi,\coordj}]}\right)\,.
\end{align}

For instance, in the case when \(\Kr=2\), for the {\rp}
\(\raisebox{-.1cm}{\includegraphics[scale=.3]{figYdiag3}}\), 
it is easy to write explicitly
\(\pg[\Mnothing][{\substack{~\\\includegraphics[scale=.15,angle=0]{figYdiag3}}}]\)
in the basis \eqref{eq:SymBasis}. Then if we write \eqref{eq:GenRep}
for a small \(\epsilon\) (keeping only the terms linear in \(\epsilon\)),
we deduce the following expression of the generators of {\GL 2} in the
{\rp}
\(\raisebox{-.1cm}{\includegraphics[scale=.3]{figYdiag3}}\)
\begin{align}
\label{eq:GenSym22}
  \pe[e_{1,1}][{\substack{~\\\includegraphics[scale=.15,angle=0]{figYdiag3}}}]
=&
\left(  \begin{array}{lll}     2&0&0\\0&1&0\\0&0&0   \end{array}
\right)
&
  \pe[e_{1,2}][{\substack{~\\\includegraphics[scale=.15,angle=0]{figYdiag3}}}]
=&
\left(  \begin{array}{lll}     0&\sqrt 2&0\\0&0&\sqrt 2\\0&0&0   \end{array}
\right)
\\
  \pe[e_{2,1}][{\substack{~\\\includegraphics[scale=.15,angle=0]{figYdiag3}}}]
=&
\left(  \begin{array}{lll}     0&0&0\\\sqrt 2&0&0\\0&\sqrt 2&0   \end{array}
\right)
&
  \pe[e_{2,2}][{\substack{~\\\includegraphics[scale=.15,angle=0]{figYdiag3}}}]
=&
\left(  \begin{array}{lll}     0&0&0\\0&1&0\\0&0&2   \end{array}
\right)\,.
\label{eq:GenSym22b}
\end{align}
As expected 
\(\pe[e_{{\coordi},{\coordj}}][{\substack{~\\\includegraphics[scale=.15,angle=0]{figYdiag3}}}]\)
obeys the same commutation relation as \(e_{{\coordi},{\coordj}}\):
\begin{align}
\label{eq:commgenGLK}  \comm{
  \pe[e_{\coordi,\coordj}][{\substack{~\\\includegraphics[scale=.15,angle=0]{figYdiag3}}}]}
{
 \pe[e_{\coordk,\coordk}][{\substack{~\\\includegraphics[scale=.15,angle=0]{figYdiag3}}}]}
=&
\delta_{\coordj,\coordk}
{
 \pe[e_{\coordi,\coordl}][{\substack{~\\\includegraphics[scale=.15,angle=0]{figYdiag3}}}]}
-
\delta_{\coordl,\coordi}
{
 \pe[e_{\coordk,\coordl}][{\substack{~\\\includegraphics[scale=.15,angle=0]{figYdiag3}}}]}\,.
\end{align}
One should nevertheless note that the relation 
\( e_{\coordi,\coordj} e_{\coordk,\coordl}
=
\delta_{\coordj,\coordk} e_{\coordi,\coordl}\) does not generalize to
the generators
\(\pe[e_{\coordi,\coordl}][{\substack{~\\\includegraphics[scale=.15,angle=0]{figYdiag3}}}]\). The
same result is easily obtained for an arbitrary {\yn} diagram.

In order to identify them with the generators of
\eqref{eq:DefReprSU2Gen}, we should introduce the generators
\(J^{(1)}=\frac {e_{12}+e_{21}}2\), 
\(J^{(2)}=ⅈ\frac {e_{21}-e_{12}}2\), and \(J^{(3)}=\frac
{e_{11}-e_{22}}2\). In a given {\rp}, their action becomes
\(\pe[J^{(1)}]=\frac {\pe[e_{12}]+\pe[e_{21}]}2\), 
\(\pe[J^{(2)}]=ⅈ\frac {\pe[e_{21}]-\pe[e_{12}]}2\), and \(\pe[J^{(3)}]=\frac
{\pe[e_{11}]-\pe[e_{22}]}2\), as it can be seen from
\eqref{eq:GenRep}. Then, the expressions
(\ref{eq:GenSym22},\ref{eq:GenSym22b}) 
allow to check that 
\(\pe[J^{(\coordl)}][{\substack{~\\\includegraphics[scale=.15,angle=0]{figYdiag3}}}]\)
coincides exactly with \(%
  J_1^{(\coordl)}%
\) defined in \eqref{eq:DefReprSU2Gen}. 
Thus we have shown the identification between the {\rp}
\(\raisebox{-.1cm}{\includegraphics[scale=.3]{figYdiag3}}\) and the
{\rp} with spin \(\js=1\) constructed in section \ref{sec:repr-de-su2}.

This analysis of the generators can be generalized to an arbitrary
{\Ydag}. For instance for the {\rp} 
\(\lambda=({\NN},0,0,\cdots)=\underbrace{\includegraphics[scale=.3]{figYdiag2}}_{{\NN}}\), it gives
\begin{gather}
{\pg[e_{\coordi,\coordj}][{{\includegraphics[scale=.2]{figYdiag2}}}]
}%
=
\proj_{{\includegraphics[scale=.2]{figYdiag2}}}\cdot \left(
\sum_{\spk=1}^{{\NN}} \bI^{\otimes (\spi-1)} \otimes e_{\coordi,\coordj}
\otimes \bI^{\otimes ({\NN}-\spk)}\right)\cdot
\proj_{{\includegraphics[scale=.2]{figYdiag2}}}\,,
\label{eq:=GenSym}
\end{gather}
expressed in terms of the projector \eqref{eq:DefPsym}.
Writing explicitly these generators, as in
(\ref{eq:GenSym22}-~\ref{eq:GenSym22b}), allows to prove the
identification between this {\rp} (if we restrict it to \SU
2) and the {\rp} of \SU 2 with spin \(\js={\NN}/2\).

\subsection{Characters}
\label{sec:caracteres}
\index{Representation!character}

We have seen that to each {{\Ydag}} \(\lambda\) is associated a {\rp}, and
for several examples we defined an orthonormal basis, and wrote the
morphism \(\g\mapsto\pg\).
It is then possible to compute the characters \(\cha {\lambda}\) of these {\rp}s, using
the definition 
\eqref{eq:charDef}.

To do this, the first useful relation is that for any {\rp} \(\lambda\), 
\begin{align}
  \cha {\lambda}(h\cdot {\g}\cdot h^{-1})=&\mathrm{tr}_{\lambda}\left(
\pg[h]\pg\pg[h^{-1}]\right) = \mathrm{tr}_{\lambda}\left(
\pg\pg[h^{-1}]\pg[h]\right)\nonumber \\=&
  \cha {\lambda}({\g})\,.
\label{eq:=chiClassInv}
\end{align}

This relation shows that the characters are ``class functions'', and
allows to find the character of an arbitrary \(\g\) provided we know
the characters of diagonal matrices%
.
Let us show explicitly how to express these characters %
for the symmetric {\rp}
\(\underbrace{\includegraphics[scale=.3]{figYdiag2}}_{{\NN}}\), by
starting with the particular case of a diagonal matrix \linebreak
\({\g}=\mathrm{diag}
\left(x_1,x_2,\cdots,x_\Kr\right)\in\GL\Kr\). 
In order to compute the trace of
\(\pg[][{{\includegraphics[scale=.2]{figYdiag2}}}]\), let us first note that
\begin{align}
  \pg[][{\underbrace{\includegraphics[scale=.2]{figYdiag2}}_{\NN}}]
  \raisebox{-.35cm}{\includegraphics{figYtabS}} =&\left( \prod_{\ii=1}^{\NN}
    x_{{\nn}_\ii} \right) \raisebox{-.35cm}{\includegraphics{figYtabS}}\,.
\end{align}
By summing over these basis vectors (defined in \eqref{eq:SymBasis}),
one can deduce that the character of \({\g}\) is equal to
\begin{align}
  \cha {{\includegraphics[scale=.2]{figYdiag2}}}({\g})=&\sum_{\substack{\C_1,\C_2,\cdots
    \C_\Kr\geq 0\\\C_1+\C_2+\cdots+\C_{\Kr}={\NN}}} \prod_{\jvp=1}^\Kr
    x_\jvp^{\C_\jvp}\label{eq:SymCharac}\,.
\end{align}
The relation \eqref{eq:=chiClassInv} allows to deduce the character of
an arbitrary diagonalizable matrix. Moreover the set of the diagonalizable
matrices is dense in \(\GL \Kr\), and for diagonalizable matrices the
character is a continuous function of the eigenvalues (it is even a
polynomial), hence we can deduce that for an arbitrary \(\g\in\GL\Kr\),
if we denote the eigenvalues of \(\g\) as \(x_1,x_2,\cdots,x_\Kr\), then
the character of \(\g\) is given by \eqref{eq:SymCharac}.

\paragraph{Schur polynomials and Weyl formulae}
\label{sec:schur-polyn-weyl}

For an arbitrary {\Ydag} \(\lambda\), the same procedure allows to
compute the character of any group element \(\g\in\GL\Kr\). Because of
the relation \eqref{eq:=chiClassInv}, that gives a symmetric function
of the eigenvalues of {\g}. Moreover, for any {\Ydag}, %
this character is a
polynomial function of the eigenvalues of {\g}, because the matrix
elements of the {\op} \(\pg\) are polynomial functions of the matrix
elements of {\g}. These polynomials, which are symmetric functions of
\(\Kr\) variables are called ``Schur polynomials''.\index{Schur polynomials}

Performing this analysis for an arbitrary {\Ydag} \(\lambda\) (identified to a set
of integers, as in 
\eqref{eq:DefYoung}),
yields the following expression of characters as ratios of
determinants:
\begin{align}
\label{eq:Weyl1}
  \cha \lambda({\g})=&\frac{
\Det{x_\jvp^{\lambda_\ii+\Kr-\ii}}{1\leq \ii,\jvp\leq\Kr}
}{\Det{ x_\jvp^{\Kr-\ii}}{1\leq
      \ii,\jvp\leq\Kr}}\,\,,
\end{align}
where the \(x_\jvp\) still denote the eigenvalues of \({\g}\), and this
expression holds even if {\g} is not diagonalizable.
This formula is sometimes called the ``first Weyl formula''.

Another relation holds for these characters, which can be obtained
from \eqref{eq:Weyl1}. It reads
\begin{align}
  \label{eq:Weyl2}
  \cha \lambda({\g})=&
\Det{\cha {\underbrace{\includegraphics[scale=.2]{figYdiag2}}_{\lambda_\ii+\jvp-\ii}}(\g)}{1\leq \ii,\jvp\leq|\lambda|}\,, 
\end{align}
where \(|\lambda|\) denotes the number of rows in the {\Ydag}
\(\lambda\) ({\idest} the largest integer \(a\) such that 
\(\lambda_a>0\)).
The formula \eqref{eq:Weyl2} is sometimes called the ``second Weyl
formula'', and if we insert the expression  \eqref{eq:SymCharac}
into it, it
allows to recover the expression \eqref{eq:Weyl1}.

Moreover, the expression \eqref{eq:SymCharac} can be very conveniently
recast into the following generating series:
\begin{align}
\label{eq:DefWz}
  w(z)\equiv&\sum_{s=0}^{\infty} z^s \chs s(\g)=\prod_{\jvp=1}^\Kr \frac 1
  {1-x_\jvp z}=\det \frac 1 {1- {\g}~ z}\\
\where&\chs s\equiv \cha {\underbrace{\includegraphics[scale=.2]{figYdiag2}}_s}\,.
\end{align}
This expression will be convenient to write the character of symmetric
representations, in order to plug it into the ``second Weyl Formula''
\eqref{eq:Weyl2}.

\section{
Generalization to the {\sugr} \GL{{\Kr}\texorpdfstring{\ensuremath{|}}{|}{\Mr}}}
\label{sec:gener-glkens}

\index{Supergroups!GL(KM)@GL({\Kr}\textbar {\Mr})}
\index{GL(KM)@GL({\Kr}\textbar {\Mr})}

\paragraph{Introduction to {\GLKM}}
\label{sec:introduction-glkm}

The group {\GL{\Kr\ensuremath{|}{\Mr}}} is obtained by introducing some
matrices such that
\(  \left(A\otimes \bI\right)\cdot\left(\bI\otimes B\right)=-
  \left(\bI\otimes B\right)\cdot\left(A\otimes \bI\right)\).
Two matrices  \(A\) and \(B\)  such that \(  \left(A\otimes \bI\right)\cdot\left(\bI\otimes B\right)=-
  \left(\bI\otimes B\right)\cdot\left(A\otimes \bI\right)\)  are said
  to be anti-commuting, and associated to the grading 
\(\gr{A}=\gr{B}=1\in~\bZ/2\bZ\). On the other hand, if \(A\) is such that for all \(B\), 
\(\left(A\otimes \bI\right)\cdot\left(\bI\otimes B\right)=
  \left(\bI\otimes B\right)\cdot\left(A\otimes \bI\right)\), then \(A\)
  is said to be commuting, and associated to the grading
  \(\gr{A}=0\in~\bZ/2\bZ\). We can see that with these definitions, if both \(A\) and
  \(B\) have a well-defined grading, then 
\begin{align}
\label{eq:anticomMatr}
\left(A\otimes \bI\right)\cdot\left(\bI\otimes B\right)=(-1)^{\gr{A}\gr{B}}
  \left(\bI\otimes B\right)\cdot\left(A\otimes \bI\right)\,%
  .
\end{align}

Similarly to the section \ref{sec:notat-tens-prod}, it is convenient
to introduce coordinates for {\ops}, as well as for vectors. %
In %
coordinates, the relation \eqref{eq:anticomMatr} means
that \(A^{\coordi_1}_{~\coordj_1}B^{\coordi_2}_{~\coordj_2}=\pm
B^{\coordi_2}_{~\coordj_2}A^{\coordi_1}_{~\coordj_1}\), which means that the
coordinates of matrices can be commuting or anti-commuting
variables. In what follows, we will introduce these coordinates in
such a way that the matrix multiplications and tensor products have
the same expression in terms of coordinates as for usual matrices in
{\GL \Kr}.

Let us then define a basis of vectors \(\nket{1}, \nket{2},\ldots,\nket
{\Kr+\Mr}\), and define a {\her} product such that the basis is
orthonormal, {\idest} such that \(\nbranket\mm\nn=\delta_{\mm,\nn}\). 
{\GL{\Kr\ensuremath{|}{\Mr}}} is defined by choosing an arbitrary
grading \({\ii}\mapsto \gr {\ii}\) (where \({\ii}\in\ninter 1 {\Kr+\Mr}\) and \(\gr {\ii}
\in \bZ/2\bZ\)) taking \(\Kr\) times the value \(0\) and \(\Mr\) times the
value \(1\). For instance one can choose
\begin{align}
  \gr{\coordi}=&0&\If&\coordi\in\ninter 1 \Kr\\
  \gr{\coordi}=&1&\If&\coordi\in\ninter {\Kr+1} {\Mr}\,,
\end{align}
Then we will say that \(\nbra \nn\) and \(\nket \nn\) have the grading
\(\gr \nn\), and the rule \eqref{eq:anticomMatr} implies that
\begin{gather}
  \left(\nbra{{\ii}}\otimes\nbra{{\jj}}\right)\cdot\left(\nket{{\kk}}\otimes\nket{{\lL}}\right)
  = (-1)^{\gr {{\jj}}\gr {\kk}} \delta_{{\ii},{\kk}}\delta_{{\jj},{\lL}}
\end{gather}

Let us denote by \(\vv=
\nket{\nn_1,\nn_2,\cdots,\nn_\lcds}\) the vector \(\nket
{\nn_1}\otimes \nket {\nn_2}\otimes \cdots \otimes \nket
{\nn_\lcds}\). We %
define
its coordinates as %
\(\vv^{{\coordi}_1,{\coordi}_2,\cdots,{\coordi}_\lcds}=\prod_{\spj=1}^\lcds
\theta_\spj^{\gr{\nn_\spj}}\delta^{\coordi_\spj}_{\nn_\spj} \), where
\((\theta_\spi)_{\spi=1\cdots\lcds}\) is a set of anti-commuting
variables, %
which obey the relation
\begin{gather}
\label{eq:acommtheta}
  \acomm{\theta_\spi}{\theta_\spj}\equiv \theta_\spi
  \theta_\spj+\theta_\spj \theta_\spi=\delta_{\spi,\spj}\,.
\end{gather}
These coordinates are designed to manipulate products of matrices and
vector with the same notation as for usual matrices, and the price
for that is that 
the coordinates of some vectors are anti-commuting objects.

For an arbitrary vector \(\ket {\vv}\), %
this definition of the coordinates %
means that
\begin{gather}
  {\vv}^{{\coordi}_1,{\coordi}_2,\cdots,{\coordi}_\lcds}\equiv 
\theta_1^{\gr {{\coordi}_1}}\theta_2^{\gr {{\coordi}_2}}\cdots
\theta_\lcds^{\gr {{\coordi}_\lcds}}
\nbraket{{\coordi}_1,{\coordi}_2,\cdots,{\coordi}_\lcds} {{\vv}}\,.
\end{gather}
In this definition,
\(\nbra{{\coordi}_1,{\coordi}_2,\cdots,{\coordi}_\lcds}\) is defined by
\begin{gather}
  \nbranket{{\coordi}_1,{\coordi}_2,\cdots,{\coordi}_\lcds}
  {{\coordj}_1,{\coordj}_2,\cdots,{\coordj}_\lcds}
  =\prod_{spk=1}^\lcds \delta_{{\coordi}_\spk,{\coordj}_\spk}\,.
\end{gather}
For instance if \(\lcds=2\) it means that
\begin{gather}
  \nbra{{\coordi}_1,{\coordi}_2} =(-1)^{\gr {{\coordi}_1}\gr {{\coordi}_2}}
  \nbra{{\coordi}_1} \otimes \nbra{{\coordi}_2}\,.
\end{gather}

We can also define the coordinates of an {\op} \(\mathcal{O}\) as 
\begin{multline}
  \mathcal{O}_{~{\coordj}_1,{\coordj}_2,\cdots,{\coordj}_\lcds}
  ^{{\coordi}_1,{\coordi}_2,\cdots,{\coordi}_\lcds} =\theta_1^{\gr
    {{\coordi}_1}}\theta_2^{\gr {{\coordi}_2}}\cdots 
\theta_\lcds^{\gr {{\coordi}_\lcds}} ~~ \theta_\lcds^{\gr
  {{\coordj}_\lcds}} \theta_{\lcds-1}^{\gr
  {{\coordj}_{\lcds-1}}}\cdots \theta_1^{\gr
  {{\coordj}_1}}%
\nbra{ {\coordi}_1,{\coordi}_2,\cdots,{\coordi}_\lcds
} \mathcal{O} \nket{ {\coordj}_1,{\coordj}_2,\cdots,{\coordj}_\lcds
}
\end{multline}
which is defined in such a way that
\begin{gather}
  \left(\mathcal{O}\ket
    {\vv}\right)^{{\coordi}_1,{\coordi}_2,\cdots,{\coordi}_\lcds} = 
  \mathcal{O}_{~{\coordk}_1,{\coordk}_2,\cdots,{\coordk}_\lcds}
  ^{{\coordi}_1,{\coordi}_2,\cdots,{\coordi}_\lcds} {\vv}^{{\coordk}_1,{\coordk}_2,\cdots,{\coordk}_\lcds}\,,
\end{gather}
which means that the manipulation of products (in terms of contracted
indices) is exactly the same as for usual groups.

Moreover one can show that\footnote{
In \eqref{eq:leftmm-nnrightc-coor}, the coordinates of \(A\) and \(B\)
are given by
\(A^{{\coordi}_1}_{~{\coordj}_1}=\theta_1^{\gr{{\coordi}_1} +
  \gr{{\coordj}_1} }
\nbra{ {\coordi}_1} A \nket{ {\coordj}_1}
 \)
and 
\(B^{{\coordi}_2}_{~{\coordj}_2}=\theta_2^{\gr{{\coordi}_2} +
  \gr{{\coordj}_2} }
\nbra{ {\coordi}_2} B \nket{ {\coordj}_2}
 \).
}
\begin{gather}
\label{eq:leftmm-nnrightc-coor}
    \left(A\otimes
      B\right)^{{\coordi}_1,{\coordi}_2}_{~{\coordj}_1,{\coordj}_2}
    = 
    A^{{\coordi}_1}_{~{\coordj}_1}
    B^{{\coordi}_2}_{~{\coordj}_2}\,.
\end{gather}

The group \GL{\Kr\ensuremath{|}{\Mr}},
is then the group of the invertible {\ops} acting on \linebreak
\({\Vect{\nket{1}, \nket{2},\ldots,\nket 
{\Kr+\Mr}}}\). By writing their coordinates as defined above,
we see that they are of the form 
\begin{align}
  {\Mnothing}=&\left(
  \begin{array}{ll}
    \mathcal{A} & \theta_\spi \mathcal{B}\\
    \theta_\spi \mathcal{C}& \mathcal{{\oD}}
  \end{array}
\right)
\label{eq:suMBlock}
\end{align}
where \(\mathcal{A}\), \(\mathcal{B}\), \(\mathcal{C}\)
and \(\mathcal{{\oD}}\) are
are complex matrices of respective size \(\Kr\times\Kr\), 
\(\Kr\times {\Mr}\), \({\Mr}\times \Kr\) and
\({\Mr}\times {\Mr}\).
The coefficients \(\theta_\spi\) are anti-commuting variables in the
sense of \eqref{eq:acommtheta}.
 We see that for a matrix of the form \eqref{eq:suMBlock}, the grading
 associated to the matrix element \({\Mnothing^\coordi}_\coordj\) of  is exactly
\({\gr{\coordi}+\gr{\coordj}}\).%

\paragraph{Representations of {\GLKM}}
\label{sec:representations-glkm}

We will call fundamental {\rp} of \GL{\Kr\ensuremath{|}{\Mr}}
the {\rp} defined by the space \(\Vect{\nket{1}, \nket{2},\ldots,\nket 
{\Kr+\Mr}}\) and by the morphism \(\pg[\Mnothing][{ }]=\Mnothing\). We will denote it
as 
{\raisebox{-.1cm}{\includegraphics[scale=.3]{figYdiag_1}}}.

For any matrix of this group, we define its ``super-trace'' and its
``super-determinant'' as
 \begin{align}
   \sutr \left(
  \begin{array}{ll}
    \mathcal{A} & \mathcal{B}\\
    \mathcal{C}& \mathcal{{\oD}}
  \end{array}
\right)=&\mathrm{tr}\mathcal{A}-\mathrm{tr}\mathcal{{\oD}}&
\sudet({\Mnothing})=e^{\sutr\left(\log {\Mnothing}\right)}
 \end{align}

With these definitions, one can generalize to these {\sugrs} the
construction given in section
\ref{sec:repr-tens-de}. 
First we should generalize the permutation {\ops} \(\perm\) which
appear in the definitions \eqref{eq:DefPsym}, \eqref{eq:DefVantisym},
\eqref{eq:DefVhook}, etc. 
For bosonic groups, an interesting property of \(\perm\) was that
the permutation {\op} commutes with \({\Mnothing}^{\otimes {\NN}}\).
This property was crucial as it allowed to prove that several vector
spaces were stable under the action of \(
\pg[\Mnothing]\), for all \({\Mnothing}\in
\GL\Kr\). For a {\sugr}, the definition \eqref{eq:Pgen} of the
permutation {\op} has to
be slightly modified. Indeed, 
one can check that if we  keep the definition \eqref{eq:Pgen} 
then  \(\perm_{\sigma}\cdot {\Mnothing}^{\NN}\) and
\({\Mnothing}^{\NN} \cdot \perm_{\sigma}\) are equal only up to a
sign. In order to make this sign disappear, one can define the
permutation operator as \index{Pa (permutation {\op})@\perm (permutation {\op})!generalized permutation}
\begin{gather}
\label{eq:PgenSug}
  \perm_{(\sigma)}\,:\,\nket{{\nn}_1,{\nn}_2,\cdots,{\nn}_{\lcds}}
  \mapsto
\prod_{\kk<\lL}\left(\frac{\sigma(\kk)-\sigma(\lL)}{|\kk-\lL|}\right)^{\gr{\nn_\kk}+\gr{\nn_\lL}}     \nket{{\nn}_{\sigma(1)},\cdots,{\nn}_{\sigma(\lcds)}}\,.
\end{gather}

With this definition of the generalized permutation, one can associate
representations of \GL{{\Kr}\ensuremath{|}{\Mr}} to {\yn} diagrams. This is done as in
section \ref{sec:repr-tens-de} by considering tensor products of the
form \(\underbrace{\raisebox{-.1cm}{\includegraphics[scale=.3]{figYdiag_1}}  ⊗ 
\raisebox{-.1cm}{\includegraphics[scale=.3]{figYdiag_1}} ⊗ 
\cdots ⊗ 
\raisebox{-.1cm}{\includegraphics[scale=.3]{figYdiag_1}}}_{\NN}\) (where
{\raisebox{-.1cm}{\includegraphics[scale=.3]{figYdiag_1}}} denotes the
fundamental {\rp}), and restricting it to the image of
combinations of the permutation {\ops} generalized according to
\eqref{eq:PgenSug}. This gives rise to a set of irreducible
representations of \GL{{\Kr}|{\Mr}}.

Unlike the \GL{{\Kr}} case, there also exist other polynomial irreducible
representations of \GL{{\Kr}|{\Mr}}, because 
the fundamental {\rp} is not
unique. Indeed, 
we have chosen to define the fundamental {\rp} %
{\raisebox{-.1cm}{\includegraphics[scale=.3]{figYdiag_1}}} 
as a set of vectors 
who
have {\Kr} coordinates with grading \(\sg \nn =+1\) and 
{\Mr} coordinates with grading \(\sg \nn =-1\)%
. But one could also
consider a set
{\raisebox{-.1cm}{\includegraphics[scale=.3]{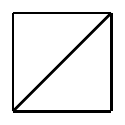}}} of vectors 
who
have {\Mr} coordinates with grading \(\sg \nn =+1\) and 
{\Kr} coordinates with grading \(\sg \nn =-1\).
Several {\rp}s
can then be built from tensor products involving both
{\raisebox{-.1cm}{\includegraphics[scale=.3]{figYdiag_1}}} and
{\raisebox{-.1cm}{\includegraphics[scale=.3]{figYdiag_1s}}}, which
makes the {\rp} theory of \GL{{\Kr}\ensuremath{|}{\Mr}} richer than for \GL{{\Kr}}. In
the present manuscript, tensor products involving
{\raisebox{-.1cm}{\includegraphics[scale=.3]{figYdiag_1s}}} will not
be considered, and we will restrict to representations described by
usual {\yn} diagrams.

For these representations, 
we can define characters (like in section \ref{sec:caracteres}, except
that the character should now be defined as a super-trace), and
one can show \cite{Baha} that 
\eqref{eq:DefWz} is generalized as follows~:
\begin{align}
    w(z)\equiv&\sum_{s=0}^{\infty} z_s \chs s=\sudet\left( \frac 1 {1-
        {\g}~ z}\right)\,
\end{align}
which allows to find the characters of arbitrary representations using
the second Weyl formula \eqref{eq:Weyl2}. On the contrary, the first
Weyl formula \eqref{eq:Weyl1} does not hold in the case of {\sugrs}.

\paragraph{``{\fat}-{\hook}'' condition}
\label{sec:{\fat}-{\hook}-condition}

For usual groups we saw that only the {\yn} diagrams with less
than {\Kr} rows gave rise to representations of %
{\GL \Kr}. It is interesting to see how this condition generalizes to
{\GL{\Kr\ensuremath{|}{\Mr}}}:
for {\sugrs}, the projector
\(\proj_{\left.\raisebox{-.55cm}{\includegraphics[scale=.2,angle=90]{figYdiag2}}\right\}{\NN}}
= \frac
1 {{{\NN}!}}\sum_{\sigma
  \in \Sgrp {\NN}} \epsilon(\sigma)~ \perm_\sigma\) (associated to
the {\rp} \(\left.\raisebox{-.55cm}{\includegraphics[scale=.2,angle=90]{figYdiag2}}\right\}{\NN}\)) does not vanish
when \({\NN}>\Kr\) because \(\perm\) itself contains a sign. This sign is such
that the indices with grading \(\sg {\ivp}=+1\) are antisymmetrized and
the indices with grading \(\sg {\ivp}=-1\) are symmetrized. It can be
shown \cite{Deguchi:1991aq,1997JPhA...30.7975T} that the Young
diagrams which give rise to representations of
{\GL{\Kr\ensuremath{|}{\Mr}}} are the diagrams such that
\(\lambda_{\Kr+1}\leq \Mr\).%
They are represented in figure \ref{fig:FatHook} page
\pageref{fig:FatHook}.

\index{Representation|)}
%
%

%%% Local Variables: ***
%%% mode:latex ***
%%% TeX-master: "english.tex" ***
%%% End: ***

%% file: diagrammatics.tex
\section{Diagrammatic expressions for {\cdrs}}
\label{sec:diagr-expr-co}

This section will explain how to explicitly compute expressions
involving {\cdrs}, mainly by using the Leibniz rule
\eqref{eq:codLeibnitz}. We will see that the repeated action of
{\cdrs}, computed through this Leibniz rule, gives rise to
diagrammatic expressions.

As indicated by expression \eqref{eq:TfromD}, we will be specifically
interested in {\cdrs} acting on characters. As it can be seen
in \eqref{eq:Weyl2}, arbitrary characters are linear combinations of
products of characters of symmetric representations. Moreover the
characters of symmetric representations are simply encoded into the
function \(w(z)\) defined in \eqref{eq:DefWz}. We will therefore focus
on {\cd} acting on \(w(z)\) or on products like \(w(x)w(y)w(z)\).

\paragraph{First properties of the {\cdrs}}
\label{sec:first-prop-coder}

Let us first remind a few simple properties of the {\cdrs}, defined in
chapter \ref{part:qoperatorsspin}. This {\cd} is defined by 
\Pv{\begin{empheq}[box=\fbox]{align}
  \hD \otimes f(\g)&\equiv \parDer {\phi^{^t}} \otimes \left.f\left(e^{\phie%
      } \g\right)\right|_{\phi=0}%
\label{eq:DefDg}\\
&=\sum_{\alpha,\beta}e_{\alpha,\beta}\otimes  \left.\left(
\parDer {\phi_{\beta,\alpha}}f\left(e^{\sum_{\gamma,\delta}
 e_{\gamma,\delta}\phi_{\gamma,\delta}}{\g}\right)
\right)\right|_{\phi\to 0}\,,
\end{empheq}}
where \(\phi\in \Mat \Kr %
\) is a %
\(\Kr\times \Kr\) 
matrix and
{\phie}
denotes \(\sum_{\alpha,\beta} e_{\alpha,\beta}\phi_{\alpha,\beta}\),
where the  \(e_{\alpha,\beta}\) are the generators of \(GL(\Kr)\),
introduced in the appendix \ref{sec:repr-tens-de}
(they are matrices with one single non-zero coefficient at position \(\alpha,\beta\)).

From this definition, 
let us show how to compute \(\hD w(z)\).
As a first step, the simplest explicit computation one can do is the {\cd} of \(\g\)
itself :
\begin{align}
  \hD\otimes \g=&\sum_{\alpha,\beta}
  e_{\alpha,\beta}\otimes\left(\left.\parDer
      {\phi_{\beta,\alpha}}\left(1+\phie+\frac {\phie^2}
        2+\cdots\right).\g\right|_{\phi\to 0}\right)\nonumber\\
=&\sum_{\alpha,\beta}
  e_{\alpha,\beta}\otimes \left(e_{\beta,\alpha}\cdot \g\right)
=\perm_{1,2}\cdot (\bI \otimes \g)\,.
\label{eq:=sum_-beta-e_alpha}
\end{align}

The next simplest thing that we can compute is \( \hD\otimes \g^{\nn}\),
and in order to write it, we need to use the Leibniz rule
\eqref{eq:codLeibnitz}
 {\below} :
\begin{align}
\hD \otimes \left(f_{1}({\g}) \cdot f_{2}({\g})\right)=&\cb{\hD \otimes
  f_{1}(\g)} \cdot \left(\bI \otimes f_{2}(\g)\right) + 
\left(\bI \otimes f_{1}(\g)\right) \cdot \cb{\hD \otimes
  f_{2}(\g) }\,.
\end{align}
From the expression \eqref{eq:=sum_-beta-e_alpha}, this Leibniz rule
allows to deduce
iteratively that
\begin{align}
  \hD\otimes \g^{\nn}=&\perm_{1,2}\cdot \left(\sum_{{\kk}=1}^{\nn} \g^{{\nn}-{\kk}}\otimes \g^{\kk}\right)\,.
\end{align}
As the trace is linear, we can %
also easily compute
\begin{align}
  \hDt \mathrm{tr}(\g^\nn)=&\mathrm{tr}_2\left(\hD \otimes \g^\nn\right)=
\nn~ %
\g^\nn%
  \,,
\end{align}
because \(\mathrm{tr}_2\left(\perm_{1,2}\cdot\left(A\otimes
    B\right)\right)=%
  B\cdot A
\). Then we get
\begin{align}
  \hDt \mathrm{tr}\left(\log (1-\g~z)\right)=&-\sum_{\nn\geq 1}\hDt
  \mathrm{tr}
  \frac{\left(\g~z\right)^\nn}{\nn}=-\sum_{\nn\geq 1}
  {\left(\g~z\right)^\nn}%
=-\frac {\g~z}{1-\g~z}
\end{align}
And finally, we can compute the derivative of
\(w(z)=e^{-\mathrm{tr}\left(\log (1-\g~z)\right)}\) :
  \begin{align}
    \hDt w(z)=&-\left[ \hDt \mathrm{tr}\left(\log
        (1-\g~z)\right)\right]\cdot e^{-\mathrm{tr}\left(\log
        (1-\g~z)\right)} = \frac {\g~z}{1-\g~z} w(z)\,.
\label{eq:Dw=fw}
  \end{align}

Now, let us see the effect of multiple successive {\cdrs},
using the Leibniz rule \eqref{eq:codLeibnitz}:
\begin{align}
  \hD \otimes \hDt w(z)=&\hD \otimes \left( \frac {\g~z}{1-\g~z} w(z)
  \right) \\=& \cb{\hD \otimes \frac {\g~z}{1-\g~z}} w(z)
+ \cb{\hDt w(z)\otimes \bI}\cdot\left(\bI \otimes \frac
  {\g~z}{1-\g~z}\right)\\=&
\left( \frac {\g \, z}{1- \g \, z} \otimes \frac {\g \, z}{1- \g
  \, z} +\perm_{1,2}\cdot\left(\frac {1}{1- \g \, z} \otimes \frac
{\g \, z}{1- \g \, z}\right)\right)w(z)
\label{eq:DDw=ff+Pff}
\end{align}
where we used 
\begin{align}
  {\hD \otimes \frac {\g~z}{1-\g~z}}&=\sum _{\nn\geq 1}\hD\otimes
  \left(\g~z\right)^\nn
=\perm_{1,2}\cdot\sum _{\substack {{\mm}\geq 0\\p\geq 1}}\g^{\mm}\otimes \g^p\\
=&\perm_{1,2}\cdot\left(\frac{1}{1-\g~z}\otimes
  \frac{\g~z}{1-\g~z}\right)\,.
\label{eq:=Dfrac}
\end{align}

\paragraph{Expression of  \(\hD^{\otimes \lcds} w(z)\) from \(\hD\)-diagrams}
\label{sec:hd-diagrams}

Let us now write the
relations \eqref{eq:Dw=fw} and \eqref{eq:DDw=ff+Pff} at the {\level} of
coordinates, and introduce diagrams summarizing these relations :
\begin{align}
\label{eq:Diag1}  \left(\hDt w(z)\right)_{~{\coordj}}^{{\coordi}}=&\left(\frac
    {\g~z}{1-\g~z}\right)_{~{\coordj}}^{{\coordi}} w(z)\equiv \raisebox{-15pt}{\MyLine[{jj1/ii1}][{}][{}]}~w(z)\\
 \left(\hD \otimes \hDt
   w(z)\right)_{~{\coordj}_1,{\coordj}_2}^{{\coordi}_1,{\coordi}_2}=&
\left(\left(\frac
    {\g~z}{1-\g~z}\right)_{~{\coordj}_1}^{{\coordi}_1} \left(\frac
    {\g~z}{1-\g~z}\right)_{~{\coordj}_2}^{{\coordi}_2}\right.\nonumber\\
&~~ +\left.
\left(\frac
    {1}{1-\g~z}\right)_{~{\coordj}_1}^{{\coordi}_2} \left(\frac
    {\g~z}{1-\g~z}\right)_{~{\coordj}_2}^{{\coordi}_1}
\right) w(z)\\
\equiv&
\left(\raisebox{-15pt}{\MyTwoNodes[{jj1/ii1, jj2/ii2}][{}][{}]}+
\raisebox{-15pt}{\MyTwoNodes[{jj2/ii1}][{jj1/ii2}][{}]}
\right)w(z)\,.
\label{eq:Diag2}
\end{align}
In this notation, the dots are {\lbd} by the indices \({\coordi}_{\spk}\) and
\({\coordj}_{\spk}\), and solid lines connecting them correspond to the {\op} \(\frac
    {\g~z}{1-\g~z}\), while dashed line correspond to the {\op} \(\frac
    {1}{1-\g~z}\). %

We will call \(\hD\)-diagrams these pictures which stand for {\ops} :
for instance the \(\hD\)-diagram \(\MyTwoNodes[{jj2/ii1}][{jj1/ii2}]\)
will stand for the {\op} \(\perm_{1,2}\cdot\left(\frac{1}{1-\g~z}\otimes
  \frac{\g~z}{1-\g~z}\right)\) which has coordinates 
\(\raisebox{-15pt}{\MyTwoNodes[{jj2/ii1}][{jj1/ii2}][{}]}\equiv \left(\frac
    {1}{1-\g~z}\right)_{~{\coordj}_1}^{{\coordi}_2} \left(\frac
    {\g~z}{1-\g~z}\right)_{~{\coordj}_2}^{{\coordi}_1}\).

As we will see, these \(\hD\)-diagrams actually allow to generalize the
results \eqref{eq:Diag1} and \eqref{eq:Diag2} to an arbitrary number
of spins. For instance we will get, for 3 spins, 
\begin{align}
\label{eq:Diag3}
\lefteqn{\left(\hD^{\otimes
    3}w(z)\right)_{~{\coordj}_1,{\coordj}_2,{\coordj}_3}^{{\coordi}_1,{\coordi}_2,{\coordi}_3}  }~~\nonumber\\&=\left(
\raisebox{-15pt}{\MyThreeNodes[{jj1/ii1,jj2/ii2,jj3/ii3}][{}][{}]}
    +
    \raisebox{-15pt}{\MyThreeNodes[{jj1/ii1,jj3/ii2}][{jj2/ii3}][{}]}
+\raisebox{-15pt}{\MyThreeNodes[{jj3/ii3,jj2/ii1}][{jj1/ii2}][{}]}
+\raisebox{-15pt}{\MyThreeNodes[{jj2/ii2,jj3/ii1}][{jj1/ii3}][{}]}
+\raisebox{-15pt}{\MyThreeNodes[{jj3/ii2,jj2/ii1}][{jj1/ii3}][{}]}
+\raisebox{-15pt}{\MyThreeNodes[{jj3/ii1}][{jj1/ii2,jj2/ii3}][{}]}
\right) w(z)\,,
\end{align}
from where one easily finds the generalization to {\lcds} spins :
\(\hD^{\otimes \lcds}w(z)\) contains \(\lcds !\) terms, corresponding to the
\(\lcds !\) permutations \(\sigma\in\Sgrp \lcds\), where 
\(\Sgrp \lcds\) denotes the
\index{S (symmetric group)@\ensuremath{\mathcal{S}} (symmetric group)}
  set of all permutations of \(\ninter 1 \lcds\).
For each given permutation, the corresponding \(\hD\)-diagram is obtained by
connecting \({\coordj}_\coordk\) to \({\coordi}_{\sigma(\coordk)}\)
 through a dashed line if \(\sigma(\coordk)>\coordk\) and through a
 solid line otherwise. The corresponding expression
 \begin{align}
\label{eq:DLw}
\framedline{   {\left(\hD^{\otimes
    \lcds} w(z)\right)_{~{\coordj}_1,{\coordj}_2,\cdots,{\coordj}_\lcds}^{{\coordi}_1,{\coordi}_2,\cdots,{\coordi}_\lcds}
}=}%
{\sum_{\sigma\in \Sgrp \lcds} \prod_{\spk=1}^\lcds
\left(\frac{\left(\g~z\right)^{\theta({\spk}-\sigma(\spk))}}{1-\g~z}\right)_{\coordj_\spk}^{\coordi_{\sigma(\spk)}}
w(z)}\\
\where \theta(\nn)=&\left\{
  \begin{array}{ll}
    1&\If \nn\geq 0\\
    0&\mathrm{otherwise}
  \end{array}
\right.
 \end{align}
can be proven by recurrence over {\lcds}.%

\begin{proof}
  To perform this recurrence, one assumes \eqref{eq:DLw} and length
  \(\lcds\)  and gets
  from \(\lcds\) to \(\lcds+1\) using the Leibniz rule to describe the
  action of a {\cd}  :
  \begin{align}
    \lefteqn{ \hD_{~\coordj_1}^{\coordi_1} {\left(\hD^{\otimes \lcds}
          w(z)\right)_{~{\coordj}_2,{\coordj}_3,\cdots,{\coordj}_{\lcds+1}}^{{\coordi}_2,{\coordi}_3,\cdots,{\coordi}_{\lcds+1}}
      }}~~\nonumber\\&=\hD_{~\coordj_1}^{\coordi_1} \sum_{\sigma\in
      \Sgrp \lcds} \prod_{\spk=1}^\lcds
    \left(\frac{\left(\g~z\right)^{\theta(\spk-\sigma(\spk))}}{1-\g~z}\right)_{\coordj_{\spk+1}}^{\coordi_{\sigma(\spk)+1}}
    w(z)\\
    &= \left(\frac{\g~z}{1-\g~z}\right) _{~\coordj_1}^{\coordi_1} \sum_{\sigma\in
      \Sgrp \lcds} \prod_{\spk=1}^\lcds
    \left(\frac{\left(\g~z\right)^{\theta(\spk-\sigma(\spk))}}{1-\g~z}\right)_{\coordj_{\spk+1}}^{\coordi_{\sigma(\spk)+1}}
    w(z)\nonumber\\
    &~+\sum_{\sigma\in \Sgrp \lcds} \sum_{\spk=1}^\lcds
    \left(\frac{\g~z}{1-\g~z}\right)_{\coordj_{\spk+1}}^{\coordi_1}
    \left(\frac{1}{1-\g~z}\right)_{\coordj_{1}}^{\coordi_{\sigma(\spl)+1}}
    \prod_{\spl\neq\spk}
    \left(\frac{\left(\g~z\right)^{\theta(\spl-\sigma(\spl))}}{1-\g~z}\right)_{\coordj_{\spl+1}}^{\coordi_{\sigma(\spl)+1}}
    w(z)
    \label{eq:recDLwO3}\,,
  \end{align}
where we also use the notation
\( \hD_{~\coordj_1}^{\coordi_1} \hD_{~\coordj_2}^{\coordi_2}\cdots
\hD_{~\coordj_\lcds}^{\coordi_\lcds} f(\g)\equiv \left(\hD^{\otimes
    \lcds} f(\g)\right) _{~{\coordj}_1,{\coordj}_2,\cdots,{\coordj}_{\lcds}}^{{\coordi}_1,{\coordi}_2,\cdots,{\coordi}_{\lcds}}
   \).

  In \eqref{eq:recDLwO3}, the first line %
  is the contribution of
  the {\cd} to the left acting %
  on \(w(z)\). It can be rewritten as
  \[\sum_{\substack{\sigma\in \Sgrp {\lcds+1}\\\sigma(1)=1}}
  \prod_{\spk=1}^{\lcds+1}
  \left(\frac{\left(\g~z\right)^{\theta({\spk}-\sigma(\spk))}}{1-\g~z}\right)_{\coordj_\spk}^{\coordi_{\sigma(\spk)}}
  w(z).\]
  The second line is the contribution of the {\cd} to the
  left acting on the \(\spk^{\mathrm{th}}\) factor of \eqref{eq:DLw}. It
  is obtained by noticing that 
\(\hD \frac 1 {1-\g~z}=\hD \frac {\g~z}
  {1-\g~z}\), and by writing \eqref{eq:=Dfrac} in coordinates. This term
  is equal to
  \[\sum_{\spk=1}^{\lcds}\sum_{\substack{\sigma\in
      \Sgrp {\lcds+1}\\\sigma(\spk+1)=1}}
  \prod_{\spk=1}^{\lcds+1}
  \left(\frac{\left(\g~z\right)^{\theta({\spk}-\sigma(\spk))}}{1-\g~z}\right)_{\coordj_\spk}^{\coordi_{\sigma(\spk)}}
  w(z).\]
 Grouping these terms together, we get \eqref{eq:DLw} for
  \(\lcds\to\lcds+1\), which proves the relation by a recurrence which
  starts from the most case \(\lcds=0\).
\end{proof}

\paragraph{generalization to \(\su_\spi\neq 0\)}
\label{sec:gener-us_sp-0}

A first generalization of this result is to express the {\op} 
\(\left[%
\DLt
w(z)\right]\) : this is not complicated and for instance at
\(\lcds=2\), it is obtained as \((\su_1+\hD)\otimes
(\su_2+\hD)w(z)=\su_1 \su_2 \bI w(z) +\su_1 \bI\otimes
\left[\hDt w(z)\right]+\su_2 
\left[\hDt w(z)\right]\otimes \bI+\left[\hD\otimes\hDt w(z)\right]\).

For arbitrary \(\lcds\), we get the expression 
\begin{align}
\label{eq:su+DLw}
  \left[\DLt %
w(z)\right]=&\sum_{\sigma\in \Sgrp \lcds} \prod_{\spk=1}^\lcds\left(
{\su}_\spk\delta^{\sigma(\spk)}_{\spk}\delta^{\coordi_\spk}_{\coordj_\spk}+\left(\frac{\left(\g~z\right)^{\theta({\spk}-\sigma(\spk))}}{1-\g~z}\right)_{\coordj_\spk}^{\coordi_{\sigma(\spk)}}\right)
w(z)\,.
\end{align}

In particular, in the case \(\forall \spi, ~ {\su}_\spi=1\), %
this simplifies to 
\begin{align}
\label{eq:1+DLw}
  \left[\left(1+\hD\right) ^{\otimes\lcds}
w(z)\right]=&\sum_{\sigma\in \Sgrp \lcds} \prod_{\spk=1}^\lcds%
\left(\frac{\left(\g~z\right)^{\theta({\spk}-\sigma(\spk)-1)}}{1-\g~z}\right)_{\coordj_\spk}^{\coordi_{\sigma(\spk)}}%
w(z)\,,
\end{align}
which means, in terms of \(\hD\)-diagrams
(\ref{eq:Diag1},~\ref{eq:Diag2},~\ref{eq:Diag3}), that the vertical
lines become dashed instead of solid.

\paragraph{Generalization to \(\left[
\DLt
 w(z_1)\cdots w(z_\nn)\right]\)}
\label{sec:gener-left-dlt}

Let us now
generalize %
the equation \eqref{eq:su+DLw}%
to
the {\op}
\({\Wt[z_1,\cdots,z_\nn]\equiv}{\left[
\DLt
 w(z_1)\cdots w(z_\nn)\right]}\). First, if \(\lcds=1\), we
get
\begin{align}
  \left[\left(\su_1+\hD\right)~~ w(z_1)\cdots w(z_\nn)\right]=&
\left(\su_1+\sum_{\kk=1}^{\nn} \frac{\g~z_\kk}{1-\g~z_\kk}\right)w(z_1)\cdots w(z_\nn)\,,
\end{align}
by using a Leibniz rule where the {\cd} can either act on
\(w(z_1)\) or on \(w(z_2)\) or on any other \(w(z_\kk)\). This can be
represented diagrammatically as 
\begin{align}
  \left[\left(\su_1+\hD\right)~~ w(z_1)\cdots w(z_\nn)\right]=&
\left(
\MyLine[{}][{}][\forget][{jj1/ii1}] +
\sum_{\kk=1}^{\nn} 
\raisebox{-13pt}{\MyLine[{jj1/ii1}][{}][\forget][][{jj1.south/\kk}]}
\right)w(z_1)\cdots w(z_\nn)\,,
\end{align}
where
\(\raisebox{-13pt}{\MyLine[{jj1/ii1}][{}][\forget][][{jj1.south/\kk}]}\)
denotes the {\op} \(\frac{\g~z_\kk}{1-\g~z_\kk}\), and 
 a line \(\MyLine[{}][{}][\forget][{jj1/ii1}]\) at position \(\spi\)
denotes the {\op} \(\su_\spi \bI\) (here \(\spi=1\)).
For two spins the action of the next {\cd} gives
\begin{align}
\lefteqn{  \left[\DLt[][][2]
~~ w(z_1)\cdots w(z_\nn)\right]}\hspace{1cm}\nonumber\\
=&
\left(
\MyTwoNodes[{}][{}][\forget][{jj1/ii1,jj2/ii2}]
+\sum_{\kk=1}^\nn
\left(\raisebox{-13pt}{\MyTwoNodes[{jj1/ii1}][{}][\forget][{jj2/ii2}][{jj1.south/\kk}]}
+ \raisebox{-13pt}{\MyTwoNodes[{jj2/ii2}][{}][\forget][{jj1/ii1}][{jj2.south/\kk}]}
\right)
+\sum_{1\leq \kk,\kk'\leq \nn}
\raisebox{-13pt}{\MyTwoNodes[{jj1/ii1,jj2/ii2}][{}][\forget][{}][{jj1.south/\kk,jj2.south/\kk'}]}
+\sum_{\kk=1}^\nn
\raisebox{-13pt}{\MyTwoNodes[{jj2/ii1}][{jj1/ii2}][\forget][{}][{jj1.south/\kk,jj2.south/\kk}]}
\right)w(z_1)\cdots w(z_\nn)\,,
\label{eq:diag2Wtlots}
\end{align}
where the last term arises from \(\left[\hD \otimes 
\sum_{\kk=1}^{\nn} 
\raisebox{-13pt}{\MyLine[{jj1/ii1}][{}][\forget][][{jj1.south/\kk}]}
\right]\), using the relation \({\left[\hD \otimes 
\raisebox{-13pt}{\MyLine[{jj1/ii1}][{}][\forget][][{jj1.south/\kk}]}
\right]=\raisebox{-13pt}{\MyTwoNodes[{jj2/ii1}][{jj1/ii2}][\forget][{}][{jj1.south/\kk,jj2.south/\kk}]}}\)
where
\(\raisebox{-13pt}{\MyLine[{}][{jj1/ii1}][\forget][][{jj1.south/\kk}]}\)stands
for the {\op} \(\frac{1}{1-\g~z_\kk}\). On the other hand, the first
terms correspond to \(\left(
\MyLine[{}][{}][\forget][{jj1/ii1}] +
\sum_{\kk=1}^{\nn} 
\raisebox{-13pt}{\MyLine[{jj1/ii1}][{}][\forget][][{jj1.south/\kk}]}
\right)\otimes \left(
\MyLine[{}][{}][\forget][{jj1/ii1}] +
\sum_{\kk=1}^{\nn} 
\raisebox{-13pt}{\MyLine[{jj1/ii1}][{}][\forget][][{jj1.south/\kk}]}
\right)w(z_1)\cdots w(z_\nn)\), which arises if all {\cdrs} act
directly on \(w(z_1)\cdots w(z_\nn)\). In this notation, the \(\hD\)-diagram
\(\raisebox{-13pt}{\MyTwoNodes[{jj1/ii1}][{}][\forget][{jj2/ii2}][{jj1.south/\kk}]}\)
(for instance) denotes the {\op} \(\frac{\g~z_\kk}{1- \g~z_\kk}\otimes
(\su_2 \bI)\).

This expression can be generalized for \(\lcds\) spin, where it is
expressed as a sum of \(\hD\)-diagrams. We saw above that the
expression \(%
  \hD^{\otimes \lcds} w(z)%
\) can be written as a
sum of \(\hD\) diagrams obeying certain rules (there is one
\(\hD\)-diagram for each permutation, \linebreak and a  rule says what line is
dashed or solid). This sum can also be explicitly \linebreak expressed  more
mathematically by the expression \eqref{eq:DLw}. For the {\op}
\linebreak
 \({\Wt[z_1,\cdots,z_\nn]\equiv}{\left[
\DLt
 w(z_1)\cdots w(z_\nn)\right]}\), it is more complicated to write an
explicit expression like \eqref{eq:DLw}, but it is easy to describe
what \(\hD\)-diagrams should be summed (this description is convenient
for instance in order to write these expressions for arbitrary \(\lcds\)
on a computer). 
To compute \(\Wt[z_1,\cdots,z_\nn]\), one should sum all the %
\(\hD\)-diagrams %
such that :
\begin{itemize}
\item All vertical lines are either double lines (associated to
  \(\su_{\spi}\bI\)) or solid lines
  \(\raisebox{-13pt}{\MyLine[{jj1/ii1}][{}][\forget][][{jj1.south/\kk}]}\)
  for a given value of \(\kk\).
\item The slant lines are solid if they go up to the left, or dashed
  if they go up to the right. They are associated to a given value of
  \(\kk\).
\item The permutation\footnote{We remind here that in the case of the
    expression \(%
  \hD^{\otimes \lcds} w(z)%
\), one single diagram was associated to each permutation (see for
instance \eqref{eq:Diag3}). By contrast, in the present case, several
different diagrams may be associated to the same permutation.
} \(\sigma\) can be decomposed into ``cycles'', 
  which are the minimal subsets of \(\ninter 1 \lcds\) stable under
  \(\sigma\).
  For instance, for two spins (\(\lcds=2\)), the identity permutation \(\bo\) has two
  cycles \((1)\) and \((2)\), whereas the permutation \(\tau_{[1,2]}\) has
  one single cycle \((1,2)\).%

  From the point of view of the Leibniz rule, the cycle \((1,2)\)
  arises in \eqref{eq:diag2Wtlots} if the derivative
  \(\hD_{~\coordj_2}^{\coordi_2}\) (associated to the second site of the
  {\cds}) acts on a given \(w(z_\kk)\) to give rise to
  \(\left(\frac{\g~z_\kk}{1-\g~z_\kk}\right)_{~{\coordj}_2}^{{\coordi}_2}\), and then the
  derivative \(\hD_{~\coordj_1}^{\coordi_1}\) (associated to the first site of the
  {\cds}) acts on this to produce 
  \(\left(\frac
    {1}{1-\g~z_\kk}\right)_{~{\coordj}_1}^{{\coordi}_2} \left(\frac
    {\g~z_\kk}{1-\g~z_\kk}\right)_{~{\coordj}_2}^{{\coordi}_1}
  \), which diagrammatically corresponds to two lines associated
  to the same value of \(\kk\). As a consequence, the \(\hD\)-diagram \(
\raisebox{-13pt}{\MyTwoNodes[{jj2/ii1}][{jj1/ii2}][\forget][{}][{jj1.south/\kk,jj2.south/\kk'}]}\)
can only arise for \(\kk=\kk'\).

  The generalization of these constraints for %
  an arbitrary number \(\lcds\) of spins is that for every cycle, all lines
  should be associated to the same \(\kk\), because in the Leibniz rule,
  they correspond to derivatives acting on the derivative of a single
  \(w(z)\) factor.
\end{itemize}

The rules above %
allow to compute explicitly the {\op} 
\(\Wt[z_1,\cdots,z_\nn]\), as it can be 
proven by the same recurrence as in the proof of
\eqref{eq:su+DLw}.
This expression is
suitable to analyze its analyticity properties (poles structure), or to
do explicit computations on a computer.

Moreover these rules can be generalized to 
\({  \left[\DLt[][][\lcds]
~~ w(z_1)^{\alpha_1}\cdots w(z_\nn)^{\alpha_\nn}\right]}\). The only
difference with the discussion above is that each \(\hD\)-diagram is
then multiplied by the factor \(\prod_{\kk=1}^{\nn} (\alpha_\kk)^{\nn_\kk}\)
where \(\nn_\kk\) is the number of cycles containing lines associated to
the label \(\kk\).

\section{Identities involving {\cdrs}}
\label{sec:ident-involv-co}

From the \(\hD\)-diagrams introduced above, one can deduce identities such as
the identity \eqref{eq:DD.1+DD=1+DD.DD}, proven in section
\ref{sec:proof-cbr-formula}.

In the next paragraphs, we will explain how to deduce \su-dependent versions
of equations like \eqref{eq:DD.1+DD=1+DD.DD}, and for that  we will
also use the following nice consequence of the 
Leibniz rule (and of the relation\footnote{The relation
  \(\hDt \det \g = \bI \det \g\) can be proven by the same elementary
  methods as the equations
  (\ref{eq:=sum_-beta-e_alpha}~\ref{eq:Dw=fw}) proven at the beginning
  of this section. } \(\hDt \det \g = \bI \det \g\)) :
\begin{align}
  \left[\hD^{\otimes \lcds} A(\g) \det \g\right]=&\det \g
  \left[\left(1+\hD\right)^{\otimes \lcds} A(\g)\right]\,,
\label{eq:detshift}
\end{align}
which holds for an arbitrary function \(A(\g)\).

\subsection{Yang-Baxter Equation}
\label{sec:yang-baxter-equation}

First, it can be instructing to rewrite the Yang-Baxter equation
\eqref{eq:YBlambda} in terms of {\cdrs} : indeed, as we saw that
the {\Rop}-matrices can be expressed in terms of {\cdrs}, we can
expect that the Yang-Baxter identity itself is nothing but an identity
on {\cdrs}.

The Yang-Baxter equation \eqref{eq:YBlambda} can be rewritten as
\begin{gather}
\Rop_{\spi,\spj}(\su_\spj-\su_\spi)\Rop_{\spj,\lambda}(\su_\spj)\Rop_{\spi,\lambda}(\su_\spi)= \Rop_{\spi,\lambda}(\su_\spi)\Rop_{\spj,\lambda}(\su_\spj)\Rop_{\spi,\spj}(\su_\spj-\su_\spi) \,,
\end{gather}
or, equivalently\footnote{%
    The equivalence of these equations relies on
  the relation
  \(\perm_{\spi,\spj}\perm_{\spi,\lambda}=\perm_{\spj,\lambda}\perm_{\spi,\spj}\).}
\begin{gather}
\Rop_{\spi,\spj}(\su_\spj-\su_\spi)\perm_{\spi,\spj}\Rop_{\spi,\lambda}(\su_\spj)\Rop_{\spj,\lambda}(\su_\spi)=  \Rop_{\spi,\lambda}(\su_\spi)\Rop_{\spj,\lambda}(\su_\spj)\Rop_{\spi,\spj}(\su_\spj-\su_\spi)\perm_{\spi,\spj},.
\end{gather}

Using the relation \eqref{eq:LDef=DDD} between {\cdrs} and
{\Roprs}, this relation becomes (for \(\spj<\spk\))
\begin{multline}
 \label{eq:YB1}
 {(1+(\su_\spk-\su_\spj)\perm_{\spj,\spk}) \cdot  \left[\DLt 
\cha \lambda(\g)\right]}\\
=\left[\DLt[{\su}_{\tau_{[\spj,\spk]}(\spi)}+\hD]
\cha \lambda(\g)\right]\cdot (1+(\su_\spk-\su_\spj)\perm_{\spj,\spk})\,.
\end{multline}
where the transposition \(\tau_{[\spj,\spk]} : \spj\leftrightarrow\spk\)
was defined in \eqref{eq:DefTAUij0}.

As the Yang-Baxter relation \eqref{eq:YBlambda} was proven for an
arbitrary {\rp} \(\lambda\), the relation \eqref{eq:YB1} holds
for the character \(\cha \lambda(\g)\) in an arbitrary {\rp}
\(\lambda\). By linearity, 
\begin{multline}
 \label{eq:YB2}
{%
(1+(\su_\spk-\su_\spj)\perm_{\spj,\spk}) \cdot 
 \left[\DLt %
A(\g)\right]}%
\\
=\left[\DLt[{\su}_{\tau_{[\spj,\spk]}(\spi)}+\hD] %
A(\g)\right]\cdot (1+(\su_\spk-\su_\spj)\perm_{\spj,\spk})
\end{multline}
holds as soon as \(A(\g)\) is a linear combination of characters. 
In particular it holds when \(A=w(z)\) and even when \(A\) is a product
of  \(w\) functions\footnote{One should remember that \(w(z)=\sum_s z^s
  \chs s\) is the generating series of the characters of symmetric
  {\rp}.}  (like \(w(x)w(y)w(z)\)). Indeed, 
it is shown in the main text (see the ``second proof'' of the {\MID}
in section \ref{sec:mid}), that 
this product
can be written as a linear combination of
the characters associated to different {\yn} diagrams.
  Moreover, the relation \eqref{eq:YB2} implies for instance
\begin{multline}
 \label{eq:YB22}
{(1+(\su_\spk-\su_\spj)\perm_{\spj,\spk}) \cdot %
  \left[\DLt A(\g)\right]}
\cdot  \left[\DLt B(\g)\right]
=\\ \left[\DLt[{\su}_{\tau_{[\spj,\spk]}(\spi)}+\hD] A(\g)\right]\cdot
\left[\DLt[{\su}_{\tau_{[\spj,\spk]}(\spi)}+\hD] B(\g)\right] \cdot
(1+(\su_\spk-\su_\spj)\perm_{\spj,\spk}) %
\,, 
\end{multline}
which will be useful to prove the relations of the next section.

\subsection{Bilinear identities}
\label{sec:bilinear-identities}

In order to show how strongly-constrained
the internal structure of {\cdrs} is, 
  let us prove a nice statement which we will apply to find the
  bilinear identity \eqref{eq:MID0Nest}, used to derive the CBR
  formula in section \ref{sec:hirota-equat-cher}.

  \begin{statmt}
Let us consider a set of \(2\kk\) arbitrary %
functions
\((A_\jj(\g))_{1\leq \jj\leq \kk}\) and \linebreak \((B_\jj(\g))_{1\leq \jj \leq
  \kk}\) %
such that
\begin{gather}
  \forall \lcds\in \bN, \g\in
\GL \Kr, \qquad \label{eq:ulemmaHyp}
\sum_\jj 
\left[\hD ^{\otimes \lcds}A_\jj(\g)\right] \cdot
\left[\hD^{\otimes \lcds} B_\jj(\g)\right]=0 \,,
\end{gather}
Then these functions also obey the stronger relation
\begin{gather}
\sum_\jj 
\left[\DLt %
  A_\jj(\g)\right] \cdot
\left[\DLt %
  B_\jj(\g)\right]=0
\nonumber\\
\hspace{3cm}\forall \lcds\in \bN, ({\su}_1, {\su}_2,\cdots,{\su}_\lcds)\in \mathbb{C}^{\lcds}, \g\in
\GL \Kr\,.
\label{eq:ulemmaConcl}
\end{gather}
  \end{statmt}

\begin{proof}

The quantity \(\mathcal A \equiv  \sum_\jj 
 \left[\DLt%
   A_\jj(\g)\right] \cdot
 \left[\DLt%
   B_\jj(\g)\right]\) is (by hypothesis)
 equal to zero if 
 \(\su_1=\su_2=\cdots=\su_{\lcds%
 }=0\). A first step is to prove that
 it is still equal to zero if  \(\su_1=\su_2=\cdots=\su_{\lcds%
   -1}=0\)
 with \(\su_{\lcds%
 }\neq 0\). In that case we can expand \(\mathcal{A}\)
 with respect to \(\su_{\lcds%
 }\) and we see that the coefficient of
 degree two in \(\su_{\lcds%
 }\) is equal to 
\(\sum_\jj 
 \left[ \hD^{\otimes \lcds%
   -1}\otimes\bI A_\jj(\g)\right] \cdot
 \left[  \hD^{\otimes \lcds%
    -1}\otimes\bI B_\jj(\g)\right]\), which is equal to zero by
hypothesis (see \eqref{eq:ulemmaHyp}). To show that the term of degree
one in \(\su_{\lcds%
 }\) is also equal to zero, we can use the Yang-Baxter equation
 \eqref{eq:YB22} to write
 \begin{multline}
\label{eq:ToBilinCoderShift}
   \sum_\jj (1+\su ~\perm_{\lcds-1,\lcds})\cdot \left[
     \hD^{\otimes \lcds-1} \otimes(\su+\hD) ~~A_\jj(\g)\right]
   \cdot \left[
     \hD^{\otimes \lcds-1} \otimes(\su+\hD) ~~B_\jj(\g)\right]\\
 -
\sum_\jj  \left[
     \hD^{\otimes \lcds-2} \otimes(\su+\hD)\otimes \hD ~~A_\jj(\g)\right]
   \cdot \left[
     \hD^{\otimes \lcds-2} \otimes(\su+\hD)\otimes \hD ~~B_\jj(\g)\right]
\cdot
   (1+\su ~\perm_{\lcds-1,\lcds}) \\ =0\,.
 \end{multline}
Then, the coefficient of the term of degree one \(\su\) (in
\eqref{eq:ToBilinCoderShift}) contains the following terms~:
\begin{itemize}
\item The terms where \(\su\) is kept in \((1+\su
  ~\perm_{\lcds-1,\lcds})\) (and set to zero in the other factors) is
equal to 
\begin{equation}
\label{eq:ToToToBilinCoderShift}\perm_{\lcds-1,\lcds} \cdot \sum_\jj  \left[\hD ^{\otimes \lcds}
  A_\jj(\g)\right] \cdot 
\left[\hD^{\otimes \lcds} B_\jj(\g)\right] 
- \sum_\jj  \left[\hD ^{\otimes \lcds}A_\jj(\g)\right] \cdot
\left[\hD^{\otimes \lcds} B_\jj(\g)\right]
\cdot\perm_{\lcds-1,\lcds}
\,,
\end{equation}
which is equal to zero by hypothesis (see \eqref{eq:ulemmaHyp}).

\item The other terms in the first line are exactly the coefficient of
  \(\mathcal{A}\) with degree one in \(\su_\lcds\)
(and degree zero in \(\su_1\), \(\su_2\), \(\cdots\) and \(\su_{\lcds-1}\)).%

\item The other terms in the second line are exactly the coefficient of
  \(\mathcal{A}\) with degree one in \(\su_{\lcds-1}\) 
(and degree zero in the other variables \(\su_{\spi}\)).
\end{itemize}

Therefore, in order to prove that the coefficient of \(\mathcal{A}\)
with degree one in \(\su_\lcds\) (and degree zero in the other variables
\(\su_{\spi}\))  vanishes, %
it is sufficient to prove that the coefficient of \(\mathcal{A}\)
with degree one in \(\su_{\lcds-1}\) vanishes. %
By iterating this argument, we simply have to prove that the
coefficient of \(\mathcal{A}\) 
with degree one in \(\su_1\) 
(and degree zero in the other variables
\(\su_{\spi}\)) is equal to zero.
But this coefficient is equal to 
\begin{multline}
  \sum_\jj  \left[\hD ^{\otimes \lcds}
  A_\jj(\g)\right] \cdot 
\left[\bI\otimes\hD^{\otimes \lcds-1} B_\jj(\g)\right]  + 
 \sum_\jj  \left[\bI\otimes\hD ^{\otimes \lcds-1}
  A_\jj(\g)\right] \cdot 
\left[\hD^{\otimes \lcds} B_\jj(\g)\right] \\
=  \hD \otimes  \sum_\jj  \left[\hD ^{\otimes \lcds-1}
  A_\jj(\g)\right] \cdot 
\left[\hD^{\otimes \lcds-1} B_\jj(\g)\right]\,,
\label{eq:LeibnToBilinShift}
\end{multline}
where the {\rhs} is equal to zero (see \eqref{eq:ulemmaHyp}), and is
equal to the {\lhs} due to the Leibniz rule \eqref{eq:codLeibnitz}.
This concludes the proof that \(\mathcal{A}\) is equal to zero when \(\su_1=\su_2=\cdots=\su_{\lcds%
   -1}=0\), even 
 with \(\su_{\lcds}\neq 0\).

Next, one can easily see that this first result allows to show that
\(\mathcal{A}\) is equal to zero when
\(\su_1=\su_2=\cdots=\su_{\lcds-2}=0\). Indeed, if we expand it with
respect to \(\su_{\lcds-1}\), then the coefficient of degree two (in
\(\su_{\lcds-1}\)) corresponds 
to the identity \eqref{eq:ulemmaConcl} for a chain of length \(\lcds-1\)
where all the variables \(\su_\spi\) are equal to zero except the last
one (hence this coefficient is equal to zero as we have just
shown). The term of degree one in \(\su_{\lcds-1}\) is also equal to
zero, as it can be shown by repeating the arguments above\footnote{
More precisely, one has
to write the Yang-Baxter equation 
 \begin{multline}
\label{eq:ToBilinCoderShift2}
   \sum_\jj (1+\su ~\perm_{\lcds-2,\lcds-1})\cdot \left[
     \hD^{\otimes \lcds-2} \otimes(\su+\hD)\otimes(\su_\lcds+\hD)
     ~~A_\jj(\g)\right]
\cdot \left[
     \hD^{\otimes \lcds-2} \otimes(\su+\hD)\otimes(\su_\lcds+\hD) ~~B_\jj(\g)\right]\\
 -
\sum_\jj 
\left[
     \hD^{\otimes \lcds-3} \otimes(\su+\hD)\otimes \hD^{\otimes 2} ~~A_\jj(\g)\right]
   \cdot
\left[
     \hD^{\otimes \lcds-3} \otimes(\su+\hD)\otimes \hD^{\otimes 2} ~~B_\jj(\g)\right]
\cdot
   (1+\su ~\perm_{\lcds-2,\lcds-1}) \\ =0\,.
 \end{multline}
and to keep only the terms of degree one in \(\su\).
Exactly like in \eqref{eq:ToToToBilinCoderShift}, the terms containing
the permutation \(\perm_{\lcds-2,\lcds-1}\) vanish, and by iterations the
problem is reduced to showing that the coefficient of \(\mathcal{A}\)
with degree one in \(\su_1\) (and degree zero in the variables \(\su_{\spi}\)
when \(1<{\spi}<\lcds\)) vanishes. Then one sees that this coefficient
is zero by using the Leibniz rule, exactly like in
\eqref{eq:LeibnToBilinShift}. 
 }.

Repeating this argument allows to show that \eqref{eq:ulemmaConcl}
still holds if we only assume that
\(\su_1=\su_2=\cdots=\su_{\lcds-3}=0\), and by iterations, it holds for
arbitrary \({({\su}_1, {\su}_2,\cdots,{\su}_\lcds)\in \mathbb{C}^{\lcds}}\).
\end{proof}

This property is very interesting because we have seen in section
\ref{sec:proof-cbr-formula} that the formula \eqref{eq:DD.1+DD=1+DD.DD}
holds when \(\lcds\geq 1\). This statement can be rewritten as :
\begin{align}
 \forall \lcds\geq 0,~~& {\left[\hD^{\otimes \lcds} z w(z)\det \g\right]
  \cdot  \left[\hD^{\otimes \lcds} w(y)\right]}\nonumber\\
&~-
  \left[\hD^{\otimes \lcds} w(z)\right]
  \cdot  \left[\hD^{\otimes \lcds} y  w(y) \det \g\right]\nonumber
\\&~-\left[\hD^{\otimes \lcds} \left((z-y) w(z) w(y) \det
    \g\right)\right]\cdot
\left[\hD^{\otimes \lcds}~~ 1\right]=0\,,
\label{eq:3tem1+DD.DD=DD.1+DD}
\end{align}
which is exactly of the form \eqref{eq:ulemmaHyp}. The first two terms
of this relation are exactly the terms of equation
\eqref{eq:DD.1+DD=1+DD.DD}, up to the rewriting \eqref{eq:detshift}.
It was also necessary to add a third term, involving
\(\left[\hD^{\otimes \lcds}~~ 1\right]\) which is zero if \(\lcds\geq
1\). If \(\lcds=0\), the factor
\(\left[\hD^{\otimes \lcds} \left((z-y) w(z) w(y) \det
    \g\right)\right]\)
is fitted so that the identity
\eqref{eq:3tem1+DD.DD=DD.1+DD} is true even when \(\lcds=0\). This condition
is important to initialize the recurrence proving the property
\eqref{eq:ulemmaConcl}.

Then the property \eqref{eq:ulemmaConcl} allows to conclude that for
arbitrary \(\su_\spi\)'s, 
\begin{align}
 \forall \lcds\geq 0,~~\mathrlap {z \left[
\DLt[{\su}_\spi+1+\hD]
w(z)\right]
  \cdot  \left[\DLt %
    w(y)\right]}
\qquad\nonumber\\
&=
  y \left[\DLt %
 w(z)\right]
  \cdot  \left[\DLt[{\su}_\spi+1+\hD]
w(y) \right]\nonumber
\\&~~+(z-y)\left[\DLt[{\su}_\spi+1+\hD] %
  w(z) w(y)\right]\cdot
\left[\prod_{\spi=1}^\lcds \su_\spi\right]\,.
\label{eq:MINoNest}
\end{align}
This equation is an important step to derive the CBR formula in
section \ref{sec:hirota-equat-cher}.

\section{Co-derivatives and eigenvalues}
\label{sec:coder-eigenv}

In this appendix, we will prove the equivalence between the
definitions  \eqref{eq:WN01DefDL} and \eqref{eq:WN01DefDLwNest}. This
equivalence means that one can freely move a factor \(1-z~x_\jvp\) from
the left of \(\DL\) to its right. Of course, such a result would not be
expected in general, and it holds only because in these definitions,
there is a prefactor \((1-\g~t)^{\otimes \lcds}\). Indeed, some
eigenvalues of this prefactor are zero when \(t\to 1/x_{\jvp}\), which makes
a few terms vanish.

To do this, the first question which arises is actually ``what is the
definition of \(\hD x_{\jvp}\)~?''. In order to use the definition
\eqref{eq:DefD1} of \(\hD\), \(x_{\jvp}\) should be defined as a function of
\(\g\), which can be computed at the point \(e^{\phi\cdot e}\g\). 
The most natural definition is based on the fact that
\(x_{\jvp}\) is the \({\jvp}^\mathrm{th}\) eigenvalue of \(\g\), or
in other words the \({\jvp}^\mathrm{th}\) root of its characteristic
polynomial. In this sense, \(x_{{\jvp}}\) is a function of the group element 
\(\g\): \(x_{{\jvp}}=x_{{\jvp}}(\g)\). In particular, \(x_{{\jvp}}(\Omega \g \Omega^{-1})=x_{{\jvp}}(\g)\) 
for any similarity transformation. 

With this definition, we can now show that 
\begin{align}
\label{eq:hDxj}
  \hDt x_\jvp=&
\proj _{\jvp} x_{\jvp}
\end{align}
where \(\proj_{\jvp}\) denotes the projector into the eigenspace of \(\g\)
associated to the eigenvalue \(x_\jvp\).

\begin{proof}
If \(\g\) is a diagonal matrix, %
the contribution of the non-diagonal-elements of the matrix
 \(e^{\phie}\g\) to 
 the characteristic polynomial \(\det\left(\lambda \bI-e^{\phie}\g
\right) \) 
is at least  
quadratic in \(\phi\). This means that at the point \(e^{\phie}\g\),
\(x_{\jvp}\) is equal to \(\left(e^{\phie}\g\right)^{\jvp}_\jvp\) to the first order in \(\phi\). As a consequence, we get 
\(\hD_{{\coordj}_1}^{{\coordi}_1} x_{\jvp}=\hD_{{\coordj}_1}^{{\coordi}_1} {\g}_{\jvp}^{\jvp}=\delta^{{\coordi}_1}_{{\jvp}}\delta^{{\jvp}}_{{\coordi}_1}
x_{\jvp}\), so that \(\hD x_{\jvp}=\proj _{\jvp} x_{\jvp}\), where the projector to the eigenspace 
 for the \({\jvp}\)-th eigenvalue \(x_{{\jvp}}\) is
 \(\proj_{{\jvp}}=e_{\jvp \jvp}\) in this case.

More generally, if \(\g=\Omega^{-1} \tilde \g \Omega\) where \(\tilde \g\) is
diagonal and \(\Omega\) is an arbitrary similarity transformation,
then
we obtain 
\begin{align}
 \begin{split}
    \hD ~ x_{\jvp}&= \frac{\partial}{\partial\phi} 
  \left.x_{\jvp}\left(e^{\phie}\Omega^{-1} \tilde \g \Omega\right)\right|_{\phi=0}
  =\frac{\partial}{\partial\phi} 
  \left.\left(\Omega e^{\phie}\Omega^{-1} \tilde \g
  \right)^{\jvp}_\jvp\right|_{\phi=0}
\\[5pt]
  &=\sum_{{\coordi}_{1}, {\coordj}_{1}}e_{{\coordi}_{1} {\coordj}_{1}} \Omega^{{\jvp}}_{{\coordj}_{1}} (\Omega^{-1})^{{\coordi}_{1}}_{{\jvp}} x_{{\jvp}} 
  =\sum_{{\coordi}_{1}, {\coordj}_{1}}e_{{\coordi}_{1} {\coordj}_{1}}
  (\Omega^{-1}e_{\jvp \jvp}\Omega )^{{\coordi}_{1}}_{{\coordj}_{1}} x_{{\jvp}} .
  \end{split}
\label{eq:Dre1}
\end{align}

This exactly means that for a non-diagonal matrix \(\g\), \(\hD ~ x_{\jvp} =
\proj _{\jvp} x_{\jvp}\), where the projector to the eigenspace for \(x_{{\jvp}}\) has the form 
\(\proj_{{\jvp}}=\Omega^{-1} e_{\jvp\jvp} \Omega\).
\end{proof}

Now that we have defined \(\hDt x_\jvp\), we can investigate the
equivalence between the
definitions  \eqref{eq:WN01DefDL} and \eqref{eq:WN01DefDLwNest}. We
can start with the \(\lcds=1\) case, where we easily show how to commute
\(x_\jvp\) through the {\cd} :
\begin{align}
\lim_{t\to\frac 1 {x_{\jvp}}}  \left(1-{\g} t\right)
\comm{
(\su +\hD)}{x_{{\jvp}} \bI 
}
=&
\left(1-\frac {\g} {x_{\jvp}}\right) \cdot \left[ \hDt x_{\jvp} \right]
\nonumber\\ =& \left(1-\frac {\g} {x_{\jvp}}\right)
x_{\jvp} \proj_{{\jvp}} =  0 \,,
\label{eq:1spCm1}
\end{align}
where we see that a key point is the
multiplication by
\(\lim_{t\to\frac 1 {x_{\jvp}}} (1-{\g} t)= (1-{\g}/x_{\jvp})\), which makes the  \(\hDt
x_{\jvp}=x_{\jvp} \proj_{{\jvp}}\) vanish,
due to the property \( (1-{\g}/x_{\jvp})~ \proj_{{\jvp}}=0\). 

For larger values of {\lcds}, we have to prove that 
\begin{gather}
\label{eq:RecToMoveXj} 
 C_{\spm,\lcds}=0 , \qquad\qquad 
 \where \quad 
 C_{\spm,\lcds}\equiv \left(1-{\g}
     /x_{\jvp}\right)^{\otimes (\lcds)} B_{\spm,\lcds}\,, \\
 B_{\spm,\lcds}\equiv
 \left(\DL[][1][\spm]\right)\otimes 
 \comm{\su_{\spm+1} +\hD}{x_{{\jvp}} \bI}
  \otimes
 \left(\DL[][\spm+2]\right)\,, \nonumber
\end{gather}
where \(0\leq\spm\leq \lcds-1\).
This statement \eqref{eq:RecToMoveXj} is exactly the statement that
\(x_\jvp\) can be commuted through the {\cdrs}, to give the
equivalence of \eqref{eq:WN01DefDL} and \eqref{eq:WN01DefDLwNest}. We
will prove it by recurrence over \(\spm\), keeping \(\lcds-\spm\) constant:

\begin{proof}[Proof of \eqref{eq:RecToMoveXj}]

For 
\(\spm=0\), \eqref{eq:RecToMoveXj} follows from \eqref{eq:1spCm1}. 
Let us  show how \(C_{\spm+1,\lcds+1}\) vanishes under the assumption that
\(C_{\spm,\lcds}=0\) for all \({\g} \in GL({\Kr})\) 
 and any \(\{\su_{{\ivp}}\} \in {\bC}^{\lcds}\).
Then for  any 
 \(\su_{0} \in {\bC} \), one can calculate: 
 \begin{align}
  0=&\left((1-{\g}/x_{\jvp})\otimes \bI^{\otimes (\lcds)}\right)\cdot\left(
    (\su_{0}+\hD) \otimes 
    C_{\spm,\lcds}\right) \nonumber \\
=&C_{\spm+1,{\lcds+1}}^{\prime }+\left((1-{\g}/x_{\jvp})\otimes \bI^{\otimes \lcds} \right)\cdot 
\left[
\hD  \otimes (1-{\g} /x_{\jvp})^{\otimes (\lcds)}
\right]
\cdot 
  (\bI \otimes B_{\spm,\lcds}),
 \label{eq:NspCm2}\\
&\!\!\!\!\!\!\where 
C_{\spm+1,{\lcds+1}}^{\prime} \equiv \left(1-{\g} /x_{\jvp}\right)^{\otimes (\lcds+1)} B_{\spm+1,\lcds+1}^{\prime}, \\
&\!\!\!\!\!\!B_{\spm+1,\lcds+1}^{\prime}\equiv \left(\DL[][0][\spm]\right)\otimes 
 \comm{\su_{\spm+1} +\hD}{x_{{\jvp}} \bI}
  \otimes
 \left(\DL[][\spm+2]\right)\,.
\end{align}

This expression \eqref{eq:NspCm2} is obtained 
by computing \((\su_{0}+\hD) \otimes 
    C_{\spm,\lcds}\) using the Leibniz
rule :
 \(\hD\otimes
  \left( \left(1-{\g}/x_{\jvp}\right)^{\otimes \lcds} \cdot B_{\spm,\lcds}\right) =
  \left[\hD\otimes 
  \left(1-{\g}/x_{\jvp}\right)^{\otimes \lcds}\right]\cdot \left(\bI\otimes
    B_{\spm,\lcds}\right) + 
\left(\bI\otimes \left(1-{\g}/x_{\jvp}\right)^{\otimes \lcds}\right) \cdot
  \left[\hD\otimes B_{\spm,\lcds}\right]
\).

Using the relation \(\hD\otimes {\g}/x_{\jvp}=\perm_{1,2}\cdot(1\otimes
{\g}/x_{\jvp})-\proj _{\jvp} \otimes {\g}/x_{\jvp}\), the second term in
\eqref{eq:NspCm2} 
 can be expanded to get
\begin{multline}
0=C_{\spm+1,{\lcds+1}}^{\prime }\\+\sum_{{\spk}=1}^{\spm+1}
\left(1-{\g}/x_{\jvp}\right)_{0,\cdots, {\spk}-1}\cdot\perm_{0,{\spk}}\cdot \left(\frac{{\g}
  }{x_{\jvp}}\right)_{\spk}\cdot
\left(1-{\g}/x_{\jvp}\right)_{{\spk}+1,\cdots , \lcds}
\cdot \left(\bI\otimes B_{\spm,\lcds}\right) 
\\
-\sum_{{\spk}=1}^{\spm+1}
\left(1-{\g}/x_{\jvp}\right)_{0,\cdots, {\spk}-1}\cdot\left(\proj_{\jvp}\frac{{\g}
  }{x_{\jvp}}\right)_{\spk}\cdot
\left(1-{\g}/x_{\jvp}\right)_{{\spk}+1,\cdots , \lcds}  \cdot \left(\bI\otimes B_{\spm,\lcds}\right)  \,
\label{eq:ComXjStep2}
\end{multline}
\begin{gather}
  \where \quad \left(\proj_{\jvp}\frac{{\g}
  }{x_{\jvp}}\right)_{\spk} \equiv \bI^{\otimes {\spk}}\otimes \left(\proj_{\jvp} \frac{{\g}
  }{x_{\jvp}}\right) \otimes \bI^{\otimes \lcds-{\spk}},\\
\left(\frac{{\g}
  }{x_{\jvp}}\right)_{\spk} \equiv \bI^{\otimes {\spk}}\otimes \frac{{\g}
  }{x_{\jvp}} \otimes \bI^{\otimes \lcds-{\spk}},
 \qquad  
\left(1-{\g}/x_{\jvp}\right)_{a,\cdots,b}\equiv \prod_{{\spk}=a}^b \left(\bI-\left(\frac{{\g}
  }{x_{\jvp}}\right)_{\spk} \right)\,.
\end{gather}
By commuting \(\perm_{0,\kk}\) to the left of the other terms, the first
sum of \eqref{eq:ComXjStep2} can
be written 
\(\sum_{{\spk}=1}^{\spm+1} \perm_{0,{\spk}}\cdot \left(\frac{{\g}
  }{x_{\jvp}}\right)_{\spk}\cdot \bI\otimes\left((1-{\g}/x_{\jvp})^{\otimes \lcds}
\cdot B_{\spm,\lcds}\right)\), which is zero because it contains
\(C_{\spm,\lcds}\). 
The second term is also zero because it contains \((1-{\g}/x_{{\jvp}}) \proj_{\jvp}\).

This completes the proof of the fact that 
\(C_{\spm+1,{\lcds+1}}^{\prime}=0\),
 from which 
\(C_{\spm+1,{\lcds+1}}=0\) follows. 
  \end{proof}

 As a consequence, we can indeed commute the factor
\(1-x_\jvp~z\)
to the right of all {\cdrs} in \eqref{eq:WN01DefDL} to get \eqref{eq:WN01DefDLwNest}.

%
%
%
%
%
%

%
%
%
%
%

%%% Local Variables: ***
%%% mode:latex ***
%%% eval: (find-file "english.tex") ***
%%% TeX-master: "english.tex" ***
%%% End: ***